\documentclass[longauth]{aa} 
\usepackage{graphicx}
\usepackage{subcaption}
\usepackage{txfonts}
\usepackage{siunitx}
\usepackage{multirow}
\usepackage{amssymb}
\usepackage{hyperref}

\usepackage{orcidlink}
\usepackage{booktabs}

\graphicspath{{./}{figures/}}

\newcounter{pta}
\renewcommand{\thepta}{\Roman{pta}} 

\DeclareRobustCommand{\defpta}[1]{%
   \refstepcounter{pta}%
   \thepta\label{#1}%
}

\newcommand{\refpta}[1]{\ref{#1}}

\DeclareSIUnit\gauss{G}
\DeclareSIUnit\erg{erg}
\DeclareSIUnit\parsec{pc}
\DeclareSIUnit\lightyear{ly}
\DeclareSIUnit\year{yr}
\DeclareSIUnit\milliarcsecond{mas}
\DeclareSIUnit\lightsecond{lt-s}
\DeclareSIUnit\kilometer{km}
\DeclareSIUnit\second{s}

\DeclareSIUnit\rydberg{Ry}
\DeclareSIUnit\magnitude{mag}
\DeclareSIUnit\jansky{Jy}
\DeclareSIUnit\h{$h$}
\DeclareSIUnit\hseven{$h$_7}

\newcommand{\tempotwo}{\textsc{tempo2}}
\newcommand{\psrchive}{\textsc{psrchive}}
\newcommand{\tnest}{\textsc{temponest}}

\def\wm{1}

\begin{document}
   \title{The second data release from the European Pulsar Timing Array}

   \subtitle{I. The dataset and timing analysis}

\author{
    J.~Antoniadis\orcidlink{0000-0003-4453-776}\inst{\ref{forth},\ref{mpifr}\ifnum\wm>1,\refpta{epta}\fi},
    \ifnum\wm>1 P.~Arumugam\orcidlink{0000-0001-9264-8024}\inst{\ref{IITR}\ifnum\wm>1,\refpta{inpta}\fi},\fi
    \ifnum\wm>1 S.~Arumugam\orcidlink{0009-0001-3587-6622}\inst{\ref{IITH_El}\ifnum\wm>1,\refpta{inpta}\fi},\fi
    S.~Babak\orcidlink{0000-0001-7469-4250}\inst{\ref{apc}\ifnum\wm>1,\refpta{epta}\fi},
    \ifnum\wm>1 M.~Bagchi\orcidlink{0000-0001-8640-8186}\inst{\ref{IMSc},\ref{HBNI}\ifnum\wm>1,\refpta{inpta}\fi},\fi
    A.-S.~Bak~Nielsen\orcidlink{ 0000-0002-1298-9392}\inst{\ref{mpifr},\ref{unibi}\ifnum\wm>1,\refpta{epta}\fi},
    C.~G.~Bassa\orcidlink{0000-0002-1429-9010}\inst{\ref{astron}\ifnum\wm>1,\refpta{epta}\fi},
    \ifnum\wm>1 A.~Bathula\orcidlink{0000-0001-7947-6703} \inst{\ref{IISERM}\ifnum\wm>1,\refpta{inpta}\fi},\fi
    A.~Berthereau\inst{\ref{lpc2e},\ref{nancay}\ifnum\wm>1,\refpta{epta}\fi},
    M.~Bonetti\orcidlink{0000-0001-7889-6810}\inst{\ref{unimib},\ref{infn-unimib},\ref{inaf-brera}\ifnum\wm>1,\refpta{epta}\fi},
    E.~Bortolas\inst{\ref{unimib},\ref{infn-unimib},\ref{inaf-brera}\ifnum\wm>1,\refpta{epta}\fi},
    P.~R.~Brook\orcidlink{0000-0003-3053-6538}\inst{\ref{unibir}\ifnum\wm>1,\refpta{epta}\fi},
    M.~Burgay\orcidlink{0000-0002-8265-4344}\inst{\ref{inaf-oac}\ifnum\wm>1,\refpta{epta}\fi},
    R.~N.~Caballero\orcidlink{0000-0001-9084-9427}\inst{\ref{HOU}\ifnum\wm>1,\refpta{epta}\fi},
    A.~Chalumeau\orcidlink{0000-0003-2111-1001}\inst{\ref{unimib}\ifnum\wm>1,\refpta{epta}\fi}\ifnum\wm=2\thanks{aurelien.chalumeau@unimib.it}\fi,
    D.~J.~Champion\orcidlink{0000-0003-1361-7723}\inst{\ref{mpifr}\ifnum\wm>1,\refpta{epta}\fi},
    S.~Chanlaridis\orcidlink{0000-0002-9323-9728}\inst{\ref{forth}\ifnum\wm>1,\refpta{epta}\fi},
    S.~Chen\orcidlink{0000-0002-3118-5963}\inst{\ref{kiaa}\ifnum\wm>1,\refpta{epta}\fi}\ifnum\wm=3\thanks{sychen@pku.edu.cn}\fi,
    I.~Cognard\orcidlink{0000-0002-1775-9692}\inst{\ref{lpc2e},\ref{nancay}\ifnum\wm>1,\refpta{epta}\fi},
    \ifnum\wm>1 S.~Dandapat\orcidlink{0000-0003-4965-9220}\inst{\ref{TIFR}\ifnum\wm>1,\refpta{inpta}\fi},\fi
    \ifnum\wm>1 D.~Deb\orcidlink{0000-0003-4067-5283}\inst{\ref{IMSc}\ifnum\wm>1,\refpta{inpta}\fi},      \fi
    \ifnum\wm>1 S.~Desai\orcidlink{0000-0002-0466-3288}\inst{\ref{IITH_Ph}\ifnum\wm>1,\refpta{inpta}\fi},\fi
    G.~Desvignes\orcidlink{0000-0003-3922-4055}\inst{\ref{mpifr}\ifnum\wm>1,\refpta{epta}\fi},
    \ifnum\wm>1 N.~Dhanda-Batra \inst{\ref{UoD}\ifnum\wm>1,\refpta{inpta}\fi},\fi
    \ifnum\wm>1 C.~Dwivedi\orcidlink{0000-0002-8804-650X}\inst{\ref{IIST}\ifnum\wm>1,\refpta{inpta}\fi},\fi
    M.~Falxa\inst{\ref{apc},\ref{lpc2e}\ifnum\wm>1,\refpta{epta}\fi}
    R.~D.~Ferdman\inst{\ref{uea}\ifnum\wm>1,\refpta{epta}\fi},
    A.~Franchini\orcidlink{0000-0002-8400-0969}\inst{\ref{unimib},\ref{infn-unimib}\ifnum\wm>1,\refpta{epta}\fi},
    J.~R.~Gair\orcidlink{0000-0002-1671-3668}\inst{\ref{aei}\ifnum\wm>1,\refpta{epta}\fi},
    B.~Goncharov\orcidlink{0000-0003-3189-5807}\inst{\ref{gssi},\ref{lngs}\ifnum\wm>1,\refpta{epta}\fi}
    \ifnum\wm>1 A.~Gopakumar\orcidlink{0000-0003-4274-4369}\inst{\ref{TIFR}\ifnum\wm>1,\refpta{inpta}\fi},\fi
    E.~Graikou\inst{\ref{mpifr}\ifnum\wm>1,\refpta{epta}\fi},
    J.-M.~Grie{\ss}meier\orcidlink{0000-0003-3362-7996}\inst{\ref{lpc2e},\ref{nancay}\ifnum\wm>1,\refpta{epta}\fi},
    L.~Guillemot\orcidlink{0000-0002-9049-8716}\inst{\ref{lpc2e},\ref{nancay}\ifnum\wm>1,\refpta{epta}\fi},
    Y.~J.~Guo\inst{\ref{mpifr}\ifnum\wm>1,\refpta{epta}\fi}\ifnum\wm=3\thanks{yjguo@mpifr-bonn.mpg.de}\fi,
    \ifnum\wm>1 Y.~Gupta\orcidlink{0000-0001-5765-0619}\inst{\ref{NCRA}\ifnum\wm>1,\refpta{inpta}\fi},\fi
    \ifnum\wm>1 S.~Hisano\orcidlink{0000-0002-7700-3379}\inst{\ref{KU_J}\ifnum\wm>1,\refpta{inpta}\fi},\fi
    H.~Hu\orcidlink{0000-0002-3407-8071}\inst{\ref{mpifr}\ifnum\wm>1,\refpta{epta}\fi}, 
    F.~Iraci\inst{\ref{unica}\ref{inaf-oac}\ifnum\wm>1,\refpta{epta}\fi},
    D.~Izquierdo-Villalba\orcidlink{0000-0002-6143-1491}\inst{\ref{unimib},\ref{infn-unimib}\ifnum\wm>1,\refpta{epta}\fi},
    J.~Jang\orcidlink{0000-0003-4454-0204}\inst{\ref{mpifr}\ifnum\wm>1,\refpta{epta}\fi}\ifnum\wm=1\thanks{jjang@mpifr-bonn.mpg.de}\fi,
    J.~Jawor\orcidlink{0000-0003-3391-0011}\inst{\ref{mpifr}\ifnum\wm>1,\refpta{epta}\fi},
    G.~H.~Janssen\orcidlink{0000-0003-3068-3677}\inst{\ref{astron},\ref{imapp}\ifnum\wm>1,\refpta{epta}\fi},
    A.~Jessner\orcidlink{0000-0001-6152-9504}\inst{\ref{mpifr}\ifnum\wm>1,\refpta{epta}\fi},
    \ifnum\wm>1 B.~C.~Joshi\orcidlink{0000-0002-0863-7781}\inst{\ref{NCRA},\ref{IITR}\ifnum\wm>1,\refpta{inpta}\fi},\fi
    \ifnum\wm>1 F.~Kareem\orcidlink{0000-0003-2444-838X} \inst{\ref{IISERK},\ref{CESSI}\ifnum\wm>1,\refpta{inpta}\fi},\fi
    R.~Karuppusamy\orcidlink{0000-0002-5307-2919}\inst{\ref{mpifr}\ifnum\wm>1,\refpta{epta}\fi},
    E.~F.~Keane\orcidlink{0000-0002-4553-655X}\inst{\ref{tcd}\ifnum\wm>1,\refpta{epta}\fi},
    M.~J.~Keith\orcidlink{0000-0001-5567-5492}\inst{\ref{jbca}\ifnum\wm>1,\refpta{epta}\fi}\ifnum\wm=2\thanks{michael.keith@manchester.ac.uk}\fi,
    \ifnum\wm>1 D.~Kharbanda\orcidlink{0000-0001-8863-4152}\inst{\ref{IITH_Ph}\ifnum\wm>1,\refpta{inpta}\fi},\fi
    \ifnum\wm>1 T.~Kikunaga\orcidlink{0000-0002-5016-3567} \inst{\ref{KU_J}\ifnum\wm>1,\refpta{inpta}\fi},\fi
    \ifnum\wm>1 N.~Kolhe\orcidlink{0000-0003-3528-9863} \inst{\ref{XCM}\ifnum\wm>1,\refpta{inpta}\fi},\fi
    M.~Kramer\inst{\ref{mpifr},\ref{jbca}\ifnum\wm>1,\refpta{epta}\fi},
    M.~A.~Krishnakumar\orcidlink{0000-0003-4528-2745}\inst{\ref{mpifr},\ref{unibi}\ifnum\wm>1,\refpta{epta}\fi\ifnum\wm>1,\refpta{inpta}\fi},
    K.~Lackeos\orcidlink{0000-0002-6554-3722}\inst{\ref{mpifr}\ifnum\wm>1,\refpta{epta}\fi},
    K.~J.~Lee\inst{3,8,\ref{mpifr}\ifnum\wm>1,\refpta{epta}\fi},
    K.~Liu\inst{\ref{mpifr}\ifnum\wm>1,\refpta{epta}\fi}\ifnum\wm=1\thanks{kliu@mpifr-bonn.mpg.de}\fi,
    Y.~Liu\orcidlink{0000-0001-9986-9360}\inst{\ref{naoc}, \ref{unibi}\ifnum\wm>1,\refpta{epta}\fi},
    A.~G.~Lyne\inst{\ref{jbca}\ifnum\wm>1,\refpta{epta}\fi},
    J.~W.~McKee\orcidlink{0000-0002-2885-8485}\inst{\ref{milne},\ref{daim}\ifnum\wm>1,\refpta{epta}\fi},
    \ifnum\wm>1 Y.~Maan\inst{\ref{NCRA}\ifnum\wm>1,\refpta{inpta}\fi},\fi
    R.~A.~Main\inst{\ref{mpifr}\ifnum\wm>1,\refpta{epta}\fi},
    M.~B.~Mickaliger\orcidlink{0000-0001-6798-5682}\inst{\ref{jbca}\ifnum\wm>1,\refpta{epta}\fi},
    I.~C.~Ni\c{t}u\orcidlink{0000-0003-3611-3464}\inst{\ref{jbca}\ifnum\wm>1,\refpta{epta}\fi},
    \ifnum\wm>1 K.~Nobleson\orcidlink{0000-0003-2715-4504}\inst{\ref{BITS}\ifnum\wm>1,\refpta{inpta}\fi},\fi
    \ifnum\wm>1 A.~K.~Paladi\orcidlink{0000-0002-8651-9510}\inst{\ref{IISc}\ifnum\wm>1,\refpta{inpta}\fi},\fi
    A.~Parthasarathy\orcidlink{0000-0002-4140-5616}\inst{\ref{mpifr}\ifnum\wm>1,\refpta{epta}\fi}\ifnum\wm=2\thanks{aparthas@mpifr-bonn.mpg.de}\fi,
    B.~B.~P.~Perera\orcidlink{0000-0002-8509-5947}\inst{\ref{arecibo}\ifnum\wm>1,\refpta{epta}\fi},
    D.~Perrodin\orcidlink{0000-0002-1806-2483}\inst{\ref{inaf-oac}\ifnum\wm>1,\refpta{epta}\fi},
    A.~Petiteau\orcidlink{0000-0002-7371-9695}\inst{\ref{irfu},\ref{apc}\ifnum\wm>1,\refpta{epta}\fi},
    N.~K.~Porayko\inst{\ref{unimib},\ref{mpifr}\ifnum\wm>1,\refpta{epta}\fi},
    A.~Possenti\inst{\ref{inaf-oac}\ifnum\wm>1,\refpta{epta}\fi},
    \ifnum\wm>1 T.~Prabu\inst{\ref{RRI}\ifnum\wm>1,\refpta{inpta}\fi},\fi
    H.~Quelquejay~Leclere\inst{\ref{apc}\ifnum\wm>1,\refpta{epta}\fi}
    \ifnum\wm>1 P.~Rana\orcidlink{0000-0001-6184-5195}\inst{\ref{TIFR}\ifnum\wm>1,\refpta{inpta}\fi},\fi
    A.~Samajdar\orcidlink{0000-0002-0857-6018}\inst{\ref{uni-potsdam}\ifnum\wm>1,\refpta{epta}\fi},
    S.~A.~Sanidas\inst{\ref{jbca}\ifnum\wm>1,\refpta{epta}\fi},
    A.~Sesana\inst{\ref{unimib},\ref{infn-unimib},\ref{inaf-brera}\ifnum\wm>1,\refpta{epta}\fi},
    G.~Shaifullah\orcidlink{0000-0002-8452-4834}\inst{\ref{unimib},\ref{infn-unimib},\ref{inaf-oac}\ifnum\wm>1,\refpta{epta}\fi}\ifnum\wm=1\thanks{golam.shaifullah@unimib.it}\fi,
    \ifnum\wm>1 J.~Singha\orcidlink{0000-0002-1636-9414}\inst{\ref{IITR}\ifnum\wm>1,\refpta{inpta}\fi},\fi
    L.~Speri\orcidlink{0000-0002-5442-7267}\inst{\ref{aei}\ifnum\wm>1,\refpta{epta}\fi},
    R.~Spiewak\inst{\ref{jbca}\ifnum\wm>1,\refpta{epta}\fi},
    \ifnum\wm>1 A.~Srivastava\orcidlink{0000-0003-3531-7887} \inst{\ref{IITH_Ph}\ifnum\wm>1,\refpta{inpta}\fi},\fi
    B.~W.~Stappers\inst{\ref{jbca}\ifnum\wm>1,\refpta{epta}\fi},
    \ifnum\wm>1 M.~Surnis\orcidlink{0000-0002-9507-6985}\inst{\ref{IISERB}\ifnum\wm>1,\refpta{inpta}\fi},\fi
    S.~C.~Susarla\orcidlink{0000-0003-4332-8201}\inst{\ref{uog}\ifnum\wm>1,\refpta{epta}\fi},
    \ifnum\wm>1 A.~Susobhanan\orcidlink{0000-0002-2820-0931}\inst{\ref{CGCA}\ifnum\wm>1,\refpta{inpta}\fi},\fi
    \ifnum\wm>1 K.~Takahashi\orcidlink{0000-0002-3034-5769}\inst{\ref{KU_J1},\ref{KU_J2}\ifnum\wm>1,\refpta{inpta}\fi} \fi
    \ifnum\wm>1 P.~Tarafdar\orcidlink{0000-0001-6921-4195}\inst{\ref{IMSc}\ifnum\wm>1,\refpta{inpta}\fi}\fi
    G.~Theureau\orcidlink{0000-0002-3649-276X}\inst{\ref{lpc2e}, \ref{nancay}, \ref{luth}\ifnum\wm>1,\refpta{epta}\fi},
    C.~Tiburzi\inst{\ref{inaf-oac}\ifnum\wm>1,\refpta{epta}\fi},
    E.~van~der~Wateren\orcidlink{0000-0003-0382-8463}\inst{\ref{astron},\ref{imapp}\ifnum\wm>1,\refpta{epta}\fi},
    A.~Vecchio\orcidlink{0000-0002-6254-1617}\inst{\ref{unibir}\ifnum\wm>1,\refpta{epta}\fi},
    V.~Venkatraman~Krishnan\orcidlink{0000-0001-9518-9819}\inst{\ref{mpifr}\ifnum\wm>1,\refpta{epta}\fi},
    J.~P.~W.~Verbiest\orcidlink{0000-0002-4088-896X}\inst{\ref{FSI},\ref{unibi},\ref{mpifr}\ifnum\wm>1,\refpta{epta}\fi},
    J.~Wang\orcidlink{0000-0003-1933-6498}\inst{\ref{unibi}, \ref{airub}, \ref{bnuz}\ifnum\wm>1,\refpta{epta}\fi},
    L.~Wang\inst{\ref{jbca}\ifnum\wm>1,\refpta{epta}\fi} and 
    Z.~Wu\orcidlink{0000-0002-1381-7859}\inst{\ref{naoc},\ref{unibi}\ifnum\wm>1,\refpta{epta}\fi}.
    }

\institute{
{Institute of Astrophysics, FORTH, N. Plastira 100, 70013, Heraklion, Greece\label{forth}}\and 
{Max-Planck-Institut f{\"u}r Radioastronomie, Auf dem H{\"u}gel 69, 53121 Bonn, Germany\label{mpifr}}\and
\ifnum\wm>1{Department of Physics, Indian Institute of Technology Roorkee, Roorkee-247667, India\label{IITR}}\and \fi
\ifnum\wm>1{Department of Electrical Engineering, IIT Hyderabad, Kandi, Telangana 502284, India \label{IITH_El}}\and\fi
{Universit{\'e} Paris Cit{\'e}, CNRS, Astroparticule et Cosmologie, 75013 Paris, France\label{apc}}\and
\ifnum\wm>1{The Institute of Mathematical Sciences, C. I. T. Campus, Taramani, Chennai 600113, India \label{IMSc}}\and\fi
\ifnum\wm>1{Homi Bhabha National Institute, Training School Complex, Anushakti Nagar, Mumbai 400094, India \label{HBNI}}\and\fi
{Fakult{\"a}t f{\"u}r Physik, Universit{\"a}t Bielefeld, Postfach 100131, 33501 Bielefeld, Germany\label{unibi}}\and
{ASTRON, Netherlands Institute for Radio Astronomy, Oude Hoogeveensedijk 4, 7991 PD, Dwingeloo, The Netherlands\label{astron}}\and
\ifnum\wm>1{Department of Physical Sciences, Indian Institute of Science Education and Research, Mohali, Punjab 140306, India \label{IISERM}}\and\fi
{Laboratoire de Physique et Chimie de l'Environnement et de l'Espace, Universit\'e d'Orl\'eans / CNRS, 45071 Orl\'eans Cedex 02, France \label{lpc2e}}\and
{Observatoire Radioastronomique de Nan\c{c}ay, Observatoire de Paris, Universit\'e PSL, Université d'Orl\'eans, CNRS, 18330 Nan\c{c}ay, France\label{nancay}}\and
{Dipartimento di Fisica ``G. Occhialini", Universit{\'a} degli Studi di Milano-Bicocca, Piazza della Scienza 3, I-20126 Milano, Italy\label{unimib}}\and
{INFN, Sezione di Milano-Bicocca, Piazza della Scienza 3, I-20126 Milano, Italy\label{infn-unimib}}\and
{INAF - Osservatorio Astronomico di Brera, via Brera 20, I-20121 Milano, Italy\label{inaf-brera}}\and
{Institute for Gravitational Wave Astronomy and School of Physics and Astronomy, University of Birmingham, Edgbaston, Birmingham B15 2TT, UK\label{unibir}}\and
{INAF - Osservatorio Astronomico di Cagliari, via della Scienza 5, 09047 Selargius (CA), Italy\label{inaf-oac}}\and
{Hellenic Open University, School of Science and Technology, 26335 Patras, Greece\label{HOU}}\and
{Kavli Institute for Astronomy and Astrophysics, Peking University, Beijing 100871, P. R. China\label{kiaa}}\and
\ifnum\wm>1{Department of Astronomy and Astrophysics, Tata Institute of Fundamental Research, Homi Bhabha Road, Navy Nagar, Colaba, Mumbai 400005, India \label{TIFR}}\and\fi
\ifnum\wm>1{Department of Physics, IIT Hyderabad, Kandi, Telangana 502284, India \label{IITH_Ph}}\and\fi
\ifnum\wm>1{Department of Physics and Astrophysics, University of Delhi, Delhi 110007, India \label{UoD}}\and\fi
\ifnum\wm>1{Department of Earth and Space Sciences, Indian Institute of Space Science and Technology, Valiamala, Thiruvananthapuram, Kerala 695547,India \label{IIST}}\and\fi
{School of Physics, Faculty of Science, University of East Anglia, Norwich NR4 7TJ, UK\label{uea}}\and
{Max Planck Institute for Gravitational Physics (Albert Einstein Institute), Am Mu{\"u}hlenberg 1, 14476 Potsdam, Germany\label{aei}}\and
{Gran Sasso Science Institute (GSSI), I-67100 L'Aquila, Italy \label{gssi}}\and
{INFN, Laboratori Nazionali del Gran Sasso, I-67100 Assergi, Italy \label{lngs}}\and 
\ifnum\wm>1{National Centre for Radio Astrophysics, Pune University Campus, Pune 411007, India \label{NCRA}}\and\fi
\ifnum\wm>1{Kumamoto University, Graduate School of Science and Technology, Kumamoto, 860-8555, Japan \label{KU_J}}\and\fi
{Universit{\'a} di Cagliari, Dipartimento di Fisica, S.P. Monserrato-Sestu Km 0,700 - 09042 Monserrato (CA), Italy\label{unica}}\and
{Department of Astrophysics/IMAPP, Radboud University Nijmegen, P.O. Box 9010, 6500 GL Nijmegen, The Netherlands\label{imapp}}\and
\ifnum\wm>1{Department of Physical Sciences,Indian Institute of Science Education and Research Kolkata, Mohanpur, 741246, India \label{IISERK}}\and\fi
\ifnum\wm>1{Center of Excellence in Space Sciences India, Indian Institute of Science Education and Research Kolkata, 741246, India \label{CESSI}}\and \fi
{School of Physics, Trinity College Dublin, College Green, Dublin 2, D02 PN40, Ireland\label{tcd}}\and
{Jodrell Bank Centre for Astrophysics, Department of Physics and Astronomy, University of Manchester, Manchester M13 9PL, UK\label{jbca}}\and
\ifnum\wm>1{Department of Physics, St. Xavier’s College (Autonomous), Mumbai 400001, India \label{XCM}}\and\fi
{National Astronomical Observatories, Chinese Academy of Sciences, Beijing 100101, P. R. China\label{naoc}}\and
{E.A. Milne Centre for Astrophysics, University of Hull, Cottingham Road, Kingston-upon-Hull, HU6 7RX, UK\label{milne}}\and
{Centre of Excellence for Data Science, Artificial Intelligence and Modelling (DAIM), University of Hull, Cottingham Road, Kingston-upon-Hull, HU6 7RX, UK\label{daim}}\and
\ifnum\wm>1{Department of Physics, BITS Pilani Hyderabad Campus, Hyderabad 500078, Telangana, India \label{BITS}}\and\fi
\ifnum\wm>1{Joint Astronomy Programme, Indian Institute of Science, Bengaluru, Karnataka, 560012, India \label{IISc}}\and\fi
{Arecibo Observatory, HC3 Box 53995, Arecibo, PR 00612, USA\label{arecibo}}\and
{IRFU, CEA, Université Paris-Saclay, F-91191 Gif-sur-Yvette, France \label{irfu}}\and
\ifnum\wm>1{Raman Research Institute India, Bengaluru, Karnataka, 560080, India \label{RRI}}\and\fi
{Institut f\"{u}r Physik und Astronomie, Universit\"{a}t Potsdam, Haus 28, Karl-Liebknecht-Str. 24/25, 14476, Potsdam, Germany\label{uni-potsdam}}\and
\ifnum\wm>1{Department of Physics, IISER Bhopal, Bhopal Bypass Road, Bhauri, Bhopal 462066, Madhya Pradesh, India \label{IISERB}}\and\fi
{Ollscoil na Gaillimhe --- University of Galway, University Road, Galway, H91 TK33, Ireland\label{uog}}\and
\ifnum\wm>1{Center for Gravitation, Cosmology, and Astrophysics, University of Wisconsin-Milwaukee, Milwaukee, WI 53211, USA \label{CGCA}}\and\fi
\ifnum\wm>1{Division of Natural Science, Faculty of Advanced Science and Technology, Kumamoto University, 2-39-1 Kurokami, Kumamoto 860-8555, Japan \label{KU_J1}}\and\fi
\ifnum\wm>1{International Research Organization for Advanced Science and Technology, Kumamoto University, 2-39-1 Kurokami, Kumamoto 860-8555, Japan \label{KU_J2}}\and\fi
{Laboratoire Univers et Th{\'e}ories LUTh, Observatoire de Paris, Universit{\'e} PSL, CNRS, Universit{\'e} de Paris, 92190 Meudon, France\label{luth}}\and
{Florida Space Institute, University of Central Florida, 12354 Research Parkway, Partnership 1 Building, Suite 214, Orlando, 32826-0650, FL, USA\label{FSI}}\and
{Ruhr University Bochum, Faculty of Physics and Astronomy, Astronomical Institute (AIRUB), 44780 Bochum, Germany \label{airub}}\and
{Advanced Institute of Natural Sciences, Beijing Normal University, Zhuhai 519087, China \label{bnuz}}\\
\ifnum\wm>1\\ {\defpta{epta} : The European Pulsar Timing Array}\fi
\ifnum\wm>1\\ {\defpta{inpta} : The Indian Pulsar Timing Array}\fi
}

      \authorrunning{the EPTA collaboration}
      \titlerunning{EPTA-DR2}
   \date{Received ; accepted }

  \abstract
   {Pulsar timing arrays offer a probe of the low-frequency gravitational wave spectrum ($1-100$ nanohertz), which is intimately connected to a number of markers that can uniquely trace the formation and evolution of the Universe. 
   We present the dataset and the results of the timing analysis from the second data release of the European Pulsar Timing Array (EPTA). The dataset contains high-precision pulsar timing data from 25 millisecond pulsars collected with the five largest radio telescopes in Europe, as well as the Large European Array for Pulsars. The dataset forms the foundation for the search for gravitational waves by the EPTA, presented in associated papers. 
   We describe the dataset and present the results of the frequentist and Bayesian pulsar timing analysis for individual millisecond pulsars that have been observed over the last $\sim$25 years. We discuss the improvements to the individual pulsar parameter estimates, as well as new measurements of the physical properties of these pulsars and their companions.
   This data release extends the dataset from EPTA Data Release 1 up to the beginning of 2021, with individual pulsar datasets with timespans ranging from 14 to 25 years. These lead to improved constraints on annual parallaxes, secular variation of the orbital period, and Shapiro delay for a number of sources. Based on these results, we derived astrophysical parameters that include distances, transverse velocities, binary pulsar masses, and annual orbital parallaxes.}


   \maketitle
%
\section{Introduction}\label{sec:intro}
  Pulsar timing arrays \citep[PTAs;][]{saz78, det79} search for
  gravitational waves (GWs) at nanohertz (nHz) frequencies by
  observing a suite of stable, rapidly rotating millisecond pulsars
  \citep[MSPs][]{bac93} over decadal timescales
  \citep{fb90,dcl+16,vlh+16}. These GWs may be produced by
  inspiralling supermassive  black holes binaries \citep[SMBHBs; see
    e.g. ][]{sesana2004,sesana2013}, cosmic strings
  \citep[][]{dv01b,gri05,bb08}, phase transitions in the early
  Universe \citep{schwaller2015}, quantum fluctuations in the
  primordial gravitational field \citep{gri05, lms+16}, or from a
  primordial magnetic field \citep[see ][and references
    therein]{cf18}.

  Several PTA experiments are currently operational: the
  European Pulsar Timing Array \citep[EPTA;][]{dcl+16}, the North
  American Nanohertz Observatory for Gravitational Waves
  \citep[NANOGrav;][]{mcl13}, the Parkes Pulsar Timing Array
  \citep[PPTA;][]{mhb+13}, the Indian Pulsar Timing Array
  \citep[InPTA]{bgp+22}, the MeerKAT Pulsar Timing Array (MPTA) based in
  South Africa \citep[][]{msb23}, and the Chinese Pulsar Timing
  Array \citep[CPTA;][]{lee16}. These regional PTAs are also organised
  under the International Pulsar Timing Array \citep[IPTA;][]{vlh+16,
    pdd+19}. Although individual GW sources remain an exciting
  prospect for detection with PTAs \citep[see e.g. ][ and references
    therein]{bps+13, fbb+23} it is likely that the first signal
  detected by PTAs will be a stochastic GW background (GWB) arising
  from multiple overlapping sources \citep{rsg15}.

  The recent detection of a common red noise process by EPTA,
  NANOGrav, PPTA and IPTA 
  \citep[][ respectively]{ccg+21, aab+21a, gsr+22, aab+22} 
  suggests that the characteristic spatial quadrupole correlation of the GWB 
  \citep{hd83} can become detectable with a
  modest expansion of PTA datasets.  In this article, we present the
  latest EPTA data release, henceforth referred to as EPTA DR2. The
  release contains high-precision time-of-arrival (TOA) data for 25 MSPs,
  as well as the corresponding timing models. 
  We also make available the full suite of software libraries required to reproduce these timing models. This EPTA data release paper is accompanied by   
  two papers \citep{wm2,wm3}, reporting the modelling of stochastic noise
processes present in individual pulsar datasets and the search for 
  GW signals, respectively. 
  
  The article is organised as follows; in Section \ref{sec:obs} we
  introduce the MSP selection process and provide details on the EPTA observing systems and observations. In
  Section \ref{sec:combination} we describe the data curation and combination process, and in Section~\ref{sec:result} 
  we present the timing solutions for the 25 pulsars in our dataset. 
  In Section \ref{sec:discussion} we discuss the implications of these results. We conclude with a
  brief summary in Section \ref{sec:conclusions}.
 
\section{Observations and data processing}\label{sec:obs}
The EPTA uses data from six European radio telescopes: the
Effelsberg \SI{100}{\m} radio telescope (EFF) in Germany, the
\SI{76}{\m} Lovell telescope at Jodrell Bank Observatory (JBO) in the
United Kingdom, the large radio telescope operated by the Nan\c{c}ay Radio Observatory
(NRT) in France, the \SI{64}{\m} Sardinia Radio Telescope
(SRT) operated by the Italian National Institute for Astrophysics
(INAF) through the Astronomical Observatory of Cagliari (OAC), and the
Westerbork Synthesis Radio Telescope (WSRT) operated by ASTRON, the
Netherlands Institute for Radio Astronomy. In addition, these
telescopes regularly operate as the Large European Array for Pulsars
(LEAP), which offers an equivalent diameter of up to \SI{194}{\m} \citep{bjk+16}. In this data release, we also incorporated, for the first time, data from the Mark\,II telescope at JBO.

The first EPTA data release \citep[henceforth DR1;][]{dcl+16} only included data from
legacy data recording systems. Most of these
made use of the incoherent dedispersion scheme implemented
on custom-built hardware, placing a fundamental limit on
achievable timing precision. In EPTA DR2, we added
data from next-generation coherent dedispersion recording systems that offer a significant
increase in bandwidth and sensitivity at each telescope.
These new backends use hardware based on field-programmable gate arrays
(FPGAs) to carry out the conversion of the electrical signal into a
digital data stream and to apply polyphase filterbanks and, in some cases,
some pre-filtering of the recorded band to reject known radio
frequency interference (RFI). For most of the next-generation EPTA recording backends, 
this data processing step is implemented on the second
generation of the Reconfigurable Open Architecture Computing Hardware
(ROACH) platform, developed by the Collaboration for Astronomy Signal
Processing and Electronics Research
\citep[CASPER;][]{hickish2016decade}. 
In the following sub-sections, we describe the pulsar recording systems and data processing 
schemes for each telescope in more detail (see Table~\ref{tab:flags} for an overview).

\begin{table*}[!htbp]
\caption{Summary of telescopes, backends, and their operating frequencies. \protect\tempotwo{} flags are constructed from these values using the telescope abbreviations, the backend names and the central observing frequency for characterising system noise properties through the `group' flag or the sub-band centre frequencies for determining phase offsets or JUMPS with `system' flags. For example, for a TOA from an observation made at the Nan\c{c}ay Radio Telescope with the NUPPI backend, from the sub-band \SI{1420}{\mega\Hz} when the backend was acquiring its full bandwith around a central frequency of \SI{1854}{\mega\Hz}, the system flag is \textit{NRT.NUPPI.1420} and the group flag is \textit{NRT.NUPPI.1484}. A dash-symbol indicates that no sub-bands were created and the system and group flags are identical.}\label{tab:flags}
\begin{center}
\begin{tabular}{@{}p{2.7cm}p{1.7cm}Sp{2.5cm}ll@{}}
\hline
\textbf{Telescope \quad\quad(Abbreviation)} & \textbf{Receiver or Backend}  & \textbf{Centre Frequency (\si{\mega\hertz})} & \textbf{Sub-bands (\si{\mega\hertz})}  & \textbf{Category}& \textbf{Polarisation}\\
\hline\noalign {\smallskip}
\multirow{5}{3.0cm}{Effelsberg 100-m Radio Telescope (EFF)} & EBPP & \SIlist{1360;1410;2639}{} & -& Legacy& Full Stokes\\
& P200 & \SIlist{1380}{}& \SIlist{1365;1425}{}&Modern& Full Stokes\\
& P217 & \SIlist{1380}{}& \SIlist{1365;1425}{}&Modern& Full Stokes\\
& S110 & \SIlist{2487}{}& -&Modern& Full Stokes\\
& S60 & \SIlist{4857}{}& -&Modern & Full Stokes\\
\noalign{\vskip 2mm}    
{Jodrell Bank Observatory (JBO)}&&&&\\
\noalign{\vskip 1.6mm}
\multirow{2}{3.0cm}{+ Lovell Telescope } & DFB & \SIlist{1400;1520}{} & - &Legacy& Full Stokes\\
& ROACH & \SIlist{1520}{} & \SIlist{1420;1620}{} &Modern& Full Stokes\\
\noalign{\vskip 1.6mm}
{+ Mark II} & MK2 & \SIlist{1520}{} & - &Modern& Full Stokes\\
\noalign{\vskip 4mm}    
\multirow{3}{3.0cm}{Nan\c{c}ay Radio Telescope (NRT)} & BON & \SIlist{1400;1600;2000}{} & - &Legacy& Full Stokes\\
& NUPPI & \SIlist{1484;1854;2154;2539}{} & \SIlist{1292;1420;1548;1676}{}; \SIlist{1662;1790;1918;2046}{}; \SIlist{1962;2090;2218;2346}{}; \SIlist{2411;2667}{}&Modern& Full Stokes\\
\noalign{\vskip 2mm}    
\multirow{2}{3.0cm}{Westerbork Synthesis Radio Telescope (WSRT)} &  PuMaI  & \SIlist{323;328;367;382;840;1380;2273}{} & - &Legacy& Dual\\
& PuMaII & \SIlist{350;1380;2273}{}& - &Modern& Full Stokes\\
\noalign{\vskip 6mm}    
The Large European Array for Pulsars (LEAP)& LEAP & \SIlist{1396}{}& -&Modern\\
\noalign{\vskip 2mm}    
\textit{Sardinia Radio Telescope (SRT)\textsuperscript{$\dag$}} &  ROACH & \SIlist{357;1396}{} & - &Modern& - \\
\hline
\noalign{\vskip \baselineskip}
\end{tabular}
\end{center}
{\footnotesize
\begin{itemize}
    \item[$\dag$]Data from SRT were only included as part of the LEAP mode of observations for this data release.
\end{itemize}
}
\end{table*}
\subsection{Effelsberg Radio Telescope}\label{ssec:eff}
The Effelsberg \SI{100}{\m} single-dish radio telescope located near
Bad Munstereifel is maintained and operated by the Max Planck Institut
f\"ur Radioastronomie. The Effelsberg Berkeley Pulsar Processor
\citep[EBPP;][]{bdz+97} was used to record observations up to 2009. 
A ROACH-based backend became the main EFF pulsar backend in 2011 \citep{lkg+16}. 
This system is capable of coherently dedispersing and folding 
dual-polarisation data streams over a total bandwidth of up to $\sim$\SI{500}{\mega\hertz}.

The observations in the Effelsberg Radio Telescope were conducted in $L$, $S$, and $C$ bands \footnote{See \url{https://eff100mwiki.mpifr-bonn.mpg.de/doku.php?id=information_for_astronomers:rx:p200mm} for a detailed description of the observing systems with the Effelsberg Radio Telescope.}. Depending on the central frequency, different receivers were used, as summarised in \cite{lkg+16}. $L-$band observations were performed with the P200mm and P217mm receivers of the telescope. These offer bandwidths of \SIlist{140}{\mega\hertz} 
and \SIlist{240}{\mega\hertz}, respectively. 

$S-$band observations around \SI{2600}{\mega\hertz} were made with the
S110mm receiver. 
The receiver frequency coverage increased
from \SIlist{80} {\mega\hertz} to \SIlist{300}{\mega\hertz} after the
second half of 2018. $C-$band observations were made with the
S60mm receiver, which offers \SIlist{500}{\mega\hertz} of bandwidth.

The typical integration length per source was \SI{30}{\min}, which included 
a \SIlist{2}{\min} scan of the noise diode for polarisation
calibration. Data reduction of the folded coherently dedispersed pulsar archives, including flux and polarisation calibrations, was 
carried out via the \textsc{CoastGuard} pipeline, which is part of the  
\textsc{toaster} software library \citep[see][for a detailed
  description]{lkg+16}. The removal of RFI was carried out with the
\texttt{pazi} command in \psrchive{}. For each pulsar, we
produced a profile template at each observing band by adding the
three observations with the highest signal-to-noise ratio (S/N). Data taken with the P200mm and P217mm
receivers were analysed with the same template. Each observation was
integrated over frequency, time, and polarisation to produce integrated
pulse profiles. For $L-$band data, observations were divided into two
subbands, except for PSRs~J1738+0333, J1843$-$1113 and J2322+2057, 
which were typically weak and required integration over the entire 
band. TOAs were then produced by cross-correlating the pulse profiles
with the respective pulse-profile templates \citep[see][for details]{lkg+16}.

\subsection{Lovell Telescope}\label{ssec:jbo}
The Lovell Telescope is located at the JBO, 
in Cheshire, UK. It is a \SI{76}{\m} (\SI{250}{ft}) parabolic dish on
an altitude-azimuth mount. The telescope is operated by the Jodrell Bank
Centre for Astrophysics at the University of Manchester. Since January
2009, pulsar data have been processed using a clone of the Digital
Filter Bank (DFB) developed by the Australian National Telescope
Facility. From April 2011 onwards, data were
simultaneously processed using the DFB and a ROACH board.

Although the DFB initially operated over a bandwidth of
\SI{128}{\mega\hertz} centred on \SI{1400}{\mega\hertz}, from
September 2009 onwards, the observing frequency coverage increased to 
\SI{512}{\mega\hertz} centred at 
\SI{1520}{\mega\hertz}, of which the central \SI{384}{\mega\hertz} were
typically used. ROACH observations covered a \SI{512}{\mega\hertz} band
centred on \SI{1532}{\mega\hertz}. The edges of the band were masked,
leaving a total of \SI{400}{\mega\hertz} of usable bandwidth.

The typical integration time per source was varied depending on the pulsar
and epoch, with the median observation times per pulsar ranging
between $\sim$\SIrange{10}{55}{\min}. These observations were
time-stamped using an on-site hydrogen maser clock and then corrected
to Coordinated Universal Time (UTC) using recorded offsets between
local time kept by the hydrogen maser and GPS time. At JBO, the
observations were typically not flux or polarisation calibrated.

For the ROACH backend, data streams from two orthogonal polarisations
were sampled at the Nyquist rate and digitised as 8-bit numbers. The
\SI{512}{\mega\hertz} band was split into \SI{16}{\mega\hertz}
sub-bands by a 32-channel polyphase filter. The signal in each 
\SI{16}{\mega\hertz} sub-band was then dedispersed and folded into
1024 bins at the pulse period in real-time. Each of the
\SI{16}{\mega\hertz} sub-bands was further divided into 64 channels,
each \SI{0.25}{\mega\hertz} wide and recorded on disk in
\SI{10}{\second} long sub-integrations. The lowest and highest
\SI{56}{\mega\hertz} of the band were discarded due to persistent and
known contamination by RFI sources at these frequencies. Therefore,
the total bandwidth recorded on disk was \SI{400}{\mega\hertz}. For the
DFB, the two orthogonal polarisation data streams, covering
\SI{512}{\mega\hertz}, were incoherently dedispersed and folded into
1024 bins at the pulse period using \SI{0.5}{\mega\hertz} wide
channels and \SI{10}{\second} sub-integrations. In this case, the
lowest and highest \SI{64}{\mega\hertz} of the total band were
discarded due to contamination by RFI.

The mitigation of RFI in both DFB and ROACH data was carried out using a
median filtering algorithm, which was followed by manual
inspection. Since November 2011, ROACH data have also been cleaned in
real-time using a spectral kurtosis method \citep{morello22}. For each
pulsar observation, the pipeline produced archive files with various
frequency and time resolutions. These were fully 
frequency-averaged archives with full time resolution (\SI{10}{\second} 
long sub-integrations); fully time-averaged profiles with full frequency 
resolution (\SI{0.25}{\mega\hertz} wide and \SI{0.50}{\mega\hertz}
wide channels for the ROACH and DFB, respectively); archives that were 
partially averaged in time and frequency (\SI{1}{\minute} long 
sub-integrations and \SI{8}{\mega\hertz}-wide and \SI{12}{\mega\hertz}-wide 
channels for the ROACH and DFB, respectively) and finally; profiles that were fully averaged in time and frequency.

TOAs were formed for each observation using the \texttt{pat} command 
in \textsc{psrchive} \citep{psrchive}, which employs the \texttt{FDM} 
algorithm \citep{vlh+16} using the template profile for the appropriate 
frequency. In the work presented here, for each epoch, we divided the 
partially (time-)averaged files to produce two archives, one for each 
subband centred on \SIlist{1420; 1620}{\mega\hertz}, respectively. These 
files were then averaged in frequency and time individually, to finally 
produce two TOAs spanning the full observation duration.

\subsection{Mark II Telescope}\label{ssec:jbo_mkii} The Mark II Telescope is also located at the JBO, in Cheshire, UK. It is an altitude-azimuth mount instrument, with an elliptical dish with a major axis of \SI{38.1}{\m} (\SI{125}{ft}) and a minor axis of \SI{25.4}{\m} (\SI{83.3}{ft}). From August 2017 onwards data have also been recorded for PSR~J1713+0747 using this telescope with the ROACH-board-based backend described above. The data were processed in the same way as for the Lovell Telescope.

\subsection{Nan\c{c}ay Radio Telescope}\label{ssec:nrt}
Regular EPTA timing observations were conducted with the NRT from late 2004. These observations
were made using the $L-$ and $S-$band receivers of the telescope,
with a frequency coverage of \SIrange{1.1}{1.8}{\giga\hertz}, and \SIrange{1.7}{3.5}{\giga\hertz},
respectively. From late 2004 until early 2014, the legacy
Berkeley-Orl\'eans-Nan\c{c}ay (BON) backend \citep{ct06} was used to
record the pulsar timing data included in the EPTA DR1 
dataset \citep[for a detailed description, see][]{dcl+16}. 
Starting in August 2011, the Nan\c{c}ay Ultimate
Pulsar Processing Instrument (NUPPI) became the primary pulsar timing
backend \citep{ctg+13,lgi+20}. The EPTA DR2 dataset includes data 
collected with the NUPPI backend between August 2011 and early 2021. 
NUPPI observations have durations ranging from \SIrange{20}{80}{\minute}, and 
cover a frequency bandwidth of \SI{512}{\mega\hertz} channelised into 128 channels 
that are coherently dedispersed in real-time. Observations with the 
$L-$band receiver were made at a central frequency of \SI{1484}{\mega\hertz}. Those 
with the $S-$band receiver were generally centred on \SI{2539}{\mega\hertz} and 
occasionally on \num{1854} and \SI{2154}{\mega\hertz}. 
Data collected before MJD\,57924 were time-stamped using a local rubidium clock, and later
corrected to GPS standard time stamped using recorded offsets between
the clock and the Paris Observatory Universal Time. Data
collected after this epoch were directly stamped with the GPS time
standard, as the backend was locked to the GPS signal. 
Data were calibrated for polarisation using a short scan 
on a reference noise diode, conducted prior to each observation, with the 
\texttt{SingleAxis}\footnote{\url{http://psrchive.sourceforge.net/manuals/pac/}} 
method of \psrchive{}, to correct for the differential phase and 
amplitude between the two polarisations. Since late 2019, this simple calibration scheme was further complemented by regular observations of bright polarised pulsars in a mode where the horn rotated by $\sim$\SI{180}{\degree} across the \SI{1}{\hour} observation, enabling a better determination of the polarimetric response of the NRT at the epoch of the observations. These observations and the procedure followed for analysing them are described in \citet{Guillemot2023}.  Automatic RFI mitigation was performed on polarisation-averaged 
archives with the full frequency and time resolution available (typically,
\SI{4}{\mega\hertz} and \SIrange{5}{60}{\second} time resolution), using a Python 
script based on the \texttt{surgical} RFI-cleaning algorithm of the 
\textsc{CoastGuard} software package \citep{lkg+16}. Observations corrupted by strong RFI, calibration issues, incidental 
backend faults, or those that contained no visible pulsar signal, were discarded. 
We formed template profiles with the four Stokes parameters and with four 
frequency sub-bands by integrating the eight highest S/N observations, 
smoothing the average using a wavelet smoothing method. Finally, 
we used these polarimetric profiles and the matrix template matching (MTM) 
technique implemented in \texttt{pat} \citep{vanStraten2006} to 
extract TOAs from the NUPPI observations. By modelling the transformation 
between polarised light curves, the MTM method corrects potential 
polarisation artefacts (caused by, e.g., calibration issues),  
yielding more accurate TOAs. The procedure developed at Nan\c{c}ay 
to improve the polarisation calibration of the NUPPI data will be presented in 
detail in \citet{Guillemot2023}.

\subsection{Sardinia Radio Telescope}\label{ssec:srt}

The Sardinia Radio Telescope is a 64\,m parabolic dish with an
altitude-azimuth mount. The receiver that is used primarily for EPTA
observations is the dual-band $L$/$P$ receiver (original $P-$band =
\SIrange{305}{410}{\mega\hertz} and $L$-band = $1.3 - 1.8$\,GHz). Data
during the EPTA observations were mainly acquired with a DFB as well as a
ROACH backend, the latter of which is a copy of those installed at JBO
and Effelsberg. Both backends are capable of performing real-time
folding of the incoming data, each with a bandwidth of
\SI{500}{\mega\hertz}. Additionally, an 8-node CPU cluster installed with ROACH
allows the baseband recording of the full LEAP bandwidth
(\SI{128}{\mega\hertz}). The data were time-stamped using a local
hydrogen maser. EPTA observations at the SRT started in March 2014 and went on 
until July 2016, and were resumed from May 2018 onwards. The intermediate gap
was due to repairs of the $L$/$P$ receiver, refurbishment of the
active surface of the dish, and relocation of the control room along
with the digital instruments to a permanent structure. Most of these
observations were conducted as part of the LEAP observing programme.
For the work in this paper, these data were directly integrated into the LEAP data products,
rather than being included as independent pulsar timing data (see Section~\ref{ssec:leap}).

\subsection{Westerbork Synthesis Radio Telescope}\label{ssec:wsrt}
The WSRT was an equatorial mount linear array consisting of \num{14}
dishes of \SI{25}{\m} diameter which were coherently added to form a
\SI{93}{\m} equivalent dish. Pulsar observations formed a key science
operation and were carried out using the Pulsar Machine \citep[PumaI;
][]{str99} backend from 1999 until mid-2010. From 2007 to mid-2015,
pulsar observations were performed using the second generation Pulsar
Machine \citep[PuMaII; ][]{kbw08} which was a coherent dedispersion
backend capable of simultaneously recording data across a total
bandwidth of \SIrange[]{64}{128}{\mega\hertz}. For both PuMa backends,
the WSRT was equipped with the turret-mounted Multi-Frequency
FrontEnds \cite[MFFEs; ][]{tan91}, which allowed for rapid changes in
observing frequency. From 2013 onwards, the effective size of WSRT for
pulsar observations gradually decreased due to dishes being
removed from the array as the infrastructure was converted to adapt
the telescopes for the new phased array feeds; the APERture Tile In
Focus \citep[APERTIF; ][]{cov22}. To account for the decrease in
sensitivity, the observing time per pulsar was steadily increased
during this transition period, which ended on
15 June 2015 when the WSRT officially ended
its multiband operations.

The PumaI backend was used in conjunction with the MFFEs to observe
pulsars at \SIlist{328;382;800;1420;2200} {\mega\hertz}, with the data 
recorded in a custom format. TOAs were created from the PuMaI backend
by first converting the observations to ASCII total intensity profiles
and then cross-correlating them with templates as described in
\cite{dcl+16}.  The PuMaII backend operated across a total bandwidth
of \SI{80}{\mega\hertz} at \SI{350}{\mega\hertz}, and
\SI{160}{\mega\hertz} at \SIlist{1380;2273}{\mega\hertz}. The total
bandwidth was subdivided into an overlapped polyphase filter bank
scheme with individual channels of \SIlist{10;20}{\mega\hertz} each with an
overlap of \SIlist{1.25;0}{\mega\hertz} at \SIlist{350;1380}{\mega\hertz}, 
respectively. Besides multiband operation, PuMaII was also operated in 
LEAP mode (see Section \ref{ssec:leap}) centred at \SI{1398}{\mega\hertz},
\SI{16.25}{\mega\hertz} of each \SI{20}{\mega\hertz} used to match
other telescopes; with a total bandwidth for timing observations set
at \SI{130}{\mega\hertz}.

The duration of pulsar observations at the WSRT varied with pulsar and
epoch and lasted between \SIlist{15;45}{\min}. However, at later
epochs, the duration per pulsar was extended to account for the loss of
sensitivity because fewer dishes were available. These observations
were time-stamped to UTC at the observatory using a GPS-referenced
local hydrogen maser.  Daily averages of the local maser clock were
also compared with other observatory masers as part of regular, very
large baseline interferometry (VLBI) recordings. The local clock at
WSRT was affected by a discrete offset of \SI{10}{\second} between
$\sim$\SIrange{56808}{56903}{MJD} due to a failure of the local
Network Time Protocol \citep{ntp} server.

To achieve phase coherence between the dishes, standard
interferometric coherent phasing of the array was performed at the
start of each observing run. By observing an astronomical calibrator,
the signal amplitudes for each polarisation were equalised for all
individual antennas. This maximised the vector sum of each polarised
data stream, leading to very stable polarisation behaviour, especially
at \SIlist{1380;2273}{\mega\hertz} which was stable for at least a
couple of days.  No further calibration (flux or polarisation) was
performed, although the polarisation response was cross-validated
against calibrated data from the other telescopes when the WSRT was
operated in LEAP mode.

The PuMAII backend recorded raw data in
\texttt{psrdada}\footnote{\url{http://psrdada.sourceforge.net/}} format,
which were dedispersed using DM values stored in
\texttt{TEMPO}-style ephemeris files. PuMaII produced
\psrchive{} archives for each band with a final sub-integration
length of \SI{10}{\second} and 64 channels of
\SIlist{10;20}{\mega\hertz}, depending on the observing band. The data
were Nyquist sampled for each channel at a resolution of
\SI{25}{\nano\second} leading to a variable number of bins for the
pulsars; thus, the fastest pulsars have 256 bins, while the slowest
have 8192.

The excision of the RFI in WSRT was performed using a custom
tool based on \psrchive{}, followed by manual inspection.  To
compensate for decreased amplitude response at the band edges and
obtain continuous frequency coverage, files per frequency band were
split and re-added, providing effective bandwidths of
\SI{70}{\mega\hertz} at \SI{350}{\mega\hertz}, \SI{160}{\mega\hertz}
at \SI{1380}{\mega\hertz}, \SI{130}{\mega\hertz} at
\SI{1396}{\mega\hertz} (LEAP mode) and \SI{150}{\mega\hertz} at
\SI{2200}{\mega\hertz}.

For each of the band-averaged datasets, individual observations were
further reprocessed for RFI removal using
\texttt{ITERATIVE\_CLEANER}\footnote{\url{https://github.com/larskuenkel/iterative\_cleaner}},
inspired by algorithms from the \textsc{CoastGuard} \citep{lkg+16}
Pulsar Processing Suite. Each observation was then averaged over
frequency and time to produce the pulse profiles.
Templates for each pulsar were produced by averaging the smallest
number of profiles that contribute $\sim$90\% of the S/N of the entire
dataset, and TOAs were generated following the recommendations in
Appendix A of \citep{vlh+16}, by using \texttt{pat} with the
\texttt{FDM} algorithm.

\subsection{The Large European Array for Pulsars (LEAP)}\label{ssec:leap}
LEAP performs simultaneous monthly observations of more than twenty
MSPs at 1.4\,GHz with the  five EPTA telescopes mentioned above \citep{bjk+16}. During each observation, the telescopes switch
between the pulsar and a nearby phase calibrator. The typical exposure
time per pair of pulsar and phase calibrator is $45-60$ and
$2-3$\,min, respectively. All phase calibrators were selected from
the VLBA calibrator catalogue\footnote{\url{http://www.vlba.nrao.edu/astro/calib/vlbaCalib.txt}}. During the observations,
baseband data were recorded at the Nyquist rate with 8-bit sampling and
8$\times$16\,MHz subbands. The data were later assembled at JBO where
they were correlated, calibrated for polarisation, and coherently combined with a
dedicated software correlator \citep{sbj+17}, 
using Effelsberg as a geometrical and time reference. 
Narrowband RFI from a particular telescope was also removed before the data were 
combined. The combined baseband data of each pulsar were then
coherently dedispersed and folded to form 10\,s sub-integrations with a
1\,MHz frequency resolution, using the \textsc{dspsr} software
package. Data were visually inspected to remove any remaining impulse
RFI. Next, for each pulsar, the data from each individual observation
were averaged in time and frequency and used to calculate the TOA of
the integrated profile with the MCMC implementation of the canonical
template matching scheme \citep{tay92a,vlh+16}. The template was
created using the observation with the highest S/N (which was eventually
not included in the timing dataset) using a wavelet smoothing scheme
to remove the radiometer noise on top of the pulsar signal. Most of
the processing of the dedispersed, folded data was carried out with the
\textsc{psrchive} software package.

\section{Data preparation, combination, and timing analysis}\label{sec:combination}
The EPTA DR2 includes \tempotwo{}-compatible pulsar ephemerides
and TOA files, with the latter produced as described in the previous section and references therein. 
Following the customised processing steps at
individual telescopes, described in Section \ref{sec:obs}, the final
TOA sets were transferred to a central repository, for which a continuous
integration scheme was developed, providing quick-look plots of the
timing residuals for inspection. The data from the central repository
were then combined in parallel, using standard manual steps as
presented in \cite{vlh+16,dcl+16} as well as a semi-automated
combination scheme. In both cases, the results were manually inspected
and cross-verified for consistency, with the final data release
containing data from the manual process. In the following paragraphs,
we describe source selection, combination steps, and the
resulting combined dataset in more detail.

\subsection{Source selection}
\begin{figure}
\centering 
\includegraphics[scale=0.6]{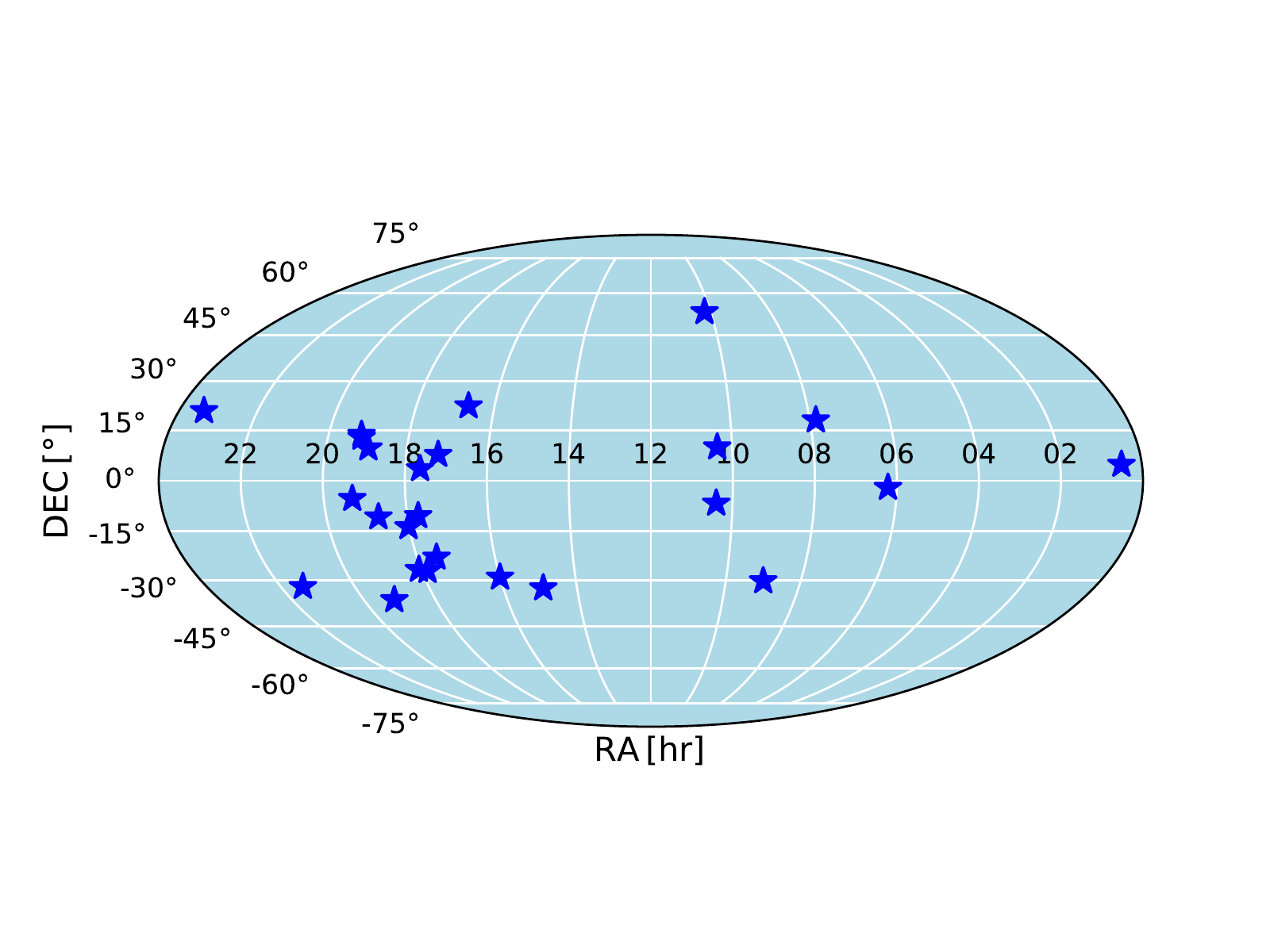}
\vspace*{-1.5cm}
  \caption{Sky projection of the 25 pulsars included in the EPTA DR2 dataset.  \label{fig:hammer}}
\end{figure}

Due to the large number of telescope and back-end combinations across EPTA observatories, 
the heterogeneous recording and processing schemes, and the complex RFI environment in most telescopes,
 curation and vetting of EPTA data require long lead
times.  
Furthermore, the noise modelling for 
individual pulsars is computationally expensive, requiring a detailed and 
iterative analysis of the possible noise models that may be applicable
for each pulsar dataset. 
For these reasons, we adopted a source selection scheme which maximises the detectability of a stochastic GWB through the ${\rm S/N_A^2}$ statistic of \citet{ rsg15,spf+23}, 
\begin{equation} 
\label{eq:signal-to-noise-A}
    \text{S/N}_\text{A}^2  = 2\sum_{a>b} \int \frac{
    \Gamma^2 _{ab} \, S^2 (f) \, T_{ab}
    }{
    P_a (f) P_b (f)
    } d f\, ,
\end{equation}
taking into account the fact that each pulsar contributes differently to the PTA response, due to its inherently distinct noise properties, $P_j(f)$. Here $\Gamma_{ab}$ represents the overlap reduction
function that translates the mean spectral density of an isotropic stochastic red-noise process, $S(f)$, to cross-correlation power between pulsars $a$ and $b$ that have been observed for a common duration of $T_{ab}$. The simple ranking produced through this scheme was then improved by applying the coupling matrix formalism introduced in \citet{roebber2019} as adapted by \citet{spf+23} to prioritise pulsars that maximise the response to HD-like correlations 
while maintaining the ability to distinguish between competing dipolar and monopolar signals.  
Using this methodology, we found that a subset of 25 pulsars out of the 42 included in the DR1, were sufficient to recover at least \SIrange{90}{98}\%
of the full array sensitivity to a simulated stochastic GW background with an amplitude of $3\times 10^{-15}$ at a frequency of 1\,yr$^{-1}$,  
and a spectral index of $\gamma=13/3$. The same subset of pulsars would also recover at least 95\% of the total sensitivity to possible individually resolvable, monochromatic gravitational wave sources across all frequencies.
These 25 pulsars comprise the EPTA DR2. 
Their distribution on the sky can be seen in Figure~\ref{fig:hammer}. 

\subsection{Combination of the dataset}
We followed the timing and combination steps described in \citet{vlh+16}
and \cite{dcl+16} to combine the data across telescopes. 
For each pulsar, data from different telescopes were
combined using \tempotwo{} to form the joint dataset, starting with
parameters from \citet{dcl+16} and using a summary TOA file, following 
 \cite{pdd+19}, \cite{vlh+16} and \cite{dcl+16}. To align the data from different observing systems, we fitted for an arbitrary phase offset (commonly referred to as \texttt{jump}) for each sub-band and backend combination, using the NUPPI sub-band centred at \SI{1420}{\mega\hertz} as our reference dataset. For a small number of individual backend datasets,  discrete time offsets were detected and estimated using multi-pulsar
analysis. These were also removed using the \tempotwo{} \texttt{TIME} 
keyword. During certain observing runs, data were collected using
both legacy and new backends, or in both single-telescope and LEAP
modes. As these observations represented the same signal and noise, we
eliminated the older backend and non-LEAP data. However, we kept the
data outside the LEAP bands to better constrain dispersion-measure (DM) 
variations. The overlapping data were removed after the \texttt{jump} values
were determined. With this final set of curated TOAs, we then produced initial timing solutions for each pulsar using \tempotwo{}. For these solutions, the timing parameters 
were fitted for iteratively using \tempotwo{}, until the linearised timing solution converged.
For each of the pulsars, we then investigated
the likelihood of introducing new timing parameters that were not fitted for in DR1, using a 5$\sigma$
detection threshold, as well as a number of $F-$statistic and information criteria based tests.

All initial timing models include 
the spin frequency and its derivative, DM
and its first and second derivative, the astrometric parameters
(position, proper motions, and in several cases the annual
parallax). For binary pulsars, we included fits for five Keplerian
parameters and a selection of post-Keplerian (pK) parameters,
depending on the pulsar. The full set of parameters included in each timing model is listed in
Tables~\ref{tab:ephem1} to \ref{tab:ephem7}. 
We note that we used equatorial coordinates to
fit for the positions of most pulsars, except for PSRs\,J0030+0451,
J1022+1001 and J1730$-$2304, for which we used ecliptic coordinates, as 
their ecliptic latitude is less than \SI{1}{\deg}. We used the DE440
version of the JPL solar system ephemeris \citep{DE440} and
TT(BIPM2021) \citep{petit09}\footnote{\protect\url{https://www.bipm.org/en/time-ftp/tt-bipm-}} as
our reference clock standard. We also applied the default spherical of
the Solar Wind electron density model implemented in \tempotwo{} 
to correct for solar wind-related DM variations, fixing the
average density in the ecliptic plane at \SI{1}{\astronomicalunit} to
\SI{7.9}{\per\cm\cubed}\citep{mca+19} following \citet{tsb+21}, except
for PSR~J1022+1001, as described in Section \ref{ssec:tnest}.

The combination scheme described above produced the full EPTA DR2
dataset, an overview of which can be found in Figure~\ref{fig:overview}. 
This dataset is used in the pulsar timing
analysis presented below, as well as in associated work, namely the 
single-pulsar noise modelling in \cite{wm2}, and the search for GWs in \cite{wm3}. Based on the full DR2 dataset, we also produced additional dataset versions for GW searches. Details for these versions can be found in Appendix~\ref{app:datasets}, as well as in \cite{wm2} and \cite{wm3}.

 \subsection{Outlier analysis} 
 The EPTA DR2 dataset was checked for outliers using the following procedures. The first step to eliminate outliers was performed when
 compiling single telescope data, either by custom automated data
 flagging or manual inspection. After initial combination, outstanding
 outliers, such as TOAs with residuals offset by more than 10 times
 the root mean square (rms) of the timing solutions, were flagged and
 the observation archives reinspected. We found that these were
 typically associated with low S/N and were therefore
 removed. Additionally, we removed observations with known calibration
 issues as well as those displaying corrupted polarisation profiles or
 systematic trends in single-epoch timing residuals. A similar
 analysis was carried out using the semi-automated analysis, including
 tests from expected correlations such as excess contribution from the
 Solar Wind or epoch-wise offsets. Finally, we inspected the whitened
 residuals using the results of the noise modelling analysis (see the next subsection), which
 revealed that all remaining TOAs were within 5$\sigma$ of the whitened
 timing residuals, indicating that no additional outlier removal was needed.

\begin{figure*}
\centering
  \centering
  \vspace*{-.5\baselineskip}
  \includegraphics[trim={0 0 0 2mm},scale=0.95]{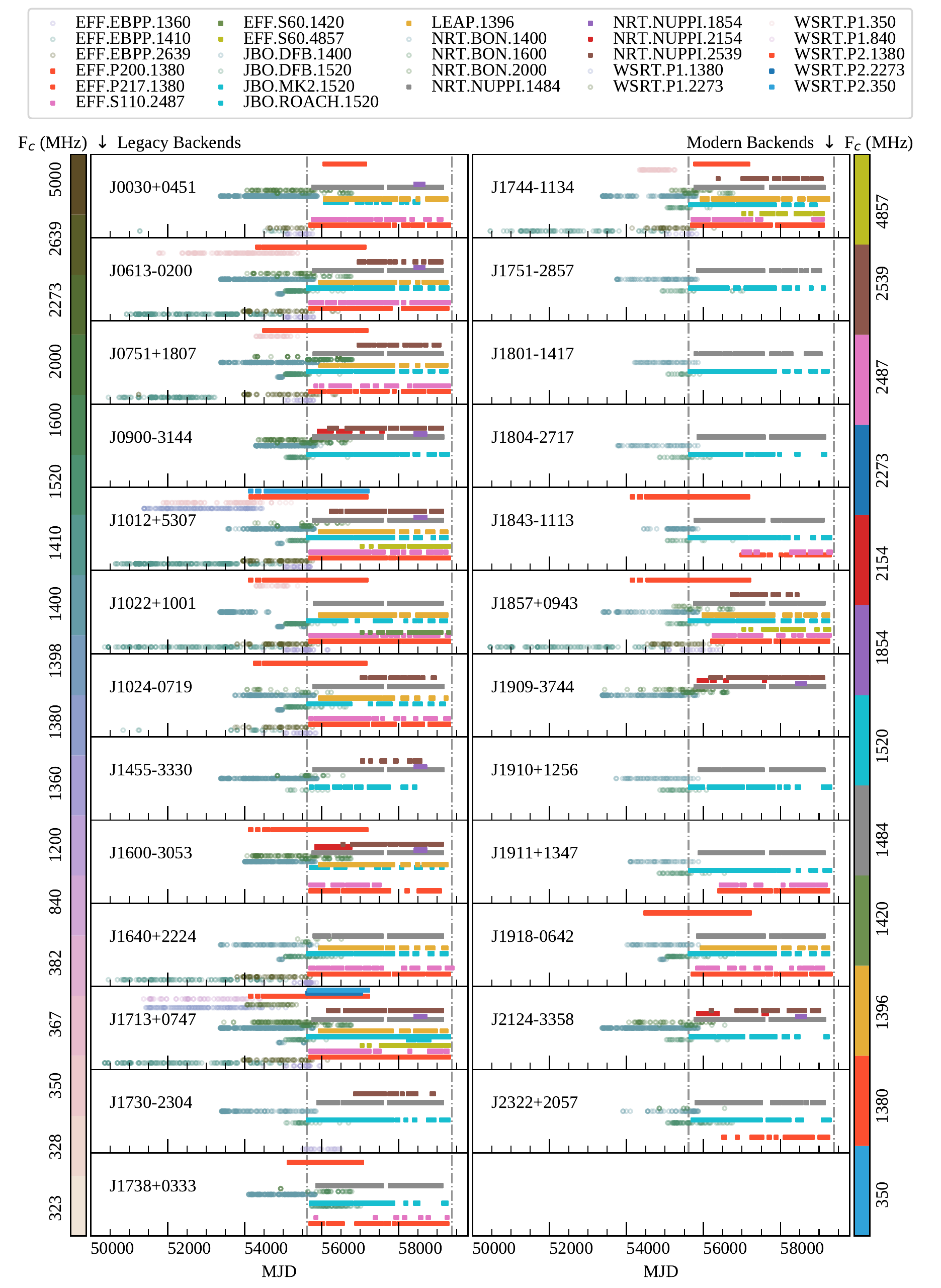}
  \vspace*{-.6\baselineskip}
  \caption{Overview of the full EPTA DR2 dataset. Empty circles denote
    legacy data, while filled squares show new EPTA backends. Vertical
    lines bound the range of the `DR2new' dataset. Readers can refer to Sec.~\ref{sec:combination} and \ref{app:datasets} for details.\label{fig:overview}}
\end{figure*}

\subsection{Bayesian timing analysis}\label{ssec:tnest}
To obtain the final timing solution for each pulsar, we performed a Bayesian timing analysis on the combined dataset using the \tnest{} toolkit \citep{lah+14}. For this step, the timing measurements obtained with \tempotwo{} were provided as initial guesses. \tnest{} is based on the software packages \tempotwo{} \citep{hem06} and \textsc{Multinest} \citep{fhb+09}. It explores the parameter space of a non-linear pulsar timing models using nested sampling
\citep{skilling2004} based on Bayesian inference \citep{lah+14}, to
provide robust estimates for the timing parameters.

For the analysis of each pulsar, in addition to the deterministic pulsar timing parameters, we included a set of stochastic parameters to characterise the noise components in the dataset. Within our adopted framework, the white noise is described by a multiplicative factor $E_{\rm f}$
(\texttt{EFAC}) that accounts for the possible underestimation of TOA uncertainties, and a factor $E_{\rm q}$ that is added in quadrature 
(\texttt{EQUAD}) to model any other possible source of noise, such as pulse phase jitter and systematics \citep[e.g.][]{lvk+11,lkl+12}. These are related to the uncertainty of the TOA measurement, $\sigma_{\rm r}$, as follows:
\begin{equation}
    \sigma=\sqrt{E^2_{\rm q}+E^2_{\rm f}\sigma^2_{\rm r}}.
    \label{eq:whitenoise}
\end{equation}
Such pairs of noise parameters were introduced for each back-end/frontent combination. These combinations correspond to different 
\texttt{-group} flags in the TOA files (see Table~\ref{tab:flags}). 
The long-term red-noise processes in the data were described by two types of models
that account for both the chromatic and achromatic noise components, respectively. In both cases, the noise components were modelled as a stationary, stochastic process with a power-law spectrum given by:
\begin{equation}
S(f)=\frac{A^2}{C_{\rm 0}}\left(\frac{f}{f_{\rm r}}\right)^{-\gamma},
\label{eq:rednoise}
\end{equation}
where $S(f)$, $A$, $\gamma$ and $f_{\rm r}$ correspond to the spectral
density of the power as a function of frequency $f$, the spectral
amplitude, the spectral index, and the reference frequency (set at
\SI{1}{\per\year}), respectively. The spectrum in both cases has a low-frequency cutoff, determined by the inverse of the total time span of the dataset. The constant $C_{\rm 0}$ was set to $1$ for the chromatic noise component and to $12\pi^2$ for the achromatic term. The spectral amplitude of the chromatic noise component is proportional to the observing radio frequency as $A\propto\nu^{-\alpha}$, where the chromatic index 
$\alpha$ is either $2$ for modelling the features induced by the DM variations, or $4$ for capturing scattering variation in the data. For each pulsar, we used a customised noise model obtained by following the procedures
outlined in \cite{cbp+22}, which employs a Bayesian model-selection framework to determine the components in the noise model and the number of frequency bins used to sample the spectrum of each component. For 21 pulsars, the noise model includes a chromatic component to model DM variation, while for 10 pulsars it includes an achromatic component. Only one pulsar (PSR~J1600$-$3053) favours the inclusion of a chromatic component for modelling scattering variation. J2322+2057 is the only pulsar in our sample that does not show any evidence for the presence of red noise. The noise components for each pulsar are summarised in the tables of Appendix~\ref{app:eph}. More details of our customised noise models can be found in the accompanying paper on noise-modelling in this series \citep{wm2}.

During the analysis with \tnest{}, both timing and noise parameters were sampled simultaneously, while the time
offsets were analytically marginalised. All model parameters were
sampled with uniform priors, except for EQUAD and the amplitude of the
red-noise processes, which used a log-uniform prior. For PSR\,J1022+1001, whose line of sight lies very close to the ecliptic plane, we also fitted the solar wind density at the orbit distance of the Earth, instead of fixing its value to \SI{7.9}{\per\cm\cubed} as was done for the other pulsars. This helped to obtain an estimate of the timing parallax consistent with other pulsar timing and VLBI measurements (see Section~\ref{ssec:J1022} for more details).

\section{Results} \label{sec:result}
Table\,\ref{tab:source} provides a summary of the EPTA DR2 dataset, while the full timing solutions are listed in Tables~\ref{tab:ephem1} through \ref{tab:ephem7}. Our timeseries have time spans of 13.6 to 24.5 years and contain $\sim$800 to $\sim$6000 TOAs. With the exception of PSR\,J1909$-$3744, all timing solutions use data from at least two telescopes. In the following, we describe some new timing measurements in DR2 and compare them with the results by \cite{dcl+16,pdd+19,aab+20a,rsc+21}; henceforth referred to as EPTA DR1, IPTA DR2, NG12 and PPTA DR2, respectively. Unless otherwise noted, quoted values and uncertainties represent the median and 68.3\% confidence intervals (C.I.) of the one-dimensional marginalised posterior distribution function. 

\begin{table*}
	\begin{center}
		\caption[]{Overview of the EPTA DR2 25 pulsar data set.
			The columns represent the name of the pulsar, the telescopes that have collected the data, the number of TOAs, the timespan $T_{\rm span}$ of the data set, the median $\sigma_{\rm TOA}$ in each frequency band, the weighted rms (wrms) of the timing residuals, and the weighted rms after subtracting both red and DM noise (i.e. of the whitened residuals). The corresponding frequency coverage for the P, L, S, and C bands is 0.3--1.0, 1.0--2.0, 2.0--4.0, and 4.0--8.0\,GHz, respectively. The asterisk next to the pulsar name denotes that this pulsar is also included in the InPTA data release (also see Appendix~\ref{app:datasets}). \label{tab:dataoverview}}
  \begin{tabular}{cccccccccccc}
  \hline
  \multirow{2}{*}{Pulsar Jname} & \multirow{2}{*}{Telescopes} & \multirow{2}{*}{MJD range} & $T_{\rm span}$ &\multirow{2}{*}{$N_{\rm TOA}$} &\multicolumn{4}{c}{Median $\sigma_{\rm TOA}$ ($\mu$s)} & wrms & wrms, \\
  \cline{6-9}
& & & (yr) & &P &L &S &C & ($\mu$s) & whitened ($\mu$s) \\
\hline \noalign {\smallskip}
J0030+0451 & EFF, LT, NRT, WSRT, LEAP &51275--59294 & 22.0 & 4069 & --- & 3.40 & 6.07 & --- & 2.85 & 2.30\\
J0613$-$0200$^{\ast}$ & EFF, LT, NRT, WSRT, LEAP &50931--59293 & 22.9  & 2909 & 7.40 & 1.43 & 6.16 & --- & 2.47 & 1.06 \\
J0751+1807$^{\ast}$ & EFF, LT, NRT, WSRT, LEAP &50460--59294 & 24.2 & 3613 & --- & 2.22 & 4.43 & --- & 2.12 & 1.50 \\
J0900$-$3144 & LT, NRT &54286--59269& 13.6 & 6064 & --- & 2.95 & 8.92 & --- & 4.28 & 2.60 \\
J1012+5307$^{\ast}$ & EFF, LT, NRT, WSRT, LEAP &50647--59295 &23.7 & 5325 & 3.77 & 1.76 & 5.59 & 4.78 & 1.28 & 1.02\\
J1022+1001$^{\ast}$ & EFF, LT, NRT, WSRT, LEAP &50361--59294 &24.5 & 2445 & 11.3 & 2.19 & 4.42 & 2.99 & 1.78 & 1.56\\
J1024$-$0719 & EFF, LT, NRT, WSRT, LEAP &50841--59294 &23.1 & 2522 & --- & 2.38 & 7.64 & --- & 1.10 & 1.02\\
J1455$-$3330 & LT, NRT &53375--59117 &15.7 & 2815 & --- & 7.23 & 17.8 & --- & 2.52 & 2.46 \\
J1600$-$3053$^{\ast}$ & EFF, LT, NRT, WSRT, LEAP &53998--59230 &14.3 & 2982 & --- & 0.48 & 1.50 & --- & 2.68 & 0.37\\
J1640+2224 & EFF, LT, NRT, WSRT, LEAP &50459--59385 & 24.4 & 2006 & --- & 3.57 & 7.65 & --- & 1.13 & 1.10\\
J1713+0747$^{\ast}$ & EFF, LT, NRT, WSRT, LEAP &50360--59295 & 24.5 & 5003 & 3.13 & 0.32 & 0.54 & 0.67 & 0.43 & 0.20 \\
J1730$-$2304 & EFF, LT, NRT &53397--59279 & 16.1 & 1315 & --- & 1.57 & 7.26 & --- & 1.00 & 0.83 \\
J1738+0333 & EFF, LT, NRT, WSRT &54103--59259 & 14.1 & 1019 & --- & 4.31 & --- & --- & 2.90 & 2.33 \\
J1744$-$1134$^{\ast}$ & EFF, LT, NRT, WSRT, LEAP &50460--59230 & 24.0 & 1946 & 3.60 & 0.90 & 2.29 & 1.15 & 1.01 & 0.56 \\
J1751$-$2857 & LT, NRT &53746--59111 & 14.7 & 398 & --- & 3.17 & --- & --- & 3.73 & 2.34 \\
J1801$-$1417 & LT, NRT &54206--59214 & 13.7 & 449 & --- & 4.09 & --- & --- & 3.94 & 2.46 \\
J1804$-$2717 & LT, NRT &53766--59145& 14.7 & 723 & --- & 5.94 & --- & --- & 1.80 & 1.63 \\
J1843$-$1113 & EFF, LT, NRT, WSRT &53156--59293 & 16.8 & 893 & --- & 1.37 & 2.64 & --- & 3.47 & 0.81 \\
J1857+0943$^{\ast}$ & EFF, LT, NRT, WSRT, LEAP &50458--59258 & 24.1 & 1540 & --- & 1.70 & 4.05 & 6.85 & 1.38 & 1.05 \\
J1909$-$3744$^{\ast}$ & NRT &53368--59115 & 15.7 & 2503 & --- & 0.29 & 0.57 & --- & 0.73 & 0.14 \\
J1910+1256 & LT, NRT &53725--59282 & 15.2 & 538 & --- & 2.65 & --- & --- & 2.03 & 1.77 \\
J1911+1347 & EFF, LT, NRT &54095--59282 & 14.2 & 882 & --- & 1.22 & 1.36 & --- & 1.06 & 0.75 \\
J1918$-$0642 & EFF, LT, NRT, WSRT, LEAP &52095--59294 & 19.7 & 1361 & --- & 2.04 & 4.14 & --- & 1.78 & 1.31 \\
J2124$-$3358$^{\ast}$ & LT, NRT &53365--59213 & 16.0 & 2018 & --- & 3.70 & 12.6 & --- & 2.24 & 2.17 \\
J2322+2057 & EFF, LT, NRT &53905--59268 & 14.7 & 804 & --- & 10.9 & --- & --- & 4.08 & 4.08 \\
			\hline \noalign {\smallskip}
			\label{tab:source}
		\end{tabular}
	\end{center}
\end{table*}

For 12 of the 25 pulsars in DR2, thanks to the extended data and
increased sensitivity of the next-generation receivers, we measure several new timing parameters in comparison to DR1. Some of these measurements are also reported here for the first time. These results are discussed in detail in the following subsections. In addition, DR2 includes improved timing solutions for the binary systems J0751$+$1807, J0900$-$3144, J1024$-$0719, J1455$-$3330, J1713$+$0747,
J1857$+$0943, J1909$-$3744 and J1910$+$1256, as well as for the
isolated pulsars J0030$+$0451, J1744$-$1134, J1843$-$1113,
J2124$-$3358, and J2322$+$2057, even though no additional timing parameters were
measured compared to EPTA DR1. For PSR~J1024$-$0719, where we only fit the second
spin frequency derivative as in previous work \citep{bjs+16}, we
performed additional investigations to explore the measurability of higher order spin frequency derivatives (see
Section~\ref{subsec:high_order}). For PSR~J1857$+$0943, we omitted the fit for secular variation of the orbital projected semi-major axis ($\dot{x}$) which was poorly constrained from the timing
analysis. Meanwhile, a rough estimate of this parameter may still be
inferred by searching for the annual orbital parallax (Section~\ref{subsec:annual-orbital-parallax}).

\subsection{PSR\,J0613$-$0200}
PSR~J0613$-$0200 is a binary pulsar with a white dwarf (WD) companion \citep{bac+16}. The new timing solution includes three pK parameters, namely the secular change of the orbital period, $\dot{P}_{\rm b}$, the third harmonic of the Shapiro delay (SD), $h_3$, and the ratio between the third and fourth SD harmonics, $\zeta$ \citep{fw10}. The uncertainty of the measurement on $\dot{P}_{\rm b}$ has improved by a factor of three compared to EPTA DR1, while estimates for the SD parameters with EPTA data are reported here for the first time. The astrometric solution includes a significant measurement of the timing parallax. These measurements agree with the constraints in NG12 and PPTA DR2, but are in slight tension with previous EPTA estimates in DR1. The results are summarised in Table~\ref{tab:ephem1}.

\subsection{PSR\,J1012$+$5307}
PSR\,J1012$+$5307 is a 1.8\,M$_{\odot}$ pulsar in a 14.4\,h orbit around a low-mass WD companion \citep{vbk96,lwj+09,ato+16,ddf+20,siv+20}.
The DR2 dataset for this pulsar now includes more than 5000\,TOAs with an average precision of $\sim$5\,$\mathrm{\mu s}$.
Previous mass estimates \citep{ato+16} for the system indicate that the expected SD amplitude is now larger than the EPTA timing precision for the pulsar (5\,$\mathrm{\mu s}\times 5000^{-1/2}\simeq0.06$\,$\mathrm{\mu s}$). This prompted us to include the orthometric SD parameters in our fit, resulting in a significant detection of $h_3$ and $h_4$ for the first time (see Table~\ref{tab:ephem2}). The updated timing solution also includes a highly significant measurement of $\dot{P}_{\rm b}$ and an improved estimate of the parallax, which is consistent with previous timing estimates \citep{lwj+09,dcl+16,pdd+19}, but in tension with the estimates from VLBI and Gaia \citep{ddf+20,ant20,ant21}. Possible reasons for this discrepancy may be related to systematics between dynamical and kinematic reference frames \citep{lza+23}, or covariance between the parallax, DM variability, and solar-wind timing signatures. These will be investigated in a future publication.

\subsection{PSR\,J1022$+$1001} \label{ssec:J1022}
In the EPTA DR1 dataset, the only measurable pK parameter for this system was the advance of periastron, $\dot{\omega}$. The updated estimate for this parameter is in agreement with the previous value, with an uncertainty improved by five times. 
The new solution includes three additional pK parameters, namely the two SD parameters and $\dot{P}_{\rm b}$, all constrained with a significance greater than $3\sigma$ (see Table~\ref{tab:ephem2}). In particular, $\dot{P}_{\rm b}$ is measured for the first time in this system. Both the SD and solar wind amplitude measurements are in agreement with those reported in \citet{rsc+21}. The updated timing parallax is slightly lower than the VLBI estimate for the system \citep{dgb+19}. 

\subsection{PSR~J1455$-$3330}
This MSP is in a 76\,d binary system with a WD companion. The measured $\dot{x}$ is $-1.98(6) 
\times 10^{-14}$\,ls\,s$^{-1}$ (see Table~\ref{tab:ephem2}), which is in agreement with the NG12 result. In addition, we report a tentative measurement of $\dot{P}_{\rm b}$ for the first time. Further timing analysis may lead to a clear detection of this parameter. The measured parallax agrees with IPTA DR2 and EPTA DR1. 

\subsection{PSR\,J1600$-$3053}
PSR\,J1600$-$3053 is a binary MSP in orbit with a WD companion of an orbital period of 14.3\,day. Using the EPTA DR2 data, we obtain a measurement of the pK parameter $\dot{P}_{\rm b}$ (see Table~\ref{tab:ephem3}) for the first time. The pK parameters measured in this system, including $\dot{P}_{\rm b}$, $\dot{x}$, $\dot{\omega}$ and SD, are all consistent with those reported by NG12 and PPTA DR2.

\subsection{PSR~J1640$+$2224}
PSR\,J1640$+$2224 is an MSP with a WD companion in a 175-day orbit \citep{zzx+20}. The new timing solution was derived using the DDH model in \textsc{tempo2} and includes significant measurements of the timing parallax, SD parameters, and the first derivative of the orbital period. The SD parameters suggest that the pulsar has a low mass, although the measurement uncertainties are still too large to provide a stringent constraint. The parallax estimate is consistent with that derived with VLBI and the timing solutions by other PTAs (see Section~\ref{subsec:parallax}). The secular orbital period change can be attributed to kinematic effects, although at face value our estimate is in slight tension with expectations, given the position and proper motion of the pulsar (see Section~\ref{ssec:pbdot}). The uncertainty for the $\dot{x}$ estimate is improved by three (five) times compared to EPTA DR1 (IPTA DR2). 
The timing parameters are listed in Table~\ref{tab:ephem3}. 

\subsection{PSR~J1730$-$2304}
The proper motion along the ecliptic latitude was not measured in either EPTA DR1, IPTA DR2 or PPTA DR2, due to the proximity of this isolated pulsar to the ecliptic plane. Here, we report the first tentative measurement of this parameter, $-4.4(1.8)$\,mas\,yr$^{-1}$. The revised timing solution is explained in Table~\ref{tab:ephem3}.

\subsection{PSR~J1738$+$0333}
A pulsar timing analysis of PSR~J1738$+$0333 based on data collected with the Arecibo and EPTA telescopes resulted in a significant detection of the timing parallax and first orbital period derivative \citep{fwe+12}. The latter is consistent with the general relativity prediction given the masses of the pulsar and its companion, which have been measured using optical spectroscopy \citep{avk+12}. Here, we report for the first time a $4\sigma$ measurement of $\dot{P}_{\rm b}=-3.0(7)\times 10^{-14}$, using EPTA-only data. This estimate is approximately $2\sigma$ higher than that of \citet{fwe+12} ($-1.7(3)\times10^{14}$), which could be due to systematics introduced by the long integration times typically used in EPTA observations ($\sim 0.5-1$\,hour). The latter correspond to a significant fraction of the 7.5\,h binary orbit and may result in smearing of the $\dot{P}_{\rm b}$ timing signature.

\subsection{PSR~J1751$-$2857}
This 110-day period binary pulsar \citep{sfl+05} is only monitored by the EPTA. Compared to EPTA DR1, the new solution (Table~\ref{tab:ephem4}) provides significantly improved estimates for all timing parameters. We also report a marginal detection of the timing parallax; $\varpi=1.1(4)$\,mas. 

\subsection{PSR~J1801$-$1417}
PSR~J1801$-$1417 is an isolated pulsar and only monitored by the EPTA. Here, we report a marginal detection of the timing parallax signature at approximately $3\sigma$ (see Tables~\ref{tab:ephem4} and \ref{table:parallax}). 

\subsection{PSR~J1804$-$2717}
PSR~J1804$-$2717 is a binary MSP with a WD companion of an orbital period of 11.1\,day. It is currently only monitored by the EPTA. The new timing solution includes a significant measurement of the timing parallax, as well as proper motion parameters that are seven times more precise compared to the EPTA DR1 solution (see Tables~\ref{table:parallax} and \ref{tab:ephem5}).

\subsection{PSR~J1911$+$1347}
PSR~J1911$+$1347 is a 4.63-ms isolated MSP. The astrometric solution for this pulsar now includes a parallax measurement (see Table~\ref{table:parallax}. This is also the first parallax measurement reported in this pulsar. The new timing solution is consistent with EPTA DR1 and IPTA DR2, with a significant improvement in the measurement precision of all timing parameters.  

\subsection{PSR~J1918$-$0642}
We report an improved measurement of the SD signature, with $h_{3}=8.3(2)\times 10^{-7}$ $\mu$s, and $\zeta=0.908(9)$, which agrees well with previous estimates. Compared to EPTA DR1, the timing solution now also includes a measured timing parallax (as seen in Table~\ref{tab:ephem6}) that is consistent with the value reported in NG12, as well as the VLBI parallax reported by \cite{dds+23}, as can be seen in Table~\ref{table:parallax}. 

\section{Discussion}\label{sec:discussion}

\subsection{Parallaxes and distances}
\label{subsec:parallax}

\begin{table*}
\begin{center}
\caption{Summary of the timing parallax measurements in the EPTA DR2. The $\varpi_{\rm mes}$ column gives the median and 68.3 percentiles of the marginalised posteriod distribution. $S_{1400}$ gives the pulsar flux densities that were used to infer the bias-corrected distances, $D_{\rm corr}$ (all taken from the ATNF catalogue v1.69 \protect\citet{hmth04}; see Section~\ref{subsec:parallax} for details on parallax inversion). The last two columns list the magnitude of the proper motion and the inferred transverse velocity ($V_{\rm trans}=4.74\mu D_{\rm corr}$) of the system, respectively.}

\renewcommand{\arraystretch}{1.4}
\begin{tabular}{ c c c c  c c c c}
\hline
Pulsar name & $\varpi_{\rm mes}$ & $S_{1400}$  & $D_{\rm corr}$ & $\varpi_{\rm ref}$ & Reference & $\mu$  & $V_{\rm trans}$  \\
 &  (mas) &  (mJy)  & (kpc) & (mas) & & (mas\,yr$^{-1}$) &  (km\,s$^{-1}$) \\
\hline

J0030$+$0451 & 3.09 $\pm$ 0.06 & 1.09 & 0.323$_{-0.006}^{+0.006}$ & 3.02$^{+0.07}_{-0.07}$ & \cite{dds+23} & $6.33_{-0.08}^{+0.09}$  & 9.7$_{-0.2}^{+0.2}$  \\
 
J0613$-$0200 & 1.00 $\pm$ 0.05 & 2.25 & 0.99$_{-0.05}^{+0.05}$ & 1.25$^{+0.13}_{-0.13}$ & DR1 & $10.521_{-0.005}^{+0.005}$  & 50$_{-3}^{+3}$  \\

J0751$+$1807 & 0.85 $\pm$ 0.04 & 1.35 & 1.17$_{-0.05}^{+0.06}$ & 0.82$^{+0.17}_{-0.17}$ & DR1 & $13.543_{-0.071}^{+0.068}$  & 75.1$_{-3.5}^{+3.6}$  \\ 

J1012$+$5307 & 0.90 $\pm$ 0.08 & 3.8 & 1.07$_{-0.08}^{+0.10}$ & 1.17$^{+0.04}_{-0.05}$ & \cite{dds+23} & $25.622_{-0.004}^{+0.004}$  & 131$_{-11}^{+11}$  \\ 

J1022$+$1001 & 1.16 $\pm$ 0.08 & 3.9 & 0.85$_{-0.05}^{+0.06}$ & 1.39$^{+0.04}_{-0.03}$ & \cite{dgb+19} & $23_{-3}^{+3}$  & 94$_{-13}^{+15}$  \\

J1024$-$0719 & 1.01 $\pm$ 0.04 & 1.5 & 0.98$_{-0.04}^{+0.04}$ & 0.94$^{+0.06}_{-0.06}$ & \cite{dds+23} & $59.75_{-0.01}^{+0.01}$  & 279$_{-11}^{+11}$  \\ 

J1455$-$3330 & 1.3 $\pm$ 0.1 & 0.73 & 0.76$_{-0.05}^{+0.06}$ & 1.04$^{+0.35}_{-0.35}$ & DR1 & $8.097_{-0.010}^{+0.011}$  & 29$_{-2}^{+2}$   \\ 

J1600$-$3053 & 0.72 $\pm$ 0.02 & 2.44 & 1.39$_{-0.04}^{+0.04}$ & 0.53$^{+0.06}_{-0.06}$ & \cite{rsc+21} & $6.984_{-0.010}^{+0.011}$  & 45.9$_{-1.3}^{+1.3}$  \\ 

J1640$+$2224 & 0.8 $\pm$ 0.2 & 0.46 & 1.08$_{-0.19}^{+0.28}$ & 0.68$^{+0.08}_{-0.08}$ & \cite{dds+23} & $11.526_{-0.007}^{+0.006}$  & 62$_{-12}^{+14}$  \\ 

J1713$+$0747 & 0.88 $\pm$ 0.01 & 8.3 & 1.136$_{-0.013}^{+0.013}$ & 0.90$^{+0.03}_{-0.03}$ & DR1 &  $6.292_{-0.001}^{+0.001}$  & 33.9$_{-0.4}^{+0.4}$   \\ 

J1730$-$2304 & 2.08 $\pm$ 0.06 & 4.0 & 0.48$_{-0.01}^{+0.01}$ & 1.57$^{+0.18}_{-0.18}$ & \cite{dds+23} & $20.7_{-0.3}^{+0.5}$  & 47.3$_{-1.6}^{+1.7}$  \\ 

J1738+0333 & --- & 0.34 & --- & 0.68$^{+0.05}_{-0.05}$  & \cite{fwe+12} & $8.713_{-0.023}^{+0.023}$  & 60.129$_{-4.264}^{+4.494}$  \\ 

J1744$-$1134 & 2.58 $\pm$ 0.03 & 2.6 & 0.388$_{-0.004}^{+0.005}$ & 2.38$^{+0.08}_{-0.08}$ & DR1 & $21.018_{-0.004}^{+0.004}$  & 38.6$_{-0.4}^{+0.5}$  \\ 

J1751$-$2857 & 1.1 $\pm$ 0.4 & 0.46 & 0.79$_{-0.21}^{+0.43}$ &  ---  & --- &  $8.67_{-0.19}^{+0.20}$  & 38$_{-12}^{+15}$  \\ 

J1801$-$1417 & 0.8 $\pm$ 0.3 & 1.54 & 1.00$_{-0.25}^{+0.46}$ &  ---  & --- & $11.01_{-0.06}^{+0.06}$  & 59$_{-17}^{+21}$  \\ 

J1804$-$2717 & 1.1 $\pm$ 0.3 & 0.4 & 0.8$_{-0.2}^{+0.3}$ & ---  & --- & $17.1_{-0.4}^{+0.4}$  & 74$_{-18}^{+23}$  \\ 

J1857$+$0943 & 0.89 $\pm$ 0.06 & 5.0 & 1.11$_{-0.07}^{+0.08}$ & 0.70$^{+0.26}_{-0.26}$ & DR1 & $6.050_{-0.005}^{+0.005}$  & 32$_{-2}^{+2}$  \\ 

J1909$-$3744 & 0.94 $\pm$ 0.02 & 1.8 & 1.06$_{-0.02}^{+0.02}$ & 0.86$^{+0.01}_{-0.01}$ & \cite{lgi+20} & $37.026_{-0.004}^{+0.004}$  & 186.6$_{-3.9}^{+4.0}$ \\ 

J1911$+$1347 & 0.40 $\pm$ 0.09 & 0.9 & 2.2$_{-0.4}^{+0.6}$ & ---  & --- & $4.683_{-0.007}^{+0.007}$  & 52.2$_{-9.9}^{+11.6}$ \\ 

J1918$-$0642 & 0.75 $\pm$ 0.07 & 0.58 & 1.3$_{-0.1}^{+0.1}$ & 0.6$^{+0.1}_{-0.1}$ & \cite{dds+23} & $9.29_{-0.01}^{+0.01}$  & 58.2$_{-5.1}^{+5.6}$  \\ 

J2124$-$3358 & 2.1 $\pm$ 0.1 & 4.5 & 0.47$_{-0.02}^{+0.02}$ & 2.50$^{+0.36}_{-0.36}$ & DR1 & $52.25_{-0.02}^{+0.02}$  & 117.6$_{-5.5}^{+5.6}$  \\

\hline
\end{tabular}
\label{table:parallax}
\end{center}
\end{table*}

We report timing parallaxes for 20 of the systems contained in this data release. For PSRs~ J1640$+$2224, J1751$-$2857, J1801$-$1417, J1804$-$2717, J1911$+$1347, and J1918$-$0642 the parallax was not measured in EPTA DR1, while for PSRs~J0900$-$3144, J1738$+$0333, J1843$-$1113, J1910$+$1256, J2322$+$2057, we still do not detect a parallax signature. In what follows we briefly compare our new results with EPTA DR1 and the estimates obtained by NANOGrav and VLBI \citep{vdk+18, dgb+19, aab+21a,dds+23}. 

For sources monitored by both EPTA and NANOGrav, we find that the parallax estimates are in agreement, with the EPTA generally being more precise, due to the improved sampling resulting from the larger number of observations. Unlike NG12, we do not find a significant parallax signature for PSR~J2322$+$2057. 

To obtain distance estimates we invert the parallaxes using Lutz-Kelker-like volume and luminosity priors \citep{lk73,Binney}, as described in \cite{vlm10,vwc+12,ivc16}\footnote{The code to correct for the Lutz-Kelker bias is available here: \url{http://psrpop.phys.wvu.edu/LKbias/}}. More specifically we use a volumetric prior given by, 

\begin{equation}
P_{\rm D}(\varpi) \propto \frac{R^{1.9}}{\varpi^{4}} \exp\left({-\frac{|\sin{b}|}{\varpi E}-5\frac{R-R_{0}}{R_{0}}}\right),
\end{equation}
where $E$ is the scale height, assumed to be \SI{500}{\parsec} for MSPs, $b$ is the Galactic latitude of the source, $R_{0}$ is the distance of the sun from the galactic centre (\SI{8.5}{\kilo\parsec}), and $R$ is the distance between the source and the galactic centre as, 
\begin{equation}
R=\sqrt{ R_{0}^{2}+\frac{\cos{b}}{\varpi}-2R_{0}\frac{\cos{b}\cos{l}}{\varpi}},
\end{equation}
where $l$ is the Galactic longitude of the source.
For the luminosity prior we adopt,
\begin{equation}
P_{\rm L}(\varpi) \propto \frac{1}{\varpi} \exp\left(-0.5\left[\frac{\log{S_{1400}}+1.1-2\log{\varpi}}{0.9}\right]^{2}\right),
\end{equation}
where $S_{1400}$ is the mean flux density of the pulsar at 1400 MHz. 
The use of these priors ensures that the posterior distribution function of the distance is well behaved, so that statistical moments can be defined even in the presence of large uncertainties.

A summary of the distance estimates obtained with the method described above is given in
Table~\ref{table:parallax}. Four MSPs in our samples, PSRs~J1640+2224, J1751$-$2857, J1801$-$1417 and J1804$-$2717,  have a poorly constrained parallax and thus the corresponding distance estimates depend sensitively on the prior assumptions and are highly asymmetric. Transverse velocities, calculated using the bias-corrected distances, and the measured proper motions, are also listed in Table~\ref{table:parallax}. The proper motions derived here are in most cases consistent with VLBI estimates, but have approximately three times smaller uncertainties \cite{dgb+19}. For PSR~J1022+1001, where numerous authors have reported inconsistent parallax determinations from pulsar timing and VLBI campaigns, we find a strong correlation between the parallax and the adopted model for the solar wind electron density. This will be explored in detail by a future work (Liu et al. in prep.).

\subsection{Pulsar mass measurements} \label{ssec:mass}
\begin{figure}[!h]
\centering
\includegraphics[trim={1.8cm 3cm 1cm 4cm},clip,width=0.5\textwidth]{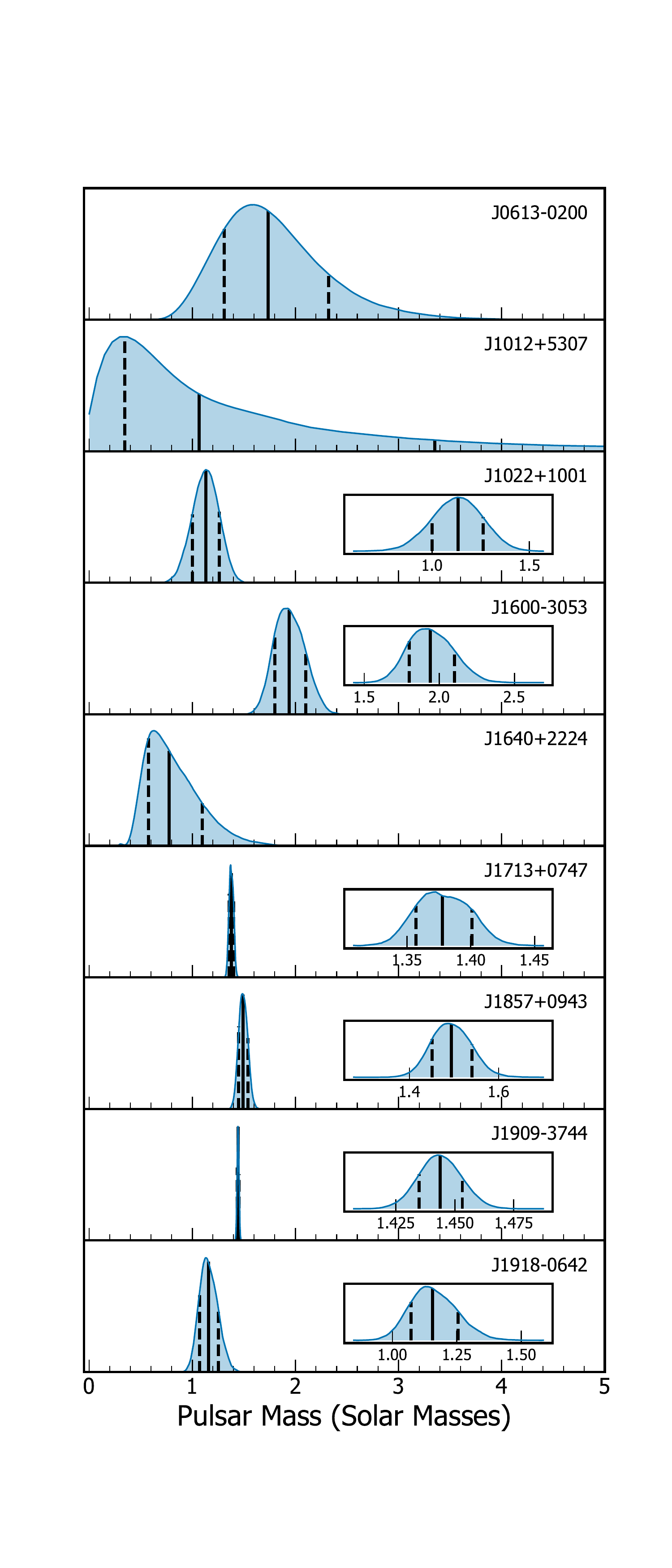}
  \caption{Posterior probability distributions for $m_{\rm p}$, for the 9 systems with significant SD measurements. Solid and dashed lines represent the median and 68.3\% C.I. of the marginalised posterior distribution function. In each panel, the inset provides a zoomed view around the peak of the distribution.}
  \label{fig:pulsar_masses}
\end{figure}

We detect the SD signature with a significance greater than $3\sigma$ in nine systems.  Figure~\ref{fig:pulsar_masses} shows the marginalised posterior probability functions for the masses of the corresponding pulsars. These estimates are based solely on SD posteriors and do not take into account all available information, such as additional pK parameters 
or independent measurements from optical spectroscopy. A more robust analysis of the astrophysical 
parameters for EPTA DR2 pulsars will be presented in a future publication. 

We find that the mass estimates derived from the SD posteriors are generally consistent with the 
measurements obtained with previous iterations of the EPTA data, as well as those derived with independent datasets \citep[e.g. see][]{dcl+16,vlh+16,psb+18,pdd+19,lgi+20,rsc+21,aab+21b}. 
In most cases, the EPTA DR2 constraints are significantly more precise.  

An SD detection for PSRs~1012+5307 and J1022+1001 is reported for the first time. 
For the former system, the SD parameter posteriors are not yet constrained with sufficient 
precision to provide informative constraints on the pulsar mass. If one considers the mass ratio 
of the system, $q \equiv m_{\rm p} / m_{\rm c} = 10.44(11)$ \citep{siv+20}, and discards the SD 
of posterior samples that correspond to physically meaningless values for the companion mass (e.g. 
$m_{\rm c}/{\mathrm M_{\odot}} \lesssim 0.14$ or $\gtrsim 0.3)$, the pulsar mass is constrained to be in the $1.7-2.0$\,M$_{\odot}$ range. As the precision of $h_{3}$ and $h_{4}$ improves with more data, the SD signature will ultimately provide an independent measurement of the component masses. Combined with the parallax, $\dot{P}_{\rm b}$, and WD spectroscopic estimates, this will significantly improve the constraints on gravitational dipole radiation, also providing a valuable test for low-mass WD cooling models. 

For PSR\,J1022+1001, the posterior distribution of the SD parameters suggests that the pulsar mass is relatively low. For the corresponding constraints on the component masses, the expected magnitude for the relativistic periastron advance is $\dot{\omega}_{\rm GR} \simeq $\SI{0.01}{\degree\per\year}, 
which is $\sim$2.5$\sigma$ higher than the measured value for this parameter. This can be explained either by an extremely low pulsar mass ($\sim$1\,M$_{\odot}$), or by systematics and non-relativistic contributions to the measured $\dot{\omega}$.

\subsection{Secular variation of the orbital period}
\label{ssec:pbdot}

\begin{table*}
\caption{Measured orbital period change and the expected contribution from extrinsic effects. Columns \numrange{2}{5} list the estimates for the Shklovskii effect, the contribution due to Galactic acceleration, the intrinsic change in the orbital period, and that due to GW damping. The final column shows the calculated kinematic distance. The superscripts indicate special cases where we assumed or cited measurements from literature: ${a}$ - Mass and parallax taken from literature; ${*}$ - Pulsar mass is assumed to be 1.4 M$_{\odot}$; ${\dag}$ - Parallax value used to determine external orbital period variation effects taken from literature. See \ref{ssec:pbdot} for more details. 
}
\centering
\renewcommand{\arraystretch}{1.4}
\begin{tabular}{ c c c c c c c c  }
\hline
Pulsar name & $\dot{P}_{\rm b,Obs}$ ($10^{-15}$) & $\dot{P}_{\rm b,Shk}$ ($10^{-15}$) & $\dot{P}_{\rm b,Gal}$ ($10^{-15}$) & $\dot{P}_{\rm b,int}$ ($10^{-15}$) &$\dot{P}_{\rm b,GW}$ ($10^{-15}$) & $D_{\rm kin}$ (Kpc) \\

\hline

\hline

J0613$-$0200 & $35^{+2}_{-2}$ & $28^{+1}_{-1}$ & $3.3^{+0.2}_{-0.2}$ & $3.6^{+2.8}_{-2.8}$ & $-3.7^{+1.2}_{-1.8}$ & $1.27^{+0.11}_{-0.10}$ & \\ 

J0751+1807$^{*}$ & $-35^{+0.5}_{-0.5}$ & $11.9^{+0.6}_{-0.5}$ & $0.66^{+0.06}_{-0.07}$ & $-47.6^{+0.7}_{-0.8}$ & $-34.2^\ast$ & --- & \\ 

J1012+5307 & $54.6^{+0.6}_{-0.6}$ & $90.3^{+7.9}_{-7.4}$ & $-4.13^{+0.08}_{-0.06}$ & $-31.6^{+7.4}_{-7.9}$ & $-6^{+5}_{-32}$ & $0.78^{+0.38}_{-0.06}$ & \\ 

J1022+1001 & $218^{+9}_{-9}$ & $753^{+243}_{-186}$ & $-66^{+1}_{-1}$ & $-469^{+187}_{-243}$ & $-0.52^{+0.08}_{-0.08}$ & $0.32^{+0.10}_{-0.08}$ & \\ 

J1455$-$3330 & $4593^{+2202}_{-2235}$ & $805^{+62}_{-59}$ & $-209^{+13}_{-13}$ & $3997^{+2203}_{-2246}$ & --- & $4.6^{+2.1}_{-2.1}$ & \\ 

J1600$-$3053$^{\dag}$ & $365^{+32}_{-31}$ & $287^{+39}_{-33}$ & $52.0^{+4.8}_{-4.9}$ & $25^{+46}_{-49}$ & $-0.11^{+0.01}_{-0.01}$ & $2.1^{+0.2}_{-0.2}$ & \\ 

J1640+2224 & $9501^{+1929}_{-1948}$ & $5556^{+1255}_{-1090}$ & $-2296^{+330}_{-382}$ & $6207^{+2285}_{-2334}$ & $-0.0006^{+0.0002}_{-0.0003}$ & --- & \\ 

J1713+0747 & $264^{+73}_{-72}$ & $642^{+7}_{-7}$ & $-333^{+3}_{-3}$ & $-45^{+73}_{-72}$ & $-0.0067^{+0.0001}_{-0.0002}$ & $1.06^{+0.13}_{-0.13}$ & \\ 

J1738+0333$^{a}$ & $-30^{+7}_{-7}$ & $6.44^{+0.08}_{-0.08}$ & $-0.367^{+0.005}_{-0.005}$ & $-36^{+7}_{-7}$ & $-28^{+1}_{-1}$ & --- & \\ 

J1909$-$3744$^{\dag}$ & $509.2^{+0.9}_{-0.9}$ & $513.3^{+8.8}_{-9.0}$ & $2.58^{+0.07}_{-0.07}$ & $-7^{+9}_{-9}$ & $-2.67^{+0.02}_{-0.02}$ & $1.154^{+0.002}_{-0.002}$ & \\
\hline

\hline
\end{tabular}
\label{table:pbdot}
\end{table*}

Using EPTA DR2, we measured binary orbital period derivatives for a total of ten pulsars. The measurements in PSRs~J1022$+$1001, J1600$-$3053, J1640$+$2224, J1713+0747 (all with $\gtrsim3\sigma$ significance), PSR~J1455$-$3330 (marginal), and J1738+0333 are new compared to those obtained with the EPTA DR1 dataset. Overall, the measured values reported here are consistent with EPTA DR1 (see
Table~\ref{table:pbdot}), although with improved uncertainties.

GW emission causes a change in the orbital period in binary systems \citep{pet64}. The observed $\dot{P}_{\rm b}$ however, may also include extrinsic contributions such as distance- and location-dependent kinematic effects. We determined these extrinsic contributions to the observed $\dot{P}_{\rm b}$ using the bias-corrected distances given in Table~\ref{table:parallax}. For PSRs~J1600$-$3053, J1909$-$3744 and J1738+0333, we used the reference parallax values from \citet{rsc+21}, \citet{lgi+20} and \citet{fwe+12}, respectively (shown in Table~\ref{table:parallax}). We consider two kinematic contributions. The first is the so-called Shklovskii effect \citep{shk70}, which scales with the proper motion of a binary 
\begin{equation}
    \frac{\dot{P}_{\rm b,Shk}}{P_{\rm b}}=\frac{\mu^{2}d}{c},
\end{equation}
where $d$ is the distance, $\mu$ is the proper measured motion, and $c$ is the speed of light, respectively.
The second is caused by differences in the acceleration by the galactic gravitational potential at the solar system and binary pulsar locations \citep{dt91}. To correct for this, we used the Milky Way potential model of \citet{mp17}, which considers the gas disc and the halo density profile, and is calibrated against Galactic maser sources. 

We then calculated the intrinsic orbital period change ($\dot{P}_{\rm b,int}$) by subtracting the kinematic contributions from the observed value,
\begin{equation}
    \dot{P}_{\rm b,Int}=\dot{P}_{\rm b,Obs}-\dot{P}_{\rm b,Shk}-\dot{P}_{\rm b,Gal}.
\end{equation}
From Table~\ref{table:pbdot}, it can be seen that for all pulsars except PSRs~J0751$+$1807, J1012+5307, J1640+2224 and J1738+0333, the derived $\dot{P}_{\rm b,int}$ is consistent with zero within 2$\sigma$.

We also calculated the change in the orbital period by GW damping ($\dot{P}_{\rm b,GW}$) using the mass estimates presented in Section~\ref{ssec:mass}, where available. We note that here we used the measurement of the SD parameter $h_3$ and assumed a 1.4\,M$_{\odot}$ pulsar mass to estimate $\dot{P}_{\rm b,GW}$ for PSR~J0751+1807. The masses of PSR~J1738$+$0333 were directly taken from \cite{avk+12}.  For PSRs~J0751+1807, J1012+5307 and J1738+0333, the derived $\dot{P}_{\rm b,int}$ is consistent with the predicted $\dot{P}_{\rm b,GW}$, suggesting that gravitational damping can account for the non-zero intrinsic secular variation of orbital period.

For PSR J1640+2224, the timing parallax and corresponding distance estimate agree well with those inferred using VLBI measurements \citep{dds+23}. Therefore, the marginal $\sim 2.7\sigma$ detection of $\dot{P}_{\rm b,int}$, which cannot be accounted for by GW damping, is unlikely to be due to an incorrect distance estimate. Although the most probable explanation for the discrepancy is the low-significance measurement of $\dot{P}_{\rm b,Obs}$ ($\sim 4.9\sigma$), another intriguing explanation could be that the system experiences an additional acceleration due to a nearby object. 

For pulsars with a $\dot{P}_{\rm b,int}$ consistent with either zero or anticipated contribution from GW damping, we calculated the kinematic distances $\dot{P}_{\rm b,obs}$, by assuming a net zero $\dot{P}_{\rm b}$ \citep{bb96}. More specifically, we estimate the kinematic distances as
\begin{equation}
D_{\rm kin}=\frac{c \left(\dot{P}_{\rm b,Obs}-\dot{P}_{\rm b,Gal}-\dot{P}_{\rm b,GW}\right)}{\mu^{2} P_{\rm b}}. 
\end{equation}
For PSR~J1455$-$3330, the masses were not measured with our data and the kinetic distance is determined without considering the orbital period derivative GW damping (which should nonetheless be negligible considering its 76-day orbital period). The results are summarised in Table~\ref{table:pbdot}. Comparison between the kinematic distance and the bias-corrected parallax distance indicates that they are, in general, consistent with each other. 

\subsection{Searches for annual orbital parallax signatures}
\label{subsec:annual-orbital-parallax}
For binary pulsars, the proper motion of the system changes the viewing geometry of the orbit with respect to the Earth. This effect induces an apparent secular variation in the projected semi-major axis of the binary orbit  \citep{kop96}\footnote{Here we used the astronomical convention as in \citet{ehm06}, where $\Omega$ and $i$ are different from those in \cite{kop95,kop96} following: $\Omega = \pi/2 - \Omega_{\rm K96}$, $i=\pi-i_{\rm K96}$. This applies to Eq.~\ref{eq:x_PM}--\ref{eq:x_AOP}.}:
\begin{equation} \label{eq:x_PM}
    x=x_{\rm int}\left[1+\cot i(\mu_{\alpha}\cos\Omega-\mu_{\delta}\sin\Omega)(t-t_0)\right],
\end{equation}
which gives
\begin{equation} \label{eq:xdot_PM}
    \frac{\dot{x}}{x_{\rm int}}=\mu\cot i \cos (\theta_{\mu}+\Omega).
\end{equation}
Here $\mu\equiv\sqrt{\mu^2_{\alpha}+\mu^2_{\delta}}$\footnote{We note that here $\mu_{\alpha}$, $\mu_{\delta}$ follows the notation as stated in Table~\ref{tab:ephem1}.}, $\Omega$ is the longitude of the ascending node of the orbit and $\theta_{\mu}$ is the position angle of the proper motion on the sky. It can be seen that if the proper motion and orbital inclination angle are measured, a measurement of $\dot{x}$ can then be used to determine $\Omega$, with an ambiguity of $\cos(-\theta_{\mu}-\Omega)=\cos (\theta_{\mu}+\Omega)$. Typically, the orbital inclination is obtained from the SD parameter $\sin i$, with a $\pi-i$ ambiguity. Therefore, the determination of $\Omega$ can have four possible pairs of solutions for the orbital inclination and ascending node \citep[e.g.][]{lgi+20}. Meanwhile, for binary pulsars, the annual motion of the Earth around the Sun also introduces an apparently periodic variation in the viewing geometry of the orbit, an effect known as the annual orbital parallax. The variation in the projected semi-major axis as a result of this effect can be written as \citep{kop95}:
\begin{equation} \label{eq:x_AOP}
x=x_{\rm int}\left[1-\frac{\cot i}{d}(\Delta_{I_0}\cos\Omega-\Delta_{J_0}\sin\Omega)\right],
\end{equation}
where
\begin{eqnarray}
\Delta_{I_0}&=&-X\sin\alpha+Y\cos\alpha, \\
\Delta_{J_0}&=&-X\sin\delta\cos\alpha-Y\sin\delta\sin\alpha+Z\cos\delta.
\end{eqnarray}
Here, $\Vec{r}=\{X, Y, Z\}$ is the position vector of the Earth with respect to the barycentre of the solar system, and ($\alpha$, $\delta$) are the spherical coordinates of the barycentre of the binary system. For nearby binary pulsars in a 
wide orbit, the annual orbital parallax can be measurable, giving a unique pair of solutions for the inclination and ascending node (e.g. for PSR~J1713+0747). We investigated sources with measured $\dot{x}$ and SD parameters to search for 
this annual orbital parallax signatures. These sources include PSRs~J1600$-$3053, J1857+0943\footnote{Although there is still no measured $\dot{x}$ for J1857+0943, we include it here as \cite{dcl+16} give tentative constraints on $\Omega$.}, 
J1640+2224, J1022+1001, J1012+5307. We used the T2 binary model in \tempotwo{}, where the annual orbital parallax effect is described with the \texttt{KOM} and \texttt{KIN} parameters, corresponding to the longitude of the ascending node 
($\Omega$) and inclination angle ($i$), respectively. The SD parameter $s\equiv\sin i$ is then treated as a function of \texttt{KIN}. We mapped the \texttt{KOM}-\texttt{KIN} space with the \tnest{} toolkit, following a scheme similar to 
described in \citet{dcl+16}. In detail, we fixed the set of white noise parameters to their maximum likelihood values and chose to analytically marginalise over the spin and astrometric parameters. We set customised uniform linear priors for \texttt{KOM}, \texttt{KIN} and $M_{\rm c}$. For \texttt{KIN} and $M_{\rm c}$ the prior ranges are set to include all possible values of these parameters that are allowed from the SD measurements (see tables in Appendix~\ref{app:eph}). For \texttt{KOM} the prior range was set from \SIrange{0}{360}{\degree}. The sampling was conducted with the constant efficiency option turned off, in order to map the multimodal parameter space more effectively.

For PSR~J1600$-$3053, four possible solutions were found in the mapping of \texttt{KIN} and \texttt{KOM} using DR1 dataset \citep{dcl+16}, which was consistent with a negligible annual orbital parallax parallax. With the EPTA DR2 data, we are now able to single out one solution for these two parameters: $\Omega=82^{+12}_{-8}$\,deg, $i=110.4^{+0.9}_{-0.7}$\,deg. Based on the number of samples ($\sim337$k) in the posterior distribution, the logarithmic likelihood ratio of this solution to the others is $\gtrsim 5.5$. This means that the signature of the annual orbital parallax is clearly detected in this binary pulsar system. 

\begin{figure*}[!htbp]
\centering
  \begin{subfigure}[b]{0.45\textwidth}
         \centering
         \includegraphics[trim={0 0 0 1cm},clip,scale=0.5]{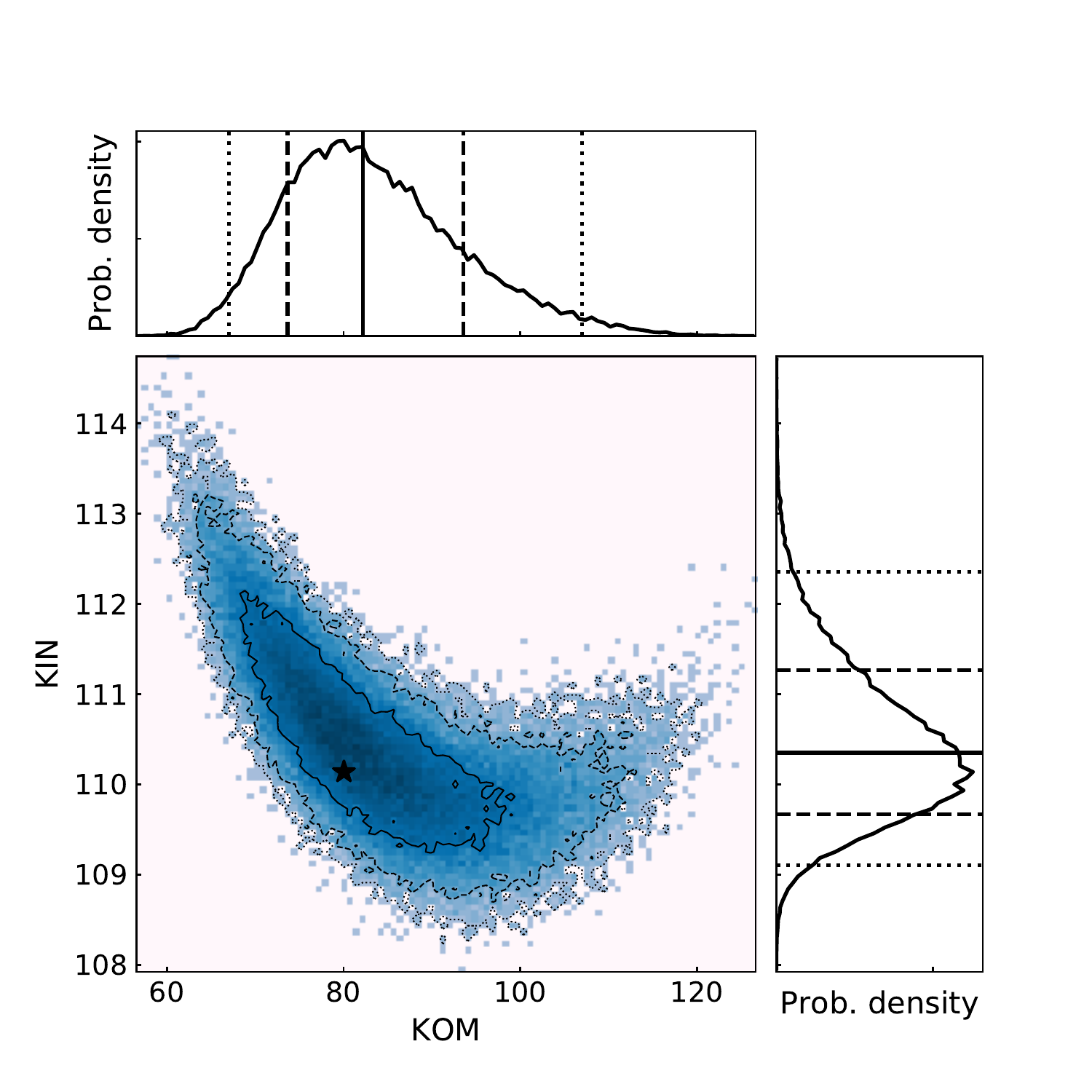}
         \caption{PSR~J1600$-$3053}
         \label{fig:j1600-3053}
  \end{subfigure}
  \begin{subfigure}[b]{0.45\textwidth}
         \centering
         \includegraphics[trim={0 0 0 1cm},clip,scale=0.5]{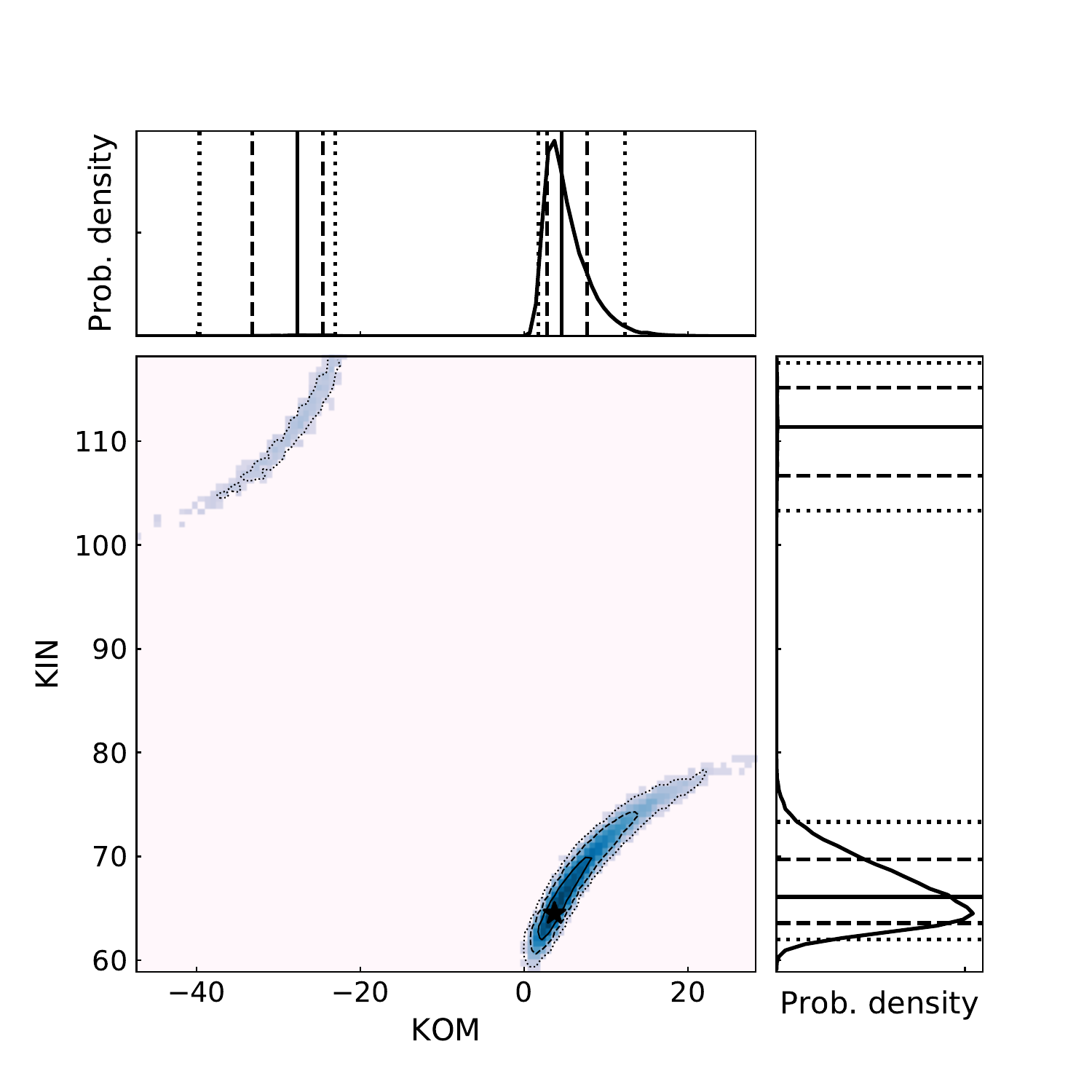}
         \caption{PSR~J1640$+$2224}
         \label{fig:j1640+2224}
  \end{subfigure}
  \begin{subfigure}[b]{0.45\textwidth}
         \centering
         \includegraphics[trim={0 0 0 1cm},clip,scale=0.5]{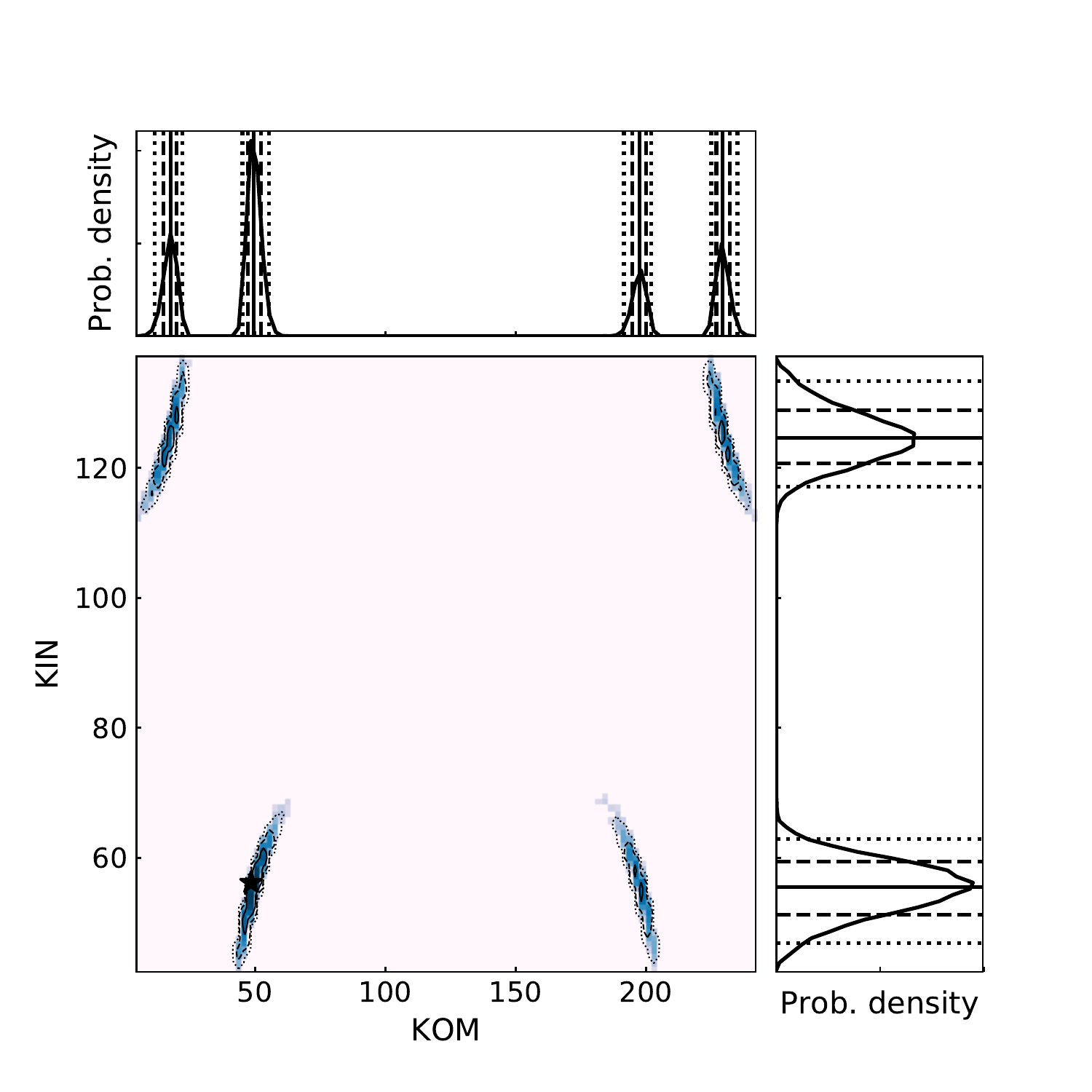}
         \caption{PSR~J1022$+$1001}
         \label{fig:j1022-1001}
  \end{subfigure}
  \begin{subfigure}[b]{0.45\textwidth}
         \centering
         \includegraphics[trim={0 0 0 1cm},clip,scale=0.5]{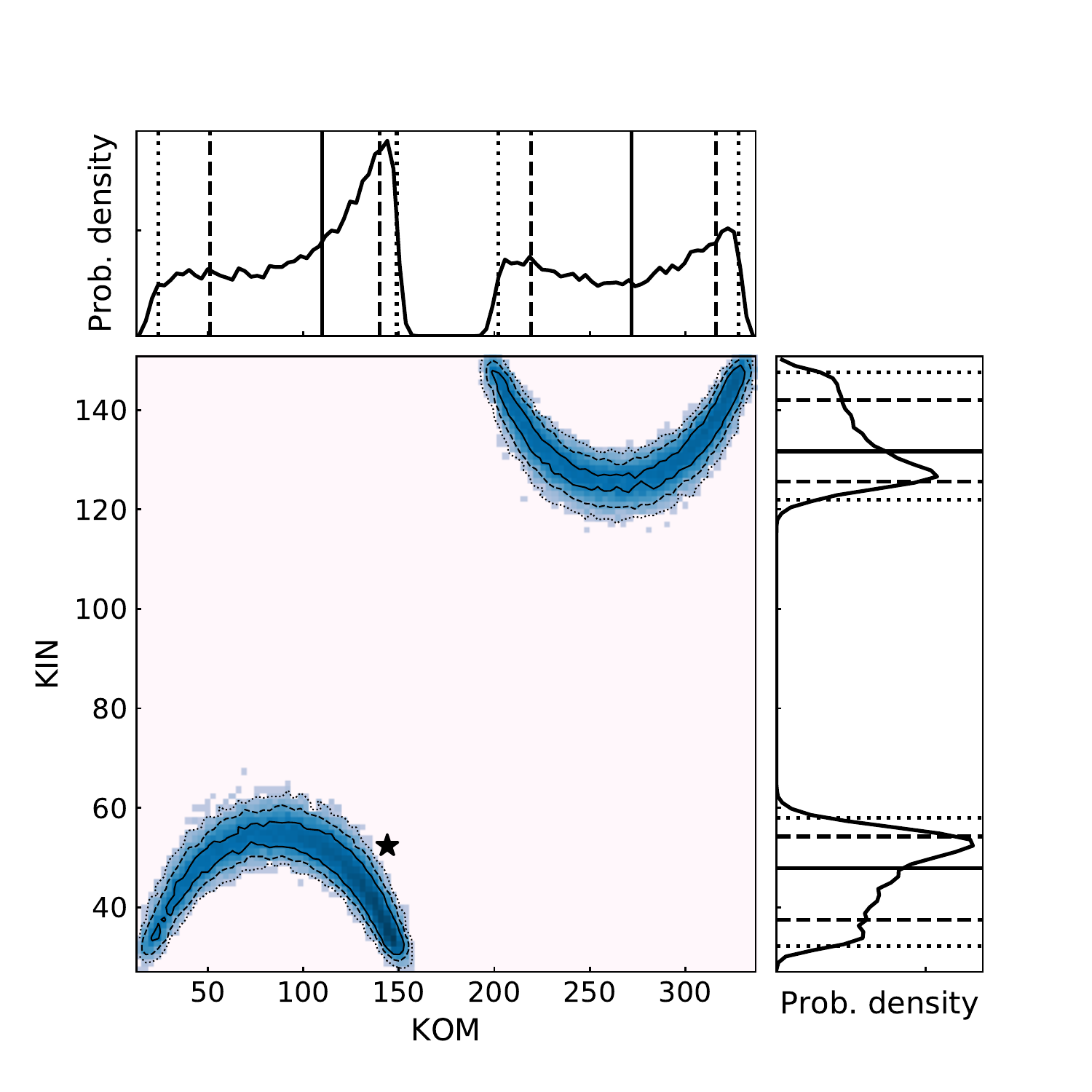}
         \caption{PSR~J1012$+$5307}
         \label{fig:j1012+5307}
  \end{subfigure}
  \begin{subfigure}[b]{0.45\textwidth}
         \centering
         \includegraphics[trim={0 0 0 1cm},clip,scale=0.5]{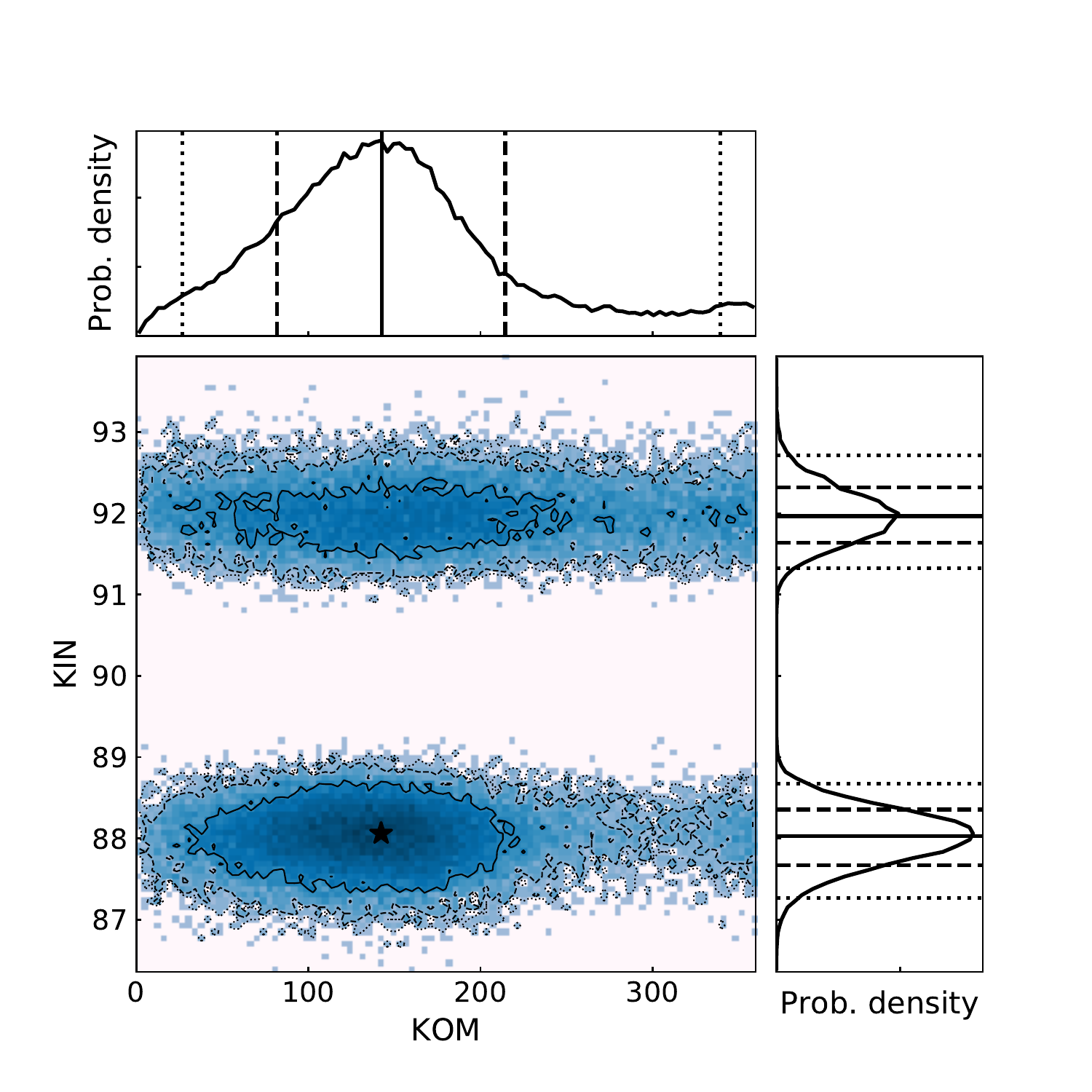}
         \caption{PSR~J1857$+$0943}
         \label{fig:j1857+0943}
  \end{subfigure}
  \caption{Posterior distributions of \texttt{KOM} and \texttt{KIN} from the search for annual-orbital parallax effect, for PSRs~J1600$-$3053 (top left), J1640+2224 (top right), J1022+1001 (middle left), J1012+5307 (middle right),  and J1857+0943 (bottom). The solid, dashed and dotted lines in the side panels stand for the median, 1$\sigma$ confidence interval and 2$\sigma$ confident interval of the distribution. The stars in the plots remark the maximum likelihood values of the two parameters each obtained from the posterior distribution in the side panel. 
  \label{fig:KK-mapping}}
\end{figure*}

For PSR~J1640+2224, two possible solutions for \texttt{KOM} and \texttt{KIN} have been found;
$\Omega=5^{+3}_{-2}$\,deg, $i=111^{+4}_{-5}$\,deg, and $\Omega=-332^{+3}_{-5}$\,deg, $i=66^{+4}_{-3}$\,deg, respectively. The logarithmic likelihood ratio of these two islands is approximately 2.1, which favours the former solution. Thus, there is weak evidence for the detection of the annual orbital parallax signal. 

In PSR~J1012+5307, from the \texttt{KOM}-\texttt{KIN} mapping, the longitude of the ascending node and the orbital inclination are restricted to two regions. The 2$\sigma$ range of $\Omega$ in these two regions, are 47--142 and 218--318\,deg, respectively. Although the most likely value of $\Omega$ is around 150\,deg, the preference for this area of the solution is unclear. 

For PSR~J1022+1001, four solutions in \texttt{KOM} and \texttt{KOM} have been found. These are 1). $\Omega=17^{+2}_{-3}$\,deg, $i=56^{+4}_{-4}$\,deg, 2). $\Omega=50^{+3}_{-2}$\,deg, $i=56^{+4}_{-4}$\,deg, 3). $\Omega=197^{+2}_{-3}$\,deg, $i=125^{+4}_{-4}$\,deg, 4). $\Omega=229^{+3}_{-2}$\,deg, $i=125^{+4}_{-4}$\,deg. Although two of them are slightly preferred in probability, they cannot be distinguished with high confidence, which means that the signal of annual orbital parallax is not detected.

\cite{dcl+16} did not report measurable secular variation of the projected semi-major axis in J1857+0943, and with the new data set we also do not measure significant $\dot{x}$. Still, as in \cite{dcl+16}, we nonetheless attempted to search for an annual orbital parallax to see if the constraint on the longtidue of the ascending node can be improved. In \cite{dcl+16}, there was only a tentative constraint on the ascending node. With the new dataset, there is now a tentative preference for $\Omega$ in the range of 100 to 200\,deg. For the orbital inclination, the posterior space around 88\,deg is slightly preferred, compared to that peaked around 92\,deg. However, there is no clear evidence for neither of the two solutions. 

The analysis of PSR~J1713+0747 yields $\Omega=91.1\pm0.5$\,deg and $i=71.3\pm0.2$\,deg as discussed in Seciont~\ref{sec:result}. These are consistent with previous analyses \citep{zdw+19,aab+21a,rsc+21}, as well as with the results based on EPTA DR1 \citep{dcl+16}.

\subsection{High order spin frequency derivatives in J1024$-$0719}
\label{subsec:high_order}
There is a significant measurement of the second spin frequency derivative in PSR~J1024$-$0719 \citep{bjs+16,kkn+16}. Based on these results, it is anticipated that this pulsar is in orbit with a main-sequence star named 2MASS J10243869$-$0719190, of an orbital period longer than 200\,yr \citep{bjs+16}. Here with the new EPTA dataset, in addition to our timing analysis presented in Table~\ref{tab:ephem2}, we conducted another round of analysis including modelling of the third, fourth and fifth spin frequency derivatives. These derivatives, if measurable, can be used to derive the properties of the Keplerian orbit of the pulsar \citep{bjs+16}. Table~\ref{tab:J1024} shows the results of this analysis, and a comparison with previous work. As can be seen, there is a tentative change in the measured second spin frequency derivative compared to previous results (even when the reference epoch of the spin frequency and the number of modelled spin frequency derivatives are both the same). This is expected given that the EPTA DR2 dataset covers a slightly longer timeline / orbit phase of the binary system. It is also intriguing to see that with the new and more sensitive data, there is now a tentative detection of the fifth spin frequency derivative. This though needs to be confirmed by future observations. 

\begin{table*}[]
    \centering
    \begin{tabular}{ccccc}
    \hline
    \hline
         & \cite{bjs+16}  & \cite{kkn+16}  &EPTA DR2, F0-2 &EPTA DR2, F0-5 \\
         \hline
Epoch of spin frequency (MJD) & 55000 & 56236 & 55000 & 55000 \\
Spin frequency, $\nu$ (s$^{-1}$) &193.7156834485468(7)  &193.7156863778085(8) & 193.715683448548(2) & 193.715683448549(3) \\
First derivative of $\nu$ (s$^{-2}$) &$-6.95893(15)\times10^{-16}$ & $-6.9638(4)\times10^{-16}$  & $-6.9593(2)\times10^{-16}$ & $-6.9598(2)\times10^{-16}$ \\
Second derivative of $\nu$ (s$^{-3}$) &$-3.92(2)\times10^{-27}$ & $-4.1(10)\times10^{-27}$  &  $-3.57(7)\times10^{-27}$  & $-2.7(4)\times10^{-27}$ \\
Third derivative of $\nu$ (s$^{-4}$) &$<2.7\times10^{-36}$  & 1.1(7)$\times10^{-34}$ & --- & $-3(3)\times10^{-36}$ \\
Forth derivative of $\nu$ (s$^{-5}$) &$<4.5\times10^{-44}$  & ---    & --- & $-1.7(7)\times10^{-43}$ \\
Fifth derivative of $\nu$ (s$^{-6}$) & ---   & ---   & --- & 2.2(9)$\times10^{-51}$ \\
\hline
    \end{tabular}
    \caption{Spin frequency and its derivatives for PSR~J1024$-$0719, measured from previous work and analysis in this paper. }
    \label{tab:J1024}
\end{table*}
\section{Conclusions} \label{sec:conclusions}
In this paper, we presented the EPTA DR2 dataset and results from a
combined frequentist and Bayesian timing analyses. This dataset
contains high-precision timing data for 25 MSPs, collected with the
five largest radio telescopes in Europe along with LEAP. The DR2
dataset combines data from EPTA DR1 \citep{dcl+16}
with those recorded by a new generation of data acquisition
systems. The dataset extends to the beginning of 2021 and has baselines ranging from 14 to 25 years. 

We conducted timing analysis of the dataset
based on a Bayesian framework to measure the timing parameters of the
pulsars. This has yielded a collection of new measurements including
annual parallaxes, secular variation of orbital period, Shapiro delay
and so forth in some pulsars. We also derived a group of astrophysical
parameters of these pulsars, including distances, transverse
velocities, binary masses, and annual orbital parallaxes. The
DR2 dataset builds the foundation for searching for GW 
signals. The results of this search are reported in an accompanying publication \citep{wm3}.


\begin{acknowledgements}
The European Pulsar Timing Array (EPTA) is a collaboration between
European and partner institutes, namely ASTRON (NL), INAF/Osservatorio
di Cagliari (IT), Max-Planck-Institut f\"{u}r Radioastronomie (GER),
Nan\c{c}ay/Paris Observatory (FRA), the University of Manchester (UK),
the University of Birmingham (UK), the University of East Anglia (UK),
the University of Bielefeld (GER), the University of Paris (FRA), the
University of Milan-Bicocca (IT), the Foundation for Research and 
Technology, Hellas (GR), and Peking University (CHN), with the
aim to provide high-precision pulsar timing to work towards the direct
detection of low-frequency gravitational waves. An Advanced Grant of
the European Research Council allowed to implement the Large European Array
for Pulsars (LEAP) under Grant Agreement Number 227947 (PI M. Kramer). 
The EPTA is part of the
International Pulsar Timing Array (IPTA); we thank our
IPTA colleagues for their support and help with this paper and the external Detection Committee members for their work on the Detection Checklist.

Part of this work is based on observations with the 100-m telescope of
the Max-Planck-Institut f\"{u}r Radioastronomie (MPIfR) at Effelsberg
in Germany. Pulsar research at the Jodrell Bank Centre for
Astrophysics and the observations using the Lovell Telescope are
supported by a Consolidated Grant (ST/T000414/1) from the UK's Science
and Technology Facilities Council (STFC). ICN is also supported by the
STFC doctoral training grant ST/T506291/1. The Nan{\c c}ay radio
Observatory is operated by the Paris Observatory, associated with the
French Centre National de la Recherche Scientifique (CNRS), and
partially supported by the Region Centre in France. We acknowledge
financial support from ``Programme National de Cosmologie and
Galaxies'' (PNCG), and ``Programme National Hautes Energies'' (PNHE)
funded by CNRS/INSU-IN2P3-INP, CEA and CNES, France. We acknowledge
financial support from Agence Nationale de la Recherche
(ANR-18-CE31-0015), France. The Westerbork Synthesis Radio Telescope
is operated by the Netherlands Institute for Radio Astronomy (ASTRON)
with support from the Netherlands Foundation for Scientific Research
(NWO). The Sardinia Radio Telescope (SRT) is funded by the Department
of University and Research (MIUR), the Italian Space Agency (ASI), and
the Autonomous Region of Sardinia (RAS) and is operated as a National
Facility by the National Institute for Astrophysics (INAF).

The work is supported by the National SKA programme of China
(2020SKA0120100), Max-Planck Partner Group, NSFC 11690024, CAS
Cultivation Project for FAST Scientific. This work is also supported
as part of the ``LEGACY'' MPG-CAS collaboration on low-frequency
gravitational wave astronomy. JA acknowledges support from the
European Commission (Grant Agreement number: 101094354). JA and SCha 
were partially supported by the Stavros
Niarchos Foundation (SNF) and the Hellenic Foundation for Research and
Innovation (H.F.R.I.) under the 2nd Call of the ``Science and Society --
Action Always strive for excellence -- Theodoros Papazoglou''
(Project Number: 01431). AC acknowledges support from the Paris
\^{I}le-de-France Region. AC, AF, ASe, ASa, EB, DI, GMS, MBo acknowledge
financial support provided under the European Union's H2020 ERC
Consolidator Grant ``Binary Massive Black Hole Astrophysics'' (B
Massive, Grant Agreement: 818691). GD, KLi, RK and MK acknowledge support
from European Research Council (ERC) Synergy Grant ``BlackHoleCam'', 
Grant Agreement Number 610058. This work is supported by the ERC 
Advanced Grant ``LEAP'', Grant Agreement Number 227947 (PI M. Kramer). 
AV and PRB are supported by the UK's Science
and Technology Facilities Council (STFC; grant ST/W000946/1). AV also acknowledges
the support of the Royal Society and Wolfson Foundation. JPWV acknowledges
support by the Deutsche Forschungsgemeinschaft (DFG) through thew
Heisenberg programme (Project No. 433075039) and by the NSF through
AccelNet award \#2114721. NKP is funded by the Deutsche
Forschungsgemeinschaft (DFG, German Research Foundation) --
Projektnummer PO 2758/1--1, through the Walter--Benjamin
programme. ASa thanks the Alexander von Humboldt foundation in
Germany for a Humboldt fellowship for postdoctoral researchers. APo, DP
and MBu acknowledge support from the research grant “iPeska”
(P.I. Andrea Possenti) funded under the INAF national call
Prin-SKA/CTA approved with the Presidential Decree 70/2016
(Italy). RNC acknowledges financial support from the Special Account
for Research Funds of the Hellenic Open University (ELKE-HOU) under
the research programme ``GRAVPUL'' (grant agreement 319/10-10-2022).
EvdW, CGB and GHJ acknowledge support from the Dutch National Science
Agenda, NWA Startimpuls – 400.17.608.
BG is supported by the Italian Ministry of Education, University and 
Research within the PRIN 2017 Research Program Framework, n. 2017SYRTCN. LS acknowledges the use of the HPC system Cobra at the Max Planck Computing and Data Facility.

\ifnum\wm>1 The Indian Pulsar Timing Array (InPTA) is an Indo-Japanese
collaboration that routinely employs TIFR's upgraded Giant Metrewave
Radio Telescope for monitoring a set of IPTA pulsars.  BCJ, YG, YM,
SD, AG and PR acknowledge the support of the Department of Atomic
Energy, Government of India, under Project Identification \# RTI 4002.
BCJ, YG and YM acknowledge support of the Department of Atomic Energy,
Government of India, under project No. 12-R\&D-TFR-5.02-0700 while SD,
AG and PR acknowledge support of the Department of Atomic Energy,
Government of India, under project no. 12-R\&D-TFR-5.02-0200.  KT is
partially supported by JSPS KAKENHI Grant Numbers 20H00180, 21H01130,
and 21H04467, Bilateral Joint Research Projects of JSPS, and the ISM
Cooperative Research Program (2021-ISMCRP-2017). AS is supported by
the NANOGrav NSF Physics Frontiers Center (awards \#1430284 and
2020265).  AKP is supported by CSIR fellowship Grant number
09/0079(15784)/2022-EMR-I.  SH is supported by JSPS KAKENHI Grant
Number 20J20509.  KN is supported by the Birla Institute of Technology
\& Science Institute fellowship.  AmS is supported by CSIR fellowship
Grant number 09/1001(12656)/2021-EMR-I and T-641 (DST-ICPS).  TK is
partially supported by the JSPS Overseas Challenge Program for Young
Researchers.  We acknowledge the National Supercomputing Mission (NSM)
for providing computing resources of ‘PARAM Ganga’ at the Indian
Institute of Technology Roorkee as well as `PARAM Seva' at IIT
Hyderabad, which is implemented by C-DAC and supported by the Ministry
of Electronics and Information Technology (MeitY) and Department of
Science and Technology (DST), Government of India. DD acknowledges the 
support from the Department of Atomic Energy, Government of India 
through Apex Project - Advance Research and Education in Mathematical 
Sciences at IMSc. \fi

The work presented here is a culmination of many years of data
analysis as well as software and instrument development. In particular,
we thank Drs. N.~D'Amico, P.~C.~C.~Freire, R.~van Haasteren, 
C.~Jordan, K.~Lazaridis, P.~Lazarus, L.~Lentati, O.~L\"{o}hmer and 
R.~Smits for their past contributions. We also
thank Dr. N. Wex for supporting the calculations of the
galactic acceleration as well as the related discussions.
The EPTA is also grateful
to staff at its observatories and telescopes who have made the
continued observations possible. 
\linebreak\linebreak\textit{Author contributions.}
The EPTA is a multi-decade effort and all authors have
contributed through conceptualisation, funding acquisition,
data-curation, methodology, software and hardware
 developments as well as (aspects of) the continued running of
the observational campaigns, which includes writing and
proofreading observing proposals, evaluating observations
and observing systems, mentoring students, developing
science cases. All authors also helped in (aspects of)
verification of the data, analysis and results as well as
in finalising the paper draft. Specific contributions from individual 
EPTA members are listed in the CRediT\footnote{\url{https://credit.niso.org/}} format below.

InPTA members contributed in uGMRT observations and data reduction to
create the InPTA data set which is employed while assembling the
\texttt{DR2full+} and \texttt{DR2new+} data sets. 

\ifnum\wm=1

JJan, KLi, GMS equally share the correspondence of the paper.

\linebreak\linebreak\textit{CRediT statement:}\newline
Conceptualisation: APa, APo, AV, BWS, CGB, CT, GHJ, GMS, GT, IC, JA, JJan, JPWV, JW, JWM, KJL, KLi, MK.\\
Methodology: APa, AV, DJC, GMS, IC, JA, JJan, JPWV, JWM, KJL, KLi, LG, MK.\\
Software: AC, AJ, APa, CGB, DJC, GMS, IC, JA, JJan, JJaw, JPWV, KJL, KLi, LG, MJK, RK.\\
Validation: AC, APa, CGB, CT, GMS, GT, IC, JA, JJan, JPWV, JWM, KLi, LG.\\
Formal Analysis: APa, CGB, DJC, DP, EvdW, GHJ, GMS, JA, JJan, JPWV, JWM, KLi.\\
Investigation: APa, APo, BWS, CGB, DJC, DP, GMS, GT, IC, JA, JJan, JPWV, JWM, KLi, LG, MBM, MBu, MJK, RK.\\
Resources: APa, APe, APo, BWS, GHJ, GMS, GT, HH, IC, JA, JJan, JPWV, JWM, KJL, KLi, LG, MJK, MK, RK.\\
Data Curation: AC, AJ, APa, BWS, CGB, DJC, DP, EG, EvdW, GHJ, GMS, GT, HH, IC, JA, JJan, JPWV, JWM, KLi, LG, MBM, MBu, MJK, MK, NKP, RK, SChe, YJG.\\
Writing – Original Draft: APa, GMS, JA, JJan, KLi, LG.\\
Writing – Review \& Editing: AC, AF, APa, APo, DJC, EB, EFK, GHJ, GMS, GT, JA, JJan, JPWV, JWM, KLi, MK, SChe, VVK.\\
Visualisation: APa, GMS, JA, JJan, KLi.\\
Supervision: APo, ASe, AV, BWS, CGB, DJC, EFK, GHJ, GMS, GT, IC, JA, JPWV, KJL, KLi, LG, MJK, MK, VVK.\\
Project Administration: APo, ASe, AV, BWS, CGB, CT, GHJ, GMS, GT, IC, JJan, JPWV, JWM, KLi, LG, MK.\\
Funding Acquisition: APe, APo, ASe, BWS, GHJ, GT, IC, JA, JJan, LG, MJK, MK.\\

\fi

\ifnum\wm=2
InPTA members contributed to the discussions that probed the impact of 
including InPTA data on single pulsar noise analysis. Furthermore, they 
provided quantitative comparisons of various noise models, wrote a brief 
description of the underlying \texttt{Tensiometer} package, and helped 
with the related interpretations.

APa, AC, MJK equally share the correspondence of the paper. 

\linebreak\linebreak\textit{CRediT statement:}\newline
Conceptualisation: AC, APa, APo, AV, BWS, CT, GMS, GT, JPWV, JWM, KJL, KLi, MJK, MK.\\
Methodology: AC, APa, AV, DJC, GMS, IC, JWM, KJL, KLi, LG, MJK, MK, SB, SChe, VVK.\\
Software: AC, AJ, APa, APe, GD, GMS, KJL, KLi, MJK, RK, SChe, VVK.\\
Validation: AC, APa, BG, GMS, IC, JPWV, JWM, KLi, LG, MJK.\\
Formal Analysis: AC, APa, BG, EvdW, GHJ, GMS, JWM, KLi, MJK.\\
Investigation: AC, APa, APo, BWS, CGB, DJC, DP, GMS, IC, JPWV, JWM, KLi, LG, MBM, MBu, MJK, RK, VVK.\\
Resources: AC, APa, APe, APo, BWS, GHJ, GMS, GT, IC, JPWV, JWM, KJL, KLi, LG, MJK, MK, RK.\\
Data Curation: AC, AJ, APa, BWS, CGB, DJC, DP, EvdW, GHJ, GMS, JA, JWM, KLi, MBM, MJK, MK, NKP, RK, SChe.\\
Writing – Original Draft: AC, APa, GMS, MJK.\\
Writing – Review \& Editing: AC, AF, APa, APo, BG, EB, EFK, GMS, GT, JA, JPWV, JWM, KLi, MJK, MK, SChe, VVK.\\
Visualisation: AC, APa, GMS, KLi, MJK.\\
Supervision: AC, APo, ASe, AV, BWS, CGB, DJC, EFK, GHJ, GT, JPWV, KJL, LG, MJK, MK, VVK.\\
Project Administration: AC, APo, ASe, AV, BWS, CGB, CT, GHJ, GMS, GT, JPWV, JWM, LG, MJK, MK.\\
Funding Acquisition: APe, APo, ASe, BWS, GHJ, GT, IC, LG, MJK, MK.\\
\fi

\ifnum\wm=3

Additionally, InPTA members contributed to GWB search efforts with 
\texttt{DR2full+} and \texttt{DR2new+} data sets and their interpretations. 
Further, they provided quantitative comparisons of GWB posteriors that 
arise from these data sets and multiple pipelines.

For this work specifically, SChen and YJG equally share the 
correspondence of the paper. 

\linebreak\linebreak\textit{CRediT statement:}\newline
Conceptualisation: AC, APa, APe, APo, ASe, AV, BG, CT, GMS, GT, IC, JA, JPWV, JWM, KJL, KLi, MK.\\
Methodology: AC, APa, ASe, AV, DJC, GMS, JWM, KJL, KLi, LS, MK, SChe.\\
Software: AC, AJ, APa, APe, GD, GMS, KJL, KLi, MJK, RK, SChe, VVK.\\
Validation: AC, APa, ASe, AV, BG, GMS, HQL, JPWV, JWM, KLi, LS, SChe, YJG.\\
Formal Analysis: AC, APa, ASe, AV, BG, EvdW, GMS, HQL, JWM, KLi, LS, MF, NKP, PRB, SChe, YJG.\\
Investigation: APa, APo, ASe, AV, BWS, CGB, DJC, DP, GMS, JWM, KLi, LS, MBM, MBu, MF, PRB, RK, SB, SChe, YJG.\\
Resources: AC, APa, APe, APo, ASe, AV, BWS, GHJ, GMS, GT, IC, JPWV, JWM, KJL, KLi, LG, LS, MJK, MK, RK.\\
Data Curation: AC, AJ, APa, BWS, CGB, DJC, DP, EvdW, GMS, JA, JWM, KLi, MBM, MJK, MK, RK, SChe.\\
Writing – Original Draft: AC, APa, BG, DJC, GMS, JA, KLi, SB, SChe, YJG.\\
Writing – Review \& Editing: AC, AF, APa, APo, ASe, AV, BG, DJC, EB, EFK, GMS, GT, JA, JPWV, JWM, KLi, LS, MBo, MK, NKP, PRB, SChe, VVK, YJG.\\
Visualisation: APa, BG, GMS, KLi, MF, PRB, SChe.\\
Supervision: APo, ASe, AV, BWS, CGB, DJC, EFK, GHJ, GMS, GT, JPWV, KJL, MK, SB.\\
Project Administration: APo, ASe, AV, BWS, CGB, CT, GHJ, GMS, GT, JPWV, JWM, LG, MK, SChe.\\
Funding Acquisition: APe, APo, ASe, AV, BWS, GHJ, GT, IC, JA, LG, MJK, MK, SB.\\
\fi

\newline
\textit{Data Availability}
Files containing the TOAs and pulsar timing models can be downloaded from \url{https://epta.pages.in2p3.fr/epta-dr2/}. The same site also provides the software environments necessary to recreate the timing analysis presented in this article. 
\end{acknowledgements}

\bibliographystyle{bibtex/aa} 
\bibliography{bibfiles/journals,bibfiles/eptadr2,bibfiles/crossrefs} 

@proceedings{cg04,
  editor = {F. Camilo and B. M. Gaensler},
  title = {Young Neutron Stars and Their Environments, {IAU} Symposium 218},
  booktitle = {Young Neutron Stars and Their Environments, {IAU} Symposium 218},
  year = {2004},
  publisher = {Astronomical Society of the Pacific},
  address = {San Francisco}
}



@ARTICLE{Guillemot2023,
       author = {{Guillemot}, L. and {Cognard}, I. and {van Straten}, W. and {Theureau}, G. and {G\’erard}, E.}, 
        title = "{Improving pulsar polarization and timing measurements with the Nan\c{c}ay Radio Telescope}",
      journal = {submitted to A\&A},
     keywords = {},
         year = 2023,
        month = {},
       volume = {},
       number = {},
        pages = {},
          doi = {},
archivePrefix = {},
       eprint = {},
 primaryClass = {},
       adsurl = {},
      adsnote = {}
}

@ARTICLE{mhb+13,
       author = {{Manchester}, R.~N. and {Hobbs}, G. and {Bailes}, M. and {Coles}, W.~A. and {van Straten}, W. and {Keith}, M.~J. and {Shannon}, R.~M. and {Bhat}, N.~D.~R. and {Brown}, A. and {Burke-Spolaor}, S.~G. and {Champion}, D.~J. and {Chaudhary}, A. and {Edwards}, R.~T. and {Hampson}, G. and {Hotan}, A.~W. and {Jameson}, A. and {Jenet}, F.~A. and {Kesteven}, M.~J. and {Khoo}, J. and {Kocz}, J. and {Maciesiak}, K. and {Oslowski}, S. and {Ravi}, V. and {Reynolds}, J.~R. and {Sarkissian}, J.~M. and {Verbiest}, J.~P.~W. and {Wen}, Z.~L. and {Wilson}, W.~E. and {Yardley}, D. and {Yan}, W.~M. and {You}, X.~P.},
        title = "{The Parkes Pulsar Timing Array Project}",
      journal = {\pasa},
     keywords = {gravitational waves, instrumentation: miscellaneous, methods: observational, pulsars: general, Astrophysics - Instrumentation and Methods for Astrophysics, Astrophysics - High Energy Astrophysical Phenomena},
         year = 2013,
        month = jan,
       volume = {30},
          eid = {e017},
        pages = {e017},
          doi = {10.1017/pasa.2012.017},
archivePrefix = {arXiv},
       eprint = {1210.6130},
 primaryClass = {astro-ph.IM},
       adsurl = {https://ui.adsabs.harvard.edu/abs/2013PASA...30...17M},
      adsnote = {Provided by the SAO/NASA Astrophysics Data System}
}

@ARTICLE{ivc16,
       author = {{Igoshev}, Andrei and {Verbunt}, Frank and {Cator}, Eric},
        title = "{Distance and luminosity probability distributions derived from parallax and flux with their measurement errors. With application to the millisecond pulsar PSR J0218+4232}",
      journal = {\aap},
     keywords = {methods: statistical, stars: luminosity function, mass function, pulsars: general, pulsars: individual: PSR J0218+4232, Astrophysics - High Energy Astrophysical Phenomena},
         year = 2016,
        month = jun,
       volume = {591},
          eid = {A123},
        pages = {A123},
          doi = {10.1051/0004-6361/201527471},
archivePrefix = {arXiv},
       eprint = {1604.08452},
 primaryClass = {astro-ph.HE},
       adsurl = {https://ui.adsabs.harvard.edu/abs/2016A&A...591A.123I},
      adsnote = {Provided by the SAO/NASA Astrophysics Data System}
}



@article{bb96,
  author = {Bell, J. F. and Bailes, M.},
  title = { A New Method for Obtaining Binary Pulsar Distances and its Implications for Tests of General Relativity. },
  journal = {\apj},
  volume = {456},
  pages = {L33-L36},
  year = 1996,
  keyword = {Shklovskii effect, Distances}
}

@ARTICLE{roebber2019,
       author = {{Roebber}, Elinore},
        title = "{Ephemeris Errors and the Gravitational-wave Signal: Harmonic Mode Coupling in Pulsar Timing Array Searches}",
      journal = {\apj},
     keywords = {ephemerides, gravitational waves, methods: data analysis, pulsars: general, time, Astrophysics - High Energy Astrophysical Phenomena, Astrophysics - Cosmology and Nongalactic Astrophysics},
         year = 2019,
        month = may,
       volume = {876},
       number = {1},
          eid = {55},
        pages = {55},
          doi = {10.3847/1538-4357/ab100e},
archivePrefix = {arXiv},
       eprint = {1901.05468},
 primaryClass = {astro-ph.HE},
       adsurl = {https://ui.adsabs.harvard.edu/abs/2019ApJ...876...55R},
      adsnote = {Provided by the SAO/NASA Astrophysics Data System}
}


@article{pet64,
  author = {Peters, P. C.},
  title = {Gravitational Radiation and the Motion of Two Point Masses},
  journal = physrev,
  volume = 136,
  pages = {1224-1232},
  year = 1964,
  keyword = {neutron stars}
}

@article{vbk96,
  author = {van Kerkwijk, M. H. and Bergeron, P. and Kulkarni, S. R.},
  title = {{The masses of the millisecond pulsar J1012+5307 and its  white-dwarf companion}},
  journal = {\apjl},
  year = {1996},
  volume = {467},
  pages = {L89-L92},
  keyword = {binaries: close --- pulsars: individual (PSR J1012+5307) ---   stars: neutron --- white dwarfs}
}

@article{tay92a,
  author = {Taylor, J. H.},
  title = {Pulsar Timing and Relativistic Gravity},
  journal = ptrsa,
  volume = {341},
  pages = {117-134},
  year = {1992}
}



@ARTICLE{lza+23,
       author = {{Liu}, N. and {Zhu}, Z. and {Antoniadis}, J. and {Liu}, J. -C. and {Zhang}, H. and {Jiang}, N.},
        title = "{Comparison of dynamical and kinematic reference frames via pulsar positions from timing, Gaia, and interferometric astrometry}",
      journal = {\aap},
     keywords = {reference systems, astrometry, pulsars: general, techniques: interferometric, ephemerides, Astrophysics - Instrumentation and Methods for Astrophysics},
         year = 2023,
        month = feb,
       volume = {670},
          eid = {A173},
        pages = {A173},
          doi = {10.1051/0004-6361/202243614},
archivePrefix = {arXiv},
       eprint = {2212.07178},
 primaryClass = {astro-ph.IM},
       adsurl = {https://ui.adsabs.harvard.edu/abs/2023A&A...670A.173L},
      adsnote = {Provided by the SAO/NASA Astrophysics Data System}
}



@article{ant21,
  title = {Gaia Pulsars and Where to Find Them},
  author = {Antoniadis, John},
  year = {2021},
  month = jan,
  journal = {Monthly Notices of the Royal Astronomical Society},
  volume = {501},
  number = {1},
  pages = {1116--1126},
  issn = {0035-8711},
  doi = {10.1093/mnras/staa3595}
}




@ARTICLE{rsg15,
       author = {{Rosado}, Pablo A. and {Sesana}, Alberto and {Gair}, Jonathan},
        title = "{Expected properties of the first gravitational wave signal detected with pulsar timing arrays}",
      journal = {\mnras},
     keywords = {black hole physics, gravitation, gravitational waves, methods: data analysis, pulsars: general, galaxies: evolution, Astrophysics - High Energy Astrophysical Phenomena, General Relativity and Quantum Cosmology},
         year = 2015,
        month = aug,
       volume = {451},
       number = {3},
        pages = {2417-2433},
          doi = {10.1093/mnras/stv1098},
archivePrefix = {arXiv},
       eprint = {1503.04803},
 primaryClass = {astro-ph.HE},
       adsurl = {https://ui.adsabs.harvard.edu/abs/2015MNRAS.451.2417R},
      adsnote = {Provided by the SAO/NASA Astrophysics Data System}
}

@ARTICLE{aab+22,
       author = {{Antoniadis}, J. and {Arzoumanian}, Z. and {Babak}, S. and {Bailes}, M. and {Bak Nielsen}, A. -S. and {Baker}, P.~T. and {Bassa}, C.~G. and {B{\'e}csy}, B. and {Berthereau}, A. and {Bonetti}, M. and {Brazier}, A. and {Brook}, P.~R. and {Burgay}, M. and {Burke-Spolaor}, S. and {Caballero}, R.~N. and {Casey-Clyde}, J.~A. and {Chalumeau}, A. and {Champion}, D.~J. and {Charisi}, M. and {Chatterjee}, S. and {Chen}, S. and {Cognard}, I. and {Cordes}, J.~M. and {Cornish}, N.~J. and {Crawford}, F. and {Cromartie}, H.~T. and {Crowter}, K. and {Dai}, S. and {DeCesar}, M.~E. and {Demorest}, P.~B. and {Desvignes}, G. and {Dolch}, T. and {Drachler}, B. and {Falxa}, M. and {Ferrara}, E.~C. and {Fiore}, W. and {Fonseca}, E. and {Gair}, J.~R. and {Garver-Daniels}, N. and {Goncharov}, B. and {Good}, D.~C. and {Graikou}, E. and {Guillemot}, L. and {Guo}, Y.~J. and {Hazboun}, J.~S. and {Hobbs}, G. and {Hu}, H. and {Islo}, K. and {Janssen}, G.~H. and {Jennings}, R.~J. and {Johnson}, A.~D. and {Jones}, M.~L. and {Kaiser}, A.~R. and {Kaplan}, D.~L. and {Karuppusamy}, R. and {Keith}, M.~J. and {Kelley}, L.~Z. and {Kerr}, M. and {Key}, J.~S. and {Kramer}, M. and {Lam}, M.~T. and {Lamb}, W.~G. and {Lazio}, T.~J.~W. and {Lee}, K.~J. and {Lentati}, L. and {Liu}, K. and {Luo}, J. and {Lynch}, R.~S. and {Lyne}, A.~G. and {Madison}, D.~R. and {Main}, R.~A. and {Manchester}, R.~N. and {McEwen}, A. and {McKee}, J.~W. and {McLaughlin}, M.~A. and {Mickaliger}, M.~B. and {Mingarelli}, C.~M.~F. and {Ng}, C. and {Nice}, D.~J. and {Os{\l}owski}, S. and {Parthasarathy}, A. and {Pennucci}, T.~T. and {Perera}, B.~B.~P. and {Perrodin}, D. and {Petiteau}, A. and {Pol}, N.~S. and {Porayko}, N.~K. and {Possenti}, A. and {Ransom}, S.~M. and {Ray}, P.~S. and {Reardon}, D.~J. and {Russell}, C.~J. and {Samajdar}, A. and {Sampson}, L.~M. and {Sanidas}, S. and {Sarkissian}, J.~M. and {Schmitz}, K. and {Schult}, L. and {Sesana}, A. and {Shaifullah}, G. and {Shannon}, R.~M. and {Shapiro-Albert}, B.~J. and {Siemens}, X. and {Simon}, J. and {Smith}, T.~L. and {Speri}, L. and {Spiewak}, R. and {Stairs}, I.~H. and {Stappers}, B.~W. and {Stinebring}, D.~R. and {Swiggum}, J.~K. and {Taylor}, S.~R. and {Theureau}, G. and {Tiburzi}, C. and {Vallisneri}, M. and {van der Wateren}, E. and {Vecchio}, A. and {Verbiest}, J.~P.~W. and {Vigeland}, S.~J. and {Wahl}, H. and {Wang}, J.~B. and {Wang}, J. and {Wang}, L. and {Witt}, C.~A. and {Zhang}, S. and {Zhu}, X.~J.},
        title = "{The International Pulsar Timing Array second data release: Search for an isotropic gravitational wave background}",
      journal = {\mnras},
     keywords = {gravitational waves, methods: data analysis, pulsars: general, Astrophysics - High Energy Astrophysical Phenomena, Astrophysics - Instrumentation and Methods for Astrophysics},
         year = 2022,
        month = mar,
       volume = {510},
       number = {4},
        pages = {4873-4887},
          doi = {10.1093/mnras/stab3418},
archivePrefix = {arXiv},
       eprint = {2201.03980},
 primaryClass = {astro-ph.HE},
       adsurl = {https://ui.adsabs.harvard.edu/abs/2022MNRAS.510.4873A},
      adsnote = {Provided by the SAO/NASA Astrophysics Data System}
}

@ARTICLE{psb+18,
       author = {{Perera}, B.~B.~P. and {Stappers}, B.~W. and {Babak}, S. and {Keith}, M.~J. and {Antoniadis}, J. and {Bassa}, C.~G. and {Caballero}, R.~N. and {Champion}, D.~J. and {Cognard}, I. and {Desvignes}, G. and {Graikou}, E. and {Guillemot}, L. and {Janssen}, G.~H. and {Karuppusamy}, R. and {Kramer}, M. and {Lazarus}, P. and {Lentati}, L. and {Liu}, K. and {Lyne}, A.~G. and {McKee}, J.~W. and {Os{\l}owski}, S. and {Perrodin}, D. and {Sanidas}, S.~A. and {Sesana}, A. and {Shaifullah}, G. and {Theureau}, G. and {Verbiest}, J.~P.~W. and {Taylor}, S.~R.},
        title = "{Improving timing sensitivity in the microhertz frequency regime: limits from PSR J1713+0747 on gravitational waves produced by supermassive black hole binaries}",
      journal = {\mnras},
     keywords = {gravitational waves, binaries: general, stars: black holes, stars: neutron, pulsars: individual: PSR J1713+0747, quasars: supermassive black holes, Astrophysics - High Energy Astrophysical Phenomena},
         year = 2018,
        month = jul,
       volume = {478},
       number = {1},
        pages = {218-227},
          doi = {10.1093/mnras/sty1116},
archivePrefix = {arXiv},
       eprint = {1804.10571},
 primaryClass = {astro-ph.HE},
       adsurl = {https://ui.adsabs.harvard.edu/abs/2018MNRAS.478..218P},
      adsnote = {Provided by the SAO/NASA Astrophysics Data System}
}





@ARTICLE{morello22,
       author = {{Morello}, V. and {Rajwade}, K.~M. and {Stappers}, B.~W.},
        title = "{IQRM: real-time adaptive RFI masking for radio transient and pulsar searches}",
      journal = {\mnras},
     keywords = {methods: data analysis, pulsars: general, fast radio bursts, Astrophysics - Instrumentation and Methods for Astrophysics},
         year = 2022,
        month = feb,
       volume = {510},
       number = {1},
        pages = {1393-1403},
          doi = {10.1093/mnras/stab3493},
archivePrefix = {arXiv},
       eprint = {2108.12434},
 primaryClass = {astro-ph.IM},
       adsurl = {https://ui.adsabs.harvard.edu/abs/2022MNRAS.510.1393M},
      adsnote = {Provided by the SAO/NASA Astrophysics Data System}
}


@ARTICLE{siv+20,
       author = {{Mata S{\'a}nchez}, D. and {Istrate}, A.~G. and {van Kerkwijk}, M.~H. and {Breton}, R.~P. and {Kaplan}, D.~L.},
        title = "{PSR J1012+5307: a millisecond pulsar with an extremely low-mass white dwarf companion}",
      journal = {\mnras},
     keywords = {stars: evolution, stars: neutron, pulsars: individual: PSRJ1012+5307, white dwarfs, Astrophysics - High Energy Astrophysical Phenomena, Astrophysics - Solar and Stellar Astrophysics},
         year = 2020,
        month = may,
       volume = {494},
       number = {3},
        pages = {4031-4042},
          doi = {10.1093/mnras/staa983},
archivePrefix = {arXiv},
       eprint = {2004.02901},
 primaryClass = {astro-ph.HE},
       adsurl = {https://ui.adsabs.harvard.edu/abs/2020MNRAS.494.4031M},
      adsnote = {Provided by the SAO/NASA Astrophysics Data System}
}


@misc{hickish2016decade,
      title={A Decade of Developing Radio-Astronomy Instrumentation using CASPER Open-Source Technology}, 
      author={Jack Hickish and Zuhra Abdurashidova and Zaki Ali and Kaushal D. Buch and Sandeep C. Chaudhari and Hong Chen and Matthew Dexter and Rachel Simone Domagalski and John Ford and Griffin Foster and David George and Joe Greenberg and Lincoln Greenhill and Adam Isaacson and Homin Jiang and Glenn Jones and Francois Kapp and Henno Kriel and Rich Lacasse and Andrew Lutomirski and David MacMahon and Jason Manley and Andrew Martens and Randy McCullough and Mekhala V. Muley and Wesley New and Aaron Parsons and Daniel C. Price and Rurik A. Primiani and Jason Ray and Andrew Siemion and Verees'e Van Tonder and Laura Vertatschitsch and Mark Wagner and Jonathan Weintroub and Dan Werthimer},
      year={2016},
      eprint={1611.01826},
      archivePrefix={arXiv},
      primaryClass={astro-ph.IM}
}


@INPROCEEDINGS{ntp,
       author = {{Mills}, D.~L.},
        title = "{The Network Computer as Precision Timekeeper}",
    booktitle = {Proceedings of the 28th Annual Precise Time and Time Interval (PTTI) Applications and Planning Meeting. Editorial Committee Chairman},
         year = 1997,
        month = jan,
        pages = {97-107},
       adsurl = {https://ui.adsabs.harvard.edu/abs/1997ptti.conf...97M},
      adsnote = {Provided by the SAO/NASA Astrophysics Data System}
}


@article{petit09, 
         title={Atomic time scales TAI and TT(BIPM): present performances and prospects}, 
         volume={5}, 
         DOI={10.1017/S1743921310008896}, 
         number={H15}, 
         journal={Proceedings of the International Astronomical Union}, 
         publisher={Cambridge University Press}, 
         author={Petit, Gérard}, 
         year={2009}, 
         pages={220–221}}

@ARTICLE{gsr+22,
       author = {{Goncharov}, Boris and {Thrane}, Eric and {Shannon}, Ryan M. and {Harms}, Jan and {Bhat}, N.~D. Ramesh and {Hobbs}, George and {Kerr}, Matthew and {Manchester}, Richard N. and {Reardon}, Daniel J. and {Russell}, Christopher J. and {Zhu}, Xing-Jiang and {Zic}, Andrew},
        title = "{Consistency of the Parkes Pulsar Timing Array Signal with a Nanohertz Gravitational-wave Background}",
      journal = {\apjl},
     keywords = {Gravitational waves, Millisecond pulsars, Pulsar timing method, Astronomy data analysis, Bayesian statistics, Importance sampling, Supermassive black holes, Gravitational wave astronomy, Hierarchical models, High energy astrophysics, Astronomical methods, 678, 1062, 1305, 1858, 1900, 1892, 1663, 675, 1925, 739, 1043, General Relativity and Quantum Cosmology, Astrophysics - High Energy Astrophysical Phenomena, Astrophysics - Instrumentation and Methods for Astrophysics},
         year = 2022,
        month = jun,
       volume = {932},
       number = {2},
          eid = {L22},
        pages = {L22},
          doi = {10.3847/2041-8213/ac76bb},
archivePrefix = {arXiv},
       eprint = {2206.03766},
 primaryClass = {gr-qc},
       adsurl = {https://ui.adsabs.harvard.edu/abs/2022ApJ...932L..22G},
      adsnote = {Provided by the SAO/NASA Astrophysics Data System}
}

@ARTICLE{tarafdar+22,
       author = {{Tarafdar}, Pratik and {Nobleson}, K. and {Rana}, Prerna and {Singha}, Jaikhomba and {Krishnakumar}, M.~A. and {Joshi}, Bhal Chandra and {Paladi}, Avinash Kumar and {Kolhe}, Neel and {Batra}, Neelam Dhanda and {Agarwal}, Nikita and {Bathula}, Adarsh and {Dandapat}, Subhajit and {Desai}, Shantanu and {Dey}, Lankeswar and {Hisano}, Shinnosuke and {Ingale}, Prathamesh and {Kato}, Ryo and {Kharbanda}, Divyansh and {Kikunaga}, Tomonosuke and {Marmat}, Piyush and {Pandian}, B. Arul and {Prabu}, T. and {Srivastava}, Aman and {Surnis}, Mayuresh and {Susarla}, Sai Chaitanya and {Susobhanan}, Abhimanyu and {Takahashi}, Keitaro and {Arumugam}, P. and {Bagchi}, Manjari and {Banik}, Sarmistha and {De}, Kishalay and {Girgaonkar}, Raghav and {Gopakumar}, A. and {Gupta}, Yashwant and {Maan}, Yogesh and {Manoharan}, P.~K. and {Naidu}, Arun and {Pathak}, Dhruv},
        title = "{The Indian Pulsar Timing Array: First data release}",
      journal = {\pasa},
     keywords = {radio telescopes, radio astronomy, astronomy data analysis, pulsar timing method, millisecond pulsars, Astrophysics - Instrumentation and Methods for Astrophysics, Astrophysics - High Energy Astrophysical Phenomena},
         year = 2022,
        month = oct,
       volume = {39},
          eid = {e053},
        pages = {e053},
          doi = {10.1017/pasa.2022.46},
archivePrefix = {arXiv},
       eprint = {2206.09289},
 primaryClass = {astro-ph.IM},
       adsurl = {https://ui.adsabs.harvard.edu/abs/2022PASA...39...53T},
      adsnote = {Provided by the SAO/NASA Astrophysics Data System}
}

@ARTICLE{DE440,
       author = {{Park}, Ryan S. and {Folkner}, William M. and {Williams}, James G. and {Boggs}, Dale H.},
        title = "{The JPL Planetary and Lunar Ephemerides DE440 and DE441}",
      journal = {\aj},
     keywords = {Celestial mechanics, Orbital motion, Orbits, Solar system planets, Solar system, The Sun, The Moon, Earth-moon system, Solar system astronomy, Inner planets, Outer planets, Pluto, 211, 1179, 1184, 1260, 1528, 1693, 1692, 436, 1529, 1267},
         year = 2021,
        month = mar,
       volume = {161},
       number = {3},
          eid = {105},
        pages = {105},
          doi = {10.3847/1538-3881/abd414},
       adsurl = {https://ui.adsabs.harvard.edu/abs/2021AJ....161..105P},
      adsnote = {Provided by the SAO/NASA Astrophysics Data System}
}



@ARTICLE{bac+16,
       author = {{Bassa}, C.~G. and {Antoniadis}, J. and {Camilo}, F. and {Cognard}, I. and {Koester}, D. and {Kramer}, M. and {Ransom}, S.~R. and {Stappers}, B.~W.},
        title = "{Cool white dwarf companions to four millisecond pulsars}",
      journal = {\mnras},
     keywords = {binaries: close, stars: individual: PSR J0613-0200, stars: individual: PSR J1231-1411, stars: individual: PSR J2017+0603, pulsars: general, white dwarfs, Astrophysics - High Energy Astrophysical Phenomena, Astrophysics - Solar and Stellar Astrophysics},
         year = 2016,
        month = feb,
       volume = {455},
       number = {4},
        pages = {3806-3813},
          doi = {10.1093/mnras/stv2607},
archivePrefix = {arXiv},
       eprint = {1511.01319},
 primaryClass = {astro-ph.HE},
       adsurl = {https://ui.adsabs.harvard.edu/abs/2016MNRAS.455.3806B},
      adsnote = {Provided by the SAO/NASA Astrophysics Data System}
}


@ARTICLE{wm1,
       author = {{the EPTA Collaboration}},
        title = "{The second data release from the European Pulsar Timing Array. I. The dataset and timing analysis}",
      journal = {\aa},
         year = 2023,
        month = mar,
       volume = {this Volume},
       number = {},
          eid = {},
        pages = {},
}


@ARTICLE{wm2,
       author = {{the EPTA and InPTA Collaborations}},
        title = "{The second data release from the European Pulsar Timing Array. II. Customised pulsar noise models for gravitational wave background searches}",
      journal = {\aa},
         year = 2023,
        month = mar,
       volume = {this Volume},
       number = {},
          eid = {},
        pages = {},
}

@ARTICLE{wm3,
       author = {{the EPTA and InPTA Collaborations}},
        title = "{The second data release from the European Pulsar Timing Array. III. Search for gravitational wave signals}",
      journal = {\aa},
         year = 2023,
        month = mar,
       volume = {this Volume},
       number = {},
          eid = {},
        pages = {},
}


@INPROCEEDINGS{skilling2004,
       author = {{Skilling}, John},
        title = "{Nested Sampling}",
     keywords = {02.50.Tt, Inference methods},
    booktitle = {Bayesian Inference and Maximum Entropy Methods in Science and Engineering: 24th International Workshop on Bayesian Inference and Maximum Entropy Methods in Science and Engineering},
         year = 2004,
       editor = {{Fischer}, Rainer and {Preuss}, Roland and {Toussaint}, Udo Von},
       series = {American Institute of Physics Conference Series},
       volume = {735},
        month = nov,
        pages = {395-405},
          doi = {10.1063/1.1835238},
       adsurl = {https://ui.adsabs.harvard.edu/abs/2004AIPC..735..395S},
      adsnote = {Provided by the SAO/NASA Astrophysics Data System}
}

@ARTICLE{tsb+21,
       author = {{Tiburzi}, C. and {Shaifullah}, G.~M. and {Bassa}, C.~G. and {Zucca}, P. and {Verbiest}, J.~P.~W. and {Porayko}, N.~K. and {van der Wateren}, E. and {Fallows}, R.~A. and {Main}, R.~A. and {Janssen}, G.~H. and {Anderson}, J.~M. and {Bak Nielsen}, A. -S. and {Donner}, J.~Y. and {Keane}, E.~F. and {K{\"u}nsem{\"o}ller}, J. and {Os{\l}owski}, S. and {Grie{\ss}meier}, J. -M. and {Serylak}, M. and {Br{\"u}ggen}, M. and {Ciardi}, B. and {Dettmar}, R. -J. and {Hoeft}, M. and {Kramer}, M. and {Mann}, G. and {Vocks}, C.},
        title = "{The impact of solar wind variability on pulsar timing}",
      journal = {\aap},
     keywords = {pulsars: general, solar wind, ISM: general, gravitational waves, Astrophysics - High Energy Astrophysical Phenomena},
         year = 2021,
        month = mar,
       volume = {647},
          eid = {A84},
        pages = {A84},
          doi = {10.1051/0004-6361/202039846},
archivePrefix = {arXiv},
       eprint = {2012.11726},
 primaryClass = {astro-ph.HE},
       adsurl = {https://ui.adsabs.harvard.edu/abs/2021A&A...647A..84T},
      adsnote = {Provided by the SAO/NASA Astrophysics Data System}
}

@ARTICLE{spf+23,
       author = {{Speri}, Lorenzo and {Porayko}, Nataliya K. and {Falxa}, Mikel and {Chen}, Siyuan and {Gair}, Jonathan R. and {Sesana}, Alberto and {Taylor}, Stephen R.},
        title = "{Quality over quantity: Optimizing pulsar timing array analysis for stochastic and continuous gravitational wave signals}",
      journal = {\mnras},
     keywords = {gravitational waves, methods: data analysis, pulsars: general, Astrophysics - High Energy Astrophysical Phenomena, General Relativity and Quantum Cosmology},
         year = 2023,
        month = jan,
       volume = {518},
       number = {2},
        pages = {1802-1817},
          doi = {10.1093/mnras/stac3237},
archivePrefix = {arXiv},
       eprint = {2211.03201},
 primaryClass = {astro-ph.HE},
       adsurl = {https://ui.adsabs.harvard.edu/abs/2023MNRAS.518.1802S},
      adsnote = {Provided by the SAO/NASA Astrophysics Data System}
}

@ARTICLE{ant20,
       author = {{Antoniadis}, John},
        title = "{Gaia Pulsars and Where to Find Them in EDR3}",
      journal = {Research Notes of the American Astronomical Society},
     keywords = {Binary pulsars, Millisecond pulsars, Optical pulsars, Radio pulsars, Rotation powered pulsars, Sky surveys, 153, 1062, 1173, 1353, 1408, 1464, Astrophysics - High Energy Astrophysical Phenomena},
         year = 2020,
        month = dec,
       volume = {4},
       number = {12},
          eid = {223},
        pages = {223},
          doi = {10.3847/2515-5172/abd189},
archivePrefix = {arXiv},
       eprint = {2012.06335},
 primaryClass = {astro-ph.HE},
       adsurl = {https://ui.adsabs.harvard.edu/abs/2020RNAAS...4..223A},
      adsnote = {Provided by the SAO/NASA Astrophysics Data System}
}



@ARTICLE{ato+16,
       author = {{Antoniadis}, John and {Tauris}, Thomas M. and {Ozel}, Feryal and {Barr}, Ewan and {Champion}, David J. and {Freire}, Paulo C.~C.},
        title = "{The millisecond pulsar mass distribution: Evidence for bimodality and constraints on the maximum neutron star mass}",
      journal = {arXiv e-prints},
     keywords = {Astrophysics - High Energy Astrophysical Phenomena, Astrophysics - Solar and Stellar Astrophysics, Nuclear Theory},
         year = 2016,
        month = may,
          eid = {arXiv:1605.01665},
        pages = {arXiv:1605.01665},
          doi = {10.48550/arXiv.1605.01665},
archivePrefix = {arXiv},
       eprint = {1605.01665},
 primaryClass = {astro-ph.HE},
       adsurl = {https://ui.adsabs.harvard.edu/abs/2016arXiv160501665A},
      adsnote = {Provided by the SAO/NASA Astrophysics Data System}
}


@ARTICLE{avk+12,
       author = {{Antoniadis}, J. and {van Kerkwijk}, M.~H. and {Koester}, D. and {Freire}, P.~C.~C. and {Wex}, N. and {Tauris}, T.~M. and {Kramer}, M. and {Bassa}, C.~G.},
        title = "{The relativistic pulsar-white dwarf binary PSR J1738+0333 - I. Mass determination and evolutionary history}",
      journal = {\mnras},
     keywords = {binaries: close, stars: neutron, pulsars: general, white dwarfs, stars: individual: PSR J1738+0333, Astrophysics - High Energy Astrophysical Phenomena, Astrophysics - Solar and Stellar Astrophysics, General Relativity and Quantum Cosmology},
         year = 2012,
        month = jul,
       volume = {423},
       number = {4},
        pages = {3316-3327},
          doi = {10.1111/j.1365-2966.2012.21124.x},
archivePrefix = {arXiv},
       eprint = {1204.3948},
 primaryClass = {astro-ph.HE},
       adsurl = {https://ui.adsabs.harvard.edu/abs/2012MNRAS.423.3316A},
      adsnote = {Provided by the SAO/NASA Astrophysics Data System}
}


@ARTICLE{cov22,
       author = {{van Cappellen}, W.~A. and {Oosterloo}, T.~A. and {Verheijen}, M.~A.~W. and {Adams}, E.~A.~K. and {Adebahr}, B. and {Braun}, R. and {Hess}, K.~M. and {Holties}, H. and {van der Hulst}, J.~M. and {Hut}, B. and {Kooistra}, E. and {van Leeuwen}, J. and {Loose}, G.~M. and {Morganti}, R. and {Moss}, V.~A. and {Orr{\'u}}, E. and {Ruiter}, M. and {Schoenmakers}, A.~P. and {Vermaas}, N.~J. and {Wijnholds}, S.~J. and {van Amesfoort}, A.~S. and {Arts}, M.~J. and {Attema}, J.~J. and {Bakker}, L. and {Bassa}, C.~G. and {Bast}, J.~E. and {Benthem}, P. and {Beukema}, R. and {Blaauw}, R. and {de Blok}, W.~J.~G. and {Bouwhuis}, M. and {van den Brink}, R.~H. and {Connor}, L. and {Coolen}, A.~H.~W.~M. and {Damstra}, S. and {van Diepen}, G.~N.~J. and {de Goei}, R. and {D{\'e}nes}, H. and {Drost}, M. and {Ebbendorf}, N. and {Frank}, B.~S. and {Gardenier}, D.~W. and {Gerbers}, M. and {Grange}, Y.~G. and {Grit}, T. and {Gunst}, A.~W. and {Gupta}, N. and {Ivashina}, M.~V. and {J{\'o}zsa}, G.~I.~G. and {Janssen}, G.~H. and {Koster}, A. and {Kruithof}, G.~H. and {Kuindersma}, S.~J. and {Kutkin}, A. and {Lucero}, D.~M. and {Maan}, Y. and {Maccagni}, F.~M. and {van der Marel}, J. and {Mika}, A. and {Morawietz}, J. and {Mulder}, H. and {Mulder}, E. and {Norden}, M.~J. and {Offringa}, A.~R. and {Oostrum}, L.~C. and {Overeem}, R.~E. and {Paragi}, Z. and {Pepping}, H.~J. and {Petroff}, E. and {Pisano}, D.~J. and {Polatidis}, A.~G. and {Prasad}, P. and {de Reijer}, J.~P.~R. and {Romein}, J.~W. and {Schaap}, J. and {Schoonderbeek}, G.~W. and {Schulz}, R. and {van der Schuur}, D. and {Sclocco}, A. and {Sluman}, J.~J. and {Smits}, R. and {Stappers}, B.~W. and {Straal}, S.~M. and {Stuurwold}, K.~J.~C. and {Verstappen}, J. and {Vohl}, D. and {Wierenga}, K.~J. and {Woestenburg}, E.~E.~M. and {Zanting}, A.~W. and {Ziemke}, J.},
        title = "{Apertif: Phased array feeds for the Westerbork Synthesis Radio Telescope. System overview and performance characteristics}",
      journal = {\aap},
     keywords = {telescopes, instrumentation: interferometers, surveys, Astrophysics - Instrumentation and Methods for Astrophysics},
         year = 2022,
        month = feb,
       volume = {658},
          eid = {A146},
        pages = {A146},
          doi = {10.1051/0004-6361/202141739},
archivePrefix = {arXiv},
       eprint = {2109.14234},
 primaryClass = {astro-ph.IM},
       adsurl = {https://ui.adsabs.harvard.edu/abs/2022A&A...658A.146V},
      adsnote = {Provided by the SAO/NASA Astrophysics Data System}
}

@INPROCEEDINGS{tan91,
       author = {{Tan}, G.~H.},
        title = "{The multi frequency front end - A new type of front end for the Westerbork Synthesis Radio Telescope}",
     keywords = {Cryogenic Cooling, Frequency Ranges, High Electron Mobility Transistors, Radio Astronomy, Radio Telescopes, Circular Polarization, Feasibility Analysis, Noise Intensity, Astronomy, CRYOGENIC COOLING, FREQUENCY RANGES, HIGH ELECTRON MOBILITY TRANSISTORS, RADIO ASTRONOMY, RADIO TELESCOPES, CIRCULAR POLARIZATION, FEASIBILITY ANALYSIS, NOISE INTENSITY},
    booktitle = {IAU Colloq. 131: Radio Interferometry. Theory, Techniques, and Applications},
         year = 1991,
       editor = {{Cornwell}, T.~J. and {Perley}, R.~A.},
       series = {Astronomical Society of the Pacific Conference Series},
       volume = {19},
        month = jan,
        pages = {42-46},
       adsurl = {https://ui.adsabs.harvard.edu/abs/1991ASPC...19...42T},
      adsnote = {Provided by the SAO/NASA Astrophysics Data System}
}


@article{kbw08,
doi = {10.1086/528699},
url = {https://dx.doi.org/10.1086/528699},
year = {2008},
month = {feb},
publisher = {University of Chicago Press},
volume = {120},
number = {864},
pages = {191},
author = {Ramesh Karuppusamy and Ben Stappers and Willem van Straten},
title = {PuMa-II: A Wide Band Pulsar Machine for the Westerbork Synthesis Radio Telescope},
journal = {Publications of the Astronomical Society of the Pacific},
abstract = {The Pulsar Machine II (PuMa-II) is the new flexible pulsar processing back-end system at the Westerbork Synthesis Radio Telescope (WSRT), specifically designed to take advantage of the upgraded WSRT. The instrument is based on a computer cluster running the Linux operating system, with minimal custom hardware. A maximum of 160 MHz analog bandwidth sampled as 8 × 20 MHz subbands with 8-bit resolution can be recorded on disks attached to separate computer nodes. Processing of the data is done in the additional 32 nodes allowing near real time coherent dedispersion for most pulsars observed at the WSRT. This has doubled the bandwidth for pulsar observations in general, and has enabled the use of coherent dedispersion over a bandwidth 8 times larger than was previously possible at the WSRT. PuMa-II is one of the widest bandwidth coherent dedispersion machines currently in use and has a maximum time resolution of 50 ns. The system is now routinely used for high-precision pulsar timing studies, polarization studies, single pulse work, and a variety of other observational work.}
}

@INPROCEEDINGS{str99,
       author = {{Strom}, R.~G.},
        title = "{The NFRA Pulsar Machine PuMA}",
    booktitle = {The Universe at Low Radio Frequencies},
         year = 2002,
       editor = {{Pramesh Rao}, A. and {Swarup}, G. and {Gopal-Krishna}},
       volume = {199},
        month = jan,
        pages = {383},
       adsurl = {https://ui.adsabs.harvard.edu/abs/2002IAUS..199..383S},
      adsnote = {Provided by the SAO/NASA Astrophysics Data System}
}

@ARTICLE{bjs+16,
       author = {{Bassa}, C.~G. and {Janssen}, G.~H. and {Stappers}, B.~W. and {Tauris}, T.~M. and {Wevers}, T. and {Jonker}, P.~G. and {Lentati}, L. and {Verbiest}, J.~P.~W. and {Desvignes}, G. and {Graikou}, E. and {Guillemot}, L. and {Freire}, P.~C.~C. and {Lazarus}, P. and {Caballero}, R.~N. and {Champion}, D.~J. and {Cognard}, I. and {Jessner}, A. and {Jordan}, C. and {Karuppusamy}, R. and {Kramer}, M. and {Lazaridis}, K. and {Lee}, K.~J. and {Liu}, K. and {Lyne}, A.~G. and {McKee}, J. and {Os{\l}owski}, S. and {Perrodin}, D. and {Sanidas}, S. and {Shaifullah}, G. and {Smits}, R. and {Theureau}, G. and {Tiburzi}, C. and {Zhu}, W.~W.},
        title = "{A millisecond pulsar in an extremely wide binary system}",
      journal = {\mnras},
     keywords = {binaries: general, stars: individual: PSR J1024-0719, stars: neutron, supernovae: general, Astrophysics - High Energy Astrophysical Phenomena, Astrophysics - Solar and Stellar Astrophysics},
         year = 2016,
        month = aug,
       volume = {460},
       number = {2},
        pages = {2207-2222},
          doi = {10.1093/mnras/stw1134},
archivePrefix = {arXiv},
       eprint = {1604.00129},
 primaryClass = {astro-ph.HE},
       adsurl = {https://ui.adsabs.harvard.edu/abs/2016MNRAS.460.2207B},
      adsnote = {Provided by the SAO/NASA Astrophysics Data System}
}

@ARTICLE{ccg+21,
       author = {{Chen}, S. and {Caballero}, R.~N. and {Guo}, Y.~J. and {Chalumeau}, A. and {Liu}, K. and {Shaifullah}, G. and {Lee}, K.~J. and {Babak}, S. and {Desvignes}, G. and {Parthasarathy}, A. and {Hu}, H. and {van der Wateren}, E. and {Antoniadis}, J. and {Bak Nielsen}, A. -S. and {Bassa}, C.~G. and {Berthereau}, A. and {Burgay}, M. and {Champion}, D.~J. and {Cognard}, I. and {Falxa}, M. and {Ferdman}, R.~D. and {Freire}, P.~C.~C. and {Gair}, J.~R. and {Graikou}, E. and {Guillemot}, L. and {Jang}, J. and {Janssen}, G.~H. and {Karuppusamy}, R. and {Keith}, M.~J. and {Kramer}, M. and {Liu}, X.~J. and {Lyne}, A.~G. and {Main}, R.~A. and {McKee}, J.~W. and {Mickaliger}, M.~B. and {Perera}, B.~B.~P. and {Perrodin}, D. and {Petiteau}, A. and {Porayko}, N.~K. and {Possenti}, A. and {Samajdar}, A. and {Sanidas}, S.~A. and {Sesana}, A. and {Speri}, L. and {Stappers}, B.~W. and {Theureau}, G. and {Tiburzi}, C. and {Vecchio}, A. and {Verbiest}, J.~P.~W. and {Wang}, J. and {Wang}, L. and {Xu}, H.},
        title = "{Common-red-signal analysis with 24-yr high-precision timing of the European Pulsar Timing Array: inferences in the stochastic gravitational-wave background search}",
      journal = {\mnras},
     keywords = {gravitational waves, methods: data analysis, pulsars: general, Astrophysics - High Energy Astrophysical Phenomena, Astrophysics - Cosmology and Nongalactic Astrophysics},
         year = 2021,
        month = dec,
       volume = {508},
       number = {4},
        pages = {4970-4993},
          doi = {10.1093/mnras/stab2833},
archivePrefix = {arXiv},
       eprint = {2110.13184},
 primaryClass = {astro-ph.HE},
       adsurl = {https://ui.adsabs.harvard.edu/abs/2021MNRAS.508.4970C},
      adsnote = {Provided by the SAO/NASA Astrophysics Data System}
}


@ARTICLE{fbb+23,
       author = {{Falxa}, M. and {Babak}, S. and {Baker}, P.~T. and {B{\'e}csy}, B. and {Chalumeau}, A. and {Chen}, S. and {Chen}, Z. and {Cornish}, N.~J. and {Guillemot}, L. and {Hazboun}, J.~S. and {Mingarelli}, C.~M.~F. and {Parthasarathy}, A. and {Petiteau}, A. and {Pol}, N.~S. and {Sesana}, A. and {Spolaor}, S.~B. and {Taylor}, S.~R. and {Theureau}, G. and {Vallisneri}, M. and {Vigeland}, S.~J. and {Witt}, C.~A. and {Zhu}, X. and {Antoniadis}, J. and {Arzoumanian}, Z. and {Bailes}, M. and {Bhat}, N.~D.~R. and {Blecha}, L. and {Brazier}, A. and {Brook}, P.~R. and {Caballero}, N. and {Cameron}, A.~D. and {Casey-Clyde}, J.~A. and {Champion}, D. and {Charisi}, M. and {Chatterjee}, S. and {Cognard}, I. and {Cordes}, J.~M. and {Crawford}, F. and {Cromartie}, H.~T. and {Crowter}, K. and {Dai}, S. and {DeCesar}, M.~E. and {Demorest}, P.~B. and {Desvignes}, G. and {Dolch}, T. and {Drachler}, B. and {Feng}, Y. and {Ferrara}, E.~C. and {Fiore}, W. and {Fonseca}, E. and {Garver-Daniels}, N. and {Glaser}, J. and {Goncharov}, B. and {Good}, D.~C. and {Griessmeier}, J. and {Guo}, Y.~J. and {G{\"u}ltekin}, K. and {Hobbs}, G. and {Hu}, H. and {Islo}, K. and {Jang}, J. and {Jennings}, R.~J. and {Johnson}, A.~D. and {Jones}, M.~L. and {Kaczmarek}, J. and {Kaiser}, A.~R. and {Kaplan}, D.~L. and {Keith}, M. and {Kelley}, L.~Z. and {Kerr}, M. and {Key}, J.~S. and {Laal}, N. and {Lam}, M.~T. and {Lamb}, W.~G. and {Lazio}, T.~J.~W. and {Liu}, K. and {Liu}, T. and {Luo}, J. and {Lynch}, R.~S. and {Madison}, D.~R. and {Main}, R. and {Manchester}, R. and {McEwen}, A. and {McKee}, J. and {McLaughlin}, M.~A. and {Ng}, C. and {Nice}, D.~J. and {Ocker}, S. and {Olum}, K.~D. and {Os{\l}owski}, S. and {Pennucci}, T.~T. and {Perera}, B.~B.~P. and {Perrodin}, D. and {Porayko}, N. and {Possenti}, A. and {Quelquejay-Leclere}, H. and {Ransom}, S.~M. and {Ray}, P.~S. and {Reardon}, D.~J. and {Russell}, C.~J. and {Samajdar}, A. and {Sarkissian}, J. and {Schult}, L. and {Shaifullah}, G. and {Shannon}, R.~M. and {Shapiro-Albert}, B.~J. and {Siemens}, X. and {Simon}, J.~J. and {Siwek}, M. and {Smith}, T.~L. and {Speri}, L. and {Spiewak}, R. and {Stairs}, I.~H. and {Stappers}, B. and {Stinebring}, D.~R. and {Swiggum}, J.~K. and {Tiburzi}, C. and {Turner}, J. and {Vecchio}, A. and {Verbiest}, J.~P.~W. and {Wahl}, H. and {Wang}, S.~Q. and {Wang}, J. and {Wang}, J. and {Wu}, Z. and {Zhang}, L. and {Zhang}, S.},
        title = "{Searching for continuous Gravitational Waves in the second data release of the International Pulsar Timing Array}",
      journal = {\mnras},
     keywords = {gravitational waves, methods:data analysis, pulsars:general, General Relativity and Quantum Cosmology, Astrophysics - Instrumentation and Methods for Astrophysics},
         year = 2023,
        month = mar,
          doi = {10.1093/mnras/stad812},
archivePrefix = {arXiv},
       eprint = {2303.10767},
 primaryClass = {gr-qc},
       adsurl = {https://ui.adsabs.harvard.edu/abs/2023MNRAS.tmp..817F},
      adsnote = {Provided by the SAO/NASA Astrophysics Data System}
}

@ARTICLE{cbp+22,
       author = {{Chalumeau}, A. and {Babak}, S. and {Petiteau}, A. and {Chen}, S. and {Samajdar}, A. and {Caballero}, R.~N. and {Theureau}, G. and {Guillemot}, L. and {Desvignes}, G. and {Parthasarathy}, A. and {Liu}, K. and {Shaifullah}, G. and {Hu}, H. and {van der Wateren}, E. and {Antoniadis}, J. and {Bak Nielsen}, A. -S. and {Bassa}, C.~G. and {Berthereau}, A. and {Burgay}, M. and {Champion}, D.~J. and {Cognard}, I. and {Falxa}, M. and {Ferdman}, R.~D. and {Freire}, P.~C.~C. and {Gair}, J.~R. and {Graikou}, E. and {Guo}, Y.~J. and {Jang}, J. and {Janssen}, G.~H. and {Karuppusamy}, R. and {Keith}, M.~J. and {Kramer}, M. and {Lee}, K.~J. and {Liu}, X.~J. and {Lyne}, A.~G. and {Main}, R.~A. and {McKee}, J.~W. and {Mickaliger}, M.~B. and {Perera}, B.~B.~P. and {Perrodin}, D. and {Porayko}, N.~K. and {Possenti}, A. and {Sanidas}, S.~A. and {Sesana}, A. and {Speri}, L. and {Stappers}, B.~W. and {Tiburzi}, C. and {Vecchio}, A. and {Verbiest}, J.~P.~W. and {Wang}, J. and {Wang}, L. and {Xu}, H.},
        title = "{Noise analysis in the European Pulsar Timing Array data release 2 and its implications on the gravitational-wave background search}",
      journal = {\mnras},
     keywords = {gravitational waves, methods: data analysis, pulsars: general, Astrophysics - High Energy Astrophysical Phenomena, Astrophysics - Instrumentation and Methods for Astrophysics},
         year = 2022,
        month = feb,
       volume = {509},
       number = {4},
        pages = {5538-5558},
          doi = {10.1093/mnras/stab3283},
archivePrefix = {arXiv},
       eprint = {2111.05186},
 primaryClass = {astro-ph.HE},
       adsurl = {https://ui.adsabs.harvard.edu/abs/2022MNRAS.509.5538C},
      adsnote = {Provided by the SAO/NASA Astrophysics Data System}
}




@ARTICLE{bgp+22,
       author = {{Joshi}, Bhal Chandra and {Gopakumar}, Achamveedu and {Pandian}, Arul and {Prabu}, Thiagaraj and {Dey}, Lankeswar and {Bagchi}, Manjari and {Desai}, Shantanu and {Tarafdar}, Pratik and {Rana}, Prerna and {Maan}, Yogesh and {Batra}, Neelam Dhanda and {Girgaonkar}, Raghav and {Agarwal}, Nikita and {Arumugam}, Paramasivan and {Basu}, Avishek and {Bathula}, Adarsh and {Dandapat}, Subhajit and {Gupta}, Yashwant and {Hisano}, Shinnosuke and {Kato}, Ryo and {Kharbanda}, Divyansh and {Kikunaga}, Tomonosuke and {Kolhe}, Neel and {Krishnakumar}, M.~A. and {Manoharan}, P.~K. and {Marmat}, Piyush and {Naidu}, Arun and {Banik}, Sarmistha and {Nobleson}, K. and {Paladi}, Avinash Kumar and {Pathak}, Dhruv and {Singha}, Jaikhomba and {Srivastava}, Aman and {Surnis}, Mayuresh and {Susarla}, Sai Chaitanya and {Susobhanan}, Abhimanyu and {Takahashi}, Keitaro},
        title = "{Nanohertz gravitational wave astronomy during SKA era: An InPTA perspective}",
      journal = {Journal of Astrophysics and Astronomy},
     keywords = {Gravitational waves, pulsars: general, stars: neutron, ISM: General, Astrophysics - High Energy Astrophysical Phenomena, Astrophysics - Instrumentation and Methods for Astrophysics, Astrophysics - Solar and Stellar Astrophysics},
         year = 2022,
        month = dec,
       volume = {43},
       number = {2},
          eid = {98},
        pages = {98},
          doi = {10.1007/s12036-022-09869-w},
archivePrefix = {arXiv},
       eprint = {2207.06461},
 primaryClass = {astro-ph.HE},
       adsurl = {https://ui.adsabs.harvard.edu/abs/2022JApA...43...98J},
      adsnote = {Provided by the SAO/NASA Astrophysics Data System}
}

@ARTICLE{sesana2013,
       author = {{Sesana}, A.},
        title = "{Systematic investigation of the expected gravitational wave signal from  supermassive black hole binaries in the pulsar timing band.}",
      journal = {\mnras},
     keywords = {black hole physics, gravitational waves, pulsars: general, galaxies: evolution, Astrophysics - Cosmology and Nongalactic Astrophysics, General Relativity and Quantum Cosmology},
         year = 2013,
        month = jun,
       volume = {433},
        pages = {L1-L5},
          doi = {10.1093/mnrasl/slt034},
archivePrefix = {arXiv},
       eprint = {1211.5375},
 primaryClass = {astro-ph.CO},
       adsurl = {https://ui.adsabs.harvard.edu/abs/2013MNRAS.433L...1S},
      adsnote = {Provided by the SAO/NASA Astrophysics Data System}
}

@ARTICLE{sesana2004,
       author = {{Sesana}, Alberto and {Haardt}, Francesco and {Madau}, Piero and {Volonteri}, Marta},
        title = "{Low-Frequency Gravitational Radiation from Coalescing Massive Black Hole Binaries in Hierarchical Cosmologies}",
      journal = {\apj},
     keywords = {Black Hole Physics, Cosmology: Theory, Cosmology: Early Universe, Gravitational Waves, Relativity, Astrophysics},
         year = 2004,
        month = aug,
       volume = {611},
       number = {2},
        pages = {623-632},
          doi = {10.1086/422185},
archivePrefix = {arXiv},
       eprint = {astro-ph/0401543},
 primaryClass = {astro-ph},
       adsurl = {https://ui.adsabs.harvard.edu/abs/2004ApJ...611..623S},
      adsnote = {Provided by the SAO/NASA Astrophysics Data System}
}

@ARTICLE{lms+16,
       author = {{Lasky}, Paul D. and {Mingarelli}, Chiara M.~F. and {Smith}, Tristan L. and {Giblin}, John T. and {Thrane}, Eric and {Reardon}, Daniel J. and {Caldwell}, Robert and {Bailes}, Matthew and {Bhat}, N.~D. Ramesh and {Burke-Spolaor}, Sarah and {Dai}, Shi and {Dempsey}, James and {Hobbs}, George and {Kerr}, Matthew and {Levin}, Yuri and {Manchester}, Richard N. and {Os{\l}owski}, Stefan and {Ravi}, Vikram and {Rosado}, Pablo A. and {Shannon}, Ryan M. and {Spiewak}, Ren{\'e}e and {van Straten}, Willem and {Toomey}, Lawrence and {Wang}, Jingbo and {Wen}, Linqing and {You}, Xiaopeng and {Zhu}, Xingjiang},
        title = "{Gravitational-Wave Cosmology across 29 Decades in Frequency}",
      journal = {Physical Review X},
     keywords = {Astrophysics - Cosmology and Nongalactic Astrophysics, General Relativity and Quantum Cosmology},
         year = 2016,
        month = jan,
       volume = {6},
       number = {1},
          eid = {011035},
        pages = {011035},
          doi = {10.1103/PhysRevX.6.011035},
archivePrefix = {arXiv},
       eprint = {1511.05994},
 primaryClass = {astro-ph.CO},
       adsurl = {https://ui.adsabs.harvard.edu/abs/2016PhRvX...6a1035L},
      adsnote = {Provided by the SAO/NASA Astrophysics Data System}
}


@ARTICLE{schwaller2015,
       author = {{Schwaller}, Pedro},
        title = "{Gravitational Waves from a Dark Phase Transition}",
      journal = {\prl},
     keywords = {04.30.-w, 12.60.-i, 95.35.+d, Gravitational waves: theory, Models beyond the standard model, Dark matter, High Energy Physics - Phenomenology, Astrophysics - Cosmology and Nongalactic Astrophysics},
         year = 2015,
        month = oct,
       volume = {115},
       number = {18},
          eid = {181101},
        pages = {181101},
          doi = {10.1103/PhysRevLett.115.181101},
archivePrefix = {arXiv},
       eprint = {1504.07263},
 primaryClass = {hep-ph},
       adsurl = {https://ui.adsabs.harvard.edu/abs/2015PhRvL.115r1101S},
      adsnote = {Provided by the SAO/NASA Astrophysics Data System}
}


@ARTICLE{dv01b,
       author = {{Damour}, Thibault and {Vilenkin}, Alexander},
        title = "{Gravitational wave bursts from cusps and kinks on cosmic strings}",
      journal = {\prd},
     keywords = {04.30.Db, 95.85.Sz, 97.60.Gb, 98.80.Cq, Wave generation and sources, Gravitational radiation magnetic fields and other observations, Pulsars, Particle-theory and field-theory models of the early Universe, Astrophysics, General Relativity and Quantum Cosmology},
         year = 2001,
        month = sep,
       volume = {64},
       number = {6},
          eid = {064008},
        pages = {064008},
          doi = {10.1103/PhysRevD.64.064008},
archivePrefix = {arXiv},
       eprint = {gr-qc/0104026},
 primaryClass = {astro-ph},
       adsurl = {https://ui.adsabs.harvard.edu/abs/2001PhRvD..64f4008D},
      adsnote = {Provided by the SAO/NASA Astrophysics Data System}
}

@ARTICLE{bps+13,
       author = {{Babak}, S. and {Petiteau}, A. and {Sesana}, A. and {Brem}, P. and {Rosado}, P.~A. and {Taylor}, S.~R. and {Lassus}, A. and {Hessels}, J.~W.~T. and {Bassa}, C.~G. and {Burgay}, M. and {Caballero}, R.~N. and {Champion}, D.~J. and {Cognard}, I. and {Desvignes}, G. and {Gair}, J.~R. and {Guillemot}, L. and {Janssen}, G.~H. and {Karuppusamy}, R. and {Kramer}, M. and {Lazarus}, P. and {Lee}, K.~J. and {Lentati}, L. and {Liu}, K. and {Mingarelli}, C.~M.~F. and {Os{\l}owski}, S. and {Perrodin}, D. and {Possenti}, A. and {Purver}, M.~B. and {Sanidas}, S. and {Smits}, R. and {Stappers}, B. and {Theureau}, G. and {Tiburzi}, C. and {van Haasteren}, R. and {Vecchio}, A. and {Verbiest}, J.~P.~W.},
        title = "{European Pulsar Timing Array limits on continuous gravitational waves from individual supermassive black hole binaries}",
      journal = {\mnras},
     keywords = {black hole physics, gravitational waves, pulsars: general, Astrophysics - Cosmology and Nongalactic Astrophysics, Astrophysics - Astrophysics of Galaxies, Astrophysics - Instrumentation and Methods for Astrophysics, General Relativity and Quantum Cosmology},
         year = 2016,
        month = jan,
       volume = {455},
       number = {2},
        pages = {1665-1679},
          doi = {10.1093/mnras/stv2092},
archivePrefix = {arXiv},
       eprint = {1509.02165},
 primaryClass = {astro-ph.CO},
       adsurl = {https://ui.adsabs.harvard.edu/abs/2016MNRAS.455.1665B},
      adsnote = {Provided by the SAO/NASA Astrophysics Data System}
}

@ARTICLE{msb23,
       author = {{Miles}, M.~T. and {Shannon}, R.~M. and {Bailes}, M. and {Reardon}, D.~J. and {Keith}, M.~J. and {Cameron}, A.~D. and {Parthasarathy}, A. and {Shamohammadi}, M. and {Spiewak}, R. and {van Straten}, W. and {Buchner}, S. and {Camilo}, F. and {Geyer}, M. and {Karastergiou}, A. and {Kramer}, M. and {Serylak}, M. and {Theureau}, G. and {Venkatraman Krishnan}, V.},
        title = "{The MeerKAT Pulsar Timing Array: first data release}",
      journal = {\mnras},
     keywords = {gravitational waves, methods: data analysis, methods: observational, (stars:) pulsars: general, Astrophysics - High Energy Astrophysical Phenomena, Astrophysics - Instrumentation and Methods for Astrophysics},
         year = 2023,
        month = mar,
       volume = {519},
       number = {3},
        pages = {3976-3991},
          doi = {10.1093/mnras/stac3644},
archivePrefix = {arXiv},
       eprint = {2212.04648},
 primaryClass = {astro-ph.HE},
       adsurl = {https://ui.adsabs.harvard.edu/abs/2023MNRAS.519.3976M},
      adsnote = {Provided by the SAO/NASA Astrophysics Data System}
}


@ARTICLE{mcl13,
       author = {{McLaughlin}, M.~A.},
        title = "{The North American Nanohertz Observatory for Gravitational Waves}",
      journal = {Classical and Quantum Gravity},
     keywords = {Astrophysics - Instrumentation and Methods for Astrophysics, Astrophysics - High Energy Astrophysical Phenomena, Astrophysics - Solar and Stellar Astrophysics},
         year = 2013,
        month = nov,
       volume = {30},
       number = {22},
          eid = {224008},
        pages = {224008},
          doi = {10.1088/0264-9381/30/22/224008},
archivePrefix = {arXiv},
       eprint = {1310.0758},
 primaryClass = {astro-ph.IM},
       adsurl = {https://ui.adsabs.harvard.edu/abs/2013CQGra..30v4008M},
      adsnote = {Provided by the SAO/NASA Astrophysics Data System}
}

@ARTICLE{hob13,
       author = {{Hobbs}, G.},
        title = "{The Parkes Pulsar Timing Array}",
      journal = {Classical and Quantum Gravity},
     keywords = {Astrophysics - Instrumentation and Methods for Astrophysics, Astrophysics - High Energy Astrophysical Phenomena, Astrophysics - Solar and Stellar Astrophysics},
         year = 2013,
        month = nov,
       volume = {30},
       number = {22},
          eid = {224007},
        pages = {224007},
          doi = {10.1088/0264-9381/30/22/224007},
archivePrefix = {arXiv},
       eprint = {1307.2629},
 primaryClass = {astro-ph.IM},
       adsurl = {https://ui.adsabs.harvard.edu/abs/2013CQGra..30v4007H},
      adsnote = {Provided by the SAO/NASA Astrophysics Data System}
}

@ARTICLE{cf18,
       author = {{Caprini}, Chiara and {Figueroa}, Daniel G.},
        title = "{Cosmological backgrounds of gravitational waves}",
      journal = {Classical and Quantum Gravity},
     keywords = {Astrophysics - Cosmology and Nongalactic Astrophysics, General Relativity and Quantum Cosmology, High Energy Physics - Phenomenology},
         year = 2018,
        month = aug,
       volume = {35},
       number = {16},
          eid = {163001},
        pages = {163001},
          doi = {10.1088/1361-6382/aac608},
archivePrefix = {arXiv},
       eprint = {1801.04268},
 primaryClass = {astro-ph.CO},
       adsurl = {https://ui.adsabs.harvard.edu/abs/2018CQGra..35p3001C},
      adsnote = {Provided by the SAO/NASA Astrophysics Data System}
}

@INPROCEEDINGS{lee16,
       author = {{Lee}, K.~J.},
        title = "{Prospects of Gravitational Wave Detection Using Pulsar Timing Array for Chinese Future Telescopes}",
    booktitle = {Frontiers in Radio Astronomy and FAST Early Sciences Symposium 2015},
         year = 2016,
       editor = {{Qain}, L. and {Li}, D.},
       series = {Astronomical Society of the Pacific Conference Series},
       volume = {502},
        month = feb,
        pages = {19},
       adsurl = {https://ui.adsabs.harvard.edu/abs/2016ASPC..502...19L},
      adsnote = {Provided by the SAO/NASA Astrophysics Data System}
}


@misc{psrchive,
      title={Pulsar data analysis with PSRCHIVE}, 
      author={Willem van Straten and Paul Demorest and Stefan Osłowski},
      year={2012},
      eprint={1205.6276},
      archivePrefix={arXiv},
      primaryClass={astro-ph.IM}
}

@ARTICLE{aab+21a,
       author = {{Alam}, Md F. and {Arzoumanian}, Zaven and {Baker}, Paul T. and {Blumer}, Harsha and {Bohler}, Keith E. and {Brazier}, Adam and {Brook}, Paul R. and {Burke-Spolaor}, Sarah and {Caballero}, Keeisi and {Camuccio}, Richard S. and {Chamberlain}, Rachel L. and {Chatterjee}, Shami and {Cordes}, James M. and {Cornish}, Neil J. and {Crawford}, Fronefield and {Cromartie}, H. Thankful and {Decesar}, Megan E. and {Demorest}, Paul B. and {Dolch}, Timothy and {Ellis}, Justin A. and {Ferdman}, Robert D. and {Ferrara}, Elizabeth C. and {Fiore}, William and {Fonseca}, Emmanuel and {Garcia}, Yhamil and {Garver-Daniels}, Nathan and {Gentile}, Peter A. and {Good}, Deborah C. and {Gusdorff}, Jordan A. and {Halmrast}, Daniel and {Hazboun}, Jeffrey S. and {Islo}, Kristina and {Jennings}, Ross J. and {Jessup}, Cody and {Jones}, Megan L. and {Kaiser}, Andrew R. and {Kaplan}, David L. and {Kelley}, Luke Zoltan and {Key}, Joey Shapiro and {Lam}, Michael T. and {Lazio}, T. Joseph W. and {Lorimer}, Duncan R. and {Luo}, Jing and {Lynch}, Ryan S. and {Madison}, Dustin R. and {Maraccini}, Kaleb and {McLaughlin}, Maura A. and {Mingarelli}, Chiara M.~F. and {Ng}, Cherry and {Nguyen}, Benjamin M.~X. and {Nice}, David J. and {Pennucci}, Timothy T. and {Pol}, Nihan S. and {Ramette}, Joshua and {Ransom}, Scott M. and {Ray}, Paul S. and {Shapiro-Albert}, Brent J. and {Siemens}, Xavier and {Simon}, Joseph and {Spiewak}, Ren{\'e}e and {Stairs}, Ingrid H. and {Stinebring}, Daniel R. and {Stovall}, Kevin and {Swiggum}, Joseph K. and {Taylor}, Stephen R. and {Tripepi}, Michael and {Vallisneri}, Michele and {Vigeland}, Sarah J. and {Witt}, Caitlin A. and {Zhu}, Weiwei and {Nanograv Collaboration}},
        title = "{The NANOGrav 12.5 yr Data Set: Wideband Timing of 47 Millisecond Pulsars}",
      journal = {\apjs},
     keywords = {Millisecond pulsars, Pulsar timing method, Rotation powered pulsars, Binary pulsars, Radio pulsars, Pulsars, Gravitational waves, Astronomy data analysis, Interstellar medium, 1062, 1305, 1408, 153, 1353, 1306, 678, 1858, 847, Astrophysics - High Energy Astrophysical Phenomena, Astrophysics - Instrumentation and Methods for Astrophysics},
         year = 2021,
        month = jan,
       volume = {252},
       number = {1},
          eid = {5},
        pages = {5},
          doi = {10.3847/1538-4365/abc6a1},
archivePrefix = {arXiv},
       eprint = {2005.06495},
 primaryClass = {astro-ph.HE},
       adsurl = {https://ui.adsabs.harvard.edu/abs/2021ApJS..252....5A},
      adsnote = {Provided by the SAO/NASA Astrophysics Data System}
}

@ARTICLE{aab+21b,
       author = {{Alam}, Md F. and {Arzoumanian}, Zaven and {Baker}, Paul T. and {Blumer}, Harsha and {Bohler}, Keith E. and {Brazier}, Adam and {Brook}, Paul R. and {Burke-Spolaor}, Sarah and {Caballero}, Keeisi and {Camuccio}, Richard S. and {Chamberlain}, Rachel L. and {Chatterjee}, Shami and {Cordes}, James M. and {Cornish}, Neil J. and {Crawford}, Fronefield and {Cromartie}, H. Thankful and {Decesar}, Megan E. and {Demorest}, Paul B. and {Dolch}, Timothy and {Ellis}, Justin A. and {Ferdman}, Robert D. and {Ferrara}, Elizabeth C. and {Fiore}, William and {Fonseca}, Emmanuel and {Garcia}, Yhamil and {Garver-Daniels}, Nathan and {Gentile}, Peter A. and {Good}, Deborah C. and {Gusdorff}, Jordan A. and {Halmrast}, Daniel and {Hazboun}, Jeffrey S. and {Islo}, Kristina and {Jennings}, Ross J. and {Jessup}, Cody and {Jones}, Megan L. and {Kaiser}, Andrew R. and {Kaplan}, David L. and {Kelley}, Luke Zoltan and {Key}, Joey Shapiro and {Lam}, Michael T. and {Lazio}, T. Joseph W. and {Lorimer}, Duncan R. and {Luo}, Jing and {Lynch}, Ryan S. and {Madison}, Dustin R. and {Maraccini}, Kaleb and {McLaughlin}, Maura A. and {Mingarelli}, Chiara M.~F. and {Ng}, Cherry and {Nguyen}, Benjamin M.~X. and {Nice}, David J. and {Pennucci}, Timothy T. and {Pol}, Nihan S. and {Ramette}, Joshua and {Ransom}, Scott M. and {Ray}, Paul S. and {Shapiro-Albert}, Brent J. and {Siemens}, Xavier and {Simon}, Joseph and {Spiewak}, Ren{\'e}e and {Stairs}, Ingrid H. and {Stinebring}, Daniel R. and {Stovall}, Kevin and {Swiggum}, Joseph K. and {Taylor}, Stephen R. and {Tripepi}, Michael and {Vallisneri}, Michele and {Vigeland}, Sarah J. and {Witt}, Caitlin A. and {Zhu}, Weiwei and {Nanograv Collaboration}},
        title = "{The NANOGrav 12.5 yr Data Set: Observations and Narrowband Timing of 47 Millisecond Pulsars}",
      journal = {\apjs},
     keywords = {Millisecond pulsars, Pulsar timing method, Radio pulsars, Pulsars, Gravitational waves, Binary pulsars, Astrometry, Astronomy databases, Radio astronomy, Rotation powered pulsars, Time series analysis, Flux calibration, 1062, 1305, 1353, 1306, 678, 153, 80, 83, 1338, 1408, 1916, 544, Astrophysics - High Energy Astrophysical Phenomena, Astrophysics - Instrumentation and Methods for Astrophysics},
         year = 2021,
        month = jan,
       volume = {252},
       number = {1},
          eid = {4},
        pages = {4},
          doi = {10.3847/1538-4365/abc6a0},
archivePrefix = {arXiv},
       eprint = {2005.06490},
 primaryClass = {astro-ph.HE},
       adsurl = {https://ui.adsabs.harvard.edu/abs/2021ApJS..252....4A},
      adsnote = {Provided by the SAO/NASA Astrophysics Data System}
}

@ARTICLE{mca+19,
       author = {{Madison}, D.~R. and {Cordes}, J.~M. and {Arzoumanian}, Z. and {Chatterjee}, S. and {Crowter}, K. and {DeCesar}, M.~E. and {Demorest}, P.~B. and {Dolch}, T. and {Ellis}, J.~A. and {Ferdman}, R.~D. and {Ferrara}, E.~C. and {Fonseca}, E. and {Gentile}, P.~A. and {Jones}, G. and {Jones}, M.~L. and {Lam}, M.~T. and {Levin}, L. and {Lorimer}, D.~R. and {Lynch}, R.~S. and {McLaughlin}, M.~A. and {Mingarelli}, C.~M.~F. and {Ng}, C. and {Nice}, D.~J. and {Pennucci}, T.~T. and {Ransom}, S.~M. and {Ray}, P.~S. and {Spiewak}, R. and {Stairs}, I.~H. and {Stovall}, K. and {Swiggum}, J.~K. and {Zhu}, W.~W.},
        title = "{The NANOGrav 11 yr Data Set: Solar Wind Sounding through Pulsar Timing}",
      journal = {\apj},
     keywords = {ISM: structure, pulsars: general, solar wind, Astrophysics - Solar and Stellar Astrophysics},
         year = 2019,
        month = feb,
       volume = {872},
       number = {2},
          eid = {150},
        pages = {150},
          doi = {10.3847/1538-4357/ab01fd},
archivePrefix = {arXiv},
       eprint = {1808.07078},
 primaryClass = {astro-ph.SR},
       adsurl = {https://ui.adsabs.harvard.edu/abs/2019ApJ...872..150M},
      adsnote = {Provided by the SAO/NASA Astrophysics Data System}
}

@article{vlm10,
    author = {Verbiest, J. P. W. and Lorimer, D. R. and McLaughlin, M. A.},
    title = "{Lutz–Kelker bias in pulsar parallax measurements}",
    journal = {\mnras},
    volume = {405},
    number = {1},
    pages = {564-572},
    year = {2010},
    month = {06},
    abstract = "{Lutz and Kelker showed that parallax measurements are systematically overestimated because they do not properly account for the larger volume of space that is sampled at smaller parallax values. We apply their analysis to neutron stars, incorporating the bias introduced by the intrinsic radio luminosity function and a realistic Galactic population model for neutron stars. We estimate the bias for all published neutron star parallax measurements and find that measurements with less than ∼95 per cent certainty are likely to be significantly biased. Through inspection of historic parallax measurements, we confirm the described effects in optical and radio measurements as well as in distance estimates based on interstellar dispersion measures. The potential impact on future tests of relativistic gravity through pulsar timing and on X-ray-based estimates of neutron star radii is briefly discussed.}",
    issn = {0035-8711},
    doi = {10.1111/j.1365-2966.2010.16488.x},
    url = {https://doi.org/10.1111/j.1365-2966.2010.16488.x},
    eprint = {https://academic.oup.com/mnras/article-pdf/405/1/564/3385712/mnras0405-0564.pdf},
}

@ARTICLE{vwc+12,
       author = {{Verbiest}, J.~P.~W. and {Weisberg}, J.~M. and {Chael}, A.~A. and {Lee}, K.~J. and {Lorimer}, D.~R.},
        title = "{On Pulsar Distance Measurements and Their Uncertainties}",
      journal = {\apj},
     keywords = {astrometry, pulsars: general, Astrophysics - Galaxy Astrophysics},
         year = 2012,
        month = aug,
       volume = {755},
       number = {1},
          eid = {39},
        pages = {39},
          doi = {10.1088/0004-637X/755/1/39},
archivePrefix = {arXiv},
       eprint = {1206.0428},
 primaryClass = {astro-ph.GA},
       adsurl = {https://ui.adsabs.harvard.edu/abs/2012ApJ...755...39V},
      adsnote = {Provided by the SAO/NASA Astrophysics Data System}
}


@BOOK{Binney,
       author = {{Binney}, James and {Merrifield}, Michael},
        title = "{Galactic Astronomy}",
         year = 1998,
         publisher = "{Princeton Univesity Press}",
       adsurl = {https://ui.adsabs.harvard.edu/abs/1998gaas.book.....B},
      adsnote = {Provided by the SAO/NASA Astrophysics Data System}
}




@article{abc08,
  author = {{Anholm}, M. and {Ballmer}, S. and {Creighton}, J.~D.~E. and {Price}, L.~R. and {Siemens}, X.},
  title = {Optimal strategies for gravitational wave stochastic background searches in pulsar timing data},
  journal = "ArXiv e-prints",
  archivePrefix = "arXiv",
  eprint = {0809.0701},
  year = 2008,
  month = sep
}

@ARTICLE{dgb+19,
       author = {{Deller}, A.~T. and {Goss}, W.~M. and {Brisken}, W.~F. and {Chatterjee}, S. and {Cordes}, J.~M. and {Janssen}, G.~H. and {Kovalev}, Y.~Y. and {Lazio}, T.~J.~W. and {Petrov}, L. and {Stappers}, B.~W. and {Lyne}, A.},
        title = "{Microarcsecond VLBI Pulsar Astrometry with PSR{\ensuremath{\pi}} II. Parallax Distances for 57 Pulsars}",
      journal = {\apj},
     keywords = {astrometry, galaxies: ISM, pulsars: general, stars: neutron, techniques: high angular resolution, Astrophysics - Instrumentation and Methods for Astrophysics, Astrophysics - High Energy Astrophysical Phenomena, Astrophysics - Solar and Stellar Astrophysics},
         year = 2019,
        month = apr,
       volume = {875},
       number = {2},
          eid = {100},
        pages = {100},
          doi = {10.3847/1538-4357/ab11c7},
archivePrefix = {arXiv},
       eprint = {1808.09046},
 primaryClass = {astro-ph.IM},
       adsurl = {https://ui.adsabs.harvard.edu/abs/2019ApJ...875..100D},
      adsnote = {Provided by the SAO/NASA Astrophysics Data System}
}

@article{stob14,
    author = "Shibata, Masaru and Taniguchi, Keisuke and Okawa, Hirotada and Buonanno, Alessandra",
    title = "{Coalescence of binary neutron stars in a scalar-tensor theory of gravity}",
    eprint = "1310.0627",
    archivePrefix = "arXiv",
    primaryClass = "gr-qc",
    doi = "10.1103/PhysRevD.89.084005",
    journal = \prd,
    volume = "89",
    number = "8",
    pages = "084005",
    year = "2014"
}

@article{sha14,
    author = "Shao, Lijing",
    title = "{Tests of local Lorentz invariance violation of gravity in the standard model extension with pulsars}",
    eprint = "1402.6452",
    archivePrefix = "arXiv",
    primaryClass = "gr-qc",
    doi = "10.1103/PhysRevLett.112.111103",
    journal = \prl,
    volume = "112",
    pages = "111103",
    year = "2014"
}

@article{asr+09,
  author = {{Archibald}, A.~M. and {Stairs}, I.~H. and {Ransom}, S.~M. and {Kaspi}, V.~M. and {Kondratiev}, V.~I. and {Lorimer}, D.~R. and {McLaughlin}, M.~A. and {Boyles}, J. and {Hessels}, J.~W.~T. and {Lynch}, R. and {van Leeuwen}, J. and {Roberts}, M.~S.~E. and {Jenet}, F. and {Champion}, D.~J. and {Rosen}, R. and {Barlow}, B.~N. and {Dunlap}, B.~H. and {Remillard}, R.~A.},
  title = "{A Radio Pulsar/X-ray Binary Link}",
  journal = \sci,
  archivePrefix = "arXiv",
  eprint = {0905.3397},
  year = 2009,
  month = jun,
  volume = 324,
  pages = {1411-}
}

@article{bb08,
  author = {{Boyle}, L.~A. and {Buonanno}, A.},
  title = {Relating gravitational wave constraints from primordial nucleosynthesis, pulsar timing, laser interferometers, and the CMB: Implications for the early universe},
  journal = "Physical Review D",
  archivePrefix = "arXiv",
  eprint = {0708.2279},
  year = 2008,
  month = aug,
  volume = 78,
  number = 4,
  pages = {043531-+}
}

@article{bbv08,
  author = {Bhat, N.~D.~R. and Bailes, M. and Verbiest, J.~P.~W.},
  title = {Gravitational-radiation losses from the pulsar--white-dwarf binary PSR J1141--6545},
  journal = \prd,
  volume = {77},
  pages = {124017},
  eprint = {arXiv:astro-ph/0804.0956},
  year = {2008}
}

@article{bgrt05,
  author = {Bacon, D. J. and Goldberg, D. M. and Rowe, B. T. P. and Taylor, A. N.},
  title = {Weak Gravitational Flexion},
  journal = {\mnras},
  volume = {},
  pages = {},
  year = {2005},
  note = {submitted (astro-ph/0504478)}
}

@article{bjd+06,
  author = {Burgay, M. and Joshi, B.~C. and {D'Amico}, N. and Possenti, A. and Lyne, A.~G. and Manchester, R.~N. and McLaughlin, M.~A. and Kramer, M. and Camilo, F. and Freire, P.~C.~C.},
  title = {The Parkes High-Latitude pulsar survey},
  journal = {\mnras},
  keywords = {methods: observational, pulsars: general},
  year = 2006,
  month = may,
  volume = 368,
  pages = {283-292}
}

@article{bkj+05,
  author = {Becker, W. and Kramer, M. and Jessner, A. and Taam, R. E. and Jia, J. J. and Cheng, K. S. and Mignani, R. and Pellizzoni, A. and de Luca, A. and Slowikowska, A. and Caraveo, P.},
  title = {A multi-wavelength study of the pulsar {PSR B1929+10} and its {X-ray} trail},
  journal = {\apj},
  volume = {},
  pages = {},
  year = {2005},
  note = {submitted (astro-ph/0506545)}
}

@article{bkk+08,
  author = {Breton, R.~P. and Kaspi, V.~M. and Kramer, M. and McLaughlin, M.~A. and Lyutikov, M. and Ransom, S.~M. and Stairs, I.~H. and Ferdman, R.~D. and Camilo, F. and Possenti, A.},
  title = {Relativistic Spin Precession in the Double Pulsar},
  journal = \sci,
  volume = {321},
  pages = {104--},
  year = {2008},
  eprint = {arXiv:astro-ph/0807.2644}
}

@book{bm98,
  author = {Binney, J. and Merrifield, M.},
  title = {Galactic astronomy},
  year = 1998,
  publisher = {Princeton University Press},
  address = {Princeton, NJ}
}

@book{bra00,
  author = {R. Bracewell},
  title = {The Fourier Transform and Its Applications},
  year = {2000},
  publisher = {McGraw--Hill},
  address = {New York},
  edition = {Third}
}

@article{cbl+96,
  author = {Cognard, I. and Bourgois, G. and Lestrade, J.-F. and Biraud, F. and Aubry, D. and Darchy, B. Drouhin, J.-P.},
  title = {High-precision timing observations of the millisecond pulsar PSR 1821$-$24 at Nancay.},
  journal = {\aa},
  keywords = {PULSARS: GENERAL, PULSARS: INDIVIDUAL: PSR 1821-24, TIME},
  year = {1996},
  month = jul,
  volume = {311},
  pages = {179--188}
}

@article{cbv+09,
  author = {Chatterjee, S. and Brisken, W.~F. and Vlemmings, W.~H.~T. and Goss, W.~M. and Lazio, T.~J.~W. and Cordes, J.~M. and Thorsett, S.~E. and Fomalont, E.~B. and Lyne, A.~G. and Kramer, M.},
  title = {Precision Astrometry with the Very Long Baseline Array: Parallaxes and Proper Motions for 14 Pulsars},
  journal = {\apj},
  archivePrefix = "arXiv",
  eprint = {0901.1436},
  year = 2009,
  month = jun,
  volume = 698,
  pages = {250-265}
}

@inproceedings{cc05,
  author = {Charles, P. A. and Coe, M. J.},
  title = {Optical, Ultraviolet, and Infrared Observations of {X}-ray Binaries},
  booktitle = {Compact Stellar {X}-ray Sources},
  editor = {Lewin, W. H. G. and van der Klis, M.},
  publisher = {Cambridge University Press},
  year = {2005},
  note = {in press, astro-ph/0308020}
}

@article{cf84,
  author = {Coles, W.~A. and Filice, J.~P.},
  title = {Dynamic spectra of interplanetary scintillations},
  journal = {\nat},
  year = 1984,
  month = nov,
  volume = 312,
  pages = {251-254}
}

@article{cfl+05,
  author = {Cordes, J.~M. and Freire, P.~C.~C. and Lorimer, D.~R. and Camilo, F.   and Champion, D.~J.  and Nice, D.~J.  and Ramachandran, R.   and Hessels, J.~W.~T.  and Vlemmings, W.  and van Leeuwen, J.   and Ransom, S.~M.  and Bhat, N.~D.~R.  and Arzoumanian, Z.   and McLaughlin, M.~A.  and Kaspi, V.~M.  and Kasian, L.  and Deneva, J.~S.   and Reid, B.  and Han, J.~L.  and Backer, D.~C.  and Stairs, I.~H.   and Deshpande, A.~A.  and Faucher-Gigu\`ere, C.-A.},
  title = {{Arecibo Pulsar Survey Using ALFA. I. Survey Strategy  and First Discoveries}},
  journal = {\apj},
  note = {submitted},
  year = 2005
}

@article{crh+06,
  title = {Transient pulsed radio emission from a magnetar},
  author = {{Camilo}, F. and {Ransom}, S. M. and {Halpern}, J. P. and {Reynolds}, J. and {Helfand}, D. J. and {Zimmerman}, N. and {Sarkissian}, J.},
  year = 2006,
  journal = {\nat},
  volume = 442,
  pages = {892-895}
}

@article{crh+06a,
  author = {{Crawford}, F. and {Roberts}, M.~S.~E. and {Hessels}, J.~W.~T. and {Ransom}, S.~M. and {Livingstone}, M. and {Tam}, C.~R. and {Kaspi}, V.~M.},
  title = "{A Survey of 56 Midlatitude EGRET Error Boxes for Radio Pulsars}",
  journal = {\apj},
  eprint = {arXiv:astro-ph/0608225},
  keywords = {Gamma Rays: Observations, Stars: Pulsars: General},
  year = 2006,
  month = dec,
  volume = 652,
  pages = {1499-1507}
}

@article{crl+08,
  author = {{Champion}, D.~J. and {Ransom}, S.~M. and {Lazarus},
                  P. and {Camilo}, F. and {Bassa}, C. and {Kaspi},
                  V.~M. and {Nice}, D.~J. and {Freire}, P.~C.~C. and
                  {Stairs}, I.~H. and {van Leeuwen}, J. and
                  {Stappers}, B.~W. and {Cordes}, J.~M. and {Hessels},
                  J.~W.~T. and {Lorimer}, D.~R. and {Arzoumanian},
                  Z. and {Backer}, D.~C. and {Bhat}, N.~D.~R. and
                  {Chatterjee}, S. and {Cognard}, I. and {Deneva},
                  J.~S. and {Faucher-Gigu{\`e}re}, C.-A. and
                  {Gaensler}, B.~M. and {Han}, J. and {Jenet},
                  F.~A. and {Kasian}, L. and {Kondratiev}, V.~I. and
                  {Kramer}, M. and {Lazio}, J. and {McLaughlin},
                  M.~A. and {Venkataraman}, A. and {Vlemmings}, W.},
  title = "{An Eccentric Binary Millisecond Pulsar in the Galactic Plane}",
  journal = sci,
  archivePrefix = "arXiv",
  eprint = {0805.2396},
  year = 2008,
  month = jun,
  volume = 320,
  pages = {1309-}
}

@article{dbt09,
  author = {Deller, A.~T. and Bailes, M. and Tingay, S.~J.},
  title = {Implications of a VLBI Distance to the Double Pulsar J0737$-$3039A/B},
  journal = {Science},
  archivePrefix = "arXiv",
  eprint = {0902.0996},
  year = 2009,
  month = mar,
  volume = 323,
  pages = {1327-}
}

@article{dtbr09,
  author = {Deller, A.~T. and Tingay, S.~J.  and Bailes, M. and Reynolds, J.~E.},
  title = {Precision Southern Hemisphere VLBI Pulsar Astrometry. II. Measurement of Seven Parallaxes},
  journal = {\apj},
  archivePrefix = "arXiv",
  eprint = {0906.3897},
  keywords = {astrometry, pulsars: general, techniques: interferometric},
  year = 2009,
  month = aug,
  volume = 701,
  pages = {1243-1257}
}

@phdthesis{del09,
  author = {Deller, A.~T.},
  school = {Swinburne University of Technology},
  year = 2009,
  title = {Precision VLBI astrometry: Instrumentation, algorithms and pulsar parallax determination}
}

@article{dms05,
  title = {Earth-mass dark-matter haloes as the first structures in the early Universe},
  author = {Diemand, J. and Moore, B and Stadel, J.},
  journal = {\nat},
  eprint = {arXiv:astro-ph/0501589},
  year = 2005,
  month = jan,
  volume = 433,
  pages = {389-391}
}

@ARTICLE{abb+18,
       author = {{Arzoumanian}, Z. and {Baker}, P.~T. and {Brazier}, A. and {Burke-Spolaor}, S. and {Chamberlin}, S.~J. and {Chatterjee}, S. and {Christy}, B. and {Cordes}, J.~M. and {Cornish}, N.~J. and {Crawford}, F. and {Thankful Cromartie}, H. and {Crowter}, K. and {DeCesar}, M. and {Demorest}, P.~B. and {Dolch}, T. and {Ellis}, J.~A. and {Ferdman}, R.~D. and {Ferrara}, E. and {Folkner}, W.~M. and {Fonseca}, E. and {Garver-Daniels}, N. and {Gentile}, P.~A. and {Haas}, R. and {Hazboun}, J.~S. and {Huerta}, E.~A. and {Islo}, K. and {Jones}, G. and {Jones}, M.~L. and {Kaplan}, D.~L. and {Kaspi}, V.~M. and {Lam}, M.~T. and {Lazio}, T.~J.~W. and {Levin}, L. and {Lommen}, A.~N. and {Lorimer}, D.~R. and {Luo}, J. and {Lynch}, R.~S. and {Madison}, D.~R. and {McLaughlin}, M.~A. and {McWilliams}, S.~T. and {Mingarelli}, C.~M.~F. and {Ng}, C. and {Nice}, D.~J. and {Park}, R.~S. and {Pennucci}, T.~T. and {Pol}, N.~S. and {Ransom}, S.~M. and {Ray}, P.~S. and {Rasskazov}, A. and {Siemens}, X. and {Simon}, J. and {Spiewak}, R. and {Stairs}, I.~H. and {Stinebring}, D.~R. and {Stovall}, K. and {Swiggum}, J. and {Taylor}, S.~R. and {Vallisneri}, M. and {van Haasteren}, R. and {Vigeland}, S. and {Zhu}, W.~W. and {NANOGrav Collaboration}},
        title = "{The NANOGrav 11 Year Data Set: Pulsar-timing Constraints on the Stochastic Gravitational-wave Background}",
      journal = {\apj},
     keywords = {ephemerides, gravitational waves, inflation, methods: data analysis, pulsars: general, quasars: supermassive black holes, Astrophysics - High Energy Astrophysical Phenomena, Astrophysics - Astrophysics of Galaxies, General Relativity and Quantum Cosmology},
         year = 2018,
        month = may,
       volume = {859},
       number = {1},
          eid = {47},
        pages = {47},
          doi = {10.3847/1538-4357/aabd3b},
archivePrefix = {arXiv},
       eprint = {1801.02617},
 primaryClass = {astro-ph.HE},
       adsurl = {https://ui.adsabs.harvard.edu/abs/2018ApJ...859...47A},
      adsnote = {Provided by the SAO/NASA Astrophysics Data System}
}


@article{ aab+21,
	author = {{GRAVITY Collaboration} and {Abuter, R.} and {Amorim, A.} and {Baub\"ock, M.} and {Berger, J. P.} and {Bonnet, H.} and {Brandner, W.} and {Cl\'enet, Y.} and {Davies, R.} and {de Zeeuw, P. T.} and {Dexter, J.} and {Dallilar, Y.} and {Drescher, A.} and {Eckart, A.} and {Eisenhauer, F.} and {F\"orster Schreiber, N. M.} and {Garcia, P.} and {Gao, F.} and {Gendron, E.} and {Genzel, R.} and {Gillessen, S.} and {Habibi, M.} and {Haubois, X.} and {Hei\ss{}el, G.} and {Henning, T.} and {Hippler, S.} and {Horrobin, M.} and {Jim\'enez-Rosales, A.} and {Jochum, L.} and {Jocou, L.} and {Kaufer, A.} and {Kervella, P.} and {Lacour, S.} and {Lapeyr\`ere, V.} and {Le Bouquin, J.-B.} and {L\'ena, P.} and {Lutz, D.} and {Nowak, M.} and {Ott, T.} and {Paumard, T.} and {Perraut, K.} and {Perrin, G.} and {Pfuhl, O.} and {Rabien, S.} and {Rodr\'{\i}guez-Coira, G.} and {Shangguan, J.} and {Shimizu, T.} and {Scheithauer, S.} and {Stadler, J.} and {Straub, O.} and {Straubmeier, C.} and {Sturm, E.} and {Tacconi, L. J.} and {Vincent, F.} and {von Fellenberg, S.} and {Waisberg, I.} and {Widmann, F.} and {Wieprecht, E.} and {Wiezorrek, E.} and {Woillez, J.} and {Yazici, S.} and {Young, A.} and {Zins, G.}},
	title = {Improved GRAVITY astrometric accuracy from modeling optical aberrations},
	DOI= "10.1051/0004-6361/202040208",
	url= "https://doi.org/10.1051/0004-6361/202040208",
	journal = {A\&A},
	year = 2021,
	volume = 647,
	pages = "A59",
}

@article{adh+19,
doi = {10.3847/1538-4357/aaf648},
url = {https://dx.doi.org/10.3847/1538-4357/aaf648},
year = {2019},
month = {jan},
publisher = {The American Astronomical Society},
volume = {871},
number = {1},
pages = {120},
author = {Anna-Christina Eilers and David W. Hogg and Hans-Walter Rix and Melissa K. Ness},
title = {The Circular Velocity Curve of the Milky Way from 5 to 25 kpc},
journal = {The Astrophysical Journal},
abstract = {We measure the circular velocity curve vc(R) of the Milky Way with the highest precision to date across Galactocentric distances of 5 ≤ R ≤ 25 kpc. Our analysis draws on the six-dimensional phase-space coordinates of ≳23,000 luminous red giant stars, for which we previously determined precise parallaxes using a data-driven model that combines spectral data from APOGEE with photometric information from WISE, 2MASS, and Gaia. We derive the circular velocity curve with the Jeans equation assuming an axisymmetric gravitational potential. At the location of the Sun we determine the circular velocity with its formal uncertainty to be  =  with systematic uncertainties at the ∼2
}

@ARTICLE{hol04,
       author = {{Holmberg}, Johan and {Flynn}, Chris},
        title = "{The local surface density of disc matter mapped by Hipparcos}",
      journal = {\mnras},
     keywords = {Galaxy: kinematics and dynamics, Galaxy: structure, dark matter, Astrophysics},
         year = 2004,
        month = aug,
       volume = {352},
       number = {2},
        pages = {440-446},
          doi = {10.1111/j.1365-2966.2004.07931.x},
archivePrefix = {arXiv},
       eprint = {astro-ph/0405155},
 primaryClass = {astro-ph},
       adsurl = {https://ui.adsabs.harvard.edu/abs/2004{\mnras}.352..440H},
      adsnote = {Provided by the SAO/NASA Astrophysics Data System}
}


@article{dtb09,
  author = {Deller, A.~T. and Tingay, S.~J. and Brisken, W.},
  title = {Precision Southern Hemisphere Pulsar VLBI Astrometry: Techniques and Results for PSR J1559$-$4438},
  journal = {\apj},
  archivePrefix = "arXiv",
  eprint = {0808.1598},
  keywords = {astrometry, pulsars: general, pulsars: individual: J1559-4438, techniques: interferometric},
  year = 2009,
  month = jan,
  volume = 690,
  pages = {198-209}
}

@article{dvtb08,
  author = {Deller, A.~T. and Verbiest, J.~P.~W. and Tingay, S.~J. and Bailes, M.},
  title = {Extremely High Precision VLBI Astrometry of PSR J0437$-$4715 and Implications for Theories of Gravity},
  journal = {\apjl},
  archivePrefix = "arXiv",
  eprint = {0808.1594},
  year = 2008,
  month = sep,
  volume = 685,
  pages = {L67-L70}
}

@article{ein05,
  author = {Einstein, A.},
  title = {Zur Elektrodynamik bewegter K{\"o}rper},
  journal = {Annalen der Physik},
  year = 1905,
  volume = 322,
  pages = {891-921}
}

@article{ekl09,
  author = {Eatough, R.~P. and Keane, E.~F. and Lyne, A.~G.},
  title = {An interference removal technique for radio pulsar searches},
  journal = {\mnras},
  year = 2009,
  month = may,
  volume = 395,
  pages = {410-415}
}

@article{em68,
  author = {{Ekers}, R.~D. and {Moffet}, A.~T.},
  title = {Further Observations of Pulsating Radio Sources at 13 cm},
  journal = {\nat},
  year = 1968,
  month = nov,
  volume = 220,
  pages = {756-761}
}

@article{emk+10,
  author = {Eatough, R.~P. and Molkenthin, N. and Kramer, M. and Noutsos, A. and Keith, M.~J. and Stappers, B.~W. and Lyne, A.~G.},
  title = {Selection of radio pulsar candidates using artificial neural networks},
  year = 2010,
  month = may,
  journal = {\mnras},
  note = {Accepted}
}

@inproceedings{esp05,
  author = {{Esposito-Far\`ese}, G.},
  title = "{Binary-Pulsar Tests of Strong-Field Gravity and Gravitational Radiation Damping}",
  booktitle = {The Tenth Marcel Grossman Meeting. On recent developments in theoretical and experimental general relativity, gravitation and relativistic field theories},
  year = 2005,
  editor = "{M.~Novello, S.~Perez Bergliaffa, \& R.~Ruffini}",
  month = jan,
  pages = {647-+},
  adsurl = {http://adsabs.harvard.edu/abs/2005tmgm.meet..647E},
}

@inproceedings{esp05hack,
  author = {{Esposito-Far\`ese}, G.},
  title = "{Binary-Pulsar Tests of Strong-Field Gravity and Gravitational Radiation Damping}",
  booktitle = {The Tenth Marcel Grossman Meeting},
  year = 2005,
  editor = "{M.~Novello, S.~Perez Bergliaffa, \& R.~Ruffini}",
  month = jan,
  pages = {647-+},
  adsurl = {http://adsabs.harvard.edu/abs/2005tmgm.meet..647E},
}

@article{esp09,
  author = {{Esposito-Far\`ese}, G.},
  title = "{Motion in alternative theories of gravity}",
  journal = {astro-ph/0905.2575},
  archivePrefix = "arXiv",
  eprint = {0905.2575},
  keywords = {General Relativity and Quantum Cosmology},
  year = 2009,
  month = may
}

@article{fbw+10,
  author = {{Freire}, P.~C.~C. et.\ al.},
  title = "{On the nature and evolution of the unique binary puslar PSR J1903+0327}",
  journal = {\mnras},
  year = 2010,
  note = {Accepted}
}

@article{fk06,
  author = {{Faucher-Gigu{\`e}re}, C.~A. and Kaspi, V.~M.},
  title = {Birth and Evolution of Isolated Radio Pulsars},
  journal = {\apj},
  eprint = {arXiv:astro-ph/0512585},
  keywords = {Galaxy: Structure, Methods: Statistical, Stars: Pulsars: General, Stars: Kinematics, Stars: Neutron},
  year = 2006,
  month = may,
  volume = 643,
  pages = {332-355}
}

@article{fmlg08,
  title = {INPOP06: a new numerical planetary ephemeris},
  author = {Fienga, A. and Manche, H. and Laskar, J. and Gastineau, M.},
  journal = aa,
  year = 2008,
  month = jan,
  volume = 477,
  pages = {315-327}
}

@article{frb+08,
  title = {Eight New Millisecond Pulsars in NGC 6440 and NGC 6441},
  author = {Freire, P.~C.~C. and Ransom, S.~M. and B{\'e}gin, S. and Stairs, I.~H. and Hessels, J.~W.~T. and Frey, L.~H. and Camilo, F.},
  journal = {\apj},
  archivePrefix = "arXiv",
  eprint = {0711.0925},
  year = 2008,
  month = mar,
  volume = {675},
  pages = {670-682}
}

@article{fwvh08,
  author = {{Freire}, P.~C.~C. and {Wolszczan}, A. and {van den Berg}, M. and {Hessels}, J.~W.~T.},
  title = "{A Massive Neutron Star in the Globular Cluster M5}",
  journal = {\apj},
  archivePrefix = "arXiv",
  eprint = {0712.3826},
  year = 2008,
  month = jun,
  volume = 679,
  pages = {1433-1442}
}

@article{gf97,
  author = {Gonzalez, A.~H. and Faber, S.~M.},
  title = {Malmquist Bias and the Distance to the Virgo Cluster},
  journal = {\apj},
  eprint = {arXiv:astro-ph/9704192},
  keywords = {Cosmology: Distance scale, Galaxies: Clusters: Individual name: Virgo, Galaxies: Distances and redshifts},
  year = 1997,
  month = aug,
  volume = 485,
  pages = {80-+}
}

@article{gm10,
  author = {{Gil}, J.~A. and {Melikidze}, G.~I.},
  title = "{Do double features in averaged pulsar profiles decipher the nature of their radio emission?}",
  journal = {ArXiv e-prints},
  archivePrefix = "arXiv",
  eprint = {1005.0678},
  primaryClass = "astro-ph.SR",
  keywords = {Astrophysics - Solar and Stellar Astrophysics},
  year = 2010,
  month = may
}

@article{gmga05,
  author = {Gupta, Y. and Mitra, D. and Green, D. A. and Acharyya, A.},
  title = {{GMRT} discovery of {PSR~J1833--1034: The} pulsar associated with the supernova remnant {G21.5--0.9}},
  journal = cursci,
  volume = {},
  pages = {},
  year = {2005},
  note = {submitted (astro-ph/0508257)}
}

@article{gri05,
  author = {{Grishchuk}, L.~P.},
  title = {Relic Gravitational Waves and Cosmology},
  journal = {Phys. Uspekhi},
  eprint = {arXiv:gr-qc/0504018},
  year = 2005,
  pages = {1235-1247}
}

@article{gzf+05,
  author = {C. Granet and H. Z. Zhang and A. R. Forsyth and G. R. Graves     and P. Doherty and K. J. Greene and G. L. James and P. Sykes and     T. S. Bird and M. W. Sinclair and G. Moorey, and R. N. Manchester.},
  title = {Design, Manufacture, and Test of a Dual-Band Feed     System for the Parkes Radio Telescope},
  journal = {IEEE Antennas and Propagation Magazine},
  year = 2005,
  note = {in press}
}

@article{haa+09,
  author = {Hobbs, G. and Archibald, A. and Arzoumanian, Z. and Backer, D. and Bailes, M. and Bhat, N.~D.~R. and Burgay, M. and Burke-Spolaor, S. and Champion, D. and Cognard, I. and Coles, W. and Cordes, J. and Demorest, P. and Desvignes, G. and Ferdman, R. and Finn, L. and Freire, P. and Gonzalez, M. and Hessels, J. and Hotan, A. and Janssen, G. and Jenet, F. and Jessner, A. and Jordan, C. and Kaspi, V. and Kramer, M. and Kondratiev, V. and Lazio, J. and Lazaridis, K. and Lee, K.~J. and Levin, Y. and Lommen, A. and Lorimer, D. and Lynch, R. and Lyne, A. and Manchester, R. and McLaughlin, M. and Nice, D. and Oslowski, S. and Pilia, M. and Possenti, A. and Purver, M. and Ransom, S. and Sanidas, S. and Sarkissian, J. and Sesana, A. and Shannon, R. and Siemens, X. and Stairs, I. and Stappers, B. and Stinebring, D. and Theureau, G. and van Haasteren, R. and van Straten, W. and Verbiest, J.~P.~W. and Yardley, D.~R.~B. and You, X.~Y.},
  title = {The international pulsar timing array project: using pulsars as a gravitational wave detector},
  journal = cqg,
  year = 2009
}

@article{hbb+09,
  author = {Hobbs, G.~B. and Bailes, M. and Bhat, N.~D.~R. and Burke-Spolaor, S. and Champion, D.~J. and Coles, W. and Hotan, A. and Jenet, F. and Kedziora-Chudczer, L. and Khoo, J. and Lee, K.~J. and Lommen, A. and Manchester, R.~N. and Reynolds, J. and Sarkissian, J. and van Straten, W. and To, S. and Verbiest, J.~P.~W. and Yardley, D. and You, X.~P.},
  title = {Gravitational-Wave Detection Using Pulsars: Status of the Parkes Pulsar Timing Array Project},
  journal = pasa,
  archivePrefix = "arXiv",
  eprint = {0812.2721},
  keywords = {Keywords: gravitational waves - pulsars: general,},
  year = 2009,
  month = jun,
  volume = 26,
  pages = {103-109}
}

@article{he07,
  author = {{Hankins}, T.~H. and {Eilek}, J.~A.},
  title = "{Radio Emission Signatures in the Crab Pulsar}",
  journal = {\apj},
  archivePrefix = "arXiv",
  eprint = {0708.2505},
  keywords = {pulsars: individual (Crab Nebula pulsar), Radiation Mechanisms: Nonthermal},
  year = 2007,
  month = nov,
  volume = 670,
  pages = {693-701}
}

@article{hem05a,
  author = {Hobbs, G. B. and Edwards, R. T. and Manchester, R. N.},
  title = {\textsc{TEMPO2}, a new pulsar timing package},
  journal = chjaa,
  volume = {},
  pages = {},
  year = 2005,
  note = {submitted}
}

@article{hgb+05,
  author = {Halpern, J. P. and Gotthelf, E. V. and Becker, R. H. and Helfand, D. J. and White, R. L.},
  title = {Discovery of Radio Emission from the Transient Anomalous X-ray Pulsar {XTE~J1810--197}},
  journal = {\apjl},
  volume = 632,
  pages = {29-32},
  year = 2005
}

@inproceedings{hjl+08,
  author = {Hobbs, G. and Jenet, F. and Lommen, A. and Coles, W. and Verbiest, J.~P.~W. and Manchester, R.},
  title = {Using pulsars to limit the existence of a gravitational wave background},
  booktitle = {40 Years of Pulsars: Millisecond Pulsars, Magnetars and More},
  year = {2008},
  series = {American Institute of Physics Conference Series},
  volume = 983,
  editor = {Bassa, C. and Wang, Z. and Cumming, A. and Kaspi, V.~M.},
  month = feb,
  pages = {630-632}
}

@ARTICLE{hjl+09,
   author = {{Hobbs}, G. and {Jenet}, F. and {Lee}, K.~J. and {Verbiest}, J.~P.~W. and
    {Yardley}, D. and {Manchester}, R. and {Lommen}, A. and {Coles}, W. and
    {Edwards}, R. and {Shettigara}, C.},
    title = "{TEMPO2: a new pulsar timing package - III. Gravitational wave simulation}",
  journal = {\mnras},
archivePrefix = "arXiv",
   eprint = {0901.0592},
 primaryClass = "astro-ph.SR",
 keywords = {gravitational waves , methods: numerical , pulsars: general},
     year = 2009,
    month = apr,
   volume = 394,
    pages = {1945-1955},
      doi = {10.1111/j.1365-2966.2009.14391.x},
   adsurl = {http://adsabs.harvard.edu/abs/2009{\mnras}.394.1945H},
  adsnote = {Provided by the SAO/NASA Astrophysics Data System}
}

@article{hlk06,
  author = {Hobbs, G. and Lyne, A. and Kramer, M.},
  title = {Pulsar Timing Noise},
  journal = {Chinese Journal of Astronomy and Astrophysics Supplement},
  year = 2006,
  month = dec,
  volume = 6,
  number = 2,
  pages = {169-175}
}

@article{hlk09,
  author = {Hobbs, G. and Lyne, A. and Kramer, M.},
  journal = {\mnras},
  year = 2009,
  note = {in preparation}
}

@article{hml+06,
  author = {Han, J.~L. and Manchester, R.~N. and Lyne, A.~G. and Qiao, G.~J. and {van Straten}, W.},
  title ={Pulsar Rotation Measures and the Large-Scale Structure of the Galactic Magnetic Field},
  journal = {\apj},
  eprint = {arxiv:astro-ph/0601357},
  keywords = {Galaxies: Magnetic Fields, Galaxy: Structure, ISM: Magnetic Fields, Stars: Pulsars: General},
  year = 2006,
  month = may,
  volume = 642,
  pages = {868-881}
}

@article{hs08,
  author = {Hemberger, D.~A. and Stinebring, D.~R.},
  title = {Time Variability of Interstellar Scattering and Improvements to Pulsar Timing},
  journal = {\apjl},
  keywords = {ISM: Structure, Stars: Pulsars: General, pulsars: individual (PSR B1737+13), Scattering, Techniques: Spectroscopic},
  year = {2008},
  month = feb,
  volume = {674},
  pages = {L37-L40}
}

@article{iih+05,
  author = {Ilyasov, Y.~P. and Imae, M. and Hanado, Y. and Oreshko, V.~V. and Potapov, V.~A. and Rodin, A.~E. and Sekido, M.},
  title = {Two-frequency timing of the pulsar B1936+21 in Kalyazin and Kashima in 1997 - 2002},
  journal = asl,
  year = {2005},
  month = jan,
  volume = {31},
  pages = {30--36}
}

@inproceedings{jsk+08,
  author = {Janssen, G.~H. and Stappers, B.~W. and Kramer, M. and Purver, M. and Jessner, A. and Cognard, I.},
  title = {European Pulsar Timing Array},
  booktitle = {40 Years of Pulsars: Millisecond Pulsars, Magnetars and More},
  year = {2008},
  series = {American Institute of Physics Conference Series},
  volume = 983,
  editor = {Bassa, C. and Wang, Z. and Cumming, A. and Kaspi, V.~M.},
  month = feb,
  pages = {633-635}
}

@article{jbo+07,
  author = {Jacoby, B.~A. and Bailes, M. and Ord, S.~M. and Knight, H.~S. and Hotan, A.~W.},
  title = {Discovery of Five Recycled Pulsars in a High Galactic Latitude Survey},
  journal = {\apj},
  eprint = {arXiv:astro-ph/0609448},
  year = {2007},
  month = feb,
  volume = {656},
  pages = {408--413}
}

@article{jbo+09,
  author = {{Jacoby}, B.~A. and {Bailes}, M. and {Ord}, S.~M. and {Edwards}, R.~T. and {Kulkarni}, S.~R.},
  title = "{A Large-Area Survey for Radio Pulsars at High Galactic Latitudes}",
  journal = {\apj},
  keywords = {pulsars: general, stars: neutron, surveys},
  year = 2009,
  month = jul,
  volume = 699,
  pages = {2009-2016}
}

@article{jhv+05,
  author = {Johnston, S. and Hobbs, G. and Vigeland, S. and Kramer, M. and Weisberg, J. M. and Lyne, A. G.},
  title = {Evidence for alignment of the rotation and velocity vectors in pulsars},
  journal = {\mnras},
  note = {In Press},
  year = 2005
}

@inproceedings{jng08,
  author = {Jin, C.~J. and Nan, R.~D. and Gan, H.~Q.},
  title = {The FAST telescope and its possible contribution to high precision astrometry},
  keywords = {telescopes, masers, pulsars: general, gravitational waves, reference systems},
  booktitle = {IAU Symposium},
  year = 2008,
  series = {IAU Symposium},
  volume = 248,
  editor = {Jin, W.~J. and Platais, I. and Perryman, M.~A.~C.},
  month = jul,
  pages = {178-181}
}

@inproceedings{jon07,
  author = {Jonas, J.},
  title = {MeerKAT science and technology},
  booktitle = {From Planets to Dark Energy: the Modern Radio Universe. October 1-5 2007, The University of Manchester, UK. Published online at SISSA, Proceedings of Science, p.7},
  year = 2007,
}

@article{jsb+10,
  author = {{Janssen}, G.~H. and {Stappers}, B.~W. and {Bassa}, C.~G. and {Cognard}, I. and {Kramer}, M. and {Theureau}, G.},
  title = "{Long-term timing of four millisecond pulsars}",
  journal = aap,
  year = 2010,
  month = may,
  volume = 514,
  pages = {A74+}
}

@article{kb04,
   author = {{Krasinsky}, G.~A. and {Brumberg}, V.~A.},
    title = "{Secular increase of astronomical unit from analysis of the major planet motions, and its interpretation}",
  journal = {Celestial Mechanics and Dynamical Astronomy},
     year = 2004,
    month = nov,
   volume = 90,
    pages = {267-288},
      doi = {10.1007/s10569-004-0633-z},
   adsurl = {http://adsabs.harvard.edu/abs/2004CeMDA..90..267K},
  adsnote = {Provided by the Smithsonian/NASA Astrophysics Data System}
}

@article{kbm+06,
   author = {{Knight}, H.~S. and {Bailes}, M. and {Manchester}, R.~N. and
    {Ord}, S.~M.},
    title = "{A Study of Giant Pulses from PSR J1824-2452A}",
  journal = {\apj},
   eprint = {astro-ph/0608155},
 keywords = {Stars: Pulsars: General, Stars: Pulsars: Individual: Alphanumeric: PSR B1821-24, pulsars: individual (PSR J1824-2452), pulsars: individual (PSR J1824-2452A)},
     year = 2006,
    month = dec,
   volume = 653,
    pages = {580-586},
      doi = {10.1086/508253},
   adsurl = {http://adsabs.harvard.edu/abs/2006ApJ...653..580K},
  adsnote = {Provided by the SAO/NASA Astrophysics Data System}
}

@article{kel+09,
  author = {Keith, M.~J. and Eatough, R.~P. and Lyne, A.~G. and Kramer, M. and Possenti, A. and Camilo, F. and Manchester, R.~N.},
  title = {Discovery of 28 pulsars using new techniques for sorting pulsar candidates},
  journal = {\mnras},
  year = 2009,
  month = may,
  volume = 395,
  pages = {837-846}
}

@article{kir05,
  author = {Kirk, J.},
  title = {Dissipation in pulsar winds},
  journal = {Adv. in Space Res.},
  volume = {},
  pages = {},
  year = {2005},
  note = {in press (astro-ph/0504410)}
}

@article{klo+06,
  author = {Kramer, M. and Lyne, A.~G. and O'Brien, J.~T. and Jordan, C.~A. and Lorimer, D.~R.},
  title = {A Periodically Active Pulsar Giving Insight into Magnetospheric Physics},
  journal = sci,
  eprint = {arXiv:astro-ph/0604605},
  year = 2006,
  month = apr,
  volume = 312,
  pages = {549-551}
}

@inproceedings{krh05,
  author = {Kaspi, V. M. and Roberts, M. S. E. and Harding, A. K.},
  title = {Isolated neutron stars},
  booktitle = {Compact Stellar X-ray Sources},
  editor = {W. H. G. Lewin and M. van der Klis},
  publisher = {CUP},
  address = {Cambridge},
  pages = { },
  year = {2005},
  note = {in press (astro-ph/0402136)}
}

@article{kva07,
  author = {Kaplan, D.~L. and {van Kerkwijk}, M.~H. and Anderson, J.},
  title = "{The Distance to the Isolated Neutron Star RX J0720.4--3125}",
  journal = {\apj},
  eprint = {arXiv:astro-ph/0703343},
  keywords = {Astrometry, Stars: Individual: Alphanumeric: RX J0720.4--3125, Stars: Neutron, X-rays: individual (RX J0720.4--3125)},
  year = 2007,
  month = may,
  volume = 660,
  pages = {1428-1443}
}

@article{kkn16,
	doi = {10.3847/0004-637x/826/1/86},
  
	url = {https://doi.org/10.3847
  
	year = 2016,
	month = {jul},
  
	publisher = {American Astronomical Society},
  
	volume = {826},
  
	number = {1},
  
	pages = {86},
  
	author = {David L. Kaplan and Thomas Kupfer and David J. Nice and Andreas Irrgang and Ulrich Heber and Zaven Arzoumanian and Elif Beklen and Kathryn Crowter and Megan E. DeCesar and Paul B. Demorest and Timothy Dolch and Justin A. Ellis and Robert D. Ferdman and Elizabeth C. Ferrara and Emmanuel Fonseca and Peter A. Gentile and Glenn Jones and Megan L. Jones and Simon Kreuzer and Michael T. Lam and Lina Levin and Duncan R. Lorimer and Ryan S. Lynch and Maura A. McLaughlin and Adam A. Miller and Cherry Ng and Timothy T. Pennucci and Tom A. Prince and Scott M. Ransom and Paul S. Ray and Renee Spiewak and Ingrid H. Stairs and Kevin Stovall and Joseph Swiggum and Weiwei Zhu},
  
	title = {{PSR} J1024{\textendash}0719: A {MILLISECOND} {PULSAR} {IN} {AN} {UNUSUAL} {LONG}-{PERIOD} {ORBIT}},
	journal = {\apj}
}

@article{lbm+07,
  author = {Lorimer, D.~R. and Bailes, M. and McLaughlin, M.~A. and Narkevic, D.~J. and Crawford, F.},
  title = {A Bright Millisecond Radio Burst of Extragalactic Origin},
  journal = sci,
  archivePrefix = "arXiv",
  eprint = {0709.4301},
  year = 2007,
  month = nov,
  volume = 318,
  pages = {777-}
}

@article{lfl+06,
  author = {Lorimer, D.~R. and Faulkner, A.~J. and Lyne, A.~G. and Manchester, R.~N. and Kramer, M. and McLaughlin, M.~A. and Hobbs, G. and Possenti, A. and Stairs, I.~H. and Camilo, F. and Burgay, M. and {D'Amico}, N. and Corongiu, A. and Crawford, F.},
  title = {The Parkes Multibeam Pulsar Survey - VI. Discovery and timing of 142 pulsars and a Galactic population analysis},
  journal = {\mnras},
  eprint = {arXiv:astro-ph/0607640},
  keywords = {methods: statistical, stars: neutron, pulsars: general},
  year = 2006,
  month = oct,
  volume = 372,
  pages = {777-800}
}

@article{ljp08,
  author = {Lee, K.~J. and Jenet, F.~A. and Price, R.~H.},
  title = {Pulsar Timing as a Probe of Non-Einsteinian Polarizations of Gravitational Waves},
  journal = {\apj},
  year = 2008,
  month = oct,
  volume = 685,
  pages = {1304-1319}
}

@article{lk73,
  author = {Lutz, T.~E. and Kelker, D.~H.},
  title = {On the Use of Trigonometric Parallaxes for the Calibration of Luminosity Systems: Theory},
  journal = pasp,
  year = 1973,
  month = oct,
  volume = 85,
  pages = {573-+}
}

@article{lkg+07,
  author = {Livingstone, M.~A. and Kaspi, V.~M. and Gavriil, F.~P. and Manchester, R.~N. and Gotthelf, E.~V.~G. and Kuiper, L.},
  title = {New phase-coherent measurements of pulsar braking indices},
  journal = apspsci,
  eprint = {arXiv:astro-ph/0702196},
  year = 2007,
  month = apr,
  volume = 308,
  pages = {317-323}
}

@article{lkn+06,
  author = {{Lommen}, A.~N. and {Kipphorn}, R.~A. and {Nice}, D.~J. and {Splaver}, E.~M. and {Stairs}, I.~H. and {Backer}, D.~C.},
  title = {The Parallax and Proper Motion of PSR J0030+0451},
  journal = {\apj},
  eprint = "arXiv:astro-ph/0601521",
  keywords = {Stars: Binaries: General, pulsars: individual (J0030+0451), Galaxy: Solar Neighborhood, Sun: Solar Wind, Stars: Distances},
  year = 2006,
  month = may,
  volume = 642,
  pages = {1012-1017}
}

@article{lmcs07,
  author = {{Lorimer}, D.~R. and {McLaughlin}, M.~A. and {Champion}, D.~J. and {Stairs}, I.~H.},
  title = "{PSR J1453+1902 and the radio luminosities of solitary versus binary millisecond pulsars}",
  journal = {\mnras},
  archivePrefix = "arXiv",
  eprint = {0705.0685},
  keywords = {pulsars: general, pulsars: individual: PSR J1453+1902},
  year = 2007,
  month = jul,
  volume = 379,
  pages = {282-288}
}

@article{lp07,
  author = {{Lattimer}, J.~M. and {Prakash}, M.},
  title = "{Neutron star observations: Prognosis for equations of state constraints}",
  journal = physrep,
  eprint = {arXiv:astro-ph/0612440},
  year = 2007,
  month = apr,
  volume = 442,
  pages = {109,165}
}

@article{lut79,
  author = {Lutz, T.~E.},
  title = {On the use of trigonometric parallaxes for the calibration of luminosity systems. II},
  journal = {\mnras},
  keywords = {calibrating, parallax, stellar luminosity, astronomical models, stellar magnitude, trigonometry},
  year = 1979,
  month = oct,
  volume = 189,
  pages = {273-278}
}

@article{lwj+09,
  author = {Lazaridis, K. and Wex, N. and Jessner, A. and Kramer, M. and Stappers, B.~W. and Janssen, G.~H. and Desvignes, G. and Purver, M.~B. and Cognard, I. and Theureau, G. and Lyne, A.~G. and Jordan, C.~A. and Zensus, J.~A.},
  title = {Generic tests of the existence of the gravitational dipole radiation and the variation of the gravitational constant},
  journal = {\mnras},
  archivePrefix = "arXiv",
  eprint = {0908.0285},
  keywords = {binaries: general, pulsars: general, pulsars: individual: PSR J1012+5307},
  year = 2009,
  month = dec,
  volume = 400,
  pages = {805-814}
}


@article{man06b,
  author = {{Manchester}, R.~N.},
  title = {The Parkes Pulsar Timing Array},
  journal = chjaa,
  volume = {6},
  pages = {139-147},
  year = {2006}
}

@inproceedings{man08,
  author = {Manchester, R.~N.},
  title = {The {P}arkes {P}ulsar {T}iming {A}rray Project},
  booktitle = {40 Years of Pulsars: Millisecond Pulsars, Magnetars and More},
  year = {2008},
  series = {American Institute of Physics Conference Series},
  volume = {983},
  editor = {{Bassa}, C. and {Wang}, Z. and {Cumming}, A. and {Kaspi}, V.~M.},
  month = feb,
  pages = {584--592}
}

@article{mll+06,
  title = {Transient radio bursts from rotating neutron stars},
  author = {McLaughlin, M. A. and Lyne, A. G. and Lorimer, D. R. and Kramer, M. and Faulkner, A. J. and Manchester, R. N. and Cordes, J. M. and Camilo, F. and Possenti, A. and Stairs, I. H. and Hobbs, G. and D'Amico, N. and Burgay, M. and O'Brien, J. T.},
  year = 2006,
  journal = {\nat},
  volume = 439,
  pages = {817-820}
}

@article{mpw08,
  author = {{Melatos}, A. and {Peralta}, C. and {Whyithe}, J.~S.~B.},
  title = "{Avalanche Dynamics of Radio Pulsar Glitches}",
  journal = {\apj},
  archivePrefix = "arXiv",
  eprint = {0710.1021},
  keywords = {Dense Matter, Stars: Pulsars: General, Stars: Interiors, Stars: Neutron, Stars: Rotation},
  year = 2008,
  month = jan,
  volume = 672,
  pages = {1103-1118}
}

@article{nan06,
  author = {{Nan}, R.},
  title = {Five hundred meter aperture spherical radio telescope (FAST)},
  journal = {Science in China G: Physics and Astronomy},
  year = 2006,
  month = mar,
  volume = 49,
  pages = {129-148}
}

@article{nic06,
  author = {Nice, David~J.},
  title = {Neutron star masses derived from relativistic measurements of radio pulsars},
  journal = asr,
  volume  = {38},
  pages = {2721--2724},
  year = {2006}
}

@article{nrb+07,
  author = {Ng, C.~Y. and Romani, R.~W. and Brisken, W.~F. and Chatterjee, S. and Kramer, M.},
  title = {The Origin and Motion of PSR J0538+2817 in S147},
  journal = {\apj},
  eprint = {arXiv:astro-ph/0611068},
  keywords = {Astrometry, Stars: Pulsars: Individual: Alphanumeric: PSR J0538+2817, Stars: Kinematics, supernovae: individual (S147)},
  year = 2007,
  month = jan,
  volume = 654,
  pages = {487-493}
}

@article{ojhb06,
  author = {Ord, S.~M. and Jacoby, B.~A. and Hotan, A.~W. and Bailes, M.},
  title = {High-precision timing of PSR J1600--3053},
  journal = {\mnras},
  year = {2006},
  month = sep,
  volume = {371},
  pages = {337--342}
}

@article{ojk+08,
  author = {{O'Brien}, J.~T. and {Johnston}, S. and {Kramer}, M. and {Lyne}, A.~G. and {Bailes}, M. and {Possenti}, A. and {Burgay}, M. and {Lorimer}, D.~R. and {McLaughlin}, M.~A. and {Hobbs}, G. and {Parent}, D. and {Guillemot}, L.},
  title = "{PSR J1410-6132: a young, energetic pulsar associated with the EGRET source 3EG J1410-6147}",
  journal = {\mnras},
  archivePrefix = "arXiv",
  eprint = {0806.0431},
  keywords = {methods: observational, pulsars: general, pulsars: individual: J1410-6132},
  year = 2008,
  month = jul,
  volume = 388,
  pages = {L1-L5}
}

@article{ojs07,
  author = {{Ord}, S.~M. and {Johnston}, S. and {Sarkissian}, J.},
  title = {The Magnetic Field of the Solar Corona from Pulsar Observations},
  journal = solphys,
  archivePrefix = "arXiv",
  eprint = {0705.1869},
  year = 2007,
  month = sep,
  volume = 245,
  pages = {109-120}
}

@article{pit05,
  author = {Pitjeva, E.~V.},
  title = {High-Precision Ephemerides of Planets{\mdash}EPM and Determination of Some Astronomical Constants},
  journal = {Solar System Research},
  year = 2005,
  month = may,
  volume = 39,
  pages = {176-186}
}

@article{plps04,
  author = {Page, D. and Lattimer, J.~M. and Prakash, M. and Steiner, A.~W.},
  title = {Minimal Cooling of Neutron Stars: A New Paradigm},
  journal = \apjs,
  eprint = {arXiv:astro-ph/0403657},
  keywords = {Dense Matter, Equation of State, Neutrinos, Stars: Neutron},
  year = 2004,
  month = dec,
  volume = 155,
  pages = {623-650}
}

@article{pns+01,
  author = {Peng, B. and Nan, R. and Su, Y. and Qiu, Y. and Zhu, L. and Zhu, W.},
  title = {Five-hundred-meter Aperture Spherical Telescope project},
  journal = ass,
  year = 2001,
  volume = 278,
  pages = {219-224}
}

@book{pri00,
  author = {Prialnik, D.},
  title = {An Introduction to the Theory of Stellar Structure and Evolution},
  publisher = {Cambridge University Press},
  year = 2000
}

@article{rdb+06,
  author = {Ramachandran, R. and Demorest, P. and Backer, D.~C. and Cognard, I. and Lommen, A.},
  title = {Interstellar Plasma Weather Effects in Long-Term Multifrequency Timing of Pulsar B1937+21},
  journal = {\apj},
  eprint = {arXiv:astro-ph/0601242},
  year = 2006,
  month = jul,
  volume = 645,
  pages = {303-313}
}

@article{rt91,
  author = {Ryba, M.~F. and Taylor, J.~H.},
  title = {High-precision timing of millisecond pulsars. I - Astrometry and masses of the PSR 1855+09 system},
  journal = {\apj},
  year = 1991,
  month = apr,
  volume = 371,
  pages = {739-748}
}

@book{sch93b,
  author = {Schutz, B.~F.},
  title = {A first course in general relativity},
  publisher = {Cambridge University Press},
  year = 1993
}

@article{sgs05,
  author = {Seward, F. D. and Gorenstein, P. and Smith, R. K.},
  title = {Chandra observations of the X-ray halo around the {Crab Nebula}},
  journal = {\apj},
  volume = {},
  pages = {},
  year = {2005},
  note = {submitted}
}

@article{sgs06,
  author = {{Smirnova}, T.~V. and {Gwinn}, C.~R. and {Shishov}, V.~I.},
  title = {Interstellar scintillation of PSR J0437-4715},
  journal = aap,
  eprint = {arXiv:astro-ph/0603490},
  keywords = {turbulence, pulsars: individual: PSR J0437-4715, scattering},
  year = 2006,
  month = jul,
  volume = 453,
  pages = {601-607}
}

@article{smi03b,
  author = {Smith, H.},
  title = {Is there really a Lutz-Kelker bias? Reconsidering calibration with trigonometric parallaxes},
  journal = {\mnras},
  keywords = {methods:statistical, astrometry, stars: distances, stars: fundamental parameters},
  year = 2003,
  month = feb,
  volume = 338,
  pages = {891-902}
}

@article{spg+05,
  author = {{Sasaki}, M. and {Plucinsky}, P.~P. and {Gaetz}, T.~J. and {Smith}, R.~K. and   {Edgar}, R.~J. and {Slane}, P.~O.},
  title = {{XMM-Newton observations of the Galactic Supernova Remnant CTB 109 (G109.1-1.0)}},
  journal = {\apj},
  year = {2005},
  note = {in press (astro-ph/0408290)}
}

@article{sti07,
  author = {{Stinebring}, D.},
  title = "{Using pulsar scintillation to probe AU-size structure in the interstellar medium}",
  journal = {Astronomical and Astrophysical Transactions},
  year = 2007,
  month = dec,
  volume = 26,
  pages = {517-524}
}

@article{svc08,
  author = {Sesana, A. and Vecchio, A. and Colacino, C.~N.},
  title = {The stochastic gravitational-wave background from massive black hole binary systems: implications for observations with Pulsar Timing Arrays},
  journal = {\mnras},
  archivePrefix = "arXiv",
  eprint = {0804.4476},
  year = 2008,
  month = oct,
  volume = 390,
  pages = {192-209}
}

@article{swb96,
  author = {{Staveley-Smith}, L. and {Wilson}, W.~E. and {Bird}, T.~S. and {Disney}, M.~J. and {Ekers}, R.~D. and {Freeman}, K.~C. and {Haynes}, R.~F. and {Sinclair}, M.~W. and {Vaile}, R.~A. and {Webster}, R.~L. and {Wright}, A.~E.},
  title = "{The Parkes 21 CM multibeam receiver}",
journal = {Publications of the Astronomical Society of Australia},
year = 1996,
month = nov,
volume = 13,
pages = {243-248}
}

@INPROCEEDINGS{tay93,
   author = {{Taylor}, J.~H.},
    title = "{Probing relativistic gravity with pulsar timing experiments}",
booktitle = {Particle Astrophysics},
     year = 1993,
   editor = {{Fontaine}, G. and {Tran Thanh van}, J.},
    pages = {367-+},
   adsurl = {http://adsabs.harvard.edu/abs/1993paas.conf..367T},
  adsnote = {Provided by the Smithsonian/NASA Astrophysics Data System}
}

@article{tlmr07,
   author = {{Torres}, R.~M. and {Loinard}, L. and {Mioduszewski}, A.~J. and
   {Rodriguez}, L.~F.},
    title = "{VLBA determination of the distance to nearby star-forming regions II. Hubble 4 and HDE 283572 in Taurus}",
  journal = {ArXiv e-prints},
   eprint = {0708.4403},
     year = 2007,
    month = aug,
   volume = 708
}

@inproceedings{val07,
  author = {{Valls-Gabaud}, D.},
  title = "{The Distance to the Pleiades Revisited}",
  booktitle = {IAU Symposium},
  year = 2007,
  series = {IAU Symposium},
  volume = 240,
  month = aug,
  pages = {281-289}
}

@ARTICLE{vb10,
   author = {{van Straten}, W. and {Bailes}, M.},
    title = "{DSPSR: Digital Signal Processing Software for Pulsar Astronomy}",
  journal = pasa,
archivePrefix = "arXiv",
   eprint = {1008.3973},
 primaryClass = "astro-ph.IM",
 keywords = {methods: data analysis, polarisation, pulsars: general, techniques: polarimetric},
     year = 2011,
    month = jan,
   volume = 28,
    pages = {1-14},
      doi = {10.1071/AS10021},
   adsurl = {http://adsabs.harvard.edu/abs/2011PASA...28....1V},
  adsnote = {Provided by the SAO/NASA Astrophysics Data System}
}

@article{vbb+10,
  author = {Verbiest, J.~P.~W. Bailes, M. and Bhat, N.~D.~R. and Burke-Spolaor, S. and Champion, D.~J. and Coles, W. and Hobbs, G.~B. and Hotan, A.~W. and Jenet, F. and Khoo, J. and Lee, K.~J. and Lommen, A. and Manchester, R.~N. and Oslowsk, S. and Reynolds, J. and Sarkissian, J. and van Straten, W. and Yardley, D.~R.~B. and You, X.~P.},
  title = {Status update of the Parkes pulsar timing array},
  journal = {Classical and Quantum Gravity},
  archivePrefix = "arXiv",
  eprint = {0912.2692},
  primaryClass = "astro-ph.GA",
  year = 2010,
  month = apr,
  volume = 27,
  number = 8,
  pages = {084015-+}
}

@article{vbc+09,
  author = {{Verbiest}, J.~P.~W. and {Bailes}, M. and {Coles}, W.~A. and {Hobbs}, G.~B. and {van Straten}, W. and {Champion}, D.~J. and {Jenet}, F.~A. and {Manchester}, R.~N. and {Bhat}, N.~D.~R. and {Sarkissian}, J.~M. and {Yardley}, D. and {Burke-Spolaor}, S. and {Hotan}, A.~W. and {You}, X.~P.},
  title = {Timing stability of millisecond pulsars and prospects for gravitational-wave detection},
  journal = {\mnras},
  archivePrefix = "arXiv",
  eprint = {0908.0244},
  keywords = {gravitational waves, pulsars: general},
  year = 2009,
  month = dec,
  volume = 400,
  pages = {951-968}
}

@article{vbv+08,
  author = {{Verbiest}, J.~P.~W. and {Bailes}, M. and {van Straten}, W. and {Hobbs}, G.~B. and {Edwards}, R.~T. and {Manchester}, R.~N. and {Bhat}, N.~D.~R. and {Sarkissian}, J.~M. and {Jacoby}, B.~A. and {Kulkarni}, S.~R.},
  title = {Precision Timing of PSR J0437--4715: An Accurate Pulsar Distance, a High Pulsar Mass, and a Limit on the Variation of Newton's Gravitational Constant},
  journal = {\apj},
  eprint = {arXiv:0801.2589},
  year = {2008},
  month = may,
  volume = {679},
  pages = {675--680}
}

@article{vk07,
  author = {{van Kerkwijk}, M.~H. and Kaplan, D.~L.},
  title = "{Isolated neutron stars: magnetic fields, distances, and spectra}",
  journal = apss,
  eprint = {arXiv:astro-ph/0607320},
  year = 2007,
  month = apr,
  volume = 308,
  pages = {191-201}
}

@article{vlml08,
  author = {van Haasteren, R. and Levin, Y. and McDonald, P. and Lu, T.},
  title = {On measuring the gravitational-wave background using Pulsar Timing Arrays},
  journal = {ArXiv e-prints},
  archivePrefix = "arXiv",
  eprint = {0809.0791},
  keywords = {Astrophysics, General Relativity and Quantum Cosmology},
  year = 2008,
  month = sep
}

@article{wei08,
  author = {Weise, W.},
  title = {Overview and Perspectives in Nuclear Physics},
  journal = {Nuclear Physics A},
  archivePrefix = "arXiv",
  eprint = {0801.1619},
  year = 2008,
  month = jun,
  volume = 805,
  pages = {115-126}
}

@article{wk05,
  author = {{Willems}, B. and {Kalogera}, V.},
  title = {{Comment on "Origin of the binary pulsar J0737-3039B".}},
  journal = {submitted to Physical Review Letters},
  year = 2005
}

@article{wksv08,
  author = {Walker, M.~A. and Koopmans, L.~V.~E. and Stinebring, D.~R. and {van Straten}, W.},
  title = {Interstellar holography},
  journal = {\mnras},
  archivePrefix = "arXiv",
  eprint = {0801.4183},
  keywords = {scattering, turbulence, techniques: interferometric, pulsars: general, pulsars: individual: B0834+06, ISM: structure},
  year = {2008},
  month = aug,
  volume = {388},
  pages = {1214-1222}
}

@article{wmsz04,
  author = {{Walker}, M.~A. and {Melrose}, D.~B. and {Stinebring}, D.~R. and {Zhang}, C.~M.},
  title = {Interpretation of parabolic arcs in pulsar secondary spectra},
  journal = {\mnras},
  eprint = {arXiv:astro-ph/0403587},
  keywords = {pulsars: general, ISM: general},
  year = 2004,
  month = oct,
  volume = 354,
  pages = {43-54}
}

@article{yhj+09,
  author = {Yardley, D.~R.~B. and Hobbs, G.~B. and Jenet, F.~A. and Wen, Z.~L. and Verbiest, J.~P.~W. and Manchester, R.~N. and Coles, W.~A. and Bailes, M. and Bhat, N.~D.~R. and Burke-Spolaor, S. and Champion, D.~J. and Hotan, A.~W. and Sarkissian, J.~M. and {van Straten}, W.},
  title = {The Sensitivity of the Parkes Pulsar Timing Array to Sinusoidal Sources of Gravitational Waves},
  journal = {\mnras},
  year = 2009,
  note = {to be submitted}
}

@article{yh05,
  author = {You, X.-P. and Han, J.-L.},
  title = {Circular polarization in pulsar integrated profiles: Updates},
  journal = chjaa,
  year = 2005,
  note = {In Press}
}

@article{yhc+07,
  author = {You, X.-P. and Hobbs, G. and Coles, W. and Manchester, R.~N. and Edwards, R. and Bailes, M. and Sarkissian, J. and Verbiest, J.~P.~W. and {van Straten}, W. and Hotan, A. and Ord, S. and Jenet, F. and Bhat, N.~D.~R. and Teoh, A.},
  title = {Dispersion Measure Variations and their Effect on Precision Pulsar Timing},
  journal = {\mnras},
  eprint = {arXiv:astro-ph/0702366},
  volume = 378,
  pages = {493-506},
  year = {2007},
  month = jun
}

@article{yhc+07b,
  author = {You, X.~P. and Hobbs, G.~B. and Coles, W.~A. and Manchester, R.~N. and Han, J.~L.},
  title = {An Improved Solar Wind Electron Density Model for Pulsar Timing},
  journal = {\apj},
  archivePrefix = "arXiv",
  eprint = {0709.0135},
  year = 2007,
  month = dec,
  volume = 671,
  pages = {907-911}
}

@techreport{sac+07,
  author = {{Schilizzi}, R.~T. and {Alexander}, P. and {Cordes}, J.~M. and {Dewdney}, P.~E. and {Ekers}, R.~D. and {Faulkner}, A.~J. and {Gaensler}, B.~M. and {Hall}, P.~J. and {Jonas}, J.~L. and {Kellerman}, K.~I.},
  title = {Preliminary Specifications for the Square Kilometre Array},
  year = 2007,
  month = 12,
  type = "Memo",
  institution = {SKA Program Development Office},
  number = 100
}

@book{mw70,
  author =  {{Mathews}, J. and {Walker}, R.~L.},
  title = {Mathematical methods of physics},
  year = 1970,
  publisher = {Addison-Wesley World Student Series},
  address = {Menlo Park, Ca.},
}

@article{cs10,
  author = {{Cordes}, J.~M. and {Shannon}, R.~M.},
  title = {A measurement model for precision pulsar timing},
  journal = {astro-ph/1107.3086},
  archivePrefix = "arXiv",
  eprint = {1010.3785},
  year = 2010,
  month = Oct,
}

@article{van11,
  author = {{van Straten}, W.},
  title = {},
  journal = {in preparation},
  year = 2011,
}

@ARTICLE{g58,
   author = {{Garfinkel}, B.},
    title = "{On the motion of a satellite of an oblate planet}",
  journal = aj,
     year = 1958,
    month = mar,
   volume = 63,
    pages = {88-+},
      doi = {10.1086/107697},
   adsurl = {http://adsabs.harvard.edu/abs/1958AJ.....63...88G},
  adsnote = {Provided by the SAO/NASA Astrophysics Data System}
}

@ARTICLE{g59,
   author = {{Garfinkel}, B.},
    title = "{The orbit of a satellite of an oblate planet}",
  journal = aj,
     year = 1959,
    month = nov,
   volume = 64,
    pages = {353-+},
      doi = {10.1086/107956},
   adsurl = {http://adsabs.harvard.edu/abs/1959AJ.....64..353G},
  adsnote = {Provided by the SAO/NASA Astrophysics Data System}
}

@ARTICLE{g60,
   author = {{Garfinkel}, B.},
    title = "{On the motion of a satellite in the vicinity of the critical inclination}",
  journal = {\aj},
     year = 1960,
    month = dec,
   volume = 65,
    pages = {624-+},
      doi = {10.1086/108308},
   adsurl = {http://adsabs.harvard.edu/abs/1960AJ.....65..624G},
  adsnote = {Provided by the SAO/NASA Astrophysics Data System}
}

@ARTICLE{s57,
   author = {{Sterne}, T.~F.},
    title = "{Analytical orbits of close satellites of oblate planets.}",
  journal = {\aj},
     year = 1957,
    month = may,
   volume = 62,
    pages = {96-+},
      doi = {10.1086/107471},
   adsurl = {http://adsabs.harvard.edu/abs/1957AJ.....62...96S},
  adsnote = {Provided by the SAO/NASA Astrophysics Data System}
}

@book{b91,
  author = {Brumberg, V. A.},
  title = {Essential Relativistic Celestial Mechanics},
  year = 1991,
  publisher = {Bristol: Hilger},
}

@ARTICLE{kw09,
   author = {{Kramer}, M and {Wex}, N},
    title = "{TOPICAL REVIEW:  The double pulsar system: a unique laboratory for gravity}",
  journal = {Classical and Quantum Gravity},
     year = 2009,
    month = apr,
   volume = 26,
   number = 7,
      eid = {073001},
    pages = {073001},
      doi = {10.1088/0264-9381/26/7/073001},
   adsurl = {http://adsabs.harvard.edu/abs/2009CQGra..26g3001K},
  adsnote = {Provided by the SAO/NASA Astrophysics Data System}
}

@ARTICLE{kop94,
   author = {{Kopeikin}, S. M.},
  journal = {\apj},
     year = 1994,
   volume = 434,
    pages = {L67-+},
}

@ARTICLE{ppr05,
   author = {{Pfahl}, E., and {Podsiadlowski}, P., and {Rappaport}, S.},
  journal = {\apj},
     year = 2005,
   volume = 628,
    pages = 343,
}

@ARTICLE{khm93,
   author = {{Kulkarni}, S. R., and {Hut}, R., and {McMillan}, S.},
  journal = {Nature},
     year = 1993,
   volume = 364,
    pages = 421,
}

@ARTICLE{mo09,
   author = {{Mandel}, I., and {O'Shaughnessy}, R.},
  journal = {preprint},
     year = 2009,
      doi = {arXiv:0912.1074v1},
}

@ARTICLE{ckl+04,
   author = {{Cordes}, J.~M. and {Kramer}, M. and {Lazio}, T.~J.~W. and {Stappers}, B.~W. and {Backer}, D.~C. and {Johnston}, S.},
    title = "{Pulsars as tools for fundamental physics {\amp} astrophysics}",
  journal = nar,
   eprint = {arXiv:astro-ph/0505555},
     year = 2004,
    month = dec,
   volume = 48,
    pages = {1413-1438},
      doi = {10.1016/j.newar.2004.09.040},
   adsurl = {http://adsabs.harvard.edu/abs/2004NewAR..48.1413C},
  adsnote = {Provided by the SAO/NASA Astrophysics Data System}
}

@article{lkl+11,
   author = {{Liu}, K. and {Keane}, E.~F. and {Lee}, K.~J. and {Kramer}, M. and
    {Cordes}, J.~M. and {Purver}, M.~B.},
    title = "{Profile-shape stability and phase-jitter analyses of millisecond pulsars}",
  journal = {\mnras},
archivePrefix = "arXiv",
   eprint = {1110.4759},
 primaryClass = "astro-ph.HE",
 keywords = {methods: data analysis, pulsars: general},
     year = 2012,
    month = feb,
   volume = 420,
    pages = {361-368},
      doi = {10.1111/j.1365-2966.2011.20041.x},
   adsurl = {http://adsabs.harvard.edu/abs/2012{\mnras}.420..361L},
  adsnote = {Provided by the SAO/NASA Astrophysics Data System}
}

@ARTICLE{gsw+08,
   author = {{Ghez}, A.~M. and {Salim}, S. and {Weinberg}, N.~N. and {Lu}, J.~R. and
    {Do}, T. and {Dunn}, J.~K. and {Matthews}, K. and {Morris}, M.~R. and
    {Yelda}, S. and {Becklin}, E.~E. and {Kremenek}, T. and {Milosavljevic}, M. and
    {Naiman}, J.},
    title = "{Measuring Distance and Properties of the Milky Way's Central Supermassive Black Hole with Stellar Orbits}",
  journal = {\apj},
archivePrefix = "arXiv",
   eprint = {0808.2870},
 keywords = {Black Hole Physics, Galaxy: Center, Galaxy: Kinematics and Dynamics, Infrared: Stars, Techniques: High Anular Resolution},
     year = 2008,
    month = dec,
   volume = 689,
    pages = {1044-1062},
      doi = {10.1086/592738},
   adsurl = {http://adsabs.harvard.edu/abs/2008ApJ...689.1044G},
  adsnote = {Provided by the SAO/NASA Astrophysics Data System}
}

@ARTICLE{get+09,
   author = {{Gillessen}, S. and {Eisenhauer}, F. and {Trippe}, S. and {Alexander}, T. and
    {Genzel}, R. and {Martins}, F. and {Ott}, T.},
    title = "{Monitoring Stellar Orbits Around the Massive Black Hole in the Galactic Center}",
  journal = {\apj},
archivePrefix = "arXiv",
   eprint = {0810.4674},
 keywords = {black hole physics, astrometry, Galaxy: center, infrared: stars},
     year = 2009,
    month = feb,
   volume = 692,
    pages = {1075-1109},
      doi = {10.1088/0004-637X/692/2/1075},
   adsurl = {http://adsabs.harvard.edu/abs/2009ApJ...692.1075G},
  adsnote = {Provided by the SAO/NASA Astrophysics Data System}
}

@ARTICLE{zcc97,
   author = {{Zhang}, S.~N. and {Cui}, W. and {Chen}, W.},
    title = "{Black Hole Spin in X-Ray Binaries: Observational Consequences}",
  journal = {ApJL},
   eprint = {arXiv:astro-ph/9704072},
 keywords = {BLACK HOLE PHYSICS, X-RAYS: STARS},
     year = 1997,
    month = jun,
   volume = 482,
    pages = {L155+},
      doi = {10.1086/310705},
   adsurl = {http://adsabs.harvard.edu/abs/1997ApJ...482L.155Z},
  adsnote = {Provided by the SAO/NASA Astrophysics Data System}
}

@ARTICLE{frs+89,
   author = {{Fabian}, A.~C. and {Rees}, M.~J. and {Stella}, L. and {White}, N.~E.
    },
    title = "{X-ray fluorescence from the inner disc in Cygnus X-1}",
  journal = {\mnras},
 keywords = {ACCRETION DISKS, BLACK HOLES (ASTRONOMY), CYGNUS CONSTELLATION, X RAY FLUORESCENCE, X RAY SPECTRA, BLACK BODY RADIATION, EMISSION SPECTRA, LINE SPECTRA, RED SHIFT},
     year = 1989,
    month = may,
   volume = 238,
    pages = {729-736},
   adsurl = {http://adsabs.harvard.edu/abs/1989{\mnras}.238..729F},
  adsnote = {Provided by the SAO/NASA Astrophysics Data System}
}

@ARTICLE{srm+11,
   author = {{Steiner}, J.~F. and {Reis}, R.~C. and {McClintock}, J.~E. and
    {Narayan}, R. and {Remillard}, R.~A. and {Orosz}, J.~A. and
    {Gou}, L. and {Fabian}, A.~C. and {Torres}, M.~A.~P.},
    title = "{The spin of the black hole microquasar XTE J1550-564 via the continuum-fitting and Fe-line methods}",
  journal = {\mnras},
archivePrefix = "arXiv",
   eprint = {1010.1013},
 primaryClass = "astro-ph.HE",
 keywords = {accretion, accretion discs, black hole physics, stars: individual: XTE J1550-564, X-rays: binaries},
     year = 2011,
    month = sep,
   volume = 416,
    pages = {941-958},
      doi = {10.1111/j.1365-2966.2011.19089.x},
   adsurl = {http://adsabs.harvard.edu/abs/2011{\mnras}.416..941S},
  adsnote = {Provided by the SAO/NASA Astrophysics Data System}
}

@ARTICLE{t80,
   author = {{Thorne}, K.~S.},
    title = "{Multipole expansions of gravitational radiation}",
  journal = {Reviews of Modern Physics},
     year = 1980,
    month = apr,
   volume = 52,
    pages = {299-340},
      doi = {10.1103/RevModPhys.52.299},
   adsurl = {http://adsabs.harvard.edu/abs/1980RvMP...52..299T},
  adsnote = {Provided by the SAO/NASA Astrophysics Data System}
}

@ARTICLE{bb11,
   author = {{Bambi}, C. and {Barausse}, E.},
    title = "{Constraining the Quadrupole Moment of Stellar-mass Black Hole Candidates with the Continuum Fitting Method}",
  journal = {\apj},
archivePrefix = "arXiv",
   eprint = {1012.2007},
 primaryClass = "gr-qc",
 keywords = {accretion, accretion disks, black hole physics, gravitation, X-rays: binaries},
     year = 2011,
    month = apr,
   volume = 731,
      eid = {121},
    pages = {121},
      doi = {10.1088/0004-637X/731/2/121},
   adsurl = {http://adsabs.harvard.edu/abs/2011ApJ...731..121B},
  adsnote = {Provided by the SAO/NASA Astrophysics Data System}
}

@book{w84,
  author = {Wald, R.~M.},
  title = {General Relativity},
  year = 1984,
  publisher = {University of Chicago Press, Chicago},
}

@ARTICLE{wnt10,
   author = {{Weisberg}, J.~M. and {Nice}, D.~J. and {Taylor}, J.~H.},
    title = "{Timing Measurements of the Relativistic Binary Pulsar PSR B1913+16}",
  journal = {\apj},
archivePrefix = "arXiv",
   eprint = {1011.0718},
 primaryClass = "astro-ph.GA",
 keywords = {binaries: close, gravitation, pulsars: individual: PSR B1913+16, stars: kinematics and dynamics},
     year = 2010,
    month = oct,
   volume = 722,
    pages = {1030-1034},
      doi = {10.1088/0004-637X/722/2/1030},
   adsurl = {http://adsabs.harvard.edu/abs/2010ApJ...722.1030W},
  adsnote = {Provided by the SAO/NASA Astrophysics Data System}
}

@ARTICLE{yp98,
   author = {{Yungelson}, L. and {Portegies Zwart}, S.~F.},
    title = "{Evolution of Close Binaries: Formation and Merger of Neutron Star Binaries}",
  journal = {arXiv:astro-ph/9801127},
   eprint = {arXiv:astro-ph/9801127},
 keywords = {Astrophysics},
     year = 1998,
    month = jan,
   adsurl = {http://adsabs.harvard.edu/abs/1998astro.ph..1127Y},
  adsnote = {Provided by the SAO/NASA Astrophysics Data System}
}

@ARTICLE{vt03,
   author = {{Voss}, R. and {Tauris}, T.~M.},
    title = "{Galactic distribution of merging neutron stars and black holes - prospects for short gamma-ray burst progenitors and LIGO/VIRGO}",
  journal = {\mnras},
   eprint = {arXiv:astro-ph/0303227},
 keywords = {black hole physics, gravitational waves, methods: numerical, binaries: close, stars: neutron, gamma-rays: bursts},
     year = 2003,
    month = jul,
   volume = 342,
    pages = {1169-1184},
      doi = {10.1046/j.1365-8711.2003.06616.x},
   adsurl = {http://adsabs.harvard.edu/abs/2003{\mnras}.342.1169V},
  adsnote = {Provided by the SAO/NASA Astrophysics Data System}
}

@ARTICLE{spn04,
   author = {{Sipior}, M.~S. and {Portegies Zwart}, S. and {Nelemans}, G.
    },
    title = "{Recycled pulsars with black hole companions: the high-mass analogues of PSR B2303+46}",
  journal = {\mnras},
   eprint = {arXiv:astro-ph/0407268},
 keywords = {binaries: close, pulsars: general},
     year = 2004,
    month = nov,
   volume = 354,
    pages = {L49-L53},
      doi = {10.1111/j.1365-2966.2004.08373.x},
   adsurl = {http://adsabs.harvard.edu/abs/2004{\mnras}.354L..49S},
  adsnote = {Provided by the SAO/NASA Astrophysics Data System}
}

@ARTICLE{fl11,
   author = {{Faucher-Gigu{\`e}re}, C.-A. and {Loeb}, A.},
    title = "{Pulsar-black hole binaries in the Galactic Centre}",
  journal = {\mnras},
archivePrefix = "arXiv",
   eprint = {1012.0573},
 primaryClass = "astro-ph.HE",
 keywords = {black hole physics, binaries: general, stars: neutron, pulsars: general, Galaxy: centre},
     year = 2011,
    month = aug,
   volume = 415,
    pages = {3951-3961},
      doi = {10.1111/j.1365-2966.2011.19019.x},
   adsurl = {http://adsabs.harvard.edu/abs/2011{\mnras}.415.3951F},
  adsnote = {Provided by the SAO/NASA Astrophysics Data System}
}


@ARTICLE{dcm+07,
   author = {{Devecchi}, B. and {Colpi}, M. and {Mapelli}, M. and {Possenti}, A.
    },
    title = "{Millisecond pulsars around intermediate-mass black holes in globular clusters}",
  journal = {\mnras},
archivePrefix = "arXiv",
   eprint = {0706.1656},
 keywords = {black hole physics, stellar dynamics, stars: neutron, pulsars: general, globular clusters: general},
     year = 2007,
    month = sep,
   volume = 380,
    pages = {691-702},
      doi = {10.1111/j.1365-2966.2007.12160.x},
   adsurl = {http://adsabs.harvard.edu/abs/2007{\mnras}.380..691D},
  adsnote = {Provided by the SAO/NASA Astrophysics Data System}
}

@ARTICLE{pcf+05,
   author = {{Patruno}, A. and {Colpi}, M. and {Faulkner}, A. and {Possenti}, A.
    },
    title = "{Radio pulsars around intermediate-mass black holes in superstellar clusters}",
  journal = {\mnras},
   eprint = {arXiv:astro-ph/0507229},
 keywords = {accretion, accretion discs, black hole physics, X-rays: binaries, X-rays: galaxies},
     year = 2005,
    month = nov,
   volume = 364,
    pages = {344-352},
      doi = {10.1111/j.1365-2966.2005.09568.x},
   adsurl = {http://adsabs.harvard.edu/abs/2005{\mnras}.364..344P},
  adsnote = {Provided by the SAO/NASA Astrophysics Data System}
}

@ARTICLE{lba05,
   author = {{Lipunov}, V.~M. and {Bogomazov}, A.~I. and {Abubekerov}, M.~K.
    },
    title = "{How abundant is the population of binary radio pulsars with black holes?}",
  journal = {\mnras},
   eprint = {arXiv:astro-ph/0503341},
 keywords = {black hole physics, stars: abundances, binaries: close, binaries: general, pulsars: general, X-rays: binaries},
     year = 2005,
    month = jun,
   volume = 359,
    pages = {1517-1523},
      doi = {10.1111/j.1365-2966.2005.08997.x},
   adsurl = {http://adsabs.harvard.edu/abs/2005{\mnras}.359.1517L},
  adsnote = {Provided by the SAO/NASA Astrophysics Data System}
}

@ARTICLE{mkf+10,
   author = {{Macquart}, {J.-P.} and {Kanekar}, N. and {Frail}, D.~A. and
    {Ransom}, S.~M.},
    title = "{A High-frequency Search for Pulsars within the Central Parsec of Sgr A*}",
  journal = {\apj},
archivePrefix = "arXiv",
   eprint = {1004.1643},
 primaryClass = "astro-ph.GA",
 keywords = {Galaxy: center, pulsars: general},
     year = 2010,
    month = jun,
   volume = 715,
    pages = {939-946},
      doi = {10.1088/0004-637X/715/2/939},
   adsurl = {http://adsabs.harvard.edu/abs/2010ApJ...715..939M},
  adsnote = {Provided by the SAO/NASA Astrophysics Data System}
}

@ARTICLE{bbf10,
   author = {{Barsdell}, B.~R. and {Barnes}, D.~G. and {Fluke}, C.~J.},
    title = "{Analysing astronomy algorithms for graphics processing units and beyond}",
  journal = {\mnras},
archivePrefix = "arXiv",
   eprint = {1007.1660},
 primaryClass = "astro-ph.IM",
 keywords = {gravitational lensing: micro, methods: data analysis, pulsars: general},
     year = 2010,
    month = nov,
   volume = 408,
    pages = {1936-1944},
      doi = {10.1111/j.1365-2966.2010.17257.x},
   adsurl = {http://adsabs.harvard.edu/abs/2010{\mnras}.408.1936B},
  adsnote = {Provided by the SAO/NASA Astrophysics Data System}
}

@ARTICLE{kac+10,
   author = {{Knispel}, B. and {Allen}, B. and {Cordes}, J.~M. and {Deneva}, J.~S. and
    {Anderson}, D. and {Aulbert}, C. and {Bhat}, N.~D.~R. and {Bock}, O. and
    {Bogdanov}, S. and {Brazier}, A. and {Camilo}, F. and {Champion}, D.~J. and
    {Chatterjee}, S. and {Crawford}, F. and {Demorest}, P.~B. and
    {Fehrmann}, H. and {Freire}, P.~C.~C. and {Gonzalez}, M.~E. and
    {Hammer}, D. and {Hessels}, J.~W.~T. and {Jenet}, F.~A. and {Kasian}, L. and {Kaspi}, V.~M. and {Kramer}, M. and {Lazarus}, P. and
    {van Leeuwen}, J. and {Lorimer}, D.~R. and {Lyne}, A.~G. and
    {Machenschalk}, B. and {McLaughlin}, M.~A. and {Messenger}, C. and
    {Nice}, D.~J. and {Papa}, M.~A. and {Pletsch}, H.~J. and {Prix}, R. and
    {Ransom}, S.~M. and {Siemens}, X. and {Stairs}, I.~H. and {Stappers}, B.~W. and
    {Stovall}, K. and {Venkataraman}, A.},
    title = "{Pulsar Discovery by Global Volunteer Computing}",
  journal = {Science},
archivePrefix = "arXiv",
   eprint = {1008.2172},
 primaryClass = "astro-ph.GA",
     year = 2010,
    month = sep,
   volume = 329,
    pages = {1305-},
      doi = {10.1126/science.1195253},
   adsurl = {http://adsabs.harvard.edu/abs/2010Sci...329.1305K},
  adsnote = {Provided by the SAO/NASA Astrophysics Data System}
}

@ARTICLE{skm+11,
   author = {{Simonetti}, J.~H. and {Kavic}, M. and {Minic}, D. and {Surani}, U. and
    {Vijayan}, V.},
    title = "{A Precision Test for an Extra Spatial Dimension Using Black-hole-Pulsar Binaries}",
  journal = {\apjl},
archivePrefix = "arXiv",
   eprint = {1010.5245},
 primaryClass = "astro-ph.HE",
 keywords = {binaries: close, black hole physics, gravitation, pulsars: general},
     year = 2011,
    month = aug,
   volume = 737,
      eid = {L28},
    pages = {L28},
      doi = {10.1088/2041-8205/737/2/L28},
   adsurl = {http://adsabs.harvard.edu/abs/2011ApJ...737L..28S},
  adsnote = {Provided by the SAO/NASA Astrophysics Data System}
}

@ARTICLE{wk07,
   author = {{Wex}, N. and {Kramer}, M.},
    title = "{A characteristic observable signature of preferred-frame effects in relativistic binary pulsars}",
  journal = {\mnras},
archivePrefix = "arXiv",
   eprint = {0706.2382},
 keywords = {gravitation, pulsars: general, pulsars: individual: PSR J0737-3039},
     year = 2007,
    month = sep,
   volume = 380,
    pages = {455-465},
      doi = {10.1111/j.1365-2966.2007.12093.x},
   adsurl = {http://adsabs.harvard.edu/abs/2007{\mnras}.380..455W},
  adsnote = {Provided by the SAO/NASA Astrophysics Data System}
}

@INPROCEEDINGS{rpr01,
   author = {{Rappaport}, S. and {Pfahl}, E. and {Rasio}, F.~A. and {Podsiadlowski}, P.
    },
    title = "{Formation of Compact Binaries in Globular Clusters}",
booktitle = {Evolution of Binary and Multiple Star Systems},
     year = 2001,
   series = {Astronomical Society of the Pacific Conference Series},
   volume = 229,
   eprint = {arXiv:astro-ph/0101535},
   editor = "{P.~Podsiadlowski, S.~Rappaport, A.~R.~King, F.~D'Antona, \& L.~Burderi
    }",
    pages = {409-+},
   adsurl = {http://adsabs.harvard.edu/abs/2001ASPC..229..409R},
  adsnote = {Provided by the SAO/NASA Astrophysics Data System}
}

@ARTICLE{hcmc10,
   author = {{Hobbs}, G. and {Coles}, W. and {Manchester}, R. and {Chen}, D.
    },
    title = "{Developing a pulsar-based timescale}",
  journal = {arXiv:1011.5285},
archivePrefix = "arXiv",
   eprint = {1011.5285},
 primaryClass = "astro-ph.SR",
 keywords = {Astrophysics - Solar and Stellar Astrophysics},
     year = 2010,
    month = nov,
   adsurl = {http://adsabs.harvard.edu/abs/2010arXiv1011.5285H},
  adsnote = {Provided by the SAO/NASA Astrophysics Data System}
}

@ARTICLE{mamw10,
   author = {{Merritt}, D. and {Alexander}, T. and {Mikkola}, S. and {Will}, C.~M.
    },
    title = "{Testing properties of the Galactic center black hole using stellar orbits}",
  journal = {\prd},
archivePrefix = "arXiv",
   eprint = {0911.4718},
 primaryClass = "astro-ph.GA",
 keywords = {Experimental tests of gravitational theories},
     year = 2010,
    month = mar,
   volume = 81,
   number = 6,
    pages = {062002-+},
      doi = {10.1103/PhysRevD.81.062002},
   adsurl = {http://adsabs.harvard.edu/abs/2010PhRvD..81f2002M},
  adsnote = {Provided by the SAO/NASA Astrophysics Data System}
}

@ARTICLE{lhk+10,
   author = {{Lyne}, A. and {Hobbs}, G. and {Kramer}, M. and {Stairs}, I. and
    {Stappers}, B.},
    title = "{Switched Magnetospheric Regulation of Pulsar Spin-Down}",
  journal = {Science},
archivePrefix = "arXiv",
   eprint = {1006.5184},
 primaryClass = "astro-ph.GA",
     year = 2010,
    month = jul,
   volume = 329,
    pages = {408-},
      doi = {10.1126/science.1186683},
   adsurl = {http://adsabs.harvard.edu/abs/2010Sci...329..408L},
  adsnote = {Provided by the SAO/NASA Astrophysics Data System}
}

@ARTICLE{hlk10,
   author = {{Hobbs}, G. and {Lyne}, A.~G. and {Kramer}, M.},
    title = "{An analysis of the timing irregularities for 366 pulsars}",
  journal = {\mnras},
archivePrefix = "arXiv",
   eprint = {0912.4537},
 primaryClass = "astro-ph.GA",
 keywords = {pulsars: general},
     year = 2010,
    month = feb,
   volume = 402,
    pages = {1027-1048},
      doi = {10.1111/j.1365-2966.2009.15938.x},
   adsurl = {http://adsabs.harvard.edu/abs/2010{\mnras}.402.1027H},
  adsnote = {Provided by the SAO/NASA Astrophysics Data System}
}

@ARTICLE{agpp04,
   author = {{Aschenbach}, B. and {Grosso}, N. and {Porquet}, D. and {Predehl}, P.
    },
    title = "{X-ray flares reveal mass and angular momentum of the Galactic Center black hole}",
  journal = {\aa},
   eprint = {arXiv:astro-ph/0401589},
 keywords = {Galaxy: center, X-rays: individuals: Sgr A*, radiation mechanisms: general, relativity, Galaxy: nucleus, Galaxy: fundamental parameters},
     year = 2004,
    month = apr,
   volume = 417,
    pages = {71-78},
      doi = {10.1051/0004-6361:20035883},
   adsurl = {http://adsabs.harvard.edu/abs/2004A
  adsnote = {Provided by the SAO/NASA Astrophysics Data System}
}

@ARTICLE{asc10,
   author = {{Aschenbach}, B.},
    title = "{Mass and spin of the Sgr A* supermassive black hole determined from flare lightcurves and flare start times}",
  journal = {\memsai},
 keywords = {black hole physics, Galaxy: center, infrared: general, X-rays: general, X-rays: individual (Sgr A*)},
     year = 2010,
   volume = 81,
    pages = {319-+},
   adsurl = {http://adsabs.harvard.edu/abs/2010MmSAI..81..319A},
  adsnote = {Provided by the SAO/NASA Astrophysics Data System}
}

@ARTICLE{btdg+06,
   author = {{B{\'e}langer}, G. and {Terrier}, R. and {de Jager}, O.~C. and
    {Goldwurm}, A. and {Melia}, F.},
    title = "{Periodic Modulations in an X-ray Flare from Sagittarius A*}",
  journal = {Journal of Physics Conference Series},
   eprint = {arXiv:astro-ph/0604337},
     year = 2006,
    month = dec,
   volume = 54,
    pages = {420-426},
      doi = {10.1088/1742-6596/54/1/066},
   adsurl = {http://adsabs.harvard.edu/abs/2006JPhCS..54..420B},
  adsnote = {Provided by the SAO/NASA Astrophysics Data System}
}

@ARTICLE{dcl09,
   author = {{Deneva}, J.~S. and {Cordes}, J.~M. and {Lazio}, T.~J.~W.},
    title = "{Discovery of Three Pulsars from a Galactic Center Pulsar Population}",
  journal = {ApJL},
archivePrefix = "arXiv",
   eprint = {0908.1331},
 primaryClass = "astro-ph.SR",
 keywords = {pulsars: general, pulsars: individual: J1746{\ndash}2850I J1746{\ndash}2850II J1745{\ndash}2910},
     year = 2009,
    month = sep,
   volume = 702,
    pages = {L177-L181},
      doi = {10.1088/0004-637X/702/2/L177},
   adsurl = {http://adsabs.harvard.edu/abs/2009ApJ...702L.177D},
  adsnote = {Provided by the SAO/NASA Astrophysics Data System}
}

@ARTICLE{jkl+06,
   author = {{Johnston}, S. and {Kramer}, M. and {Lorimer}, D.~R. and {Lyne}, A.~G. and
    {McLaughlin}, M. and {Klein}, B. and {Manchester}, R.~N.},
    title = "{Discovery of two pulsars towards the Galactic Centre}",
  journal = {\mnras},
   eprint = {arXiv:astro-ph/0606465},
 keywords = {pulsars: general, pulsars: individual: J1745-2912, pulsars: individual: J1746-2856},
     year = 2006,
    month = nov,
   volume = 373,
    pages = {L6-L10},
      doi = {10.1111/j.1745-3933.2006.00232.x},
   adsurl = {http://adsabs.harvard.edu/abs/2006{\mnras}.373L...6J},
  adsnote = {Provided by the SAO/NASA Astrophysics Data System}
}

@ARTICLE{gsoe+03,
   author = {{Genzel}, R. and {Sch{\"o}del}, R. and {Ott}, T. and {Eisenhauer}, F. and
    {Hofmann}, R. and {Lehnert}, M. and {Eckart}, A. and {Alexander}, T. and
    {Sternberg}, A. and {Lenzen}, R. and {Cl{\'e}net}, Y. and {Lacombe}, F. and
    {Rouan}, D. and {Renzini}, A. and {Tacconi-Garman}, L.~E.},
    title = "{The Stellar Cusp around the Supermassive Black Hole in the Galactic Center}",
  journal = {\apj},
   eprint = {arXiv:astro-ph/0305423},
 keywords = {Black Hole Physics, Galaxies: Nuclei, Galaxy: Center, Stars: Formation},
     year = 2003,
    month = sep,
   volume = 594,
    pages = {812-832},
      doi = {10.1086/377127},
   adsurl = {http://adsabs.harvard.edu/abs/2003ApJ...594..812G},
  adsnote = {Provided by the SAO/NASA Astrophysics Data System}
}

@ARTICLE{osm+11,
   author = {{Orosz}, J.~A. and {Steiner}, J.~F. and {McClintock}, J.~E. and
    {Torres}, M.~A.~P. and {Remillard}, R.~A. and {Bailyn}, C.~D. and
    {Miller}, J.~M.},
    title = "{An Improved Dynamical Model for the Microquasar XTE J1550-564}",
  journal = {ArXiv e-prints:1101.2499},
archivePrefix = "arXiv",
   eprint = {1101.2499},
 primaryClass = "astro-ph.SR",
 keywords = {Astrophysics - Solar and Stellar Astrophysics},
     year = 2011,
    month = jan,
   adsurl = {http://adsabs.harvard.edu/abs/2011arXiv1101.2499O},
  adsnote = {Provided by the SAO/NASA Astrophysics Data System}
}


@ARTICLE{lvk+11,
   author = {{Liu}, K. and {Verbiest}, J.~P.~W. and {Kramer}, M. and {Stappers}, B.~W. and
    {van Straten}, W. and {Cordes}, J.~M.},
    title = "{Prospects for high-precision pulsar timing}",
  journal = {\mnras},
archivePrefix = "arXiv",
   eprint = {1107.3086},
 primaryClass = "astro-ph.HE",
 keywords = {methods: data analysis, pulsars: individual: PSR J0437-4715, ISM: general},
     year = 2011,
   volume = 417,
    month = sep,
    pages = {2916-2926},
      doi = {10.1111/j.1365-2966.2011.19452.x},
   adsurl = {http://adsabs.harvard.edu/abs/2011{\mnras}.tmp.1442L},
  adsnote = {Provided by the SAO/NASA Astrophysics Data System}
}

@ARTICLE{mldr11,
   author = {{Messenger}, C. and {Lommen}, A. and {Demorest}, P. and {Ransom}, S.
    },
    title = "{A Bayesian parameter estimation approach to pulsar time-of-arrival analysis}",
  journal = {Classical and Quantum Gravity},
archivePrefix = "arXiv",
   eprint = {1103.0518},
 primaryClass = "astro-ph.HE",
     year = 2011,
    month = mar,
   volume = 28,
   number = 5,
    pages = {055001-+},
      doi = {10.1088/0264-9381/28/5/055001},
   adsurl = {http://adsabs.harvard.edu/abs/2011CQGra..28e5001M},
  adsnote = {Provided by the SAO/NASA Astrophysics Data System}
}

@ARTICLE{hlj+11,
   author = {{van Haasteren}, R. and {Levin}, Y. and {Janssen}, G.~H. and
    {Lazaridis}, K. and {Kramer}, M. and {Stappers}, B.~W. and {Desvignes}, G. and
    {Purver}, M.~B. and {Lyne}, A.~G. and {Ferdman}, R.~D. and {Jessner}, A. and
    {Cognard}, I. and {Theureau}, G. and {D'Amico}, N. and {Possenti}, A. and
    {Burgay}, M. and {Corongiu}, A. and {Hessels}, J.~W.~T. and
    {Smits}, R. and {Verbiest}, J.~P.~W.},
    title = "{Placing limits on the stochastic gravitational-wave background using European Pulsar Timing Array data}",
  journal = {\mnras},
archivePrefix = "arXiv",
   eprint = {1103.0576},
 primaryClass = "astro-ph.CO",
 keywords = {gravitational waves, methods: data analysis, pulsars: general},
     year = 2011,
    month = jul,
   volume = 414,
    pages = {3117-3128},
      doi = {10.1111/j.1365-2966.2011.18613.x},
   adsurl = {http://adsabs.harvard.edu/abs/2011{\mnras}.414.3117V},
  adsnote = {Provided by the SAO/NASA Astrophysics Data System}
}


@ARTICLE{ych+11,
   author = {{Yardley}, D.~R.~B. and {Coles}, W.~A. and {Hobbs}, G.~B. and
    {Verbiest}, J.~P.~W. and {Manchester}, R.~N. and {van Straten}, W. and
    {Jenet}, F.~A. and {Bailes}, M. and {Bhat}, N.~D.~R. and {Burke-Spolaor}, S. and
    {Champion}, D.~J. and {Hotan}, A.~W. and {Oslowski}, S. and
    {Reynolds}, J.~E. and {Sarkissian}, J.~M.},
    title = "{On detection of the stochastic gravitational-wave background using the Parkes pulsar timing array}",
  journal = {\mnras},
archivePrefix = "arXiv",
   eprint = {1102.2230},
 primaryClass = "astro-ph.GA",
 keywords = {gravitational waves, methods: data analysis, stars: pulsars: general},
     year = 2011,
    month = jun,
   volume = 414,
    pages = {1777-1787},
      doi = {10.1111/j.1365-2966.2011.18517.x},
   adsurl = {http://adsabs.harvard.edu/abs/2011{\mnras}.414.1777Y},
  adsnote = {Provided by the SAO/NASA Astrophysics Data System}
}


@ARTICLE{yhj+10,
   author = {{Yardley}, D.~R.~B. and {Hobbs}, G.~B. and {Jenet}, F.~A. and
    {Verbiest}, J.~P.~W. and {Wen}, Z.~L. and {Manchester}, R.~N. and
    {Coles}, W.~A. and {van Straten}, W. and {Bailes}, M. and {Bhat}, N.~D.~R. and
    {Burke-Spolaor}, S. and {Champion}, D.~J. and {Hotan}, A.~W. and
    {Sarkissian}, J.~M.},
    title = "{The sensitivity of the Parkes Pulsar Timing Array to individual sources of gravitational waves}",
  journal = {\mnras},
archivePrefix = "arXiv",
   eprint = {1005.1667},
 primaryClass = "astro-ph.GA",
 keywords = {gravitational waves, methods: data analysis, pulsars: general, galaxies: evolution},
     year = 2010,
    month = sep,
   volume = 407,
    pages = {669-680},
      doi = {10.1111/j.1365-2966.2010.16949.x},
   adsurl = {http://adsabs.harvard.edu/abs/2010{\mnras}.407..669Y},
  adsnote = {Provided by the SAO/NASA Astrophysics Data System}
}

@ARTICLE{wjy+11,
   author = {{Wen}, Z.~L. and {Jenet}, F.~A. and {Yardley}, D. and {Hobbs}, G.~B. and
    {Manchester}, R.~N.},
    title = "{Constraining the Coalescence Rate of Supermassive Black-hole Binaries Using Pulsar Timing}",
  journal = {\apj},
archivePrefix = "arXiv",
   eprint = {1103.2808},
 primaryClass = "astro-ph.CO",
 keywords = {early universe, galaxies: statistics, gravitational waves, methods: data analysis, pulsars: general},
     year = 2011,
    month = mar,
   volume = 730,
    pages = {29-+},
      doi = {10.1088/0004-637X/730/1/29},
   adsurl = {http://adsabs.harvard.edu/abs/2011ApJ...730...29W},
  adsnote = {Provided by the SAO/NASA Astrophysics Data System}
}

@ARTICLE{opr+10,
   author = {{{\"O}zel}, F. and {Psaltis}, D. and {Ransom}, S. and {Demorest}, P. and
    {Alford}, M.},
    title = "{The Massive Pulsar PSR J1614-2230: Linking Quantum Chromodynamics, Gamma-ray Bursts, and Gravitational Wave Astronomy}",
  journal = {\apjl},
archivePrefix = "arXiv",
   eprint = {1010.5790},
 primaryClass = "astro-ph.HE",
 keywords = {gamma-ray burst: general, pulsars: individual: PSR J1614{\ndash}2230, stars: neutron},
     year = 2010,
    month = dec,
   volume = 724,
    pages = {L199-L202},
      doi = {10.1088/2041-8205/724/2/L199},
   adsurl = {http://adsabs.harvard.edu/abs/2010ApJ...724L.199O},
  adsnote = {Provided by the SAO/NASA Astrophysics Data System}
}

@ARTICLE{dpr+10,
   author = {{Demorest}, P.~B. and {Pennucci}, T. and {Ransom}, S.~M. and
    {Roberts}, M.~S.~E. and {Hessels}, J.~W.~T.},
    title = "{A two-solar-mass neutron star measured using Shapiro delay}",
  journal = na,
archivePrefix = "arXiv",
   eprint = {1010.5788},
 primaryClass = "astro-ph.HE",
     year = 2010,
    month = oct,
   volume = 467,
    pages = {1081-1083},
      doi = {10.1038/nature09466},
   adsurl = {http://adsabs.harvard.edu/abs/2010Natur.467.1081D},
  adsnote = {Provided by the SAO/NASA Astrophysics Data System}
}

@ARTICLE{lwk+11,
   author = {{Lee}, K.~J. and {Wex}, N. and {Kramer}, M. and {Stappers}, B.~W. and
    {Bassa}, C.~G. and {Janssen}, G.~H. and {Karuppusamy}, R. and
    {Smits}, R.},
    title = "{Gravitational wave astronomy of single sources with a pulsar timing array}",
  journal = {\mnras},
archivePrefix = "arXiv",
   eprint = {1103.0115},
 primaryClass = "astro-ph.HE",
 keywords = {gravitational waves, pulsars: general},
     year = 2011,
    month = jul,
   volume = 414,
    pages = {3251-3264},
      doi = {10.1111/j.1365-2966.2011.18622.x},
   adsurl = {http://adsabs.harvard.edu/abs/2011{\mnras}.414.3251L},
  adsnote = {Provided by the SAO/NASA Astrophysics Data System}
}


@ARTICLE{ljp+10,
   author = {{Lee}, K. and {Jenet}, F.~A. and {Price}, R.~H. and {Wex}, N. and
    {Kramer}, M.},
    title = "{Detecting Massive Gravitons Using Pulsar Timing Arrays}",
  journal = {\apj},
archivePrefix = "arXiv",
   eprint = {1008.2561},
 primaryClass = "astro-ph.HE",
 keywords = {elementary particles, gravitational waves, pulsars: general},
     year = 2010,
    month = oct,
   volume = 722,
    pages = {1589-1597},
      doi = {10.1088/0004-637X/722/2/1589},
   adsurl = {http://adsabs.harvard.edu/abs/2010ApJ...722.1589L},
  adsnote = {Provided by the SAO/NASA Astrophysics Data System}
}

@ARTICLE{fl10b,
   author = {{Finn}, L.~S. and {Lommen}, A.~N.},
    title = "{Detection, Localization, and Characterization of Gravitational Wave Bursts in a Pulsar Timing Array}",
  journal = {\apj},
archivePrefix = "arXiv",
   eprint = {1004.3499},
 primaryClass = "astro-ph.IM",
 keywords = {gravitational waves, methods: data analysis, methods: statistical},
     year = 2010,
    month = aug,
   volume = 718,
    pages = {1400-1415},
      doi = {10.1088/0004-637X/718/2/1400},
   adsurl = {http://adsabs.harvard.edu/abs/2010ApJ...718.1400F},
  adsnote = {Provided by the SAO/NASA Astrophysics Data System}
}

@ARTICLE{cor97,
   author = {{Cordes}, J.~M.},
    title = "{Pulsar PSR 1919+21 - Notches, drifting subpulses, microstructure, and other emission}",
  journal = {\apj},
 keywords = {ASTRONOMICAL MODELS, PULSARS, SPECTRUM ANALYSIS, STELLAR STRUCTURE, DATA ACQUISITION, DATA REDUCTION, ENERGY SPECTRA, MICROSTRUCTURE, RADIO SPECTRA, STELLAR SPECTRA},
     year = 1975,
    month = jan,
   volume = 195,
    pages = {193-202},
      doi = {10.1086/153318},
   adsurl = {http://adsabs.harvard.edu/abs/1975ApJ...195..193C},
  adsnote = {Provided by the SAO/NASA Astrophysics Data System}
}

@INPROCEEDINGS{pms11,
   author = {{Palliyaguru}, N. and {McLaughlin}, M. and {Stinebring}, D.},
    title = "{Pulsar Phase Jitter and Cyclic Spectroscopy Derived Arrival Times}",
booktitle = {American Astronomical Society Meeting Abstracts {\#}217},
     year = 2011,
   series = {Bulletin of the American Astronomical Society},
   volume = 43,
    month = jan,
    pages = {{\#}139.02-+},
   adsurl = {http://adsabs.harvard.edu/abs/2011AAS...21713902P},
  adsnote = {Provided by the SAO/NASA Astrophysics Data System}
}

@ARTICLE{mks+10,
   author = {{Manchester}, R.~N. and {Kramer}, M. and {Stairs}, I.~H. and
    {Burgay}, M. and {Camilo}, F. and {Hobbs}, G.~B. and {Lorimer}, D.~R. and
    {Lyne}, A.~G. and {McLaughlin}, M.~A. and {McPhee}, C.~A. and
    {Possenti}, A. and {Reynolds}, J.~E. and {van Straten}, W.},
    title = "{Observations and Modeling of Relativistic Spin Precession in PSR J1141-6545}",
  journal = {\apj},
archivePrefix = "arXiv",
   eprint = {1001.1483},
 primaryClass = "astro-ph.GA",
 keywords = {pulsars: individual: PSR J1141{\ndash}6545, radiation mechanisms: non-thermal, relativistic processes},
     year = 2010,
    month = feb,
   volume = 710,
    pages = {1694-1709},
      doi = {10.1088/0004-637X/710/2/1694},
   adsurl = {http://adsabs.harvard.edu/abs/2010ApJ...710.1694M},
  adsnote = {Provided by the SAO/NASA Astrophysics Data System}
}

@ARTICLE{cor75,
   author = {{Cordes}, J.~M.},
    title = "{Pulsar PSR 1919+21 - Notches, drifting subpulses, microstructure, and other emission}",
  journal = {\apj},
 keywords = {ASTRONOMICAL MODELS, PULSARS, SPECTRUM ANALYSIS, STELLAR STRUCTURE, DATA ACQUISITION, DATA REDUCTION, ENERGY SPECTRA, MICROSTRUCTURE, RADIO SPECTRA, STELLAR SPECTRA},
     year = 1975,
    month = jan,
   volume = 195,
    pages = {193-202},
      doi = {10.1086/153318},
   adsurl = {http://adsabs.harvard.edu/abs/1975ApJ...195..193C},
  adsnote = {Provided by the SAO/NASA Astrophysics Data System}
}

@ARTICLE{wmj07,
   author = {{Wang}, N. and {Manchester}, R.~N. and {Johnston}, S.},
    title = "{Pulsar nulling and mode changing}",
  journal = {\mnras},
   eprint = {arXiv:astro-ph/0703241},
 keywords = {radiation mechanisms: non-thermal, pulsars: general},
     year = 2007,
    month = may,
   volume = 377,
    pages = {1383-1392},
      doi = {10.1111/j.1365-2966.2007.11703.x},
   adsurl = {http://adsabs.harvard.edu/abs/2007{\mnras}.377.1383W},
  adsnote = {Provided by the SAO/NASA Astrophysics Data System}
}

@ARTICLE{fvb+10,
   author = {{Ferdman}, R.~D. and {van Haasteren}, R. and {Bassa}, C.~G. and
    {Burgay}, M. and {Cognard}, I. and {Corongiu}, A. and {D'Amico}, N. and
    {Desvignes}, G. and {Hessels}, J.~W.~T. and {Janssen}, G.~H. and
    {Jessner}, A. and {Jordan}, C. and {Karuppusamy}, R. and {Keane}, E.~F. and
    {Kramer}, M. and {Lazaridis}, K. and {Levin}, Y. and {Lyne}, A.~G. and
    {Pilia}, M. and {Possenti}, A. and {Purver}, M. and {Stappers}, B. and
    {Sanidas}, S. and {Smits}, R. and {Theureau}, G.},
    title = "{The European Pulsar Timing Array: current efforts and a LEAP toward the future}",
  journal = {Classical and Quantum Gravity},
archivePrefix = "arXiv",
   eprint = {1003.3405},
 primaryClass = "astro-ph.HE",
     year = 2010,
    month = apr,
   volume = 27,
   number = 8,
    pages = {084014-+},
      doi = {10.1088/0264-9381/27/8/084014},
   adsurl = {http://adsabs.harvard.edu/abs/2010CQGra..27h4014F},
  adsnote = {Provided by the SAO/NASA Astrophysics Data System}
}

@ARTICLE{zio08,
   author = {{Zi{\'o}{\l}kowski}, J.},
    title = "{Masses of Black Holes in the Universe}",
  journal = {Chinese Journal of Astronomy and Astrophysics Supplement},
archivePrefix = "arXiv",
   eprint = {0808.0435},
 keywords = {black holes: mass determination, stars: black holes, intermediate mass black holes, supermassive black holes},
     year = 2008,
    month = oct,
   volume = 8,
    pages = {273-280},
   adsurl = {http://adsabs.harvard.edu/abs/2008ChJAS...8..273Z},
  adsnote = {Provided by the SAO/NASA Astrophysics Data System}
}

@ARTICLE{sf08,
   author = {{Silverman}, J.~M. and {Filippenko}, A.~V.},
    title = "{On IC 10 X-1, the Most Massive Known Stellar-Mass Black Hole}",
  journal = {\apjl},
archivePrefix = "arXiv",
   eprint = {0802.2716},
 keywords = {Black Hole Physics, Galaxies: Starburst, Stars: Wolf-Rayet, X-Rays: Binaries},
     year = 2008,
    month = may,
   volume = 678,
    pages = {L17-L20},
      doi = {10.1086/588096},
   adsurl = {http://adsabs.harvard.edu/abs/2008ApJ...678L..17S},
  adsnote = {Provided by the SAO/NASA Astrophysics Data System}
}

@ARTICLE{bbf+10,
   author = {{Belczynski}, K. and {Bulik}, T. and {Fryer}, C.~L. and {Ruiter}, A. and
    {Valsecchi}, F. and {Vink}, J.~S. and {Hurley}, J.~R.},
    title = "{On the Maximum Mass of Stellar Black Holes}",
  journal = {\apj},
archivePrefix = "arXiv",
   eprint = {0904.2784},
 primaryClass = "astro-ph.SR",
 keywords = {binaries: close, black hole physics, gravitational waves, stars: evolution, stars: neutron},
     year = 2010,
    month = may,
   volume = 714,
    pages = {1217-1226},
      doi = {10.1088/0004-637X/714/2/1217},
   adsurl = {http://adsabs.harvard.edu/abs/2010ApJ...714.1217B},
  adsnote = {Provided by the SAO/NASA Astrophysics Data System}
}

@ARTICLE{bgf+06,
   author = {{Bower}, G.~C. and {Goss}, W.~M. and {Falcke}, H. and {Backer}, D.~C. and
    {Lithwick}, Y.},
    title = "{The Intrinsic Size of Sagittarius A* from 0.35 to 6 cm}",
  journal = {ApJL},
   eprint = {arXiv:astro-ph/0608004},
 keywords = {Galaxies: Active, Galaxy: Center, Scattering, Techniques: Interferometric},
     year = 2006,
    month = sep,
   volume = 648,
    pages = {L127-L130},
      doi = {10.1086/508019},
   adsurl = {http://adsabs.harvard.edu/abs/2006ApJ...648L.127B},
  adsnote = {Provided by the SAO/NASA Astrophysics Data System}
}

@ARTICLE{lke+11,
   author = {{Lu}, R.-S. and {Krichbaum}, T.~P. and {Eckart}, A. and {K{\"o}nig}, S. and
    {Kunneriath}, D. and {Witzel}, G. and {Witzel}, A. and {Zensus}, J.~A.
    },
    title = "{Multiwavelength VLBI observations of Sagittarius A*}",
  journal = {\aa},
archivePrefix = "arXiv",
   eprint = {1010.1287},
 primaryClass = "astro-ph.GA",
 keywords = {Galaxy: center, galaxies: individual: Sgr A*, scattering, techniques: interferometric, galaxies: nuclei},
     year = 2011,
    month = jan,
   volume = 525,
    pages = {A76+},
      doi = {10.1051/0004-6361/200913807},
   adsurl = {http://adsabs.harvard.edu/abs/2011A
  adsnote = {Provided by the SAO/NASA Astrophysics Data System}
}

@ARTICLE{rfr+90,
   author = {{Reich}, W. and {Fuerst}, E. and {Reich}, P. and {Reif}, K.},
    title = "{A radio continuum survey of the Galactic Plane at 11 CM wavelength. II - The area L = 358-76 deg, B = -5 to 5 deg. III}",
  journal = {\aaps},
 keywords = {CENTIMETER WAVES, CONTINUOUS RADIATION, GALACTIC RADIO WAVES, MILKY WAY GALAXY, SKY SURVEYS (ASTRONOMY), ANGULAR RESOLUTION, FLUX DENSITY, TELESCOPES, Galactic radio emission, radio continuum survey},
     year = 1990,
    month = oct,
   volume = 85,
    pages = {633-690},
   adsurl = {http://adsabs.harvard.edu/abs/1990A
  adsnote = {Provided by the SAO/NASA Astrophysics Data System}
}

@ARTICLE{hl11,
   author = {{Hartnett}, J.~G. and {Luiten}, A.~N.},
    title = "{Colloquium: Comparison of astrophysical and terrestrial frequency standards}",
  journal = {Reviews of Modern Physics},
archivePrefix = "arXiv",
   eprint = {1004.0115},
 primaryClass = "astro-ph.IM",
 keywords = {Time and frequency, Auxiliary and recording instruments; clocks and frequency standards, Pulsations, oscillations, and stellar seismology, Pulsars},
     year = 2011,
    month = jan,
   volume = 83,
    pages = {1-9},
      doi = {10.1103/RevModPhys.83.1},
   adsurl = {http://adsabs.harvard.edu/abs/2011RvMP...83....1H},
  adsnote = {Provided by the SAO/NASA Astrophysics Data System}
}

@phdthesis{ver09,
  author = {Verbiest, J.~P.~W.},
  school = {Swinburne University of Technology},
  year = 2009,
  title = {Long-Term Timing of Millisecond Pulsars and Gravitational Wave Detection}
}

@manual{hb08,
  author = {{Hampson}, G., and {Brown}, A.},
  year = 2008,
  title = {A 1 GHz Pulsar Digital Filter Bank and RFI Mitigation System, unpublished manual,
  Australia Telescope National Facility - CSIRO, Corner Vimiera and Pembroke Roads, Marsfield, NSW, 2122}
}

@ARTICLE{fw10,
   author = {{Freire}, P.~C.~C. and {Wex}, N.},
    title = "{The orthometric parametrization of the Shapiro delay and an improved test of general relativity with binary pulsars}",
  journal = {\mnras},
archivePrefix = "arXiv",
   eprint = {1007.0933},
 primaryClass = "astro-ph.IM",
 keywords = {gravitation, methods: observational, stars: neutron, pulsars: general},
     year = 2010,
    month = nov,
   volume = 409,
    pages = {199-212},
      doi = {10.1111/j.1365-2966.2010.17319.x},
   adsurl = {http://adsabs.harvard.edu/abs/2010{\mnras}.409..199F},
  adsnote = {Provided by the SAO/NASA Astrophysics Data System}
}

@ARTICLE{ych+09,
   author = {{Yuan}, Y.-F. and {Cao}, X. and {Huang}, L. and {Shen}, Z.-Q.
    },
    title = "{Images of the Radiatively Inefficient Accretion Flow Surrounding a Kerr Black Hole: Application in Sgr A*}",
  journal = {\apj},
archivePrefix = "arXiv",
   eprint = {0904.4090},
 primaryClass = "astro-ph.HE",
 keywords = {black hole physics, Galaxy: center, radiative transfer, submillimeter},
     year = 2009,
    month = jul,
   volume = 699,
    pages = {722-731},
      doi = {10.1088/0004-637X/699/1/722},
   adsurl = {http://adsabs.harvard.edu/abs/2009ApJ...699..722Y},
  adsnote = {Provided by the SAO/NASA Astrophysics Data System}
}

@ARTICLE{bbb+01,
   author = {{Baganoff}, F.~K. and {Bautz}, M.~W. and {Brandt}, W.~N. and
    {Chartas}, G. and {Feigelson}, E.~D. and {Garmire}, G.~P. and
    {Maeda}, Y. and {Morris}, M. and {Ricker}, G.~R. and {Townsley}, L.~K. and
    {Walter}, F.},
    title = "{Rapid X-ray flaring from the direction of the supermassive black hole at the Galactic Centre}",
  journal = {\nat},
   eprint = {arXiv:astro-ph/0109367},
     year = 2001,
    month = sep,
   volume = 413,
    pages = {45-48},
      doi = {10.1038/35092510},
   adsurl = {http://adsabs.harvard.edu/abs/2001Natur.413...45B},
  adsnote = {Provided by the SAO/NASA Astrophysics Data System}
}

@ARTICLE{zed+11,
   author = {{Zamaninasab}, M. and {Eckart}, A. and {Dov{\v c}iak}, M. and
    {Karas}, V. and {Sch{\"o}del}, R. and {Witzel}, G. and {Sabha}, N. and
    {Garc{\'{\i}}a-Mar{\'{\i}}n}, M. and {Kunneriath}, D. and {Mu{\v z}i{\'c}}, K. and
    {Straubmeier}, C. and {Valencia-S}, M. and {Zensus}, J.~A.},
    title = "{Near-infrared polarimetry as a tool for testing properties of accreting supermassive black holes}",
  journal = {\mnras},
archivePrefix = "arXiv",
   eprint = {1102.0775},
 primaryClass = "astro-ph.HE",
 keywords = {accretion, accretion discs, black hole physics, Galaxy: centre, Galaxy: nucleus, infrared: general},
     year = 2011,
    month = may,
   volume = 413,
    pages = {322-332},
      doi = {10.1111/j.1365-2966.2010.18139.x},
   adsurl = {http://adsabs.harvard.edu/abs/2011{\mnras}.413..322Z},
  adsnote = {Provided by the SAO/NASA Astrophysics Data System}
}

@ARTICLE{nlj+11,
   author = {{Nan}, R. and {Li}, D. and {Jin}, C. and {Wang}, Q. and {Zhu}, L. and
    {Zhu}, W. and {Zhang}, H. and {Yue}, Y. and {Qian}, L.},
    title = "{The Five-Hundred Aperture Spherical Radio Telescope (fast) Project}",
  journal = {International Journal of Modern Physics D},
archivePrefix = "arXiv",
   eprint = {1105.3794},
 primaryClass = "astro-ph.IM",
 keywords = {Radio telescope, active main reflector, HI 21cm line, pulsar},
     year = 2011,
   volume = 20,
    pages = {989-1024},
      doi = {10.1142/S0218271811019335},
   adsurl = {http://adsabs.harvard.edu/abs/2011IJMPD..20..989N},
  adsnote = {Provided by the SAO/NASA Astrophysics Data System}
}

@ARTICLE{sks+09,
   author = {{Smits}, R. and {Kramer}, M. and {Stappers}, B. and {Lorimer}, D.~R. and
    {Cordes}, J. and {Faulkner}, A.},
    title = "{Pulsar searches and timing with the square kilometre array}",
  journal = aap,
 keywords = {stars: neutron, stars: pulsars: general, telescopes},
     year = 2009,
    month = jan,
   volume = 493,
    pages = {1161-1170},
      doi = {10.1051/0004-6361:200810383},
   adsurl = {http://adsabs.harvard.edu/abs/2009A
  adsnote = {Provided by the SAO/NASA Astrophysics Data System}
}

@ARTICLE{slk+09,
   author = {{Smits}, R. and {Lorimer}, D.~R. and {Kramer}, M. and {Manchester}, R. and
    {Stappers}, B. and {Jin}, C.~J. and {Nan}, R.~D. and {Li}, D.
    },
    title = "{Pulsar science with the Five hundred metre Aperture Spherical Telescope}",
  journal = {\aa},
archivePrefix = "arXiv",
   eprint = {0908.1689},
 primaryClass = "astro-ph.IM",
 keywords = {stars: neutron, stars: pulsars: general, telescopes},
     year = 2009,
    month = oct,
   volume = 505,
    pages = {919-926},
      doi = {10.1051/0004-6361/200911939},
   adsurl = {http://adsabs.harvard.edu/abs/2009A
  adsnote = {Provided by the SAO/NASA Astrophysics Data System}
}

@ARTICLE{lwk+12,
   author = {{Liu}, K. and {Wex}, N. and {Kramer}, M. and {Cordes}, J.~M. and
    {Lazio}, T.~J.~W.},
    title = "{Prospects for Probing the Spacetime of Sgr A* with Pulsars}",
  journal = {\apj},
archivePrefix = "arXiv",
   eprint = {1112.2151},
 primaryClass = "astro-ph.HE",
 keywords = {black hole physics, Galaxy: center, pulsars: general},
     year = 2012,
    month = mar,
   volume = 747,
      eid = {1},
    pages = {1},
      doi = {10.1088/0004-637X/747/1/1},
   adsurl = {http://adsabs.harvard.edu/abs/2012ApJ...747....1L},
  adsnote = {Provided by the SAO/NASA Astrophysics Data System}
}


@article{heu98,
  title    = {Stationary Black Holes: Uniqueness and Beyond},
  author   = {Markus Heusler},
  journal  = {Living Reviews in Relativity 1},
  year     = {1998},
  volume   = {6. URL (cited on 2011/7/22): http://www.livingreviews.org/lrr-1998-6},
  keywords = {self-gravitating classical fields, uniqueness theorems, black holes},
  url      = {http://www.livingreviews.org/lrr-1998-6}
}


@ARTICLE{pri72a,
   author = {{Price}, R.~H.},
    title = "{Nonspherical Perturbations of Relativistic Gravitational Collapse. I. Scalar and Gravitational Perturbations}",
  journal = prevd,
     year = {1972},
    month = may,
   volume = 5,
    pages = {2419-2438},
      doi = {10.1103/PhysRevD.5.2419},
   adsurl = {http://adsabs.harvard.edu/abs/1972PhRvD...5.2419P},
  adsnote = {Provided by the SAO/NASA Astrophysics Data System}
}

@ARTICLE{pri72b,
   author = {{Price}, R.~H.},
    title = "{Nonspherical Perturbations of Relativistic Gravitational Collapse. II. Integer-Spin, Zero-Rest-Mass Fields}",
  journal = prevd,
     year = {1972},
    month = may,
   volume = 5,
    pages = {2439-2454},
      doi = {10.1103/PhysRevD.5.2439},
   adsurl = {http://adsabs.harvard.edu/abs/1972PhRvD...5.2439P},
  adsnote = {Provided by the SAO/NASA Astrophysics Data System}
}

@INPROCEEDINGS{pen79,
   author = {{Penrose}, R.},
    title = "{Singularities and time-asymmetry.}",
 keywords = {Cosmological Models:Relativity Theory, Cosmological Models:Singularities},
booktitle = {General Relativity: An Einstein centenary survey},
     year = 1979,
   editor = "{S.~W.~Hawking \& W.~Israel}",
   volume = {1},
     publisher = {Cambridge; New York: Cambridge University Press},
    pages = {581-638},
   adsurl = {http://adsabs.harvard.edu/abs/1979grec.conf..581P},
  adsnote = {Provided by the SAO/NASA Astrophysics Data System}
}

@ARTICLE{han74,
   author = {{Hansen}, R. O.},
  journal = {Journal of Mathematical Physics},
     year = 1974,
   volume = 15,
    pages = {46},
}

@ARTICLE{sw11,
   author = {{Sadeghian}, L. and {Will}, C.~M.},
    title = "{Testing the black hole no-hair theorem at the galactic center: Perturbing effects of stars in the surrounding cluster}",
  journal = {arXiv:astro-ph/1106.5056},
archivePrefix = "arXiv",
   eprint = {1106.5056},
 primaryClass = "gr-qc",
 keywords = {General Relativity and Quantum Cosmology},
     year = 2011,
    month = jun,
   adsurl = {http://adsabs.harvard.edu/abs/2011arXiv1106.5056S},
  adsnote = {Provided by the SAO/NASA Astrophysics Data System}
}

@ARTICLE{els+11,
   author = {{Espinoza}, C.~M. and {Lyne}, A.~G. and {Stappers}, B.~W. and
    {Kramer}, M.},
    title = "{A study of 315 glitches in the rotation of 102 pulsars}",
  journal = {\mnras},
archivePrefix = "arXiv",
   eprint = {1102.1743},
 primaryClass = "astro-ph.HE",
 keywords = {stars: neutron, pulsars: general},
     year = 2011,
    month = jun,
   volume = 414,
    pages = {1679-1704},
      doi = {10.1111/j.1365-2966.2011.18503.x},
   adsurl = {http://adsabs.harvard.edu/abs/2011{\mnras}.414.1679E},
  adsnote = {Provided by the SAO/NASA Astrophysics Data System}
}

@ARTICLE{ywm+10,
   author = {{Yuan}, J.~P. and {Wang}, N. and {Manchester}, R.~N. and {Liu}, Z.~Y.
    },
    title = "{29 glitches detected at Urumqi Observatory}",
  journal = {\mnras},
archivePrefix = "arXiv",
   eprint = {1001.1471},
 primaryClass = "astro-ph.GA",
 keywords = {methods: data analysis, stars: neutron, pulsars: general},
     year = 2010,
    month = may,
   volume = 404,
    pages = {289-304},
      doi = {10.1111/j.1365-2966.2010.16272.x},
   adsurl = {http://adsabs.harvard.edu/abs/2010{\mnras}.404..289Y},
  adsnote = {Provided by the SAO/NASA Astrophysics Data System}
}

@ARTICLE{pop08,
   author = {{Popov}, S.~B.},
    title = "{Tkachenko waves, glitches and precession in neutron stars}",
  journal = {\apss},
archivePrefix = "arXiv",
   eprint = {0808.3040},
 keywords = {Neutron stars, Pulsars},
     year = 2008,
    month = oct,
   volume = 317,
    pages = {175-179},
      doi = {10.1007/s10509-008-9871-y},
   adsurl = {http://adsabs.harvard.edu/abs/2008Ap
  adsnote = {Provided by the SAO/NASA Astrophysics Data System}
}

@ARTICLE{lyx07,
   author = {{Liu}, K. and {Yue}, Y.~L. and {Xu}, R.~X.},
    title = "{PSR B1828-11: a precession pulsar torqued by a quark planet?}",
  journal = {\mnras},
   eprint = {arXiv:astro-ph/0611729},
 keywords = {gravitational waves , planetary systems , pulsars: individual: PSR B1828-11},
     year = 2007,
    month = oct,
   volume = 381,
    pages = {L1-L5},
      doi = {10.1111/j.1745-3933.2007.00337.x},
   adsurl = {http://adsabs.harvard.edu/abs/2007{\mnras}.381L...1L},
  adsnote = {Provided by the SAO/NASA Astrophysics Data System}
}

@ARTICLE{qxx+03,
   author = {{Qiao}, G.~J. and {Xue}, Y.~Q. and {Xu}, R.~X. and {Wang}, H.~G. and
    {Xiao}, B.~W.},
    title = "{An accretion disk model for periodic timing variations of pulsars}",
  journal = {\aa},
   eprint = {arXiv:astro-ph/0306489},
 keywords = {stars: neutron, accretion, accretion disks},
     year = 2003,
    month = aug,
   volume = 407,
    pages = {L25-L28},
      doi = {10.1051/0004-6361:20031055},
   adsurl = {http://adsabs.harvard.edu/abs/2003A
  adsnote = {Provided by the SAO/NASA Astrophysics Data System}
}

@ARTICLE{cs08,
   author = {{Cordes}, J.~M. and {Shannon}, R.~M.},
    title = "{Rocking the Lighthouse: Circumpulsar Asteroids and Radio Intermittency}",
  journal = {\apj},
   eprint = {arXiv:astro-ph/0605145},
 keywords = {Acceleration of Particles, Accretion, Accretion Disks, Stars: Pulsars: General, pulsars: individual (B0656+14), pulsars: individual (B1931+24), Stars: Neutron},
     year = 2008,
    month = aug,
   volume = 682,
    pages = {1152-1165},
      doi = {10.1086/589425},
   adsurl = {http://adsabs.harvard.edu/abs/2008ApJ...682.1152C},
  adsnote = {Provided by the SAO/NASA Astrophysics Data System}
}

@ARTICLE{lor08,
   author = {{Lorimer}, D.~R.},
    title = "{Binary and Millisecond Pulsars}",
  journal = {Living Reviews in Relativity},
archivePrefix = "arXiv",
   eprint = {0811.0762},
 keywords = {pulsars},
     year = 2008,
    month = nov,
   volume = 11,
    pages = {8-+},
   adsurl = {http://adsabs.harvard.edu/abs/2008LRR....11....8L},
  adsnote = {Provided by the SAO/NASA Astrophysics Data System}
}

@ARTICLE{frg07,
   author = {{Freire}, P.~C.~C. and {Ransom}, S.~M. and {Gupta}, Y.},
    title = "{Timing the Eccentric Binary Millisecond Pulsar in NGC 1851}",
  journal = {\apj},
   eprint = {arXiv:astro-ph/0703051},
 keywords = {Stars: Binaries: General, Galaxy: Globular Clusters: General, Galaxy: Globular Clusters: Individual: NGC Number: NGC 1851, Stars: Pulsars: General, Stars: Pulsars: Individual: Alphanumeric: PSR J0514-4002A},
     year = 2007,
    month = jun,
   volume = 662,
    pages = {1177-1182},
      doi = {10.1086/517904},
   adsurl = {http://adsabs.harvard.edu/abs/2007ApJ...662.1177F},
  adsnote = {Provided by the SAO/NASA Astrophysics Data System}
}

@ARTICLE{fbw+11,
   author = {{Freire}, P.~C.~C. and {Bassa}, C.~G. and {Wex}, N. and {Stairs}, I.~H. and
    {Champion}, D.~J. and {Ransom}, S.~M. and {Lazarus}, P. and
    {Kaspi}, V.~M. and {Hessels}, J.~W.~T. and {Kramer}, M. and
    {Cordes}, J.~M. and {Verbiest}, J.~P.~W. and {Podsiadlowski}, P. and
    {Nice}, D.~J. and {Deneva}, J.~S. and {Lorimer}, D.~R. and {Stappers}, B.~W. and
    {McLaughlin}, M.~A. and {Camilo}, F.},
    title = "{On the nature and evolution of the unique binary pulsar J1903+0327}",
  journal = {\mnras},
archivePrefix = "arXiv",
   eprint = {1011.5809},
 primaryClass = "astro-ph.GA",
 keywords = {equation of state, binaries: spectroscopic, stars: neutron, pulsars: general, pulsars: individual: PSR J1903+0327},
     year = 2011,
    month = apr,
   volume = 412,
    pages = {2763-2780},
      doi = {10.1111/j.1365-2966.2010.18109.x},
   adsurl = {http://adsabs.harvard.edu/abs/2011{\mnras}.412.2763F},
  adsnote = {Provided by the SAO/NASA Astrophysics Data System}
}

@ARTICLE{gcm01,
   author = {{Greiner}, J. and {Cuby}, J.~G. and {McCaughrean}, M.~J.},
    title = "{An unusually massive stellar black hole in the Galaxy}",
  journal = {\nat},
   eprint = {arXiv:astro-ph/0111538},
     year = 2001,
    month = nov,
   volume = 414,
    pages = {522-525},
   adsurl = {http://adsabs.harvard.edu/abs/2001Natur.414..522G},
  adsnote = {Provided by the SAO/NASA Astrophysics Data System}
}

@ARTICLE{omn+07,
   author = {{Orosz}, J.~A. and {McClintock}, J.~E. and {Narayan}, R. and
    {Bailyn}, C.~D. and {Hartman}, J.~D. and {Macri}, L. and {Liu}, J. and
    {Pietsch}, W. and {Remillard}, R.~A. and {Shporer}, A. and {Mazeh}, T.
    },
    title = "{A 15.65-solar-mass black hole in an eclipsing binary in the nearby spiral galaxy M 33}",
  journal = {\nat},
archivePrefix = "arXiv",
   eprint = {0710.3165},
     year = 2007,
    month = oct,
   volume = 449,
    pages = {872-875},
      doi = {10.1038/nature06218},
   adsurl = {http://adsabs.harvard.edu/abs/2007Natur.449..872O},
  adsnote = {Provided by the SAO/NASA Astrophysics Data System}
}

@ARTICLE{mw09,
   author = {{Melatos}, A. and {Warszawski}, L.},
    title = "{Superfluid Vortex Unpinning as a Coherent Noise Process, and the Scale Invariance of Pulsar Glitches}",
  journal = {\apj},
archivePrefix = "arXiv",
   eprint = {0904.2998},
 primaryClass = "astro-ph.SR",
 keywords = {dense matter, hydrodynamics, stars: interiors, stars: neutron, stars: rotation},
     year = 2009,
    month = aug,
   volume = 700,
    pages = {1524-1540},
      doi = {10.1088/0004-637X/700/2/1524},
   adsurl = {http://adsabs.harvard.edu/abs/2009ApJ...700.1524M},
  adsnote = {Provided by the SAO/NASA Astrophysics Data System}
}

@ARTICLE{fbw+11,
   author = {{Freire}, P.~C.~C. and {Bassa}, C.~G. and {Wex}, N. and {Stairs}, I.~H. and
    {Champion}, D.~J. and {Ransom}, S.~M. and {Lazarus}, P. and
    {Kaspi}, V.~M. and {Hessels}, J.~W.~T. and {Kramer}, M. and
    {Cordes}, J.~M. and {Verbiest}, J.~P.~W. and {Podsiadlowski}, P. and
    {Nice}, D.~J. and {Deneva}, J.~S. and {Lorimer}, D.~R. and {Stappers}, B.~W. and
    {McLaughlin}, M.~A. and {Camilo}, F.},
    title = "{On the nature and evolution of the unique binary pulsar J1903+0327}",
  journal = {\mnras},
archivePrefix = "arXiv",
   eprint = {1011.5809},
 primaryClass = "astro-ph.GA",
 keywords = {equation of state, binaries: spectroscopic, stars: neutron, pulsars: general, pulsars: individual: PSR J1903+0327},
     year = 2011,
    month = apr,
   volume = 412,
    pages = {2763-2780},
      doi = {10.1111/j.1365-2966.2010.18109.x},
   adsurl = {http://adsabs.harvard.edu/abs/2011{\mnras}.412.2763F},
  adsnote = {Provided by the SAO/NASA Astrophysics Data System}
}

@ARTICLE{psa08,
   author = {{Psaltis}, D.},
    title = "{Probes and Tests of Strong-Field Gravity with Observations in the Electromagnetic Spectrum}",
  journal = {Living Reviews in Relativity 11},
archivePrefix = "arXiv",
   eprint = {0806.1531},
 keywords = {Neutron stars, Black holes, Tests of relativistic gravity},
     year = 2008,
    month = nov,
  volume  = {9. URL (cited on 2011/7/22: http://www.livingreviews.org/lrr-2008-9},
   adsurl = {http://adsabs.harvard.edu/abs/2008LRR....11....9P},
  adsnote = {Provided by the SAO/NASA Astrophysics Data System}
}

@ARTICLE{pj10,
   author = {{Johannsen}, T. and {Psaltis}, D.},
    title = "{Testing the No-hair Theorem with Observations in the Electromagnetic Spectrum. III. Quasi-periodic Variability}",
  journal = {\apj},
archivePrefix = "arXiv",
   eprint = {1010.1000},
 primaryClass = "astro-ph.HE",
 keywords = {accretion, accretion disks, black hole physics, gravitation, hydrodynamics, stars: individual: Sgr Aast, X-rays: binaries},
     year = 2011,
    month = jan,
   volume = 726,
    pages = {11-+},
      doi = {10.1088/0004-637X/726/1/11},
   adsurl = {http://adsabs.harvard.edu/abs/2011ApJ...726...11J},
  adsnote = {Provided by the SAO/NASA Astrophysics Data System}
}

@ARTICLE{wil08,
   author = {{Will}, C.~M.},
    title = "{Testing the General Relativistic ``No-Hair'' Theorems Using the Galactic Center Black Hole Sagittarius A*}",
  journal = {ApJL},
archivePrefix = "arXiv",
   eprint = {0711.1677},
 keywords = {Black Hole Physics, Galaxy: Center, Relativity},
     year = 2008,
    month = feb,
   volume = 674,
    pages = {L25-L28},
      doi = {10.1086/528847},
   adsurl = {http://adsabs.harvard.edu/abs/2008ApJ...674L..25W},
  adsnote = {Provided by the SAO/NASA Astrophysics Data System}
}

@INPROCEEDINGS{epb+09,
   author = {{Eisenhauer}, F. and {Perrin}, G. and {Brandner}, W. and {Straubmeier}, C. and
    {B{\"o}hm}, A. and {Baumeister}, H. and {Cassaing}, F. and {Cl{\'e}net}, Y. and
    {Dodds-Eden}, K. and {Eckart}, A. and {Gendron}, E. and {Genzel}, R. and
    {Gillessen}, S. and {Gr{\"a}ter}, A. and {Gueriau}, C. and {Hamaus}, N. and
    {Haubois}, X. and {Haug}, M. and {Henning}, T. and {Hippler}, S. and
    {Hofmann}, R. and {Hormuth}, F. and {Houairi}, K. and {Kellner}, S. and
    {Kervella}, P. and {Klein}, R. and {Kolmeder}, J. and {Laun}, W. and
    {L{\'e}na}, P. and {Lenzen}, R. and {Marteaud}, M. and {Naranjo}, V. and
    {Neumann}, U. and {Paumard}, T. and {Rabien}, S. and {Ramos}, J.~R. and
    {Reess}, J.~M. and {Rohloff}, R.-R. and {Rouan}, D. and {Rousset}, G. and
    {Ruyet}, B. and {Sevin}, A. and {Thiel}, M. and {Ziegleder}, J. and
    {Ziegler}, D.},
    title = "{GRAVITY: Microarcsecond Astrometry and Deep Interferometric Imaging with the VLT}",
booktitle = {Science with the VLT in the ELT Era},
     year = 2009,
   editor = "{A.~Moorwood}",
   volume = {Astrophysics and Space Science Proceedings},
publisher = {Springer Netherlands},
    pages = {361-365},
      doi = {10.1007/978-1-4020-9190-2_61},
   adsurl = {http://adsabs.harvard.edu/abs/2009svlt.conf..361E},
  adsnote = {Provided by the SAO/NASA Astrophysics Data System}
}

@ARTICLE{asm10,
   author = {{Ang{\'e}lil}, R. and {Saha}, P. and {Merritt}, D.},
    title = "{Toward Relativistic Orbit Fitting of Galactic Center Stars and Pulsars}",
  journal = {\apj},
archivePrefix = "arXiv",
   eprint = {1007.0007},
 primaryClass = "astro-ph.GA",
 keywords = {Galaxy: nucleus, gravitation, stars: kinematics and dynamics},
     year = 2010,
    month = sep,
   volume = 720,
    pages = {1303-1310},
      doi = {10.1088/0004-637X/720/2/1303},
   adsurl = {http://adsabs.harvard.edu/abs/2010ApJ...720.1303A},
  adsnote = {Provided by the SAO/NASA Astrophysics Data System}
}

@ARTICLE{wcp+09,
   author = {{Wang}, Y. and {Creighton}, T. and {Price}, R.~H. and {Jenet}, F.~A.
   },
    title = "{Strong Field Effects on Pulsar Arrival Times: General Orientations}",
  journal = {\apj},
archivePrefix = "arXiv",
   eprint = {0909.2709},
 primaryClass = "astro-ph.GA",
 keywords = {black hole physics, pulsars: general},
     year = 2009,
    month = nov,
   volume = 705,
    pages = {1252-1259},
      doi = {10.1088/0004-637X/705/2/1252},
   adsurl = {http://adsabs.harvard.edu/abs/2009ApJ...705.1252W},
  adsnote = {Provided by the SAO/NASA Astrophysics Data System}
}

@ARTICLE{wjc+09,
   author = {{Wang}, Y. and {Jenet}, F.~A. and {Creighton}, T. and {Price}, R.~H.
    },
    title = "{Strong Field Effects on Pulsar Arrival Times: Circular Orbits and Equatorial Beams}",
  journal = {\apj},
archivePrefix = "arXiv",
   eprint = {0812.2302},
 keywords = {black hole physics, pulsars: general},
     year = 2009,
    month = may,
   volume = 697,
    pages = {237-246},
      doi = {10.1088/0004-637X/697/1/237},
   adsurl = {http://adsabs.harvard.edu/abs/2009ApJ...697..237W},
  adsnote = {Provided by the SAO/NASA Astrophysics Data System}
}

@ARTICLE{kg05,
   author = {{K{\"o}nigsd{\"o}rffer}, C. and {Gopakumar}, A.},
    title = "{Post-Newtonian accurate parametric solution to the dynamics of spinning compact binaries in eccentric orbits: The leading order spin-orbit interaction}",
  journal = {\prd},
   eprint = {arXiv:gr-qc/0501011},
 keywords = {Post-Newtonian approximation; perturbation theory; related approximations, Wave generation and sources, Neutron stars, Black holes},
     year = 2005,
    month = jan,
   volume = 71,
   number = 2,
    pages = {024039-+},
      doi = {10.1103/PhysRevD.71.024039},
   adsurl = {http://adsabs.harvard.edu/abs/2005PhRvD..71b4039K},
  adsnote = {Provided by the SAO/NASA Astrophysics Data System}
}

@ARTICLE{gso+03,
   author = {{Genzel}, R. and {Sch{\"o}del}, R. and {Ott}, T. and {Eckart}, A. and
    {Alexander}, T. and {Lacombe}, F. and {Rouan}, D. and {Aschenbach}, B.
    },
    title = "{Near-infrared flares from accreting gas around the supermassive black hole at the Galactic Centre}",
  journal = {\nat},
   eprint = {arXiv:astro-ph/0310821},
     year = 2003,
    month = oct,
   volume = 425,
    pages = {934-937},
      doi = {10.1038/nature02065},
   adsurl = {http://adsabs.harvard.edu/abs/2003Natur.425..934G},
  adsnote = {Provided by the SAO/NASA Astrophysics Data System}
}

@ARTICLE{pmk+10,
   author = {{Perera}, B.~B.~P. and {McLaughlin}, M.~A. and {Kramer}, M. and
    {Stairs}, I.~H. and {Ferdman}, R.~D. and {Freire}, P.~C.~C. and
    {Possenti}, A. and {Breton}, R.~P. and {Manchester}, R.~N. and
    {Burgay}, M. and {Lyne}, A.~G. and {Camilo}, F.},
    title = "{The Evolution of PSR J0737-3039B and a Model for Relativistic Spin Precession}",
  journal = {\apj},
archivePrefix = "arXiv",
   eprint = {1008.1097},
 primaryClass = "astro-ph.GA",
 keywords = {pulsars: general, stars: neutron},
     year = 2010,
    month = oct,
   volume = 721,
    pages = {1193-1205},
      doi = {10.1088/0004-637X/721/2/1193},
   adsurl = {http://adsabs.harvard.edu/abs/2010ApJ...721.1193P},
  adsnote = {Provided by the SAO/NASA Astrophysics Data System}
}

@article{pdd+19,
    author = {Perera, B B P and DeCesar, M E and Demorest, P B and Kerr, M and Lentati, L and Nice, D J and Osłowski, S and Ransom, S M and Keith, M J and Arzoumanian, Z and Bailes, M and Baker, P T and Bassa, C G and Bhat, N D R and Brazier, A and Burgay, M and Burke-Spolaor, S and Caballero, R N and Champion, D J and Chatterjee, S and Chen, S and Cognard, I and Cordes, J M and Crowter, K and Dai, S and Desvignes, G and Dolch, T and Ferdman, R D and Ferrara, E C and Fonseca, E and Goldstein, J M and Graikou, E and Guillemot, L and Hazboun, J S and Hobbs, G and Hu, H and Islo, K and Janssen, G H and Karuppusamy, R and Kramer, M and Lam, M T and Lee, K J and Liu, K and Luo, J and Lyne, A G and Manchester, R N and McKee, J W and McLaughlin, M A and Mingarelli, C M F and Parthasarathy, A P and Pennucci, T T and Perrodin, D and Possenti, A and Reardon, D J and Russell, C J and Sanidas, S A and Sesana, A and Shaifullah, G and Shannon, R M and Siemens, X and Simon, J and Spiewak, R and Stairs, I H and Stappers, B W and Swiggum, J K and Taylor, S R and Theureau, G and Tiburzi, C and Vallisneri, M and Vecchio, A and Wang, J B and Zhang, S B and Zhang, L and Zhu, W W and Zhu, X J},
    title = "{The International Pulsar Timing Array: second data release}",
    journal = {Monthly Notices of the Royal Astronomical Society},
    volume = {490},
    number = {4},
    pages = {4666-4687},
    year = {2019},
    month = {10},
    abstract = "{In this paper, we describe the International Pulsar Timing Array second data release, which includes recent pulsar timing data obtained by three regional consortia: the European Pulsar Timing Array, the North American Nanohertz Observatory for Gravitational Waves, and the Parkes Pulsar Timing Array. We analyse and where possible combine high-precision timing data for 65 millisecond pulsars which are regularly observed by these groups. A basic noise analysis, including the processes which are both correlated and uncorrelated in time, provides noise models and timing ephemerides for the pulsars. We find that the timing precisions of pulsars are generally improved compared to the previous data release, mainly due to the addition of new data in the combination. The main purpose of this work is to create the most up-to-date IPTA data release. These data are publicly available for searches for low-frequency gravitational waves and other pulsar science.}",
    issn = {0035-8711},
    doi = {10.1093/mnras/stz2857},
    url = {https://doi.org/10.1093/mnras/stz2857},
    eprint = {https://academic.oup.com/mnras/article-pdf/490/4/4666/30489619/stz2857.pdf},
}


@article{vdk+18,
doi = {10.3847/1538-4357/aaaa73},
url = {https://dx.doi.org/10.3847/1538-4357/aaaa73},
year = {2018},
month = {mar},
publisher = {The American Astronomical Society},
volume = {855},
number = {2},
pages = {122},
author = {Sarah J. Vigeland and Adam T. Deller and David L. Kaplan and Alina G. Istrate and Benjamin W. Stappers and Thomas M. Tauris},
title = {Reconciling Optical and Radio Observations of the Binary Millisecond Pulsar PSR J1640+2224},
journal = {The Astrophysical Journal},
abstract = {Previous optical and radio observations of the binary millisecond pulsar PSR J1640+2224 have come to inconsistent conclusions about the identity of its companion, with some observations suggesting that the companion is a low-mass helium-core (He-core) white dwarf (WD), while others indicate that it is most likely a high-mass carbon–oxygen (CO) WD. Binary evolution models predict PSR J1640+2224 most likely formed in a low-mass X-ray binary based on the pulsar’s short spin period and long-period, low-eccentricity orbit, in which case its companion should be a He-core WD with mass about 0.35–0.39 M⊙, depending on metallicity. If instead it is a CO WD, it would suggest that the system has an unusual formation history. In this paper we present the first astrometric parallax measurement for this system from observations made with the Very Long Baseline Array (VLBA), from which we determine the distance to be . We use this distance and a reanalysis of archival optical observations originally taken in 1995 with the Wide Field Planetary Camera 2 on the Hubble Space Telescope (HST) to measure the WD’s mass. We also incorporate improvements in calibration, extinction model, and WD cooling models. We find that the existing observations are not sufficient to tightly constrain the companion mass, but we conclude the WD mass is &gt;0.4 M⊙ with &gt;90
}

@article{kop97b,
  author = {Kopeikin, S. M.},
  journal = {Journal of Mathematical Physics},
  year = {1997},
  volume = {38},
  pages = {2587}
}

@ARTICLE{sog+02,
   author = {{Sch{\"o}del}, R. and {Ott}, T. and {Genzel}, R. and {Hofmann}, R. and
    {Lehnert}, M. and {Eckart}, A. and {Mouawad}, N. and {Alexander}, T. and
    {Reid}, M.~J. and {Lenzen}, R. and {Hartung}, M. and {Lacombe}, F. and
    {Rouan}, D. and {Gendron}, E. and {Rousset}, G. and {Lagrange}, A.-M. and
    {Brandner}, W. and {Ageorges}, N. and {Lidman}, C. and {Moorwood}, A.~F.~M. and
    {Spyromilio}, J. and {Hubin}, N. and {Menten}, K.~M.},
    title = "{A star in a 15.2-year orbit around the supermassive black hole at the centre of the Milky Way}",
  journal = {\nat},
   eprint = {arXiv:astro-ph/0210426},
     year = 2002,
    month = oct,
   volume = 419,
    pages = {694-696},
      doi = {10.1038/nature01121},
   adsurl = {http://adsabs.harvard.edu/abs/2002Natur.419..694S},
  adsnote = {Provided by the SAO/NASA Astrophysics Data System}
}

@article{k+11,
  author = {{Kramer et~al.}},
  journal = {private communication},
  year = {2011}
}

@article{lan32,
  author = {{Landau}, L.~D.},
  journal = {Phys. Z. Sowjetunion},
  year = {1932},
  volume = {1},
  pages = {285}
}

@article{cha32,
  author = {{Chadwick}, L.},
  journal = {\nat},
  year = {1932},
  volume = {129},
  pages = {312}
}

@ARTICLE{lx11,
   author = {{Lai}, X.-Y. and {Xu}, R.-X.},
    title = "{A note on the discovery of a 2M_{odot} pulsar}",
  journal = {Research in Astronomy and Astrophysics},
archivePrefix = "arXiv",
   eprint = {1011.0526},
 primaryClass = "astro-ph.SR",
     year = 2011,
    month = jun,
   volume = 11,
    pages = {687-691},
      doi = {10.1088/1674-4527/11/6/008},
   adsurl = {http://adsabs.harvard.edu/abs/2011RAA....11..687L},
  adsnote = {Provided by the SAO/NASA Astrophysics Data System}
}

@ARTICLE{afo86,
   author = {{Alcock}, C. and {Farhi}, E. and {Olinto}, A.},
    title = "{Strange stars}",
  journal = {\apj},
 keywords = {NEUTRON STARS, QUARKS, STELLAR EVOLUTION, STELLAR PHYSICS, COOLING, EQUATIONS OF STATE, GROUND STATE, STELLAR STRUCTURE},
     year = 1986,
    month = nov,
   volume = 310,
    pages = {261-272},
      doi = {10.1086/164679},
   adsurl = {http://adsabs.harvard.edu/abs/1986ApJ...310..261A},
  adsnote = {Provided by the SAO/NASA Astrophysics Data System}
}

@ARTICLE{xu02,
   author = {{Xu}, R.~X.},
    title = "{A Thermal Featureless Spectrum: Evidence for Bare Strange Stars?}",
  journal = {ApJL},
   eprint = {arXiv:astro-ph/0202365},
 keywords = {Dense Matter, Elementary Particles, Stars: Pulsars: General, Stars: Neutron},
     year = 2002,
    month = may,
   volume = 570,
    pages = {L65-L68},
      doi = {10.1086/340993},
   adsurl = {http://adsabs.harvard.edu/abs/2002ApJ...570L..65X},
  adsnote = {Provided by the SAO/NASA Astrophysics Data System}
}

@article{ho65,
  author = {Hewish, A. and Okoye, S.E.},
  title = {Evidence of an unusual source of high radio brightness temperature in the Crab Nebula},
  journal = {\nat},
  volume = {207},
  pages = {59},
  year = {1965}
}

@ARTICLE{ovh+11,
   author = {{Os{\l}owski}, S. and {van Straten}, W. and {Hobbs}, G.~B. and
    {Bailes}, M. and {Demorest}, P.},
    title = "{High signal-to-noise ratio observations and the ultimate limits of precision pulsar timing}",
  journal = {\mnras},
archivePrefix = "arXiv",
   eprint = {1108.0812},
 primaryClass = "astro-ph.GA",
 keywords = {pulsars: general, pulsars: individual: PSR J0437-4715},
     year = 2011,
    month = dec,
   volume = 418,
    pages = {1258-1271},
      doi = {10.1111/j.1365-2966.2011.19578.x},
   adsurl = {http://adsabs.harvard.edu/abs/2011{\mnras}.418.1258O},
  adsnote = {Provided by the SAO/NASA Astrophysics Data System}
}

@ARTICLE{crg+10,
   author = {{Coles}, W.~A. and {Rickett}, B.~J. and {Gao}, J.~J. and {Hobbs}, G. and
    {Verbiest}, J.~P.~W.},
    title = "{Scattering of Pulsar Radio Emission by the Interstellar Plasma}",
  journal = {\apj},
archivePrefix = "arXiv",
   eprint = {1005.4914},
 primaryClass = "astro-ph.GA",
 keywords = {ISM: general, pulsars: general},
     year = 2010,
    month = jul,
   volume = 717,
    pages = {1206-1221},
      doi = {10.1088/0004-637X/717/2/1206},
   adsurl = {http://adsabs.harvard.edu/abs/2010ApJ...717.1206C},
  adsnote = {Provided by the SAO/NASA Astrophysics Data System}
}

@ARTICLE{dem11,
   author = {{Demorest}, P.},
    title = "{Cyclic Spectral Analysis of Radio Pulsars}",
  journal = {ArXiv e-prints:1106.3345},
archivePrefix = "arXiv",
   eprint = {1106.3345},
 primaryClass = "astro-ph.IM",
 keywords = {Astrophysics - Instrumentation and Methods for Astrophysics},
     year = 2011,
    month = jun,
   adsurl = {http://adsabs.harvard.edu/abs/2011arXiv1106.3345D},
  adsnote = {Provided by the SAO/NASA Astrophysics Data System}
}

@phdthesis{eat09,
  author = {Eatough, R.~P.},
  school = {University of Manchester},
  year = 2009,
  title = {A search for relativistic binary pulsars in the galactic plane}
}

@book{pw00,
  author = {Percival, D.~B. and Walden, A.~T.},
  title = {Wavelet Methods for Time Series Analysis},
  year = 2000,
  publisher = {Cambridge University Press}
}

@ARTICLE{srr+11,
   author = {{Sesana}, A. and {Roedig}, C. and {Reynolds}, M.~T. and {Dotti}, M.
    },
    title = "{Multimessenger astronomy with pulsar timing and X-ray observations of massive black hole binaries}",
  journal = {astro-ph/1107.2927},
archivePrefix = "arXiv",
   eprint = {1107.2927},
 primaryClass = "astro-ph.CO",
 keywords = {Astrophysics - Cosmology and Extragalactic Astrophysics},
     year = 2011,
    month = jul,
   adsurl = {http://adsabs.harvard.edu/abs/2011arXiv1107.2927S},
  adsnote = {Provided by the SAO/NASA Astrophysics Data System}
}

@ARTICLE{kkl+11,
   author = {{Keane}, E.~F. and {Kramer}, M. and {Lyne}, A.~G. and {Stappers}, B.~W. and
    {McLaughlin}, M.~A.},
    title = "{Rotating Radio Transients: new discoveries, timing solutions and musings}",
  journal = {\mnras},
 keywords = {surveys, ephemerides, stars: neutron, pulsars: general, Galaxy: stellar content},
     year = 2011,
    month = aug,
   volume = 415,
    pages = {3065-3080},
      doi = {10.1111/j.1365-2966.2011.18917.x},
   adsurl = {http://adsabs.harvard.edu/abs/2011{\mnras}.415.3065K},
  adsnote = {Provided by the SAO/NASA Astrophysics Data System}
}

@phdthesis{esp10,
  author = {Espinoza, C.~M.},
  school = {University of Manchester},
  year = 2010,
  title = {The rotational history of young pulsars},
}

@ARTICLE{gmr+11,
   author = {{Gou}, L. and {McClintock}, J.~E. and {Reid}, M.~J. and {Orosz}, J.~A. and
    {Steiner}, J.~F. and {Narayan}, R. and {Xiang}, J. and {Remillard}, R.~A. and
    {Arnaud}, K.~A. and {Davis}, S.~W.},
    title = "{The Extreme Spin of the Black Hole in Cygnus X-1}",
  journal = {\apj},
archivePrefix = "arXiv",
   eprint = {1106.3690},
 primaryClass = "astro-ph.HE",
 keywords = {accretion, accretion disks, black hole physics, stars: individual: Cygnus X-1, X-rays: binaries},
     year = 2011,
    month = dec,
   volume = 742,
      eid = {85},
    pages = {85},
      doi = {10.1088/0004-637X/742/2/85},
   adsurl = {http://adsabs.harvard.edu/abs/2011ApJ...742...85G},
  adsnote = {Provided by the SAO/NASA Astrophysics Data System}
}


@ARTICLE{par09,
   author = {{Paredes}, J.~M.},
    title = "{Black Holes in the Galaxy}",
  journal = {ArXiv e-prints},
archivePrefix = "arXiv",
   eprint = {0907.3602},
 primaryClass = "astro-ph.HE",
 keywords = {Astrophysics - High Energy Astrophysical Phenomena, Astrophysics - Solar and Stellar Astrophysics},
     year = 2009,
    month = jul,
   adsurl = {http://adsabs.harvard.edu/abs/2009arXiv0907.3602P},
  adsnote = {Provided by the SAO/NASA Astrophysics Data System}
}

@INPROCEEDINGS{de08,
   author = {{de Luca}, A.},
    title = "{Central Compact Objects in Supernova Remnants}",
 keywords = {Supernova remnants, Neutron stars, X-ray sources; X-ray bursts, Pulsars},
booktitle = {40 Years of Pulsars: Millisecond Pulsars, Magnetars and More},
     year = 2008,
   series = {American Institute of Physics Conference Series},
   volume = 983,
archivePrefix = "arXiv",
   eprint = {0712.2209},
   editor = "{C.~Bassa, Z.~Wang, A.~Cumming, \& V.~M.~Kaspi}",
    month = feb,
    pages = {311-319},
      doi = {10.1063/1.2900173},
   adsurl = {http://adsabs.harvard.edu/abs/2008AIPC..983..311D},
  adsnote = {Provided by the SAO/NASA Astrophysics Data System}
}

@ARTICLE{sch60,
   author = {{Schiff}, L.~I.},
    title = "{Possible New Experimental Test of General Relativity Theory}",
  journal = {Physical Review Letters},
     year = 1960,
    month = mar,
   volume = 4,
    pages = {215-217},
      doi = {10.1103/PhysRevLett.4.215},
   adsurl = {http://adsabs.harvard.edu/abs/1960PhRvL...4..215S},
  adsnote = {Provided by the SAO/NASA Astrophysics Data System}
}

@ARTICLE{pug59,
   author = {{Pugh}, G.~E.},
    title = "{Proposal for a satellite test of the Coriolis prediction of general relativity}",
  journal = {WSEG research Memorandom No.11, Weapons Systems Evaluation Group, The Pentagon, Washington},
     year = 1959,
}

@phdthesis{pur10,
  author = {Purver, M.~B.},
  school = {University of Manchester},
  year = 2010,
  title = {High-precision pulsar timing: The stability of integrated pulse profiles and their representation by analytic templates}
}

@INPROCEEDINGS{ks10,
   author = {{Kramer}, M. and {Stappers}, B.},
    title = "{LOFAR, LEAP and beyond: Using next generation telescopes for pulsar astrophysics}",
booktitle = {ISKAF2010 Science Meeting},
     year = 2010,
archivePrefix = "arXiv",
   eprint = {1009.1938},
 primaryClass = "astro-ph.IM",
   adsurl = {http://adsabs.harvard.edu/abs/2010iska.meetE..34K},
  adsnote = {Provided by the SAO/NASA Astrophysics Data System}
}

@INPROCEEDINGS{hm11,
   author = {{Hopman}, C. and {Madigan}, A.},
    title = "{Mass Segregation in the Galactic Centre}",
booktitle = {Astronomical Society of the Pacific Conference Series},
     year = 2011,
   series = {Astronomical Society of the Pacific Conference Series},
   volume = 439,
archivePrefix = "arXiv",
   eprint = {1002.1220},
 primaryClass = "astro-ph.GA",
   editor = "{M.~R.~Morris, Q.~D.~Wang, \& F.~Yuan}",
    month = may,
    pages = {180-+},
   adsurl = {http://adsabs.harvard.edu/abs/2011ASPC..439..180H},
  adsnote = {Provided by the SAO/NASA Astrophysics Data System}
}

@ARTICLE{fak06,
   author = {{Freitag}, M. and {Amaro-Seoane}, P. and {Kalogera}, V.},
    title = "{Models of mass segregation at the Galactic Centre}",
  journal = {Journal of Physics Conference Series},
   eprint = {arXiv:astro-ph/0607001},
     year = 2006,
    month = dec,
   volume = 54,
    pages = {252-258},
      doi = {10.1088/1742-6596/54/1/040},
   adsurl = {http://adsabs.harvard.edu/abs/2006JPhCS..54..252F},
  adsnote = {Provided by the SAO/NASA Astrophysics Data System}
}

@ARTICLE{okl09,
   author = {{O'Leary}, R.~M. and {Kocsis}, B. and {Loeb}, A.},
    title = "{Gravitational waves from scattering of stellar-mass black holes in galactic nuclei}",
  journal = {\mnras},
archivePrefix = "arXiv",
   eprint = {0807.2638},
 keywords = {black hole physics , gravitational waves , galaxies: kinematics and dynamics , galaxies: nuclei},
     year = 2009,
    month = jun,
   volume = 395,
    pages = {2127-2146},
      doi = {10.1111/j.1365-2966.2009.14653.x},
   adsurl = {http://adsabs.harvard.edu/abs/2009{\mnras}.395.2127O},
  adsnote = {Provided by the SAO/NASA Astrophysics Data System}
}

@ARTICLE{kha09,
   author = {{Keshet}, U. and {Hopman}, C. and {Alexander}, T.},
    title = "{Analytic Study of Mass Segregation Around a Massive Black Hole}",
  journal = {ApJL},
archivePrefix = "arXiv",
   eprint = {0901.4343},
 primaryClass = "astro-ph.GA",
 keywords = {black hole physics, Galaxy: kinematics and dynamics, stellar dynamics},
     year = 2009,
    month = jun,
   volume = 698,
    pages = {L64-L67},
      doi = {10.1088/0004-637X/698/1/L64},
   adsurl = {http://adsabs.harvard.edu/abs/2009ApJ...698L..64K},
  adsnote = {Provided by the SAO/NASA Astrophysics Data System}
}

@ARTICLE{kt11,
   author = {{Kocsis}, B. and {Tremaine}, S.},
    title = "{Resonant relaxation and the warp of the stellar disc in the Galactic Centre}",
  journal = {\mnras},
archivePrefix = "arXiv",
   eprint = {1006.0001},
 primaryClass = "astro-ph.GA",
 keywords = {celestial mechanics, Galaxy: centre, Galaxy: nucleus},
     year = 2011,
    month = mar,
   volume = 412,
    pages = {187-207},
      doi = {10.1111/j.1365-2966.2010.17897.x},
   adsurl = {http://adsabs.harvard.edu/abs/2011{\mnras}.412..187K},
  adsnote = {Provided by the SAO/NASA Astrophysics Data System}
}

@ARTICLE{vh10,
   author = {{Vigeland}, S.~J. and {Hughes}, S.~A.},
    title = "{Spacetime and orbits of bumpy black holes}",
  journal = {\prd},
archivePrefix = "arXiv",
   eprint = {0911.1756},
 primaryClass = "gr-qc",
 keywords = {Post-Newtonian approximation; perturbation theory; related approximations, Wave generation and sources, Classical black holes},
     year = 2010,
    month = jan,
   volume = 81,
   number = 2,
    pages = {024030-+},
      doi = {10.1103/PhysRevD.81.024030},
   adsurl = {http://adsabs.harvard.edu/abs/2010PhRvD..81b4030V},
  adsnote = {Provided by the SAO/NASA Astrophysics Data System}
}

@article{psa11,
    Author = {Psaltis, Dimitrios},
    Date-Added = {2011-10-04 14:56:53 +0000},
    Date-Modified = {2011-10-04 14:59:08 +0000},
    Journal = {in prep.},
    Year = {2011}
    }

@ARTICLE{vlj11,
   author = {{van Haasteren}, R. and {Levin}, Y. and {Janssen}, G.~H. and
    {Lazaridis}, K. and {Kramer}, M. and {Stappers}, B.~W. and {Desvignes}, G. and
    {Purver}, M.~B. and {Lyne}, A.~G. and {Ferdman}, R.~D. and {Jessner}, A. and
    {Cognard}, I. and {Theureau}, G. and {D'Amico}, N. and {Possenti}, A. and
    {Burgay}, M. and {Corongiu}, A. and {Hessels}, J.~W.~T. and
    {Smits}, R. and {Verbiest}, J.~P.~W.},
    title = "{Placing limits on the stochastic gravitational-wave background using European Pulsar Timing Array data}",
  journal = {\mnras},
archivePrefix = "arXiv",
   eprint = {1103.0576},
 primaryClass = "astro-ph.CO",
 keywords = {gravitational waves, methods: data analysis, pulsars: general},
     year = 2011,
    month = jul,
   volume = 414,
    pages = {3117-3128},
      doi = {10.1111/j.1365-2966.2011.18613.x},
   adsurl = {http://adsabs.harvard.edu/abs/2011{\mnras}.414.3117V},
  adsnote = {Provided by the SAO/NASA Astrophysics Data System}
}

@ARTICLE{dfg+13,
   author = {{Demorest}, P.~B. and {Ferdman}, R.~D. and {Gonzalez}, M.~E. and
    {Nice}, D. and {Ransom}, S. and {Stairs}, I.~H. and {Arzoumanian}, Z. and
    {Brazier}, A. and {Burke-Spolaor}, S. and {Chamberlin}, S.~J. and
    {Cordes}, J.~M. and {Ellis}, J. and {Finn}, L.~S. and {Freire}, P. and
    {Giampanis}, S. and {Jenet}, F. and {Kaspi}, V.~M. and {Lazio}, J. and
    {Lommen}, A.~N. and {McLaughlin}, M. and {Palliyaguru}, N. and
    {Perrodin}, D. and {Shannon}, R.~M. and {Siemens}, X. and {Stinebring}, D. and
    {Swiggum}, J. and {Zhu}, W.~W.},
    title = "{Limits on the Stochastic Gravitational Wave Background from the North American Nanohertz Observatory for Gravitational Waves}",
  journal = {\apj},
archivePrefix = "arXiv",
   eprint = {1201.6641},
 primaryClass = "astro-ph.CO",
 keywords = {gravitational waves, methods: data analysis, pulsars: general, pulsars: individual: J0030+0451 J0613-0200 J1012+5307 J1455-3330 J1600-3053 J1640+2224 J1643-1224 J1713+0747 J1744-1134 J1853+1308 B1855+09 J1909-3744 J1910+1256 J1918-0642 B1953+29 J2145-0750 J2317+1439},
     year = 2013,
    month = jan,
   volume = 762,
      eid = {94},
    pages = {94},
      doi = {10.1088/0004-637X/762/2/94},
   adsurl = {http://adsabs.harvard.edu/abs/2013ApJ...762...94D},
  adsnote = {Provided by the SAO/NASA Astrophysics Data System}
}

@ARTICLE{des13,
   author = {{Desvignes}, G.},
     year = 2013,
  journal = {private communication},
}

@ARTICLE{fre13,
   author = {{Freire}, P.~C.~C.},
     year = 2013,
  journal = {private communication},
}

@phdthesis{liu12,
  author = {Liu, K.},
  school = {University of Manchester},
  year = 2012,
  title = {EXPLORING PULSAR{BLACK HOLE BINARIES USING THE NEXT GENERATION OF RADIO TELESCOPES}
}
}

@ARTICLE{fwe+12,
   author = {{Freire}, P.~C.~C. and {Wex}, N. and {Esposito-Far{\`e}se}, G. and
    {Verbiest}, J.~P.~W. and {Bailes}, M. and {Jacoby}, B.~A. and
    {Kramer}, M. and {Stairs}, I.~H. and {Antoniadis}, J. and {Janssen}, G.~H.
    },
    title = "{The relativistic pulsar-white dwarf binary PSR J1738+0333 - II. The most stringent test of scalar-tensor gravity}",
  journal = {\mnras},
archivePrefix = "arXiv",
   eprint = {1205.1450},
 primaryClass = "astro-ph.GA",
 keywords = {gravitation, gravitational waves, pulsars: individual: PSR J1738+0333},
     year = 2012,
    month = jul,
   volume = 423,
    pages = {3328-3343},
      doi = {10.1111/j.1365-2966.2012.21253.x},
   adsurl = {http://adsabs.harvard.edu/abs/2012{\mnras}.423.3328F},
  adsnote = {Provided by the SAO/NASA Astrophysics Data System}
}

@ARTICLE{hpk81,
   author = {{Haensel}, P. and {Proszynski}, M. and {Kutschera}, M.},
    title = "{Uncertainty in the saturation density of nuclear matter and neutron star models}",
  journal = \aap,
 keywords = {NEUTRON STARS, NUCLEAR PARTICLES, PARTICLE DENSITY (CONCENTRATION), STELLAR MASS, STELLAR MODELS, EQUATIONS OF STATE, FIELD THEORY (PHYSICS), MATTER (PHYSICS), NUCLEAR ASTROPHYSICS, NUCLEONS},
     year = 1981,
    month = oct,
   volume = 102,
    pages = {299-302},
   adsurl = {http://adsabs.harvard.edu/abs/1981A
  adsnote = {Provided by the SAO/NASA Astrophysics Data System}
}

@ARTICLE{sc12,
   author = {{Shannon}, R.~M. and {Cordes}, J.~M.},
    title = "{Pulse Intensity Modulation and the Timing Stability of Millisecond Pulsars: A Case Study of PSR J1713+0747}",
  journal = {\apj},
archivePrefix = "arXiv",
   eprint = {1210.7021},
 primaryClass = "astro-ph.SR",
 keywords = {gravitational waves, methods: statistical, pulsars: general, pulsars: individual: PSR J1713+0747 },
     year = 2012,
    month = dec,
   volume = 761,
      eid = {64},
    pages = {64},
      doi = {10.1088/0004-637X/761/1/64},
   adsurl = {http://adsabs.harvard.edu/abs/2012ApJ...761...64S},
  adsnote = {Provided by the SAO/NASA Astrophysics Data System}
}

@ARTICLE{ovd+13,
   author = {{Os{\l}owski}, S. and {van Straten}, W. and {Demorest}, P. and
    {Bailes}, M.},
    title = "{Improving the precision of pulsar timing through polarization statistics}",
  journal = {\mnras},
archivePrefix = "arXiv",
   eprint = {1301.2374},
 primaryClass = "astro-ph.IM",
 keywords = {methods: statistical, pulsars: general, pulsars: individual: PSR J0437-4715},
     year = 2013,
    month = mar,
   volume = 430,
    pages = {416-424},
      doi = {10.1093/mnras/sts662},
   adsurl = {http://adsabs.harvard.edu/abs/2013{\mnras}.430..416O},
  adsnote = {Provided by the SAO/NASA Astrophysics Data System}
}

@INPROCEEDINGS{dam09,
   author = {{Damour}, T.},
    title = "{Binary Systems as Test-Beds of Gravity Theories}",
booktitle = {Astrophysics and Space Science Library},
     year = 2009,
   series = {Astrophysics and Space Science Library},
   volume = 359,
   editor = {{Colpi}, M. and {Casella}, P. and {Gorini}, V. and {Moschella}, U. and
    {Possenti}, A.},
    pages = {1-4020},
      doi = {10.1007/978-1-4020-9264-0_1},
   adsurl = {http://adsabs.harvard.edu/abs/2009ASSL..359....1D},
  adsnote = {Provided by the SAO/NASA Astrophysics Data System}
}

@ARTICLE{kcs+13,
   author = {{Keith}, M.~J. and {Coles}, W. and {Shannon}, R.~M. and {Hobbs}, G.~B. and
    {Manchester}, R.~N. and {Bailes}, M. and {Bhat}, N.~D.~R. and
    {Burke-Spolaor}, S. and {Champion}, D.~J. and {Chaudhary}, A. and
    {Hotan}, A.~W. and {Khoo}, J. and {Kocz}, J. and {Os{\l}owski}, S. and
    {Ravi}, V. and {Reynolds}, J.~E. and {Sarkissian}, J. and {van Straten}, W. and
    {Yardley}, D.~R.~B.},
    title = "{Measurement and correction of variations in interstellar dispersion in high-precision pulsar timing}",
  journal = {\mnras},
archivePrefix = "arXiv",
   eprint = {1211.5887},
 primaryClass = "astro-ph.GA",
 keywords = {methods: data analysis, pulsars: general, ISM: structure},
     year = 2013,
    month = mar,
   volume = 429,
    pages = {2161-2174},
      doi = {10.1093/mnras/sts486},
   adsurl = {http://adsabs.harvard.edu/abs/2013{\mnras}.429.2161K},
  adsnote = {Provided by the SAO/NASA Astrophysics Data System}
}

@ARTICLE{sti09,
   author = {{Stinebring}, D.~R.},
     year = 2009,
  journal = {private communication},
}

@ARTICLE{van13,
   author = {{van Straten}, W.},
    title = "{High-fidelity Radio Astronomical Polarimetry Using a Millisecond Pulsar as a Polarized Reference Source}",
  journal = {\apjss},
archivePrefix = "arXiv",
   eprint = {1212.3446},
 primaryClass = "astro-ph.IM",
 keywords = {methods: data analysis, polarization, pulsars: general, pulsars: individual: PSR J1022+1001 PSR J0437-4715, techniques: polarimetric},
     year = 2013,
    month = jan,
   volume = 204,
      eid = {13},
    pages = {13},
      doi = {10.1088/0067-0049/204/1/13},
   adsurl = {http://adsabs.harvard.edu/abs/2013ApJS..204...13V},
  adsnote = {Provided by the SAO/NASA Astrophysics Data System}
}

@ARTICLE{ksv08,
   author = {{Karuppusamy}, R. and {Stappers}, B. and {van Straten}, W.},
    title = "{PuMa-II: A Wide Band Pulsar Machine for the Westerbork Synthesis Radio Telescope}",
  journal = pasp,
 keywords = {Astronomical Instrumentation},
     year = 2008,
    month = feb,
   volume = 120,
    pages = {191-202},
      doi = {10.1086/528699},
   adsurl = {http://adsabs.harvard.edu/abs/2008PASP..120..191K},
  adsnote = {Provided by the SAO/NASA Astrophysics Data System}
}

@ARTICLE{vdo12,
   author = {{van Straten}, W. and {Demorest}, P. and {Oslowski}, S.},
    title = "{Pulsar Data Analysis with PSRCHIVE}",
  journal = {Astronomical Research and Technology},
archivePrefix = "arXiv",
   eprint = {1205.6276},
 primaryClass = "astro-ph.IM",
 keywords = {pulsar, data analysis, software},
     year = 2012,
    month = jul,
   volume = 9,
    pages = {237-256},
   adsurl = {http://adsabs.harvard.edu/abs/2012AR
  adsnote = {Provided by the SAO/NASA Astrophysics Data System}
}

@ARTICLE{vb11,
   author = {{van Straten}, W. and {Bailes}, M.},
    title = "{DSPSR: Digital Signal Processing Software for Pulsar Astronomy}",
  journal = pasa,
archivePrefix = "arXiv",
   eprint = {1008.3973},
 primaryClass = "astro-ph.IM",
 keywords = {methods: data analysis, polarisation, pulsars: general, techniques: polarimetric},
     year = 2011,
    month = jan,
   volume = 28,
    pages = {1-14},
      doi = {10.1071/AS10021},
   adsurl = {http://adsabs.harvard.edu/abs/2011PASA...28....1V},
  adsnote = {Provided by the SAO/NASA Astrophysics Data System}
}

@ARTICLE{afw+13,
   author = {{Antoniadis}, J. and {Freire}, P.~C.~C. and {Wex}, N. and {Tauris}, T.~M. and
    {Lynch}, R.~S. and {van Kerkwijk}, M.~H. and {Kramer}, M. and
    {Bassa}, C. and {Dhillon}, V.~S. and {Driebe}, T. and {Hessels}, J.~W.~T. and
    {Kaspi}, V.~M. and {Kondratiev}, V.~I. and {Langer}, N. and
    {Marsh}, T.~R. and {McLaughlin}, M.~A. and {Pennucci}, T.~T. and
    {Ransom}, S.~M. and {Stairs}, I.~H. and {van Leeuwen}, J. and
    {Verbiest}, J.~P.~W. and {Whelan}, D.~G.},
    title = "{A Massive Pulsar in a Compact Relativistic Binary}",
  journal = {Science},
archivePrefix = "arXiv",
   eprint = {1304.6875},
 primaryClass = "astro-ph.HE",
 keywords = {Pulsars, Neutron Stars, General relativity,  Tests of General relativity, Gravitational Radiation,  Stellar evolution},
     year = 2013,
    month = apr,
   volume = 340,
    pages = {448},
      doi = {10.1126/science.1233232},
   adsurl = {http://cdsads.u-strasbg.fr/abs/2013Sci...340..448A},
  adsnote = {Provided by the SAO/NASA Astrophysics Data System}
}

@ARTICLE{gjv+12,
   author = {{Guillemot}, L. and {Johnson}, T.~J. and {Venter}, C. and {Kerr}, M. and
    {Pancrazi}, B. and {Livingstone}, M. and {Janssen}, G.~H. and
    {Jaroenjittichai}, P. and {Kramer}, M. and {Cognard}, I. and
    {Stappers}, B.~W. and {Harding}, A.~K. and {Camilo}, F. and
    {Espinoza}, C.~M. and {Freire}, P.~C.~C. and {Gargano}, F. and
    {Grove}, J.~E. and {Johnston}, S. and {Michelson}, P.~F. and
    {Noutsos}, A. and {Parent}, D. and {Ransom}, S.~M. and {Ray}, P.~S. and
    {Shannon}, R. and {Smith}, D.~A. and {Theureau}, G. and {Thorsett}, S.~E. and
    {Webb}, N.},
    title = "{Pulsed Gamma Rays from the Original Millisecond and Black Widow Pulsars: A Case for Caustic Radio Emission?}",
  journal = {\apj},
archivePrefix = "arXiv",
   eprint = {1110.1271},
 primaryClass = "astro-ph.HE",
 keywords = {gamma rays: general, pulsars: general, pulsars: individual: PSR B1937+21 PSR B1957+20, radiation mechanisms: non-thermal},
     year = 2012,
    month = jan,
   volume = 744,
      eid = {33},
    pages = {33},
      doi = {10.1088/0004-637X/744/1/33},
   adsurl = {http://adsabs.harvard.edu/abs/2012ApJ...744...33G},
  adsnote = {Provided by the SAO/NASA Astrophysics Data System}
}

@ARTICLE{man13,
   author = {{Manchester}, R.~N.},
    title = "{The International Pulsar Timing Array}",
  journal = {Class. Quant Grav. in press, astro-ph/1309.7392},
archivePrefix = "arXiv",
   eprint = {1309.7392},
 primaryClass = "astro-ph.IM",
 keywords = {Astrophysics - Instrumentation and Methods for Astrophysics},
     year = 2013,
    month = sep,
   adsurl = {http://adsabs.harvard.edu/abs/2013arXiv1309.7392M},
  adsnote = {Provided by the SAO/NASA Astrophysics Data System}
}

@article{l+13,
  author = {{Lee et~al.}},
  journal = {in prep.},
  year = {2013}
}

@ARTICLE{aaa+13,
   author = {{Abdo}, A.~A. and {Ajello}, M. and {Allafort}, A. and {Baldini}, L. and
    {Ballet}, J. and {Barbiellini}, G. and {Baring}, M.~G. and {Bastieri}, D. and
    {Belfiore}, A. and {Bellazzini}, R. and et al.},
    title = "{The Second Fermi Large Area Telescope Catalog of Gamma-Ray Pulsars}",
  journal = \apjs,
archivePrefix = "arXiv",
   eprint = {1305.4385},
 primaryClass = "astro-ph.HE",
 keywords = {catalogs, pulsars: general, stars: neutron },
     year = 2013,
    month = oct,
   volume = 208,
      eid = {17},
    pages = {17},
      doi = {10.1088/0067-0049/208/2/17},
   adsurl = {http://cdsads.u-strasbg.fr/abs/2013ApJS..208...17A},
  adsnote = {Provided by the SAO/NASA Astrophysics Data System}
}

@ARTICLE{kcs+13,
   author = {{Keith}, M.~J. and {Coles}, W. and {Shannon}, R.~M. and {Hobbs}, G.~B. and
    {Manchester}, R.~N. and {Bailes}, M. and {Bhat}, N.~D.~R. and
    {Burke-Spolaor}, S. and {Champion}, D.~J. and {Chaudhary}, A. and
    {Hotan}, A.~W. and {Khoo}, J. and {Kocz}, J. and {Os{\l}owski}, S. and
    {Ravi}, V. and {Reynolds}, J.~E. and {Sarkissian}, J. and {van Straten}, W. and
    {Yardley}, D.~R.~B.},
    title = "{Measurement and correction of variations in interstellar dispersion in high-precision pulsar timing}",
  journal = {\mnras},
archivePrefix = "arXiv",
   eprint = {1211.5887},
 primaryClass = "astro-ph.GA",
 keywords = {methods: data analysis, pulsars: general, ISM: structure},
     year = 2013,
    month = mar,
   volume = 429,
    pages = {2161-2174},
      doi = {10.1093/mnras/sts486},
   adsurl = {http://cdsads.u-strasbg.fr/abs/2013{\mnras}.429.2161K},
  adsnote = {Provided by the SAO/NASA Astrophysics Data System}
}

@INPROCEEDINGS{ct06,
   author = {{Cognard}, I. and {Theureau}, G.},
    title = "{The Nan{\c c}ay Pulsar Instrumentation : The BON Coherent Dedispersor}",
booktitle = {IAU Joint Discussion},
     year = 2006,
   series = {IAU Joint Discussion},
   volume = 2,
    month = aug,
   adsurl = {http://adsabs.harvard.edu/abs/2006IAUJD...2E..36C},
  adsnote = {Provided by the SAO/NASA Astrophysics Data System}
}

@INPROCEEDINGS{rdf+09,
   author = {{Ransom}, S.~M. and {Demorest}, P. and {Ford}, J. and {McCullough}, R. and
    {Ray}, J. and {DuPlain}, R. and {Brandt}, P.},
    title = "{GUPPI: Green Bank Ultimate Pulsar Processing Instrument}",
booktitle = {American Astronomical Society Meeting Abstracts 214},
     year = 2009,
   series = {American Astronomical Society Meeting Abstracts},
   volume = 214,
    month = dec,
    pages = {605.08},
   adsurl = {http://adsabs.harvard.edu/abs/2009AAS...21460508R},
  adsnote = {Provided by the SAO/NASA Astrophysics Data System}
}

@INPROCEEDINGS{ctg+13,
   author = {{Cognard}, I. and {Theureau}, G. and {Guillemot}, L. and {Liu}, K. and
    {Lassus}, A. and {Desvignes}, G.},
    title = "{Nan{\c c}ay contribution to the worldwide pulsar programs}",
 keywords = {pulsar, Nan$\backslash$c cay, high energies, gravitational waves},
booktitle = {SF2A-2013: Proceedings of the Annual meeting of the French Society of Astronomy and Astrophysics},
     year = 2013,
   editor = {{Cambresy}, L. and {Martins}, F. and {Nuss}, E. and {Palacios}, A.
    },
    month = nov,
    pages = {327-330},
   adsurl = {http://adsabs.harvard.edu/abs/2013sf2a.conf..327C},
  adsnote = {Provided by the SAO/NASA Astrophysics Data System}
}

@ARTICLE{pdr14,
   author = {{Pennucci}, T.~T. and {Demorest}, P.~B. and {Ransom}, S.~M.},
    title = "{Elementary Wideband Timing of Radio Pulsars}",
  journal = {astro-ph/1402.1672},
archivePrefix = "arXiv",
   eprint = {1402.1672},
 primaryClass = "astro-ph.IM",
 keywords = {Astrophysics - Instrumentation and Methods for Astrophysics, Astrophysics - Solar and Stellar Astrophysics},
     year = 2014,
    month = feb,
   adsurl = {http://adsabs.harvard.edu/abs/2014arXiv1402.1672P},
  adsnote = {Provided by the SAO/NASA Astrophysics Data System}
}

@INPROCEEDINGS{ldl+97,
   author = {{Lorimer}, D.~R. and {D'Amico}, N. and {Lyne}, A.~G. and {Seiradakis}, J.~H. and
    {Athanasopoulos}, A. and {Camilo}, F. and {Jessner}, A. and
    {Kramer}, M. and {Wielebinski}, R. and {Xilouris}, K.~M.},
    title = "{The European Pulsar Network Pulse Profile Database}",
booktitle = {Joint European and National Astronomical Meeting},
     year = 1997,
   editor = {{Hadjidemetrioy}, J.~D. and {Seiradakis}, J.~H.},
   adsurl = {http://adsabs.harvard.edu/abs/1997jena.confE.236L},
  adsnote = {Provided by the SAO/NASA Astrophysics Data System}
}

@ARTICLE{dbl+13,
   author = {{Deller}, A.~T. and {Boyles}, J. and {Lorimer}, D.~R. and {Kaspi}, V.~M. and
    {McLaughlin}, M.~A. and {Ransom}, S. and {Stairs}, I.~H. and
    {Stovall}, K.},
    title = "{VLBI Astrometry of PSR J2222-0137: A Pulsar Distance Measured to 0.4\% Accuracy}",
  journal = {\apj},
archivePrefix = "arXiv",
   eprint = {1305.4865},
 primaryClass = "astro-ph.SR",
 keywords = {astrometry, pulsars: general, pulsars: individual: J2222{\ndash}0137, techniques: interferometric},
     year = 2013,
    month = jun,
   volume = 770,
      eid = {145},
    pages = {145},
      doi = {10.1088/0004-637X/770/2/145},
   adsurl = {http://adsabs.harvard.edu/abs/2013ApJ...770..145D},
  adsnote = {Provided by the SAO/NASA Astrophysics Data System}
}

@ARTICLE{stw+11,
   author = {{Smits}, R. and {Tingay}, S.~J. and {Wex}, N. and {Kramer}, M. and
    {Stappers}, B.},
    title = "{Prospects for accurate distance measurements of pulsars with the Square Kilometre Array: Enabling fundamental physics}",
  journal = aass,
archivePrefix = "arXiv",
   eprint = {1101.5971},
 primaryClass = "astro-ph.IM",
 keywords = {parallaxes, pulsars: general, stars: neutron, telescopes, astrometry},
     year = 2011,
    month = apr,
   volume = 528,
      eid = {A108},
    pages = {A108},
      doi = {10.1051/0004-6361/201016141},
   adsurl = {http://adsabs.harvard.edu/abs/2011A
  adsnote = {Provided by the SAO/NASA Astrophysics Data System}
}

@ARTICLE{sha+11,
   author = {{Stappers}, B.~W. and {Hessels}, J.~W.~T. and {Alexov}, A. and
    {Anderson}, K. and {Coenen}, T. and {Hassall}, T. and {Karastergiou}, A. and
    {Kondratiev}, V.~I. and {Kramer}, M. and {van Leeuwen}, J. and
    {Mol}, J.~D. and {Noutsos}, A. and {Romein}, J.~W. and {Weltevrede}, P. and
    {Fender}, R. and {Wijers}, R.~A.~M.~J. and {B{\"a}hren}, L. and
    {Bell}, M.~E. and {Broderick}, J. and {Daw}, E.~J. and {Dhillon}, V.~S. and
    {Eisl{\"o}ffel}, J. and {Falcke}, H. and {Griessmeier}, J. and
    {Law}, C. and {Markoff}, S. and {Miller-Jones}, J.~C.~A. and
    {Scheers}, B. and {Spreeuw}, H. and {Swinbank}, J. and {Ter Veen}, S. and
    {Wise}, M.~W. and {Wucknitz}, O. and {Zarka}, P. and {Anderson}, J. and
    {Asgekar}, A. and {Avruch}, I.~M. and {Beck}, R. and {Bennema}, P. and
    {Bentum}, M.~J. and {Best}, P. and {Bregman}, J. and {Brentjens}, M. and
    {van de Brink}, R.~H. and {Broekema}, P.~C. and {Brouw}, W.~N. and
    {Br{\"u}ggen}, M. and {de Bruyn}, A.~G. and {Butcher}, H.~R. and
    {Ciardi}, B. and {Conway}, J. and {Dettmar}, R.-J. and {van Duin}, A. and
    {van Enst}, J. and {Garrett}, M. and {Gerbers}, M. and {Grit}, T. and
    {Gunst}, A. and {van Haarlem}, M.~P. and {Hamaker}, J.~P. and
    {Heald}, G. and {Hoeft}, M. and {Holties}, H. and {Horneffer}, A. and
    {Koopmans}, L.~V.~E. and {Kuper}, G. and {Loose}, M. and {Maat}, P. and
    {McKay-Bukowski}, D. and {McKean}, J.~P. and {Miley}, G. and
    {Morganti}, R. and {Nijboer}, R. and {Noordam}, J.~E. and {Norden}, M. and
    {Olofsson}, H. and {Pandey-Pommier}, M. and {Polatidis}, A. and
    {Reich}, W. and {R{\"o}ttgering}, H. and {Schoenmakers}, A. and
    {Sluman}, J. and {Smirnov}, O. and {Steinmetz}, M. and {Sterks}, C.~G.~M. and
    {Tagger}, M. and {Tang}, Y. and {Vermeulen}, R. and {Vermaas}, N. and
    {Vogt}, C. and {de Vos}, M. and {Wijnholds}, S.~J. and {Yatawatta}, S. and
    {Zensus}, A.},
    title = "{Observing pulsars and fast transients with LOFAR}",
  journal = aass,
archivePrefix = "arXiv",
   eprint = {1104.1577},
 primaryClass = "astro-ph.IM",
 keywords = {telescopes, pulsars: general, instrumentation: interferometers, methods: observational, stars: neutron, ISM: general},
     year = 2011,
    month = jun,
   volume = 530,
      eid = {A80},
    pages = {A80},
      doi = {10.1051/0004-6361/201116681},
   adsurl = {http://adsabs.harvard.edu/abs/2011A
  adsnote = {Provided by the SAO/NASA Astrophysics Data System}
}

@ARTICLE{bcg+13,
   author = {{Berti}, E. and {Cardoso}, V. and {Gualtieri}, L. and {Horbatsch}, M. and
    {Sperhake}, U.},
    title = "{Numerical simulations of single and binary black holes in scalar-tensor theories: Circumventing the no-hair theorem}",
  journal = {Phys.~Rev.~D},
archivePrefix = "arXiv",
   eprint = {1304.2836},
 primaryClass = "gr-qc",
 keywords = {Modified theories of gravity, Numerical studies of black holes and black-hole binaries, Wave generation and sources},
     year = 2013,
    month = jun,
   volume = 87,
   number = 12,
      eid = {124020},
    pages = {124020},
      doi = {10.1103/PhysRevD.87.124020}
}

@INPROCEEDINGS{dam88,
   author = {{Damour}, T.},
    title = "{Strong-field tests of general relativity and the binary pulsar.}",
 keywords = {Binary Pulsars:General Relativity, Binary Pulsars:Gravitation Theory, General Relativity:Binary Pulsars, Gravitation Theory:Binary Pulsars},
booktitle = {Proceedings of the 2nd Canadian Conference on General Relativity and Relativistic Astrophysics},
     year = 1988,
   editor = {{Coley}, A. and {Dyer}, C. and {Tupper}, T.},
    pages = {315-334}
}

@INPROCEEDINGS{dam09,
   author = {{Damour}, T.},
    title = "{Binary Systems as Test-Beds of Gravity Theories}",
booktitle = {Astrophysics and Space Science Library},
     year = 2009,
   series = {Astrophysics and Space Science Library},
   volume = 359,
   editor = {{Colpi}, M. and {Casella}, P. and {Gorini}, V. and {Moschella}, U. and
    {Possenti}, A.},
    pages = {1-4020},
      doi = {10.1007/978-1-4020-9264-0_1}
}

@ARTICLE{esp09,
   author = {{Esposito-Far{\`e}se}, G.},
    title = "{Motion in alternative theories of gravity}",
  journal = {ArXiv e-prints: 0905.2575},
archivePrefix = "arXiv",
   eprint = {0905.2575},
 primaryClass = "gr-qc",
 keywords = {General Relativity and Quantum Cosmology},
     year = 2009,
}

@ARTICLE{haw72,
   author = {{Hawking}, S.~W.},
    title = "{Black holes in the Brans-Dicke: Theory of gravitation}",
  journal = {Communications in Mathematical Physics},
     year = 1972,
    month = jun,
   volume = 25,
    pages = {167-171},
      doi = {10.1007/BF01877518}
}

@ARTICLE{gmr+13,
   author = {{Gou}, L. and {McClintock}, J.~E. and {Remillard}, R.~A. and
    {Steiner}, J.~F. and {Reid}, M.~J. and {Orosz}, J.~A. and {Narayan}, R. and
    {Hanke}, M. and {Garc{\'{\i}}a}, J.},
    title = "{Confirmation Via the Continuum-Fitting Method that the Spin of the Black Hole in Cygnus X-1 is Extreme}",
 primaryClass = "astro-ph.HE",
 keywords = {Astrophysics - High Energy Astrophysical Phenomena, Astrophysics - Solar and Stellar Astrophysics, High Energy Physics - Theory},
   journal = {astro-ph/1308.4760},
     year = 2013,
    month = aug
}

@article{jac99,
  title = {Primordial Black Hole Evolution in Tensor-Scalar Cosmology},
  author = {Jacobson, Ted},
  journal = {Phys. Rev. Lett.},
  volume = {83},
  issue = {14},
  pages = {2699--2702},
  year = {1999},
  month = {Oct},
}

@ARTICLE{laz13,
   author = {{Lazio}, T.~J.~W.},
    title = "{The Square Kilometre Array pulsar timing array}",
  journal = {Class.~Quantum~Grav.},
     year = 2013,
    month = nov,
   volume = 30,
   number = 22,
      eid = {224011},
    pages = {224011},
      doi = {10.1088/0264-9381/30/22/224011},
}

@article{mw13,
  title = {Compact binary systems in scalar-tensor gravity: Equations of motion to 2.5 post-Newtonian order},
  author = {Mirshekari, Saeed and Will, Clifford M.},
  journal = {Phys.~Rev.~D},
  volume = {87},
  issue = {8},
  pages = {084070},
  numpages = {27},
  year = {2013},
  month = {Apr},
  publisher = {American Physical Society},
  doi = {10.1103/PhysRevD.87.084070}
}

@ARTICLE{mpa87,
   author = {{M{\"u}ther}, H. and {Prakash}, M. and {Ainsworth}, T.~L.},
    title = "{The nuclear symmetry energy in relativistic Brueckner-Hartree-Fock calculations}",
  journal = {Physics Letters B},
     year = 1987,
    month = dec,
   volume = 199,
    pages = {469-474},
      doi = {10.1016/0370-2693(87)91611-X},
}

@ARTICLE{nm13,
   author = {{Narayan}, R. and {McClintock}, J.~E.},
    title = "{Observational Evidence for Black Holes}",
  journal = {ArXiv e-prints: 1312.6698},
archivePrefix = "arXiv",
   eprint = {1312.6698},
 primaryClass = "astro-ph.HE",
 keywords = {Astrophysics - High Energy Astrophysical Phenomena},
     year = 2013,
    month = dec,
   adsurl = {http://adsabs.harvard.edu/abs/2013arXiv1312.6698N},
  adsnote = {Provided by the SAO/NASA Astrophysics Data System}
}

@BOOK{will93,
   author = {{Will}, C.~M.},
    title = "{Theory and Experiment in Gravitational Physics}",
booktitle = {Theory and Experiment in Gravitational Physics, by Clifford M.~Will, pp.~396.~ISBN 0521439736.~Cambridge, UK: Cambridge University Press, March 1993.},
     year = 1993
}

@ARTICLE{k+14,
   author = {{Karuppusamy}, R. and {Karuppusamy}, R. and {Karuppusamy}, R. and {Karuppusamy}, R. and {Karuppusamy}, R. and {Karuppusamy}, R. and {Karuppusamy}, R. and {Karuppusamy}, R. and {Karuppusamy}, R. and {Karuppusamy}, R. and {Karuppusamy}, R. and {Karuppusamy}, R. and {Karuppusamy}, R.},
  journal = {in prep.},
     year = 2014,
}

@ARTICLE{dv01,
   author = {{Dhurandhar}, S.~V. and {Vecchio}, A.},
    title = "{Searching for continuous gravitational wave sources in binary systems}",
  journal = prd,
   eprint = {gr-qc/0011085},
 keywords = {Gravitational wave detectors and experiments, Gravitational radiation detectors, mass spectrometers, and other instrumentation and techniques, Mathematical procedures and computer techniques, Pulsars},
     year = 2001,
    month = jun,
   volume = 63,
   number = 12,
      eid = {122001},
    pages = {122001},
      doi = {10.1103/PhysRevD.63.122001},
   adsurl = {http://adsabs.harvard.edu/abs/2001PhRvD..63l2001D},
  adsnote = {Provided by the SAO/NASA Astrophysics Data System}
}

@ARTICLE{ekl+13,
   author = {{Eatough}, R.~P. and {Kramer}, M. and {Lyne}, A.~G. and {Keith}, M.~J.
    },
    title = "{A coherent acceleration search of the Parkes multibeam pulsar survey - techniques and the discovery and timing of 16 pulsars}",
  journal = {\mnras},
archivePrefix = "arXiv",
   eprint = {1301.6346},
 primaryClass = "astro-ph.IM",
 keywords = {methods: data analysis, stars: neutron, pulsars: general},
     year = 2013,
    month = may,
   volume = 431,
    pages = {292-307},
      doi = {10.1093/mnras/stt161},
   adsurl = {http://adsabs.harvard.edu/abs/2013{\mnras}.431..292E},
  adsnote = {Provided by the SAO/NASA Astrophysics Data System}
}

@ARTICLE{kjv+10,
   author = {{Keith}, M.~J. and {Jameson}, A. and {van Straten}, W. and {Bailes}, M. and
    {Johnston}, S. and {Kramer}, M. and {Possenti}, A. and {Bates}, S.~D. and
    {Bhat}, N.~D.~R. and {Burgay}, M. and {Burke-Spolaor}, S. and
    {D'Amico}, N. and {Levin}, L. and {McMahon}, P.~L. and {Milia}, S. and
    {Stappers}, B.~W.},
    title = "{The High Time Resolution Universe Pulsar Survey - I. System configuration and initial discoveries}",
  journal = {\mnras},
archivePrefix = "arXiv",
   eprint = {1006.5744},
 primaryClass = "astro-ph.HE",
 keywords = {pulsars: general},
     year = 2010,
    month = dec,
   volume = 409,
    pages = {619-627},
      doi = {10.1111/j.1365-2966.2010.17325.x},
   adsurl = {http://adsabs.harvard.edu/abs/2010{\mnras}.409..619K},
  adsnote = {Provided by the SAO/NASA Astrophysics Data System}
}

@ARTICLE{cai04,
   author = {{Cairns}, I.~H.},
    title = "{Properties and Interpretations of Giant Micropulses and Giant Pulses from Pulsars}",
  journal = {\apj},
   eprint = {astro-ph/0404174},
 keywords = {Methods: Statistical, Plasmas, Stars: Pulsars: General, Stars: Pulsars: Individual: Name: Vela Pulsar, Radiation Mechanisms: Nonthermal, Stars: Neutron},
     year = 2004,
    month = aug,
   volume = 610,
    pages = {948-955},
      doi = {10.1086/421756},
   adsurl = {http://adsabs.harvard.edu/abs/2004ApJ...610..948C},
  adsnote = {Provided by the SAO/NASA Astrophysics Data System}
}

@ARTICLE{pet11,
   author = {{P{\'e}tri}, J.},
    title = "{A unified polar cap/striped wind model for pulsed radio and gamma-ray emission in pulsars}",
  journal = {\mnras},
archivePrefix = "arXiv",
   eprint = {1011.3431},
 primaryClass = "astro-ph.HE",
 keywords = {radiation mechanisms: non-thermal, relativistic processes, pulsars: general, stars: winds, outflows, gamma-rays: stars},
     year = 2011,
    month = apr,
   volume = 412,
    pages = {1870-1880},
      doi = {10.1111/j.1365-2966.2010.18023.x},
   adsurl = {http://cdsads.u-strasbg.fr/abs/2011{\mnras}.412.1870P},
  adsnote = {Provided by the SAO/NASA Astrophysics Data System}
}

@ARTICLE{drh04,
   author = {{Dyks}, J. and {Rudak}, B. and {Harding}, A.~K.},
    title = "{On the Methods of Determining the Radio Emission Geometry in Pulsar Magnetospheres}",
  journal = {\apj},
   eprint = {astro-ph/0307251},
 keywords = {Stars: Pulsars: General, Radiation Mechanisms: Nonthermal},
     year = 2004,
    month = jun,
   volume = 607,
    pages = {939-948},
      doi = {10.1086/383587},
   adsurl = {http://cdsads.u-strasbg.fr/abs/2004ApJ...607..939D},
  adsnote = {Provided by the SAO/NASA Astrophysics Data System}
}

@ARTICLE{lbj+14,
   author = {{Lee}, K.~J. and {Bassa}, C.~G. and {Janssen}, G.~H. and {Karuppusamy}, R. and
    {Kramer}, M. and {Liu}, K. and {Perrodin}, D. and {Smits}, R. and
    {Stappers}, B.~W. and {van Haasteren}, R. and {Lentati}, L.},
    title = "{Model-based asymptotically optimal dispersion measure correction for pulsar timing}",
  journal = {{\mnras} in press},
   eprint = {1404.2084},
 primaryClass = "astro-ph.IM",
 keywords = {Astrophysics - Instrumentation and Methods for Astrophysics},
     year = 2014,
    month = apr,
    volume = {astro-ph/1404.2084},
   adsurl = {http://adsabs.harvard.edu/abs/2014arXiv1404.2084L},
  adsnote = {Provided by the SAO/NASA Astrophysics Data System}
}

@INPROCEEDINGS{wle+13,
   author = {{Wex}, N. and {Liu}, K. and {Eatough}, R.~P. and {Kramer}, M. and
    {Cordes}, J.~M. and {Lazio}, T.~J.~W.},
    title = "{Prospects for probing strong gravity with a pulsar-black hole system}",
 keywords = {pulsars: general, black hole physics, relativity},
booktitle = {IAU Symposium},
     year = 2013,
   series = {IAU Symposium},
   volume = 291,
   editor = {{van Leeuwen}, J.},
    month = mar,
    pages = {171-176},
      doi = {10.1017/S1743921312023538},
}

@ARTICLE{kek+13,
   author = {{Knispel}, B. and {Eatough}, R.~P. and {Kim}, H. and {Keane}, E.~F. and
    {Allen}, B. and {Anderson}, D. and {Aulbert}, C. and {Bock}, O. and
    {Crawford}, F. and {Eggenstein}, H.-B. and {Fehrmann}, H. and
    {Hammer}, D. and {Kramer}, M. and {Lyne}, A.~G. and {Machenschalk}, B. and
    {Miller}, R.~B. and {Papa}, M.~A. and {Rastawicki}, D. and {Sarkissian}, J. and
    {Siemens}, X. and {Stappers}, B.~W.},
    title = "{Einstein@Home Discovery of 24 Pulsars in the Parkes Multi-beam Pulsar Survey}",
  journal = {\apj},
archivePrefix = "arXiv",
   eprint = {1302.0467},
 primaryClass = "astro-ph.GA",
 keywords = {methods: data analysis, pulsars: general, stars: neutron},
     year = 2013,
    month = sep,
   volume = 774,
      eid = {93},
    pages = {93},
      doi = {10.1088/0004-637X/774/2/93},
   adsurl = {http://adsabs.harvard.edu/abs/2013ApJ...774...93K},
  adsnote = {Provided by the SAO/NASA Astrophysics Data System}
}

@ARTICLE{blw13,
   author = {{Bagchi}, M. and {Lorimer}, D.~R. and {Wolfe}, S.},
    title = "{On the detectability of eccentric binary pulsars}",
  journal = {\mnras},
archivePrefix = "arXiv",
   eprint = {1302.4914},
 primaryClass = "astro-ph.SR",
 keywords = {methods: analytical, methods: numerical, binaries: general, stars: neutron, pulsars: general},
     year = 2013,
    month = jun,
   volume = 432,
    pages = {1303-1314},
      doi = {10.1093/mnras/stt559},
   adsurl = {http://adsabs.harvard.edu/abs/2013{\mnras}.432.1303B},
  adsnote = {Provided by the SAO/NASA Astrophysics Data System}
}

@ARTICLE{lrf+11,
   author = {{Lynch}, R.~S. and {Ransom}, S.~M. and {Freire}, P.~C.~C. and
    {Stairs}, I.~H.},
    title = "{Six New Recycled Globular Cluster Pulsars Discovered with the Green Bank Telescope}",
  journal = {\apj},
archivePrefix = "arXiv",
   eprint = {1101.1467},
 primaryClass = "astro-ph.SR",
 keywords = {globular clusters: individual: M22 NGC 6517, pulsars: individual: J1836{\ndash}2354A J1836{\ndash}2354B J1801{\ndash}0857A J1801{\ndash}0857B J1801{\ndash}0857C J1801{\ndash}0857D},
     year = 2011,
    month = jun,
   volume = 734,
      eid = {89},
    pages = {89},
      doi = {10.1088/0004-637X/734/2/89},
   adsurl = {http://adsabs.harvard.edu/abs/2011ApJ...734...89L},
  adsnote = {Provided by the SAO/NASA Astrophysics Data System}
}

@ARTICLE{dtm+13,
   author = {{Dewdney}, P.~E. and {Turner}, W. and {Millenaar}, and {McCool}, R. and {Lazio}, J. and {Cornwell}, T.~J.},
   journal = {SKA-TEL-SKO-DD-001},
      year = 2013
}

@INPROCEEDINGS{yln13,
   author = {{Yue}, Y. and {Li}, D. and {Nan}, R.},
    title = "{FAST low frequency pulsar survey}",
 keywords = {pulsars: general},
booktitle = {IAU Symposium},
     year = 2013,
   series = {IAU Symposium},
   volume = 291,
archivePrefix = "arXiv",
   eprint = {1211.0748},
 primaryClass = "astro-ph.IM",
   editor = {{van Leeuwen}, J.},
    month = mar,
    pages = {577-579},
      doi = {10.1017/S174392131300001X},
   adsurl = {http://adsabs.harvard.edu/abs/2013IAUS..291..577Y},
  adsnote = {Provided by the SAO/NASA Astrophysics Data System}
}

@ARTICLE{csc14,
   author = {{Clausen}, D. and {Sigurdsson}, S. and {Chernoff}, D.~F.},
    title = "{Dynamically formed black hole+millisecond pulsar binaries in globular clusters}",
  journal = {astro-ph/1404.7502},
archivePrefix = "arXiv",
   eprint = {1404.7502},
 primaryClass = "astro-ph.HE",
 keywords = {Astrophysics - High Energy Astrophysical Phenomena},
     year = 2014,
    month = apr,
   adsurl = {http://adsabs.harvard.edu/abs/2014arXiv1404.7502C},
  adsnote = {Provided by the SAO/NASA Astrophysics Data System}
}

@ARTICLE{lah+14,
   author = {{Lentati}, L. and {Alexander}, P. and {Hobson}, M.~P. and {Feroz}, F. and
    {van Haasteren}, R. and {Lee}, K.~J. and {Shannon}, R.~M.},
    title = "{TEMPONEST: a Bayesian approach to pulsar timing analysis}",
  journal = {\mnras},
archivePrefix = "arXiv",
   eprint = {1310.2120},
 primaryClass = "astro-ph.IM",
 keywords = {methods: data analysis, pulsars: general, pulsars: individual: B1937+21},
     year = 2014,
    month = jan,
   volume = 437,
    pages = {3004-3023},
      doi = {10.1093/mnras/stt2122},
   adsurl = {http://adsabs.harvard.edu/abs/2014/mnras.437.3004L},
  adsnote = {Provided by the SAO/NASA Astrophysics Data System}
}

@ARTICLE{src+13,
   author = {{Shannon}, R.~M. and {Ravi}, V. and {Coles}, W.~A. and {Hobbs}, G. and
    {Keith}, M.~J. and {Manchester}, R.~N. and {Wyithe}, J.~S.~B. and
    {Bailes}, M. and {Bhat}, N.~D.~R. and {Burke-Spolaor}, S. and
    {Khoo}, J. and {Levin}, Y. and {Oslowski}, S. and {Sarkissian}, J.~M. and
    {van Straten}, W. and {Verbiest}, J.~P.~W. and {Want}, J.-B.
    },
    title = "{Gravitational-wave limits from pulsar timing constrain supermassive black hole evolution.}",
  journal = {Science},
archivePrefix = "arXiv",
   eprint = {1310.4569},
 primaryClass = "astro-ph.CO",
     year = 2013,
    month = oct,
   volume = 342,
    pages = {334-337},
   adsurl = {http://adsabs.harvard.edu/abs/2013Sci...342..334S},
  adsnote = {Provided by the SAO/NASA Astrophysics Data System}
}

@ARTICLE{mks+11,
   author = {{Magro}, A. and {Karastergiou}, A. and {Salvini}, S. and {Mort}, B. and
    {Dulwich}, F. and {Zarb Adami}, K.},
    title = "{Real-time, fast radio transient searches with GPU de-dispersion}",
  journal = {\mnras},
archivePrefix = "arXiv",
   eprint = {1107.2516},
 primaryClass = "astro-ph.IM",
 keywords = {instrumentation: miscellaneous},
     year = 2011,
    month = nov,
   volume = 417,
    pages = {2642-2650},
      doi = {10.1111/j.1365-2966.2011.19426.x},
   adsurl = {http://adsabs.harvard.edu/abs/2011{\mnras}.417.2642M},
  adsnote = {Provided by the SAO/NASA Astrophysics Data System}
}

@ARTICLE{bbb+12,
   author = {{Barsdell}, B.~R. and {Bailes}, M. and {Barnes}, D.~G. and {Fluke}, C.~J.
    },
    title = "{Accelerating incoherent dedispersion}",
  journal = {\mnras},
archivePrefix = "arXiv",
   eprint = {1201.5380},
 primaryClass = "astro-ph.IM",
 keywords = {methods: data analysis, pulsars: general},
     year = 2012,
    month = may,
   volume = 422,
    pages = {379-392},
      doi = {10.1111/j.1365-2966.2012.20622.x},
   adsurl = {http://adsabs.harvard.edu/abs/2012{\mnras}.422..379B},
  adsnote = {Provided by the SAO/NASA Astrophysics Data System}
}

@INPROCEEDINGS{akg+12,
   author = {{Armour}, W. and {Karastergiou}, A. and {Giles}, M. and {Williams}, C. and
    {Magro}, A. and {Zagkouris}, K. and {Roberts}, S. and {Salvini}, S. and
    {Dulwich}, F. and {Mort}, B.},
    title = "{A GPU-based Survey for Millisecond Radio Transients Using ARTEMIS}",
booktitle = {Astronomical Data Analysis Software and Systems XXI},
     year = 2012,
   series = {Astronomical Society of the Pacific Conference Series},
   volume = 461,
archivePrefix = "arXiv",
   eprint = {1111.6399},
 primaryClass = "astro-ph.IM",
   editor = {{Ballester}, P. and {Egret}, D. and {Lorente}, N.~P.~F.},
    month = sep,
    pages = {33},
   adsurl = {http://adsabs.harvard.edu/abs/2012ASPC..461...33A},
  adsnote = {Provided by the SAO/NASA Astrophysics Data System}
}

@PHDTHESIS{hes07,
   author = {{Hessels}, J.~W.~T.},
    title = "{Discovery and study of exotic radio pulsars}",
 keywords = {Exotic radio pulsars, Pulsar discovery, Globular clusters, Pulsar physics, Ultradense matter, Strong gravity},
   school = {McGill University, Canada},
     year = 2007,
   adsurl = {http://adsabs.harvard.edu/abs/2007PhDT........28H},
  adsnote = {Provided by the SAO/NASA Astrophysics Data System}
}

@ARTICLE{n+14,
   author = {{Ng}, C. and {Karuppusamy}, R. and {Karuppusamy}, R. and {Karuppusamy}, R. and {Karuppusamy}, R. and {Karuppusamy}, R. and {Karuppusamy}, R. and {Karuppusamy}, R. and {Karuppusamy}, R. and {Karuppusamy}, R. and {Karuppusamy}, R. and {Karuppusamy}, R. and {Karuppusamy}, R.},
  journal = {in prep.},
     year = 2014,
}

@ARTICLE{dsm+13,
   author = {{Deneva}, J.~S. and {Stovall}, K. and {McLaughlin}, M.~A. and
    {Bates}, S.~D. and {Freire}, P.~C.~C. and {Martinez}, J.~G. and
    {Jenet}, F. and {Bagchi}, M.},
    title = "{Goals, Strategies and First Discoveries of AO327, the Arecibo All-sky 327 MHz Drift Pulsar Survey}",
  journal = {\apj},
archivePrefix = "arXiv",
   eprint = {1307.8142},
 primaryClass = "astro-ph.SR",
 keywords = {pulsars: general, stars: neutron},
     year = 2013,
    month = sep,
   volume = 775,
      eid = {51},
    pages = {51},
      doi = {10.1088/0004-637X/775/1/51},
   adsurl = {http://adsabs.harvard.edu/abs/2013ApJ...775...51D},
  adsnote = {Provided by the SAO/NASA Astrophysics Data System}
}

@ARTICLE{lfr+12,
   author = {{Lynch}, R.~S. and {Freire}, P.~C.~C. and {Ransom}, S.~M. and
    {Jacoby}, B.~A.},
    title = "{The Timing of Nine Globular Cluster Pulsars}",
  journal = {\apj},
archivePrefix = "arXiv",
   eprint = {1112.2612},
 primaryClass = "astro-ph.HE",
 keywords = {globular clusters: individual: M62 NGC 6544 NGC 6624, pulsars: individual: J1701-3006D J1701-3006E J1701-3006F J1807-2459A J1807-2500B J1823-3021D J1823-3021E J1823-3021F},
     year = 2012,
    month = feb,
   volume = 745,
      eid = {109},
    pages = {109},
      doi = {10.1088/0004-637X/745/2/109},
   adsurl = {http://cdsads.u-strasbg.fr/abs/2012ApJ...745..109L},
  adsnote = {Provided by the SAO/NASA Astrophysics Data System}
}

@ARTICLE{fkl09,
   author = {{Freire}, P.~C. and {Kramer}, M. and {Lyne}, A.~G.},
    title = "{Erratum: Determination of the orbital parameters of binary pulsars}",
  journal = {\mnras},
 keywords = {errata, addenda , binaries: general , pulsars: general , globular clusters: individual: 47 Tucanae},
     year = 2009,
    month = may,
   volume = 395,
    pages = {1775-1775},
      doi = {10.1111/j.1365-2966.2009.14609.x},
   adsurl = {http://adsabs.harvard.edu/abs/2009{\mnras}.395.1775F},
  adsnote = {Provided by the SAO/NASA Astrophysics Data System}
}

@ARTICLE{ovb+14,
   author = {{Os{\l}owski}, S. and {van Straten}, W. and {Bailes}, M. and
    {Jameson}, A. and {Hobbs}, G.},
    title = "{Timing, polarimetry and physics of the bright, nearby millisecond pulsar PSR J0437-4715 - a single-pulse perspective}",
  journal = {\mnras},
archivePrefix = "arXiv",
   eprint = {1405.2638},
 primaryClass = "astro-ph.SR",
 keywords = {pulsars: general, pulsars: individual (PSR J0437-4715)},
     year = 2014,
    month = jul,
   volume = 441,
    pages = {3148-3160},
      doi = {10.1093/mnras/stu804},
   adsurl = {http://adsabs.harvard.edu/abs/2014{\mnras}.441.3148O},
  adsnote = {Provided by the SAO/NASA Astrophysics Data System}
}

@ARTICLE{rya97,
   author = {{Ryan}, F.~D.},
    title = "{Spinning boson stars with large self-interaction}",
  journal = {\prd},
 keywords = {Relativistic stars: structure stability and oscillations, Stellar rotation},
     year = 1997,
    month = may,
   volume = 55,
    pages = {6081-6091},
      doi = {10.1103/PhysRevD.55.6081},
   adsurl = {http://adsabs.harvard.edu/abs/1997PhRvD..55.6081R},
  adsnote = {Provided by the SAO/NASA Astrophysics Data System}
}

@ARTICLE{wwj12,
   author = {{Weltevrede}, P. and {Wright}, G. and {Johnston}, S.},
  journal = {\mnras},
     year = 2012,
    month = aug,
   volume = 424,
    pages = {843-854}
}

@ARTICLE{ymv+11,
   author = {{Yan}, W.~M. and {Manchester}, R.~N. and {van Straten}, W. and
    {Reynolds}, J.~E. and {Hobbs}, G. and {Wang}, N. and {Bailes}, M. and
    {Bhat}, N.~D.~R. and {Burke-Spolaor}, S. and {Champion}, D.~J. and
    {Coles}, W.~A. and {Hotan}, A.~W. and {Khoo}, J. and {Oslowski}, S. and
    {Sarkissian}, J.~M. and {Verbiest}, J.~P.~W. and {Yardley}, D.~R.~B.
    },
    title = "{Polarization observations of 20 millisecond pulsars}",
  journal = {\mnras},
archivePrefix = "arXiv",
   eprint = {1102.2274},
 primaryClass = "astro-ph.SR",
 keywords = {polarization, pulsars: general, radio continuum: stars},
     year = 2011,
    month = jul,
   volume = 414,
    pages = {2087-2100},
      doi = {10.1111/j.1365-2966.2011.18522.x},
   adsurl = {http://adsabs.harvard.edu/abs/2011{\mnras}.414.2087Y},
  adsnote = {Provided by the SAO/NASA Astrophysics Data System}
}

@ARTICLE{sod+14,
   author = {{Shannon}, R.~M. and {Os{\l}owski}, S. and {Dai}, S. and {Bailes}, M. and
    {Hobbs}, G. and {Manchester}, R.~N. and {van Straten}, W. and
    {Raithel}, C.~A. and {Ravi}, V. and {Toomey}, L. and {Bhat}, N.~D.~R. and
    {Burke-Spolaor}, S. and {Coles}, W.~A. and {Keith}, M.~J. and
    {Kerr}, M. and {Levin}, Y. and {Sarkissian}, J.~M. and {Wang}, J.-B. and
    {Wen}, L. and {Zhu}, X.-J.},
    title = "{Limitations in timing precision due to single-pulse shape variability in millisecond pulsars}",
  journal = {\mnras},
archivePrefix = "arXiv",
   eprint = {1406.4716},
 primaryClass = "astro-ph.SR",
 keywords = {methods: data analysis, stars: neutron, pulsars: general},
     year = 2014,
    month = sep,
   volume = 443,
    pages = {1463-1481},
      doi = {10.1093/mnras/stu1213},
   adsurl = {http://adsabs.harvard.edu/abs/2014{\mnras}.443.1463S},
  adsnote = {Provided by the SAO/NASA Astrophysics Data System}
}

@ARTICLE{lkl+15,
   author = {{Liu}, K. and {Karuppusamy}, R. and {Lee}, K.~J. and {Stappers}, B.~W. and
    {Kramer}, M. and {Smits}, R. and {Purver}, M.~B. and {Janssen}, G.~H. and
    {Perrodin}, D.},
    title = "{Single-pulse and profile-variability study of PSR J1022+1001}",
  journal = {\mnras},
archivePrefix = "arXiv",
   eprint = {1502.06785},
 primaryClass = "astro-ph.HE",
 keywords = {methods: data analysis, pulsars: individual: PSR J1022+1001},
     year = 2015,
    month = may,
   volume = 449,
    pages = {1158-1169},
      doi = {10.1093/mnras/stv397},
   adsurl = {http://adsabs.harvard.edu/abs/2015{\mnras}.449.1158L},
  adsnote = {Provided by the SAO/NASA Astrophysics Data System}
}

@article{ard+06,
doi = {10.1086/501067},
url = {https://dx.doi.org/10.1086/501067},
year = {2006},
month = {may},
publisher = {},
volume = {642},
number = {2},
pages = {1012},
author = {Andrea N. Lommen and Richard A. Kipphorn and David J. Nice and Eric M. Splaver and Ingrid H. Stairs and Donald C. Backer},
title = {The Parallax and Proper Motion of PSR J0030+0451},
journal = {The Astrophysical Journal},
abstract = {We report the parallax and proper motion of millisecond pulsar J0030+0451, one of 13 known isolated millisecond pulsars in the disk of the Galaxy. We obtained more than 6 years of monthly data from the 305 m Arecibo telescope at 430 and 1410 MHz. We measure the parallax of PSR J0030+0451 to be 3.3 ± 0.9 mas, corresponding to a distance of 300 ± 90 pc. The Cordes and Lazio model of Galactic electron distribution yields a dispersion-measure-derived distance of 317 pc, which agrees with our measurement. We place the pulsar's transverse space velocity in the range of 8-17 km s-1, making this pulsar one of the slowest known. We perform a brief census of velocities of isolated versus binary millisecond pulsars. We find that the velocities of the two populations are indistinguishable. However, the scale height of the binary population is twice that of the isolated population, and the luminosity functions of the two populations are different. We suggest that the scale height difference may be an artifact of the luminosity difference.}
}

@ARTICLE{bjd+06,
       author = {{Burgay}, M. and {Joshi}, B.~C. and {D'Amico}, N. and {Possenti}, A. and {Lyne}, A.~G. and {Manchester}, R.~N. and {McLaughlin}, M.~A. and {Kramer}, M. and {Camilo}, F. and {Freire}, P.~C.~C.},
        title = "{The Parkes High-Latitude pulsar survey}",
      journal = {\mnras},
     keywords = {methods: observational, pulsars: general},
         year = 2006,
        month = may,
       volume = {368},
       number = {1},
        pages = {283-292},
          doi = {10.1111/j.1365-2966.2006.10100.x},
       adsurl = {https://ui.adsabs.harvard.edu/abs/2006MNRAS.368..283B},
      adsnote = {Provided by the SAO/NASA Astrophysics Data System}
}

@ARTICLE{wes06,
   author = {{Weltevrede}, P. and {Edwards}, R.~T. and {Stappers}, B.~W.},
    title = "{The subpulse modulation properties of pulsars at 21 cm}",
  journal = aa,
   eprint = {astro-ph/0507282},
 keywords = {stars: pulsars: general},
     year = 2006,
    month = jan,
   volume = 445,
    pages = {243-272},
      doi = {10.1051/0004-6361:20053088},
   adsurl = {http://adsabs.harvard.edu/abs/2006A
  adsnote = {Provided by the SAO/NASA Astrophysics Data System}
}

@ARTICLE{dlc+14,
   author = {{Dolch}, T. and {Lam}, M.~T. and {Cordes}, J. and {Chatterjee}, S. and
    {Bassa}, C. and {Bhattacharyya}, B. and {Champion}, D.~J. and
    {Cognard}, I. and {Crowter}, K. and {Demorest}, P.~B. and {Hessels}, J.~W.~T. and
    {Janssen}, G. and {Jenet}, F.~A. and {Jones}, G. and {Jordan}, C. and
    {Karuppusamy}, R. and {Keith}, M. and {Kondratiev}, V. and {Kramer}, M. and
    {Lazarus}, P. and {Lazio}, T.~J.~W. and {Lee}, K.~J. and {McLaughlin}, M.~A. and
    {Roy}, J. and {Shannon}, R.~M. and {Stairs}, I. and {Stovall}, K. and
    {Verbiest}, J.~P.~W. and {Madison}, D.~R. and {Palliyaguru}, N. and
    {Perrodin}, D. and {Ransom}, S. and {Stappers}, B. and {Zhu}, W.~W. and
    {Dai}, S. and {Desvignes}, G. and {Guillemot}, L. and {Liu}, K. and
    {Lyne}, A. and {Perera}, B.~B.~P. and {Petroff}, E. and {Rankin}, J.~M. and
    {Smits}, R.},
    title = "{A 24 Hr Global Campaign to Assess Precision Timing of the Millisecond Pulsar J1713+0747}",
  journal = {\apj},
archivePrefix = "arXiv",
   eprint = {1408.1694},
 primaryClass = "astro-ph.HE",
 keywords = {gravitational waves, ISM: structure, pulsars: individual: PSR J1713+0747},
     year = 2014,
    month = oct,
   volume = 794,
      eid = {21},
    pages = {21},
      doi = {10.1088/0004-637X/794/1/21},
   adsurl = {http://adsabs.harvard.edu/abs/2014ApJ...794...21D},
  adsnote = {Provided by the SAO/NASA Astrophysics Data System}
}

@ARTICLE{mgm09,
   author = {{Mitra}, D. and {Gil}, J. and {Melikidze}, G.~I.},
    title = "{Unraveling the Nature of Coherent Pulsar Radio Emission}",
  journal = {\apjl},
archivePrefix = "arXiv",
   eprint = {0903.3023},
 primaryClass = "astro-ph.HE",
 keywords = {pulsars: general, radiation mechanisms: non-thermal},
     year = 2009,
    month = may,
   volume = 696,
    pages = {L141-L145},
      doi = {10.1088/0004-637X/696/2/L141},
   adsurl = {http://adsabs.harvard.edu/abs/2009ApJ...696L.141M},
  adsnote = {Provided by the SAO/NASA Astrophysics Data System}
}

@ARTICLE{tjb+13,
   author = {{Tiburzi}, C. and {Johnston}, S. and {Bailes}, M. and {Bates}, S.~D. and
    {Bhat}, N.~D.~R. and {Burgay}, M. and {Burke-Spolaor}, S. and
    {Champion}, D. and {Coster}, P. and {D'Amico}, N. and {Keith}, M.~J. and
    {Kramer}, M. and {Levin}, L. and {Milia}, S. and {Ng}, C. and
    {Possenti}, A. and {Stappers}, B.~W. and {Thornton}, D. and
    {van Straten}, W.},
    title = "{The High Time Resolution Universe survey - IX. Polarimetry of long-period pulsars}",
  journal = {\mnras},
archivePrefix = "arXiv",
   eprint = {1310.1823},
 primaryClass = "astro-ph.SR",
 keywords = {magnetic fields, polarization, methods: observational, pulsars: general},
     year = 2013,
    month = dec,
   volume = 436,
    pages = {3557-3572},
      doi = {10.1093/mnras/stt1834},
   adsurl = {http://adsabs.harvard.edu/abs/2013{\mnras}.436.3557T},
  adsnote = {Provided by the SAO/NASA Astrophysics Data System}
}

@INPROCEEDINGS{elc+15,
   author = {{Eatough}, R. and {Lazio}, T.~J.~W. and {Casanellas}, J. and
    {Chatterjee}, S. and {Cordes}, J.~M. and {Demorest}, P.~B. and
    {Kramer}, M. and {Lee}, K.~J. and {Liu}, K. and {Ransom}, S.~M. and
    {Wex}, N.},
    title = "{Observing Radio Pulsars in the Galactic Centre with the Square Kilometre Array}",
booktitle = {Advancing Astrophysics with the Square Kilometre Array (AASKA14)},
     year = 2015,
archivePrefix = "arXiv",
   eprint = {1501.00281},
 primaryClass = "astro-ph.IM",
      eid = {45},
    pages = {45},
   adsurl = {http://adsabs.harvard.edu/abs/2015aska.confE..45E},
  adsnote = {Provided by the SAO/NASA Astrophysics Data System}
}

@ARTICLE{ltm+15,
   author = {{Lentati}, L. and {Taylor}, S.~R. and {Mingarelli}, C.~M.~F. and
    {Sesana}, A. and {Sanidas}, S.~A. and {Vecchio}, A. and {Caballero}, R.~N. and
    {Lee}, K.~J. and {van Haasteren}, R. and {Babak}, S. and {Bassa}, C.~G. and
    {Brem}, P. and {Burgay}, M. and {Champion}, D.~J. and {Cognard}, I. and
    {Desvignes}, G. and {Gair}, J.~R. and {Guillemot}, L. and {Hessels}, J.~W.~T. and
    {Janssen}, G.~H. and {Karuppusamy}, R. and {Kramer}, M. and
    {Lassus}, A. and {Lazarus}, P. and {Liu}, K. and {Os{\l}owski}, S. and
    {Perrodin}, D. and {Petiteau}, A. and {Possenti}, A. and {Purver}, M.~B. and
    {Rosado}, P.~A. and {Smits}, R. and {Stappers}, B. and {Theureau}, G. and
    {Tiburzi}, C. and {Verbiest}, J.~P.~W.},
    title = "{European Pulsar Timing Array limits on an isotropic stochastic gravitational-wave background}",
  journal = {\mnras},
archivePrefix = "arXiv",
   eprint = {1504.03692},
 keywords = {gravitational waves, methods: data analysis, pulsars: general},
     year = 2015,
    month = nov,
   volume = 453,
    pages = {2576-2598},
      doi = {10.1093/mnras/stv1538},
   adsurl = {http://adsabs.harvard.edu/abs/2015{\mnras}.453.2576L},
  adsnote = {Provided by the SAO/NASA Astrophysics Data System}
}

@ARTICLE{ick+15,
   author = {{Imgrund}, M. and {Champion}, D.~J. and {Kramer}, M. and {Lesch}, H.
    },
    title = "{A Bayesian method for pulsar template generation}",
  journal = {\mnras},
archivePrefix = "arXiv",
   eprint = {1501.03497},
 primaryClass = "astro-ph.IM",
 keywords = {methods: analytical, methods: statistical, pulsars: general, pulsars: individual: B1133+16, pulsars: individual: B0329+54, pulsars: individual: J1713+0747},
     year = 2015,
    month = jun,
   volume = 449,
    pages = {4162-4183},
      doi = {10.1093/mnras/stv449},
   adsurl = {http://adsabs.harvard.edu/abs/2015{\mnras}.449.4162I},
  adsnote = {Provided by the SAO/NASA Astrophysics Data System}
}

@PHDTHESIS{bil12,
   author = {{Bilous}, A.~V.},
    title = "{Single-pulse study of radio pulsars}",
   school = {University of Virginia},
     year = 2012,
   adsurl = {http://adsabs.harvard.edu/abs/2012PhDT.......231B},
  adsnote = {Provided by the SAO/NASA Astrophysics Data System}
}

@ARTICLE{bpd+15,
   author = {{Bilous}, A.~V. and {Pennucci}, T.~T. and {Demorest}, P. and
    {Ransom}, S.~M.},
    title = "{A Broadband Radio Study of the Average Profile and Giant Pulses from PSR B1821-24A}",
  journal = {\apj},
archivePrefix = "arXiv",
   eprint = {1412.7629},
 primaryClass = "astro-ph.SR",
 keywords = {pulsars: individual: B1821{\ndash}24A},
     year = 2015,
    month = apr,
   volume = 803,
      eid = {83},
    pages = {83},
      doi = {10.1088/0004-637X/803/2/83},
   adsurl = {http://adsabs.harvard.edu/abs/2015ApJ...803...83B},
  adsnote = {Provided by the SAO/NASA Astrophysics Data System}
}

@ARTICLE{zps+13,
   author = {{Zhuravlev}, V.~I. and {Popov}, M.~V. and {Soglasnov}, V.~A. and
    {Kondrat'ev}, V.~I. and {Kovalev}, Y.~Y. and {Bartel}, N. and
    {Ghigo}, F.},
    title = "{Statistical and polarization properties of giant pulses of the millisecond pulsar B1937+21}",
  journal = {\mnras},
archivePrefix = "arXiv",
   eprint = {1301.5134},
 primaryClass = "astro-ph.HE",
 keywords = {radiation mechanisms: non-thermal, scattering, methods: data analysis, pulsars: general, pulsars: individual: B1937+21},
     year = 2013,
    month = apr,
   volume = 430,
    pages = {2815-2821},
      doi = {10.1093/mnras/stt094},
   adsurl = {http://adsabs.harvard.edu/abs/2013{\mnras}.430.2815Z},
  adsnote = {Provided by the SAO/NASA Astrophysics Data System}
}

@ARTICLE{kni07,
   author = {{Knight}, H.~S.},
    title = "{A Parkes radio telescope study of giant pulses from PSR J1823-3021A}",
  journal = {\mnras},
 keywords = {pulsars: individual: J1823-3021A},
     year = 2007,
    month = jun,
   volume = 378,
    pages = {723-729},
      doi = {10.1111/j.1365-2966.2007.11810.x},
   adsurl = {http://adsabs.harvard.edu/abs/2007{\mnras}.378..723K},
  adsnote = {Provided by the SAO/NASA Astrophysics Data System}
}

@ARTICLE{kbm+06b,
   author = {{Knight}, H.~S. and {Bailes}, M. and {Manchester}, R.~N. and
    {Ord}, S.~M. and {Jacoby}, B.~A.},
    title = "{Green Bank Telescope Studies of Giant Pulses from Millisecond Pulsars}",
  journal = {\apj},
   eprint = {astro-ph/0512341},
 keywords = {Stars: Pulsars: General, Stars: Pulsars: Individual: Alphanumeric: PSR J0218+4232, Stars: Pulsars: Individual: Alphanumeric: PSR J1012+5307, pulsars: individual (PSR J1843-1113), Stars: Pulsars: Individual: Alphanumeric: PSR B1957+20},
     year = 2006,
    month = apr,
   volume = 640,
    pages = {941-949},
      doi = {10.1086/500292},
   adsurl = {http://adsabs.harvard.edu/abs/2006ApJ...640..941K},
  adsnote = {Provided by the SAO/NASA Astrophysics Data System}
}

@ARTICLE{zsd+15,
   author = {{Zhu}, W.~W. and {Stairs}, I.~H. and {Demorest}, P.~B. and {Nice}, D.~J. and
    {Ellis}, J.~A. and {Ransom}, S.~M. and {Arzoumanian}, Z. and
    {Crowter}, K. and {Dolch}, T. and {Ferdman}, R.~D. and {Fonseca}, E. and
    {Gonzalez}, M.~E. and {Jones}, G. and {Jones}, M.~L. and {Lam}, M.~T. and
    {Levin}, L. and {McLaughlin}, M.~A. and {Pennucci}, T. and {Stovall}, K. and
    {Swiggum}, J.},
    title = "{Testing Theories of Gravitation Using 21-Year Timing of Pulsar Binary J1713+0747}",
  journal = {\apj},
archivePrefix = "arXiv",
   eprint = {1504.00662},
 primaryClass = "astro-ph.SR",
 keywords = {binaries: general, gravitation, parallaxes, pulsars: individual: PSR J1713+0747, stars: neutron},
     year = 2015,
    month = aug,
   volume = 809,
      eid = {41},
    pages = {41},
      doi = {10.1088/0004-637X/809/1/41},
   adsurl = {http://adsabs.harvard.edu/abs/2015ApJ...809...41Z},
  adsnote = {Provided by the SAO/NASA Astrophysics Data System}
}

@ARTICLE{bkb+14,
   author = {{Brook}, P.~R. and {Karastergiou}, A. and {Buchner}, S. and
    {Roberts}, S.~J. and {Keith}, M.~J. and {Johnston}, S. and {Shannon}, R.~M.
    },
    title = "{Evidence of an Asteroid Encountering a Pulsar}",
  journal = {\apjl},
archivePrefix = "arXiv",
   eprint = {1311.3541},
 primaryClass = "astro-ph.HE",
 keywords = {pulsars: general, pulsars: individual: PSR J0738-4042, stars: neutron},
     year = 2014,
    month = jan,
   volume = 780,
      eid = {L31},
    pages = {L31},
      doi = {10.1088/2041-8205/780/2/L31},
   adsurl = {http://adsabs.harvard.edu/abs/2014ApJ...780L..31B},
  adsnote = {Provided by the SAO/NASA Astrophysics Data System}
}

@ARTICLE{gil85,
   author = {{Gil}, J.},
    title = "{Fluctuations of pulsar emission with sub-microsecond time-scales}",
  journal = {\apjss},
 keywords = {Emission Spectra, Pulsars, Stellar Models, Stellar Oscillations, Temporal Distribution, Coherent Radiation, Radiant Flux Density},
     year = 1985,
    month = mar,
   volume = 110,
    pages = {293-296},
      doi = {10.1007/BF00653674},
   adsurl = {http://adsabs.harvard.edu/abs/1985Ap
  adsnote = {Provided by the SAO/NASA Astrophysics Data System}
}

@ARTICLE{es04,
   author = {{Edwards}, R.~T. and {Stappers}, B.~W.},
    title = "{Ellipticity and deviations from orthogonality in the polarization modes of PSR B0329+54}",
  journal = aa,
   eprint = {astro-ph/0404092},
 keywords = {plasmas, polarization, stars: pulsars: individual: PSR B0329+54, waves},
     year = 2004,
    month = jul,
   volume = 421,
    pages = {681-691},
      doi = {10.1051/0004-6361:20040228},
   adsurl = {http://adsabs.harvard.edu/abs/2004A
  adsnote = {Provided by the SAO/NASA Astrophysics Data System}
}

@ARTICLE{jg03,
   author = {{Jenet}, F.~A. and {Gil}, J.},
    title = "{Using the Intensity Modulation Index to Test Pulsar Radio Emission Models}",
  journal = {ApJL},
   eprint = {astro-ph/0306031},
 keywords = {Stars: Pulsars: General},
     year = 2003,
    month = oct,
   volume = 596,
    pages = {L215-L218},
      doi = {10.1086/379501},
   adsurl = {http://adsabs.harvard.edu/abs/2003ApJ...596L.215J},
  adsnote = {Provided by the SAO/NASA Astrophysics Data System}
}

@ARTICLE{bjk+16,
   author = {{Bassa}, C.~G. and {Janssen}, G.~H. and {Karuppusamy}, R. and
    {Kramer}, M. and {Lee}, K.~J. and {Liu}, K. and {McKee}, J. and
    {Perrodin}, D. and {Purver}, M. and {Sanidas}, S. and {Smits}, R. and
    {Stappers}, B.~W.},
    title = "{LEAP: the Large European Array for Pulsars}",
  journal = {\mnras},
archivePrefix = "arXiv",
   eprint = {1511.06597},
 primaryClass = "astro-ph.IM",
 keywords = {gravitational waves, methods: data analysis, techniques: interferometric, pulsars: general},
     year = 2016,
    month = feb,
   volume = 456,
    pages = {2196-2209},
      doi = {10.1093/mnras/stv2755},
   adsurl = {http://adsabs.harvard.edu/abs/2016{\mnras}.456.2196B},
  adsnote = {Provided by the SAO/NASA Astrophysics Data System}
}

@ARTICLE{pfb+13,
   author = {{Papitto}, A. and {Ferrigno}, C. and {Bozzo}, E. and {Rea}, N. and
    {Pavan}, L. and {Burderi}, L. and {Burgay}, M. and {Campana}, S. and
    {di Salvo}, T. and {Falanga}, M. and {Filipovi{\'c}}, M.~D. and
    {Freire}, P.~C.~C. and {Hessels}, J.~W.~T. and {Possenti}, A. and
    {Ransom}, S.~M. and {Riggio}, A. and {Romano}, P. and {Sarkissian}, J.~M. and
    {Stairs}, I.~H. and {Stella}, L. and {Torres}, D.~F. and {Wieringa}, M.~H. and
    {Wong}, G.~F.},
    title = "{Swings between rotation and accretion power in a binary millisecond pulsar}",
  journal = {\nat},
archivePrefix = "arXiv",
   eprint = {1305.3884},
 primaryClass = "astro-ph.HE",
     year = 2013,
    month = sep,
   volume = 501,
    pages = {517-520},
      doi = {10.1038/nature12470},
   adsurl = {http://adsabs.harvard.edu/abs/2013Natur.501..517P},
  adsnote = {Provided by the SAO/NASA Astrophysics Data System}
}

@ARTICLE{ssw09,
   author = {{Serylak}, M. and {Stappers}, B.~W. and {Weltevrede}, P.},
    title = "{S2DFS: analysis of temporal changes of drifting subpulses}",
  journal = aa,
archivePrefix = "arXiv",
   eprint = {0909.2488},
 primaryClass = "astro-ph.SR",
 keywords = {methods: analytical, pulsars: general},
     year = 2009,
    month = nov,
   volume = 506,
    pages = {865-874},
      doi = {10.1051/0004-6361/200912453},
   adsurl = {http://adsabs.harvard.edu/abs/2009A
  adsnote = {Provided by the SAO/NASA Astrophysics Data System}
}

@ARTICLE{mar15,
   author = {{Mitra}, D. and {Arjunwadkar}, M. and {Rankin}, J.~M.},
    title = "{Polarized Quasiperiodic Structures in Pulsar Radio Emission Reflect Temporal Modulations of Non-stationary Plasma Flow}",
  journal = {\apj},
archivePrefix = "arXiv",
   eprint = {1502.06897},
 primaryClass = "astro-ph.HE",
 keywords = {magnetohydrodynamics: MHD, plasmas, pulsars: general, radiation mechanism: non-thermal},
     year = 2015,
    month = jun,
   volume = 806,
      eid = {236},
    pages = {236},
      doi = {10.1088/0004-637X/806/2/236},
   adsurl = {http://adsabs.harvard.edu/abs/2015ApJ...806..236M},
  adsnote = {Provided by the SAO/NASA Astrophysics Data System}
}

@ARTICLE{jhm+15,
   author = {{Janssen}, G. and {Hobbs}, G. and {McLaughlin}, M. and {Bassa}, C. and
    {Deller}, A. and {Kramer}, M. and {Lee}, K. and {Mingarelli}, C. and
    {Rosado}, P. and {Sanidas}, S. and {Sesana}, A. and {Shao}, L. and
    {Stairs}, I. and {Stappers}, B. and {Verbiest}, J.~P.~W.},
    title = "{Gravitational Wave Astronomy with the SKA}",
  journal = {Advancing Astrophysics with the Square Kilometre Array (AASKA14)},
archivePrefix = "arXiv",
   eprint = {1501.00127},
 primaryClass = "astro-ph.IM",
     year = 2015,
      eid = {37},
    pages = {37},
   adsurl = {http://adsabs.harvard.edu/abs/2015aska.confE..37J},
  adsnote = {Provided by the SAO/NASA Astrophysics Data System}
}


@ARTICLE{van10,
   author = {{van Straten}, W.},
    title = "{Erratum: ''The Statistics of Radio Astronomical Polarimetry: Bright Sources and High Time Resolution'' <A href=''/abs/2009ApJ...694.1413V''>(2009, {\apj}, 694, 1413)</A>}",
  journal = {\apj},
     year = 2010,
    month = aug,
   volume = 719,
    pages = {985},
      doi = {10.1088/0004-637X/719/1/985},
   adsurl = {http://adsabs.harvard.edu/abs/2010ApJ...719..985V},
  adsnote = {Provided by the SAO/NASA Astrophysics Data System}
}

@ARTICLE{gmg03,
   author = {{Gil}, J. and {Melikidze}, G.~I. and {Geppert}, U.},
    title = "{Drifting subpulses and inner accelerationregions in radio pulsars}",
  journal = aa,
   eprint = {astro-ph/0305463},
 keywords = {stars: pulsars: individual: PSRs B0943+10, B0809+74, B0826-34, B2303+30, B2319+60, B0031-07},
     year = 2003,
    month = aug,
   volume = 407,
    pages = {315-324},
      doi = {10.1051/0004-6361:20030854},
   adsurl = {http://adsabs.harvard.edu/abs/2003A
  adsnote = {Provided by the SAO/NASA Astrophysics Data System}
}

@ARTICLE{qlz+04,
   author = {{Qiao}, G.~J. and {Lee}, K.~J. and {Zhang}, B. and {Xu}, R.~X. and
    {Wang}, H.~G.},
    title = "{A Model for the Challenging ``Bi-drifting'' Phenomenon in PSR J0815+09}",
  journal = {\apjl},
   eprint = {astro-ph/0410479},
 keywords = {Elementary Particles, Stars: Pulsars: General, Stars: Pulsars: Individual: Alphanumeric: PSR J0815+09, Radiation Mechanisms: Nonthermal, Stars: Neutron},
     year = 2004,
    month = dec,
   volume = 616,
    pages = {L127-L130},
      doi = {10.1086/426862},
   adsurl = {http://adsabs.harvard.edu/abs/2004ApJ...616L.127Q},
  adsnote = {Provided by the SAO/NASA Astrophysics Data System}
}

@BOOK{lg06,
   author = {{Lyne}, A.~G. and {Graham-Smith}, F.},
    title = "{Pulsar Astronomy}",
booktitle = {Pulsar astronomy, 3rd ed., by A.G.~Lyne and F.~Graham-Smith.~Cambridge astrophysics series.~Cambridge, UK: Cambridge University Press, 2006 ISBN 0521839548.},
     year = 2006,
    month = feb,
   adsurl = {http://adsabs.harvard.edu/abs/2006puas.book.....L},
  adsnote = {Provided by the SAO/NASA Astrophysics Data System}
}

@ARTICLE{vlh+16,
   author = {{Verbiest}, J.~P.~W. and {Lentati}, L. and {Hobbs}, G. and {van Haasteren}, R. and
    {Demorest}, P.~B. and {Janssen}, G.~H. and {Wang}, J.-B. and
    {Desvignes}, G. and {Caballero}, R.~N. and {Keith}, M.~J. and
    {Champion}, D.~J. and {Arzoumanian}, Z. and {Babak}, S. and
    {Bassa}, C.~G. and {Bhat}, N.~D.~R. and {Brazier}, A. and {Brem}, P. and
    {Burgay}, M. and {Burke-Spolaor}, S. and {Chamberlin}, S.~J. and
    {Chatterjee}, S. and {Christy}, B. and {Cognard}, I. and {Cordes}, J.~M. and
    {Dai}, S. and {Dolch}, T. and {Ellis}, J.~A. and {Ferdman}, R.~D. and
    {Fonseca}, E. and {Gair}, J.~R. and {Garver-Daniels}, N.~E. and
    {Gentile}, P. and {Gonzalez}, M.~E. and {Graikou}, E. and {Guillemot}, L. and
    {Hessels}, J.~W.~T. and {Jones}, G. and {Karuppusamy}, R. and
    {Kerr}, M. and {Kramer}, M. and {Lam}, M.~T. and {Lasky}, P.~D. and
    {Lassus}, A. and {Lazarus}, P. and {Lazio}, T.~J.~W. and {Lee}, K.~J. and
    {Levin}, L. and {Liu}, K. and {Lynch}, R.~S. and {Lyne}, A.~G. and
    {Mckee}, J. and {McLaughlin}, M.~A. and {McWilliams}, S.~T. and
    {Madison}, D.~R. and {Manchester}, R.~N. and {Mingarelli}, C.~M.~F. and
    {Nice}, D.~J. and {Os{\l}owski}, S. and {Palliyaguru}, N.~T. and
    {Pennucci}, T.~T. and {Perera}, B.~B.~P. and {Perrodin}, D. and
    {Possenti}, A. and {Petiteau}, A. and {Ransom}, S.~M. and {Reardon}, D. and
    {Rosado}, P.~A. and {Sanidas}, S.~A. and {Sesana}, A. and {Shaifullah}, G. and
    {Shannon}, R.~M. and {Siemens}, X. and {Simon}, J. and {Smits}, R. and
    {Spiewak}, R. and {Stairs}, I.~H. and {Stappers}, B.~W. and
    {Stinebring}, D.~R. and {Stovall}, K. and {Swiggum}, J.~K. and
    {Taylor}, S.~R. and {Theureau}, G. and {Tiburzi}, C. and {Toomey}, L. and
    {Vallisneri}, M. and {van Straten}, W. and {Vecchio}, A. and
    {Wang}, Y. and {Wen}, L. and {You}, X.~P. and {Zhu}, W.~W. and
    {Zhu}, X.-J.},
    title = "{The International Pulsar Timing Array: First data release}",
  journal = {\mnras},
archivePrefix = "arXiv",
   eprint = {1602.03640},
 primaryClass = "astro-ph.IM",
 keywords = {methods: data analysis, pulsars: general},
     year = 2016,
    month = may,
   volume = 458,
    pages = {1267-1288},
      doi = {10.1093/mnras/stw347},
   adsurl = {http://adsabs.harvard.edu/abs/2016{\mnras}.458.1267V},
  adsnote = {Provided by the SAO/NASA Astrophysics Data System}
}

@ARTICLE{wel16,
   author = {{Weltevrede}, P.},
    title = "{Investigation of the bi-drifting subpulses of radio pulsar B1839-04 utilising the open-source data-analysis project PSRSALSA}",
  journal = aa,
archivePrefix = "arXiv",
   eprint = {1605.06413},
 primaryClass = "astro-ph.HE",
 keywords = {Astrophysics - High Energy Astrophysical Phenomena, Astrophysics - Solar and Stellar Astrophysics},
     year = 2016,
   volume = 590,
    pages = {A109},
}

@ARTICLE{srl+15,
   author = {{Shannon}, R.~M. and {Ravi}, V. and {Lentati}, L.~T. and {Lasky}, P.~D. and
    {Hobbs}, G. and {Kerr}, M. and {Manchester}, R.~N. and {Coles}, W.~A. and
    {Levin}, Y. and {Bailes}, M. and {Bhat}, N.~D.~R. and {Burke-Spolaor}, S. and
    {Dai}, S. and {Keith}, M.~J. and {Os{\l}owski}, S. and {Reardon}, D.~J. and
    {van Straten}, W. and {Toomey}, L. and {Wang}, J.-B. and {Wen}, L. and
    {Wyithe}, J.~S.~B. and {Zhu}, X.-J.},
    title = "{Gravitational waves from binary supermassive black holes missing in pulsar observations}",
  journal = {Science},
archivePrefix = "arXiv",
   eprint = {1509.07320},
     year = 2015,
    month = sep,
   volume = 349,
    pages = {1522-1525},
      doi = {10.1126/science.aab1910},
   adsurl = {http://adsabs.harvard.edu/abs/2015Sci...349.1522S},
  adsnote = {Provided by the SAO/NASA Astrophysics Data System}
}

@ARTICLE{abb+16,
   author = {{Arzoumanian}, Z. and {Brazier}, A. and {Burke-Spolaor}, S. and
    {Chamberlin}, S.~J. and {Chatterjee}, S. and {Christy}, B. and
    {Cordes}, J.~M. and {Cornish}, N.~J. and {Crowter}, K. and {Demorest}, P.~B. and
    {Deng}, X. and {Dolch}, T. and {Ellis}, J.~A. and {Ferdman}, R.~D. and
    {Fonseca}, E. and {Garver-Daniels}, N. and {Gonzalez}, M.~E. and
    {Jenet}, F. and {Jones}, G. and {Jones}, M.~L. and {Kaspi}, V.~M. and
    {Koop}, M. and {Lam}, M.~T. and {Lazio}, T.~J.~W. and {Levin}, L. and
    {Lommen}, A.~N. and {Lorimer}, D.~R. and {Luo}, J. and {Lynch}, R.~S. and
    {Madison}, D.~R. and {McLaughlin}, M.~A. and {McWilliams}, S.~T. and
    {Mingarelli}, C.~M.~F. and {Nice}, D.~J. and {Palliyaguru}, N. and
    {Pennucci}, T.~T. and {Ransom}, S.~M. and {Sampson}, L. and
    {Sanidas}, S.~A. and {Sesana}, A. and {Siemens}, X. and {Simon}, J. and
    {Stairs}, I.~H. and {Stinebring}, D.~R. and {Stovall}, K. and
    {Swiggum}, J. and {Taylor}, S.~R. and {Vallisneri}, M. and {van Haasteren}, R. and
    {Wang}, Y. and {Zhu}, W.~W. and {The NANOGrav Collaboration}
    },
    title = "{The NANOGrav Nine-year Data Set: Limits on the Isotropic Stochastic Gravitational Wave Background}",
  journal = {\apj},
archivePrefix = "arXiv",
   eprint = {1508.03024},
 keywords = {gravitational waves, methods: data analysis, pulsars: general},
     year = 2016,
    month = apr,
   volume = 821,
      eid = {13},
    pages = {13},
      doi = {10.3847/0004-637X/821/1/13},
   adsurl = {http://adsabs.harvard.edu/abs/2016ApJ...821...13A},
  adsnote = {Provided by the SAO/NASA Astrophysics Data System}
}

@ARTICLE{dcl+16,
   author = {{Desvignes}, G. and {Caballero}, R.~N. and {Lentati}, L. and
    {Verbiest}, J.~P.~W. and {Champion}, D.~J. and {Stappers}, B.~W. and
    {Janssen}, G.~H. and {Lazarus}, P. and {Os{\l}owski}, S. and
    {Babak}, S. and {Bassa}, C.~G. and {Brem}, P. and {Burgay}, M. and
    {Cognard}, I. and {Gair}, J.~R. and {Graikou}, E. and {Guillemot}, L. and
    {Hessels}, J.~W.~T. and {Jessner}, A. and {Jordan}, C. and {Karuppusamy}, R. and
    {Kramer}, M. and {Lassus}, A. and {Lazaridis}, K. and {Lee}, K.~J. and
    {Liu}, K. and {Lyne}, A.~G. and {McKee}, J. and {Mingarelli}, C.~M.~F. and
    {Perrodin}, D. and {Petiteau}, A. and {Possenti}, A. and {Purver}, M.~B. and
    {Rosado}, P.~A. and {Sanidas}, S. and {Sesana}, A. and {Shaifullah}, G. and
    {Smits}, R. and {Taylor}, S.~R. and {Theureau}, G. and {Tiburzi}, C. and
    {van Haasteren}, R. and {Vecchio}, A.},
    title = "{High-precision timing of 42 millisecond pulsars with the European Pulsar Timing Array}",
  journal = {\mnras},
archivePrefix = "arXiv",
   eprint = {1602.08511},
 primaryClass = "astro-ph.HE",
 keywords = {proper motions, stars: distances, pulsars: general},
     year = 2016,
    month = may,
   volume = 458,
    pages = {3341-3380},
      doi = {10.1093/mnras/stw483},
   adsurl = {http://adsabs.harvard.edu/abs/2016{\mnras}.458.3341D},
  adsnote = {Provided by the SAO/NASA Astrophysics Data System}
}

@ARTICLE{abb+15,
   author = {{The NANOGrav Collaboration} and {Arzoumanian}, Z. and {Brazier}, A. and
    {Burke-Spolaor}, S. and {Chamberlin}, S. and {Chatterjee}, S. and
    {Christy}, B. and {Cordes}, J.~M. and {Cornish}, N. and {Crowter}, K. and
    {Demorest}, P.~B. and {Dolch}, T. and {Ellis}, J.~A. and {Ferdman}, R.~D. and
    {Fonseca}, E. and {Garver-Daniels}, N. and {Gonzalez}, M.~E. and
    {Jenet}, F.~A. and {Jones}, G. and {Jones}, M.~L. and {Kaspi}, V.~M. and
    {Koop}, M. and {Lam}, M.~T. and {Lazio}, T.~J.~W. and {Levin}, L. and
    {Lommen}, A.~N. and {Lorimer}, D.~R. and {Luo}, J. and {Lynch}, R.~S. and
    {Madison}, D. and {McLaughlin}, M.~A. and {McWilliams}, S.~T. and
    {Nice}, D.~J. and {Palliyaguru}, N. and {Pennucci}, T.~T. and
    {Ransom}, S.~M. and {Siemens}, X. and {Stairs}, I.~H. and {Stinebring}, D.~R. and
    {Stovall}, K. and {Swiggum}, J.~K. and {Vallisneri}, M. and
    {van Haasteren}, R. and {Wang}, Y. and {Zhu}, W.},
    title = "{The NANOGrav Nine-year Data Set: Observations, Arrival Time Measurements, and Analysis of 37 Millisecond Pulsars}",
  journal = {\apj},
archivePrefix = "arXiv",
   eprint = {1505.07540},
 primaryClass = "astro-ph.IM",
 keywords = {gravitational waves, methods: data analysis, pulsars: general},
     year = 2015,
    month = nov,
   volume = 813,
      eid = {65},
    pages = {65},
      doi = {10.1088/0004-637X/813/1/65},
   adsurl = {http://adsabs.harvard.edu/abs/2015ApJ...813...65T},
  adsnote = {Provided by the SAO/NASA Astrophysics Data System}
}

@ARTICLE{wws+06,
   author = {{Weltevrede}, P. and {Wright}, G.~A.~E. and {Stappers}, B.~W. and
    {Rankin}, J.~M.},
    title = "{The bright spiky emission of pulsar B0656+14}",
  journal = aa,
   eprint = {astro-ph/0608023},
 keywords = {stars: pulsars: individual: PSR B0656+14, stars: pulsars: general, radiation mechanisms: non-thermal},
     year = 2006,
    month = oct,
   volume = 458,
    pages = {269-283},
      doi = {10.1051/0004-6361:20065572},
   adsurl = {http://adsabs.harvard.edu/abs/2006A
  adsnote = {Provided by the SAO/NASA Astrophysics Data System}
}

@ARTICLE{jkw06,
       author = {{Johnston}, Simon and {Karastergiou}, Aris and {Willett}, Kyle},
        title = "{High-frequency observations of southern pulsars}",
      journal = {\mnras},
     keywords = {pulsars: general, Astrophysics},
         year = "2006",
        month = "Jul",
       volume = {369},
       number = {4},
        pages = {1916-1928},
          doi = {10.1111/j.1365-2966.2006.10440.x},
archivePrefix = {arXiv},
       eprint = {astro-ph/0604404},
 primaryClass = {astro-ph},
       adsurl = {https://ui.adsabs.harvard.edu/abs/2006{\mnras}.369.1916J},
      adsnote = {Provided by the SAO/NASA Astrophysics Data System}
}

@ARTICLE{naa+11,
       author = {{Noutsos}, A. and {Abdo}, A.~A. and {Ackermann}, M. and {Ajello}, M. and
         {Ballet}, J. and {Barbiellini}, G. and {Baring}, M.~G. and
         {Bastieri}, D. and {Bechtol}, K. and {Bellazzini}, R. and
         {Berenji}, B. and {Bonamente}, E. and {Borgland}, A.~W. and
         {Bregeon}, J. and {Brez}, A. and {Brigida}, M. and {Bruel}, P. and
         {Buehler}, R. and {Busetto}, G. and {Caliandro}, G.~A. and
         {Cameron}, R.~A. and {Camilo}, F. and {Caraveo}, P.~A. and {Casand
        jian}, J.~M. and {Cecchi}, C. and {{\c{C}}elik}, {\"O}. and {Chaty}, S. and
         {Chekhtman}, A. and {Chiang}, J. and {Ciprini}, S. and {Claus}, R. and
         {Cognard}, I. and {Cohen-Tanugi}, J. and {Colafrancesco}, S. and
         {Cutini}, S. and {Dermer}, C.~D. and {de Palma}, F. and {Drell}, P.~S. and
         {Dumora}, D. and {Ea}, C.~M. and {Favuzzi}, C. and {Ferrara}, E.~C. and
         {Focke}, W.~B. and {Frailis}, M. and {Freire}, P.~C.~C. and
         {Fukazawa}, Y. and {Funk}, S. and {Fusco}, P. and {Gargano}, F. and
         {Germani}, S. and {Giglietto}, N. and {Giordano}, F. and
         {Giroletti}, M. and {Godfrey}, G. and {Grandi}, P. and
         {Grenier}, I.~A. and {Grove}, J.~E. and {Guillemot}, L. and
         {Guiriec}, S. and {Harding}, A.~K. and {Hughes}, R.~E. and
         {Jackson}, M.~S. and {J{\'o}hannesson}, G. and {Johnson}, A.~S. and
         {Johnson}, T.~J. and {Johnson}, W.~N. and {Johnston}, S. and
         {Kamae}, T. and {Katagiri}, H. and {Kataoka}, J. and
         {Kn{\"o}dlseder}, J. and {Kramer}, M. and {Kuss}, M. and {Lande}, J. and
         {Lee}, S. -H. and {Longo}, F. and {Loparco}, F. and
         {Lovellette}, M.~N. and {Lubrano}, P. and {Lyne}, A.~G. and
         {Makeev}, A. and {Marelli}, M. and {Mazziotta}, M.~N. and
         {McEnery}, J.~E. and {Mehault}, J. and {Michelson}, P.~F. and
         {Mizuno}, T. and {Monte}, C. and {Monzani}, M.~E. and {Morselli}, A. and
         {Moskalenko}, I.~V. and {Murgia}, S. and {Naumann-Godo}, M. and
         {Nolan}, P.~L. and {Nuss}, E. and {Ohsugi}, T. and {Okumura}, A. and
         {Omodei}, N. and {Orlando}, E. and {Ormes}, J.~F. and {Panetta}, J.~H. and
         {Parent}, D. and {Pelassa}, V. and {Pepe}, M. and {Persic}, M. and
         {Pesce-Rollins}, M. and {Piron}, F. and {Porter}, T.~A. and
         {Rain{\`o}}, S. and {Ray}, P.~S. and {Razzano}, M. and {Reimer}, A. and
         {Reimer}, O. and {Reposeur}, T. and {Romani}, R.~W. and
         {Sadrozinski}, H.~F. -W. and {Sander}, A. and {Saz Parkinson}, P.~M. and
         {Sgr{\`o}}, C. and {Siskind}, E.~J. and {Smith}, D.~A. and
         {Smith}, P.~D. and {Spandre}, G. and {Spinelli}, P. and
         {Stappers}, B.~W. and {Strickman}, M.~S. and {Suson}, D.~J. and
         {Takahashi}, H. and {Tanaka}, T. and {Theureau}, G. and
         {Thompson}, D.~J. and {Thorsett}, S.~E. and {Tibolla}, O. and
         {Torres}, D.~F. and {Tramacere}, A. and {Usher}, T.~L. and {Vand
        enbroucke}, J. and {Vianello}, G. and {Vilchez}, N. and {Villata}, M. and
         {Vitale}, V. and {von Kienlin}, A. and {Waite}, A.~P. and {Wang}, P. and
         {Watters}, K. and {Weltevrede}, P. and {Winer}, B.~L. and
         {Wood}, K.~S. and {Ziegler}, M.},
        title = "{Radio and {\ensuremath{\gamma}}-ray Constraints on the Emission Geometry and Birthplace of PSR J2043+2740}",
      journal = {\apj},
     keywords = {gamma rays: stars, pulsars: individual: PSR J2043 + 2740, Astrophysics - Astrophysics of Galaxies},
         year = "2011",
        month = "Feb",
       volume = {728},
       number = {2},
          eid = {77},
        pages = {77},
          doi = {10.1088/0004-637X/728/2/77},
archivePrefix = {arXiv},
       eprint = {1012.4658},
 primaryClass = {astro-ph.GA},
       adsurl = {https://ui.adsabs.harvard.edu/abs/2011ApJ...728...77N},
      adsnote = {Provided by the SAO/NASA Astrophysics Data System}
}

@ARTICLE{hje15,
       author = {{Hankins}, T.~H. and {Jones}, G. and {Eilek}, J.~A.},
        title = "{The Crab Pulsar at Centimeter Wavelengths. I. Ensemble Characteristics}",
      journal = {\apj},
     keywords = {pulsars: general, pulsars: individual: Crab pulsar, Astrophysics - High Energy Astrophysical Phenomena},
         year = "2015",
        month = "Apr",
       volume = {802},
       number = {2},
          eid = {130},
        pages = {130},
          doi = {10.1088/0004-637X/802/2/130},
archivePrefix = {arXiv},
       eprint = {1502.00677},
 primaryClass = {astro-ph.HE},
       adsurl = {https://ui.adsabs.harvard.edu/abs/2015ApJ...802..130H},
      adsnote = {Provided by the SAO/NASA Astrophysics Data System}
}

@ARTICLE{tek+15,
       author = {{Torne}, P. and {Eatough}, R.~P. and {Karuppusamy}, R. and {Kramer}, M. and
         {Paubert}, G. and {Klein}, B. and {Desvignes}, G. and
         {Champion}, D.~J. and {Wiesemeyer}, H. and {Kramer}, C. and
         {Spitler}, L.~G. and {Thum}, C. and {Gusten}, R. and {Schuster}, K.~F. and
         {Cognard}, I.},
        title = "{Simultaneous multifrequency radio observations of the Galactic Centre magnetar SGR J1745-2900.}",
      journal = {\mnras},
     keywords = {radiation mechanisms: non-thermal, stars: magnetars, stars: neutron, pulsars: general, pulsars: individual: SGR J1745-2900, Galaxy: centre, Astrophysics - High Energy Astrophysical Phenomena},
         year = "2015",
        month = "Jul",
       volume = {451},
        pages = {L50-L54},
          doi = {10.1093/mnrasl/slv063},
archivePrefix = {arXiv},
       eprint = {1504.07241},
 primaryClass = {astro-ph.HE},
       adsurl = {https://ui.adsabs.harvard.edu/abs/2015{\mnras}.451L..50T},
      adsnote = {Provided by the SAO/NASA Astrophysics Data System}
}

@ARTICLE{kjl+11,
       author = {{Keith}, M.~J. and {Johnston}, S. and {Levin}, L. and {Bailes}, M.},
        title = "{17- and 24-GHz observations of southern pulsars}",
      journal = {\mnras},
     keywords = {pulsars: general, pulsars: individual: PSR J0437-4715, pulsars: individual: PSR J0835-4510, pulsars: individual: PSR J1622-4950, Astrophysics - Solar and Stellar Astrophysics},
         year = "2011",
        month = "Sep",
       volume = {416},
       number = {1},
        pages = {346-354},
          doi = {10.1111/j.1365-2966.2011.19041.x},
archivePrefix = {arXiv},
       eprint = {1105.3961},
 primaryClass = {astro-ph.SR},
       adsurl = {https://ui.adsabs.harvard.edu/abs/2011{\mnras}.416..346K},
      adsnote = {Provided by the SAO/NASA Astrophysics Data System}
}

@ARTICLE{cfp+07,
       author = {{Camilo}, F. and {Ransom}, S.~M. and {Pe{\~n}alver}, J. and
         {Karastergiou}, A. and {van Kerkwijk}, M.~H. and {Durant}, M. and
         {Halpern}, J.~P. and {Reynolds}, J. and {Thum}, C. and {Helfand
        }, D.~J. and {Zimmerman}, N. and {Cognard}, I.},
        title = "{The Variable Radio-to-X-Ray Spectrum of the Magnetar XTE J1810-197}",
      journal = {\apj},
     keywords = {Stars: Pulsars: Individual: Alphanumeric: XTE J1810-197, Stars: Neutron, Astrophysics},
         year = "2007",
        month = "Nov",
       volume = {669},
       number = {1},
        pages = {561-569},
          doi = {10.1086/521548},
archivePrefix = {arXiv},
       eprint = {0705.4095},
 primaryClass = {astro-ph},
       adsurl = {https://ui.adsabs.harvard.edu/abs/2007ApJ...669..561C},
      adsnote = {Provided by the SAO/NASA Astrophysics Data System}
}

@ARTICLE{hej16,
       author = {{Hankins}, T.~H. and {Eilek}, J.~A. and {Jones}, G.},
        title = "{The Crab Pulsar at Centimeter Wavelengths. II. Single Pulses}",
      journal = {\apj},
     keywords = {pulsars: general, pulsars: individual: Crab pulsar, Astrophysics - High Energy Astrophysical Phenomena, Astrophysics - Solar and Stellar Astrophysics},
         year = "2016",
        month = "Dec",
       volume = {833},
       number = {1},
          eid = {47},
        pages = {47},
          doi = {10.3847/1538-4357/833/1/47},
archivePrefix = {arXiv},
       eprint = {1608.08881},
 primaryClass = {astro-ph.HE},
       adsurl = {https://ui.adsabs.harvard.edu/abs/2016ApJ...833...47H},
      adsnote = {Provided by the SAO/NASA Astrophysics Data System}
}

@ARTICLE{hss+19,
       author = {{Hessels}, J.~W.~T. and {Spitler}, L.~G. and {Seymour}, A.~D. and
         {Cordes}, J.~M. and {Michilli}, D. and {Lynch}, R.~S. and
         {Gourdji}, K. and {Archibald}, A.~M. and {Bassa}, C.~G. and
         {Bower}, G.~C. and {Chatterjee}, S. and {Connor}, L. and
         {Crawford}, F. and {Deneva}, J.~S. and {Gajjar}, V. and {Kaspi}, V.~M. and
         {Keimpema}, A. and {Law}, C.~J. and {Marcote}, B. and
         {McLaughlin}, M.~A. and {Paragi}, Z. and {Petroff}, E. and
         {Ransom}, S.~M. and {Scholz}, P. and {Stappers}, B.~W. and
         {Tendulkar}, S.~P.},
        title = "{FRB 121102 Bursts Show Complex Time-Frequency Structure}",
      journal = {ApJL},
     keywords = {galaxies: dwarf, radiation mechanisms: non-thermal, radio continuum: general, Astrophysics - High Energy Astrophysical Phenomena},
         year = "2019",
        month = "May",
       volume = {876},
       number = {2},
          eid = {L23},
        pages = {L23},
          doi = {10.3847/2041-8213/ab13ae},
archivePrefix = {arXiv},
       eprint = {1811.10748},
 primaryClass = {astro-ph.HE},
       adsurl = {https://ui.adsabs.harvard.edu/abs/2019ApJ...876L..23H},
      adsnote = {Provided by the SAO/NASA Astrophysics Data System}
}

@ARTICLE{agh+18,
       author = {{Archibald}, Anne M. and {Gusinskaia}, Nina V. and
         {Hessels}, Jason W.~T. and {Deller}, Adam T. and {Kaplan}, David L. and
         {Lorimer}, Duncan R. and {Lynch}, Ryan S. and {Ransom}, Scott M. and
         {Stairs}, Ingrid H.},
        title = "{Universality of free fall from the orbital motion of a pulsar in a stellar triple system}",
      journal = {\nat},
     keywords = {Astrophysics - High Energy Astrophysical Phenomena, General Relativity and Quantum Cosmology},
         year = 2018,
        month = jul,
       volume = {559},
       number = {7712},
        pages = {73-76},
          doi = {10.1038/s41586-018-0265-1},
archivePrefix = {arXiv},
       eprint = {1807.02059},
 primaryClass = {astro-ph.HE},
       adsurl = {https://ui.adsabs.harvard.edu/abs/2018Natur.559...73A},
      adsnote = {Provided by the SAO/NASA Astrophysics Data System}
}

@ARTICLE{stt13,
       author = {{Bates}, S.~D. and {Lorimer}, D.~R. and {Verbiest}, J.~P.~W.},
        title = "{The pulsar spectral index distribution}",
      journal = {\mnras},
     keywords = {methods: statistical, stars: neutron, pulsars: general, Astrophysics - Solar and Stellar Astrophysics, Astrophysics - High Energy Astrophysical Phenomena},
         year = 2013,
        month = may,
       volume = {431},
       number = {2},
        pages = {1352-1358},
          doi = {10.1093/mnras/stt257},
archivePrefix = {arXiv},
       eprint = {1302.2053},
 primaryClass = {astro-ph.SR},
       adsurl = {https://ui.adsabs.harvard.edu/abs/2013{\mnras}.431.1352B},
      adsnote = {Provided by the SAO/NASA Astrophysics Data System}
}

@ARTICLE{jvk+18,
       author = {{Jankowski}, F. and {van Straten}, W. and {Keane}, E.~F. and
         {Bailes}, M. and {Barr}, E.~D. and {Johnston}, S. and {Kerr}, M.},
        title = "{Spectral properties of 441 radio pulsars}",
      journal = {\mnras},
     keywords = {radiation mechanisms: non-thermal, methods: data analysis, pulsars: general, radio continuum: stars, Astrophysics - High Energy Astrophysical Phenomena},
         year = 2018,
        month = feb,
       volume = {473},
       number = {4},
        pages = {4436-4458},
          doi = {10.1093/mnras/stx2476},
archivePrefix = {arXiv},
       eprint = {1709.08864},
 primaryClass = {astro-ph.HE},
       adsurl = {https://ui.adsabs.harvard.edu/abs/2018{\mnras}.473.4436J},
      adsnote = {Provided by the SAO/NASA Astrophysics Data System}
}

@ARTICLE{btk08,
       author = {{Bhat}, N.~D. Ramesh and {Tingay}, Steven J. and {Knight}, Haydon S.},
        title = "{Bright Giant Pulses from the Crab Nebula Pulsar: Statistical Properties, Pulse Broadening, and Scattering Due to the Nebula}",
      journal = {\apj},
     keywords = {ISM: individual: Crab Nebula, ISM: structure, pulsars: general, pulsars: individual: Crab pulsar, scattering, Astrophysics},
         year = 2008,
        month = apr,
       volume = {676},
       number = {2},
        pages = {1200-1209},
          doi = {10.1086/528735},
archivePrefix = {arXiv},
       eprint = {0801.0334},
 primaryClass = {astro-ph},
       adsurl = {https://ui.adsabs.harvard.edu/abs/2008ApJ...676.1200B},
      adsnote = {Provided by the SAO/NASA Astrophysics Data System}
}

@ARTICLE{gcc+87,
       author = {{Gilmozzi}, R. and {Cassatella}, A. and {Clavel}, J. and {Fransson}, C. and
         {Gonzalez}, R. and {Gry}, C. and {Panagia}, N. and {Talavera}, A. and
         {Wamsteker}, W.},
        title = "{The progenitor of SN1987A}",
      journal = {Nature},
     keywords = {Astronomical Spectroscopy, Iue, Magellanic Clouds, Supernovae, B Stars, Spatial Distribution, Stellar Spectra, Astrophysics},
         year = 1987,
        month = jul,
       volume = {328},
       number = {6128},
        pages = {318-320},
          doi = {10.1038/328318a0},
       adsurl = {https://ui.adsabs.harvard.edu/abs/1987Natur.328..318G},
      adsnote = {Provided by the SAO/NASA Astrophysics Data System}
}

@ARTICLE{sak+87,
       author = {{Sonneborn}, George and {Altner}, Bruce and {Kirshner}, Robert P.},
        title = "{The Progenitor of SN 1987A: Spatially Resolved Ultraviolet Spectroscopy of the Supernova Field}",
      journal = {ApJL},
     keywords = {Astronomical Spectroscopy, Iue, Supernova 1987a, Gravitational Collapse, Spatial Resolution, Temporal Distribution, Ultraviolet Spectra, Astrophysics, ASTROMETRY, STARS: SUPERNOVAE, ULTRAVIOLET: SPECTRA},
         year = 1987,
        month = dec,
       volume = {323},
        pages = {L35},
          doi = {10.1086/185052},
       adsurl = {https://ui.adsabs.harvard.edu/abs/1987ApJ...323L..35S},
      adsnote = {Provided by the SAO/NASA Astrophysics Data System}
}

@article{sw12,
      author         = "Shao, Lijing and Wex, Norbert",
      title          = "{New tests of local Lorentz invariance of gravity with
                        small-eccentricity binary pulsars}",
      journal        = "CQGra",
      volume         = "29",
      year           = "2012",
      pages          = "215018",
      doi            = "10.1088/0264-9381/29/21/215018",
      eprint         = "1209.4503",
      archivePrefix  = "arXiv",
      primaryClass   = "gr-qc",
      SLACcitation   = "
}

@article{sck+13,
      author         = "Shao, Lijing and Caballero, R. Nicolas and Kramer,
                        Michael and Wex, Norbert and Champion, David J. and
                        Jessner, Axel",
      title          = "{A new limit on local Lorentz invariance violation of
                        gravity from solitary pulsars}",
      journal        = "CQGra",
      volume         = "30",
      year           = "2013",
      pages          = "165019",
      doi            = "10.1088/0264-9381/30/16/165019",
      eprint         = "1307.2552",
      archivePrefix  = "arXiv",
      primaryClass   = "gr-qc",
      SLACcitation   = "
}

@ARTICLE{nor87,
       author = {{Nordtvedt}, Ken},
        title = "{Probing Gravity to the Second Post-Newtonian Order and to One Part in 10 7 Using the Spin Axis of the Sun}",
      journal = {\apj},
         year = "1987",
        month = "Sep",
       volume = {320},
        pages = {871},
          doi = {10.1086/165603},
       adsurl = {https://ui.adsabs.harvard.edu/abs/1987ApJ...320..871N},
      adsnote = {Provided by the SAO/NASA Astrophysics Data System}
}

@book{wil18,
      author         = "Will, Clifford M.",
      title          = "{Theory and Experiment in Gravitational Physics}",
      publisher      = "Cambridge University Press",
      place={Cambridge},
      year           = "2018",
      ISBN           = "9781108679824, 9781107117440",
      SLACcitation   = "
}

@ARTICLE{kop96,
   author = {{Kopeikin}, S.~M.},
    title = "{Proper Motion of Binary Pulsars as a Source of Secular Variations of Orbital Parameters}",
  journal = {ApJL},
     year = 1996,
    month = aug,
   volume = 467,
    pages = {L93},
      doi = {10.1086/310201},
   adsurl = {http://adsabs.harvard.edu/abs/1996ApJ...467L..93K},
  adsnote = {Provided by the SAO/NASA Astrophysics Data System}
}

@article{fhlg13,
      author         = "Foreman-Mackey, Daniel and Hogg, David W. and Lang,
                        Dustin and Goodman, Jonathan",
      title          = "{emcee: The MCMC Hammer}",
      journal        = "PASP",
      volume         = "125",
      year           = "2013",
      pages          = "306-312",
      doi            = "10.1086/670067",
      eprint         = "1202.3665",
      archivePrefix  = "arXiv",
      primaryClass   = "astro-ph.IM",
      SLACcitation   = "
}

@incollection{wex14,
      author         = "Wex, Norbert",
      title          = "{Testing Relativistic Gravity with Radio Pulsars}",
      booktitle      = "{Frontiers in Relativistic Celestial Mechanics:
        Applications and Experiments}",
      publisher      = "Walter de Gruyter GmbH, Berlin/Boston",
      year           = "2014",
      editor         = "Kopeikin, Sergei M.",
      volume         = "2",
      pages          = "39",
      eprint         = "1402.5594",
      archivePrefix  = "arXiv",
      primaryClass   = "gr-qc",
      SLACcitation   = "
}

@article{kra16,
      author         = "Kramer, Michael",
      title          = "{Pulsars as probes of gravity and fundamental physics}",
      journal        = "IJMPD",
      volume         = "25",
      year           = "2016",
      number         = "14",
      pages          = "1630029",
      doi            = "10.1142/S0218271816300299",
      eprint         = "1606.03843",
      archivePrefix  = "arXiv",
      primaryClass   = "astro-ph.HE",
      SLACcitation   = "
}

@article{zdw+19,
       author = {{Zhu}, W.~W. and {Desvignes}, G. and {Wex}, N. and {Caballero}, R.~N. and
         {Champion}, D.~J. and {Demorest}, P.~B. and {Ellis}, J.~A. and
         {Janssen}, G.~H. and {Kramer}, M. and {Krieger}, A. and {Lentati}, L. and
         {Nice}, D.~J. and {Ransom}, S.~M. and {Stairs}, I.~H. and
         {Stappers}, B.~W. and {Verbiest}, J.~P.~W. and {Arzoumanian}, Z. and
         {Bassa}, C.~G. and {Burgay}, M. and {Cognard}, I. and {Crowter}, K. and
         {Dolch}, T. and {Ferdman}, R.~D. and {Fonseca}, E. and
         {Gonzalez}, M.~E. and {Graikou}, E. and {Guillemot}, L. and
         {Hessels}, J.~W.~T. and {Jessner}, A. and {Jones}, G. and
         {Jones}, M.~L. and {Jordan}, C. and {Karuppusamy}, R. and {Lam}, M.~T. and
         {Lazaridis}, K. and {Lazarus}, P. and {Lee}, K.~J. and {Levin}, L. and
         {Liu}, K. and {Lyne}, A.~G. and {McKee}, J.~W. and {McLaughlin}, M.~A. and
         {Os{\l}owski}, S. and {Pennucci}, T. and {Perrodin}, D. and
         {Possenti}, A. and {Sanidas}, S. and {Shaifullah}, G. and {Smits}, R. and
         {Stovall}, K. and {Swiggum}, J. and {Theureau}, G. and {Tiburzi}, C.},
        title = "{Tests of gravitational symmetries with pulsar binary J1713+0747}",
      journal = {\mnras},
     keywords = {gravitation, binaries: general, stars: neutron, pulsars: individual (PSR J1713+0747), Astrophysics - High Energy Astrophysical Phenomena},
         year = "2019",
        month = "Jan",
       volume = {482},
       number = {3},
        pages = {3249-3260},
          doi = {10.1093/mnras/sty2905},
archivePrefix = {arXiv},
       eprint = {1802.09206},
 primaryClass = {astro-ph.HE},
       adsurl = {https://ui.adsabs.harvard.edu/abs/2019{\mnras}.482.3249Z},
      adsnote = {Provided by the SAO/NASA Astrophysics Data System}
}

@book{lk05,
      author         = "Lorimer, D. R. and Kramer, M.",
      title          = "{Handbook of Pulsar Astronomy}",
      publisher      = "Cambridge University Press",
      address        = "Cambridge, England",
      year           = "2005"
}

@PHDTHESIS{ant13,
       author = {{Antoniadis}, John Ioannis},
        title = "{Multi-wavelength studies of pulsars and their companions}",
       school = {University of Bonn},
         year = "2013",
        month = "Sep",
       adsurl = {https://ui.adsabs.harvard.edu/abs/2013PhDT.......184A},
      adsnote = {Provided by the SAO/NASA Astrophysics Data System}
}

@article{sw16,
      author         = "Shao, Lijing and Wex, Norbert",
      title          = "{Tests of gravitational symmetries with radio pulsars}",
      journal        = "SCPMA",
      volume         = "59",
      year           = "2016",
      number         = "9",
      pages          = "699501",
      doi            = "10.1007/s11433-016-0087-6",
      eprint         = "1604.03662",
      archivePrefix  = "arXiv",
      primaryClass   = "gr-qc",
      SLACcitation   = "
}

@article{wil14,
      author         = "Will, Clifford M.",
      title          = "{The Confrontation between General Relativity and
                        Experiment}",
      journal        = "LRR",
      volume         = "17",
      year           = "2014",
      pages          = "4",
      doi            = "10.12942/lrr-2014-4",
      eprint         = "1403.7377",
      archivePrefix  = "arXiv",
      primaryClass   = "gr-qc",
      SLACcitation   = "
}

@INBOOK{tv06,
       author = {{Tauris}, T.~M. and {van den Heuvel}, E.~P.~J.},
        title = "{Formation and evolution of compact stellar X-ray sources}",
     keywords = {Stellar X-Ray Sources, Formation, Evolution, Astrophysics},
    booktitle = {Compact stellar X-ray sources},
         year = 2006,
       volume = {39},
        pages = {623-665},
    publisher = "{Cambridge University Press}",
       adsurl = {https://ui.adsabs.harvard.edu/abs/2006csxs.book..623T},
      adsnote = {Provided by the SAO/NASA Astrophysics Data System}
}

@INPROCEEDINGS{tau11,
       author = {{Tauris}, T.~M.},
        title = "{Five and a Half Roads to Form a Millisecond Pulsar}",
     keywords = {Astrophysics - High Energy Astrophysical Phenomena, Astrophysics - Solar and Stellar Astrophysics},
    booktitle = {Evolution of Compact Binaries},
         year = 2011,
       editor = {{Schmidtobreick}, L. and {Schreiber}, M.~R. and {Tappert}, C.},
       volume = {447},
       series = {Astronomical Society of the Pacific Conference Series},
        pages = {285},
        journal = "{ASP Conf. Ser.}",
       adsurl = {https://ui.adsabs.harvard.edu/abs/2011ASPC..447..285T},
      adsnote = {Provided by the SAO/NASA Astrophysics Data System}
}



@article{hd83,
  title = {Upper Limits on the Isotropic Gravitational Radiation Background from Pulsar Timing Analysis.},
  author = {Hellings, R. W. and Downs, G. S.},
  year = {1983},
  month = feb,
  journal = {The Astrophysical Journal},
  volume = {265},
  pages = {L39-L42},
  issn = {0004-637X},
  doi = {10.1086/183954},
  abstract = {A pulsar and the earth may be thought of as end masses of a free-mass gravitational wave antenna in which the relative motion of the masses is monitored by observing the Doppler shift of the pulse arrival times. Using timing residuals from PSR 1133 + 16, 1237 + 25, 1604-00, and 2045-16, an upper limit to the spectrum of the isotropic gravitational radiation background has been derived in the frequency band 4 x 10 to the -9th to 10 to the -7th Hz. This limit is found to be S(E) = 10 to the 21st f-cubed ergs/cu cm Hz, where S(E) is the energy density spectrum and f is the frequency in Hz. This would limit the energy density at frequencies below 10 to the -8th Hz to be 0.00014 times the critical density.},
  keywords = {Astrophysics,Background Radiation,Cosmology,Doppler Effect,Gravitation Theory,Gravitational Wave Antennas,Gravitational Waves,Isotropy,Limits (Mathematics),Perturbation Theory,Power Spectra,Pulsars,Spectral Energy Distribution,Stochastic Processes,Time Measurement},
  annotation = {ADS Bibcode: 1983ApJ...265L..39H},
  file = {/Users/jantoniadis/Zotero/storage/TGUTU9G7/Hellings and Downs - 1983 - Upper limits on the isotropic gravitational radiat.pdf}
}



@ARTICLE{itla14,
       author = {{Istrate}, A.~G. and {Tauris}, T.~M. and {Langer}, N. and
         {Antoniadis}, J.},
        title = "{The timescale of low-mass proto-helium white dwarf evolution}",
      journal = {\aa},
     keywords = {white dwarfs, stars: evolution, binaries: close, X-rays: binaries, stars: neutron, Astrophysics - Solar and Stellar Astrophysics, Astrophysics - High Energy Astrophysical Phenomena},
         year = 2014,
        month = nov,
       volume = {571},
          eid = {L3},
        pages = {L3},
          doi = {10.1051/0004-6361/201424681},
archivePrefix = {arXiv},
       eprint = {1410.5471},
 primaryClass = {astro-ph.SR},
       adsurl = {https://ui.adsabs.harvard.edu/abs/2014A&A...571L...3I},
      adsnote = {Provided by the SAO/NASA Astrophysics Data System}
}

@ARTICLE{imt+16,
       author = {{Istrate}, A.~G. and {Marchant}, P. and {Tauris}, T.~M. and
         {Langer}, N. and {Stancliffe}, R.~J. and {Grassitelli}, L.},
        title = "{Models of low-mass helium white dwarfs including gravitational settling, thermal and chemical diffusion, and rotational mixing}",
      journal = {\aa},
     keywords = {white dwarfs, binaries: general, stars: low-mass, pulsars: general, binaries: close, Astrophysics - Solar and Stellar Astrophysics, Astrophysics - High Energy Astrophysical Phenomena},
         year = 2016,
        month = oct,
       volume = {595},
          eid = {A35},
        pages = {A35},
          doi = {10.1051/0004-6361/201628874},
archivePrefix = {arXiv},
       eprint = {1606.04947},
 primaryClass = {astro-ph.SR},
       adsurl = {https://ui.adsabs.harvard.edu/abs/2016A&A...595A..35I},
      adsnote = {Provided by the SAO/NASA Astrophysics Data System}
}

@ARTICLE{Alsing:2012,
       author = {{Alsing}, Justin and {Berti}, Emanuele and {Will}, Clifford M. and
         {Zaglauer}, Helmut},
        title = "{Gravitational radiation from compact binary systems in the massive Brans-Dicke theory of gravity}",
      journal = {\prd},
     keywords = {04.50.-h, 04.25.Nx, 04.30.Db, 04.80.Cc, Higher-dimensional gravity and other theories of gravity, Post-Newtonian approximation, perturbation theory, related approximations, Wave generation and sources, Experimental tests of gravitational theories, General Relativity and Quantum Cosmology, Astrophysics - High Energy Astrophysical Phenomena, High Energy Physics - Phenomenology, High Energy Physics - Theory},
         year = "2012",
        month = "Mar",
       volume = {85},
       number = {6},
          eid = {064041},
        pages = {064041},
          doi = {10.1103/PhysRevD.85.064041},
archivePrefix = {arXiv},
       eprint = {1112.4903},
 primaryClass = {gr-qc},
       adsurl = {https://ui.adsabs.harvard.edu/abs/2012PhRvD..85f4041A},
      adsnote = {Provided by the SAO/NASA Astrophysics Data System}
}

@article{Fierz:1956,
  author =	"{Fierz}, M",
  journal =	"Helvetica Physica Acta",
  year =	"1956",
  volume =	"29",
  pages =	"128",
  issn =	"0018-0238",
  url =	"http://tinyurl.sfx.mpg.de/vc9p",
}

@ARTICLE{Brans:1961,
       author = {{Brans}, C. and {Dicke}, R.~H.},
        title = "{Mach's Principle and a Relativistic Theory of Gravitation}",
      journal = {Physical Review},
         year = "1961",
        month = "Nov",
       volume = {124},
       number = {3},
        pages = {925-935},
          doi = {10.1103/PhysRev.124.925},
       adsurl = {https://ui.adsabs.harvard.edu/abs/1961PhRv..124..925B},
      adsnote = {Provided by the SAO/NASA Astrophysics Data System}
}

@book{Jordan:1955schwerkraft,
  title={Schwerkraft und Weltall},
  author={{Jordan}, P.},
  lccn={56036377},
  series={Die Wissenschaft},
  url={https://books.google.de/books?id=VP85AQAAIAAJ},
  year={1955},
  publisher={Vieweg}
}

@ARTICLE{Mendes:2016,
       author = {{Mendes}, Raissa F.~P. and {Ortiz}, N{\'e}stor},
        title = "{Highly compact neutron stars in scalar-tensor theories of gravity: Spontaneous scalarization versus gravitational collapse}",
      journal = {\prd},
     keywords = {General Relativity and Quantum Cosmology, Astrophysics - High Energy Astrophysical Phenomena, High Energy Physics - Phenomenology},
         year = "2016",
        month = "Jun",
       volume = {93},
       number = {12},
          eid = {124035},
        pages = {124035},
          doi = {10.1103/PhysRevD.93.124035},
archivePrefix = {arXiv},
       eprint = {1604.04175},
 primaryClass = {gr-qc},
       adsurl = {https://ui.adsabs.harvard.edu/abs/2016PhRvD..93l4035M},
      adsnote = {Provided by the SAO/NASA Astrophysics Data System}
}

@ARTICLE{De_Felice:2010,
       author = {{De Felice}, Antonio and {Tsujikawa}, Shinji},
        title = "{f( R) Theories}",
      journal = {Living Reviews in Relativity},
     keywords = {inflation, dark energy, cosmological perturbations, f(R) gravity, modified gravity, General Relativity and Quantum Cosmology, Astrophysics - Cosmology and Extragalactic Astrophysics, High Energy Physics - Phenomenology, High Energy Physics - Theory},
         year = "2010",
        month = "Jun",
       volume = {13},
       number = {1},
          eid = {3},
        pages = {3},
          doi = {10.12942/lrr-2010-3},
archivePrefix = {arXiv},
       eprint = {1002.4928},
 primaryClass = {gr-qc},
       adsurl = {https://ui.adsabs.harvard.edu/abs/2010LRR....13....3D},
      adsnote = {Provided by the SAO/NASA Astrophysics Data System}
}

@BOOK{will18,
      address = {Cambridge, England},
       author = {{Will}, C.~M.},
        title = "{Theory and experiment in gravitational physics, Second Edition, Cambridge University Press}",
         year = "2018",
    publisher = {Cambridge University Press},
       adsurl = {https://ui.adsabs.harvard.edu/abs/1981tegp.book.....W},
      adsnote = {Provided by the SAO/NASA Astrophysics Data System}
}

@ARTICLE{Shao:2017,
       author = {{Shao}, Lijing and {Sennett}, Noah and {Buonanno}, Alessandra and
         {Kramer}, Michael and {Wex}, Norbert},
        title = "{Constraining Nonperturbative Strong-Field Effects in Scalar-Tensor Gravity by Combining Pulsar Timing and Laser-Interferometer Gravitational-Wave Detectors}",
      journal = {Physical Review X},
     keywords = {General Relativity and Quantum Cosmology, Astrophysics - High Energy Astrophysical Phenomena, High Energy Physics - Phenomenology},
         year = "2017",
        month = "Oct",
       volume = {7},
       number = {4},
          eid = {041025},
        pages = {041025},
          doi = {10.1103/PhysRevX.7.041025},
archivePrefix = {arXiv},
       eprint = {1704.07561},
 primaryClass = {gr-qc},
       adsurl = {https://ui.adsabs.harvard.edu/abs/2017PhRvX...7d1025S},
      adsnote = {Provided by the SAO/NASA Astrophysics Data System}
}

@PHDTHESIS{dem07,
       author = {{Demorest}, Paul B.},
        title = "{Measuring the gravitational wave background using precision pulsar timing}",
     keywords = {Gravitational wave background, Pulsar timing, Radio pulsars, Astronomy signal processors, Interstellar plasma},
       school = {University of California, Berkeley},
         year = 2007,
        month = aug,
       adsurl = {https://ui.adsabs.harvard.edu/abs/2007PhDT........14D},
      adsnote = {Provided by the SAO/NASA Astrophysics Data System}
}

@ARTICLE{ldc+14,
       author = {{Liu}, K. and {Desvignes}, G. and {Cognard}, I. and {Stappers}, B.~W. and
         {Verbiest}, J.~P.~W. and {Lee}, K.~J. and {Champion}, D.~J. and
         {Kramer}, M. and {Freire}, P.~C.~C. and {Karuppusamy}, R.},
        title = "{Measuring pulse times of arrival from broad-band pulsar observations}",
      journal = {\mnras},
     keywords = {methods: data analysis, pulsars: general, pulsars: individual (PSR J1909-3744), Astrophysics - Astrophysics of Galaxies},
         year = 2014,
        month = oct,
       volume = {443},
       number = {4},
        pages = {3752-3760},
          doi = {10.1093/mnras/stu1420},
archivePrefix = {arXiv},
       eprint = {1407.3827},
 primaryClass = {astro-ph.GA},
       adsurl = {https://ui.adsabs.harvard.edu/abs/2014{\mnras}.443.3752L},
      adsnote = {Provided by the SAO/NASA Astrophysics Data System}
}

@ARTICLE{krh+20,
       author = {{Kerr}, M. and {Reardon}, D.~J. and {Hobbs}, G. and {Shannon}, R.~M. and
         {Manchester}, R.~N. and {Dai}, S. and {Russell}, C.~J. and
         {Zhang}, S. -B. and {van Straten}, W. and {Os{\l}owski}, S. and
         {Parthasarathy}, A. and {Spiewak}, R. and {Bailes}, M. and
         {Bhat}, N.~D.~R. and {Cameron}, A.~D. and {Coles}, W.~A. and
         {Dempsey}, J. and {Deng}, X. and {Goncharov}, B. and {Kaczmarek}, J. F and
         {Keith}, M.~J. and {Lasky}, P.~D. and {Lower}, M.~E. and {Preisig}, B. and
         {Sarkissian}, J.~M. and {Toomey}, L. and {Wang}, H. and {Wang}, J. and
         {Zhang}, L. and {Zhu}, X.},
        title = "{The Parkes Pulsar Timing Array Project: Second data release}",
      journal = {arXiv e-prints},
     keywords = {Astrophysics - Instrumentation and Methods for Astrophysics, Astrophysics - High Energy Astrophysical Phenomena, General Relativity and Quantum Cosmology},
         year = 2020,
        month = mar,
          eid = {arXiv:2003.09780},
        pages = {arXiv:2003.09780},
archivePrefix = {arXiv},
       eprint = {2003.09780},
 primaryClass = {astro-ph.IM},
       adsurl = {https://ui.adsabs.harvard.edu/abs/2020arXiv200309780K},
      adsnote = {Provided by the SAO/NASA Astrophysics Data System}
}

@ARTICLE{abb+18b,
       author = {{Arzoumanian}, Z. and {Baker}, P.~T. and {Brazier}, A. and
         {Burke-Spolaor}, S. and {Chamberlin}, S.~J. and {Chatterjee}, S. and
         {Christy}, B. and {Cordes}, J.~M. and {Cornish}, N.~J. and
         {Crawford}, F. and {Thankful Cromartie}, H. and {Crowter}, K. and
         {DeCesar}, M. and {Demorest}, P.~B. and {Dolch}, T. and {Ellis}, J.~A. and
         {Ferdman}, R.~D. and {Ferrara}, E. and {Folkner}, W.~M. and
         {Fonseca}, E. and {Garver-Daniels}, N. and {Gentile}, P.~A. and
         {Haas}, R. and {Hazboun}, J.~S. and {Huerta}, E.~A. and {Islo}, K. and
         {Jones}, G. and {Jones}, M.~L. and {Kaplan}, D.~L. and {Kaspi}, V.~M. and
         {Lam}, M.~T. and {Lazio}, T.~J.~W. and {Levin}, L. and {Lommen}, A.~N. and
         {Lorimer}, D.~R. and {Luo}, J. and {Lynch}, R.~S. and {Madison}, D.~R. and
         {McLaughlin}, M.~A. and {McWilliams}, S.~T. and {Mingarelli}, C.~M.~F. and
         {Ng}, C. and {Nice}, D.~J. and {Park}, R.~S. and {Pennucci}, T.~T. and
         {Pol}, N.~S. and {Ransom}, S.~M. and {Ray}, P.~S. and {Rasskazov}, A. and
         {Siemens}, X. and {Simon}, J. and {Spiewak}, R. and {Stairs}, I.~H. and
         {Stinebring}, D.~R. and {Stovall}, K. and {Swiggum}, J. and
         {Taylor}, S.~R. and {Vallisneri}, M. and {van Haasteren}, R. and {Vigeland
        }, S. and {Zhu}, W.~W. and {NANOGrav Collaboration}},
        title = "{The NANOGrav 11 Year Data Set: Pulsar-timing Constraints on the Stochastic Gravitational-wave Background}",
      journal = {\apj},
     keywords = {ephemerides, gravitational waves, inflation, methods: data analysis, pulsars: general, quasars: supermassive black holes, Astrophysics - High Energy Astrophysical Phenomena, Astrophysics - Astrophysics of Galaxies, General Relativity and Quantum Cosmology},
         year = 2018,
        month = may,
       volume = {859},
       number = {1},
          eid = {47},
        pages = {47},
          doi = {10.3847/1538-4357/aabd3b},
archivePrefix = {arXiv},
       eprint = {1801.02617},
 primaryClass = {astro-ph.HE},
       adsurl = {https://ui.adsabs.harvard.edu/abs/2018ApJ...859...47A},
      adsnote = {Provided by the SAO/NASA Astrophysics Data System}
}

@ARTICLE{abb+18a,
       author = {{Arzoumanian}, Zaven and {Brazier}, Adam and {Burke-Spolaor}, Sarah and
         {Chamberlin}, Sydney and {Chatterjee}, Shami and {Christy}, Brian and
         {Cordes}, James M. and {Cornish}, Neil J. and {Crawford}, Fronefield and
         {Thankful Cromartie}, H. and {Crowter}, Kathryn and
         {DeCesar}, Megan E. and {Demorest}, Paul B. and {Dolch}, Timothy and
         {Ellis}, Justin A. and {Ferdman}, Robert D. and
         {Ferrara}, Elizabeth C. and {Fonseca}, Emmanuel and
         {Garver-Daniels}, Nathan and {Gentile}, Peter A. and
         {Halmrast}, Daniel and {Huerta}, E.~A. and {Jenet}, Fredrick A. and
         {Jessup}, Cody and {Jones}, Glenn and {Jones}, Megan L. and
         {Kaplan}, David L. and {Lam}, Michael T. and {Lazio}, T. Joseph W. and
         {Levin}, Lina and {Lommen}, Andrea and {Lorimer}, Duncan R. and
         {Luo}, Jing and {Lynch}, Ryan S. and {Madison}, Dustin and
         {Matthews}, Allison M. and {McLaughlin}, Maura A. and
         {McWilliams}, Sean T. and {Mingarelli}, Chiara and {Ng}, Cherry and
         {Nice}, David J. and {Pennucci}, Timothy T. and {Ransom}, Scott M. and
         {Ray}, Paul S. and {Siemens}, Xavier and {Simon}, Joseph and
         {Spiewak}, Ren{\'e}e and {Stairs}, Ingrid H. and
         {Stinebring}, Daniel R. and {Stovall}, Kevin and {Swiggum}, Joseph K. and
         {Taylor}, Stephen R. and {Vallisneri}, Michele and
         {van Haasteren}, Rutger and {Vigeland}, Sarah J. and {Zhu}, Weiwei and
         {NANOGrav Collaboration}},
        title = "{The NANOGrav 11-year Data Set: High-precision Timing of 45 Millisecond Pulsars}",
      journal = {\apjs},
     keywords = {binaries: general, gravitational waves, parallaxes, proper motions, pulsars: general, stars: neutron, Astrophysics - High Energy Astrophysical Phenomena, Astrophysics - Instrumentation and Methods for Astrophysics},
         year = 2018,
        month = apr,
       volume = {235},
       number = {2},
          eid = {37},
        pages = {37},
          doi = {10.3847/1538-4365/aab5b0},
archivePrefix = {arXiv},
       eprint = {1801.01837},
 primaryClass = {astro-ph.HE},
       adsurl = {https://ui.adsabs.harvard.edu/abs/2018ApJS..235...37A},
      adsnote = {Provided by the SAO/NASA Astrophysics Data System}
}

@ARTICLE{cgl+18,
       author = {{Caballero}, R.~N. and {Guo}, Y.~J. and {Lee}, K.~J. and {Lazarus}, P. and
         {Champion}, D.~J. and {Desvignes}, G. and {Kramer}, M. and {Plant}, K. and
         {Arzoumanian}, Z. and {Bailes}, M. and {Bassa}, C.~G. and
         {Bhat}, N.~D.~R. and {Brazier}, A. and {Burgay}, M. and
         {Burke-Spolaor}, S. and {Chamberlin}, S.~J. and {Chatterjee}, S. and
         {Cognard}, I. and {Cordes}, J.~M. and {Dai}, S. and {Demorest}, P. and
         {Dolch}, T. and {Ferdman}, R.~D. and {Fonseca}, E. and {Gair}, J.~R. and
         {Garver-Daniels}, N. and {Gentile}, P. and {Gonzalez}, M.~E. and
         {Graikou}, E. and {Guillemot}, L. and {Hobbs}, G. and {Janssen}, G.~H. and
         {Karuppusamy}, R. and {Keith}, M.~J. and {Kerr}, M. and {Lam}, M.~T. and
         {Lasky}, P.~D. and {Lazio}, T.~J.~W. and {Levin}, L. and {Liu}, K. and
         {Lommen}, A.~N. and {Lorimer}, D.~R. and {Lynch}, R.~S. and
         {Madison}, D.~R. and {Manchester}, R.~N. and {McKee}, J.~W. and
         {McLaughlin}, M.~A. and {McWilliams}, S.~T. and {Mingarelli}, C.~M.~F. and
         {Nice}, D.~J. and {Os{\l}owski}, S. and {Palliyaguru}, N.~T. and
         {Pennucci}, T.~T. and {Perera}, B.~B.~P. and {Perrodin}, D. and
         {Possenti}, A. and {Ransom}, S.~M. and {Reardon}, D.~J. and
         {Sanidas}, S.~A. and {Sesana}, A. and {Shaifullah}, G. and
         {Shannon}, R.~M. and {Siemens}, X. and {Simon}, J. and {Spiewak}, R. and
         {Stairs}, I. and {Stappers}, B. and {Stinebring}, D.~R. and
         {Stovall}, K. and {Swiggum}, J.~K. and {Taylor}, S.~R. and
         {Theureau}, G. and {Tiburzi}, C. and {Toomey}, L. and
         {van Haasteren}, R. and {van Straten}, W. and {Verbiest}, J.~P.~W. and
         {Wang}, J.~B. and {Zhu}, X.~J. and {Zhu}, W.~W.},
        title = "{Studying the Solar system with the International Pulsar Timing Array}",
      journal = {\mnras},
     keywords = {pulsars: general, methods: data analysis, methods: statistical, ephemerides, Astrophysics - Earth and Planetary Astrophysics},
         year = 2018,
        month = dec,
       volume = {481},
       number = {4},
        pages = {5501-5516},
          doi = {10.1093/mnras/sty2632},
archivePrefix = {arXiv},
       eprint = {1809.10744},
 primaryClass = {astro-ph.EP},
       adsurl = {https://ui.adsabs.harvard.edu/abs/2018{\mnras}.481.5501C},
      adsnote = {Provided by the SAO/NASA Astrophysics Data System}
}

@ARTICLE{chm+10,
       author = {{Champion}, D.~J. and {Hobbs}, G.~B. and {Manchester}, R.~N. and
         {Edwards}, R.~T. and {Backer}, D.~C. and {Bailes}, M. and
         {Bhat}, N.~D.~R. and {Burke-Spolaor}, S. and {Coles}, W. and
         {Demorest}, P.~B. and {Ferdman}, R.~D. and {Folkner}, W.~M. and
         {Hotan}, A.~W. and {Kramer}, M. and {Lommen}, A.~N. and {Nice}, D.~J. and
         {Purver}, M.~B. and {Sarkissian}, J.~M. and {Stairs}, I.~H. and
         {van Straten}, W. and {Verbiest}, J.~P.~W. and {Yardley}, D.~R.~B.},
        title = "{Measuring the Mass of Solar System Planets Using Pulsar Timing}",
      journal = {ApJL},
     keywords = {planets and satellites: general, planets and satellites: individual: Jupiter, pulsars: general, Astrophysics - Earth and Planetary Astrophysics},
         year = 2010,
        month = sep,
       volume = {720},
       number = {2},
        pages = {L201-L205},
          doi = {10.1088/2041-8205/720/2/L201},
archivePrefix = {arXiv},
       eprint = {1008.3607},
 primaryClass = {astro-ph.EP},
       adsurl = {https://ui.adsabs.harvard.edu/abs/2010ApJ...720L.201C},
      adsnote = {Provided by the SAO/NASA Astrophysics Data System}
}

@ARTICLE{hgc+20,
       author = {{Hobbs}, G. and {Guo}, L. and {Caballero}, R.~N. and {Coles}, W. and
         {Lee}, K.~J. and {Manchester}, R.~N. and {Reardon}, D.~J. and
         {Matsakis}, D. and {Tong}, M.~L. and {Arzoumanian}, Z. and
         {Bailes}, M. and {Bassa}, C.~G. and {Bhat}, N.~D.~R. and {Brazier}, A. and
         {Burke-Spolaor}, S. and {Champion}, D.~J. and {Chatterjee}, S. and
         {Cognard}, I. and {Dai}, S. and {Desvignes}, G. and {Dolch}, T. and
         {Ferdman}, R.~D. and {Graikou}, E. and {Guillemot}, L. and
         {Janssen}, G.~H. and {Keith}, M.~J. and {Kerr}, M. and {Kramer}, M. and
         {Lam}, M.~T. and {Liu}, K. and {Lyne}, A. and {Lazio}, T.~J.~W. and
         {Lynch}, R. and {McKee}, J.~W. and {McLaughlin}, M.~A. and
         {Mingarelli}, C.~M.~F. and {Nice}, D.~J. and {Os{\l}owski}, S. and
         {Pennucci}, T.~T. and {Perera}, B.~B.~P. and {Perrodin}, D. and
         {Possenti}, A. and {Russell}, C.~J. and {Sanidas}, S. and {Sesana}, A. and
         {Shaifullah}, G. and {Shannon}, R.~M. and {Simon}, J. and
         {Spiewak}, R. and {Stairs}, I.~H. and {Stappers}, B.~W. and
         {Swiggum}, J.~K. and {Taylor}, S.~R. and {Theureau}, G. and
         {Toomey}, L. and {van Haasteren}, R. and {Wang}, J.~B. and {Wang}, Y. and
         {Zhu}, X.~J.},
        title = "{A pulsar-based time-scale from the International Pulsar Timing Array}",
      journal = {\mnras},
     keywords = {time, pulsars: general, Astrophysics - Instrumentation and Methods for Astrophysics},
         year = 2020,
        month = feb,
       volume = {491},
       number = {4},
        pages = {5951-5965},
          doi = {10.1093/mnras/stz3071},
archivePrefix = {arXiv},
       eprint = {1910.13628},
 primaryClass = {astro-ph.IM},
       adsurl = {https://ui.adsabs.harvard.edu/abs/2020{\mnras}.491.5951H},
      adsnote = {Provided by the SAO/NASA Astrophysics Data System}
}

@misc{mamajek2015iau,
    title={IAU 2015 Resolution B3 on Recommended Nominal Conversion Constants for Selected Solar and Planetary Properties},
    author={E. E. Mamajek and A. Prsa and G. Torres and P. Harmanec and M. Asplund and P. D. Bennett and N. Capitaine and J. Christensen-Dalsgaard and E. Depagne and W. M. Folkner and M. Haberreiter and S. Hekker and J. L. Hilton and V. Kostov and D. W. Kurtz and J. Laskar and B. D. Mason and E. F. Milone and M. M. Montgomery and M. T. Richards and J. Schou and S. G. Stewart},
    year={2015},
    eprint={1510.07674},
    archivePrefix={arXiv},
    primaryClass={astro-ph.SR}
}

@ARTICLE{McMillan:2017,
       author = {{McMillan}, Paul J.},
        title = "{The mass distribution and gravitational potential of the Milky Way}",
      journal = {\mnras},
     keywords = {methods: statistical, Galaxy: fundamental parameters, Galaxy: kinematics and dynamics, Galaxy: structure, Astrophysics - Astrophysics of Galaxies},
         year = 2017,
        month = feb,
       volume = {465},
       number = {1},
        pages = {76-94},
          doi = {10.1093/mnras/stw2759},
archivePrefix = {arXiv},
       eprint = {1608.00971},
 primaryClass = {astro-ph.GA},
       adsurl = {https://ui.adsabs.harvard.edu/abs/2017{\mnras}.465...76M},
      adsnote = {Provided by the SAO/NASA Astrophysics Data System}
}

@ARTICLE{rhc+16,
       author = {{Reardon}, D.~J. and {Hobbs}, G. and {Coles}, W. and {Levin}, Y. and
         {Keith}, M.~J. and {Bailes}, M. and {Bhat}, N.~D.~R. and
         {Burke-Spolaor}, S. and {Dai}, S. and {Kerr}, M. and {Lasky}, P.~D. and
         {Manchester}, R.~N. and {Os{\l}owski}, S. and {Ravi}, V. and
         {Shannon}, R.~M. and {van Straten}, W. and {Toomey}, L. and {Wang}, J. and
         {Wen}, L. and {You}, X.~P. and {Zhu}, X. -J.},
        title = "{Timing analysis for 20 millisecond pulsars in the Parkes Pulsar Timing Array}",
      journal = {\mnras},
     keywords = {astrometry, ephemerides, parallaxes, proper motions, pulsars: general, Astrophysics - High Energy Astrophysical Phenomena, Astrophysics - Instrumentation and Methods for Astrophysics},
         year = 2016,
        month = jan,
       volume = {455},
       number = {2},
        pages = {1751-1769},
          doi = {10.1093/mnras/stv2395},
archivePrefix = {arXiv},
       eprint = {1510.04434},
 primaryClass = {astro-ph.HE},
       adsurl = {https://ui.adsabs.harvard.edu/abs/2016{\mnras}.455.1751R},
      adsnote = {Provided by the SAO/NASA Astrophysics Data System}
}

@ARTICLE{fhb+09,
       author = {{Feroz}, F. and {Hobson}, M.~P. and {Bridges}, M.},
        title = "{MULTINEST: an efficient and robust Bayesian inference tool for cosmology and particle physics}",
      journal = {\mnras},
     keywords = {methods: data analysis, methods: statistical, Astrophysics},
         year = 2009,
        month = oct,
       volume = {398},
       number = {4},
        pages = {1601-1614},
          doi = {10.1111/j.1365-2966.2009.14548.x},
archivePrefix = {arXiv},
       eprint = {0809.3437},
 primaryClass = {astro-ph},
       adsurl = {https://ui.adsabs.harvard.edu/abs/2009{\mnras}.398.1601F},
      adsnote = {Provided by the SAO/NASA Astrophysics Data System}
}

@ARTICLE{lcc+16,
       author = {{Lam}, M.~T. and {Cordes}, J.~M. and {Chatterjee}, S. and
         {Arzoumanian}, Z. and {Crowter}, K. and {Demorest}, P.~B. and
         {Dolch}, T. and {Ellis}, J.~A. and {Ferdman}, R.~D. and
         {Fonseca}, E.~F. and {Gonzalez}, M.~E. and {Jones}, G. and
         {Jones}, M.~L. and {Levin}, L. and {Madison}, D.~R. and
         {McLaughlin}, M.~A. and {Nice}, D.~J. and {Pennucci}, T.~T. and
         {Ransom}, S.~M. and {Siemens}, X. and {Stairs}, I.~H. and
         {Stovall}, K. and {Swiggum}, J.~K. and {Zhu}, W.~W.},
        title = "{The NANOGrav Nine-year Data Set: Noise Budget for Pulsar Arrival Times on Intraday Timescales}",
      journal = {\apj},
     keywords = {gravitational waves, pulsars: general, Astrophysics - Instrumentation and Methods for Astrophysics, Astrophysics - High Energy Astrophysical Phenomena},
         year = 2016,
        month = mar,
       volume = {819},
       number = {2},
          eid = {155},
        pages = {155},
          doi = {10.3847/0004-637X/819/2/155},
archivePrefix = {arXiv},
       eprint = {1512.08326},
 primaryClass = {astro-ph.IM},
       adsurl = {https://ui.adsabs.harvard.edu/abs/2016ApJ...819..155L},
      adsnote = {Provided by the SAO/NASA Astrophysics Data System}
}

@ARTICLE{lma+19,
       author = {{Lam}, M.~T. and {McLaughlin}, M.~A. and {Arzoumanian}, Z. and
         {Blumer}, H. and {Brook}, P.~R. and {Cromartie}, H.~T. and
         {Demorest}, P.~B. and {DeCesar}, M.~E. and {Dolch}, T. and
         {Ellis}, J.~A. and {Ferdman}, R.~D. and {Ferrara}, E.~C. and
         {Fonseca}, E. and {Garver-Daniels}, N. and {Gentile}, P.~A. and
         {Jones}, M.~L. and {Lorimer}, D.~R. and {Lynch}, R.~S. and {Ng}, C. and
         {Nice}, D.~J. and {Pennucci}, T.~T. and {Ransom}, S.~M. and
         {Spiewak}, R. and {Stairs}, I.~H. and {Stovall}, K. and
         {Swiggum}, J.~K. and {Vigeland}, S.~J. and {Zhu}, W.~W.},
        title = "{The NANOGrav 12.5 yr Data Set: The Frequency Dependence of Pulse Jitter in Precision Millisecond Pulsars}",
      journal = {\apj},
     keywords = {gravitational waves, pulsars: general, Astrophysics - High Energy Astrophysical Phenomena, Astrophysics - Instrumentation and Methods for Astrophysics},
         year = 2019,
        month = feb,
       volume = {872},
       number = {2},
          eid = {193},
        pages = {193},
          doi = {10.3847/1538-4357/ab01cd},
archivePrefix = {arXiv},
       eprint = {1809.03058},
 primaryClass = {astro-ph.HE},
       adsurl = {https://ui.adsabs.harvard.edu/abs/2019ApJ...872..193L},
      adsnote = {Provided by the SAO/NASA Astrophysics Data System}
}

@ARTICLE{cll+16,
       author = {{Caballero}, R.~N. and {Lee}, K.~J. and {Lentati}, L. and
         {Desvignes}, G. and {Champion}, D.~J. and {Verbiest}, J.~P.~W. and
         {Janssen}, G.~H. and {Stappers}, B.~W. and {Kramer}, M. and
         {Lazarus}, P. and {Possenti}, A. and {Tiburzi}, C. and {Perrodin}, D. and
         {Os{\l}owski}, S. and {Babak}, S. and {Bassa}, C.~G. and {Brem}, P. and
         {Burgay}, M. and {Cognard}, I. and {Gair}, J.~R. and {Graikou}, E. and
         {Guillemot}, L. and {Hessels}, J.~W.~T. and {Karuppusamy}, R. and
         {Lassus}, A. and {Liu}, K. and {McKee}, J. and {Mingarelli}, C.~M.~F. and
         {Petiteau}, A. and {Purver}, M.~B. and {Rosado}, P.~A. and
         {Sanidas}, S. and {Sesana}, A. and {Shaifullah}, G. and {Smits}, R. and
         {Taylor}, S.~R. and {Theureau}, G. and {van Haasteren}, R. and
         {Vecchio}, A.},
        title = "{The noise properties of 42 millisecond pulsars from the European Pulsar Timing Array and their impact on gravitational-wave searches}",
      journal = {\mnras},
     keywords = {gravitational waves, methods: data analysis, pulsars: general, Astrophysics - Instrumentation and Methods for Astrophysics, Astrophysics - Cosmology and Nongalactic Astrophysics},
         year = 2016,
        month = apr,
       volume = {457},
       number = {4},
        pages = {4421-4440},
          doi = {10.1093/mnras/stw179},
archivePrefix = {arXiv},
       eprint = {1510.09194},
 primaryClass = {astro-ph.IM},
       adsurl = {https://ui.adsabs.harvard.edu/abs/2016{\mnras}.457.4421C},
      adsnote = {Provided by the SAO/NASA Astrophysics Data System}
}


@ARTICLE{GRAVITY:2020,
       author = {{Gravity Collaboration} and {Abuter}, R. and {Amorim}, A. and
         {Baub{\"o}ck}, M. and {Berger}, J.~P. and {Bonnet}, H. and {Brand
        ner}, W. and {Cardoso}, V. and {Cl{\'e}net}, Y. and {de Zeeuw}, P.~T. and
         {Dexter}, J. and {Eckart}, A. and {Eisenhauer}, F. and
         {F{\"o}rster Schreiber}, N.~M. and {Garcia}, P. and {Gao}, F. and
         {Gendron}, E. and {Genzel}, R. and {Gillessen}, S. and {Habibi}, M. and
         {Haubois}, X. and {Henning}, T. and {Hippler}, S. and {Horrobin}, M. and
         {Jim{\'e}nez-Rosales}, A. and {Jochum}, L. and {Jocou}, L. and
         {Kaufer}, A. and {Kervella}, P. and {Lacour}, S. and
         {Lapeyr{\`e}re}, V. and {Le Bouquin}, J. -B. and {L{\'e}na}, P. and
         {Nowak}, M. and {Ott}, T. and {Paumard}, T. and {Perraut}, K. and
         {Perrin}, G. and {Pfuhl}, O. and {Rodr{\'\i}guez-Coira}, G. and
         {Shangguan}, J. and {Scheithauer}, S. and {Stadler}, J. and
         {Straub}, O. and {Straubmeier}, C. and {Sturm}, E. and
         {Tacconi}, L.~J. and {Vincent}, F. and {von Fellenberg}, S. and
         {Waisberg}, I. and {Widmann}, F. and {Wieprecht}, E. and
         {Wiezorrek}, E. and {Woillez}, J. and {Yazici}, S. and {Zins}, G.},
        title = "{Detection of the Schwarzschild precession in the orbit of the star S2 near the Galactic centre massive black hole}",
      journal = {\aa},
     keywords = {black hole physics, Galaxy: nucleus, gravitation, relativistic processes, Astrophysics - Astrophysics of Galaxies, Astrophysics - Instrumentation and Methods for Astrophysics, General Relativity and Quantum Cosmology},
         year = 2020,
        month = apr,
       volume = {636},
          eid = {L5},
        pages = {L5},
          doi = {10.1051/0004-6361/202037813},
archivePrefix = {arXiv},
       eprint = {2004.07187},
 primaryClass = {astro-ph.GA},
       adsurl = {https://ui.adsabs.harvard.edu/abs/2020A&A...636L...5G},
      adsnote = {Provided by the SAO/NASA Astrophysics Data System}
}

@ARTICLE{Reid:2004,
       author = {{Reid}, M.~J. and {Brunthaler}, A.},
        title = "{The Proper Motion of Sagittarius A*. II. The Mass of Sagittarius A*}",
      journal = {\apj},
     keywords = {Astrometry, Black Hole Physics, Galaxy: Center, Galaxy: Fundamental Parameters, Galaxy: Structure, Astrophysics},
         year = 2004,
        month = dec,
       volume = {616},
       number = {2},
        pages = {872-884},
          doi = {10.1086/424960},
archivePrefix = {arXiv},
       eprint = {astro-ph/0408107},
 primaryClass = {astro-ph},
       adsurl = {https://ui.adsabs.harvard.edu/abs/2004ApJ...616..872R},
      adsnote = {Provided by the SAO/NASA Astrophysics Data System}
}

@ARTICLE{Reid:2014,
       author = {{Reid}, M.~J. and {Menten}, K.~M. and {Brunthaler}, A. and
         {Zheng}, X.~W. and {Dame}, T.~M. and {Xu}, Y. and {Wu}, Y. and
         {Zhang}, B. and {Sanna}, A. and {Sato}, M. and {Hachisuka}, K. and
         {Choi}, Y.~K. and {Immer}, K. and {Moscadelli}, L. and
         {Rygl}, K.~L.~J. and {Bartkiewicz}, A.},
        title = "{Trigonometric Parallaxes of High Mass Star Forming Regions: The Structure and Kinematics of the Milky Way}",
      journal = {\apj},
     keywords = {Galaxy: fundamental parameters, Galaxy: kinematics and dynamics, Galaxy: structure, gravitational waves, parallaxes, stars: formation, Astrophysics - Galaxy Astrophysics},
         year = 2014,
        month = mar,
       volume = {783},
       number = {2},
          eid = {130},
        pages = {130},
          doi = {10.1088/0004-637X/783/2/130},
archivePrefix = {arXiv},
       eprint = {1401.5377},
 primaryClass = {astro-ph.GA},
       adsurl = {https://ui.adsabs.harvard.edu/abs/2014ApJ...783..130R},
      adsnote = {Provided by the SAO/NASA Astrophysics Data System}
}

@ARTICLE{Bland-Hawthorn:2016,
       author = {{Bland-Hawthorn}, Joss and {Gerhard}, Ortwin},
        title = "{The Galaxy in Context: Structural, Kinematic, and Integrated Properties}",
      journal = {\araa},
     keywords = {Astrophysics - Astrophysics of Galaxies},
         year = 2016,
        month = sep,
       volume = {54},
        pages = {529-596},
          doi = {10.1146/annurev-astro-081915-023441},
archivePrefix = {arXiv},
       eprint = {1602.07702},
 primaryClass = {astro-ph.GA},
       adsurl = {https://ui.adsabs.harvard.edu/abs/2016ARA&A..54..529B},
      adsnote = {Provided by the SAO/NASA Astrophysics Data System}
}


@ARTICLE{ymw17,
       author = {{Yao}, J.~M. and {Manchester}, R.~N. and {Wang}, N.},
        title = "{Determination of the Sun's offset from the Galactic plane using pulsars}",
      journal = {\mnras},
     keywords = {Sun: fundamental parameters, stars: distances, pulsars: general, Astrophysics - Solar and Stellar Astrophysics},
         year = 2017,
        month = jul,
       volume = {468},
       number = {3},
        pages = {3289-3294},
          doi = {10.1093/mnras/stx729},
archivePrefix = {arXiv},
       eprint = {1704.01272},
 primaryClass = {astro-ph.SR},
       adsurl = {https://ui.adsabs.harvard.edu/abs/2017{\mnras}.468.3289Y},
      adsnote = {Provided by the SAO/NASA Astrophysics Data System}
}

@software{eh+17,
  author       = {Justin Ellis and Rutger van Haasteren},
  title        = {jellis18/PTMCMCSampler: Official Release},
  month        = oct,
  year         = 2017,
  publisher    = {Zenodo},
  version      = {1.0.0},
  doi          = {10.5281/zenodo.1037579},
  url          = {https://doi.org/10.5281/zenodo.1037579}
}

@misc{evtb+19,
  author = {Ellis, J. A. and Vallisneri, M. and Taylor, S. R. and Baker, P. T.},
  title = {{ENTERPRISE: Enhanced Numerical Toolbox Enabling a Robust PulsaR Inference SuitE}},
  year = {2019},
  publisher = {GitHub},
  howpublished = {\url{http://ascl.net/1912.015}}
}

@article{vhl+12,
    author = {van Haasteren, Rutger and Levin, Yuri},
    title = "{Understanding and analysing time-correlated stochastic signals in pulsar timing}",
    journal = {Monthly Notices of the Royal Astronomical Society},
    volume = {428},
    pages = {1147-1159},
    year = {2012},
}

@ARTICLE{kilic2018,
       author = {{Kilic}, Mukremin and {Hermes}, J.~J. and {C{\'o}rsico}, A.~H. and
         {Kosakowski}, Alekzander and {Brown}, Warren R. and {Antoniadis}, John and
         {Calcaferro}, Leila M. and {Gianninas}, A. and {Althaus}, Leandro G. and
         {Green}, M.~J.},
        title = "{A refined search for pulsations in white dwarf companions to millisecond pulsars}",
      journal = {\mnras},
     keywords = {stars: oscillations, stars: variables: general, white dwarfs, pulsars: individual: PSR J1738+0333, PSR J1911- 5958A, PSR J2234+0611, Astrophysics - Solar and Stellar Astrophysics},
         year = 2018,
        month = sep,
       volume = {479},
       number = {1},
        pages = {1267-1272},
          doi = {10.1093/mnras/sty1546},
archivePrefix = {arXiv},
       eprint = {1806.03650},
 primaryClass = {astro-ph.SR},
       adsurl = {https://ui.adsabs.harvard.edu/abs/2018{\mnras}.479.1267K},
      adsnote = {Provided by the SAO/NASA Astrophysics Data System}
}

@ARTICLE{webbink1983,
       author = {{Webbink}, R.~F. and {Rappaport}, S. and {Savonije}, G.~J.},
        title = "{On the evolutionary status of bright, low-mass X-ray sources.}",
      journal = {\apj},
     keywords = {Binary Stars, Giant Stars, Stellar Evolution, Stellar Luminosity, Stellar Mass Accretion, X Ray Sources, Galactic Bulge, Neutron Stars, Stellar Mass Ejection, Stellar Models, White Dwarf Stars, X Ray Binaries, Astrophysics},
         year = 1983,
        month = jul,
       volume = {270},
        pages = {678-693},
          doi = {10.1086/161159},
       adsurl = {https://ui.adsabs.harvard.edu/abs/1983ApJ...270..678W},
      adsnote = {Provided by the SAO/NASA Astrophysics Data System}
}
@ARTICLE{Joss1987,
       author = {{Joss}, P.~C. and {Rappaport}, S. and {Lewis}, W.},
        title = "{The Core Mass--Radius Relation for Giants: A New Test of Stellar Evolution Theory}",
      journal = {\apj},
     keywords = {Giant Stars, Radii, Stellar Cores, Stellar Evolution, Stellar Interiors, Stellar Mass, Binary Stars, Computational Astrophysics, Constraints, Dwarf Stars, Late Stars, Astrophysics, STARS: EVOLUTION, STARS: INTERIORS, STARS: LATE-TYPE},
         year = 1987,
        month = aug,
       volume = {319},
        pages = {180},
          doi = {10.1086/165443},
       adsurl = {https://ui.adsabs.harvard.edu/abs/1987ApJ...319..180J},
      adsnote = {Provided by the SAO/NASA Astrophysics Data System}
}

@ARTICLE{refsdal1971,
       author = {{Refsdal}, S. and {Weigert}, A.},
        title = "{On the Production of White Dwarfs in Binary Systems of Small Mass}",
      journal = {\aa},
         year = 1971,
        month = aug,
       volume = {13},
        pages = {367},
       adsurl = {https://ui.adsabs.harvard.edu/abs/1971A&A....13..367R},
      adsnote = {Provided by the SAO/NASA Astrophysics Data System}
}
@ARTICLE{rappaport1995,
       author = {{Rappaport}, S. and {Podsiadlowski}, Ph. and {Joss}, P.~C. and
         {Di Stefano}, R. and {Han}, Z.},
        title = "{The relation between white dwarf mass and orbital period in wide binary radio pulsars}",
      journal = {\mnras},
     keywords = {BINARIES: GENERAL, STARS: EVOLUTION, STARS: NEUTRON, PULSARS: GENERAL, WHITE DWARFS},
         year = 1995,
        month = apr,
       volume = {273},
       number = {3},
        pages = {731-741},
          doi = {10.1093/mnras/273.3.731},
       adsurl = {https://ui.adsabs.harvard.edu/abs/1995{\mnras}.273..731R},
      adsnote = {Provided by the SAO/NASA Astrophysics Data System}
}
@ARTICLE{lin2011,
       author = {{Lin}, Jinrong and {Rappaport}, S. and {Podsiadlowski}, Ph. and
         {Nelson}, L. and {Paxton}, B. and {Todorov}, P.},
        title = "{LMXB and IMXB Evolution: I. The Binary Radio Pulsar PSR J1614-2230}",
      journal = {\apj},
     keywords = {accretion, accretion disks, binaries: general, pulsars: individual: PSR J1614─2230, stars: evolution, X-rays: binaries, Astrophysics - High Energy Astrophysical Phenomena},
         year = 2011,
        month = may,
       volume = {732},
       number = {2},
          eid = {70},
        pages = {70},
          doi = {10.1088/0004-637X/732/2/70},
archivePrefix = {arXiv},
       eprint = {1012.1877},
 primaryClass = {astro-ph.HE},
       adsurl = {https://ui.adsabs.harvard.edu/abs/2011ApJ...732...70L},
      adsnote = {Provided by the SAO/NASA Astrophysics Data System}
}

@ARTICLE{jia2014,
       author = {{Jia}, Kun and {Li}, X. -D.},
        title = "{Formation of Millisecond Pulsars with Low-mass Helium White Dwarf Companions in Very Compact Binaries}",
      journal = {\apj},
     keywords = {pulsars: individual: PSR J1738+0333 PSR J0348+0432, stars: evolution, stars: neutron, X-rays: binaries, Astrophysics - High Energy Astrophysical Phenomena, Astrophysics - Solar and Stellar Astrophysics},
         year = 2014,
        month = aug,
       volume = {791},
       number = {2},
          eid = {127},
        pages = {127},
          doi = {10.1088/0004-637X/791/2/127},
archivePrefix = {arXiv},
       eprint = {1407.3150},
 primaryClass = {astro-ph.HE},
       adsurl = {https://ui.adsabs.harvard.edu/abs/2014ApJ...791..127J},
      adsnote = {Provided by the SAO/NASA Astrophysics Data System}
}

@ARTICLE{paxton2011,
       author = {{Paxton}, Bill and {Bildsten}, Lars and {Dotter}, Aaron and
         {Herwig}, Falk and {Lesaffre}, Pierre and {Timmes}, Frank},
        title = "{Modules for Experiments in Stellar Astrophysics (MESA)}",
      journal = {\apjs},
     keywords = {methods: numerical, stars: evolution, stars: general, Astrophysics - Solar and Stellar Astrophysics, Astrophysics - Instrumentation and Methods for Astrophysics},
         year = 2011,
        month = jan,
       volume = {192},
       number = {1},
          eid = {3},
        pages = {3},
          doi = {10.1088/0067-0049/192/1/3},
archivePrefix = {arXiv},
       eprint = {1009.1622},
 primaryClass = {astro-ph.SR},
       adsurl = {https://ui.adsabs.harvard.edu/abs/2011ApJS..192....3P},
      adsnote = {Provided by the SAO/NASA Astrophysics Data System}
}
@ARTICLE{paxton2013,
       author = {{Paxton}, Bill and {Cantiello}, Matteo and {Arras}, Phil and
         {Bildsten}, Lars and {Brown}, Edward F. and {Dotter}, Aaron and
         {Mankovich}, Christopher and {Montgomery}, M.~H. and {Stello}, Dennis and
         {Timmes}, F.~X. and {Townsend}, Richard},
        title = "{Modules for Experiments in Stellar Astrophysics (MESA): Planets, Oscillations, Rotation, and Massive Stars}",
      journal = {\apjs},
     keywords = {asteroseismology, methods: numerical, planets and satellites: physical evolution, stars: evolution, stars: massive, stars: rotation, Astrophysics - Solar and Stellar Astrophysics, Astrophysics - Instrumentation and Methods for Astrophysics},
         year = 2013,
        month = sep,
       volume = {208},
       number = {1},
          eid = {4},
        pages = {4},
          doi = {10.1088/0067-0049/208/1/4},
archivePrefix = {arXiv},
       eprint = {1301.0319},
 primaryClass = {astro-ph.SR},
       adsurl = {https://ui.adsabs.harvard.edu/abs/2013ApJS..208....4P},
      adsnote = {Provided by the SAO/NASA Astrophysics Data System}
}
@ARTICLE{paxton2015,
       author = {{Paxton}, Bill and {Marchant}, Pablo and {Schwab}, Josiah and
         {Bauer}, Evan B. and {Bildsten}, Lars and {Cantiello}, Matteo and
         {Dessart}, Luc and {Farmer}, R. and {Hu}, H. and {Langer}, N. and
         {Townsend}, R.~H.~D. and {Townsley}, Dean M. and {Timmes}, F.~X.},
        title = "{Modules for Experiments in Stellar Astrophysics (MESA): Binaries, Pulsations, and Explosions}",
      journal = {\apjs},
     keywords = {binaries: general, methods: numerical, nuclear reactions, nucleosynthesis, abundances, shock waves, stars: evolution, stars: oscillations, Astrophysics - Solar and Stellar Astrophysics},
         year = 2015,
        month = sep,
       volume = {220},
       number = {1},
          eid = {15},
        pages = {15},
          doi = {10.1088/0067-0049/220/1/15},
archivePrefix = {arXiv},
       eprint = {1506.03146},
 primaryClass = {astro-ph.SR},
       adsurl = {https://ui.adsabs.harvard.edu/abs/2015ApJS..220...15P},
      adsnote = {Provided by the SAO/NASA Astrophysics Data System}
}
@ARTICLE{paxton2018,
       author = {{Paxton}, Bill and {Schwab}, Josiah and {Bauer}, Evan B. and
         {Bildsten}, Lars and {Blinnikov}, Sergei and {Duffell}, Paul and
         {Farmer}, R. and {Goldberg}, Jared A. and {Marchant}, Pablo and
         {Sorokina}, Elena and {Thoul}, Anne and {Townsend}, Richard H.~D. and
         {Timmes}, F.~X.},
        title = "{Modules for Experiments in Stellar Astrophysics (MESA): Convective Boundaries, Element Diffusion, and Massive Star Explosions}",
      journal = {\apjs},
     keywords = {convection, diffusion, hydrodynamics, methods: numerical, stars: evolution, supernovae: general, Astrophysics - Solar and Stellar Astrophysics, Astrophysics - High Energy Astrophysical Phenomena},
         year = 2018,
        month = feb,
       volume = {234},
       number = {2},
          eid = {34},
        pages = {34},
          doi = {10.3847/1538-4365/aaa5a8},
archivePrefix = {arXiv},
       eprint = {1710.08424},
 primaryClass = {astro-ph.SR},
       adsurl = {https://ui.adsabs.harvard.edu/abs/2018ApJS..234...34P},
      adsnote = {Provided by the SAO/NASA Astrophysics Data System}
}


@ARTICLE{althaus2001,
       author = {{Althaus}, L.~G. and {Serenelli}, A.~M. and {Benvenuto}, O.~G.},
        title = "{The impact of element diffusion on the formation and evolution of helium white dwarf stars}",
      journal = {\mnras},
     keywords = {DIFFUSION, STARS: EVOLUTION, STARS: INTERIORS, PULSARS: GENERAL, PULSARS: INDIVIDUAL: PSR B1855+09, WHITE DWARFS},
         year = 2001,
        month = jun,
       volume = {324},
       number = {3},
        pages = {617-622},
          doi = {10.1046/j.1365-8711.2001.04324.x},
       adsurl = {https://ui.adsabs.harvard.edu/abs/2001{\mnras}.324..617A},
      adsnote = {Provided by the SAO/NASA Astrophysics Data System}
}

@ARTICLE{Capano2020,
       author = {{Capano}, Collin D. and {Tews}, Ingo and {Brown}, Stephanie M. and
         {Margalit}, Ben and {De}, Soumi and {Kumar}, Sumit and
         {Brown}, Duncan A. and {Krishnan}, Badri and {Reddy}, Sanjay},
        title = "{Stringent constraints on neutron-star radii from multimessenger observations and nuclear theory}",
      journal = {Nature Astronomy},
     keywords = {Astrophysics - High Energy Astrophysical Phenomena, General Relativity and Quantum Cosmology, High Energy Physics - Phenomenology, Nuclear Theory},
         year = 2020,
        month = mar,
       volume = {4},
        pages = {625-632},
          doi = {10.1038/s41550-020-1014-6},
archivePrefix = {arXiv},
       eprint = {1908.10352},
 primaryClass = {astro-ph.HE},
       adsurl = {https://ui.adsabs.harvard.edu/abs/2020NatAs...4..625C},
      adsnote = {Provided by the SAO/NASA Astrophysics Data System}
}


@ARTICLE{ung08,
       author = {{Unglaub}, K.},
        title = "{Mass-loss and diffusion in subdwarf B stars and hot white dwarfs: do weak winds exist?}",
      journal = {\aa},
     keywords = {hydrodynamics, stars: chemically peculiar, stars: mass-loss, stars: winds, outflows, subdwarfs, white dwarfs, Astrophysics},
         year = 2008,
        month = aug,
       volume = {486},
       number = {3},
        pages = {923-940},
          doi = {10.1051/0004-6361:20078019},
archivePrefix = {arXiv},
       eprint = {0808.1072},
 primaryClass = {astro-ph},
       adsurl = {https://ui.adsabs.harvard.edu/abs/2008A&A...486..923U},
      adsnote = {Provided by the SAO/NASA Astrophysics Data System}
}

@ARTICLE{viani2018,
       author = {{Viani}, Lucas S. and {Basu}, Sarbani and {Ong J.}, M. Joel and
         {Bonaca}, Ana and {Chaplin}, William J.},
        title = "{Investigating the Metallicity-Mixing-length Relation}",
      journal = {\apj},
     keywords = {stars: fundamental parameters, stars: interiors, stars: oscillations: including pulsations, Astrophysics - Solar and Stellar Astrophysics},
         year = 2018,
        month = may,
       volume = {858},
       number = {1},
          eid = {28},
        pages = {28},
          doi = {10.3847/1538-4357/aab7eb},
archivePrefix = {arXiv},
       eprint = {1803.05924},
 primaryClass = {astro-ph.SR},
       adsurl = {https://ui.adsabs.harvard.edu/abs/2018ApJ...858...28V},
      adsnote = {Provided by the SAO/NASA Astrophysics Data System}
}



@ARTICLE{valle2019,
       author = {{Valle}, G. and {Dell'Omodarme}, M. and {Prada Moroni}, P.~G. and
         {Degl'Innocenti}, S.},
        title = "{Mixing-length calibration from field stars. An investigation on statistical errors, systematic biases, and spurious metallicity trends}",
      journal = {\aa},
     keywords = {stars: fundamental parameters, convection, methods: statistical, stars: evolution, stars: interiors, stars: low-mass, Astrophysics - Solar and Stellar Astrophysics},
         year = 2019,
        month = mar,
       volume = {623},
          eid = {A59},
        pages = {A59},
          doi = {10.1051/0004-6361/201834949},
archivePrefix = {arXiv},
       eprint = {1902.02272},
 primaryClass = {astro-ph.SR},
       adsurl = {https://ui.adsabs.harvard.edu/abs/2019A&A...623A..59V},
      adsnote = {Provided by the SAO/NASA Astrophysics Data System}
}
@ARTICLE{tayar2017,
       author = {{Tayar}, Jamie and {Somers}, Garrett and {Pinsonneault}, Marc H. and
         {Stello}, Dennis and {Mints}, Alexey and {Johnson}, Jennifer A. and
         {Zamora}, O. and {Garc{\'\i}a-Hern{\'a}ndez}, D.~A. and
         {Maraston}, Claudia and {Serenelli}, Aldo and {Allende Prieto}, Carlos and
         {Bastien}, Fabienne A. and {Basu}, Sarbani and {Bird}, J.~C. and
         {Cohen}, R.~E. and {Cunha}, Katia and {Elsworth}, Yvonne and
         {Garc{\'\i}a}, Rafael A. and {Girardi}, Leo and {Hekker}, Saskia and
         {Holtzman}, Jon and {Huber}, Daniel and {Mathur}, Savita and
         {M{\'e}sz{\'a}ros}, Szabolcs and {Mosser}, B. and {Shetrone}, Matthew and
         {Silva Aguirre}, Victor and {Stassun}, Keivan and
         {Stringfellow}, Guy S. and {Zasowski}, Gail and {Roman-Lopes}, A.},
        title = "{The Correlation between Mixing Length and Metallicity on the Giant Branch: Implications for Ages in the Gaia Era}",
      journal = {\apj},
     keywords = {stars: evolution, stars: fundamental parameters, Astrophysics - Solar and Stellar Astrophysics},
         year = 2017,
        month = may,
       volume = {840},
       number = {1},
          eid = {17},
        pages = {17},
          doi = {10.3847/1538-4357/aa6a1e},
archivePrefix = {arXiv},
       eprint = {1704.01164},
 primaryClass = {astro-ph.SR},
       adsurl = {https://ui.adsabs.harvard.edu/abs/2017ApJ...840...17T},
      adsnote = {Provided by the SAO/NASA Astrophysics Data System}
}
@ARTICLE{sonoi2019,
       author = {{Sonoi}, T. and {Ludwig}, H. -G. and {Dupret}, M. -A. and
         {Montalb{\'a}n}, J. and {Samadi}, R. and {Belkacem}, K. and
         {Caffau}, E. and {Goupil}, M. -J.},
        title = "{Calibration of mixing-length parameter {\ensuremath{\alpha}} for MLT and FST models by matching with CO$^{5}$BOLD models}",
      journal = {\aa},
     keywords = {convection, stars: late-type, stars: solar-type, Astrophysics - Solar and Stellar Astrophysics},
         year = 2019,
        month = jan,
       volume = {621},
          eid = {A84},
        pages = {A84},
          doi = {10.1051/0004-6361/201833495},
archivePrefix = {arXiv},
       eprint = {1811.05229},
 primaryClass = {astro-ph.SR},
       adsurl = {https://ui.adsabs.harvard.edu/abs/2019A&A...621A..84S},
      adsnote = {Provided by the SAO/NASA Astrophysics Data System}
}

@ARTICLE{corsico2012,
       author = {{C{\'o}rsico}, A.~H. and {Romero}, A.~D. and {Althaus}, L.~G. and
         {Hermes}, J.~J.},
        title = "{The seismic properties of low-mass He-core white dwarf stars}",
      journal = {\aa},
     keywords = {asteroseismology, stars: oscillations, white dwarfs, stars: evolution, stars: interiors, Astrophysics - Solar and Stellar Astrophysics},
         year = 2012,
        month = nov,
       volume = {547},
          eid = {A96},
        pages = {A96},
          doi = {10.1051/0004-6361/201220114},
archivePrefix = {arXiv},
       eprint = {1209.5107},
 primaryClass = {astro-ph.SR},
       adsurl = {https://ui.adsabs.harvard.edu/abs/2012A&A...547A..96C},
      adsnote = {Provided by the SAO/NASA Astrophysics Data System}
}

@ARTICLE{calcaferro2018,
       author = {{Calcaferro}, Leila M. and {C{\'o}rsico}, Alejandro H. and {Althaus}, Leand
        ro G. and {Romero}, Alejandra D. and {Kepler}, S.~O.},
        title = "{Pulsating low-mass white dwarfs in the frame of new evolutionary sequences. VI. Thin H-envelope sequences and asteroseismology of ELMV stars revisited}",
      journal = {\aa},
     keywords = {asteroseismology, stars: oscillations, white dwarfs, stars: evolution, stars: interiors, Astrophysics - Solar and Stellar Astrophysics},
         year = 2018,
        month = dec,
       volume = {620},
          eid = {A196},
        pages = {A196},
          doi = {10.1051/0004-6361/201833781},
archivePrefix = {arXiv},
       eprint = {1810.11502},
 primaryClass = {astro-ph.SR},
       adsurl = {https://ui.adsabs.harvard.edu/abs/2018A&A...620A.196C},
      adsnote = {Provided by the SAO/NASA Astrophysics Data System}
}
@ARTICLE{grootel2013,
       author = {{Van Grootel}, V. and {Fontaine}, G. and {Brassard}, P. and
         {Dupret}, M. -A.},
        title = "{The Newly Discovered Pulsating Low-mass White Dwarfs: An Extension of the ZZ Ceti Instability Strip}",
      journal = {\apj},
     keywords = {stars: oscillations, white dwarfs},
         year = 2013,
        month = jan,
       volume = {762},
       number = {1},
          eid = {57},
        pages = {57},
          doi = {10.1088/0004-637X/762/1/57},
       adsurl = {https://ui.adsabs.harvard.edu/abs/2013ApJ...762...57V},
      adsnote = {Provided by the SAO/NASA Astrophysics Data System}
}




@ARTICLE{vnvj12,
       author = {{van Haaften}, L.~M. and {Nelemans}, G. and {Voss}, R. and
         {Jonker}, P.~G.},
        title = "{Formation of the planet around the millisecond pulsar J1719-1438}",
      journal = {\aa},
     keywords = {pulsars: individual: J1719-1438, binaries: close, planets and satellites: formation, Astrophysics - Solar and Stellar Astrophysics, Astrophysics - Earth and Planetary Astrophysics},
         year = 2012,
        month = may,
       volume = {541},
          eid = {A22},
        pages = {A22},
          doi = {10.1051/0004-6361/201218798},
archivePrefix = {arXiv},
       eprint = {1203.2919},
 primaryClass = {astro-ph.SR},
       adsurl = {https://ui.adsabs.harvard.edu/abs/2012A&A...541A..22V},
      adsnote = {Provided by the SAO/NASA Astrophysics Data System}
}

@ARTICLE{ccth13,
       author = {{Chen}, Hai-Liang and {Chen}, Xuefei and {Tauris}, Thomas M. and
         {Han}, Zhanwen},
        title = "{Formation of Black Widows and Redbacks{\textemdash}Two Distinct Populations of Eclipsing Binary Millisecond Pulsars}",
      journal = {\apj},
     keywords = {binaries: eclipsing, pulsars: general, stars: evolution, stars: mass-loss, X-rays: binaries, Astrophysics - Solar and Stellar Astrophysics, Astrophysics - High Energy Astrophysical Phenomena},
         year = 2013,
        month = sep,
       volume = {775},
       number = {1},
          eid = {27},
        pages = {27},
          doi = {10.1088/0004-637X/775/1/27},
archivePrefix = {arXiv},
       eprint = {1308.4107},
 primaryClass = {astro-ph.SR},
       adsurl = {https://ui.adsabs.harvard.edu/abs/2013ApJ...775...27C},
      adsnote = {Provided by the SAO/NASA Astrophysics Data System}
}

@ARTICLE{hf00,
       author = {{Holmberg}, Johan and {Flynn}, Chris},
        title = "{The local density of matter mapped by Hipparcos}",
      journal = {\mnras},
     keywords = {GALAXY: KINEMATICS AND DYNAMICS, SOLAR NEIGHBOURHOOD, GALAXY: STRUCTURE, DARK MATTER, Astrophysics},
         year = 2000,
        month = apr,
       volume = {313},
       number = {2},
        pages = {209-216},
          doi = {10.1046/j.1365-8711.2000.02905.x},
archivePrefix = {arXiv},
       eprint = {astro-ph/9812404},
 primaryClass = {astro-ph},
       adsurl = {https://ui.adsabs.harvard.edu/abs/2000{\mnras}.313..209H},
      adsnote = {Provided by the SAO/NASA Astrophysics Data System}
}

@ARTICLE{vbv+20,
       author = {{Venkatraman Krishnan}, V. and {Bailes}, M. and {van Straten}, W. and
         {Wex}, N. and {Freire}, P.~C.~C. and {Keane}, E.~F. and
         {Tauris}, T.~M. and {Rosado}, P.~A. and {Bhat}, N.~D.~R. and
         {Flynn}, C. and {Jameson}, A. and {Os{\l}owski}, S.},
        title = "{Lense─Thirring frame dragging induced by a fast-rotating white dwarf in a binary pulsar system}",
      journal = {Science},
     keywords = {ASTRONOMY; PHYSICS, Astrophysics - High Energy Astrophysical Phenomena, Astrophysics - Solar and Stellar Astrophysics, General Relativity and Quantum Cosmology},
         year = 2020,
        month = jan,
       volume = {367},
       number = {6477},
        pages = {577-580},
          doi = {10.1126/science.aax7007},
archivePrefix = {arXiv},
       eprint = {2001.11405},
 primaryClass = {astro-ph.HE},
       adsurl = {https://ui.adsabs.harvard.edu/abs/2020Sci...367..577V},
      adsnote = {Provided by the SAO/NASA Astrophysics Data System}
}

@ARTICLE{ddf+20,
       author = {{Ding}, Hao and {Deller}, Adam T. and {Freire}, Paulo and
         {Kaplan}, David L. and {Lazio}, T. Joseph W. and {Shannon}, Ryan and
         {Stappers}, Benjamin},
        title = "{Very Long Baseline Astrometry of PSR J1012+5307 and its Implications on Alternative Theories of Gravity}",
      journal = {\apj},
     keywords = {1305, 661, 1108, 80, 571, 153, 1197, Astrophysics - High Energy Astrophysical Phenomena, General Relativity and Quantum Cosmology},
         year = 2020,
       volume = {896},
       number = {1},
          eid = {85},
        pages = {85},
          doi = {10.3847/1538-4357/ab8f27},
archivePrefix = {arXiv},
       eprint = {2004.14668},
 primaryClass = {astro-ph.HE},
       adsurl = {https://ui.adsabs.harvard.edu/abs/2020ApJ...896...85D},
      adsnote = {Provided by the SAO/NASA Astrophysics Data System}
}

@ARTICLE{bbb+18,
       author = {{Bailes}, M. and {Barr}, E. and {Bhat}, N.~D.~R. and {Brink}, J. and
         {Buchner}, S. and {Burgay}, M. and {Camilo}, F. and {Champion}, D.~J. and
         {Hessels}, J. and {Janssen}, G.~H. and {Jameson}, A. and
         {Johnston}, S. and {Karastergiou}, A. and {Karuppusamy}, R. and
         {Kaspi}, V. and {Keith}, M.~J. and {Kramer}, M. and
         {McLaughlin}, M.~A. and {Moodley}, K. and {Oslowski}, S. and
         {Possenti}, A. and {Ransom}, S.~M. and {Rasio}, F.~A. and
         {Sievers}, J. and {Serylak}, M. and {Stappers}, B.~W. and
         {Stairs}, I.~H. and {Theureau}, G. and {van Straten}, W. and
         {Weltevrede}, P. and {Wex}, N.},
        title = "{MeerTime - the MeerKAT Key Science Program on Pulsar Timing}",
      journal = {arXiv e-prints},
     keywords = {Astrophysics - Instrumentation and Methods for Astrophysics, Astrophysics - High Energy Astrophysical Phenomena},
         year = 2018,
        month = mar,
          eid = {arXiv:1803.07424},
        pages = {arXiv:1803.07424},
archivePrefix = {arXiv},
       eprint = {1803.07424},
 primaryClass = {astro-ph.IM},
       adsurl = {https://ui.adsabs.harvard.edu/abs/2018arXiv180307424B},
      adsnote = {Provided by the SAO/NASA Astrophysics Data System}
}

@article{de92b,
	doi = {10.1088/0264-9381/9/9/015},
	url = {https://doi.org/10.1088
	year = 1992,
	month = {sep},
	publisher = {{IOP} Publishing},
	volume = {9},
	number = {9},
	pages = {2093--2176},
	author = {T Damour and G Esposito-Farese},
	title = {Tensor-multi-scalar theories of gravitation},
	journal = {Classical and Quantum Gravity}
}

@ARTICLE{sjm14,
       author = {{Shannon}, R.~M. and {Johnston}, S. and {Manchester}, R.~N.},
        title = "{The kinematics and orbital dynamics of the PSR B1259-63/LS 2883 system from 23 yr of pulsar timing}",
      journal = {\mnras},
     keywords = {binaries: general, stars: kinematics and dynamics, pulsars: general, pulsars: individual: PSR B1259-63, Astrophysics - Solar and Stellar Astrophysics, Astrophysics - High Energy Astrophysical Phenomena},
         year = 2014,
        month = feb,
       volume = {437},
       number = {4},
        pages = {3255-3264},
          doi = {10.1093/mnras/stt2123},
archivePrefix = {arXiv},
       eprint = {1311.0588},
 primaryClass = {astro-ph.SR},
       adsurl = {https://ui.adsabs.harvard.edu/abs/2014{\mnras}.437.3255S},
      adsnote = {Provided by the SAO/NASA Astrophysics Data System}
}

@ARTICLE{rch+19,
       author = {{Reardon}, D.~J. and {Coles}, W.~A. and {Hobbs}, G. and {Ord}, S. and
         {Kerr}, M. and {Bailes}, M. and {Bhat}, N.~D.~R. and
         {Venkatraman Krishnan}, V.},
        title = "{Modelling annual and orbital variations in the scintillation of the relativistic binary PSR J1141-6545}",
      journal = {\mnras},
     keywords = {astrometry, scattering, pulsars: general, pulsars: individual (PSR J1141-6545), ISM: general, ISM: structure, Astrophysics - High Energy Astrophysical Phenomena, Astrophysics - Solar and Stellar Astrophysics},
         year = 2019,
        month = may,
       volume = {485},
       number = {3},
        pages = {4389-4403},
          doi = {10.1093/mnras/stz643},
archivePrefix = {arXiv},
       eprint = {1903.01990},
 primaryClass = {astro-ph.HE},
       adsurl = {https://ui.adsabs.harvard.edu/abs/2019{\mnras}.485.4389R},
      adsnote = {Provided by the SAO/NASA Astrophysics Data System}
}

@ARTICLE{cfg+17,
       author = {{Cognard}, Isma{\"e}l and {Freire}, Paulo C.~C. and {Guillemot}, Lucas and
         {Theureau}, Gilles and {Tauris}, Thomas M. and {Wex}, Norbert and
         {Graikou}, Eleni and {Kramer}, Michael and {Stappers}, Benjamin and
         {Lyne}, Andrew G. and {Bassa}, Cees and {Desvignes}, Gregory and
         {Lazarus}, Patrick},
        title = "{A Massive-born Neutron Star with a Massive White Dwarf Companion}",
      journal = {\apj},
     keywords = {binaries: close, gravitational waves, pulsars: general, pulsars: individual: J2222-0137, stars: neutron, white dwarfs, Astrophysics - High Energy Astrophysical Phenomena, Astrophysics - Solar and Stellar Astrophysics, General Relativity and Quantum Cosmology},
         year = 2017,
        month = aug,
       volume = {844},
       number = {2},
          eid = {128},
        pages = {128},
          doi = {10.3847/1538-4357/aa7bee},
archivePrefix = {arXiv},
       eprint = {1706.08060},
 primaryClass = {astro-ph.HE},
       adsurl = {https://ui.adsabs.harvard.edu/abs/2017ApJ...844..128C},
      adsnote = {Provided by the SAO/NASA Astrophysics Data System}
}

@ARTICLE{krh+20,
       author = {{Kerr}, Matthew and {Reardon}, Daniel J. and {Hobbs}, George and
         {Shannon}, Ryan M. and {Manchester}, Richard N. and {Dai}, Shi and
         {Russell}, Christopher J. and {Zhang}, Songbo and
         {van Straten}, Willem and {Os{\l}owski}, Stefan and
         {Parthasarathy}, Aditya and {Spiewak}, Renee and {Bailes}, Matthew and
         {Bhat}, N.~D. Ramesh and {Cameron}, Andrew D. and {Coles}, William A. and
         {Dempsey}, James and {Deng}, Xinping and {Goncharov}, Boris and
         {Kaczmarek}, Jane F. and {Keith}, Michael J. and {Lasky}, Paul D. and
         {Lower}, Marcus E. and {Preisig}, Brett and {Sarkissian}, John Mihran and
         {Toomey}, Lawrence and {Wang}, Hongguang and {Wang}, Jingbo and
         {Zhang}, Lei and {Zhu}, Xingjiang},
        title = "{The Parkes Pulsar Timing Array project: second data release}",
      journal = {\pasa},
     keywords = {gravitational waves, instrumentation: miscellaneous, methods: observational, pulsars: general, Astrophysics - Instrumentation and Methods for Astrophysics, Astrophysics - High Energy Astrophysical Phenomena, General Relativity and Quantum Cosmology},
         year = 2020,
        month = jun,
       volume = {37},
          eid = {e020},
        pages = {e020},
          doi = {10.1017/pasa.2020.11},
archivePrefix = {arXiv},
       eprint = {2003.09780},
 primaryClass = {astro-ph.IM},
       adsurl = {https://ui.adsabs.harvard.edu/abs/2020PASA...37...20K},
      adsnote = {Provided by the SAO/NASA Astrophysics Data System}
}

@ARTICLE{aab+20a,
       author = {{Alam}, Md F. and {Arzoumanian}, Zaven and {Baker}, Paul T. and
         {Blumer}, Harsha and {Bohler}, Keith E. and {Brazier}, Adam and
         {Brook}, Paul R. and {Burke-Spolaor}, Sarah and {Caballero}, Keeisi and
         {Camuccio}, Richard S. and {Chamberlain}, Rachel L. and
         {Chatterjee}, Shami and {Cordes}, James M. and {Cornish}, Neil J. and
         {Crawford}, Fronefield and {Cromartie}, H. Thankful and
         {DeCesar}, Megan E. and {Demorest}, Paul B. and {Dolch}, Timothy and
         {Ellis}, Justin A. and {Ferdman}, Robert D. and
         {Ferrara}, Elizabeth C. and {Fiore}, William and {Fonseca}, Emmanuel and
         {Garcia}, Yhamil and {Garver-Daniels}, Nathan and {Gentile}, Peter A. and
         {Good}, Deborah C. and {Gusdorff}, Jordan A. and {Halmrast}, Daniel and
         {Hazboun}, Jeffrey and {Islo}, Kristina and {Jennings}, Ross J. and
         {Jessup}, Cody and {Jones}, Megan L. and {Kaiser}, Andrew R. and
         {Kaplan}, David L. and {Kelley}, Luke Zoltan and {Shapiro Key}, Joey and
         {Lam}, Michael T. and {Lazio}, T. Joseph W. and {Lorimer}, Duncan R. and
         {Luo}, Jing and {Lynch}, Ryan S. and {Madison}, Dustin and
         {Maraccini}, Kaleb and {McLaughlin}, Maura A. and
         {Mingarelli}, Chiara M.~F. and {Ng}, Cherry and
         {Nguyen}, Benjamin M.~X. and {Nice}, David J. and
         {Pennucci}, Timothy T. and {Pol}, Nihan S. and {Ramette}, Joshua and
         {Ransom}, Scott M. and {Ray}, Paul S. and {Shapiro-Albert}, Brent J. and
         {Siemens}, Xavier and {Simon}, Joseph and {Spiewak}, Renee and
         {Stairs}, Ingrid H. and {Stinebring}, Daniel R. and {Stovall}, Kevin and
         {Swiggum}, Joseph K. and {Taylor}, Stephen R. and {Tripepi}, Michael and
         {Vallisneri}, Michele and {Vigeland}, Sarah J. and {Witt}, Caitlin A. and
         {Zhu}, Weiwei},
        title = "{The NANOGrav 12.5-year Data Set: Observations and Narrowband Timing of 47 Millisecond Pulsars}",
      journal = {arXiv e-prints},
     keywords = {Astrophysics - High Energy Astrophysical Phenomena, Astrophysics - Instrumentation and Methods for Astrophysics},
         year = 2020,
        month = may,
          eid = {arXiv:2005.06490},
        pages = {arXiv:2005.06490},
archivePrefix = {arXiv},
       eprint = {2005.06490},
 primaryClass = {astro-ph.HE},
       adsurl = {https://ui.adsabs.harvard.edu/abs/2020arXiv200506490A},
      adsnote = {Provided by the SAO/NASA Astrophysics Data System}
}

@ARTICLE{aab+20b,
       author = {{Alam}, Md F. and {Arzoumanian}, Zaven and {Baker}, Paul T. and
         {Blumer}, Harsha and {Bohler}, Keith E. and {Brazier}, Adam and
         {Brook}, Paul R. and {Burke-Spolaor}, Sarah and {Caballero}, Keeisi and
         {Camuccio}, Richard S. and {Chamberlain}, Rachel L. and
         {Chatterjee}, Shami and {Cordes}, James M. and {Cornish}, Neil J. and
         {Crawford}, Fronefield and {Cromartie}, H. Thankful and
         {DeCesar}, Megan E. and {Demorest}, Paul B. and {Dolch}, Timothy and
         {Ellis}, Justin A. and {Ferdman}, Robert D. and
         {Ferrara}, Elizabeth C. and {Fiore}, William and {Fonseca}, Emmanuel and
         {Garcia}, Yhamil and {Garver-Daniels}, Nathan and {Gentile}, Peter A. and
         {Good}, Deborah C. and {Gusdorff}, Jordan A. and {Halmrast}, Daniel and
         {Hazboun}, Jeffrey and {Islo}, Kristina and {Jennings}, Ross J. and
         {Jessup}, Cody and {Jones}, Megan L. and {Kaiser}, Andrew R. and
         {Kaplan}, David L. and {Kelley}, Luke Zoltan and {Shapiro Key}, Joey and
         {Lam}, Michael T. and {Lazio}, T. Joseph W. and {Lorimer}, Duncan R. and
         {Luo}, Jing and {Lynch}, Ryan S. and {Madison}, Dustin and
         {Maraccini}, Kaleb and {McLaughlin}, Maura A. and
         {Mingarelli}, Chiara M.~F. and {Ng}, Cherry and
         {Nguyen}, Benjamin M.~X. and {Nice}, David J. and
         {Pennucci}, Timothy T. and {Pol}, Nihan S. and {Ramette}, Joshua and
         {Ransom}, Scott M. and {Ray}, Paul S. and {Shapiro-Albert}, Brent J. and
         {Siemens}, Xavier and {Simon}, Joseph and {Spiewak}, Renee and
         {Stairs}, Ingrid H. and {Stinebring}, Daniel R. and {Stovall}, Kevin and
         {Swiggum}, Joseph K. and {Taylor}, Stephen R. and {Tripepi}, Michael and
         {Vallisneri}, Michele and {Vigeland}, Sarah J. and {Witt}, Caitlin A. and
         {Zhu}, Weiwei},
        title = "{The NANOGrav 12.5-year Data Set: Wideband Timing of 47 Millisecond Pulsars}",
      journal = {arXiv e-prints},
     keywords = {Astrophysics - High Energy Astrophysical Phenomena, Astrophysics - Instrumentation and Methods for Astrophysics},
         year = 2020,
        month = may,
          eid = {arXiv:2005.06495},
        pages = {arXiv:2005.06495},
archivePrefix = {arXiv},
       eprint = {2005.06495},
 primaryClass = {astro-ph.HE},
       adsurl = {https://ui.adsabs.harvard.edu/abs/2020arXiv200506495A},
      adsnote = {Provided by the SAO/NASA Astrophysics Data System}
}

@ARTICLE{jk17,
       author = {{Johnston}, Simon and {Karastergiou}, A.},
        title = "{Pulsar braking and the P-dot\{P\} diagram}",
      journal = {\mnras},
     keywords = {pulsars: general, Astrophysics - High Energy Astrophysical Phenomena, Astrophysics - Solar and Stellar Astrophysics},
         year = 2017,
        month = may,
       volume = {467},
       number = {3},
        pages = {3493-3499},
          doi = {10.1093/mnras/stx377},
archivePrefix = {arXiv},
       eprint = {1702.03616},
 primaryClass = {astro-ph.HE},
       adsurl = {https://ui.adsabs.harvard.edu/abs/2017mnras.467.3493J},
      adsnote = {Provided by the SAO/NASA Astrophysics Data System}
}

@ARTICLE{psj+19,
       author = {{Parthasarathy}, A. and {Shannon}, R.~M. and {Johnston}, S. and
         {Lentati}, L. and {Bailes}, M. and {Dai}, S. and {Kerr}, M. and
         {Manchester}, R.~N. and {Os{\l}owski}, S. and {Sobey}, C. and
         {van Straten}, W. and {Weltevrede}, P.},
        title = "{Timing of young radio pulsars - I. Timing noise, periodic modulation, and proper motion}",
      journal = {\mnras},
     keywords = {methods: data analysis, stars: neutron, pulsars: general, Astrophysics - High Energy Astrophysical Phenomena, Astrophysics - Solar and Stellar Astrophysics},
         year = 2019,
        month = nov,
       volume = {489},
       number = {3},
        pages = {3810-3826},
          doi = {10.1093/mnras/stz2383},
}

@ARTICLE{glw+16,
       author = {{Gao}, Z.~F. and {Li}, X. -D. and {Wang}, N. and {Yuan}, J.~P. and
         {Wang}, P. and {Peng}, Q.~H. and {Du}, Y.~J.},
        title = "{Constraining the braking indices of magnetars}",
      journal = {\mnras},
     keywords = {stars: magnetars, stars: magnetic field, stars: rotation, stars: winds, outflows, ISM: supernova remnants, Astrophysics - High Energy Astrophysical Phenomena, Astrophysics - Solar and Stellar Astrophysics},
         year = 2016,
        month = feb,
       volume = {456},
       number = {1},
        pages = {55-65},
          doi = {10.1093/mnras/stv2465},
archivePrefix = {arXiv},
       eprint = {1505.07013},
 primaryClass = {astro-ph.HE},
       adsurl = {https://ui.adsabs.harvard.edu/abs/2016{\mnras}.456...55G},
      adsnote = {Provided by the SAO/NASA Astrophysics Data System}
}

@ARTICLE{djw+18,
       author = {{Dai}, S. and {Johnston}, S. and {Weltevrede}, P. and {Kerr}, M. and
         {Burgay}, M. and {Esposito}, P. and {Israel}, G. and {Possenti}, A. and
         {Rea}, N. and {Sarkissian}, J.},
        title = "{Peculiar spin frequency and radio profile evolution of PSR J1119-6127 following magnetar-like X-ray bursts}",
      journal = {\mnras},
     keywords = {pulsars: general, stars: magnetars, pulsars: individual: PSR J1119-6127, Astrophysics - High Energy Astrophysical Phenomena},
         year = 2018,
        month = nov,
       volume = {480},
       number = {3},
        pages = {3584-3594},
          doi = {10.1093/mnras/sty2063},
archivePrefix = {arXiv},
       eprint = {1806.05064},
 primaryClass = {astro-ph.HE},
       adsurl = {https://ui.adsabs.harvard.edu/abs/2018{\mnras}.480.3584D},
      adsnote = {Provided by the SAO/NASA Astrophysics Data System}
}

@ARTICLE{tml+20,
       author = {{Torne}, P. and {Mac{\'\i}as-P{\'e}rez}, J. and {Ladjelate}, B. and
         {Ritacco}, A. and {S{\'a}nchez-Portal}, M. and {Berta}, S. and
         {Paubert}, G. and {Calvo}, M. and {Desvignes}, G. and
         {Karuppusamy}, R. and {Navarro}, S. and {John}, D. and
         {S{\'a}nchez}, S. and {Pe{\~n}alver}, J. and {Kramer}, M. and
         {Schuster}, K.},
        title = "{Detection of the magnetar XTE J1810-197 at 150 and 260 GHz with the NIKA2 kinetic inductance detector camera}",
      journal = {\aa},
     keywords = {stars: magnetars, pulsars: individual: XTE J1810-197, radiation mechanisms: non-thermal, instrumentation: detectors, Astrophysics - Instrumentation and Methods for Astrophysics, Astrophysics - High Energy Astrophysical Phenomena},
         year = 2020,
        month = aug,
       volume = {640},
          eid = {L2},
        pages = {L2},
          doi = {10.1051/0004-6361/202038504},
archivePrefix = {arXiv},
       eprint = {2007.02702},
 primaryClass = {astro-ph.IM},
       adsurl = {https://ui.adsabs.harvard.edu/abs/2020A&A...640L...2T},
      adsnote = {Provided by the SAO/NASA Astrophysics Data System}
}

@ARTICLE{tde+17,
       author = {{Torne}, P. and {Desvignes}, G. and {Eatough}, R.~P. and
         {Karuppusamy}, R. and {Paubert}, G. and {Kramer}, M. and {Cognard}, I. and
         {Champion}, D.~J. and {Spitler}, L.~G.},
        title = "{Detection of the magnetar SGR J1745-2900 up to 291 GHz with evidence of polarized millimetre emission}",
      journal = {\mnras},
     keywords = {radiation mechanisms: non-thermal, stars: magnetars, stars: neutron, pulsars: general, pulsars: individual: SGR J1745-2900, Astrophysics - High Energy Astrophysical Phenomena},
         year = 2017,
        month = feb,
       volume = {465},
       number = {1},
        pages = {242-247},
          doi = {10.1093/mnras/stw2757},
archivePrefix = {arXiv},
       eprint = {1610.07616},
 primaryClass = {astro-ph.HE},
       adsurl = {https://ui.adsabs.harvard.edu/abs/2017{\mnras}.465..242T},
      adsnote = {Provided by the SAO/NASA Astrophysics Data System}
}

@ARTICLE{kk16,
       author = {{Kaspi}, Victoria M. and {Kramer}, Michael},
        title = "{Radio Pulsars: The Neutron Star Population \&amp; Fundamental Physics}",
      journal = {arXiv e-prints},
     keywords = {Astrophysics - High Energy Astrophysical Phenomena},
         year = 2016,
        month = feb,
          eid = {arXiv:1602.07738},
        pages = {arXiv:1602.07738},
archivePrefix = {arXiv},
       eprint = {1602.07738},
 primaryClass = {astro-ph.HE},
       adsurl = {https://ui.adsabs.harvard.edu/abs/2016arXiv160207738K},
      adsnote = {Provided by the SAO/NASA Astrophysics Data System}
}

@ARTICLE{egk+20,
       author = {{Evans}, P.~A. and {Gropp}, J.~D. and {Kennea}, J.~A. and
         {Klingler}, N.~J. and {Laha}, S. and {Lien}, A.~Y. and {Page}, K.~L. and
         {Sakamoto}, T. and {Tohuvavohu}, A. and
         {Neil Gehrels Swift Observatory Team}},
        title = "{Swift-BAT trigger 960986: Swift detection of a new SGR Swift J1818.0-1607}",
      journal = {GRB Coordinates Network},
         year = 2020,
        month = mar,
       volume = {27373},
        pages = {1},
       adsurl = {https://ui.adsabs.harvard.edu/abs/2020GCN.27373....1E},
      adsnote = {Provided by the SAO/NASA Astrophysics Data System}
}

@ARTICLE{esy+20,
       author = {{Enoto}, Teruaki and {Sakamoto}, Takanori and {Younes}, George and
         {Hu}, Chin-Ping and {Ho}, Wynn C.~G. and {Gendreau}, Keith and
         {Arzoumanian}, Zaven and {Guver}, Tolga and {Guillot}, Sebastien and
         {Altamirano}, Diego and {Ray}, Paul S. and {Ng}, Mason and
         {Chakrabarty}, Deepto and {Jaisawal}, Gaurava K. and {Bogdanov}, Slavko},
        title = "{NICER detection of 1.36 sec periodicity from a new magnetar, Swift J1818.0-1607}",
      journal = {The Astronomer's Telegram},
     keywords = {Soft Gamma-ray Repeater},
         year = 2020,
        month = mar,
       volume = {13551},
        pages = {1},
       adsurl = {https://ui.adsabs.harvard.edu/abs/2020ATel13551....1E},
      adsnote = {Provided by the SAO/NASA Astrophysics Data System}
}

@ARTICLE{kdk+20,
       author = {{Karuppusamy}, Ramesh and {Desvignes}, Gregory and {Kramer}, Michael and
         {Porayko}, Nataliya and {Champion}, David and {Torne}, Pablo and
         {Stappers}, Ben and {van der Horst}, Alexander and
         {Kouveliotou}, Chryssa and {O'Connor}, Brendan},
        title = "{Detection of pulsed radio emission from new magnetar Swift J1818.0-1607}",
      journal = {The Astronomer's Telegram},
     keywords = {Neutron Star, Soft Gamma-ray Repeater, Star, Pulsar, Magnetar},
         year = 2020,
        month = mar,
       volume = {13553},
        pages = {1},
       adsurl = {https://ui.adsabs.harvard.edu/abs/2020ATel13553....1K},
      adsnote = {Provided by the SAO/NASA Astrophysics Data System}
}

@ARTICLE{cdj+20,
       author = {{Champion}, David and {Desvignes}, Gregory and {Jankowski}, Fabian and
         {Karuppusamy}, Ramesh and {Keith}, Michael and {Kouveliotou}, Chryssa and
         {Kramer}, Michael and {Lyne}, Andrew and {Mickaliger}, Mitchell B. and
         {O'Connor}, Brendan and {Porayko}, Nataliya and {Rajwade}, Kaustubh and
         {Stappers}, Ben and {Torne}, Pablo and {van der Horst}, Alexander and
         {Weltevrede}, Patrick},
        title = "{Spin-evolution of the new magnetar J1818.0-1607}",
      journal = {The Astronomer's Telegram},
     keywords = {Neutron Star, Soft Gamma-ray Repeater, Star, Pulsar, Magnetar},
         year = 2020,
        month = mar,
       volume = {13559},
        pages = {1},
       adsurl = {https://ui.adsabs.harvard.edu/abs/2020ATel13559....1C},
      adsnote = {Provided by the SAO/NASA Astrophysics Data System}
}

@ARTICLE{scholz20,
       author = {{Scholz}, Paul and {Chime/Frb Collaboration}},
        title = "{A bright millisecond-timescale radio burst from the direction of the Galactic magnetar SGR 1935+2154}",
      journal = {The Astronomer's Telegram},
     keywords = {Soft Gamma-ray Repeater, Magnetar},
         year = 2020,
        month = apr,
       volume = {13681},
        pages = {1},
       adsurl = {https://ui.adsabs.harvard.edu/abs/2020ATel13681....1S},
      adsnote = {Provided by the SAO/NASA Astrophysics Data System}
}

@ARTICLE{lld+19,
       author = {{Levin}, L. and {Lyne}, A.~G. and {Desvignes}, G. and {Eatough}, R.~P. and
         {Karuppusamy}, R. and {Kramer}, M. and {Mickaliger}, M. and
         {Stappers}, B.~W. and {Weltevrede}, P.},
        title = "{Spin frequency evolution and pulse profile variations of the recently re-activated radio magnetar XTE J1810-197}",
      journal = {\mnras},
     keywords = {stars: magnetars, stars: neutron, pulsars: individual: PSR J1809-1943, Astrophysics - High Energy Astrophysical Phenomena},
         year = 2019,
        month = oct,
       volume = {488},
       number = {4},
        pages = {5251-5258},
          doi = {10.1093/mnras/stz2074},
archivePrefix = {arXiv},
       eprint = {1903.02660},
 primaryClass = {astro-ph.HE},
       adsurl = {https://ui.adsabs.harvard.edu/abs/2019{\mnras}.488.5251L},
      adsnote = {Provided by the SAO/NASA Astrophysics Data System}
}

@ARTICLE{efk+13,
       author = {{Eatough}, R.~P. and {Falcke}, H. and {Karuppusamy}, R. and
         {Lee}, K.~J. and {Champion}, D.~J. and {Keane}, E.~F. and
         {Desvignes}, G. and {Schnitzeler}, D.~H.~F.~M. and {Spitler}, L.~G. and
         {Kramer}, M. and {Klein}, B. and {Bassa}, C. and {Bower}, G.~C. and
         {Brunthaler}, A. and {Cognard}, I. and {Deller}, A.~T. and
         {Demorest}, P.~B. and {Freire}, P.~C.~C. and {Kraus}, A. and
         {Lyne}, A.~G. and {Noutsos}, A. and {Stappers}, B. and {Wex}, N.},
        title = "{A strong magnetic field around the supermassive black hole at the centre of the Galaxy}",
      journal = {\nat},
     keywords = {Astrophysics - Galaxy Astrophysics, Astrophysics - High Energy Astrophysical Phenomena},
         year = 2013,
        month = sep,
       volume = {501},
       number = {7467},
        pages = {391-394},
          doi = {10.1038/nature12499},
archivePrefix = {arXiv},
       eprint = {1308.3147},
 primaryClass = {astro-ph.GA},
       adsurl = {https://ui.adsabs.harvard.edu/abs/2013Natur.501..391E},
      adsnote = {Provided by the SAO/NASA Astrophysics Data System}
}


@ARTICLE{dep+18,
       author = {{Desvignes}, G. and {Eatough}, R.~P. and {Pen}, U.~L. and {Lee}, K.~J. and
         {Mao}, S.~A. and {Karuppusamy}, R. and {Schnitzeler}, D.~H.~F.~M. and
         {Falcke}, H. and {Kramer}, M. and {Wucknitz}, O. and {Spitler}, L.~G. and
         {Torne}, P. and {Liu}, K. and {Bower}, G.~C. and {Cognard}, I. and
         {Lyne}, A.~G. and {Stappers}, B.~W.},
        title = "{Large Magneto-ionic Variations toward the Galactic Center Magnetar, PSR J1745-2900}",
      journal = {\apjl},
     keywords = {Galaxy: center, magnetic fields, pulsars: individual: J1745-2900, Astrophysics - High Energy Astrophysical Phenomena},
         year = 2018,
        month = jan,
       volume = {852},
       number = {1},
          eid = {L12},
        pages = {L12},
          doi = {10.3847/2041-8213/aaa2f8},
archivePrefix = {arXiv},
       eprint = {1711.10323},
 primaryClass = {astro-ph.HE},
       adsurl = {https://ui.adsabs.harvard.edu/abs/2018ApJ...852L..12D},
      adsnote = {Provided by the SAO/NASA Astrophysics Data System}
}

@ARTICLE{bel17,
       author = {{Beloborodov}, Andrei M.},
        title = "{A Flaring Magnetar in FRB 121102?}",
      journal = {\apjl},
     keywords = {dense matter, magnetic fields, radiation mechanisms: general, relativistic processes, stars: magnetars, supernovae: general, Astrophysics - High Energy Astrophysical Phenomena},
         year = 2017,
        month = jul,
       volume = {843},
       number = {2},
          eid = {L26},
        pages = {L26},
          doi = {10.3847/2041-8213/aa78f3},
archivePrefix = {arXiv},
       eprint = {1702.08644},
 primaryClass = {astro-ph.HE},
       adsurl = {https://ui.adsabs.harvard.edu/abs/2017ApJ...843L..26B},
      adsnote = {Provided by the SAO/NASA Astrophysics Data System}
}

@ARTICLE{cfp+07,
       author = {{Camilo}, F. and {Ransom}, S.~M. and {Pe{\~n}alver}, J. and
         {Karastergiou}, A. and {van Kerkwijk}, M.~H. and {Durant}, M. and
         {Halpern}, J.~P. and {Reynolds}, J. and {Thum}, C. and {Helfand
        }, D.~J. and {Zimmerman}, N. and {Cognard}, I.},
        title = "{The Variable Radio-to-X-Ray Spectrum of the Magnetar XTE J1810-197}",
      journal = {\apj},
     keywords = {Stars: Pulsars: Individual: Alphanumeric: XTE J1810-197, Stars: Neutron, Astrophysics},
         year = "2007",
        month = "Nov",
       volume = {669},
       number = {1},
        pages = {561-569},
          doi = {10.1086/521548},
archivePrefix = {arXiv},
       eprint = {0705.4095},
 primaryClass = {astro-ph},
       adsurl = {https://ui.adsabs.harvard.edu/abs/2007ApJ...669..561C},
      adsnote = {Provided by the SAO/NASA Astrophysics Data System}
}

@ARTICLE{hl06,
       author = {{Harding}, Alice K. and {Lai}, Dong},
        title = "{Physics of strongly magnetized neutron stars}",
      journal = {Reports on Progress in Physics},
     keywords = {Astrophysics},
         year = 2006,
        month = sep,
       volume = {69},
       number = {9},
        pages = {2631-2708},
          doi = {10.1088/0034-4885/69/9/R03},
archivePrefix = {arXiv},
       eprint = {astro-ph/0606674},
 primaryClass = {astro-ph},
       adsurl = {https://ui.adsabs.harvard.edu/abs/2006RPPh...69.2631H},
      adsnote = {Provided by the SAO/NASA Astrophysics Data System}
}

@ARTICLE{ok14,
       author = {{Olausen}, S.~A. and {Kaspi}, V.~M.},
        title = "{The McGill Magnetar Catalog}",
      journal = {\apjss},
     keywords = {catalogs, pulsars: general, stars: magnetars, stars: neutron, Astrophysics - High Energy Astrophysical Phenomena},
         year = 2014,
        month = may,
       volume = {212},
       number = {1},
          eid = {6},
        pages = {6},
          doi = {10.1088/0067-0049/212/1/6},
archivePrefix = {arXiv},
       eprint = {1309.4167},
 primaryClass = {astro-ph.HE},
       adsurl = {https://ui.adsabs.harvard.edu/abs/2014ApJS..212....6O},
      adsnote = {Provided by the SAO/NASA Astrophysics Data System}
}

@INPROCEEDINGS{kk15,
       author = {{Kaspi}, Victoria M. and {Kramer}, Michael},
        title = "{Radio Pulsars: The Neutron Star Population \&amp; Fundamental Physics}",
booktitle = {Proceedings of the 26th Solvay Conference of Physics },
     year = 2016,
   series = {World Scientific},
             eid = {arXiv:1602.07738},
        pages = {arXiv:1602.07738},
archivePrefix = {arXiv},
       eprint = {1602.07738},
 primaryClass = {astro-ph.HE},
       adsurl = {https://ui.adsabs.harvard.edu/abs/2016arXiv160207738K},
      adsnote = {Provided by the SAO/NASA Astrophysics Data System}
}

@ARTICLE{ljk+08,
       author = {{Lazaridis}, K. and {Jessner}, A. and {Kramer}, M. and
         {Stappers}, B.~W. and {Lyne}, A.~G. and {Jordan}, C.~A. and
         {Serylak}, M. and {Zensus}, J.~A.},
        title = "{Radio spectrum of the AXP J1810-197 and of its profile components}",
      journal = {\mnras},
     keywords = {stars: neutron, pulsars: general, pulsars: individual: AXP J1810-197, Astrophysics},
         year = 2008,
        month = oct,
       volume = {390},
       number = {2},
        pages = {839-846},
          doi = {10.1111/j.1365-2966.2008.13794.x},
archivePrefix = {arXiv},
       eprint = {0808.0244},
 primaryClass = {astro-ph},
       adsurl = {https://ui.adsabs.harvard.edu/abs/2008{\mnras}.390..839L},
      adsnote = {Provided by the SAO/NASA Astrophysics Data System}
}

@ARTICLE{crh+16,
       author = {{Camilo}, F. and {Ransom}, S.~M. and {Halpern}, J.~P. and
         {Alford}, J.~A.~J. and {Cognard}, I. and {Reynolds}, J.~E. and
         {Johnston}, S. and {Sarkissian}, J. and {van Straten}, W.},
        title = "{Radio Disappearance of the Magnetar XTE J1810-197 and Continued X-ray Timing}",
      journal = {\apj},
     keywords = {pulsars: individual: XTE J1810─197, PSR J1809─1943, stars: neutron, Astrophysics - Solar and Stellar Astrophysics, Astrophysics - High Energy Astrophysical Phenomena},
         year = 2016,
        month = apr,
       volume = {820},
       number = {2},
          eid = {110},
        pages = {110},
          doi = {10.3847/0004-637X/820/2/110},
archivePrefix = {arXiv},
       eprint = {1603.02170},
 primaryClass = {astro-ph.SR},
       adsurl = {https://ui.adsabs.harvard.edu/abs/2016ApJ...820..110C},
      adsnote = {Provided by the SAO/NASA Astrophysics Data System}
}

@ARTICLE{lsjb20,
       author = {{Lower}, Marcus E. and {Shannon}, Ryan M. and {Johnston}, Simon and
         {Bailes}, Matthew},
        title = "{Spectropolarimetric properties of Swift J1818.0$-$1607: a 1.4 s radio magnetar}",
      journal = {arXiv e-prints},
     keywords = {Astrophysics - High Energy Astrophysical Phenomena},
         year = 2020,
        month = apr,
          eid = {arXiv:2004.11522},
        pages = {arXiv:2004.11522},
archivePrefix = {arXiv},
       eprint = {2004.11522},
 primaryClass = {astro-ph.HE},
       adsurl = {https://ui.adsabs.harvard.edu/abs/2020arXiv200411522L},
      adsnote = {Provided by the SAO/NASA Astrophysics Data System}
}

@ARTICLE{kb17,
       author = {{Kaspi}, Victoria M. and {Beloborodov}, Andrei M.},
        title = "{Magnetars}",
      journal = {\araa},
     keywords = {Astrophysics - High Energy Astrophysical Phenomena},
         year = 2017,
        month = aug,
       volume = {55},
       number = {1},
        pages = {261-301},
          doi = {10.1146/annurev-astro-081915-023329},
archivePrefix = {arXiv},
       eprint = {1703.00068},
 primaryClass = {astro-ph.HE},
       adsurl = {https://ui.adsabs.harvard.edu/abs/2017ARA&A..55..261K},
      adsnote = {Provided by the SAO/NASA Astrophysics Data System}
}

@ARTICLE{ccc+20,
       author = {{Champion}, David and {Cognard}, Ismael and {Cruces}, Marilyn and {Desvignes}, Gregory and {Jankowski}, Fabian and {Karuppusamy}, Ramesh and {Keith}, Michael J. and {Kouveliotou}, Chryssa and {Kramer}, Michael and {Liu}, Kuo and {Lyne}, Andrew G. and {Mickaliger}, Mitchell B. and {O'Connor}, Brendan and {Parthasarathy}, Aditya and {Porayko}, Nataliya and {Rajwade}, Kaustubh and {Stappers}, Ben W. and {Torne}, Pablo and {van der Horst}, Alexander J. and {Weltevrede}, Patrick},
        title = "{High-cadence observations and variable spin behaviour of magnetar Swift J1818.0-1607 after its outburst}",
      journal = {\mnras},
     keywords = {polarization, radiation mechanisms: non-thermal, stars: magnetars, stars: pulsars: individual: Swift J1818.0-1607, Astrophysics - High Energy Astrophysical Phenomena, Astrophysics - Astrophysics of Galaxies},
         year = 2020,
        month = nov,
       volume = {498},
       number = {4},
        pages = {6044-6056},
          doi = {10.1093/mnras/staa2764},
archivePrefix = {arXiv},
       eprint = {2009.03568},
 primaryClass = {astro-ph.HE},
       adsurl = {https://ui.adsabs.harvard.edu/abs/2020{\mnras}.498.6044C},
      adsnote = {Provided by the SAO/NASA Astrophysics Data System}
}

@ARTICLE{crh+07,
       author = {{Camilo}, F. and {Ransom}, S.~M. and {Halpern}, J.~P. and {Reynolds}, J.},
        title = "{1E 1547.0-5408: A Radio-emitting Magnetar with a Rotation Period of 2 Seconds}",
      journal = {ApJL},
     keywords = {ISM: individual (G327.24-0.13), pulsars: individual (1E 1547.0-5408), pulsars: individual (PSR J1550-5418), pulsars: individual (XTE J1810-197), Stars: Neutron, Astrophysics},
         year = 2007,
        month = sep,
       volume = {666},
       number = {2},
        pages = {L93-L96},
          doi = {10.1086/521826},
archivePrefix = {arXiv},
       eprint = {0708.0002},
 primaryClass = {astro-ph},
       adsurl = {https://ui.adsabs.harvard.edu/abs/2007ApJ...666L..93C},
      adsnote = {Provided by the SAO/NASA Astrophysics Data System}
}

@ARTICLE{crp+07,
       author = {{Camilo}, F. and {Ransom}, S.~M. and {Pe{\~n}alver}, J. and {Karastergiou}, A. and {van Kerkwijk}, M.~H. and {Durant}, M. and {Halpern}, J.~P. and {Reynolds}, J. and {Thum}, C. and {Helfand}, D.~J. and {Zimmerman}, N. and {Cognard}, I.},
        title = "{The Variable Radio-to-X-Ray Spectrum of the Magnetar XTE J1810-197}",
      journal = {\apj},
     keywords = {Stars: Pulsars: Individual: Alphanumeric: XTE J1810-197, Stars: Neutron, Astrophysics},
         year = 2007,
        month = nov,
       volume = {669},
       number = {1},
        pages = {561-569},
          doi = {10.1086/521548},
archivePrefix = {arXiv},
       eprint = {0705.4095},
 primaryClass = {astro-ph},
       adsurl = {https://ui.adsabs.harvard.edu/abs/2007ApJ...669..561C},
      adsnote = {Provided by the SAO/NASA Astrophysics Data System}
}

@ARTICLE{lbb+10,
       author = {{Levin}, Lina and {Bailes}, Matthew and {Bates}, Samuel and {Bhat}, N.~D. Ramesh and {Burgay}, Marta and {Burke-Spolaor}, Sarah and {D'Amico}, Nichi and {Johnston}, Simon and {Keith}, Michael and {Kramer}, Michael and {Milia}, Sabrina and {Possenti}, Andrea and {Rea}, Nanda and {Stappers}, Ben and {van Straten}, Willem},
        title = "{A Radio-loud Magnetar in X-ray Quiescence}",
      journal = {ApJL},
     keywords = {pulsars: individual: 1E 1547.0─5408 PSR J1622─4950 XTE J1810─197, stars: magnetars, stars: neutron, Astrophysics - High Energy Astrophysical Phenomena},
         year = 2010,
        month = sep,
       volume = {721},
       number = {1},
        pages = {L33-L37},
          doi = {10.1088/2041-8205/721/1/L33},
archivePrefix = {arXiv},
       eprint = {1007.1052},
 primaryClass = {astro-ph.HE},
       adsurl = {https://ui.adsabs.harvard.edu/abs/2010ApJ...721L..33L},
      adsnote = {Provided by the SAO/NASA Astrophysics Data System}
}

@ARTICLE{cab+20,
       author = {{CHIME/FRB Collaboration} and {Andersen}, B.~C. and {Bandura}, K.~M. and {Bhardwaj}, M. and {Bij}, A. and {Boyce}, M.~M. and {Boyle}, P.~J. and {Brar}, C. and {Cassanelli}, T. and {Chawla}, P. and {Chen}, T. and {Cliche}, J. -F. and {Cook}, A. and {Cubranic}, D. and {Curtin}, A.~P. and {Denman}, N.~T. and {Dobbs}, M. and {Dong}, F.~Q. and {Fandino}, M. and {Fonseca}, E. and {Gaensler}, B.~M. and {Giri}, U. and {Good}, D.~C. and {Halpern}, M. and {Hill}, A.~S. and {Hinshaw}, G.~F. and {H{\"o}fer}, C. and {Josephy}, A. and {Kania}, J.~W. and {Kaspi}, V.~M. and {Landecker}, T.~L. and {Leung}, C. and {Li}, D.~Z. and {Lin}, H. -H. and {Masui}, K.~W. and {McKinven}, R. and {Mena-Parra}, J. and {Merryfield}, M. and {Meyers}, B.~W. and {Michilli}, D. and {Milutinovic}, N. and {Mirhosseini}, A. and {M{\"u}nchmeyer}, M. and {Naidu}, A. and {Newburgh}, L.~B. and {Ng}, C. and {Patel}, C. and {Pen}, U. -L. and {Pinsonneault-Marotte}, T. and {Pleunis}, Z. and {Quine}, B.~M. and {Rafiei-Ravandi}, M. and {Rahman}, M. and {Ransom}, S.~M. and {Renard}, A. and {Sanghavi}, P. and {Scholz}, P. and {Shaw}, J.~R. and {Shin}, K. and {Siegel}, S.~R. and {Singh}, S. and {Smegal}, R.~J. and {Smith}, K.~M. and {Stairs}, I.~H. and {Tan}, C.~M. and {Tendulkar}, S.~P. and {Tretyakov}, I. and {Vanderlinde}, K. and {Wang}, H. and {Wulf}, D. and {Zwaniga}, A.~V.},
        title = "{A bright millisecond-duration radio burst from a Galactic magnetar}",
      journal = {\nat},
     keywords = {Astrophysics - High Energy Astrophysical Phenomena},
         year = 2020,
        month = nov,
       volume = {587},
       number = {7832},
        pages = {54-58},
          doi = {10.1038/s41586-020-2863-y},
archivePrefix = {arXiv},
       eprint = {2005.10324},
 primaryClass = {astro-ph.HE},
       adsurl = {https://ui.adsabs.harvard.edu/abs/2020Natur.587...54C},
      adsnote = {Provided by the SAO/NASA Astrophysics Data System}
}

@ARTICLE{brb+20,
       author = {{Bochenek}, C.~D. and {Ravi}, V. and {Belov}, K.~V. and {Hallinan}, G. and {Kocz}, J. and {Kulkarni}, S.~R. and {McKenna}, D.~L.},
        title = "{A fast radio burst associated with a Galactic magnetar}",
      journal = {\nat},
     keywords = {Astrophysics - High Energy Astrophysical Phenomena},
         year = 2020,
        month = nov,
       volume = {587},
       number = {7832},
        pages = {59-62},
          doi = {10.1038/s41586-020-2872-x},
archivePrefix = {arXiv},
       eprint = {2005.10828},
 primaryClass = {astro-ph.HE},
       adsurl = {https://ui.adsabs.harvard.edu/abs/2020Natur.587...59B},
      adsnote = {Provided by the SAO/NASA Astrophysics Data System}
}

@ARTICLE{mpp+20,
       author = {{Majid}, Walid A. and {Pearlman}, Aaron B. and {Prince}, Thomas A. and {Naudet}, Charles J. and {Bansal}, Karishma and {Bansal}, Karishma and {Bansal}, Karishma},
        title = "{Significant Flattening of Swift J1818.0-1607's Spectral Index via Dual Radio Frequency Observations with the Deep Space Network}",
      journal = {The Astronomer's Telegram},
     keywords = {Neutron Star, Soft Gamma-ray Repeater, Star, Transient, Pulsar, Magnetar},
         year = 2020,
        month = jul,
       volume = {13898},
        pages = {1},
       adsurl = {https://ui.adsabs.harvard.edu/abs/2020ATel13898....1M},
      adsnote = {Provided by the SAO/NASA Astrophysics Data System}
}

@ARTICLE{lkc+20,
       author = {{Liu}, Kuo and {Karuppusamy}, Ramesh and {Cognard}, Ismael and {Desvignes}, Gregory and {Kramer}, Michael and {Lyne}, Andrew and {Rajwade}, Kaustubh and {Stappers}, Ben and {Torne}, Pablo},
        title = "{Polarimetric detection of the magnetar Swift J1818.0-1607 from 4 to 22 GHz with the Effelsberg 100-m Telescope}",
      journal = {The Astronomer's Telegram},
     keywords = {Neutron Star, Pulsar, Magnetar},
         year = 2020,
        month = sep,
       volume = {13997},
        pages = {1},
       adsurl = {https://ui.adsabs.harvard.edu/abs/2020ATel13997....1L},
      adsnote = {Provided by the SAO/NASA Astrophysics Data System}
}

@ARTICLE{ljs+20,
       author = {{Lower}, M.~E. and {Johnston}, S. and {Shannon}, R.~M. and {Bailes}, M. and {Camilo}, F.},
        title = "{The dynamic magnetosphere of Swift J1818.0-1607}",
      journal = {\mnras},
     keywords = {stars: magnetars, stars: neutron, pulsars: individual: PSR J1818-1607, Astrophysics - High Energy Astrophysical Phenomena},
         year = 2020,
        month = dec,
          doi = {10.1093/mnras/staa3789},
archivePrefix = {arXiv},
       eprint = {2011.12463},
 primaryClass = {astro-ph.HE},
       adsurl = {https://ui.adsabs.harvard.edu/abs/2020{\mnras}.tmp.3624L},
      adsnote = {Provided by the SAO/NASA Astrophysics Data System}
}

@ARTICLE{tww+21,
       author = {{Tong}, H. and {Wang}, P.~F. and {Wang}, H.~G. and {Yan}, Z.},
        title = "{Rotating vector model for magnetars}",
      journal = {\mnras},
     keywords = {Astrophysics - High Energy Astrophysical Phenomena},
         year = 2021,
        month = jan,
          doi = {10.1093/mnras/stab108},
archivePrefix = {arXiv},
       eprint = {2101.04504},
 primaryClass = {astro-ph.HE},
       adsurl = {https://ui.adsabs.harvard.edu/abs/2021{\mnras}.tmp..172T},
      adsnote = {Provided by the SAO/NASA Astrophysics Data System}
}

@ARTICLE{tlc+20,
       author = {{Torne}, Pablo and {Liu}, Kuo and {Cognard}, Ismael and {Desvignes}, Gregory and {Karuppusamy}, Ramesh and {Kramer}, Michael and {Paubert}, Gabriel and {Lyne}, Andrew and {Rajwade}, Kaustubh and {Stappers}, Ben and {Eatough}, Ralph and {Sanchez}, Salvador and {Macias-Perez}, Juan and {Ladjelate}, Bilal and {Berta}, Stefano and {Sanchez-Portal}, Miguel and {Navarro}, Santiago and {Bongiovanni}, Angel and {Kramer}, Carsten and {Schuster}, Karl},
        title = "{IRAM 30m telescope detection of the magnetar Swift J1818.0-1607 between 86 and 154 GHz}",
      journal = {The Astronomer's Telegram},
     keywords = {Neutron Star, Soft Gamma-ray Repeater, Transient, Pulsar, Young Stellar Object, Magnetar},
         year = 2020,
        month = sep,
       volume = {14001},
        pages = {1},
       adsurl = {https://ui.adsabs.harvard.edu/abs/2020ATel14001....1T},
      adsnote = {Provided by the SAO/NASA Astrophysics Data System}
}

@ARTICLE{dk14,
       author = {{Dib}, Rim and {Kaspi}, Victoria M.},
        title = "{16 yr of RXTE Monitoring of Five Anomalous X-Ray Pulsars}",
      journal = {\apj},
     keywords = {pulsars: individual: 1E 1841─045 RXS J170849.0─400910 1E 2259+586 4U 0142+61 1E 1048.1─5937, stars: neutron, X-rays: stars, Astrophysics - High Energy Astrophysical Phenomena},
         year = 2014,
        month = mar,
       volume = {784},
       number = {1},
          eid = {37},
        pages = {37},
          doi = {10.1088/0004-637X/784/1/37},
archivePrefix = {arXiv},
       eprint = {1401.3085},
 primaryClass = {astro-ph.HE},
       adsurl = {https://ui.adsabs.harvard.edu/abs/2014ApJ...784...37D},
      adsnote = {Provided by the SAO/NASA Astrophysics Data System}
}

@ARTICLE{ibr+21,
       author = {{Israel}, G.~L. and {Burgay}, M. and {Rea}, N. and {Esposito}, P. and {Possenti}, A. and {Dall'Osso}, S. and {Stella}, L. and {Pilia}, M. and {Tiengo}, A. and {Ridnaia}, A. and {Lien}, A.~Y. and {Frederiks}, D.~D. and {Bernardini}, F.},
        title = "{X-Ray and Radio Bursts from the Magnetar 1E 1547.0-5408}",
      journal = {\apj},
     keywords = {X-ray bursts, Magnetars, Radio transient sources, Radio bursts, Pulsars, Neutron stars, 1814, 992, 2008, 1339, 1306, 1108, Astrophysics - High Energy Astrophysical Phenomena},
         year = 2021,
        month = jan,
       volume = {907},
       number = {1},
          eid = {7},
        pages = {7},
          doi = {10.3847/1538-4357/abca95},
archivePrefix = {arXiv},
       eprint = {2011.06607},
 primaryClass = {astro-ph.HE},
       adsurl = {https://ui.adsabs.harvard.edu/abs/2021ApJ...907....7I},
      adsnote = {Provided by the SAO/NASA Astrophysics Data System}
}

@ARTICLE{Astropy2013,
   author = {{Astropy Collaboration} and {Robitaille}, T.~P. and {Tollerud}, E.~J. and
	{Greenfield}, P. and {Droettboom}, M. and {Bray}, E. and {Aldcroft}, T. and
	{Davis}, M. and {Ginsburg}, A. and {Price-Whelan}, A.~M. and
	{Kerzendorf}, W.~E. and {Conley}, A. and {Crighton}, N. and
	{Barbary}, K. and {Muna}, D. and {Ferguson}, H. and {Grollier}, F. and
	{Parikh}, M.~M. and {Nair}, P.~H. and {Unther}, H.~M. and {Deil}, C. and
	{Woillez}, J. and {Conseil}, S. and {Kramer}, R. and {Turner}, J.~E.~H. and
	{Singer}, L. and {Fox}, R. and {Weaver}, B.~A. and {Zabalza}, V. and
	{Edwards}, Z.~I. and {Azalee Bostroem}, K. and {Burke}, D.~J. and
	{Casey}, A.~R. and {Crawford}, S.~M. and {Dencheva}, N. and
	{Ely}, J. and {Jenness}, T. and {Labrie}, K. and {Lim}, P.~L. and
	{Pierfederici}, F. and {Pontzen}, A. and {Ptak}, A. and {Refsdal}, B. and
	{Servillat}, M. and {Streicher}, O.},
    title = "{Astropy: A community Python package for astronomy}",
  journal = {\aa},
archivePrefix = "arXiv",
   eprint = {1307.6212},
 primaryClass = "astro-ph.IM",
 keywords = {methods: data analysis, methods: miscellaneous, virtual observatory tools},
     year = 2013,
    month = oct,
   volume = 558,
      eid = {A33},
    pages = {A33},
      doi = {10.1051/0004-6361/201322068},
   adsurl = {http://adsabs.harvard.edu/abs/2013A
  adsnote = {Provided by the SAO/NASA Astrophysics Data System}
}

@ARTICLE{mpr+17,
   author = {{Mignani}, R.~P. and {Paladino}, R. and {Rudak}, B. and {Zajczyk}, A. and
	{Corongiu}, A. and {de Luca}, A. and {Hummel}, W. and {Possenti}, A. and
	{Geppert}, U. and {Burgay}, M. and {Marconi}, G.},
    title = "{The First Detection of a Pulsar with ALMA}",
  journal = {ApJL},
archivePrefix = "arXiv",
   eprint = {1708.02828},
 primaryClass = "astro-ph.HE",
 keywords = {pulsars: individual: Vela pulsar},
     year = 2017,
    month = dec,
   volume = 851,
      eid = {L10},
    pages = {L10},
      doi = {10.3847/2041-8213/aa9c3e},
   adsurl = {http://adsabs.harvard.edu/abs/2017ApJ...851L..10M},
  adsnote = {Provided by the SAO/NASA Astrophysics Data System}
}

@ARTICLE{pwk16,
   author = {{Psaltis}, D. and {Wex}, N. and {Kramer}, M.},
    title = "{A Quantitative Test of the No-hair Theorem with Sgr A* Using Stars, Pulsars, and the Event Horizon Telescope}",
  journal = {\apj},
archivePrefix = "arXiv",
   eprint = {1510.00394},
 primaryClass = "astro-ph.HE",
 keywords = {black hole physics, Galaxy: center, gravitation, pulsars: general, stars: general},
     year = 2016,
    month = feb,
   volume = 818,
      eid = {121},
    pages = {121},
      doi = {10.3847/0004-637X/818/2/121},
   adsurl = {http://adsabs.harvard.edu/abs/2016ApJ...818..121P},
  adsnote = {Provided by the SAO/NASA Astrophysics Data System}
}

@ARTICLE{wcc+12,
   author = {{Wharton}, R.~S. and {Chatterjee}, S. and {Cordes}, J.~M. and
	{Deneva}, J.~S. and {Lazio}, T.~J.~W.},
    title = "{Multiwavelength Constraints on Pulsar Populations in the Galactic Center}",
  journal = {\apj},
archivePrefix = "arXiv",
   eprint = {1111.4216},
 primaryClass = "astro-ph.HE",
 keywords = {Galaxy: center, pulsars: general},
     year = 2012,
    month = jul,
   volume = 753,
      eid = {108},
    pages = {108},
      doi = {10.1088/0004-637X/753/2/108},
   adsurl = {http://adsabs.harvard.edu/abs/2012ApJ...753..108W},
  adsnote = {Provided by the SAO/NASA Astrophysics Data System}
}

@ARTICLE{gfk+17,
   author = {{Goddi}, C. and {Falcke}, H. and {Kramer}, M. and {Rezzolla}, L. and
	{Brinkerink}, C. and {Bronzwaer}, T. and {Davelaar}, J.~R.~J. and
	{Deane}, R. and {de Laurentis}, M. and {Desvignes}, G. and {Eatough}, R.~P. and
	{Eisenhauer}, F. and {Fraga-Encinas}, R. and {Fromm}, C.~M. and
	{Gillessen}, S. and {Grenzebach}, A. and {Issaoun}, S. and {Jan{\ss}en}, M. and
	{Konoplya}, R. and {Krichbaum}, T.~P. and {Laing}, R. and {Liu}, K. and
	{Lu}, R.-S. and {Mizuno}, Y. and {Moscibrodzka}, M. and {M{\"u}ller}, C. and
	{Olivares}, H. and {Pfuhl}, O. and {Porth}, O. and {Roelofs}, F. and
	{Ros}, E. and {Schuster}, K. and {Tilanus}, R. and {Torne}, P. and
	{van Bemmel}, I. and {van Langevelde}, H.~J. and {Wex}, N. and
	{Younsi}, Z. and {Zhidenko}, A.},
    title = "{BlackHoleCam: Fundamental physics of the galactic center}",
  journal = {International Journal of Modern Physics D},
archivePrefix = "arXiv",
   eprint = {1606.08879},
 primaryClass = "astro-ph.HE",
 keywords = {General relativity, black holes, tests of general relativity, pulsars, high energy astrophysical phenomena},
     year = 2017,
   volume = 26,
      eid = {1730001-239},
    pages = {1730001-239},
      doi = {10.1142/S0218271817300014},
   adsurl = {http://adsabs.harvard.edu/abs/2017IJMPD..2630001G},
  adsnote = {Provided by the SAO/NASA Astrophysics Data System}
}


@phdthesis{tor16,
  author = {Torne, P.},
  school = {Bonn University},
  year = {2016}
}

@ARTICLE{van06,
   author = {{van Straten}, W.},
    title = "{Radio Astronomical Polarimetry and High-Precision Pulsar Timing}",
  journal = {\apj},
   eprint = {astro-ph/0510334},
 keywords = {Methods: Data Analysis, Polarization, Stars: Pulsars: General, Techniques: Polarimetric},
     year = 2006,
    month = may,
   volume = 642,
    pages = {1004-1011},
      doi = {10.1086/501001},
   adsurl = {http://adsabs.harvard.edu/abs/2006ApJ...642.1004V},
  adsnote = {Provided by the SAO/NASA Astrophysics Data System}
}

@ARTICLE{mcd+18,
   author = {{Matthews}, L.~D. and {Crew}, G.~B. and {Doeleman}, S.~S. and
	{Lacasse}, R. and {Saez}, A.~F. and {Alef}, W. and {Akiyama}, K. and
	{Amestica}, R. and {Anderson}, J.~M. and {Barkats}, D.~A. and
	{Baudry}, A. and {Brogui{\`e}re}, D. and {Escoffier}, R. and
	{Fish}, V.~L. and {Greenberg}, J. and {Hecht}, M.~H. and {Hiriart}, R. and
	{Hirota}, A. and {Honma}, M. and {Ho}, P.~T.~P. and {Impellizzeri}, C.~M.~V. and
	{Inoue}, M. and {Kohno}, Y. and {Lopez}, B. and {Mart{\'{\i}}-Vidal}, I. and
	{Messias}, H. and {Meyer-Zhao}, Z. and {Mora-Klein}, M. and
	{Nagar}, N.~M. and {Nishioka}, H. and {Oyama}, T. and {Pankratius}, V. and
	{Perez}, J. and {Phillips}, N. and {Pradel}, N. and {Rottmann}, H. and
	{Roy}, A.~L. and {Ruszczyk}, C.~A. and {Shillue}, B. and {Suzuki}, S. and
	{Treacy}, R.},
    title = "{The ALMA Phasing System: A Beamforming Capability for Ultra-high-resolution Science at (Sub)Millimeter Wavelengths}",
  journal = {\pasp},
archivePrefix = "arXiv",
   eprint = {1711.06770},
 primaryClass = "astro-ph.IM",
     year = 2018,
    month = jan,
   volume = 130,
   number = 1,
    pages = {015002},
      doi = {10.1088/1538-3873/aa9c3d},
   adsurl = {http://adsabs.harvard.edu/abs/2018PASP..130a5002M},
  adsnote = {Provided by the SAO/NASA Astrophysics Data System}
}

@ARTICLE{dwr+08,
   author = {{Doeleman}, S.~S. and {Weintroub}, J. and {Rogers}, A.~E.~E. and
	{Plambeck}, R. and {Freund}, R. and {Tilanus}, R.~P.~J. and
	{Friberg}, P. and {Ziurys}, L.~M. and {Moran}, J.~M. and {Corey}, B. and
	{Young}, K.~H. and {Smythe}, D.~L. and {Titus}, M. and {Marrone}, D.~P. and
	{Cappallo}, R.~J. and {Bock}, D.~C.-J. and {Bower}, G.~C. and
	{Chamberlin}, R. and {Davis}, G.~R. and {Krichbaum}, T.~P. and
	{Lamb}, J. and {Maness}, H. and {Niell}, A.~E. and {Roy}, A. and
	{Strittmatter}, P. and {Werthimer}, D. and {Whitney}, A.~R. and
	{Woody}, D.},
    title = "{Event-horizon-scale structure in the supermassive black hole candidate at the Galactic Centre}",
  journal = {\nat},
archivePrefix = "arXiv",
   eprint = {0809.2442},
     year = 2008,
    month = sep,
   volume = 455,
    pages = {78-80},
      doi = {10.1038/nature07245},
   adsurl = {http://adsabs.harvard.edu/abs/2008Natur.455...78D},
  adsnote = {Provided by the SAO/NASA Astrophysics Data System}
}

@ARTICLE{wbc+13,
   author = {{Whitney}, A.~R. and {Beaudoin}, C.~J. and {Cappallo}, R.~J. and
	{Corey}, B.~E. and {Crew}, G.~B. and {Doeleman}, S.~S. and {Lapsley}, D.~E. and
	{Hinton}, A.~A. and {McWhirter}, S.~R. and {Niell}, A.~E. and
	{Rogers}, A.~E.~E. and {Ruszczyk}, C.~A. and {Smythe}, D.~L. and
	{SooHoo}, J. and {Titus}, M.~A.},
    title = "{Demonstration of a 16 Gbps Station$^{-1}$ Broadband-RF VLBI System}",
  journal = {\pasp},
     year = 2013,
    month = feb,
   volume = 125,
    pages = {196},
      doi = {10.1086/669718},
   adsurl = {http://adsabs.harvard.edu/abs/2013PASP..125..196W},
  adsnote = {Provided by the SAO/NASA Astrophysics Data System}
}

@article{hsm04,
title = "{psrchive and psrfits: An Open Approach to Radio Pulsar Data Storage and Analysis}",
volume = 21,
doi = {10.1071/AS04022},
number = 3,
journal = {Publications of the Astronomical Society of Australia},
publisher = {Cambridge University Press},
author= {Hotan, A. W. and {van Straten}, W. and Manchester, R. N.},
year=2004,
pages={302–309}
}

@ARTICLE{gmm+19,
       author = {{Goddi}, C. and {Mart{\'\i}-Vidal}, I. and {Messias}, H. and
         {Crew}, G.~B. and {Herrero-Illana}, R. and {Impellizzeri}, V. and
         {Rottmann}, H. and {Wagner}, J. and {Fomalont}, E. and
         {Matthews}, L.~D.},
        title = "{Calibration of ALMA as a Phased Array. ALMA Observations During the 2017 VLBI Campaign}",
      journal = {\pasp},
     keywords = {Astrophysics - Instrumentation and Methods for Astrophysics},
         year = "2019",
        month = "Jul",
       volume = {131},
       number = {1001},
        pages = {075003},
          doi = {10.1088/1538-3873/ab136a},
archivePrefix = {arXiv},
       eprint = {1901.09987},
 primaryClass = {astro-ph.IM},
       adsurl = {https://ui.adsabs.harvard.edu/abs/2019PASP..131g5003G},
      adsnote = {Provided by the SAO/NASA Astrophysics Data System}
}

@ARTICLE{ddb+17,
       author = {{Dexter}, J. and {Deller}, A. and {Bower}, G.~C. and {Demorest}, P. and
         {Kramer}, M. and {Stappers}, B.~W. and {Lyne}, A.~G. and {Kerr}, M. and
         {Spitler}, L.~G. and {Psaltis}, D.},
        title = "{Locating the intense interstellar scattering towards the inner Galaxy}",
      journal = {\mnras},
     keywords = {scattering, pulsars: general, H \&lt;sc\&gt;II\&lt;/sc\&gt; regions, ISM: supernova remnants, Galaxy: centre, Astrophysics - Astrophysics of Galaxies},
         year = "2017",
        month = "Nov",
       volume = {471},
       number = {3},
        pages = {3563-3576},
          doi = {10.1093/mnras/stx1777},
archivePrefix = {arXiv},
       eprint = {1707.03842},
 primaryClass = {astro-ph.GA},
       adsurl = {https://ui.adsabs.harvard.edu/abs/2017{\mnras}.471.3563D},
      adsnote = {Provided by the SAO/NASA Astrophysics Data System}
}

@ARTICLE{eht+19,
       author = {{Event Horizon Telescope Collaboration} and {Akiyama}, Kazunori and
         {Alberdi}, Antxon and {Alef}, Walter and {Asada}, Keiichi and
         {Azulay}, Rebecca and {Baczko}, Anne-Kathrin and {Ball}, David and
         {Balokovi{\'c}}, Mislav and {Barrett}, John},
        title = "{First M87 Event Horizon Telescope Results. II. Array and Instrumentation}",
      journal = {ApJL},
     keywords = {black hole physics, galaxies: individual: M87, Galaxy: center, gravitational lensing: strong, instrumentation: interferometers, techniques: high angular resolution, Astrophysics - Instrumentation and Methods for Astrophysics, Astrophysics - Astrophysics of Galaxies, Astrophysics - High Energy Astrophysical Phenomena},
         year = "2019",
        month = "Apr",
       volume = {875},
       number = {1},
          eid = {L2},
        pages = {L2},
          doi = {10.3847/2041-8213/ab0c96},
archivePrefix = {arXiv},
       eprint = {1906.11239},
 primaryClass = {astro-ph.IM},
       adsurl = {https://ui.adsabs.harvard.edu/abs/2019ApJ...875L...2E},
      adsnote = {Provided by the SAO/NASA Astrophysics Data System}
}

@ARTICLE{sle+14,
       author = {{Spitler}, L.~G. and {Lee}, K.~J. and {Eatough}, R.~P. and {Kramer}, M. and
         {Karuppusamy}, R. and {Bassa}, C.~G. and {Cognard}, I. and
         {Desvignes}, G. and {Lyne}, A.~G. and {Stappers}, B.~W.},
        title = "{Pulse Broadening Measurements from the Galactic Center Pulsar J1745-2900}",
      journal = {ApJL},
     keywords = {Galaxy: center, pulsars: individual: J1745-2900, scattering, Astrophysics - High Energy Astrophysical Phenomena, Astrophysics - Astrophysics of Galaxies},
         year = "2014",
        month = "Jan",
       volume = {780},
       number = {1},
          eid = {L3},
        pages = {L3},
          doi = {10.1088/2041-8205/780/1/L3},
archivePrefix = {arXiv},
       eprint = {1309.4673},
 primaryClass = {astro-ph.HE},
       adsurl = {https://ui.adsabs.harvard.edu/abs/2014ApJ...780L...3S},
      adsnote = {Provided by the SAO/NASA Astrophysics Data System}
}

@ARTICLE{sbj+17,
       author = {{Smits}, R. and {Bassa}, C.~G. and {Janssen}, G.~H. and
         {Karuppusamy}, R. and {Kramer}, M. and {Lee}, K.~J. and {Liu}, K. and
         {McKee}, J. and {Perrodin}, D. and {Purver}, M. and {Sanidas}, S. and
         {Stappers}, B.~W. and {Zhu}, W.~W.},
        title = "{The beamformer and correlator for the Large European Array for Pulsars}",
      journal = {Astronomy and Computing},
     keywords = {Gravitational waves, Techniques, Interferometric, Pulsars, General, Astrophysics - Instrumentation and Methods for Astrophysics},
         year = "2017",
        month = "Apr",
       volume = {19},
        pages = {66-74},
          doi = {10.1016/j.ascom.2017.02.002},
archivePrefix = {arXiv},
       eprint = {1703.06438},
 primaryClass = {astro-ph.IM},
       adsurl = {https://ui.adsabs.harvard.edu/abs/2017A&C....19...66S},
      adsnote = {Provided by the SAO/NASA Astrophysics Data System}
}

@INPROCEEDINGS{tor2018,
       author = {{Torne}, Pablo},
        title = "{Pulsar observations at millimetre wavelengths}",
     keywords = {pulsars: general, radiation mechanisms: nonthermal, Galaxy: center, telescopes, Astrophysics - High Energy Astrophysical Phenomena, Astrophysics - Instrumentation and Methods for Astrophysics, 85-06},
    booktitle = {Pulsar Astrophysics the Next Fifty Years},
         year = "2018",
       editor = {{Weltevrede}, P. and {Perera}, B.~B.~P. and {Preston}, L.~L. and
         {Sanidas}, S.},
       series = {IAU Symposium},
       volume = {337},
        month = "Aug",
        pages = {92-95},
          doi = {10.1017/S1743921317009085},
archivePrefix = {arXiv},
       eprint = {1806.10617},
 primaryClass = {astro-ph.HE},
       adsurl = {https://ui.adsabs.harvard.edu/abs/2018IAUS..337...92T},
      adsnote = {Provided by the SAO/NASA Astrophysics Data System}
}

@ARTICLE{le17,
       author = {{Liu}, Kuo and {Eatough}, Ralph},
        title = "{Few and far between}",
      journal = {Nature Astronomy},
         year = "2017",
        month = "Dec",
       volume = {1},
        pages = {812-813},
          doi = {10.1038/s41550-017-0327-6},
       adsurl = {https://ui.adsabs.harvard.edu/abs/2017NatAs...1..812L},
      adsnote = {Provided by the SAO/NASA Astrophysics Data System}
}

@ARTICLE{ddl+20,
       author = {{Ding}, H. and {Deller}, A.~T. and {Lower}, M.~E. and {Flynn}, C. and
         {Chatterjee}, S. and {Brisken}, W. and {Hurley-Walker}, N. and
         {Camilo}, F. and {Sarkissian}, J. and {Gupta}, V.},
        title = "{A magnetar parallax}",
      journal = {\mnras},
     keywords = {parallaxes, proper motions, pulsars: individual: XTE J1810-197, radio continuum: transients, Astrophysics - Instrumentation and Methods for Astrophysics, Astrophysics - High Energy Astrophysical Phenomena},
         year = 2020,
        month = aug,
       volume = {498},
       number = {3},
        pages = {3736-3743},
          doi = {10.1093/mnras/staa2531},
archivePrefix = {arXiv},
       eprint = {2008.06438},
 primaryClass = {astro-ph.IM},
       adsurl = {https://ui.adsabs.harvard.edu/abs/2020{\mnras}.498.3736D},
      adsnote = {Provided by the SAO/NASA Astrophysics Data System}
}

@ARTICLE{aab+19,
       author = {{Gravity Collaboration} and {Abuter}, R. and {Amorim}, A. and
         {Baub{\"o}ck}, M. and {Berger}, J.~P. and {Bonnet}, H. and {Brand
        ner}, W. and {Cl{\'e}net}, Y. and {Coud{\'e} Du Foresto}, V. and
         {de Zeeuw}, P.~T. and {Dexter}, J. and {Duvert}, G. and {Eckart}, A. and
         {Eisenhauer}, F. and {F{\"o}rster Schreiber}, N.~M. and {Garcia}, P. and
         {Gao}, F. and {Gendron}, E. and {Genzel}, R. and {Gerhard}, O. and
         {Gillessen}, S. and {Habibi}, M. and {Haubois}, X. and {Henning}, T. and
         {Hippler}, S. and {Horrobin}, M. and {Jim{\'e}nez-Rosales}, A. and
         {Jocou}, L. and {Kervella}, P. and {Lacour}, S. and
         {Lapeyr{\`e}re}, V. and {Le Bouquin}, J. -B. and {L{\'e}na}, P. and
         {Ott}, T. and {Paumard}, T. and {Perraut}, K. and {Perrin}, G. and
         {Pfuhl}, O. and {Rabien}, S. and {Rodriguez Coira}, G. and
         {Rousset}, G. and {Scheithauer}, S. and {Sternberg}, A. and
         {Straub}, O. and {Straubmeier}, C. and {Sturm}, E. and
         {Tacconi}, L.~J. and {Vincent}, F. and {von Fellenberg}, S. and
         {Waisberg}, I. and {Widmann}, F. and {Wieprecht}, E. and
         {Wiezorrek}, E. and {Woillez}, J. and {Yazici}, S.},
        title = "{A geometric distance measurement to the Galactic center black hole with 0.3\% uncertainty}",
      journal = {\aa},
     keywords = {black hole physics, astrometry, Galaxy: nucleus, Astrophysics - Astrophysics of Galaxies},
         year = 2019,
        month = may,
       volume = {625},
          eid = {L10},
        pages = {L10},
          doi = {10.1051/0004-6361/201935656},
archivePrefix = {arXiv},
       eprint = {1904.05721},
 primaryClass = {astro-ph.GA},
       adsurl = {https://ui.adsabs.harvard.edu/abs/2019A&A...625L..10G},
      adsnote = {Provided by the SAO/NASA Astrophysics Data System}
}

@ARTICLE{lyw+19,
       author = {{Liu}, Kuo and {Young}, Andr{\'e} and {Wharton}, Robert and
         {Blackburn}, Lindy and {Cappallo}, Roger and {Chatterjee}, Shami and
         {Cordes}, James M. and {Crew}, Geoffrey B. and {Desvignes}, Gregory and
         {Doeleman}, Sheperd S. and {Eatough}, Ralph P. and {Falcke}, Heino and
         {Goddi}, Ciriaco and {Johnson}, Michael D. and {Johnston}, Simon and
         {Karuppusamy}, Ramesh and {Kramer}, Michael and {Matthews}, Lynn D. and
         {Ransom}, Scott M. and {Rezzolla}, Luciano and {Rottmann}, Helge and
         {Tilanus}, Remo P.~J. and {Torne}, Pablo},
        title = "{Detection of Pulses from the Vela Pulsar at Millimeter Wavelengths with Phased ALMA}",
      journal = {ApJL},
     keywords = {Interferometry, Pulsars, Millimeter astronomy, 808, 1306, 1061, Astrophysics - Astrophysics of Galaxies, Astrophysics - Instrumentation and Methods for Astrophysics},
         year = 2019,
        month = nov,
       volume = {885},
       number = {1},
          eid = {L10},
        pages = {L10},
          doi = {10.3847/2041-8213/ab4da8},
archivePrefix = {arXiv},
       eprint = {1910.07974},
 primaryClass = {astro-ph.GA},
       adsurl = {https://ui.adsabs.harvard.edu/abs/2019ApJ...885L..10L},
      adsnote = {Provided by the SAO/NASA Astrophysics Data System}
}

@ARTICLE{ccc+20,
       author = {{Champion}, David and {Cognard}, Ismael and {Cruces}, Marilyn and
         {Desvignes}, Gregory and {Jankowski}, Fabian and {Karuppusamy}, Ramesh and
         {Keith}, Michael J. and {Kouveliotou}, Chryssa and {Kramer}, Michael and
         {Liu}, Kuo and {Lyne}, Andrew G. and {Mickaliger}, Mitchell B. and
         {O'Connor}, Brendan and {Parthasarathy}, Aditya and
         {Porayko}, Nataliya and {Rajwade}, Kaustubh and {Stappers}, Ben W. and
         {Torne}, Pablo and {van der Horst}, Alexander J. and
         {Weltevrede}, Patrick},
        title = "{High-cadence observations and variable spin behaviour of magnetar Swift J1818.0-1607 after its outburst}",
      journal = {\mnras},
     keywords = {stars: magnetars, stars: pulsars: individual: Swift J1818.0-1607, polarization, radiation mechanisms: non-thermal, Astrophysics - High Energy Astrophysical Phenomena, Astrophysics - Astrophysics of Galaxies},
         year = 2020,
        month = sep,
          doi = {10.1093/mnras/staa2764},
archivePrefix = {arXiv},
       eprint = {2009.03568},
 primaryClass = {astro-ph.HE},
       adsurl = {https://ui.adsabs.harvard.edu/abs/2020{\mnras}.tmp.2597C},
      adsnote = {Provided by the SAO/NASA Astrophysics Data System}
}

@ARTICLE{klm+11,
       author = {{Kijak}, J. and {Lewandowski}, W. and {Maron}, O. and {Gupta}, Y. and
         {Jessner}, A.},
        title = "{Pulsars with gigahertz-peaked spectra}",
      journal = {\aa},
     keywords = {pulsars: general, ISM: general, Astrophysics - Galaxy Astrophysics},
         year = 2011,
        month = jul,
       volume = {531},
          eid = {A16},
        pages = {A16},
          doi = {10.1051/0004-6361/201014274},
archivePrefix = {arXiv},
       eprint = {1105.2065},
 primaryClass = {astro-ph.GA},
       adsurl = {https://ui.adsabs.harvard.edu/abs/2011A&A...531A..16K},
      adsnote = {Provided by the SAO/NASA Astrophysics Data System}
}

@ARTICLE{ar18,
       author = {{Andersen}, Bridget C. and {Ransom}, Scott M.},
        title = "{A Fourier Domain {\textquotedblleft}Jerk{\textquotedblright} Search for Binary Pulsars}",
      journal = {ApJL},
     keywords = {binaries: general, pulsars: general, pulsars: individual: J1748─2446am, stars: neutron, Astrophysics - High Energy Astrophysical Phenomena, Astrophysics - Instrumentation and Methods for Astrophysics},
         year = 2018,
        month = aug,
       volume = {863},
       number = {1},
          eid = {L13},
        pages = {L13},
          doi = {10.3847/2041-8213/aad59f},
archivePrefix = {arXiv},
       eprint = {1807.07900},
 primaryClass = {astro-ph.HE},
       adsurl = {https://ui.adsabs.harvard.edu/abs/2018ApJ...863L..13A},
      adsnote = {Provided by the SAO/NASA Astrophysics Data System}
}

@ARTICLE{wcc+19,
       author = {{Wharton}, R.~S. and {Chatterjee}, S. and {Cordes}, J.~M. and
         {Bower}, G.~C. and {Butler}, B.~J. and {Deller}, A.~T. and
         {Demorest}, P. and {Lazio}, T.~J.~W. and {Ransom}, S.~M.},
        title = "{VLA Observations of Single Pulses from the Galactic Center Magnetar}",
      journal = {\apj},
     keywords = {Galaxy: center, pulsars: individual: J1745{\ensuremath{-}}2900, Astrophysics - High Energy Astrophysical Phenomena},
         year = 2019,
        month = apr,
       volume = {875},
       number = {2},
          eid = {143},
        pages = {143},
          doi = {10.3847/1538-4357/ab100a},
archivePrefix = {arXiv},
       eprint = {1905.00632},
 primaryClass = {astro-ph.HE},
       adsurl = {https://ui.adsabs.harvard.edu/abs/2019ApJ...875..143W},
      adsnote = {Provided by the SAO/NASA Astrophysics Data System}
}

@ARTICLE{pmp+18,
       author = {{Pearlman}, Aaron B. and {Majid}, Walid A. and {Prince}, Thomas A. and
         {Kocz}, Jonathon and {Horiuchi}, Shinji},
        title = "{Pulse Morphology of the Galactic Center Magnetar PSR J1745-2900}",
      journal = {\apj},
     keywords = {Galaxy: center, pulsars: individual: PSR J1745─2900, scattering, stars: magnetars, stars: neutron, Astrophysics - High Energy Astrophysical Phenomena},
         year = 2018,
        month = oct,
       volume = {866},
       number = {2},
          eid = {160},
        pages = {160},
          doi = {10.3847/1538-4357/aade4d},
archivePrefix = {arXiv},
       eprint = {1809.02140},
 primaryClass = {astro-ph.HE},
       adsurl = {https://ui.adsabs.harvard.edu/abs/2018ApJ...866..160P},
      adsnote = {Provided by the SAO/NASA Astrophysics Data System}
}

@ARTICLE{ijb+19,
       author = {{Issaoun}, S. and {Johnson}, M.~D. and {Blackburn}, L. and
         {Brinkerink}, C.~D. and {Mo{\'s}cibrodzka}, M. and {Chael}, A. and
         {Goddi}, C. and {Mart{\'\i}-Vidal}, I. and {Wagner}, J. and
         {Doeleman}, S.~S. and {Falcke}, H. and {Krichbaum}, T.~P. and
         {Akiyama}, K. and {Bach}, U. and {Bouman}, K.~L. and {Bower}, G.~C. and
         {Broderick}, A. and {Cho}, I. and {Crew}, G. and {Dexter}, J. and
         {Fish}, V. and {Gold}, R. and {G{\'o}mez}, J.~L. and {Hada}, K. and
         {Hern{\'a}ndez-G{\'o}mez}, A. and {Jan{\ss}en}, M. and {Kino}, M. and
         {Kramer}, M. and {Loinard}, L. and {Lu}, R. -S. and {Markoff}, S. and
         {Marrone}, D.~P. and {Matthews}, L.~D. and {Moran}, J.~M. and
         {M{\"u}ller}, C. and {Roelofs}, F. and {Ros}, E. and {Rottmann}, H. and
         {Sanchez}, S. and {Tilanus}, R.~P.~J. and {de Vicente}, P. and
         {Wielgus}, M. and {Zensus}, J.~A. and {Zhao}, G. -Y.},
        title = "{The Size, Shape, and Scattering of Sagittarius A* at 86 GHz: First VLBI with ALMA}",
      journal = {\apj},
     keywords = {accretion, accretion disks, galaxies: individual: Sgr Aa, Galaxy: center, techniques: interferometric, Astrophysics - High Energy Astrophysical Phenomena},
         year = 2019,
        month = jan,
       volume = {871},
       number = {1},
          eid = {30},
        pages = {30},
          doi = {10.3847/1538-4357/aaf732},
archivePrefix = {arXiv},
       eprint = {1901.06226},
 primaryClass = {astro-ph.HE},
       adsurl = {https://ui.adsabs.harvard.edu/abs/2019ApJ...871...30I},
      adsnote = {Provided by the SAO/NASA Astrophysics Data System}
}

@ARTICLE{bdd+15,
       author = {{Bower}, Geoffrey C. and {Deller}, Adam and {Demorest}, Paul and
         {Brunthaler}, Andreas and {Falcke}, Heino and {Moscibrodzka}, Monika and
         {O'Leary}, Ryan M. and {Eatough}, Ralph P. and {Kramer}, Michael and
         {Lee}, K.~J. and {Spitler}, Laura and {Desvignes}, Gregory and
         {Rushton}, Anthony P. and {Doeleman}, Sheperd and {Reid}, Mark J.},
        title = "{The Proper Motion of the Galactic Center Pulsar Relative to Sagittarius A*}",
      journal = {\apj},
     keywords = {black hole physics, Galaxy: center, proper motions, pulsars: general, pulsars: individual: J1745─2900, Astrophysics - High Energy Astrophysical Phenomena},
         year = 2015,
        month = jan,
       volume = {798},
       number = {2},
          eid = {120},
        pages = {120},
          doi = {10.1088/0004-637X/798/2/120},
archivePrefix = {arXiv},
       eprint = {1411.0399},
 primaryClass = {astro-ph.HE},
       adsurl = {https://ui.adsabs.harvard.edu/abs/2015ApJ...798..120B},
      adsnote = {Provided by the SAO/NASA Astrophysics Data System}
}

@ARTICLE{lbh+15,
       author = {{Lazarus}, P. and {Brazier}, A. and {Hessels}, J.~W.~T. and
         {Karako-Argaman}, C. and {Kaspi}, V.~M. and {Lynch}, R. and
         {Madsen}, E. and {Patel}, C. and {Ransom}, S.~M. and {Scholz}, P. and
         {Swiggum}, J. and {Zhu}, W.~W. and {Allen}, B. and {Bogdanov}, S. and
         {Camilo}, F. and {Cardoso}, F. and {Chatterjee}, S. and
         {Cordes}, J.~M. and {Crawford}, F. and {Deneva}, J.~S. and
         {Ferdman}, R. and {Freire}, P.~C.~C. and {Jenet}, F.~A. and
         {Knispel}, B. and {Lee}, K.~J. and {van Leeuwen}, J. and
         {Lorimer}, D.~R. and {Lyne}, A.~G. and {McLaughlin}, M.~A. and
         {Siemens}, X. and {Spitler}, L.~G. and {Stairs}, I.~H. and
         {Stovall}, K. and {Venkataraman}, A.},
        title = "{Arecibo Pulsar Survey Using ALFA. IV. Mock Spectrometer Data Analysis, Survey Sensitivity, and the Discovery of 40 Pulsars}",
      journal = {\apj},
     keywords = {methods: data analysis, pulsars: general, Astrophysics - High Energy Astrophysical Phenomena},
         year = 2015,
        month = oct,
       volume = {812},
       number = {1},
          eid = {81},
        pages = {81},
          doi = {10.1088/0004-637X/812/1/81},
archivePrefix = {arXiv},
       eprint = {1504.02294},
 primaryClass = {astro-ph.HE},
       adsurl = {https://ui.adsabs.harvard.edu/abs/2015ApJ...812...81L},
      adsnote = {Provided by the SAO/NASA Astrophysics Data System}
}

@ARTICLE{eg96,
       author = {{Eckart}, A. and {Genzel}, R.},
        title = "{Observations of stellar proper motions near the Galactic Centre}",
      journal = {\nat},
         year = 1996,
        month = oct,
       volume = {383},
       number = {6599},
        pages = {415-417},
          doi = {10.1038/383415a0},
       adsurl = {https://ui.adsabs.harvard.edu/abs/1996Natur.383..415E},
      adsnote = {Provided by the SAO/NASA Astrophysics Data System}
}

@ARTICLE{gkm+98,
       author = {{Ghez}, A.~M. and {Klein}, B.~L. and {Morris}, M. and {Becklin}, E.~E.},
        title = "{High Proper-Motion Stars in the Vicinity of Sagittarius A*: Evidence for a Supermassive Black Hole at the Center of Our Galaxy}",
      journal = {\apj},
     keywords = {BLACK HOLE PHYSICS, GALAXY: CENTER, GALAXY: KINEMATICS AND DYNAMICS, INFRARED: STARS, STARS: KINEMATICS, TECHNIQUES: IMAGE PROCESSING, Black Hole Physics, Galaxy: Center, Galaxy: Kinematics and Dynamics, Infrared: Stars, Stars: Kinematics, Techniques: Image Processing, Astrophysics},
         year = 1998,
        month = dec,
       volume = {509},
       number = {2},
        pages = {678-686},
          doi = {10.1086/306528},
archivePrefix = {arXiv},
       eprint = {astro-ph/9807210},
 primaryClass = {astro-ph},
       adsurl = {https://ui.adsabs.harvard.edu/abs/1998ApJ...509..678G},
      adsnote = {Provided by the SAO/NASA Astrophysics Data System}
}

@ARTICLE{aaa+18,
       author = {{Gravity Collaboration} and {Abuter}, R. and {Amorim}, A. and
         {Baub{\"o}ck}, M. and {Berger}, J.~P. and {Bonnet}, H. and {Brand
        ner}, W. and {Cl{\'e}net}, Y. and {Coud{\'e} Du Foresto}, V. and
         {de Zeeuw}, P.~T. and {Deen}, C. and {Dexter}, J. and {Duvert}, G. and
         {Eckart}, A. and {Eisenhauer}, F. and {F{\"o}rster Schreiber}, N.~M. and
         {Garcia}, P. and {Gao}, F. and {Gendron}, E. and {Genzel}, R. and
         {Gillessen}, S. and {Guajardo}, P. and {Habibi}, M. and {Haubois}, X. and
         {Henning}, Th. and {Hippler}, S. and {Horrobin}, M. and {Huber}, A. and
         {Jim{\'e}nez-Rosales}, A. and {Jocou}, L. and {Kervella}, P. and
         {Lacour}, S. and {Lapeyr{\`e}re}, V. and {Lazareff}, B. and
         {Le Bouquin}, J. -B. and {L{\'e}na}, P. and {Lippa}, M. and {Ott}, T. and
         {Panduro}, J. and {Paumard}, T. and {Perraut}, K. and {Perrin}, G. and
         {Pfuhl}, O. and {Plewa}, P.~M. and {Rabien}, S. and
         {Rodr{\'\i}guez-Coira}, G. and {Rousset}, G. and {Sternberg}, A. and
         {Straub}, O. and {Straubmeier}, C. and {Sturm}, E. and
         {Tacconi}, L.~J. and {Vincent}, F. and {von Fellenberg}, S. and
         {Waisberg}, I. and {Widmann}, F. and {Wieprecht}, E. and
         {Wiezorrek}, E. and {Woillez}, J. and {Yazici}, S.},
        title = "{Detection of orbital motions near the last stable circular orbit of the massive black hole SgrA*}",
      journal = {\aa},
     keywords = {Galaxy: center, black hole physics, gravitation, relativistic processes, Astrophysics - Astrophysics of Galaxies},
         year = 2018,
        month = oct,
       volume = {618},
          eid = {L10},
        pages = {L10},
          doi = {10.1051/0004-6361/201834294},
archivePrefix = {arXiv},
       eprint = {1810.12641},
 primaryClass = {astro-ph.GA},
       adsurl = {https://ui.adsabs.harvard.edu/abs/2018A&A...618L..10G},
      adsnote = {Provided by the SAO/NASA Astrophysics Data System}
}

@ARTICLE{bsa+12,
       author = {{Burkert}, A. and {Schartmann}, M. and {Alig}, C. and {Gillessen}, S. and
         {Genzel}, R. and {Fritz}, T.~K. and {Eisenhauer}, F.},
        title = "{Physics of the Galactic Center Cloud G2, on Its Way toward the Supermassive Black Hole}",
      journal = {\apj},
     keywords = {galaxies: active, galaxies: ISM, Galaxy: center, Galaxy: nucleus, Astrophysics - Astrophysics of Galaxies},
         year = 2012,
        month = may,
       volume = {750},
       number = {1},
          eid = {58},
        pages = {58},
          doi = {10.1088/0004-637X/750/1/58},
archivePrefix = {arXiv},
       eprint = {1201.1414},
 primaryClass = {astro-ph.GA},
       adsurl = {https://ui.adsabs.harvard.edu/abs/2012ApJ...750...58B},
      adsnote = {Provided by the SAO/NASA Astrophysics Data System}
}

@ARTICLE{bmd+15,
       author = {{Bower}, Geoffrey C. and {Markoff}, Sera and {Dexter}, Jason and
         {Gurwell}, Mark A. and {Moran}, James M. and {Brunthaler}, Andreas and
         {Falcke}, Heino and {Fragile}, P. Chris and {Maitra}, Dipankar and
         {Marrone}, Dan and {Peck}, Alison and {Rushton}, Anthony and
         {Wright}, Melvyn C.~H.},
        title = "{Radio and Millimeter Monitoring of Sgr A*: Spectrum, Variability, and Constraints on the G2 Encounter}",
      journal = {\apj},
     keywords = {accretion, accretion disks, black hole physics, galaxies: active, galaxies: jets, Galaxy: center, Astrophysics - High Energy Astrophysical Phenomena},
         year = 2015,
        month = mar,
       volume = {802},
       number = {1},
          eid = {69},
        pages = {69},
          doi = {10.1088/0004-637X/802/1/69},
archivePrefix = {arXiv},
       eprint = {1502.06534},
 primaryClass = {astro-ph.HE},
       adsurl = {https://ui.adsabs.harvard.edu/abs/2015ApJ...802...69B},
      adsnote = {Provided by the SAO/NASA Astrophysics Data System}
}

@ARTICLE{pbj+16,
       author = {{Petroff}, E. and {Barr}, E.~D. and {Jameson}, A. and {Keane}, E.~F. and
         {Bailes}, M. and {Kramer}, M. and {Morello}, V. and {Tabbara}, D. and
         {van Straten}, W.},
        title = "{FRBCAT: The Fast Radio Burst Catalogue}",
      journal = {\pasa},
     keywords = {catalogs, methods: data analysis, telescopes, Astrophysics - High Energy Astrophysical Phenomena},
         year = 2016,
        month = sep,
       volume = {33},
          eid = {e045},
        pages = {e045},
          doi = {10.1017/pasa.2016.35},
archivePrefix = {arXiv},
       eprint = {1601.03547},
 primaryClass = {astro-ph.HE},
       adsurl = {https://ui.adsabs.harvard.edu/abs/2016PASA...33...45P},
      adsnote = {Provided by the SAO/NASA Astrophysics Data System}
}

@ARTICLE{lew+14,
       author = {{Liu}, K. and {Eatough}, R.~P. and {Wex}, N. and {Kramer}, M.},
        title = "{Pulsar-black hole binaries: prospects for new gravity tests with future radio telescopes}",
      journal = {\mnras},
     keywords = {gravitation, pulsars: general, Astrophysics - Astrophysics of Galaxies, Astrophysics - High Energy Astrophysical Phenomena},
         year = 2014,
        month = dec,
       volume = {445},
       number = {3},
        pages = {3115-3132},
          doi = {10.1093/mnras/stu1913},
archivePrefix = {arXiv},
       eprint = {1409.3882},
 primaryClass = {astro-ph.GA},
       adsurl = {https://ui.adsabs.harvard.edu/abs/2014{\mnras}.445.3115L},
      adsnote = {Provided by the SAO/NASA Astrophysics Data System}
}

@ARTICLE{cl14,
       author = {{Chennamangalam}, J. and {Lorimer}, D.~R.},
        title = "{The Galactic Centre pulsar population.}",
      journal = {\mnras},
     keywords = {methods: statistical, stars: neutron, pulsars: general, Galaxy: centre, Astrophysics - High Energy Astrophysical Phenomena, Astrophysics - Solar and Stellar Astrophysics},
         year = 2014,
        month = may,
       volume = {440},
        pages = {L86-L90},
          doi = {10.1093/mnrasl/slu025},
archivePrefix = {arXiv},
       eprint = {1311.4846},
 primaryClass = {astro-ph.HE},
       adsurl = {https://ui.adsabs.harvard.edu/abs/2014{\mnras}.440L..86C},
      adsnote = {Provided by the SAO/NASA Astrophysics Data System}
}

@ARTICLE{pge+15,
       author = {{Plewa}, P.~M. and {Gillessen}, S. and {Eisenhauer}, F. and {Ott}, T. and
         {Pfuhl}, O. and {George}, E. and {Dexter}, J. and {Habibi}, M. and
         {Genzel}, R. and {Reid}, M.~J. and {Menten}, K.~M.},
        title = "{Pinpointing the near-infrared location of Sgr A* by correcting optical distortion in the NACO imager}",
      journal = {\mnras},
     keywords = {methods: data analysis, techniques: high angular resolution, astrometry, Galaxy: centre, infrared: stars, Astrophysics - Astrophysics of Galaxies, Astrophysics - Instrumentation and Methods for Astrophysics},
         year = 2015,
        month = nov,
       volume = {453},
       number = {3},
        pages = {3234-3244},
          doi = {10.1093/mnras/stv1910},
archivePrefix = {arXiv},
       eprint = {1509.01941},
 primaryClass = {astro-ph.GA},
       adsurl = {https://ui.adsabs.harvard.edu/abs/2015{\mnras}.453.3234P},
      adsnote = {Provided by the SAO/NASA Astrophysics Data System}
}

@ARTICLE{egh+93,
       author = {{Eckart}, A. and {Genzel}, R. and {Hofmann}, R. and {Sams}, B.~J. and
         {Tacconi-Garman}, L.~E.},
        title = "{Near-Infrared 0 -8pt.15 Resolution Imaging of the Galactic Center}",
      journal = {ApJL},
     keywords = {Galactic Evolution, Galactic Nuclei, Infrared Imagery, Star Clusters, Star Formation, Autocorrelation, Black Holes (Astronomy), Data Reduction, Hot Stars, Main Sequence Stars, Near Infrared Radiation, Astrophysics, GALAXIES: NUCLEI, GALAXY: CENTER, INFRARED: STARS},
         year = 1993,
        month = apr,
       volume = {407},
        pages = {L77},
          doi = {10.1086/186810},
       adsurl = {https://ui.adsabs.harvard.edu/abs/1993ApJ...407L..77E},
      adsnote = {Provided by the SAO/NASA Astrophysics Data System}
}

@ARTICLE{gbd+03,
       author = {{Ghez}, A.~M. and {Becklin}, E. and {Duchjne}, G. and {Hornstein}, S. and
         {Morris}, M. and {Salim}, S. and {Tanner}, A.},
        title = "{Full Three Dimensional Orbits For Multiple Stars on Close Approaches to the Central Supermassive Black Hole}",
      journal = {Astronomische Nachrichten Supplement},
     keywords = {Astrophysics},
         year = 2003,
        month = sep,
       volume = {324},
       number = {1},
        pages = {527-533},
          doi = {10.1002/asna.200385103},
archivePrefix = {arXiv},
       eprint = {astro-ph/0303151},
 primaryClass = {astro-ph},
       adsurl = {https://ui.adsabs.harvard.edu/abs/2003ANS...324..527G},
      adsnote = {Provided by the SAO/NASA Astrophysics Data System}
}

@ARTICLE{ega+05,
       author = {{Eisenhauer}, F. and {Genzel}, R. and {Alexander}, T. and {Abuter}, R. and
         {Paumard}, T. and {Ott}, T. and {Gilbert}, A. and {Gillessen}, S. and
         {Horrobin}, M. and {Trippe}, S. and {Bonnet}, H. and {Dumas}, C. and
         {Hubin}, N. and {Kaufer}, A. and {Kissler-Patig}, M. and {Monnet}, G. and
         {Str{\"o}bele}, S. and {Szeifert}, T. and {Eckart}, A. and
         {Sch{\"o}del}, R. and {Zucker}, S.},
        title = "{SINFONI in the Galactic Center: Young Stars and Infrared Flares in the Central Light-Month}",
      journal = {\apj},
     keywords = {Black Hole Physics, Galaxy: Center, Galaxy: Structure, Infrared: Stars, Techniques: Spectroscopic, Astrophysics},
         year = 2005,
        month = jul,
       volume = {628},
       number = {1},
        pages = {246-259},
          doi = {10.1086/430667},
archivePrefix = {arXiv},
       eprint = {astro-ph/0502129},
 primaryClass = {astro-ph},
       adsurl = {https://ui.adsabs.harvard.edu/abs/2005ApJ...628..246E},
      adsnote = {Provided by the SAO/NASA Astrophysics Data System}
}

@ARTICLE{fma00,
       author = {{Falcke}, Heino and {Melia}, Fulvio and {Agol}, Eric},
        title = "{Viewing the Shadow of the Black Hole at the Galactic Center}",
      journal = {ApJL},
     keywords = {BLACK HOLE PHYSICS, GALAXIES: ACTIVE, GALAXY: CENTER, RELATIVITY, SUBMILLIMETER, TECHNIQUES: INTERFEROMETRIC, Black Hole Physics, Galaxies: Active, Galaxy: Center, Relativity, Submillimeter, Techniques: Interferometric, Astrophysics},
         year = 2000,
        month = jan,
       volume = {528},
       number = {1},
        pages = {L13-L16},
          doi = {10.1086/312423},
archivePrefix = {arXiv},
       eprint = {astro-ph/9912263},
 primaryClass = {astro-ph},
       adsurl = {https://ui.adsabs.harvard.edu/abs/2000ApJ...528L..13F},
      adsnote = {Provided by the SAO/NASA Astrophysics Data System}
}

@ARTICLE{bgs+16,
       author = {{Boehle}, A. and {Ghez}, A.~M. and {Sch{\"o}del}, R. and {Meyer}, L. and
         {Yelda}, S. and {Albers}, S. and {Martinez}, G.~D. and
         {Becklin}, E.~E. and {Do}, T. and {Lu}, J.~R. and {Matthews}, K. and
         {Morris}, M.~R. and {Sitarski}, B. and {Witzel}, G.},
        title = "{An Improved Distance and Mass Estimate for Sgr A* from a Multistar Orbit Analysis}",
      journal = {\apj},
     keywords = {astrometry, Galaxy: center, Galaxy: fundamental parameters, infrared: stars, quasars: supermassive black holes, techniques: high angular resolution, Astrophysics - Astrophysics of Galaxies},
         year = 2016,
        month = oct,
       volume = {830},
       number = {1},
          eid = {17},
        pages = {17},
          doi = {10.3847/0004-637X/830/1/17},
archivePrefix = {arXiv},
       eprint = {1607.05726},
 primaryClass = {astro-ph.GA},
       adsurl = {https://ui.adsabs.harvard.edu/abs/2016ApJ...830...17B},
      adsnote = {Provided by the SAO/NASA Astrophysics Data System}
}

@ARTICLE{gpe+17,
       author = {{Gillessen}, S. and {Plewa}, P.~M. and {Eisenhauer}, F. and {Sari}, R. and
         {Waisberg}, I. and {Habibi}, M. and {Pfuhl}, O. and {George}, E. and
         {Dexter}, J. and {von Fellenberg}, S. and {Ott}, T. and {Genzel}, R.},
        title = "{An Update on Monitoring Stellar Orbits in the Galactic Center}",
      journal = {\apj},
     keywords = {astrometry, black hole physics, Galaxy: center, Galaxy: fundamental parameters, techniques: high angular resolution, Astrophysics - Astrophysics of Galaxies},
         year = 2017,
        month = mar,
       volume = {837},
       number = {1},
          eid = {30},
        pages = {30},
          doi = {10.3847/1538-4357/aa5c41},
archivePrefix = {arXiv},
       eprint = {1611.09144},
 primaryClass = {astro-ph.GA},
       adsurl = {https://ui.adsabs.harvard.edu/abs/2017ApJ...837...30G},
      adsnote = {Provided by the SAO/NASA Astrophysics Data System}
}

@ARTICLE{fdb+11,
       author = {{Fish}, Vincent L. and {Doeleman}, Sheperd S. and
         {Beaudoin}, Christopher and {Blundell}, Ray and {Bolin}, David E. and
         {Bower}, Geoffrey C. and {Chamberlin}, Richard and {Freund}, Robert and
         {Friberg}, Per and {Gurwell}, Mark A. and {Honma}, Mareki and
         {Inoue}, Makoto and {Krichbaum}, Thomas P. and {Lamb}, James and
         {Marrone}, Daniel P. and {Moran}, James M. and {Oyama}, Tomoaki and
         {Plambeck}, Richard and {Primiani}, Rurik and {Rogers}, Alan E.~E. and
         {Smythe}, Daniel L. and {SooHoo}, Jason and {Strittmatter}, Peter and
         {Tilanus}, Remo P.~J. and {Titus}, Michael and {Weintroub}, Jonathan and
         {Wright}, Melvyn and {Woody}, David and {Young}, Ken H. and
         {Ziurys}, Lucy M.},
        title = "{1.3 mm Wavelength VLBI of Sagittarius A*: Detection of Time-variable Emission on Event Horizon Scales}",
      journal = {ApJL},
     keywords = {Galaxy: center, submillimeter: general, techniques: high angular resolution, techniques: interferometric, Astrophysics - Astrophysics of Galaxies},
         year = 2011,
        month = feb,
       volume = {727},
       number = {2},
          eid = {L36},
        pages = {L36},
          doi = {10.1088/2041-8205/727/2/L36},
archivePrefix = {arXiv},
       eprint = {1011.2472},
 primaryClass = {astro-ph.GA},
       adsurl = {https://ui.adsabs.harvard.edu/abs/2011ApJ...727L..36F},
      adsnote = {Provided by the SAO/NASA Astrophysics Data System}
}

@ARTICLE{lkr+18,
       author = {{Lu}, Ru-Sen and {Krichbaum}, Thomas P. and {Roy}, Alan L. and
         {Fish}, Vincent L. and {Doeleman}, Sheperd S. and
         {Johnson}, Michael D. and {Akiyama}, Kazunori and {Psaltis}, Dimitrios and
         {Alef}, Walter and {Asada}, Keiichi and {Beaudoin}, Christopher and
         {Bertarini}, Alessandra and {Blackburn}, Lindy and {Blundell}, Ray and
         {Bower}, Geoffrey C. and {Brinkerink}, Christiaan and
         {Broderick}, Avery E. and {Cappallo}, Roger and {Crew}, Geoffrey B. and
         {Dexter}, Jason and {Dexter}, Matt and {Falcke}, Heino and
         {Freund}, Robert and {Friberg}, Per and {Greer}, Christopher H. and
         {Gurwell}, Mark A. and {Ho}, Paul T.~P. and {Honma}, Mareki and
         {Inoue}, Makoto and {Kim}, Junhan and {Lamb}, James and
         {Lindqvist}, Michael and {Macmahon}, David and {Marrone}, Daniel P. and
         {Mart{\'\i}-Vidal}, Ivan and {Menten}, Karl M. and {Moran}, James M. and
         {Nagar}, Neil M. and {Plambeck}, Richard L. and {Primiani}, Rurik A. and
         {Rogers}, Alan E.~E. and {Ros}, Eduardo and {Rottmann}, Helge and
         {SooHoo}, Jason and {Spilker}, Justin and {Stone}, Jordan and
         {Strittmatter}, Peter and {Tilanus}, Remo P.~J. and {Titus}, Michael and
         {Vertatschitsch}, Laura and {Wagner}, Jan and {Weintroub}, Jonathan and
         {Wright}, Melvyn and {Young}, Ken H. and {Zensus}, J. Anton and
         {Ziurys}, Lucy M.},
        title = "{Detection of Intrinsic Source Structure at {\ensuremath{\sim}}3 Schwarzschild Radii with Millimeter-VLBI Observations of SAGITTARIUS A*}",
      journal = {\apj},
     keywords = {Galaxy: center, submillimeter: general, techniques: high angular resolution, techniques: interferometric, Astrophysics - Astrophysics of Galaxies, General Relativity and Quantum Cosmology},
         year = 2018,
        month = may,
       volume = {859},
       number = {1},
          eid = {60},
        pages = {60},
          doi = {10.3847/1538-4357/aabe2e},
archivePrefix = {arXiv},
       eprint = {1805.09223},
 primaryClass = {astro-ph.GA},
       adsurl = {https://ui.adsabs.harvard.edu/abs/2018ApJ...859...60L},
      adsnote = {Provided by the SAO/NASA Astrophysics Data System}
}

@ARTICLE{eaa+19,
       author = {{Event Horizon Telescope Collaboration} and {Akiyama}, Kazunori and
         {Alberdi}, Antxon and {Alef}, Walter and {Asada}, Keiichi and
         {Azulay}, Rebecca and {Baczko}, Anne-Kathrin and {Ball}, David and
         {Balokovi{\'c}}, Mislav and {Barrett}, John and {Bintley}, Dan and
         {Blackburn}, Lindy and {Boland}, Wilfred and {Bouman}, Katherine L. and
         {Bower}, Geoffrey C. and {Bremer}, Michael and
         {Brinkerink}, Christiaan D. and {Brissenden}, Roger and
         {Britzen}, Silke and {Broderick}, Avery E. and {Broguiere}, Dominique and
         {Bronzwaer}, Thomas and {Byun}, Do-Young and {Carlstrom}, John E. and
         {Chael}, Andrew and {Chan}, Chi-kwan and {Chatterjee}, Shami and
         {Chatterjee}, Koushik and {Chen}, Ming-Tang and {Chen}, Yongjun and
         {Cho}, Ilje and {Christian}, Pierre and {Conway}, John E. and
         {Cordes}, James M. and {Crew}, Geoffrey B. and {Cui}, Yuzhu and
         {Davelaar}, Jordy and {De Laurentis}, Mariafelicia and {Deane}, Roger and
         {Dempsey}, Jessica and {Desvignes}, Gregory and {Dexter}, Jason and
         {Doeleman}, Sheperd S. and {Eatough}, Ralph P. and {Falcke}, Heino and
         {Fish}, Vincent L. and {Fomalont}, Ed and {Fraga-Encinas}, Raquel and
         {Freeman}, William T. and {Friberg}, Per and {Fromm}, Christian M. and
         {G{\'o}mez}, Jos{\'e} L. and {Galison}, Peter and {Gammie}, Charles F. and
         {Garc{\'\i}a}, Roberto and {Gentaz}, Olivier and {Georgiev}, Boris and
         {Goddi}, Ciriaco and {Gold}, Roman and {Gu}, Minfeng and
         {Gurwell}, Mark and {Hada}, Kazuhiro and {Hecht}, Michael H. and
         {Hesper}, Ronald and {Ho}, Luis C. and {Ho}, Paul and {Honma}, Mareki and
         {Huang}, Chih-Wei L. and {Huang}, Lei and {Hughes}, David H. and
         {Ikeda}, Shiro and {Inoue}, Makoto and {Issaoun}, Sara and
         {James}, David J. and {Jannuzi}, Buell T. and {Janssen}, Michael and
         {Jeter}, Britton and {Jiang}, Wu and {Johnson}, Michael D. and
         {Jorstad}, Svetlana and {Jung}, Taehyun and {Karami}, Mansour and
         {Karuppusamy}, Ramesh and {Kawashima}, Tomohisa and
         {Keating}, Garrett K. and {Kettenis}, Mark and {Kim}, Jae-Young and
         {Kim}, Junhan and {Kim}, Jongsoo and {Kino}, Motoki and {Koay}, Jun Yi and
         {Koch}, Patrick M. and {Koyama}, Shoko and {Kramer}, Michael and
         {Kramer}, Carsten and {Krichbaum}, Thomas P. and {Kuo}, Cheng-Yu and
         {Lauer}, Tod R. and {Lee}, Sang-Sung and {Li}, Yan-Rong and
         {Li}, Zhiyuan and {Lindqvist}, Michael and {Liu}, Kuo and
         {Liuzzo}, Elisabetta and {Lo}, Wen-Ping and {Lobanov}, Andrei P. and
         {Loinard}, Laurent and {Lonsdale}, Colin and {Lu}, Ru-Sen and
         {MacDonald}, Nicholas R. and {Mao}, Jirong and {Markoff}, Sera and
         {Marrone}, Daniel P. and {Marscher}, Alan P. and
         {Mart{\'\i}-Vidal}, Iv{\'a}n and {Matsushita}, Satoki and
         {Matthews}, Lynn D. and {Medeiros}, Lia and {Menten}, Karl M. and
         {Mizuno}, Yosuke and {Mizuno}, Izumi and {Moran}, James M. and
         {Moriyama}, Kotaro and {Moscibrodzka}, Monika and
         {M{\"u}ller}, Cornelia and {Nagai}, Hiroshi and {Nagar}, Neil M. and
         {Nakamura}, Masanori and {Narayan}, Ramesh and {Narayanan}, Gopal and
         {Natarajan}, Iniyan and {Neri}, Roberto and {Ni}, Chunchong and
         {Noutsos}, Aristeidis and {Okino}, Hiroki and {Olivares}, H{\'e}ctor and
         {Ortiz-Le{\'o}n}, Gisela N. and {Oyama}, Tomoaki and
         {{\"O}zel}, Feryal and {Palumbo}, Daniel C.~M. and {Patel}, Nimesh and
         {Pen}, Ue-Li and {Pesce}, Dominic W. and {Pi{\'e}tu}, Vincent and
         {Plambeck}, Richard and {PopStefanija}, Aleksandar and {Porth}, Oliver and
         {Prather}, Ben and {Preciado-L{\'o}pez}, Jorge A. and
         {Psaltis}, Dimitrios and {Pu}, Hung-Yi and {Ramakrishnan}, Venkatessh and
         {Rao}, Ramprasad and {Rawlings}, Mark G. and {Raymond}, Alexander W. and
         {Rezzolla}, Luciano and {Ripperda}, Bart and {Roelofs}, Freek and
         {Rogers}, Alan and {Ros}, Eduardo and {Rose}, Mel and
         {Roshanineshat}, Arash and {Rottmann}, Helge and {Roy}, Alan L. and
         {Ruszczyk}, Chet and {Ryan}, Benjamin R. and {Rygl}, Kazi L.~J. and
         {S{\'a}nchez}, Salvador and {S{\'a}nchez-Arguelles}, David and
         {Sasada}, Mahito and {Savolainen}, Tuomas and {Schloerb}, F. Peter and
         {Schuster}, Karl-Friedrich and {Shao}, Lijing and {Shen}, Zhiqiang and
         {Small}, Des and {Sohn}, Bong Won and {SooHoo}, Jason and
         {Tazaki}, Fumie and {Tiede}, Paul and {Tilanus}, Remo P.~J. and
         {Titus}, Michael and {Toma}, Kenji and {Torne}, Pablo and
         {Trent}, Tyler and {Trippe}, Sascha and {Tsuda}, Shuichiro and
         {van Bemmel}, Ilse and {van Langevelde}, Huib Jan and
         {van Rossum}, Daniel R. and {Wagner}, Jan and {Wardle}, John and
         {Weintroub}, Jonathan and {Wex}, Norbert and {Wharton}, Robert and
         {Wielgus}, Maciek and {Wong}, George N. and {Wu}, Qingwen and
         {Young}, Ken and {Young}, Andr{\'e} and {Younsi}, Ziri and
         {Yuan}, Feng and {Yuan}, Ye-Fei and {Zensus}, J. Anton and
         {Zhao}, Guangyao and {Zhao}, Shan-Shan and {Zhu}, Ziyan and
         {Algaba}, Juan-Carlos and {Allardi}, Alexander and {Amestica}, Rodrigo and
         {Anczarski}, Jadyn and {Bach}, Uwe and {Baganoff}, Frederick K. and
         {Beaudoin}, Christopher and {Benson}, Bradford A. and {Berthold}, Ryan and
         {Blanchard}, Jay M. and {Blundell}, Ray and {Bustamente}, Sandra and
         {Cappallo}, Roger and {Castillo-Dom{\'\i}nguez}, Edgar and
         {Chang}, Chih-Cheng and {Chang}, Shu-Hao and {Chang}, Song-Chu and
         {Chen}, Chung-Chen and {Chilson}, Ryan and {Chuter}, Tim C. and
         {C{\'o}rdova Rosado}, Rodrigo and {Coulson}, Iain M. and
         {Crawford}, Thomas M. and {Crowley}, Joseph and {David}, John and
         {Derome}, Mark and {Dexter}, Matthew and {Dornbusch}, Sven and
         {Dudevoir}, Kevin A. and {Dzib}, Sergio A. and {Eckart}, Andreas and
         {Eckert}, Chris and {Erickson}, Neal R. and {Everett}, Wendeline B. and
         {Faber}, Aaron and {Farah}, Joseph R. and {Fath}, Vernon and
         {Folkers}, Thomas W. and {Forbes}, David C. and {Freund}, Robert and
         {G{\'o}mez-Ruiz}, Arturo I. and {Gale}, David M. and {Gao}, Feng and
         {Geertsema}, Gertie and {Graham}, David A. and {Greer}, Christopher H. and
         {Grosslein}, Ronald and {Gueth}, Fr{\'e}d{\'e}ric and {Haggard}, Daryl and
         {Halverson}, Nils W. and {Han}, Chih-Chiang and {Han}, Kuo-Chang and
         {Hao}, Jinchi and {Hasegawa}, Yutaka and {Henning}, Jason W. and
         {Hern{\'a}ndez-G{\'o}mez}, Antonio and {Herrero-Illana}, Rub{\'e}n and
         {Heyminck}, Stefan and {Hirota}, Akihiko and {Hoge}, James and
         {Huang}, Yau-De and {Impellizzeri}, C.~M. Violette and {Jiang}, Homin and
         {Kamble}, Atish and {Keisler}, Ryan and {Kimura}, Kimihiro and
         {Kono}, Yusuke and {Kubo}, Derek and {Kuroda}, John and
         {Lacasse}, Richard and {Laing}, Robert A. and {Leitch}, Erik M. and
         {Li}, Chao-Te and {Lin}, Lupin C. -C. and {Liu}, Ching-Tang and
         {Liu}, Kuan-Yu and {Lu}, Li-Ming and {Marson}, Ralph G. and
         {Martin-Cocher}, Pierre L. and {Massingill}, Kyle D. and
         {Matulonis}, Callie and {McColl}, Martin P. and
         {McWhirter}, Stephen R. and {Messias}, Hugo and {Meyer-Zhao}, Zheng and
         {Michalik}, Daniel and {Monta{\~n}a}, Alfredo and
         {Montgomerie}, William and {Mora-Klein}, Matias and {Muders}, Dirk and
         {Nadolski}, Andrew and {Navarro}, Santiago and {Neilsen}, Joseph and
         {Nguyen}, Chi H. and {Nishioka}, Hiroaki and {Norton}, Timothy and
         {Nowak}, Michael A. and {Nystrom}, George and {Ogawa}, Hideo and
         {Oshiro}, Peter and {Oyama}, Tomoaki and {Parsons}, Harriet and
         {Paine}, Scott N. and {Pe{\~n}alver}, Juan and {Phillips}, Neil M. and
         {Poirier}, Michael and {Pradel}, Nicolas and {Primiani}, Rurik A. and
         {Raffin}, Philippe A. and {Rahlin}, Alexandra S. and {Reiland}, George and
         {Risacher}, Christopher and {Ruiz}, Ignacio and
         {S{\'a}ez-Mada{\'\i}n}, Alejandro F. and {Sassella}, Remi and
         {Schellart}, Pim and {Shaw}, Paul and {Silva}, Kevin M. and
         {Shiokawa}, Hotaka and {Smith}, David R. and {Snow}, William and
         {Souccar}, Kamal and {Sousa}, Don and {Sridharan}, T.~K. and
         {Srinivasan}, Ranjani and {Stahm}, William and {Stark}, Anthony A. and
         {Story}, Kyle and {Timmer}, Sjoerd T. and {Vertatschitsch}, Laura and
         {Walther}, Craig and {Wei}, Ta-Shun and {Whitehorn}, Nathan and
         {Whitney}, Alan R. and {Woody}, David P. and {Wouterloot}, Jan G.~A. and
         {Wright}, Melvin and {Yamaguchi}, Paul and {Yu}, Chen-Yu and
         {Zeballos}, Milagros and {Zhang}, Shuo and {Ziurys}, Lucy},
        title = "{First M87 Event Horizon Telescope Results. I. The Shadow of the Supermassive Black Hole}",
      journal = {ApJL},
     keywords = {accretion, accretion disks, black hole physics, galaxies: active, galaxies: individual: M87, galaxies: jets, gravitation, Astrophysics - Astrophysics of Galaxies, Astrophysics - High Energy Astrophysical Phenomena, General Relativity and Quantum Cosmology},
         year = 2019,
        month = apr,
       volume = {875},
       number = {1},
          eid = {L1},
        pages = {L1},
          doi = {10.3847/2041-8213/ab0ec7},
archivePrefix = {arXiv},
       eprint = {1906.11238},
 primaryClass = {astro-ph.GA},
       adsurl = {https://ui.adsabs.harvard.edu/abs/2019ApJ...875L...1E},
      adsnote = {Provided by the SAO/NASA Astrophysics Data System}
}

@ARTICLE{zly14,
       author = {{Zhang}, Fupeng and {Lu}, Youjun and {Yu}, Qingjuan},
        title = "{On the Existence of Pulsars in the Vicinity of the Massive Black Hole in the Galactic Center}",
      journal = {\apj},
     keywords = {black hole physics, Galaxy: center, Galaxy: kinematics and dynamics, pulsars: general, Astrophysics - Astrophysics of Galaxies, Astrophysics - High Energy Astrophysical Phenomena, Astrophysics - Solar and Stellar Astrophysics},
         year = 2014,
        month = apr,
       volume = {784},
       number = {2},
          eid = {106},
        pages = {106},
          doi = {10.1088/0004-637X/784/2/106},
archivePrefix = {arXiv},
       eprint = {1402.2505},
 primaryClass = {astro-ph.GA},
       adsurl = {https://ui.adsabs.harvard.edu/abs/2014ApJ...784..106Z},
      adsnote = {Provided by the SAO/NASA Astrophysics Data System}
}

@ARTICLE{mk15,
       author = {{Macquart}, Jean-Pierre and {Kanekar}, Nissim},
        title = "{On Detecting Millisecond Pulsars at the Galactic Center}",
      journal = {\apj},
     keywords = {Galaxy: center, pulsars: general, Astrophysics - High Energy Astrophysical Phenomena},
         year = 2015,
        month = jun,
       volume = {805},
       number = {2},
          eid = {172},
        pages = {172},
          doi = {10.1088/0004-637X/805/2/172},
archivePrefix = {arXiv},
       eprint = {1504.02492},
 primaryClass = {astro-ph.HE},
       adsurl = {https://ui.adsabs.harvard.edu/abs/2015ApJ...805..172M},
      adsnote = {Provided by the SAO/NASA Astrophysics Data System}
}

@ARTICLE{bbe+11,
       author = {{Burke-Spolaor}, S. and {Bailes}, Matthew and {Ekers}, Ronald and
         {Macquart}, Jean-Pierre and {Crawford}, Fronefield, III},
        title = "{Radio Bursts with Extragalactic Spectral Characteristics Show Terrestrial Origins}",
      journal = {\apj},
     keywords = {atmospheric effects, plasmas, pulsars: general, radio continuum: general, Astrophysics - Cosmology and Nongalactic Astrophysics, Astrophysics - Earth and Planetary Astrophysics},
         year = 2011,
        month = jan,
       volume = {727},
       number = {1},
          eid = {18},
        pages = {18},
          doi = {10.1088/0004-637X/727/1/18},
archivePrefix = {arXiv},
       eprint = {1009.5392},
 primaryClass = {astro-ph.CO},
       adsurl = {https://ui.adsabs.harvard.edu/abs/2011ApJ...727...18B},
      adsnote = {Provided by the SAO/NASA Astrophysics Data System}
}

@INPROCEEDINGS{hmk+16,
       author = {{Huang}, Yau De (Ted) and {Morata}, Oscar and {Koch}, Patrick Michel and
         {Kemper}, Ciska and {Hwang}, Yuh-Jing and {Chiong}, Chau-Ching and
         {Ho}, Paul and {Chu}, You-Hua and {Huang}, Chi-Den and
         {Liu}, Ching-Tang and {Hsieh}, Fang-Chia and {Tseng}, Yen-Hsiang and
         {Weng}, Shou-Hsien and {Ho}, Chin-Ting and {Chiang}, Po-Han and
         {Wu}, Hsiao-Ling and {Chang}, Chih-Cheng and {Jian}, Shou-Ting and
         {Lee}, Chien-Feng and {Lee}, Yi-Wei and {Iguchi}, Satoru and
         {Asayama}, Shin'ichiro and {Iono}, Daisuke and {Gonzalez}, Alvaro and
         {Effland}, John and {Saini}, Kamaljeet and {Pospieszalski}, Marian and
         {Henke}, Doug and {Yeung}, Keith and {Finger}, Ricardo and
         {Tapia}, Valeria and {Reyes}, Nicolas},
        title = "{The Atacama Large Millimeter/sub-millimeter Array band-1 receiver}",
     keywords = {Astrophysics - Instrumentation and Methods for Astrophysics},
    booktitle = {Modeling, Systems Engineering, and Project Management for Astronomy VI},
         year = 2016,
       editor = {{Angeli}, George Z. and {Dierickx}, Philippe},
       series = {Society of Photo-Optical Instrumentation Engineers (SPIE) Conference Series},
       volume = {9911},
        month = aug,
          eid = {99111V},
        pages = {99111V},
          doi = {10.1117/12.2232193},
archivePrefix = {arXiv},
       eprint = {1612.00893},
 primaryClass = {astro-ph.IM},
       adsurl = {https://ui.adsabs.harvard.edu/abs/2016SPIE.9911E..1VH},
      adsnote = {Provided by the SAO/NASA Astrophysics Data System}
}

@ARTICLE{zew+10,
       author = {{Zamaninasab}, M. and {Eckart}, A. and {Witzel}, G. and {Dovciak}, M. and {Karas}, V. and {Sch{\"o}del}, R. and {Gie{\ss}{\"u}bel}, R. and {Bremer}, M. and {Garc{\'\i}a-Mar{\'\i}n}, M. and {Kunneriath}, D. and {Mu{\v{z}}i{\'c}}, K. and {Nishiyama}, S. and {Sabha}, N. and {Straubmeier}, C. and {Zensus}, A.},
        title = "{Near infrared flares of Sagittarius A*. Importance of near infrared polarimetry}",
      journal = {\aa},
     keywords = {black hole physics, infrared: general, accretion, accretion disks, Galaxy: center, Galaxy: nucleus, Astrophysics - Astrophysics of Galaxies},
         year = 2010,
        month = jan,
       volume = {510},
          eid = {A3},
        pages = {A3},
          doi = {10.1051/0004-6361/200912473},
archivePrefix = {arXiv},
       eprint = {0911.4659},
 primaryClass = {astro-ph.GA},
       adsurl = {https://ui.adsabs.harvard.edu/abs/2010A&A...510A...3Z},
      adsnote = {Provided by the SAO/NASA Astrophysics Data System}
}

@ARTICLE{do14,
       author = {{Dexter}, Jason and {O'Leary}, Ryan M.},
        title = "{The Peculiar Pulsar Population of the Central Parsec}",
      journal = {ApJL},
     keywords = {Galaxy: center, pulsars: general, pulsars: individual: SGR J1745─29, stars: neutron, Astrophysics - Galaxy Astrophysics, Astrophysics - High Energy Astrophysical Phenomena, Astrophysics - Solar and Stellar Astrophysics},
         year = 2014,
        month = mar,
       volume = {783},
       number = {1},
          eid = {L7},
        pages = {L7},
          doi = {10.1088/2041-8205/783/1/L7},
archivePrefix = {arXiv},
       eprint = {1310.7022},
 primaryClass = {astro-ph.GA},
       adsurl = {https://ui.adsabs.harvard.edu/abs/2014ApJ...783L...7D},
      adsnote = {Provided by the SAO/NASA Astrophysics Data System}
}

@ARTICLE{kcc21,
       author = {{Koch Ocker}, Stella and {Cordes}, James M. and {Chatterjee}, Shami},
        title = "{Constraining Galaxy Haloes from the Dispersion and Scattering of Fast Radio Bursts and Pulsars}",
      journal = {arXiv e-prints},
     keywords = {Astrophysics - Astrophysics of Galaxies, Astrophysics - High Energy Astrophysical Phenomena},
         year = 2021,
        month = jan,
          eid = {arXiv:2101.04784},
        pages = {arXiv:2101.04784},
archivePrefix = {arXiv},
       eprint = {2101.04784},
 primaryClass = {astro-ph.GA},
       adsurl = {https://ui.adsabs.harvard.edu/abs/2021arXiv210104784K},
      adsnote = {Provided by the SAO/NASA Astrophysics Data System}
}

@ARTICLE{lyu13,
       author = {{Lyutikov}, Maxim},
        title = "{Inverse Compton model of pulsar high-energy emission}",
      journal = {\mnras},
     keywords = {pulsars: individual: Crab, gamma-rays: stars, X-rays: stars, Astrophysics - High Energy Astrophysical Phenomena},
         year = 2013,
        month = may,
       volume = {431},
       number = {3},
        pages = {2580-2589},
          doi = {10.1093/mnras/stt351},
archivePrefix = {arXiv},
       eprint = {1208.5329},
 primaryClass = {astro-ph.HE},
       adsurl = {https://ui.adsabs.harvard.edu/abs/2013{\mnras}.431.2580L},
      adsnote = {Provided by the SAO/NASA Astrophysics Data System}
}

@ARTICLE{kzw+93,
       author = {{Krichbaum}, T.~P. and {Zensus}, J.~A. and {Witzel}, A. and {Mezger}, P.~G. and {Standke}, K.~J. and {Schalinski}, C.~J. and {Alberdi}, A. and {Marcaide}, J.~M. and {Zylka}, R. and {Rogers}, A.~E.~E. and {Booth}, R.~S. and {Ronnang}, B.~O. and {Colomer}, F. and {Bartel}, N. and {Shapiro}, I.~I.},
        title = "{First 43 GHz VLBI detection of the compact source SGR A in the Galactic Center.}",
      journal = {\aa},
     keywords = {The Galactic Center, Interferometry, Jets of Galaxies},
         year = 1993,
        month = jul,
       volume = {274},
        pages = {L37-L40},
       adsurl = {https://ui.adsabs.harvard.edu/abs/1993A&A...274L..37K},
      adsnote = {Provided by the SAO/NASA Astrophysics Data System}
}

@ARTICLE{dbp+11,
       author = {{Deller}, A.~T. and {Brisken}, W.~F. and {Phillips}, C.~J. and {Morgan}, J. and {Alef}, W. and {Cappallo}, R. and {Middelberg}, E. and {Romney}, J. and {Rottmann}, H. and {Tingay}, S.~J. and {Wayth}, R.},
        title = "{DiFX-2: A More Flexible, Efficient, Robust, and Powerful Software Correlator}",
      journal = {\pasp},
     keywords = {Astrophysics - Instrumentation and Methods for Astrophysics},
         year = 2011,
        month = mar,
       volume = {123},
       number = {901},
        pages = {275},
          doi = {10.1086/658907},
archivePrefix = {arXiv},
       eprint = {1101.0885},
 primaryClass = {astro-ph.IM},
       adsurl = {https://ui.adsabs.harvard.edu/abs/2011PASP..123..275D},
      adsnote = {Provided by the SAO/NASA Astrophysics Data System}
}

@ARTICLE{bdd+14,
       author = {{Bower}, Geoffrey C. and {Deller}, Adam and {Demorest}, Paul and {Brunthaler}, Andreas and {Eatough}, Ralph and {Falcke}, Heino and {Kramer}, Michael and {Lee}, K.~J. and {Spitler}, Laura},
        title = "{The Angular Broadening of the Galactic Center Pulsar SGR J1745-29: A New Constraint on the Scattering Medium}",
      journal = {ApJL},
     keywords = {black hole physics, pulsars: general, pulsars: individual: SGR 1745-2900, scattering, Astrophysics - Astrophysics of Galaxies, Astrophysics - High Energy Astrophysical Phenomena},
         year = 2014,
        month = jan,
       volume = {780},
       number = {1},
          eid = {L2},
        pages = {L2},
          doi = {10.1088/2041-8205/780/1/L2},
archivePrefix = {arXiv},
       eprint = {1309.4672},
 primaryClass = {astro-ph.GA},
       adsurl = {https://ui.adsabs.harvard.edu/abs/2014ApJ...780L...2B},
      adsnote = {Provided by the SAO/NASA Astrophysics Data System}
}

@ARTICLE{ddb+17,
       author = {{Dexter}, J. and {Deller}, A. and {Bower}, G.~C. and {Demorest}, P. and {Kramer}, M. and {Stappers}, B.~W. and {Lyne}, A.~G. and {Kerr}, M. and {Spitler}, L.~G. and {Psaltis}, D. and {Johnson}, M. and {Narayan}, R.},
        title = "{Locating the intense interstellar scattering towards the inner Galaxy}",
      journal = {\mnras},
     keywords = {scattering, pulsars: general, H $\lt$sc$\gt$II$\lt$/sc$\gt$ regions, ISM: supernova remnants, Galaxy: centre, Astrophysics - Astrophysics of Galaxies},
         year = 2017,
        month = nov,
       volume = {471},
       number = {3},
        pages = {3563-3576},
          doi = {10.1093/mnras/stx1777},
archivePrefix = {arXiv},
       eprint = {1707.03842},
 primaryClass = {astro-ph.GA},
       adsurl = {https://ui.adsabs.harvard.edu/abs/2017{\mnras}.471.3563D},
      adsnote = {Provided by the SAO/NASA Astrophysics Data System}
}

@ARTICLE{lag+99,
       author = {{Lazio}, T. Joseph W. and {Anantharamaiah}, K.~R. and {Goss}, W.~M. and {Kassim}, Namir E. and {Cordes}, James M.},
        title = "{G359.87+0.18, An FR II Radio Galaxy 15' from Sagittarius A$^{*}$: Implications for the Scattering Region in the Galactic Center}",
      journal = {\apj},
     keywords = {GALAXIES: INDIVIDUAL (G359.87+0.18), GALAXY: CENTER, RADIO CONTINUUM: GALAXIES, SCATTERING, galaxies: individual (G359.87+0.18), Galaxy: Center, Radio Continuum: Galaxies, Scattering, Astrophysics},
         year = 1999,
        month = apr,
       volume = {515},
       number = {1},
        pages = {196-205},
          doi = {10.1086/307019},
archivePrefix = {arXiv},
       eprint = {astro-ph/9811317},
 primaryClass = {astro-ph},
       adsurl = {https://ui.adsabs.harvard.edu/abs/1999ApJ...515..196L},
      adsnote = {Provided by the SAO/NASA Astrophysics Data System}
}

@ARTICLE{gaa+20,
       author = {{Gravity Collaboration} and {Abuter}, R. and {Amorim}, A. and {Baub{\"o}ck}, M. and {Berger}, J.~P. and {Bonnet}, H. and {Brandner}, W. and {Cardoso}, V. and {Cl{\'e}net}, Y. and {de Zeeuw}, P.~T. and {Dexter}, J. and {Eckart}, A. and {Eisenhauer}, F. and {F{\"o}rster Schreiber}, N.~M. and {Garcia}, P. and {Gao}, F. and {Gendron}, E. and {Genzel}, R. and {Gillessen}, S. and {Habibi}, M. and {Haubois}, X. and {Henning}, T. and {Hippler}, S. and {Horrobin}, M. and {Jim{\'e}nez-Rosales}, A. and {Jochum}, L. and {Jocou}, L. and {Kaufer}, A. and {Kervella}, P. and {Lacour}, S. and {Lapeyr{\`e}re}, V. and {Le Bouquin}, J. -B. and {L{\'e}na}, P. and {Nowak}, M. and {Ott}, T. and {Paumard}, T. and {Perraut}, K. and {Perrin}, G. and {Pfuhl}, O. and {Rodr{\'\i}guez-Coira}, G. and {Shangguan}, J. and {Scheithauer}, S. and {Stadler}, J. and {Straub}, O. and {Straubmeier}, C. and {Sturm}, E. and {Tacconi}, L.~J. and {Vincent}, F. and {von Fellenberg}, S. and {Waisberg}, I. and {Widmann}, F. and {Wieprecht}, E. and {Wiezorrek}, E. and {Woillez}, J. and {Yazici}, S. and {Zins}, G.},
        title = "{Detection of the Schwarzschild precession in the orbit of the star S2 near the Galactic centre massive black hole}",
      journal = {\aa},
     keywords = {black hole physics, Galaxy: nucleus, gravitation, relativistic processes, Astrophysics - Astrophysics of Galaxies, Astrophysics - Instrumentation and Methods for Astrophysics, General Relativity and Quantum Cosmology},
         year = 2020,
        month = apr,
       volume = {636},
          eid = {L5},
        pages = {L5},
          doi = {10.1051/0004-6361/202037813},
archivePrefix = {arXiv},
       eprint = {2004.07187},
 primaryClass = {astro-ph.GA},
       adsurl = {https://ui.adsabs.harvard.edu/abs/2020A&A...636L...5G},
      adsnote = {Provided by the SAO/NASA Astrophysics Data System}
}

@ARTICLE{dhg+19,
       author = {{Do}, Tuan and {Hees}, Aurelien and {Ghez}, Andrea and {Martinez}, Gregory D. and {Chu}, Devin S. and {Jia}, Siyao and {Sakai}, Shoko and {Lu}, Jessica R. and {Gautam}, Abhimat K. and {O'Neil}, Kelly Kosmo and {Becklin}, Eric E. and {Morris}, Mark R. and {Matthews}, Keith and {Nishiyama}, Shogo and {Campbell}, Randy and {Chappell}, Samantha and {Chen}, Zhuo and {Ciurlo}, Anna and {Dehghanfar}, Arezu and {Gallego-Cano}, Eulalia and {Kerzendorf}, Wolfgang E. and {Lyke}, James E. and {Naoz}, Smadar and {Saida}, Hiromi and {Sch{\"o}del}, Rainer and {Takahashi}, Masaaki and {Takamori}, Yohsuke and {Witzel}, Gunther and {Wizinowich}, Peter},
        title = "{Relativistic redshift of the star S0-2 orbiting the Galactic Center supermassive black hole}",
      journal = {Science},
     keywords = {ASTRONOMY; PHYSICS, Astrophysics - Astrophysics of Galaxies, General Relativity and Quantum Cosmology},
         year = 2019,
        month = aug,
       volume = {365},
       number = {6454},
        pages = {664-668},
          doi = {10.1126/science.aav8137},
archivePrefix = {arXiv},
       eprint = {1907.10731},
 primaryClass = {astro-ph.GA},
       adsurl = {https://ui.adsabs.harvard.edu/abs/2019Sci...365..664D},
      adsnote = {Provided by the SAO/NASA Astrophysics Data System}
}


@ARTICLE{ljk+08b,
       author = {{L{\"o}hmer}, O. and {Jessner}, A. and {Kramer}, M. and {Wielebinski}, R. and {Maron}, O.},
        title = "{Observations of pulsars at 9 millimetres}",
      journal = aa,
     keywords = {stars: pulsars: individual: PSR B0144+59, stars: pulsars: individual: PSR B0823+26, stars: pulsars: individual:, PSR B2022+50, radiation mechanisms: non-thermal, radio continuum: stars, Astrophysics},
         year = 2008,
        month = mar,
       volume = {480},
       number = {3},
        pages = {623-628},
          doi = {10.1051/0004-6361:20066806},
archivePrefix = {arXiv},
       eprint = {0802.1339},
 primaryClass = {astro-ph},
       adsurl = {https://ui.adsabs.harvard.edu/abs/2008A&A...480..623L},
      adsnote = {Provided by the SAO/NASA Astrophysics Data System}
}

@ARTICLE{dyp+18,
       author = {{De Laurentis}, Mariafelicia and {Younsi}, Ziri and {Porth}, Oliver and {Mizuno}, Yosuke and {Rezzolla}, Luciano},
        title = "{Test-particle dynamics in general spherically symmetric black hole spacetimes}",
      journal = {\prd},
     keywords = {General Relativity and Quantum Cosmology, Astrophysics - High Energy Astrophysical Phenomena},
         year = 2018,
        month = may,
       volume = {97},
       number = {10},
          eid = {104024},
        pages = {104024},
          doi = {10.1103/PhysRevD.97.104024},
archivePrefix = {arXiv},
       eprint = {1712.00265},
 primaryClass = {gr-qc},
       adsurl = {https://ui.adsabs.harvard.edu/abs/2018PhRvD..97j4024D},
      adsnote = {Provided by the SAO/NASA Astrophysics Data System}
}

@ARTICLE{jnp+18,
       author = {{Johnson}, Michael D. and {Narayan}, Ramesh and {Psaltis}, Dimitrios and {Blackburn}, Lindy and {Kovalev}, Yuri Y. and {Gwinn}, Carl R. and {Zhao}, Guang-Yao and {Bower}, Geoffrey C. and {Moran}, James M. and {Kino}, Motoki and {Kramer}, Michael and {Akiyama}, Kazunori and {Dexter}, Jason and {Broderick}, Avery E. and {Sironi}, Lorenzo},
        title = "{The Scattering and Intrinsic Structure of Sagittarius A* at Radio Wavelengths}",
      journal = {\apj},
     keywords = {Galaxy: nucleus, ISM: structure, radio continuum: ISM, scattering, techniques: interferometric, turbulence, Astrophysics - Astrophysics of Galaxies, Astrophysics - Instrumentation and Methods for Astrophysics},
         year = 2018,
        month = oct,
       volume = {865},
       number = {2},
          eid = {104},
        pages = {104},
          doi = {10.3847/1538-4357/aadcff},
archivePrefix = {arXiv},
       eprint = {1808.08966},
 primaryClass = {astro-ph.GA},
       adsurl = {https://ui.adsabs.harvard.edu/abs/2018ApJ...865..104J},
      adsnote = {Provided by the SAO/NASA Astrophysics Data System}
}

@ARTICLE{mgz+13,
       author = {{Mori}, Kaya and {Gotthelf}, Eric V. and {Zhang}, Shuo and {An}, Hongjun and {Baganoff}, Frederick K. and {Barri{\`e}re}, Nicolas M. and {Beloborodov}, Andrei M. and {Boggs}, Steven E. and {Christensen}, Finn E. and {Craig}, William W. and {Dufour}, Francois and {Grefenstette}, Brian W. and {Hailey}, Charles J. and {Harrison}, Fiona A. and {Hong}, Jaesub and {Kaspi}, Victoria M. and {Kennea}, Jamie A. and {Madsen}, Kristin K. and {Markwardt}, Craig B. and {Nynka}, Melania and {Stern}, Daniel and {Tomsick}, John A. and {Zhang}, William W.},
        title = "{NuSTAR Discovery of a 3.76 s Transient Magnetar Near Sagittarius A*}",
      journal = {ApJL},
     keywords = {pulsars: general, pulsars: individual: SGR J1745-29, SGR J1745-2900, PSR J1745-2900, stars: neutron, Astrophysics - High Energy Astrophysical Phenomena},
         year = 2013,
        month = jun,
       volume = {770},
       number = {2},
          eid = {L23},
        pages = {L23},
          doi = {10.1088/2041-8205/770/2/L23},
archivePrefix = {arXiv},
       eprint = {1305.1945},
 primaryClass = {astro-ph.HE},
       adsurl = {https://ui.adsabs.harvard.edu/abs/2013ApJ...770L..23M},
      adsnote = {Provided by the SAO/NASA Astrophysics Data System}
}

@ARTICLE{rep+13,
       author = {{Rea}, N. and {Esposito}, P. and {Pons}, J.~A. and {Turolla}, R. and {Torres}, D.~F. and {Israel}, G.~L. and {Possenti}, A. and {Burgay}, M. and {Vigan{\`o}}, D. and {Papitto}, A. and {Perna}, R. and {Stella}, L. and {Ponti}, G. and {Baganoff}, F.~K. and {Haggard}, D. and {Camero-Arranz}, A. and {Zane}, S. and {Minter}, A. and {Mereghetti}, S. and {Tiengo}, A. and {Sch{\"o}del}, R. and {Feroci}, M. and {Mignani}, R. and {G{\"o}tz}, D.},
        title = "{A Strongly Magnetized Pulsar within the Grasp of the Milky Way's Supermassive Black Hole}",
      journal = {ApJL},
     keywords = {Galaxy: center, stars: neutron, X-rays: individual: SGR J1745-2900, Astrophysics - Astrophysics of Galaxies, Astrophysics - High Energy Astrophysical Phenomena},
         year = 2013,
        month = oct,
       volume = {775},
       number = {2},
          eid = {L34},
        pages = {L34},
          doi = {10.1088/2041-8205/775/2/L34},
archivePrefix = {arXiv},
       eprint = {1307.6331},
 primaryClass = {astro-ph.GA},
       adsurl = {https://ui.adsabs.harvard.edu/abs/2013ApJ...775L..34R},
      adsnote = {Provided by the SAO/NASA Astrophysics Data System}
}

@ARTICLE{wk20,
       author = {{Wex}, Norbert and {Kramer}, Michael},
        title = "{Gravity Tests with Radio Pulsars}",
      journal = {Universe},
         year = 2020,
        month = sep,
       volume = {6},
       number = {9},
        pages = {156},
          doi = {10.3390/universe6090156},
       adsurl = {https://ui.adsabs.harvard.edu/abs/2020Univ....6..156W},
      adsnote = {Provided by the SAO/NASA Astrophysics Data System}
}

@ARTICLE{bs76b,
       author = {{Blandford}, R.~D. and {Scharlemann}, E.~T.},
        title = "{On the scattering and absorption of electromagnetic radiation with pulsar magnetospheres.}",
      journal = {\mnras},
     keywords = {Electromagnetic Absorption, Electromagnetic Scattering, Pulsar Magnetospheres, Pulsars, Stellar Atmospheres, Coherent Electromagnetic Radiation, Integral Equations, Lines Of Force, Relativistic Particles, Resonance Scattering, Astrophysics},
         year = 1976,
        month = jan,
       volume = {174},
        pages = {59-85},
          doi = {10.1093/mnras/174.1.59},
       adsurl = {https://ui.adsabs.harvard.edu/abs/1976{\mnras}.174...59B},
      adsnote = {Provided by the SAO/NASA Astrophysics Data System}
}

@ARTICLE{pts20,
       author = {{Philippov}, Alexander and {Timokhin}, Andrey and {Spitkovsky}, Anatoly},
        title = "{Origin of Pulsar Radio Emission}",
      journal = {\prl},
     keywords = {Astrophysics - High Energy Astrophysical Phenomena},
         year = 2020,
        month = jun,
       volume = {124},
       number = {24},
          eid = {245101},
        pages = {245101},
          doi = {10.1103/PhysRevLett.124.245101},
archivePrefix = {arXiv},
       eprint = {2001.02236},
 primaryClass = {astro-ph.HE},
       adsurl = {https://ui.adsabs.harvard.edu/abs/2020PhRvL.124x5101P},
      adsnote = {Provided by the SAO/NASA Astrophysics Data System}
}

@ARTICLE{ckl01,
       author = {{Crusius-W{\"a}tzel}, Andr{\'e} R. and {Kunzl}, Thomas and {Lesch}, Harald},
        title = "{Synchrotron Model for the Infrared, Optical, and X-Ray Emission of the Crab Pulsar}",
      journal = {\apj},
     keywords = {Stars: Pulsars: Individual: Alphanumeric: PSR 0531+21, Radiation Mechanisms: Nonthermal, Stars: Neutron, X-Rays: Stars, Astrophysics},
         year = 2001,
        month = jan,
       volume = {546},
       number = {1},
        pages = {401-405},
          doi = {10.1086/318235},
archivePrefix = {arXiv},
       eprint = {astro-ph/0009324},
 primaryClass = {astro-ph},
       adsurl = {https://ui.adsabs.harvard.edu/abs/2001ApJ...546..401C},
      adsnote = {Provided by the SAO/NASA Astrophysics Data System}
}

@ARTICLE{zwy+17,
       author = {{Zhao}, Ru-Shuang and {Wu}, Xin-Ji and {Yan}, Zhen and {Shen}, Zhi-Qiang and {Manchester}, R.~N. and {Qiao}, Guo-Jun and {Xu}, Ren-Xin and {Wu}, Ya-Jun and {Zhao}, Rong-Bing and {Li}, Bin and {Du}, Yuan-Jie and {Lee}, Ke-Jia and {Hao}, Long-Fei and {Liu}, Qing-Hui and {Lu}, Ji-Guang and {Shang}, Lun-Hua and {Wang}, Jin-Qing and {Wang}, Min and {Yuan}, Jin and {Zhi}, Qi-Jun and {Zhong}, Wei-Ye},
        title = "{TMRT Observations of 26 Pulsars at 8.6 GHz}",
      journal = {\apj},
     keywords = {pulsars: general, Astrophysics - High Energy Astrophysical Phenomena},
         year = 2017,
        month = aug,
       volume = {845},
       number = {2},
          eid = {156},
        pages = {156},
          doi = {10.3847/1538-4357/aa8170},
archivePrefix = {arXiv},
       eprint = {1707.06432},
 primaryClass = {astro-ph.HE},
       adsurl = {https://ui.adsabs.harvard.edu/abs/2017ApJ...845..156Z},
      adsnote = {Provided by the SAO/NASA Astrophysics Data System}
}

@ARTICLE{rmp+18,
       author = {{Ritacco}, A. and {Mac{\'\i}as-P{\'e}rez}, J.~F. and {Ponthieu}, N. and {Adam}, R. and {Ade}, P. and {Andr{\'e}}, P. and {Aumont}, J. and {Beelen}, A. and {Beno{\^\i}t}, A. and {Bideaud}, A. and {Billot}, N. and {Bourrion}, O. and {Bracco}, A. and {Calvo}, M. and {Catalano}, A. and {Coiffard}, G. and {Comis}, B. and {D'Addabbo}, A. and {De Petris}, M. and {D{\'e}sert}, F. -X. and {Doyle}, S. and {Goupy}, J. and {Kramer}, C. and {Lagache}, G. and {Leclercq}, S. and {Lestrade}, J. -F. and {Mauskopf}, P. and {Mayet}, F. and {Maury}, A. and {Monfardini}, A. and {Pajot}, F. and {Pascale}, E. and {Perotto}, L. and {Pisano}, G. and {Rebolo-Iglesias}, M. and {Rev{\'e}ret}, V. and {Rodriguez}, L. and {Romero}, C. and {Roussel}, H. and {Ruppin}, F. and {Schuster}, K. and {Sievers}, A. and {Siringo}, G. and {Thum}, C. and {Triqueneaux}, S. and {Tucker}, C. and {Wiesemeyer}, H. and {Zylka}, R.},
        title = "{NIKA 150 GHz polarization observations of the Crab nebula and its spectral energy distribution}",
      journal = {\aa},
     keywords = {polarization, instrumentation: high angular resolution, instrumentation: detectors, methods: observational, supernovae: general, Astrophysics - Cosmology and Nongalactic Astrophysics, Astrophysics - Instrumentation and Methods for Astrophysics},
         year = 2018,
        month = aug,
       volume = {616},
          eid = {A35},
        pages = {A35},
          doi = {10.1051/0004-6361/201731551},
archivePrefix = {arXiv},
       eprint = {1804.09581},
 primaryClass = {astro-ph.CO},
       adsurl = {https://ui.adsabs.harvard.edu/abs/2018A&A...616A..35R},
      adsnote = {Provided by the SAO/NASA Astrophysics Data System}
}

@ARTICLE{pbg+20,
       author = {{Page}, Dany and {Beznogov}, Mikhail V. and {Garibay}, Iv{\'a}n and {Lattimer}, James M. and {Prakash}, Madappa and {Janka}, Hans-Thomas},
        title = "{Ns 1987A in SN 1987A}",
      journal = {\apj},
     keywords = {Neutron stars, Supernovae, 1108, 1668, Astrophysics - High Energy Astrophysical Phenomena, Nuclear Theory},
         year = 2020,
        month = jul,
       volume = {898},
       number = {2},
          eid = {125},
        pages = {125},
          doi = {10.3847/1538-4357/ab93c2},
archivePrefix = {arXiv},
       eprint = {2004.06078},
 primaryClass = {astro-ph.HE},
       adsurl = {https://ui.adsabs.harvard.edu/abs/2020ApJ...898..125P},
      adsnote = {Provided by the SAO/NASA Astrophysics Data System}
}

@ARTICLE{gmo+21,
       author = {{Greco}, Emanuele and {Miceli}, Marco and {Orlando}, Salvatore and {Olmi}, Barbara and {Bocchino}, Fabrizio and {Nagataki}, Shigehiro and {Ono}, Masaomi and {Dohi}, Akira and {Peres}, Giovanni},
        title = "{Indication of a Pulsar Wind Nebula in the Hard X-Ray Emission from SN 1987A}",
      journal = {ApJL},
     keywords = {Supernova remnants, Compact objects, Neutron stars, X-ray astronomy, X-ray sources, X-ray point sources, X-ray observatories, Shocks, Interstellar synchrotron emission, Plasma astrophysics, 1667, 288, 1108, 1810, 1822, 1270, 1819, 2086, 856, 1261, Astrophysics - High Energy Astrophysical Phenomena},
         year = 2021,
        month = feb,
       volume = {908},
       number = {2},
          eid = {L45},
        pages = {L45},
          doi = {10.3847/2041-8213/abdf5a},
archivePrefix = {arXiv},
       eprint = {2101.09029},
 primaryClass = {astro-ph.HE},
       adsurl = {https://ui.adsabs.harvard.edu/abs/2021ApJ...908L..45G},
      adsnote = {Provided by the SAO/NASA Astrophysics Data System}
}


@ARTICLE{lkg+16,
       author = {{Lazarus}, P. and {Karuppusamy}, R. and {Graikou}, E. and {Caballero}, R.~N. and {Champion}, D.~J. and {Lee}, K.~J. and {Verbiest}, J.~P.~W. and {Kramer}, M.},
        title = "{Prospects for high-precision pulsar timing with the new Effelsberg PSRIX backend}",
      journal = {\mnras},
     keywords = {gravitational waves, stars: neutron, pulsars: general, Astrophysics - Instrumentation and Methods for Astrophysics, Astrophysics - Solar and Stellar Astrophysics},
         year = 2016,
        month = may,
       volume = {458},
       number = {1},
        pages = {868-880},
          doi = {10.1093/mnras/stw189},
archivePrefix = {arXiv},
       eprint = {1601.06194},
 primaryClass = {astro-ph.IM},
       adsurl = {https://ui.adsabs.harvard.edu/abs/2016{\mnras}.458..868L},
      adsnote = {Provided by the SAO/NASA Astrophysics Data System}
}

@INPROCEEDINGS{pan99,
       author = {{Panagia}, N.},
        title = "{Distance to SN 1987 A and the LMC}",
    booktitle = {New Views of the Magellanic Clouds},
         year = 1999,
       editor = {{Chu}, Y. -H. and {Suntzeff}, N. and {Hesser}, J. and {Bohlender}, D.},
       volume = {190},
        month = jan,
        pages = {549},
       adsurl = {https://ui.adsabs.harvard.edu/abs/1999IAUS..190..549P},
      adsnote = {Provided by the SAO/NASA Astrophysics Data System}
}

@ARTICLE{rcl+13,
       author = {{Ridley}, J.~P. and {Crawford}, F. and {Lorimer}, D.~R. and {Bailey}, S.~R. and {Madden}, J.~H. and {Anella}, R. and {Chennamangalam}, J.},
        title = "{Eight new radio pulsars in the Large Magellanic Cloud}",
      journal = {\mnras},
     keywords = {stars: neutron, pulsars: general, Magellanic Clouds, Astrophysics - Astrophysics of Galaxies, Astrophysics - High Energy Astrophysical Phenomena},
         year = 2013,
        month = jul,
       volume = {433},
       number = {1},
        pages = {138-146},
          doi = {10.1093/mnras/stt709},
archivePrefix = {arXiv},
       eprint = {1304.6412},
 primaryClass = {astro-ph.GA},
       adsurl = {https://ui.adsabs.harvard.edu/abs/2013{\mnras}.433..138R},
      adsnote = {Provided by the SAO/NASA Astrophysics Data System}
}

@ARTICLE{mfl+06,
       author = {{Manchester}, R.~N. and {Fan}, G. and {Lyne}, A.~G. and {Kaspi}, V.~M. and {Crawford}, F.},
        title = "{Discovery of 14 Radio Pulsars in a Survey of the Magellanic Clouds}",
      journal = {\apj},
     keywords = {Galaxies: Magellanic Clouds, Stars: Pulsars: General, Surveys, Astrophysics},
         year = 2006,
        month = sep,
       volume = {649},
       number = {1},
        pages = {235-242},
          doi = {10.1086/505461},
archivePrefix = {arXiv},
       eprint = {astro-ph/0604421},
 primaryClass = {astro-ph},
       adsurl = {https://ui.adsabs.harvard.edu/abs/2006ApJ...649..235M},
      adsnote = {Provided by the SAO/NASA Astrophysics Data System}
}

@ARTICLE{cmg+19,
       author = {{Cigan}, Phil and {Matsuura}, Mikako and {Gomez}, Haley L. and {Indebetouw}, Remy and {Abell{\'a}n}, Fran and {Gabler}, Michael and {Richards}, Anita and {Alp}, Dennis and {Davis}, Timothy A. and {Janka}, Hans-Thomas and {Spyromilio}, Jason and {Barlow}, M.~J. and {Burrows}, David and {Dwek}, Eli and {Fransson}, Claes and {Gaensler}, Bryan and {Larsson}, Josefin and {Bouchet}, P. and {Lundqvist}, Peter and {Marcaide}, J.~M. and {Ng}, C. -Y. and {Park}, Sangwook and {Roche}, Pat and {van Loon}, Jacco Th. and {Wheeler}, J.~C. and {Zanardo}, Giovanna},
        title = "{High Angular Resolution ALMA Images of Dust and Molecules in the SN 1987A Ejecta}",
      journal = {\apj},
     keywords = {Interstellar dust, Interstellar molecules, Supernovae, 836, 1668, 849, Astrophysics - High Energy Astrophysical Phenomena, Astrophysics - Solar and Stellar Astrophysics},
         year = 2019,
        month = nov,
       volume = {886},
       number = {1},
          eid = {51},
        pages = {51},
          doi = {10.3847/1538-4357/ab4b46},
archivePrefix = {arXiv},
       eprint = {1910.02960},
 primaryClass = {astro-ph.HE},
       adsurl = {https://ui.adsabs.harvard.edu/abs/2019ApJ...886...51C},
      adsnote = {Provided by the SAO/NASA Astrophysics Data System}
}

@ARTICLE{erl+18,
       author = {{Esposito}, Paolo and {Rea}, Nanda and {Lazzati}, Davide and {Matsuura}, Mikako and {Perna}, Rosalba and {Pons}, Jos{\'e} A.},
        title = "{Can a Bright and Energetic X-Ray Pulsar Be Hiding Amid the Debris of SN 1987A?}",
      journal = {\apj},
     keywords = {stars: neutron, supernovae: individual: SN 1987A, Astrophysics - High Energy Astrophysical Phenomena},
         year = 2018,
        month = apr,
       volume = {857},
       number = {1},
          eid = {58},
        pages = {58},
          doi = {10.3847/1538-4357/aab6b6},
archivePrefix = {arXiv},
       eprint = {1803.04692},
 primaryClass = {astro-ph.HE},
       adsurl = {https://ui.adsabs.harvard.edu/abs/2018ApJ...857...58E},
      adsnote = {Provided by the SAO/NASA Astrophysics Data System}
}

@INPROCEEDINGS{man07,
       author = {{Manchester}, R.~N.},
        title = "{Searching for a Pulsar in SN1987A}",
     keywords = {97.60.Bw, 97.60.Gb, 97.60.Jd, Supernovae, Pulsars, Neutron stars, Astrophysics},
    booktitle = {Supernova 1987A: 20 Years After: Supernovae and Gamma-Ray Bursters},
         year = 2007,
       editor = {{Immler}, Stefan and {Weiler}, Kurt and {McCray}, Richard},
       series = {American Institute of Physics Conference Series},
       volume = {937},
        month = oct,
        pages = {134-143},
          doi = {10.1063/1.3682894},
archivePrefix = {arXiv},
       eprint = {0708.2372},
 primaryClass = {astro-ph},
       adsurl = {https://ui.adsabs.harvard.edu/abs/2007AIPC..937..134M},
      adsnote = {Provided by the SAO/NASA Astrophysics Data System}
}

@ARTICLE{zdh+18,
       author = {{Zhang}, S. -B. and {Dai}, S. and {Hobbs}, G. and {Staveley-Smith}, L. and {Manchester}, R.~N. and {Russell}, C.~J. and {Zanardo}, G. and {Wu}, X. -F.},
        title = "{Search for a radio pulsar in the remnant of supernova 1987A}",
      journal = {\mnras},
     keywords = {stars: neutron, pulsars: general, supernova: individual (SN 1987A), Astrophysics - High Energy Astrophysical Phenomena, Astrophysics - Solar and Stellar Astrophysics, High Energy Physics - Phenomenology},
         year = 2018,
        month = sep,
       volume = {479},
       number = {2},
        pages = {1836-1841},
          doi = {10.1093/mnras/sty1573},
archivePrefix = {arXiv},
       eprint = {1806.04062},
 primaryClass = {astro-ph.HE},
       adsurl = {https://ui.adsabs.harvard.edu/abs/2018{\mnras}.479.1836Z},
      adsnote = {Provided by the SAO/NASA Astrophysics Data System}
}

@ARTICLE{lgw+13,
       author = {{Lyne}, Andrew and {Graham-Smith}, Francis and {Weltevrede}, Patrick and {Jordan}, Christine and {Stappers}, Ben and {Bassa}, Cees and {Kramer}, Michael},
        title = "{Evolution of the Magnetic Field Structure of the Crab Pulsar}",
      journal = {Science},
     keywords = {ASTRONOMY Astronomy, Physics, Applied-Physics, Astrophysics - High Energy Astrophysical Phenomena},
         year = 2013,
        month = nov,
       volume = {342},
       number = {6158},
        pages = {598-601},
          doi = {10.1126/science.1243254},
archivePrefix = {arXiv},
       eprint = {1311.0408},
 primaryClass = {astro-ph.HE},
       adsurl = {https://ui.adsabs.harvard.edu/abs/2013Sci...342..598L},
      adsnote = {Provided by the SAO/NASA Astrophysics Data System}
}

@ARTICLE{mf16,
       author = {{McCray}, Richard and {Fransson}, Claes},
        title = "{The Remnant of Supernova 1987A}",
      journal = {Ann. Rev. Astr. Ap.},
         year = 2016,
        month = sep,
       volume = {54},
        pages = {19-52},
          doi = {10.1146/annurev-astro-082615-105405},
       adsurl = {https://ui.adsabs.harvard.edu/abs/2016ARA&A..54...19M},
      adsnote = {Provided by the SAO/NASA Astrophysics Data System}
}

@ARTICLE{lj20,
       author = {{Lander}, S.~K. and {Jones}, D.~I.},
        title = "{Magnetar birth: rotation rates and gravitational-wave emission}",
      journal = {\mnras},
     keywords = {stars: evolution, stars: interiors, stars: magnetic field, stars: neutron, stars: rotation, Astrophysics - High Energy Astrophysical Phenomena, General Relativity and Quantum Cosmology},
         year = 2020,
        month = jun,
       volume = {494},
       number = {4},
        pages = {4838-4847},
          doi = {10.1093/mnras/staa966},
archivePrefix = {arXiv},
       eprint = {1910.14336},
 primaryClass = {astro-ph.HE},
       adsurl = {https://ui.adsabs.harvard.edu/abs/2020{\mnras}.494.4838L},
      adsnote = {Provided by the SAO/NASA Astrophysics Data System}
}

@ARTICLE{rgo+16,
       author = {{Rembiasz}, T. and {Guilet}, J. and {Obergaulinger}, M. and {Cerd{\'a}-Dur{\'a}n}, P. and {Aloy}, M.~A. and {M{\"u}ller}, E.},
        title = "{On the maximum magnetic field amplification by the magnetorotational instability in core-collapse supernovae}",
      journal = {\mnras},
     keywords = {accretion, accretion discs, instabilities, MHD, stars: magnetic field, supernovae: general, Astrophysics - Solar and Stellar Astrophysics},
         year = 2016,
        month = aug,
       volume = {460},
       number = {3},
        pages = {3316-3334},
          doi = {10.1093/mnras/stw1201},
archivePrefix = {arXiv},
       eprint = {1603.00466},
 primaryClass = {astro-ph.SR},
       adsurl = {https://ui.adsabs.harvard.edu/abs/2016{\mnras}.460.3316R},
      adsnote = {Provided by the SAO/NASA Astrophysics Data System}
}

@ARTICLE{gaa+21,
       author = {{Gravity Collaboration} and {Abuter}, R. and {Amorim}, A. and {Baub{\"o}ck}, M. and {Berger}, J.~P. and {Bonnet}, H. and {Brandner}, W. and {Cl{\'e}net}, Y. and {Davies}, R. and {de Zeeuw}, P.~T. and {Dexter}, J. and {Dallilar}, Y. and {Drescher}, A. and {Eckart}, A. and {Eisenhauer}, F. and {F{\"o}rster Schreiber}, N.~M. and {Garcia}, P. and {Gao}, F. and {Gendron}, E. and {Genzel}, R. and {Gillessen}, S. and {Habibi}, M. and {Haubois}, X. and {Hei{\ss}el}, G. and {Henning}, T. and {Hippler}, S. and {Horrobin}, M. and {Jim{\'e}nez-Rosales}, A. and {Jochum}, L. and {Jocou}, L. and {Kaufer}, A. and {Kervella}, P. and {Lacour}, S. and {Lapeyr{\`e}re}, V. and {Le Bouquin}, J. -B. and {L{\'e}na}, P. and {Lutz}, D. and {Nowak}, M. and {Ott}, T. and {Paumard}, T. and {Perraut}, K. and {Perrin}, G. and {Pfuhl}, O. and {Rabien}, S. and {Rodr{\'\i}guez-Coira}, G. and {Shangguan}, J. and {Shimizu}, T. and {Scheithauer}, S. and {Stadler}, J. and {Straub}, O. and {Straubmeier}, C. and {Sturm}, E. and {Tacconi}, L.~J. and {Vincent}, F. and {von Fellenberg}, S. and {Waisberg}, I. and {Widmann}, F. and {Wieprecht}, E. and {Wiezorrek}, E. and {Woillez}, J. and {Yazici}, S. and {Young}, A. and {Zins}, G.},
        title = "{Improved GRAVITY astrometric accuracy from modeling optical aberrations}",
      journal = {\aa},
     keywords = {Galaxy: center, Galaxy: fundamental parameters, instrumentation: interferometers, instrumentation: high angular resolution, methods: data analysis, Astrophysics - Astrophysics of Galaxies, Astrophysics - Instrumentation and Methods for Astrophysics},
         year = 2021,
        month = mar,
       volume = {647},
          eid = {A59},
        pages = {A59},
          doi = {10.1051/0004-6361/202040208},
archivePrefix = {arXiv},
       eprint = {2101.12098},
 primaryClass = {astro-ph.GA},
       adsurl = {https://ui.adsabs.harvard.edu/abs/2021A&A...647A..59G},
      adsnote = {Provided by the SAO/NASA Astrophysics Data System}
}

@ARTICLE{ww95,
       author = {{Woosley}, S.~E. and {Weaver}, Thomas A.},
        title = "{The Evolution and Explosion of Massive Stars. II. Explosive Hydrodynamics and Nucleosynthesis}",
      journal = {\apjs},
     keywords = {HYDRODYNAMICS, NUCLEAR REACTIONS, NUCLEOSYNTHESIS, ABUNDANCES, STARS: EVOLUTION, STARS: INTERIORS, STARS: SUPERNOVAE: GENERAL},
         year = 1995,
        month = nov,
       volume = {101},
        pages = {181},
          doi = {10.1086/192237},
       adsurl = {https://ui.adsabs.harvard.edu/abs/1995ApJS..101..181W},
      adsnote = {Provided by the SAO/NASA Astrophysics Data System}
}

@ARTICLE{bub05,
       author = {{Bonanno}, A. and {Urpin}, V. and {Belvedere}, G.},
        title = "{Protoneutron star dynamos and pulsar magnetism}",
      journal = {\aa},
     keywords = {magnetohydrodynamics (MHD), pulsars: general, stars: neutron, stars: magnetic fields, Astrophysics},
         year = 2005,
        month = sep,
       volume = {440},
       number = {1},
        pages = {199-205},
          doi = {10.1051/0004-6361:20042098},
archivePrefix = {arXiv},
       eprint = {astro-ph/0504328},
 primaryClass = {astro-ph},
       adsurl = {https://ui.adsabs.harvard.edu/abs/2005A&A...440..199B},
      adsnote = {Provided by the SAO/NASA Astrophysics Data System}
}

@ARTICLE{wjm13,
       author = {{Wongwathanarat}, A. and {Janka}, H. -Th. and {M{\"u}ller}, E.},
        title = "{Three-dimensional neutrino-driven supernovae: Neutron star kicks, spins, and asymmetric ejection of nucleosynthesis products}",
      journal = {\aa},
     keywords = {supernovae: general, stars: neutron, pulsars: general, hydrodynamics, neutrinos, Astrophysics - High Energy Astrophysical Phenomena, High Energy Physics - Phenomenology},
         year = 2013,
        month = apr,
       volume = {552},
          eid = {A126},
        pages = {A126},
          doi = {10.1051/0004-6361/201220636},
archivePrefix = {arXiv},
       eprint = {1210.8148},
 primaryClass = {astro-ph.HE},
       adsurl = {https://ui.adsabs.harvard.edu/abs/2013A&A...552A.126W},
      adsnote = {Provided by the SAO/NASA Astrophysics Data System}
}

@ARTICLE{uwj+19,
       author = {{Utrobin}, V.~P. and {Wongwathanarat}, A. and {Janka}, H. -Th. and {M{\"u}ller}, E. and {Ertl}, T. and {Woosley}, S.~E.},
        title = "{Three-dimensional mixing and light curves: constraints on the progenitor of supernova 1987A}",
      journal = {\aa},
     keywords = {supernovae: general, supernovae: individual: SN 1987A, hydrodynamics, Astrophysics - High Energy Astrophysical Phenomena, Astrophysics - Solar and Stellar Astrophysics},
         year = 2019,
        month = apr,
       volume = {624},
          eid = {A116},
        pages = {A116},
          doi = {10.1051/0004-6361/201834976},
archivePrefix = {arXiv},
       eprint = {1812.11083},
 primaryClass = {astro-ph.HE},
       adsurl = {https://ui.adsabs.harvard.edu/abs/2019A&A...624A.116U},
      adsnote = {Provided by the SAO/NASA Astrophysics Data System}
}

@ARTICLE{wmj15,
       author = {{Wongwathanarat}, A. and {M{\"u}ller}, E. and {Janka}, H. -Th.},
        title = "{Three-dimensional simulations of core-collapse supernovae: from shock revival to shock breakout}",
      journal = {\aa},
     keywords = {supernovae: general, hydrodynamics, stars: massive, Astrophysics - High Energy Astrophysical Phenomena, Astrophysics - Solar and Stellar Astrophysics},
         year = 2015,
        month = may,
       volume = {577},
          eid = {A48},
        pages = {A48},
          doi = {10.1051/0004-6361/201425025},
archivePrefix = {arXiv},
       eprint = {1409.5431},
 primaryClass = {astro-ph.HE},
       adsurl = {https://ui.adsabs.harvard.edu/abs/2015A&A...577A..48W},
      adsnote = {Provided by the SAO/NASA Astrophysics Data System}
}

@ARTICLE{lj18,
       author = {{Lander}, S.~K. and {Jones}, D.~I.},
        title = "{Neutron-star spindown and magnetic inclination-angle evolution}",
      journal = {\mnras},
     keywords = {stars: evolution, stars: interiors, stars: magnetic field, stars: neutron, stars: rotation, Astrophysics - High Energy Astrophysical Phenomena, Astrophysics - Solar and Stellar Astrophysics},
         year = 2018,
        month = dec,
       volume = {481},
       number = {3},
        pages = {4169-4193},
          doi = {10.1093/mnras/sty2553},
archivePrefix = {arXiv},
       eprint = {1807.01289},
 primaryClass = {astro-ph.HE},
       adsurl = {https://ui.adsabs.harvard.edu/abs/2018{\mnras}.481.4169L},
      adsnote = {Provided by the SAO/NASA Astrophysics Data System}
}

@ARTICLE{hsh+12,
       author = {{Hassall}, T.~E. and {Stappers}, B.~W. and {Hessels}, J.~W.~T. and {Kramer}, M. and {Alexov}, A. and {Anderson}, K. and {Coenen}, T. and {Karastergiou}, A. and {Keane}, E.~F. and {Kondratiev}, V.~I. and {Lazaridis}, K. and {van Leeuwen}, J. and {Noutsos}, A. and {Serylak}, M. and {Sobey}, C. and {Verbiest}, J.~P.~W. and {Weltevrede}, P. and {Zagkouris}, K. and {Fender}, R. and {Wijers}, R.~A.~M.~J. and {B{\"a}hren}, L. and {Bell}, M.~E. and {Broderick}, J.~W. and {Corbel}, S. and {Daw}, E.~J. and {Dhillon}, V.~S. and {Eisl{\"o}ffel}, J. and {Falcke}, H. and {Grie{\ss}meier}, J. -M. and {Jonker}, P. and {Law}, C. and {Markoff}, S. and {Miller-Jones}, J.~C.~A. and {Osten}, R. and {Rol}, E. and {Scaife}, A.~M.~M. and {Scheers}, B. and {Schellart}, P. and {Spreeuw}, H. and {Swinbank}, J. and {ter Veen}, S. and {Wise}, M.~W. and {Wijnands}, R. and {Wucknitz}, O. and {Zarka}, P. and {Asgekar}, A. and {Bell}, M.~R. and {Bentum}, M.~J. and {Bernardi}, G. and {Best}, P. and {Bonafede}, A. and {Boonstra}, A.~J. and {Brentjens}, M. and {Brouw}, W.~N. and {Br{\"u}ggen}, M. and {Butcher}, H.~R. and {Ciardi}, B. and {Garrett}, M.~A. and {Gerbers}, M. and {Gunst}, A.~W. and {van Haarlem}, M.~P. and {Heald}, G. and {Hoeft}, M. and {Holties}, H. and {de Jong}, A. and {Koopmans}, L.~V.~E. and {Kuniyoshi}, M. and {Kuper}, G. and {Loose}, G.~M. and {Maat}, P. and {Masters}, J. and {McKean}, J.~P. and {Meulman}, H. and {Mevius}, M. and {Munk}, H. and {Noordam}, J.~E. and {Orr{\'u}}, E. and {Paas}, H. and {Pandey-Pommier}, M. and {Pandey}, V.~N. and {Pizzo}, R. and {Polatidis}, A. and {Reich}, W. and {R{\"o}ttgering}, H. and {Sluman}, J. and {Steinmetz}, M. and {Sterks}, C.~G.~M. and {Tagger}, M. and {Tang}, Y. and {Tasse}, C. and {Vermeulen}, R. and {van Weeren}, R.~J. and {Wijnholds}, S.~J. and {Yatawatta}, S.},
        title = "{Wide-band simultaneous observations of pulsars: disentangling dispersion measure and profile variations}",
      journal = {\aa},
     keywords = {pulsars: general, ISM: general, magnetic fields, telescopes, Astrophysics - High Energy Astrophysical Phenomena},
         year = 2012,
        month = jul,
       volume = {543},
          eid = {A66},
        pages = {A66},
          doi = {10.1051/0004-6361/201218970},
archivePrefix = {arXiv},
       eprint = {1204.3864},
 primaryClass = {astro-ph.HE},
       adsurl = {https://ui.adsabs.harvard.edu/abs/2012A&A...543A..66H},
      adsnote = {Provided by the SAO/NASA Astrophysics Data System}
}

@ARTICLE{pts20,
       author = {{Philippov}, Alexander and {Timokhin}, Andrey and {Spitkovsky}, Anatoly},
        title = "{Origin of Pulsar Radio Emission}",
      journal = {PRL},
     keywords = {Astrophysics - High Energy Astrophysical Phenomena},
         year = 2020,
        month = jun,
       volume = {124},
       number = {24},
          eid = {245101},
        pages = {245101},
          doi = {10.1103/PhysRevLett.124.245101},
archivePrefix = {arXiv},
       eprint = {2001.02236},
 primaryClass = {astro-ph.HE},
       adsurl = {https://ui.adsabs.harvard.edu/abs/2020PhRvL.124x5101P},
      adsnote = {Provided by the SAO/NASA Astrophysics Data System}
}

@ARTICLE{lj18,
       author = {{Lander}, S.~K. and {Jones}, D.~I.},
        title = "{Neutron-star spindown and magnetic inclination-angle evolution}",
      journal = {\mnras},
     keywords = {stars: evolution, stars: interiors, stars: magnetic field, stars: neutron, stars: rotation, Astrophysics - High Energy Astrophysical Phenomena, Astrophysics - Solar and Stellar Astrophysics},
         year = 2018,
        month = dec,
       volume = {481},
       number = {3},
        pages = {4169-4193},
          doi = {10.1093/mnras/sty2553},
archivePrefix = {arXiv},
       eprint = {1807.01289},
 primaryClass = {astro-ph.HE},
       adsurl = {https://ui.adsabs.harvard.edu/abs/2018{\mnras}.481.4169L},
      adsnote = {Provided by the SAO/NASA Astrophysics Data System}
}

@ARTICLE{fc87,
       author = {{Fransson}, C. and {Chevalier}, R.~A.},
        title = "{Late Emission from SN 1987A}",
      journal = {ApJL},
     keywords = {Emission Spectra, Stellar Evolution, Stellar Mass Ejection, Stellar Spectrophotometry, Supernovae, Cosmic Gases, Infrared Spectra, Pulsars, Stellar Models, Visible Spectrum, Astrophysics, STARS: SUPERNOVAE},
         year = 1987,
        month = nov,
       volume = {322},
        pages = {L15},
          doi = {10.1086/185028},
       adsurl = {https://ui.adsabs.harvard.edu/abs/1987ApJ...322L..15F},
      adsnote = {Provided by the SAO/NASA Astrophysics Data System}
}

@ARTICLE{rgj+20,
       author = {{Raynaud}, Rapha{\"e}l and {Guilet}, J{\'e}r{\^o}me and {Janka}, Hans-Thomas and {Gastine}, Thomas},
        title = "{Magnetar formation through a convective dynamo in protoneutron stars}",
      journal = {Science Advances},
     keywords = {Astrophysics - High Energy Astrophysical Phenomena, Astrophysics - Solar and Stellar Astrophysics},
         year = 2020,
        month = mar,
       volume = {6},
       number = {11},
        pages = {eaay2732},
          doi = {10.1126/sciadv.aay2732},
archivePrefix = {arXiv},
       eprint = {2003.06662},
 primaryClass = {astro-ph.HE},
       adsurl = {https://ui.adsabs.harvard.edu/abs/2020SciA....6.2732R},
      adsnote = {Provided by the SAO/NASA Astrophysics Data System}
}

@ARTICLE{nbr+20,
       author = {{Nagakura}, Hiroki and {Burrows}, Adam and {Radice}, David and {Vartanyan}, David},
        title = "{A systematic study of proto-neutron star convection in three-dimensional core-collapse supernova simulations}",
      journal = {\mnras},
     keywords = {turbulence, supernovae: general, Astrophysics - High Energy Astrophysical Phenomena},
         year = 2020,
        month = mar,
       volume = {492},
       number = {4},
        pages = {5764-5779},
          doi = {10.1093/mnras/staa261},
archivePrefix = {arXiv},
       eprint = {1912.07615},
 primaryClass = {astro-ph.HE},
       adsurl = {https://ui.adsabs.harvard.edu/abs/2020{\mnras}.492.5764N},
      adsnote = {Provided by the SAO/NASA Astrophysics Data System}
}

@ARTICLE{vf14,
       author = {{Verbunt}, Frank and {Freire}, Paulo C.~C.},
        title = "{On the disruption of pulsar and X-ray binar ies in globular clusters}",
      journal = {\aa},
     keywords = {globular clusters: general, stars: neutron, pulsars: general, Astrophysics - Solar and Stellar Astrophysics, Astrophysics - High Energy Astrophysical Phenomena},
         year = 2014,
        month = jan,
       volume = {561},
          eid = {A11},
        pages = {A11},
          doi = {10.1051/0004-6361/201321177},
archivePrefix = {arXiv},
       eprint = {1310.4669},
 primaryClass = {astro-ph.SR},
       adsurl = {https://ui.adsabs.harvard.edu/abs/2014A&A...561A..11V},
      adsnote = {Provided by the SAO/NASA Astrophysics Data System}
}

@ARTICLE{hrs+06,
       author = {{Hessels}, Jason W.~T. and {Ransom}, Scott M. and {Stairs}, Ingrid H. and {Freire}, Paulo C.~C. and {Kaspi}, Victoria M. and {Camilo}, Fernando},
        title = "{A Radio Pulsar Spinning at 716 Hz}",
      journal = {Science},
     keywords = {ASTRONOMY, Astrophysics},
         year = 2006,
        month = mar,
       volume = {311},
       number = {5769},
        pages = {1901-1904},
          doi = {10.1126/science.1123430},
archivePrefix = {arXiv},
       eprint = {astro-ph/0601337},
 primaryClass = {astro-ph},
       adsurl = {https://ui.adsabs.harvard.edu/abs/2006Sci...311.1901H},
      adsnote = {Provided by the SAO/NASA Astrophysics Data System}
}

@ARTICLE{psl+17,
       author = {{Perera}, B.~B.~P. and {Stappers}, B.~W. and {Lyne}, A.~G. and {Bassa}, C.~G. and {Cognard}, I. and {Guillemot}, L. and {Kramer}, M. and {Theureau}, G. and {Desvignes}, G.},
        title = "{Evidence for an intermediate-mass black hole in the globular cluster NGC 6624}",
      journal = {\mnras},
     keywords = {black hole physics, stars: neutron, pulsars: individual: PSR B1820-30A, globular clusters: individual: NGC 6624, Astrophysics - High Energy Astrophysical Phenomena, Astrophysics - Astrophysics of Galaxies, Astrophysics - Solar and Stellar Astrophysics},
         year = 2017,
        month = jun,
       volume = {468},
       number = {2},
        pages = {2114-2127},
          doi = {10.1093/mnras/stx501},
archivePrefix = {arXiv},
       eprint = {1705.01612},
 primaryClass = {astro-ph.HE},
       adsurl = {https://ui.adsabs.harvard.edu/abs/2017{\mnras}.468.2114P},
      adsnote = {Provided by the SAO/NASA Astrophysics Data System}
}

@ARTICLE{apt+20,
       author = {{Abbate}, Federico and {Possenti}, Andrea and {Tiburzi}, Caterina and {Barr}, Ewan and {van Straten}, Willem and {Ridolfi}, Alessandro and {Freire}, Paulo},
        title = "{Constraints on the magnetic field in the Galactic halo from globular cluster pulsars}",
      journal = {Nature Astronomy},
     keywords = {Astrophysics - High Energy Astrophysical Phenomena},
         year = 2020,
        month = mar,
       volume = {4},
        pages = {704-710},
          doi = {10.1038/s41550-020-1030-6},
archivePrefix = {arXiv},
       eprint = {2003.02867},
 primaryClass = {astro-ph.HE},
       adsurl = {https://ui.adsabs.harvard.edu/abs/2020NatAs...4..704A},
      adsnote = {Provided by the SAO/NASA Astrophysics Data System}
}

@ARTICLE{ccp+20,
       author = {{Cadelano}, Mario and {Chen}, Jianxing and {Pallanca}, Cristina and {Istrate}, Alina G. and {Ferraro}, Francesco R. and {Lanzoni}, Barbara and {Freire}, Paulo C.~C. and {Salaris}, Maurizio},
        title = "{PSR J1641+3627F: A Low-mass He White Dwarf Orbiting a Possible High-mass Neutron Star in the Globular Cluster M13}",
      journal = {\apj},
     keywords = {Binary pulsars, Pulsars, Compact binary stars, Globular star clusters, Star clusters, White dwarf stars, Binary stars, Neutron stars, HST photometry, Millisecond pulsars, 153, 1306, 283, 656, 1567, 1799, 154, 1108, 756, 1062, Astrophysics - High Energy Astrophysical Phenomena},
         year = 2020,
        month = dec,
       volume = {905},
       number = {1},
          eid = {63},
        pages = {63},
          doi = {10.3847/1538-4357/abc345},
archivePrefix = {arXiv},
       eprint = {2010.09740},
 primaryClass = {astro-ph.HE},
       adsurl = {https://ui.adsabs.harvard.edu/abs/2020ApJ...905...63C},
      adsnote = {Provided by the SAO/NASA Astrophysics Data System}
}

@ARTICLE{lbj+16,
       author = {{Liu}, K. and {Bassa}, C.~G. and {Janssen}, G.~H. and {Karuppusamy}, R. and {McKee}, J. and {Kramer}, M. and {Lee}, K.~J. and {Perrodin}, D. and {Purver}, M. and {Sanidas}, S. and {Smits}, R. and {Stappers}, B.~W. and {Weltevrede}, P. and {Zhu}, W.~W.},
        title = "{Variability, polarimetry, and timing properties of single pulses from PSR J1713+0747 using the Large European Array for Pulsars}",
      journal = {\mnras},
     keywords = {methods: data analysis, pulsars: individual: PSR J1713+0747, Astrophysics - High Energy Astrophysical Phenomena, Astrophysics - Astrophysics of Galaxies},
         year = 2016,
        month = dec,
       volume = {463},
       number = {3},
        pages = {3239-3248},
          doi = {10.1093/mnras/stw2223},
archivePrefix = {arXiv},
       eprint = {1609.00188},
 primaryClass = {astro-ph.HE},
       adsurl = {https://ui.adsabs.harvard.edu/abs/2016{\mnras}.463.3239L},
      adsnote = {Provided by the SAO/NASA Astrophysics Data System}
}

@ARTICLE{msb+19,
       author = {{McKee}, J.~W. and {Stappers}, B.~W. and {Bassa}, C.~G. and {Chen}, S. and {Cognard}, I. and {Gaikwad}, M. and {Janssen}, G.~H. and {Karuppusamy}, R. and {Kramer}, M. and {Lee}, K.~J. and {Liu}, K. and {Perrodin}, D. and {Sanidas}, S.~A. and {Smits}, R. and {Wang}, L. and {Zhu}, W.~W.},
        title = "{A detailed study of giant pulses from PSR B1937+21 using the Large European Array for Pulsars}",
      journal = {\mnras},
     keywords = {pulsars:general, pulsars:individual (PSR B1937+21), pulsars:individual (PSR J1939+2134), stars:neutron, stars:rotation, radiation mechanisms: non-thermal, Astrophysics - High Energy Astrophysical Phenomena},
         year = 2019,
        month = mar,
       volume = {483},
       number = {4},
        pages = {4784-4802},
          doi = {10.1093/mnras/sty3058},
archivePrefix = {arXiv},
       eprint = {1811.02856},
 primaryClass = {astro-ph.HE},
       adsurl = {https://ui.adsabs.harvard.edu/abs/2019{\mnras}.483.4784M},
      adsnote = {Provided by the SAO/NASA Astrophysics Data System}
}

@ARTICLE{dgs16,
       author = {{De}, Kishalay and {Gupta}, Yashwant and {Sharma}, Prateek},
        title = "{Detection of Polarized Quasi-periodic Microstructure Emission in Millisecond Pulsars}",
      journal = {\apjl},
     keywords = {methods: observational, pulsars: general, pulsars: individual: J0437-4715, J2145-0750, techniques: polarimetric, Astrophysics - High Energy Astrophysical Phenomena},
         year = 2016,
        month = dec,
       volume = {833},
       number = {1},
          eid = {L10},
        pages = {L10},
          doi = {10.3847/2041-8213/833/1/L10},
archivePrefix = {arXiv},
       eprint = {1611.07330},
 primaryClass = {astro-ph.HE},
       adsurl = {https://ui.adsabs.harvard.edu/abs/2016ApJ...833L..10D},
      adsnote = {Provided by the SAO/NASA Astrophysics Data System}
}

@ARTICLE{abb+21,
       author = {{The CHIME/FRB Collaboration} and {Andersen}, Bridget C. and {Bandura}, Kevin and {Bhardwaj}, Mohit and {Boyle}, P.~J. and {Brar}, Charanjot and {Breitman}, Daniela and {Cassanelli}, Tomas and {Chatterjee}, Shami and {Chawla}, Pragya and {Cliche}, Jean-Fran{\c{c}}ois and {Cubranic}, Davor and {Curtin}, Alice P. and {Deng}, Meiling and {Dobbs}, Matt and {Dong}, Fengqiu Adam and {Fonseca}, Emmanuel and {Gaensler}, B.~M. and {Giri}, Utkarsh and {Good}, Deborah C. and {Hill}, Alex S. and {Josephy}, Alexander and {Kaczmarek}, J.~F. and {Kader}, Zarif and {Kania}, Joseph and {Kaspi}, Victoria M. and {Leung}, Calvin and {Li}, D.~Z. and {Lin}, Hsiu-Hsien and {Masui}, Kiyoshi W. and {Mckinven}, Ryan and {Mena-Parra}, Juan and {Merryfield}, Marcus and {Meyers}, B.~W. and {Michilli}, D. and {Naidu}, Arun and {Newburgh}, Laura and {Ng}, C. and {Ordog}, Anna and {Patel}, Chitrang and {Pearlman}, Aaron B. and {Pen}, Ue-Li and {Petroff}, Emily and {Pleunis}, Ziggy and {Rafiei-Ravandi}, Masoud and {Rahman}, Mubdi and {Ransom}, Scott and {Renard}, Andre and {Sanghavi}, Pranav and {Scholz}, Paul and {Shaw}, J. Richard and {Shin}, Kaitlyn and {Siegel}, Seth R. and {Singh}, Saurabh and {Smith}, Kendrick and {Stairs}, Ingrid and {Tan}, Chia Min and {Tendulkar}, Shriharsh P. and {Vanderlinde}, Keith and {Wiebe}, D.~V. and {Wulf}, Dallas and {Zwaniga}, Andrew},
        title = "{Sub-second periodicity in a fast radio burst}",
      journal = {arXiv e-prints},
     keywords = {Astrophysics - High Energy Astrophysical Phenomena},
         year = 2021,
        month = jul,
          eid = {arXiv:2107.08463},
        pages = {arXiv:2107.08463},
archivePrefix = {arXiv},
       eprint = {2107.08463},
 primaryClass = {astro-ph.HE},
       adsurl = {https://ui.adsabs.harvard.edu/abs/2021arXiv210708463T},
      adsnote = {Provided by the SAO/NASA Astrophysics Data System}
}

@ARTICLE{nhk+21b,
       author = {{Nimmo}, K. and {Hessels}, J.~W.~T. and {Kirsten}, F. and {Keimpema}, A. and {Cordes}, J.~M. and {Snelders}, M.~P. and {Hewitt}, D.~M. and {Karuppusamy}, R. and {Archibald}, A.~M. and {Bezukovs}, V. and {Bhardwaj}, M. and {Blaauw}, R. and {Buttaccio}, S.~T. and {Cassanelli}, T. and {Conway}, J.~E. and {Corongiu}, A. and {Feiler}, R. and {Fonseca}, E. and {Forssen}, O. and {Gawronski}, M. and {Giroletti}, M. and {Kharinov}, M.~A. and {Leung}, C. and {Lindqvist}, M. and {Maccaferri}, G. and {Marcote}, B. and {Masui}, K.~W. and {Mckinven}, R. and {Melnikov}, A. and {Michilli}, D. and {Mikhailov}, A. and {Ng}, C. and {Orbidans}, A. and {Ould-Boukattine}, O.~S. and {Paragi}, Z. and {Pearlman}, A.~B. and {Petroff}, E. and {Rahman}, M. and {Scholz}, P. and {Shin}, K. and {Smith}, K.~M. and {Stairs}, I.~H. and {Surcis}, G. and {Tendulkar}, S.~P. and {Vlemmings}, W. and {Wang}, N. and {Yang}, J. and {Yuan}, J.},
        title = "{Burst timescales and luminosities link young pulsars and fast radio bursts}",
      journal = {arXiv e-prints},
     keywords = {Astrophysics - High Energy Astrophysical Phenomena},
         year = 2021,
        month = may,
          eid = {arXiv:2105.11446},
        pages = {arXiv:2105.11446},
archivePrefix = {arXiv},
       eprint = {2105.11446},
 primaryClass = {astro-ph.HE},
       adsurl = {https://ui.adsabs.harvard.edu/abs/2021arXiv210511446N},
      adsnote = {Provided by the SAO/NASA Astrophysics Data System}
}

@ARTICLE{mpp+21,
       author = {{Majid}, Walid A. and {Pearlman}, Aaron B. and {Prince}, Thomas A. and {Wharton}, Robert S. and {Naudet}, Charles J. and {Bansal}, Karishma and {Connor}, Liam and {Bhardwaj}, Mohit and {Tendulkar}, Shriharsh P.},
        title = "{A Bright Fast Radio Burst from FRB 20200120E with Sub-100 Nanosecond Structure}",
      journal = {\apjl},
     keywords = {Radio transient sources, Radio bursts, 2008, 1339, Astrophysics - High Energy Astrophysical Phenomena},
         year = 2021,
        month = sep,
       volume = {919},
       number = {1},
          eid = {L6},
        pages = {L6},
          doi = {10.3847/2041-8213/ac1921},
archivePrefix = {arXiv},
       eprint = {2105.10987},
 primaryClass = {astro-ph.HE},
       adsurl = {https://ui.adsabs.harvard.edu/abs/2021ApJ...919L...6M},
      adsnote = {Provided by the SAO/NASA Astrophysics Data System}
}

@ARTICLE{lwm+20,
       author = {{Luo}, R. and {Wang}, B.~J. and {Men}, Y.~P. and {Zhang}, C.~F. and {Jiang}, J.~C. and {Xu}, H. and {Wang}, W.~Y. and {Lee}, K.~J. and {Han}, J.~L. and {Zhang}, B. and {Caballero}, R.~N. and {Chen}, M.~Z. and {Chen}, X.~L. and {Gan}, H.~Q. and {Guo}, Y.~J. and {Hao}, L.~F. and {Huang}, Y.~X. and {Jiang}, P. and {Li}, H. and {Li}, J. and {Li}, Z.~X. and {Luo}, J.~T. and {Pan}, J. and {Pei}, X. and {Qian}, L. and {Sun}, J.~H. and {Wang}, M. and {Wang}, N. and {Wen}, Z.~G. and {Xu}, R.~X. and {Xu}, Y.~H. and {Yan}, J. and {Yan}, W.~M. and {Yu}, D.~J. and {Yuan}, J.~P. and {Zhang}, S.~B. and {Zhu}, Y.},
        title = "{Diverse polarization angle swings from a repeating fast radio burst source}",
      journal = {\nat},
     keywords = {Astrophysics - High Energy Astrophysical Phenomena},
         year = 2020,
        month = oct,
       volume = {586},
       number = {7831},
        pages = {693-696},
          doi = {10.1038/s41586-020-2827-2},
archivePrefix = {arXiv},
       eprint = {2011.00171},
 primaryClass = {astro-ph.HE},
       adsurl = {https://ui.adsabs.harvard.edu/abs/2020Natur.586..693L},
      adsnote = {Provided by the SAO/NASA Astrophysics Data System}
}

@ARTICLE{nhk+21a,
       author = {{Nimmo}, K. and {Hessels}, J.~W.~T. and {Keimpema}, A. and {Archibald}, A.~M. and {Cordes}, J.~M. and {Karuppusamy}, R. and {Kirsten}, F. and {Li}, D.~Z. and {Marcote}, B. and {Paragi}, Z.},
        title = "{Highly polarized microstructure from the repeating FRB 20180916B}",
      journal = {Nature Astronomy},
     keywords = {Astrophysics - High Energy Astrophysical Phenomena},
         year = 2021,
        month = jan,
       volume = {5},
        pages = {594-603},
          doi = {10.1038/s41550-021-01321-3},
archivePrefix = {arXiv},
       eprint = {2010.05800},
 primaryClass = {astro-ph.HE},
       adsurl = {https://ui.adsabs.harvard.edu/abs/2021NatAs...5..594N},
      adsnote = {Provided by the SAO/NASA Astrophysics Data System}
}

@ARTICLE{ben77,
       author = {{Benford}, G.},
        title = "{Model for the microstructure emission of pulsars.}",
      journal = {\mnras},
     keywords = {Coherent Radiation, Microstructure, Pulsars, Stellar Models, Stellar Radiation, Dense Plasmas, Filaments, Fine Structure, Plasma Radiation, Radio Spectra, Relativistic Plasmas, Astrophysics},
         year = 1977,
        month = may,
       volume = {179},
        pages = {311-315},
          doi = {10.1093/mnras/179.3.311},
       adsurl = {https://ui.adsabs.harvard.edu/abs/1977{\mnras}.179..311B},
      adsnote = {Provided by the SAO/NASA Astrophysics Data System}
}

@INPROCEEDINGS{bf81,
       author = {{Boriakoff}, V. and {Ferguson}, D.~C.},
        title = "{Microstructure crosscorrelation in pulses simultaneously observed at frequencies separated by up to 1 GHz}",
     keywords = {Cross Correlation, Microstructure, Neutron Stars, Pulsars, Pulsed Radiation, Astronomical Observatories, Autocorrelation, Fine Structure, Radio Spectroscopy, Spectral Correlation, Ultrahigh Frequencies, Astrophysics},
    booktitle = {Pulsars: 13 Years of Research on Neutron Stars},
         year = 1981,
       editor = {{Sieber}, W. and {Wielebinski}, R.},
       volume = {95},
        month = jan,
        pages = {191-196},
       adsurl = {https://ui.adsabs.harvard.edu/abs/1981IAUS...95..191B},
      adsnote = {Provided by the SAO/NASA Astrophysics Data System}
}

@ARTICLE{msa+20,
       author = {{Main}, R.~A. and {Sanidas}, S.~A. and {Antoniadis}, J. and {Bassa}, C. and {Chen}, S. and {Cognard}, I. and {Gaikwad}, M. and {Hu}, H. and {Janssen}, G.~H. and {Karuppusamy}, R. and {Kramer}, M. and {Lee}, K.~J. and {Liu}, K. and {Mall}, G. and {McKee}, J.~W. and {Mickaliger}, M.~B. and {Perrodin}, D. and {Stappers}, B.~W. and {Tiburzi}, C. and {Wucknitz}, O. and {Wang}, L. and {Zhu}, W.~W.},
        title = "{Measuring interstellar delays of PSR J0613-0200 over 7 yr, using the Large European Array for Pulsars}",
      journal = {\mnras},
     keywords = {pulsars: general, pulsars: individual: PSR J0613-0200, Astrophysics - High Energy Astrophysical Phenomena},
         year = 2020,
        month = nov,
       volume = {499},
       number = {1},
        pages = {1468-1479},
          doi = {10.1093/mnras/staa2955},
archivePrefix = {arXiv},
       eprint = {2009.10707},
 primaryClass = {astro-ph.HE},
       adsurl = {https://ui.adsabs.harvard.edu/abs/2020{\mnras}.499.1468M},
      adsnote = {Provided by the SAO/NASA Astrophysics Data System}
}


@ARTICLE{cr04b,
       author = {{Clemens}, J. Christopher and {Rosen}, R.},
        title = "{Observations of Nonradial Pulsations in Radio Pulsars}",
      journal = {\apj},
     keywords = {Stars: Pulsars: General, Stars: Pulsars: Individual: Alphanumeric: PSR 1237+25, Stars: Pulsars: Individual: Alphanumeric: PSR 1919+21, Stars: Neutron, Stars: Oscillations, Astrophysics},
         year = 2004,
        month = jul,
       volume = {609},
       number = {1},
        pages = {340-353},
          doi = {10.1086/421013},
archivePrefix = {arXiv},
       eprint = {astro-ph/0403317},
 primaryClass = {astro-ph},
       adsurl = {https://ui.adsabs.harvard.edu/abs/2004ApJ...609..340C},
      adsnote = {Provided by the SAO/NASA Astrophysics Data System}
}

@ARTICLE{dhm+15,
       author = {{Dai}, S. and {Hobbs}, G. and {Manchester}, R.~N. and {Kerr}, M. and {Shannon}, R.~M. and {van Straten}, W. and {Mata}, A. and {Bailes}, M. and {Bhat}, N.~D.~R. and {Burke-Spolaor}, S. and {Coles}, W.~A. and {Johnston}, S. and {Keith}, M.~J. and {Levin}, Y. and {Os{\l}owski}, S. and {Reardon}, D. and {Ravi}, V. and {Sarkissian}, J.~M. and {Tiburzi}, C. and {Toomey}, L. and {Wang}, H.~G. and {Wang}, J. -B. and {Wen}, L. and {Xu}, R.~X. and {Yan}, W.~M. and {Zhu}, X. -J.},
        title = "{A study of multifrequency polarization pulse profiles of millisecond pulsars}",
      journal = {\mnras},
     keywords = {polarization, radiation mechanisms: non-thermal, pulsars: general, radio continuum: general, Astrophysics - Astrophysics of Galaxies, Astrophysics - High Energy Astrophysical Phenomena},
         year = 2015,
        month = may,
       volume = {449},
       number = {3},
        pages = {3223-3262},
          doi = {10.1093/mnras/stv508},
archivePrefix = {arXiv},
       eprint = {1503.01841},
 primaryClass = {astro-ph.GA},
       adsurl = {https://ui.adsabs.harvard.edu/abs/2015{\mnras}.449.3223D},
      adsnote = {Provided by the SAO/NASA Astrophysics Data System}
}

@ARTICLE{lgi+20,
       author = {{Liu}, K. and {Guillemot}, L. and {Istrate}, A.~G. and {Shao}, L. and {Tauris}, T.~M. and {Wex}, N. and {Antoniadis}, J. and {Chalumeau}, A. and {Cognard}, I. and {Desvignes}, G. and {Freire}, P.~C.~C. and {Kehl}, M.~S. and {Theureau}, G.},
        title = "{A revisit of PSR J1909-3744 with 15-yr high-precision timing}",
      journal = {\mnras},
     keywords = {gravitation, methods: data analysis, binaries: general, pulsars: individual (PSR J1909-3744), Astrophysics - High Energy Astrophysical Phenomena, Astrophysics - Solar and Stellar Astrophysics},
         year = 2020,
        month = dec,
       volume = {499},
       number = {2},
        pages = {2276-2291},
          doi = {10.1093/mnras/staa2993},
archivePrefix = {arXiv},
       eprint = {2009.12544},
 primaryClass = {astro-ph.HE},
       adsurl = {https://ui.adsabs.harvard.edu/abs/2020{\mnras}.499.2276L},
      adsnote = {Provided by the SAO/NASA Astrophysics Data System}
}

@article{abbs+18,
doi = {10.3847/1538-4365/aab5b0},
url = {https://dx.doi.org/10.3847/1538-4365/aab5b0},
year = {2018},
month = {apr},
publisher = {The American Astronomical Society},
volume = {235},
number = {2},
pages = {37},
author = {Zaven Arzoumanian and Adam Brazier and Sarah Burke-Spolaor and Sydney Chamberlin and Shami Chatterjee and Brian Christy and James M. Cordes and Neil J. Cornish and Fronefield Crawford and H. Thankful Cromartie and Kathryn Crowter and Megan E. DeCesar and Paul B. Demorest and Timothy Dolch and Justin A. Ellis and Robert D. Ferdman and Elizabeth C. Ferrara and Emmanuel Fonseca and Nathan Garver-Daniels and Peter A. Gentile and Daniel Halmrast and E. A. Huerta and Fredrick A. Jenet and Cody Jessup and Glenn Jones and Megan L. Jones and David L. Kaplan and Michael T. Lam and T. Joseph W. Lazio and Lina Levin and Andrea Lommen and Duncan R. Lorimer and Jing Luo and Ryan S. Lynch and Dustin Madison and Allison M. Matthews and Maura A. McLaughlin and Sean T. McWilliams and Chiara Mingarelli and Cherry Ng and David J. Nice and Timothy T. Pennucci and Scott M. Ransom and Paul S. Ray and Xavier Siemens and Joseph Simon and Renée Spiewak and Ingrid H. Stairs and Daniel R. Stinebring and Kevin Stovall and Joseph K. Swiggum and Stephen R. Taylor and Michele Vallisneri and Rutger van Haasteren and Sarah J. Vigeland and Weiwei Zhu and The NANOGrav Collaboration},
title = {The NANOGrav 11-year Data Set: High-precision Timing of 45 Millisecond Pulsars},
journal = {The Astrophysical Journal Supplement Series},
abstract = {We present high-precision timing data over time spans of up to 11 years for 45 millisecond pulsars observed as part of the North American Nanohertz Observatory for Gravitational Waves (NANOGrav) project, aimed at detecting and characterizing low-frequency gravitational waves. The pulsars were observed with the Arecibo Observatory and/or the Green Bank Telescope at frequencies ranging from 327 MHz to 2.3 GHz. Most pulsars were observed with approximately monthly cadence, and six high-timing-precision pulsars were observed weekly. All were observed at widely separated frequencies at each observing epoch in order to fit for time-variable dispersion delays. We describe our methods for data processing, time-of-arrival (TOA) calculation, and the implementation of a new, automated method for removing outlier TOAs. We fit a timing model for each pulsar that includes spin, astrometric, and (for binary pulsars) orbital parameters; time-variable dispersion delays; and parameters that quantify pulse-profile evolution with frequency. The timing solutions provide three new parallax measurements, two new Shapiro delay measurements, and two new measurements of significant orbital-period variations. We fit models that characterize sources of noise for each pulsar. We find that 11 pulsars show significant red noise, with generally smaller spectral indices than typically measured for non-recycled pulsars, possibly suggesting a different origin. A companion paper uses these data to constrain the strength of the gravitational-wave background.}
}

@ARTICLE{dds+23,
       author = {{Ding}, H. and {Deller}, A.~T. and {Stappers}, B.~W. and {Lazio}, T.~J.~W. and {Kaplan}, D. and {Chatterjee}, S. and {Brisken}, W. and {Cordes}, J. and {Freire}, P.~C.~C. and {Fonseca}, E. and {Stairs}, I. and {Guillemot}, L. and {Lyne}, A. and {Cognard}, I. and {Reardon}, D.~J. and {Theureau}, G.},
        title = "{The MSPSR{\ensuremath{\pi}} catalogue: VLBA astrometry of 18 millisecond pulsars}",
      journal = {\mnras},
     keywords = {gravitation, stars: kinematics and dynamics, pulsars: individual: PSR J0030+0451, PSR J0610-2100, PSR J0621+1002, PSR J1024-0719, PSR J1537+1155, PSR J1853+1303, PSR J1910+1256, PSR J1918-0642, PSR J1939+2134, gamma-rays: stars, radio continuum: stars, Astrophysics - High Energy Astrophysical Phenomena, General Relativity and Quantum Cosmology},
         year = 2023,
        month = mar,
       volume = {519},
       number = {4},
        pages = {4982-5007},
          doi = {10.1093/mnras/stac3725},
archivePrefix = {arXiv},
       eprint = {2212.06351},
 primaryClass = {astro-ph.HE},
       adsurl = {https://ui.adsabs.harvard.edu/abs/2023MNRAS.519.4982D},
      adsnote = {Provided by the SAO/NASA Astrophysics Data System}
}

@ARTICLE{rsc+21,
       author = {{Reardon}, D.~J. and {Shannon}, R.~M. and {Cameron}, A.~D. and {Goncharov}, B. and {Hobbs}, G.~B. and {Middleton}, H. and {Shamohammadi}, M. and {Thyagarajan}, N. and {Bailes}, M. and {Bhat}, N.~D.~R. and {Dai}, S. and {Kerr}, M. and {Manchester}, R.~N. and {Russell}, C.~J. and {Spiewak}, R. and {Wang}, J.~B. and {Zhu}, X.~J.},
        title = "{The Parkes pulsar timing array second data release: timing analysis}",
      journal = {\mnras},
     keywords = {astrometry, parallaxes, stars: neutron, pulsars: general, Astrophysics - High Energy Astrophysical Phenomena, Astrophysics - Solar and Stellar Astrophysics},
         year = 2021,
        month = oct,
       volume = {507},
       number = {2},
        pages = {2137-2153},
          doi = {10.1093/mnras/stab1990},
archivePrefix = {arXiv},
       eprint = {2107.04609},
 primaryClass = {astro-ph.HE},
       adsurl = {https://ui.adsabs.harvard.edu/abs/2021MNRAS.507.2137R},
      adsnote = {Provided by the SAO/NASA Astrophysics Data System}
}

@ARTICLE{mp17,
       author = {{McMillan}, Paul J.},
        title = "{The mass distribution and gravitational potential of the Milky Way}",
      journal = {\mnras},
     keywords = {methods: statistical, Galaxy: fundamental parameters, Galaxy: kinematics and dynamics, Galaxy: structure, Astrophysics - Astrophysics of Galaxies},
         year = 2017,
        month = feb,
       volume = {465},
       number = {1},
        pages = {76-94},
          doi = {10.1093/mnras/stw2759},
archivePrefix = {arXiv},
       eprint = {1608.00971},
 primaryClass = {astro-ph.GA},
       adsurl = {https://ui.adsabs.harvard.edu/abs/2017MNRAS.465...76M},
      adsnote = {Provided by the SAO/NASA Astrophysics Data System}
}


@article{zzx+20,
doi = {10.3847/1538-4357/ab76c4},
url = {https://dx.doi.org/10.3847/1538-4357/ab76c4},
year = {2020},
month = {mar},
publisher = {The American Astronomical Society},
volume = {892},
number = {1},
pages = {4},
author = {Zhu-Ling Deng and Zhi-Fu Gao and Xiang-Dong Li and Yong Shao},
title = {On the Formation of PSR J1640+2224: A Neutron Star Born Massive?},
journal = {The Astrophysical Journal},
abstract = {PSR J1640+2224 is a binary millisecond pulsar (BMSP) with a white dwarf (WD) companion. Recent observations indicate that the WD is very likely to be a ∼0.7 M⊙ CO WD. Thus, the BMSP should have evolved from an intermediate-mass X-ray binary (IMXB). However, previous investigations on IMXB evolution predict that the orbital periods of the resultant BMSPs are generally &lt; 40 days, in contrast with the 175 day orbital period of PSR J1640+2224. In this paper, we explore the influence of the mass of the neutron star (NS) and the chemical compositions of the companion star on the formation of BMSPs. Our results show that the final orbital period becomes longer with increasing NS mass, and the WD mass becomes larger with decreasing metallicity. In particular, to reproduce the properties of PSR J1640+2224, the NS was likely born massive (&gt;2.0 M⊙).}
}

@ARTICLE{kkn+16,
       author = {{Kaplan}, David L. and {Kupfer}, Thomas and {Nice}, David J. and {Irrgang}, Andreas and {Heber}, Ulrich and {Arzoumanian}, Zaven and {Beklen}, Elif and {Crowter}, Kathryn and {DeCesar}, Megan E. and {Demorest}, Paul B. and {Dolch}, Timothy and {Ellis}, Justin A. and {Ferdman}, Robert D. and {Ferrara}, Elizabeth C. and {Fonseca}, Emmanuel and {Gentile}, Peter A. and {Jones}, Glenn and {Jones}, Megan L. and {Kreuzer}, Simon and {Lam}, Michael T. and {Levin}, Lina and {Lorimer}, Duncan R. and {Lynch}, Ryan S. and {McLaughlin}, Maura A. and {Miller}, Adam A. and {Ng}, Cherry and {Pennucci}, Timothy T. and {Prince}, Tom A. and {Ransom}, Scott M. and {Ray}, Paul S. and {Spiewak}, Renee and {Stairs}, Ingrid H. and {Stovall}, Kevin and {Swiggum}, Joseph and {Zhu}, Weiwei},
        title = "{PSR J1024-0719: A Millisecond Pulsar in an Unusual Long-period Orbit}",
      journal = {\apj},
     keywords = {binaries: general, pulsars: individual: PSR J1024{\textendash}0719, stars: distances, Astrophysics - High Energy Astrophysical Phenomena, Astrophysics - Solar and Stellar Astrophysics},
         year = 2016,
        month = jul,
       volume = {826},
       number = {1},
          eid = {86},
        pages = {86},
          doi = {10.3847/0004-637X/826/1/86},
archivePrefix = {arXiv},
       eprint = {1604.00131},
 primaryClass = {astro-ph.HE},
       adsurl = {https://ui.adsabs.harvard.edu/abs/2016ApJ...826...86K},
      adsnote = {Provided by the SAO/NASA Astrophysics Data System}
}

@ARTICLE{vanStraten2006,
       author = {{van Straten}, W.},
        title = "{Radio Astronomical Polarimetry and High-Precision Pulsar Timing}",
      journal = {\apj},
     keywords = {Methods: Data Analysis, Polarization, Stars: Pulsars: General, Techniques: Polarimetric, Astrophysics},
         year = 2006,
        month = may,
       volume = {642},
       number = {2},
        pages = {1004-1011},
          doi = {10.1086/501001},
archivePrefix = {arXiv},
       eprint = {astro-ph/0510334},
 primaryClass = {astro-ph},
       adsurl = {https://ui.adsabs.harvard.edu/abs/2006ApJ...642.1004V},
      adsnote = {Provided by the SAO/NASA Astrophysics Data System}
}

@InProceedings{hmth04,
  author   = {{Hobbs}, G. and {Manchester}, R. and {Teoh}, A. and {Hobbs}, M.},
  title    = {The ATNF Pulsar Catalog},
  pages    = {139-140},
  crossref = {cg04},
}

@Article{det79,
  author  = {S. Detweiler},
  title   = {Pulsar Timing Measurements and the Search for Gravitational Waves},
  pages   = {1100},
  volume  = {234},
  journal = {\apj},
  year    = {1979},
}

@Article{dt91,
  author  = {T. Damour and J. H. Taylor},
  title   = {On the Orbital Period Change of the Binary Pulsar {PSR}\,1913+16},
  pages   = {501-511},
  volume  = {366},
  journal = {\apj},
  year    = {1991},
}

@Article{shk70,
  author  = {Shklovskii, I. S.},
  title   = {Possible causes of the secular increase in pulsar periods},
  pages   = {562-565},
  volume  = {13},
  journal = sa,
  keyword = {periods},
  year    = {1970},
}

@InProceedings{bac93,
  author    = {D. C. Backer},
  booktitle = {Planets around Pulsars},
  title     = {A Pulsar Timing Tutorial and {NRAO} {G}reen {B}ank Observations of {PSR~1257+12}},
  editor    = {J. A. Phillips and S. E. Thorsett and S. R. Kulkarni},
  pages     = {11-18},
  publisher = {Astron.\ Soc.\ Pac.\ Conf.\ Ser.\ Vol.\ 36},
  year      = {1993},
}

@Article{tsb+99,
  author  = {{Toscano}, M. and {Sandhu}, J. S. and {Bailes}, M. and {Manchester}, R. N. and {Britton}, M. C. and {Kulkarni}, S. R. and {Anderson}, S. B. and {Stappers}, B. W.},
  title   = {Millisecond pulsar velocities},
  pages   = {925--933},
  volume  = {307},
  journal = mnras,
  month   = aug,
  year    = {1999},
}

@Article{kop95,
  author  = {Kopeikin, S. M.},
  title   = {On possible implications of orbital parallaxes of wide orbit binary pulsars and their measurability},
  pages   = {L5-L8},
  volume  = {439},
  journal = {\apj},
  keyword = {1713+0747, 2019+2425, 1259-63},
  year    = {1995},
}

@Article{saz78,
  author  = {M. V. Sazhin},
  pages   = {36},
  volume  = {22},
  journal = sa,
  year    = {1978},
}

@Article{sfl+05,
  author  = {Stairs, I.~H. and Faulkner, A.~J. and Lyne, A.~G. and Kramer, M. and Lorimer, D.~R. and McLaughlin, M.~A. and Manchester, R.~N. and Hobbs, G.~B. and Camilo, F. and Possenti, A. and Burgay, M. and D'Amico, N. and Freire, P.~C.~C. and Gregory, P.~C. and Wex, N.},
  title   = {Discovery of three wide-orbit binary pulsars: implications for binary evolution and equivalence principles},
  pages   = {1060-1068},
  volume  = {632},
  journal = {\apj},
  year    = {2005},
}

@Article{bdz+97,
  author  = {Backer, D. C. and Dexter, M. R. and Zepka, A. and Ng. D. and Wertheimer, D. J. and Ray, P. S. and Foster, R. S.},
  title   = {A digital signal processor for pulsar research},
  pages   = {61},
  volume  = {109},
  journal = pasp,
  year    = {1997},
}

@Article{ehm06,
  author  = {Edwards, R. T. and Hobbs, G. B. and Manchester, R. N.},
  title   = {\textsc{TEMPO2}, a new pulsar timing package - {II}. The timing model and precision estimates},
  pages   = {1549--1574},
  volume  = {372},
  journal = mnras,
  month   = nov,
  year    = {2006},
}

@Article{fb90,
  author  = {R. S. Foster and D. C. Backer},
  title   = {Constructing a Pulsar Timing Array},
  pages   = {300},
  volume  = {361},
  journal = {\apj},
  year    = {1990},
}

@Article{hem06,
  author  = {{Hobbs}, G.\ B. and {Edwards}, R.\ T. and {Manchester}, R.\ N.},
  title   = {\textsc{TEMPO2}, a new pulsar-timing package - {I}. An overview},
  pages   = {655--672},
  volume  = {369},
  journal = mnras,
  month   = jun,
  year    = {2006},
}

@ARTICLE{2015MNRAS.451.2417R,
       author = {{Rosado}, Pablo A. and {Sesana}, Alberto and {Gair}, Jonathan},
        title = "{Expected properties of the first gravitational wave signal detected with pulsar timing arrays}",
      journal = {\mnras},
     keywords = {black hole physics, gravitation, gravitational waves, methods: data analysis, pulsars: general, galaxies: evolution, Astrophysics - High Energy Astrophysical Phenomena, General Relativity and Quantum Cosmology},
         year = 2015,
        month = aug,
       volume = {451},
       number = {3},
        pages = {2417-2433},
          doi = {10.1093/mnras/stv1098},
archivePrefix = {arXiv},
       eprint = {1503.04803},
 primaryClass = {astro-ph.HE},
       adsurl = {https://ui.adsabs.harvard.edu/abs/2015MNRAS.451.2417R},
      adsnote = {Provided by the SAO/NASA Astrophysics Data System}
}

@article{lkl+12,
   author = {{Liu}, K. and {Keane}, E.~F. and {Lee}, K.~J. and {Kramer}, M. and
    {Cordes}, J.~M. and {Purver}, M.~B.},
    title = "{Profile-shape stability and phase-jitter analyses of millisecond pulsars}",
  journal = {MNRAS},
archivePrefix = "arXiv",
   eprint = {1110.4759},
 primaryClass = "astro-ph.HE",
 keywords = {methods: data analysis, pulsars: general},
     year = 2012,
    month = feb,
   volume = 420,
    pages = {361-368},
      doi = {10.1111/j.1365-2966.2011.20041.x},
   adsurl = {http://adsabs.harvard.edu/abs/2012MNRAS.420..361L},
  adsnote = {Provided by the SAO/NASA Astrophysics Data System}
}


%
%
%
%
%
@string{aa="A\&A"}
@string{aap="A\&A"}
@string{aar="Astron.\ Astrophys.\ Rev."}
@string{aat="Astron.\ Astrophys.\ Trans."}
@string{aass="A\&AS"}
@string{aasup="A\&AS"}
@string{acta = "Acta Astron."}
@string{aj="AJ"}
@string{ajp="Aust. J. Phys."}
@string{ajpas="Aust. J. Phys. Astr. Supp."}
@string{anihp="Ann. Inst. H. Poincar\'e (Physique Th\'eorique)"}
@string{annrev="Ann. Rev. Astr. Ap."}
@string{annrevnucl="Ann. Rev. Nucl. Sci."}
@string{anphys="Ann. Phys. (U.S.A.)"}
@string{anyas="Ann. N. Y. Acad. Sci."}
@string{ap="Astropart.\ Phys."}
@string{apj="ApJ"}
@string{apjl="ApJ"}
@string{apjss="ApJS"}
@string{apjs="ApJS"}
@string{apl="Astrophys. Lett."}
@string{apspsci="Ap\&SS"}
@string{apss="Ap\&SS"}
@string{araa="Ann. Rev. Astr. Ap."}
@string{asl="Astron. Lett."}
@string{asr="Adv. Space Res."}
@string{ass="Astrophys. Space Sci."}
@string{astlc="Astro.\,Lett. and Communications"}
@string{baas = "BAAS"}
@string{ban="Bull. Astr. Inst. Netherlands"}
@string{basi="Bull.\ Astr.\ Soc.\ India"}
@string{bssa="Bull. Seis. Soc. Am."}
@string{chjaa="Chin.\ J.\ Astron.\ Astrophys."}
@string{cntphy="Contemp. Phys."}
@string{comas="Comm. Astrophys."}
@string{cqg="Class. Quant Grav."}
@string{crasm="Academie des Sciences Paris Comptes Rendus Ser.\,Scie.\,Math."}
@string{cursci="Curr. Sci."}
@string{fcph="Fundam. Cosmic Phys."}
@string{gjras= "Geophys. J. R. astr. soc."}
@string{grl = "Geophys. Res. Lett."}
@string{jgr = "J. Geophys. Res."}
@string{iauc="IAU Circ"}
@string{ieee="Proc. I. E. E. E."}
@string{ieeetap="I. E. E. E. Trans. Ant. Propag."}
@string{jaa="J. Astrophys. Astr."}
@string{jatp="J. Atm. Phys."}
@string{jhep="J. High Energy Phys."}
@string{jpha="J. Phys. A: Math. Gen."}
@string{jphg="J. Phys. G: Nucl. Part. Phys."}
@string{mc="Math. Comput."}
@string{memras = "MmRAS"}
@string{memsait = "Mem. della Soc. Ast. It."}
@string{mnras="MNRAS"}
@string{na="Nature"}
@string{nap="Nature Phys. Sci."}
@string{nar="New Astron. Rev."}
@string{nat="Nature"}
@string{obs="Observatory "}
@string{newast="New Astr."}
@string{pasa="PASA"}
@string{pasj="PASJ"}
@string{pasp="PASP"}
@string{physica="Physica"}
@string{physrep="Phys. Rep. "}
@string{physrev="Phys. Rev. "}
@string{phystod="Physics Today"}
@string{pla="Phys. Lett. A"}
@string{pnas="Proc. Nat. Acad. Sci."}
@string{prev="Phys. Rev."}
@string{prevb="Phys. Rev. B"}
@string{prevd="Phys. Rev. D"}
@string{prl="Phys. Rev. Lett."}
@string{pspie = "Proc. SPIE"}
@string{procspie = "Proc. SPIE"}
@string{pthps = "Prog. Theor. Phys. Suppl."}
@string{ptrsa = "Philos. Trans. Roy. Soc. London A"}
@string{qjras="QJRAS"}
@string{radsci="Rad. Sci."}
@string{reppp="Rep. Prog. Phys."}
@string{revgsp= "Reviews of Geophysics and Space Physics"}
@string{rmp= "Reviews of Modern Physics"}
@string{rpp= "Reports on Progress in Physics"}
@string{rsi= "Rev. Sci. Instrum."}
@string{rspsa="Proc. R. Soc. Lond. A"}
@string{sa="Sov. Astron."}
@string{sal="Sov. Astron. Lett."}
@string{sci="Science"}
@string{solphys="Solar Phys."}
@string{sphys="Soviet Phys. JETP "}
@string{spscrev="Space Sci. Rev."}
@string{sr="Space Res."}
@string{ssr="Space Sci. Rev."}
@string{ssphy="Solid State Phys."}
@string{vistas="Vistas Astron."}
@string{zp="Z. Phys."}
@string{znata="Z. Naturforsch. Teil A"}


\begin{thebibliography}{100}
\expandafter\ifx\csname natexlab\endcsname\relax\def\natexlab#1{#1}\fi

\bibitem[{{Alam} {et~al.}(2020){Alam}, {Arzoumanian}, {Baker}, {Blumer},
  {Bohler}, {Brazier}, {Brook}, {Burke-Spolaor}, {Caballero}, {Camuccio},
  {Chamberlain}, {Chatterjee}, {Cordes}, {Cornish}, {Crawford}, {Cromartie},
  {DeCesar}, {Demorest}, {Dolch}, {Ellis}, {Ferdman}, {Ferrara}, {Fiore},
  {Fonseca}, {Garcia}, {Garver-Daniels}, {Gentile}, {Good}, {Gusdorff},
  {Halmrast}, {Hazboun}, {Islo}, {Jennings}, {Jessup}, {Jones}, {Kaiser},
  {Kaplan}, {Kelley}, {Shapiro Key}, {Lam}, {Lazio}, {Lorimer}, {Luo}, {Lynch},
  {Madison}, {Maraccini}, {McLaughlin}, {Mingarelli}, {Ng}, {Nguyen}, {Nice},
  {Pennucci}, {Pol}, {Ramette}, {Ransom}, {Ray}, {Shapiro-Albert}, {Siemens},
  {Simon}, {Spiewak}, {Stairs}, {Stinebring}, {Stovall}, {Swiggum}, {Taylor},
  {Tripepi}, {Vallisneri}, {Vigeland}, {Witt}, \& {Zhu}}]{aab+20a}
{Alam}, M.~F., {Arzoumanian}, Z., {Baker}, P.~T., {et~al.} 2020, arXiv
  e-prints, arXiv:2005.06490

\bibitem[{{Alam} {et~al.}(2021{\natexlab{a}}){Alam}, {Arzoumanian}, {Baker},
  {Blumer}, {Bohler}, {Brazier}, {Brook}, {Burke-Spolaor}, {Caballero},
  {Camuccio}, {Chamberlain}, {Chatterjee}, {Cordes}, {Cornish}, {Crawford},
  {Cromartie}, {Decesar}, {Demorest}, {Dolch}, {Ellis}, {Ferdman}, {Ferrara},
  {Fiore}, {Fonseca}, {Garcia}, {Garver-Daniels}, {Gentile}, {Good},
  {Gusdorff}, {Halmrast}, {Hazboun}, {Islo}, {Jennings}, {Jessup}, {Jones},
  {Kaiser}, {Kaplan}, {Kelley}, {Key}, {Lam}, {Lazio}, {Lorimer}, {Luo},
  {Lynch}, {Madison}, {Maraccini}, {McLaughlin}, {Mingarelli}, {Ng}, {Nguyen},
  {Nice}, {Pennucci}, {Pol}, {Ramette}, {Ransom}, {Ray}, {Shapiro-Albert},
  {Siemens}, {Simon}, {Spiewak}, {Stairs}, {Stinebring}, {Stovall}, {Swiggum},
  {Taylor}, {Tripepi}, {Vallisneri}, {Vigeland}, {Witt}, {Zhu}, \& {Nanograv
  Collaboration}}]{aab+21a}
{Alam}, M.~F., {Arzoumanian}, Z., {Baker}, P.~T., {et~al.} 2021{\natexlab{a}},
  \apjs, 252, 5

\bibitem[{{Alam} {et~al.}(2021{\natexlab{b}}){Alam}, {Arzoumanian}, {Baker},
  {Blumer}, {Bohler}, {Brazier}, {Brook}, {Burke-Spolaor}, {Caballero},
  {Camuccio}, {Chamberlain}, {Chatterjee}, {Cordes}, {Cornish}, {Crawford},
  {Cromartie}, {Decesar}, {Demorest}, {Dolch}, {Ellis}, {Ferdman}, {Ferrara},
  {Fiore}, {Fonseca}, {Garcia}, {Garver-Daniels}, {Gentile}, {Good},
  {Gusdorff}, {Halmrast}, {Hazboun}, {Islo}, {Jennings}, {Jessup}, {Jones},
  {Kaiser}, {Kaplan}, {Kelley}, {Key}, {Lam}, {Lazio}, {Lorimer}, {Luo},
  {Lynch}, {Madison}, {Maraccini}, {McLaughlin}, {Mingarelli}, {Ng}, {Nguyen},
  {Nice}, {Pennucci}, {Pol}, {Ramette}, {Ransom}, {Ray}, {Shapiro-Albert},
  {Siemens}, {Simon}, {Spiewak}, {Stairs}, {Stinebring}, {Stovall}, {Swiggum},
  {Taylor}, {Tripepi}, {Vallisneri}, {Vigeland}, {Witt}, {Zhu}, \& {Nanograv
  Collaboration}}]{aab+21b}
{Alam}, M.~F., {Arzoumanian}, Z., {Baker}, P.~T., {et~al.} 2021{\natexlab{b}},
  \apjs, 252, 4

\bibitem[{{Antoniadis}(2020)}]{ant20}
{Antoniadis}, J. 2020, Research Notes of the American Astronomical Society, 4,
  223

\bibitem[{Antoniadis(2021)}]{ant21}
Antoniadis, J. 2021, Monthly Notices of the Royal Astronomical Society, 501,
  1116

\bibitem[{{Antoniadis} {et~al.}(2022){Antoniadis}, {Arzoumanian}, {Babak},
  {Bailes}, {Bak Nielsen}, {Baker}, {Bassa}, {B{\'e}csy}, {Berthereau},
  {Bonetti}, {Brazier}, {Brook}, {Burgay}, {Burke-Spolaor}, {Caballero},
  {Casey-Clyde}, {Chalumeau}, {Champion}, {Charisi}, {Chatterjee}, {Chen},
  {Cognard}, {Cordes}, {Cornish}, {Crawford}, {Cromartie}, {Crowter}, {Dai},
  {DeCesar}, {Demorest}, {Desvignes}, {Dolch}, {Drachler}, {Falxa}, {Ferrara},
  {Fiore}, {Fonseca}, {Gair}, {Garver-Daniels}, {Goncharov}, {Good}, {Graikou},
  {Guillemot}, {Guo}, {Hazboun}, {Hobbs}, {Hu}, {Islo}, {Janssen}, {Jennings},
  {Johnson}, {Jones}, {Kaiser}, {Kaplan}, {Karuppusamy}, {Keith}, {Kelley},
  {Kerr}, {Key}, {Kramer}, {Lam}, {Lamb}, {Lazio}, {Lee}, {Lentati}, {Liu},
  {Luo}, {Lynch}, {Lyne}, {Madison}, {Main}, {Manchester}, {McEwen}, {McKee},
  {McLaughlin}, {Mickaliger}, {Mingarelli}, {Ng}, {Nice}, {Os{\l}owski},
  {Parthasarathy}, {Pennucci}, {Perera}, {Perrodin}, {Petiteau}, {Pol},
  {Porayko}, {Possenti}, {Ransom}, {Ray}, {Reardon}, {Russell}, {Samajdar},
  {Sampson}, {Sanidas}, {Sarkissian}, {Schmitz}, {Schult}, {Sesana},
  {Shaifullah}, {Shannon}, {Shapiro-Albert}, {Siemens}, {Simon}, {Smith},
  {Speri}, {Spiewak}, {Stairs}, {Stappers}, {Stinebring}, {Swiggum}, {Taylor},
  {Theureau}, {Tiburzi}, {Vallisneri}, {van der Wateren}, {Vecchio},
  {Verbiest}, {Vigeland}, {Wahl}, {Wang}, {Wang}, {Wang}, {Witt}, {Zhang}, \&
  {Zhu}}]{aab+22}
{Antoniadis}, J., {Arzoumanian}, Z., {Babak}, S., {et~al.} 2022, \mnras, 510,
  4873

\bibitem[{{Antoniadis} {et~al.}(2016){Antoniadis}, {Tauris}, {Ozel}, {Barr},
  {Champion}, \& {Freire}}]{ato+16}
{Antoniadis}, J., {Tauris}, T.~M., {Ozel}, F., {et~al.} 2016, arXiv e-prints,
  arXiv:1605.01665

\bibitem[{{Antoniadis} {et~al.}(2012){Antoniadis}, {van Kerkwijk}, {Koester},
  {Freire}, {Wex}, {Tauris}, {Kramer}, \& {Bassa}}]{avk+12}
{Antoniadis}, J., {van Kerkwijk}, M.~H., {Koester}, D., {et~al.} 2012, \mnras,
  423, 3316

\bibitem[{{Babak} {et~al.}(2016){Babak}, {Petiteau}, {Sesana}, {Brem},
  {Rosado}, {Taylor}, {Lassus}, {Hessels}, {Bassa}, {Burgay}, {Caballero},
  {Champion}, {Cognard}, {Desvignes}, {Gair}, {Guillemot}, {Janssen},
  {Karuppusamy}, {Kramer}, {Lazarus}, {Lee}, {Lentati}, {Liu}, {Mingarelli},
  {Os{\l}owski}, {Perrodin}, {Possenti}, {Purver}, {Sanidas}, {Smits},
  {Stappers}, {Theureau}, {Tiburzi}, {van Haasteren}, {Vecchio}, \&
  {Verbiest}}]{bps+13}
{Babak}, S., {Petiteau}, A., {Sesana}, A., {et~al.} 2016, \mnras, 455, 1665

\bibitem[{Backer(1993)}]{bac93}
Backer, D.~C. 1993, in Planets around Pulsars, ed. J.~A. Phillips, S.~E.
  Thorsett, \& S.~R. Kulkarni (Astron.\ Soc.\ Pac.\ Conf.\ Ser.\ Vol.\ 36),
  11--18

\bibitem[{Backer {et~al.}(1997)Backer, Dexter, Zepka, D., Wertheimer, Ray, \&
  Foster}]{bdz+97}
Backer, D.~C., Dexter, M.~R., Zepka, A., {et~al.} 1997, PASP, 109, 61

\bibitem[{{Bassa} {et~al.}(2016{\natexlab{a}}){Bassa}, {Antoniadis}, {Camilo},
  {Cognard}, {Koester}, {Kramer}, {Ransom}, \& {Stappers}}]{bac+16}
{Bassa}, C.~G., {Antoniadis}, J., {Camilo}, F., {et~al.} 2016{\natexlab{a}},
  \mnras, 455, 3806

\bibitem[{{Bassa} {et~al.}(2016{\natexlab{b}}){Bassa}, {Janssen},
  {Karuppusamy}, {Kramer}, {Lee}, {Liu}, {McKee}, {Perrodin}, {Purver},
  {Sanidas}, {Smits}, \& {Stappers}}]{bjk+16}
{Bassa}, C.~G., {Janssen}, G.~H., {Karuppusamy}, R., {et~al.}
  2016{\natexlab{b}}, \mnras, 456, 2196

\bibitem[{{Bassa} {et~al.}(2016{\natexlab{c}}){Bassa}, {Janssen}, {Stappers},
  {Tauris}, {Wevers}, {Jonker}, {Lentati}, {Verbiest}, {Desvignes}, {Graikou},
  {Guillemot}, {Freire}, {Lazarus}, {Caballero}, {Champion}, {Cognard},
  {Jessner}, {Jordan}, {Karuppusamy}, {Kramer}, {Lazaridis}, {Lee}, {Liu},
  {Lyne}, {McKee}, {Os{\l}owski}, {Perrodin}, {Sanidas}, {Shaifullah}, {Smits},
  {Theureau}, {Tiburzi}, \& {Zhu}}]{bjs+16}
{Bassa}, C.~G., {Janssen}, G.~H., {Stappers}, B.~W., {et~al.}
  2016{\natexlab{c}}, \mnras, 460, 2207

\bibitem[{Bell \& Bailes(1996)}]{bb96}
Bell, J.~F. \& Bailes, M. 1996, \apj, 456, L33

\bibitem[{{Binney} \& {Merrifield}(1998)}]{Binney}
{Binney}, J. \& {Merrifield}, M. 1998, {Galactic Astronomy} ({Princeton
  Univesity Press})

\bibitem[{{Boyle} \& {Buonanno}(2008)}]{bb08}
{Boyle}, L.~A. \& {Buonanno}, A. 2008, Physical Review D, 78, 043531

\bibitem[{{Caprini} \& {Figueroa}(2018)}]{cf18}
{Caprini}, C. \& {Figueroa}, D.~G. 2018, Classical and Quantum Gravity, 35,
  163001

\bibitem[{{Chalumeau} {et~al.}(2022){Chalumeau}, {Babak}, {Petiteau}, {Chen},
  {Samajdar}, {Caballero}, {Theureau}, {Guillemot}, {Desvignes},
  {Parthasarathy}, {Liu}, {Shaifullah}, {Hu}, {van der Wateren}, {Antoniadis},
  {Bak Nielsen}, {Bassa}, {Berthereau}, {Burgay}, {Champion}, {Cognard},
  {Falxa}, {Ferdman}, {Freire}, {Gair}, {Graikou}, {Guo}, {Jang}, {Janssen},
  {Karuppusamy}, {Keith}, {Kramer}, {Lee}, {Liu}, {Lyne}, {Main}, {McKee},
  {Mickaliger}, {Perera}, {Perrodin}, {Porayko}, {Possenti}, {Sanidas},
  {Sesana}, {Speri}, {Stappers}, {Tiburzi}, {Vecchio}, {Verbiest}, {Wang},
  {Wang}, \& {Xu}}]{cbp+22}
{Chalumeau}, A., {Babak}, S., {Petiteau}, A., {et~al.} 2022, \mnras, 509, 5538

\bibitem[{{Chen} {et~al.}(2021){Chen}, {Caballero}, {Guo}, {Chalumeau}, {Liu},
  {Shaifullah}, {Lee}, {Babak}, {Desvignes}, {Parthasarathy}, {Hu}, {van der
  Wateren}, {Antoniadis}, {Bak Nielsen}, {Bassa}, {Berthereau}, {Burgay},
  {Champion}, {Cognard}, {Falxa}, {Ferdman}, {Freire}, {Gair}, {Graikou},
  {Guillemot}, {Jang}, {Janssen}, {Karuppusamy}, {Keith}, {Kramer}, {Liu},
  {Lyne}, {Main}, {McKee}, {Mickaliger}, {Perera}, {Perrodin}, {Petiteau},
  {Porayko}, {Possenti}, {Samajdar}, {Sanidas}, {Sesana}, {Speri}, {Stappers},
  {Theureau}, {Tiburzi}, {Vecchio}, {Verbiest}, {Wang}, {Wang}, \&
  {Xu}}]{ccg+21}
{Chen}, S., {Caballero}, R.~N., {Guo}, Y.~J., {et~al.} 2021, \mnras, 508, 4970

\bibitem[{{Cognard} \& {Theureau}(2006)}]{ct06}
{Cognard}, I. \& {Theureau}, G. 2006, in IAU Joint Discussion, Vol.~2, IAU
  Joint Discussion

\bibitem[{{Cognard} {et~al.}(2013){Cognard}, {Theureau}, {Guillemot}, {Liu},
  {Lassus}, \& {Desvignes}}]{ctg+13}
{Cognard}, I., {Theureau}, G., {Guillemot}, L., {et~al.} 2013, in SF2A-2013:
  Proceedings of the Annual meeting of the French Society of Astronomy and
  Astrophysics, ed. L.~{Cambresy}, F.~{Martins}, E.~{Nuss}, \& A.~{Palacios},
  327--330

\bibitem[{Damour \& Taylor(1991)}]{dt91}
Damour, T. \& Taylor, J.~H. 1991, \apj, 366, 501

\bibitem[{{Damour} \& {Vilenkin}(2001)}]{dv01b}
{Damour}, T. \& {Vilenkin}, A. 2001, \prd, 64, 064008

\bibitem[{{Deller} {et~al.}(2019){Deller}, {Goss}, {Brisken}, {Chatterjee},
  {Cordes}, {Janssen}, {Kovalev}, {Lazio}, {Petrov}, {Stappers}, \&
  {Lyne}}]{dgb+19}
{Deller}, A.~T., {Goss}, W.~M., {Brisken}, W.~F., {et~al.} 2019, \apj, 875, 100

\bibitem[{Deng {et~al.}(2020)Deng, Gao, Li, \& Shao}]{zzx+20}
Deng, Z.-L., Gao, Z.-F., Li, X.-D., \& Shao, Y. 2020, The Astrophysical
  Journal, 892, 4

\bibitem[{{Desvignes} {et~al.}(2016){Desvignes}, {Caballero}, {Lentati},
  {Verbiest}, {Champion}, {Stappers}, {Janssen}, {Lazarus}, {Os{\l}owski},
  {Babak}, {Bassa}, {Brem}, {Burgay}, {Cognard}, {Gair}, {Graikou},
  {Guillemot}, {Hessels}, {Jessner}, {Jordan}, {Karuppusamy}, {Kramer},
  {Lassus}, {Lazaridis}, {Lee}, {Liu}, {Lyne}, {McKee}, {Mingarelli},
  {Perrodin}, {Petiteau}, {Possenti}, {Purver}, {Rosado}, {Sanidas}, {Sesana},
  {Shaifullah}, {Smits}, {Taylor}, {Theureau}, {Tiburzi}, {van Haasteren}, \&
  {Vecchio}}]{dcl+16}
{Desvignes}, G., {Caballero}, R.~N., {Lentati}, L., {et~al.} 2016, \mnras, 458,
  3341

\bibitem[{Detweiler(1979)}]{det79}
Detweiler, S. 1979, \apj, 234, 1100

\bibitem[{{Ding} {et~al.}(2020){Ding}, {Deller}, {Freire}, {Kaplan}, {Lazio},
  {Shannon}, \& {Stappers}}]{ddf+20}
{Ding}, H., {Deller}, A.~T., {Freire}, P., {et~al.} 2020, \apj, 896, 85

\bibitem[{{Ding} {et~al.}(2023){Ding}, {Deller}, {Stappers}, {Lazio}, {Kaplan},
  {Chatterjee}, {Brisken}, {Cordes}, {Freire}, {Fonseca}, {Stairs},
  {Guillemot}, {Lyne}, {Cognard}, {Reardon}, \& {Theureau}}]{dds+23}
{Ding}, H., {Deller}, A.~T., {Stappers}, B.~W., {et~al.} 2023, \mnras, 519,
  4982

\bibitem[{Edwards {et~al.}(2006)Edwards, Hobbs, \& Manchester}]{ehm06}
Edwards, R.~T., Hobbs, G.~B., \& Manchester, R.~N. 2006, MNRAS, 372, 1549

\bibitem[{{Falxa} {et~al.}(2023){Falxa}, {Babak}, {Baker}, {B{\'e}csy},
  {Chalumeau}, {Chen}, {Chen}, {Cornish}, {Guillemot}, {Hazboun}, {Mingarelli},
  {Parthasarathy}, {Petiteau}, {Pol}, {Sesana}, {Spolaor}, {Taylor},
  {Theureau}, {Vallisneri}, {Vigeland}, {Witt}, {Zhu}, {Antoniadis},
  {Arzoumanian}, {Bailes}, {Bhat}, {Blecha}, {Brazier}, {Brook}, {Caballero},
  {Cameron}, {Casey-Clyde}, {Champion}, {Charisi}, {Chatterjee}, {Cognard},
  {Cordes}, {Crawford}, {Cromartie}, {Crowter}, {Dai}, {DeCesar}, {Demorest},
  {Desvignes}, {Dolch}, {Drachler}, {Feng}, {Ferrara}, {Fiore}, {Fonseca},
  {Garver-Daniels}, {Glaser}, {Goncharov}, {Good}, {Griessmeier}, {Guo},
  {G{\"u}ltekin}, {Hobbs}, {Hu}, {Islo}, {Jang}, {Jennings}, {Johnson},
  {Jones}, {Kaczmarek}, {Kaiser}, {Kaplan}, {Keith}, {Kelley}, {Kerr}, {Key},
  {Laal}, {Lam}, {Lamb}, {Lazio}, {Liu}, {Liu}, {Luo}, {Lynch}, {Madison},
  {Main}, {Manchester}, {McEwen}, {McKee}, {McLaughlin}, {Ng}, {Nice}, {Ocker},
  {Olum}, {Os{\l}owski}, {Pennucci}, {Perera}, {Perrodin}, {Porayko},
  {Possenti}, {Quelquejay-Leclere}, {Ransom}, {Ray}, {Reardon}, {Russell},
  {Samajdar}, {Sarkissian}, {Schult}, {Shaifullah}, {Shannon},
  {Shapiro-Albert}, {Siemens}, {Simon}, {Siwek}, {Smith}, {Speri}, {Spiewak},
  {Stairs}, {Stappers}, {Stinebring}, {Swiggum}, {Tiburzi}, {Turner},
  {Vecchio}, {Verbiest}, {Wahl}, {Wang}, {Wang}, {Wang}, {Wu}, {Zhang}, \&
  {Zhang}}]{fbb+23}
{Falxa}, M., {Babak}, S., {Baker}, P.~T., {et~al.} 2023, \mnras
  [\eprint[arXiv]{2303.10767}]

\bibitem[{{Feroz} {et~al.}(2009){Feroz}, {Hobson}, \& {Bridges}}]{fhb+09}
{Feroz}, F., {Hobson}, M.~P., \& {Bridges}, M. 2009, \mnras, 398, 1601

\bibitem[{Foster \& Backer(1990)}]{fb90}
Foster, R.~S. \& Backer, D.~C. 1990, \apj, 361, 300

\bibitem[{{Freire} \& {Wex}(2010)}]{fw10}
{Freire}, P.~C.~C. \& {Wex}, N. 2010, \mnras, 409, 199

\bibitem[{{Freire} {et~al.}(2012){Freire}, {Wex}, {Esposito-Far{\`e}se},
  {Verbiest}, {Bailes}, {Jacoby}, {Kramer}, {Stairs}, {Antoniadis}, \&
  {Janssen}}]{fwe+12}
{Freire}, P.~C.~C., {Wex}, N., {Esposito-Far{\`e}se}, G., {et~al.} 2012,
  \mnras, 423, 3328

\bibitem[{{Goncharov} {et~al.}(2022){Goncharov}, {Thrane}, {Shannon}, {Harms},
  {Bhat}, {Hobbs}, {Kerr}, {Manchester}, {Reardon}, {Russell}, {Zhu}, \&
  {Zic}}]{gsr+22}
{Goncharov}, B., {Thrane}, E., {Shannon}, R.~M., {et~al.} 2022, \apjl, 932, L22

\bibitem[{{Grishchuk}(2005)}]{gri05}
{Grishchuk}, L.~P. 2005, Phys. Uspekhi, 1235

\bibitem[{{Guillemot} {et~al.}(2023){Guillemot}, {Cognard}, {van Straten},
  {Theureau}, \& {Gerard}}]{Guillemot2023}
{Guillemot}, L., {Cognard}, I., {van Straten}, W., {Theureau}, G., \&
  {Gerard}, E. 2023, submitted to A\&A

\bibitem[{Hellings \& Downs(1983)}]{hd83}
Hellings, R.~W. \& Downs, G.~S. 1983, The Astrophysical Journal, 265, L39

\bibitem[{Hickish {et~al.}(2016)Hickish, Abdurashidova, Ali, Buch, Chaudhari,
  Chen, Dexter, Domagalski, Ford, Foster, George, Greenberg, Greenhill,
  Isaacson, Jiang, Jones, Kapp, Kriel, Lacasse, Lutomirski, MacMahon, Manley,
  Martens, McCullough, Muley, New, Parsons, Price, Primiani, Ray, Siemion,
  Tonder, Vertatschitsch, Wagner, Weintroub, \& Werthimer}]{hickish2016decade}
Hickish, J., Abdurashidova, Z., Ali, Z., {et~al.} 2016, A Decade of Developing
  Radio-Astronomy Instrumentation using CASPER Open-Source Technology

\bibitem[{{Hobbs} {et~al.}(2004){Hobbs}, {Manchester}, {Teoh}, \&
  {Hobbs}}]{hmth04}
{Hobbs}, G., {Manchester}, R., {Teoh}, A., \& {Hobbs}, M. 2004, in Young
  Neutron Stars and Their Environments, {IAU} Symposium 218, ed. F.~Camilo \&
  B.~M. Gaensler (San Francisco: Astronomical Society of the Pacific), 139--140

\bibitem[{{Hobbs} {et~al.}(2006){Hobbs}, {Edwards}, \& {Manchester}}]{hem06}
{Hobbs}, G.~B., {Edwards}, R.~T., \& {Manchester}, R.~N. 2006, MNRAS, 369, 655

\bibitem[{{Igoshev} {et~al.}(2016){Igoshev}, {Verbunt}, \& {Cator}}]{ivc16}
{Igoshev}, A., {Verbunt}, F., \& {Cator}, E. 2016, \aap, 591, A123

\bibitem[{{Joshi} {et~al.}(2022){Joshi}, {Gopakumar}, {Pandian}, {Prabu},
  {Dey}, {Bagchi}, {Desai}, {Tarafdar}, {Rana}, {Maan}, {Batra}, {Girgaonkar},
  {Agarwal}, {Arumugam}, {Basu}, {Bathula}, {Dandapat}, {Gupta}, {Hisano},
  {Kato}, {Kharbanda}, {Kikunaga}, {Kolhe}, {Krishnakumar}, {Manoharan},
  {Marmat}, {Naidu}, {Banik}, {Nobleson}, {Paladi}, {Pathak}, {Singha},
  {Srivastava}, {Surnis}, {Susarla}, {Susobhanan}, \& {Takahashi}}]{bgp+22}
{Joshi}, B.~C., {Gopakumar}, A., {Pandian}, A., {et~al.} 2022, Journal of
  Astrophysics and Astronomy, 43, 98

\bibitem[{{Kaplan} {et~al.}(2016){Kaplan}, {Kupfer}, {Nice}, {Irrgang},
  {Heber}, {Arzoumanian}, {Beklen}, {Crowter}, {DeCesar}, {Demorest}, {Dolch},
  {Ellis}, {Ferdman}, {Ferrara}, {Fonseca}, {Gentile}, {Jones}, {Jones},
  {Kreuzer}, {Lam}, {Levin}, {Lorimer}, {Lynch}, {McLaughlin}, {Miller}, {Ng},
  {Pennucci}, {Prince}, {Ransom}, {Ray}, {Spiewak}, {Stairs}, {Stovall},
  {Swiggum}, \& {Zhu}}]{kkn+16}
{Kaplan}, D.~L., {Kupfer}, T., {Nice}, D.~J., {et~al.} 2016, \apj, 826, 86

\bibitem[{Karuppusamy {et~al.}(2008)Karuppusamy, Stappers, \& van
  Straten}]{kbw08}
Karuppusamy, R., Stappers, B., \& van Straten, W. 2008, Publications of the
  Astronomical Society of the Pacific, 120, 191

\bibitem[{Kopeikin(1995)}]{kop95}
Kopeikin, S.~M. 1995, \apj, 439, L5

\bibitem[{{Kopeikin}(1996)}]{kop96}
{Kopeikin}, S.~M. 1996, ApJL, 467, L93

\bibitem[{{Lasky} {et~al.}(2016){Lasky}, {Mingarelli}, {Smith}, {Giblin},
  {Thrane}, {Reardon}, {Caldwell}, {Bailes}, {Bhat}, {Burke-Spolaor}, {Dai},
  {Dempsey}, {Hobbs}, {Kerr}, {Levin}, {Manchester}, {Os{\l}owski}, {Ravi},
  {Rosado}, {Shannon}, {Spiewak}, {van Straten}, {Toomey}, {Wang}, {Wen},
  {You}, \& {Zhu}}]{lms+16}
{Lasky}, P.~D., {Mingarelli}, C. M.~F., {Smith}, T.~L., {et~al.} 2016, Physical
  Review X, 6, 011035

\bibitem[{Lazaridis {et~al.}(2009)Lazaridis, Wex, Jessner, Kramer, Stappers,
  Janssen, Desvignes, Purver, Cognard, Theureau, Lyne, Jordan, \&
  Zensus}]{lwj+09}
Lazaridis, K., Wex, N., Jessner, A., {et~al.} 2009, \mnras, 400, 805

\bibitem[{{Lazarus} {et~al.}(2016){Lazarus}, {Karuppusamy}, {Graikou},
  {Caballero}, {Champion}, {Lee}, {Verbiest}, \& {Kramer}}]{lkg+16}
{Lazarus}, P., {Karuppusamy}, R., {Graikou}, E., {et~al.} 2016, \mnras, 458,
  868

\bibitem[{{Lee}(2016)}]{lee16}
{Lee}, K.~J. 2016, in Astronomical Society of the Pacific Conference Series,
  Vol. 502, Frontiers in Radio Astronomy and FAST Early Sciences Symposium
  2015, ed. L.~{Qain} \& D.~{Li}, 19

\bibitem[{{Lentati} {et~al.}(2014){Lentati}, {Alexander}, {Hobson}, {Feroz},
  {van Haasteren}, {Lee}, \& {Shannon}}]{lah+14}
{Lentati}, L., {Alexander}, P., {Hobson}, M.~P., {et~al.} 2014, \mnras, 437,
  3004

\bibitem[{{Liu} {et~al.}(2020){Liu}, {Guillemot}, {Istrate}, {Shao}, {Tauris},
  {Wex}, {Antoniadis}, {Chalumeau}, {Cognard}, {Desvignes}, {Freire}, {Kehl},
  \& {Theureau}}]{lgi+20}
{Liu}, K., {Guillemot}, L., {Istrate}, A.~G., {et~al.} 2020, \mnras, 499, 2276

\bibitem[{{Liu} {et~al.}(2012){Liu}, {Keane}, {Lee}, {Kramer}, {Cordes}, \&
  {Purver}}]{lkl+12}
{Liu}, K., {Keane}, E.~F., {Lee}, K.~J., {et~al.} 2012, MNRAS, 420, 361

\bibitem[{{Liu} {et~al.}(2011){Liu}, {Verbiest}, {Kramer}, {Stappers}, {van
  Straten}, \& {Cordes}}]{lvk+11}
{Liu}, K., {Verbiest}, J.~P.~W., {Kramer}, M., {et~al.} 2011, \mnras, 417, 2916

\bibitem[{{Liu} {et~al.}(2023){Liu}, {Zhu}, {Antoniadis}, {Liu}, {Zhang}, \&
  {Jiang}}]{lza+23}
{Liu}, N., {Zhu}, Z., {Antoniadis}, J., {et~al.} 2023, \aap, 670, A173

\bibitem[{Lutz \& Kelker(1973)}]{lk73}
Lutz, T.~E. \& Kelker, D.~H. 1973, PASP, 85, 573

\bibitem[{{Madison} {et~al.}(2019){Madison}, {Cordes}, {Arzoumanian},
  {Chatterjee}, {Crowter}, {DeCesar}, {Demorest}, {Dolch}, {Ellis}, {Ferdman},
  {Ferrara}, {Fonseca}, {Gentile}, {Jones}, {Jones}, {Lam}, {Levin}, {Lorimer},
  {Lynch}, {McLaughlin}, {Mingarelli}, {Ng}, {Nice}, {Pennucci}, {Ransom},
  {Ray}, {Spiewak}, {Stairs}, {Stovall}, {Swiggum}, \& {Zhu}}]{mca+19}
{Madison}, D.~R., {Cordes}, J.~M., {Arzoumanian}, Z., {et~al.} 2019, \apj, 872,
  150

\bibitem[{{Manchester} {et~al.}(2013){Manchester}, {Hobbs}, {Bailes}, {Coles},
  {van Straten}, {Keith}, {Shannon}, {Bhat}, {Brown}, {Burke-Spolaor},
  {Champion}, {Chaudhary}, {Edwards}, {Hampson}, {Hotan}, {Jameson}, {Jenet},
  {Kesteven}, {Khoo}, {Kocz}, {Maciesiak}, {Oslowski}, {Ravi}, {Reynolds},
  {Sarkissian}, {Verbiest}, {Wen}, {Wilson}, {Yardley}, {Yan}, \&
  {You}}]{mhb+13}
{Manchester}, R.~N., {Hobbs}, G., {Bailes}, M., {et~al.} 2013, \pasa, 30, e017

\bibitem[{{Mata S{\'a}nchez} {et~al.}(2020){Mata S{\'a}nchez}, {Istrate}, {van
  Kerkwijk}, {Breton}, \& {Kaplan}}]{siv+20}
{Mata S{\'a}nchez}, D., {Istrate}, A.~G., {van Kerkwijk}, M.~H., {Breton},
  R.~P., \& {Kaplan}, D.~L. 2020, \mnras, 494, 4031

\bibitem[{{McLaughlin}(2013)}]{mcl13}
{McLaughlin}, M.~A. 2013, Classical and Quantum Gravity, 30, 224008

\bibitem[{{McMillan}(2017)}]{mp17}
{McMillan}, P.~J. 2017, \mnras, 465, 76

\bibitem[{{Miles} {et~al.}(2023){Miles}, {Shannon}, {Bailes}, {Reardon},
  {Keith}, {Cameron}, {Parthasarathy}, {Shamohammadi}, {Spiewak}, {van
  Straten}, {Buchner}, {Camilo}, {Geyer}, {Karastergiou}, {Kramer}, {Serylak},
  {Theureau}, \& {Venkatraman Krishnan}}]{msb23}
{Miles}, M.~T., {Shannon}, R.~M., {Bailes}, M., {et~al.} 2023, \mnras, 519,
  3976

\bibitem[{{Mills}(1997)}]{ntp}
{Mills}, D.~L. 1997, in Proceedings of the 28th Annual Precise Time and Time
  Interval (PTTI) Applications and Planning Meeting. Editorial Committee
  Chairman, 97--107

\bibitem[{{Morello} {et~al.}(2022){Morello}, {Rajwade}, \&
  {Stappers}}]{morello22}
{Morello}, V., {Rajwade}, K.~M., \& {Stappers}, B.~W. 2022, \mnras, 510, 1393

\bibitem[{{Park} {et~al.}(2021){Park}, {Folkner}, {Williams}, \&
  {Boggs}}]{DE440}
{Park}, R.~S., {Folkner}, W.~M., {Williams}, J.~G., \& {Boggs}, D.~H. 2021,
  \aj, 161, 105

\bibitem[{Perera {et~al.}(2019)Perera, DeCesar, Demorest, Kerr, Lentati, Nice,
  Osłowski, Ransom, Keith, Arzoumanian, Bailes, Baker, Bassa, Bhat, Brazier,
  Burgay, Burke-Spolaor, Caballero, Champion, Chatterjee, Chen, Cognard,
  Cordes, Crowter, Dai, Desvignes, Dolch, Ferdman, Ferrara, Fonseca, Goldstein,
  Graikou, Guillemot, Hazboun, Hobbs, Hu, Islo, Janssen, Karuppusamy, Kramer,
  Lam, Lee, Liu, Luo, Lyne, Manchester, McKee, McLaughlin, Mingarelli,
  Parthasarathy, Pennucci, Perrodin, Possenti, Reardon, Russell, Sanidas,
  Sesana, Shaifullah, Shannon, Siemens, Simon, Spiewak, Stairs, Stappers,
  Swiggum, Taylor, Theureau, Tiburzi, Vallisneri, Vecchio, Wang, Zhang, Zhang,
  Zhu, \& Zhu}]{pdd+19}
Perera, B. B.~P., DeCesar, M.~E., Demorest, P.~B., {et~al.} 2019, Monthly
  Notices of the Royal Astronomical Society, 490, 4666

\bibitem[{{Perera} {et~al.}(2018){Perera}, {Stappers}, {Babak}, {Keith},
  {Antoniadis}, {Bassa}, {Caballero}, {Champion}, {Cognard}, {Desvignes},
  {Graikou}, {Guillemot}, {Janssen}, {Karuppusamy}, {Kramer}, {Lazarus},
  {Lentati}, {Liu}, {Lyne}, {McKee}, {Os{\l}owski}, {Perrodin}, {Sanidas},
  {Sesana}, {Shaifullah}, {Theureau}, {Verbiest}, \& {Taylor}}]{psb+18}
{Perera}, B.~B.~P., {Stappers}, B.~W., {Babak}, S., {et~al.} 2018, \mnras, 478,
  218

\bibitem[{Peters(1964)}]{pet64}
Peters, P.~C. 1964, Phys. Rev., 136, 1224

\bibitem[{Petit(2009)}]{petit09}
Petit, G. 2009, Proceedings of the International Astronomical Union, 5,
  220–221

\bibitem[{{Reardon} {et~al.}(2021){Reardon}, {Shannon}, {Cameron}, {Goncharov},
  {Hobbs}, {Middleton}, {Shamohammadi}, {Thyagarajan}, {Bailes}, {Bhat}, {Dai},
  {Kerr}, {Manchester}, {Russell}, {Spiewak}, {Wang}, \& {Zhu}}]{rsc+21}
{Reardon}, D.~J., {Shannon}, R.~M., {Cameron}, A.~D., {et~al.} 2021, \mnras,
  507, 2137

\bibitem[{{Roebber}(2019)}]{roebber2019}
{Roebber}, E. 2019, \apj, 876, 55

\bibitem[{{Rosado} {et~al.}(2015){Rosado}, {Sesana}, \& {Gair}}]{rsg15}
{Rosado}, P.~A., {Sesana}, A., \& {Gair}, J. 2015, \mnras, 451, 2417

\bibitem[{Sazhin(1978)}]{saz78}
Sazhin, M.~V. 1978, Sov. Astron., 22, 36

\bibitem[{{Schwaller}(2015)}]{schwaller2015}
{Schwaller}, P. 2015, \prl, 115, 181101

\bibitem[{{Sesana}(2013)}]{sesana2013}
{Sesana}, A. 2013, \mnras, 433, L1

\bibitem[{{Sesana} {et~al.}(2004){Sesana}, {Haardt}, {Madau}, \&
  {Volonteri}}]{sesana2004}
{Sesana}, A., {Haardt}, F., {Madau}, P., \& {Volonteri}, M. 2004, \apj, 611,
  623

\bibitem[{Shklovskii(1970)}]{shk70}
Shklovskii, I.~S. 1970, Sov. Astron., 13, 562

\bibitem[{{Skilling}(2004)}]{skilling2004}
{Skilling}, J. 2004, in American Institute of Physics Conference Series, Vol.
  735, Bayesian Inference and Maximum Entropy Methods in Science and
  Engineering: 24th International Workshop on Bayesian Inference and Maximum
  Entropy Methods in Science and Engineering, ed. R.~{Fischer}, R.~{Preuss}, \&
  U.~V. {Toussaint}, 395--405

\bibitem[{{Smits} {et~al.}(2017){Smits}, {Bassa}, {Janssen}, {Karuppusamy},
  {Kramer}, {Lee}, {Liu}, {McKee}, {Perrodin}, {Purver}, {Sanidas}, {Stappers},
  \& {Zhu}}]{sbj+17}
{Smits}, R., {Bassa}, C.~G., {Janssen}, G.~H., {et~al.} 2017, Astronomy and
  Computing, 19, 66

\bibitem[{{Speri} {et~al.}(2023){Speri}, {Porayko}, {Falxa}, {Chen}, {Gair},
  {Sesana}, \& {Taylor}}]{spf+23}
{Speri}, L., {Porayko}, N.~K., {Falxa}, M., {et~al.} 2023, \mnras, 518, 1802

\bibitem[{Stairs {et~al.}(2005)Stairs, Faulkner, Lyne, Kramer, Lorimer,
  McLaughlin, Manchester, Hobbs, Camilo, Possenti, Burgay, D'Amico, Freire,
  Gregory, \& Wex}]{sfl+05}
Stairs, I.~H., Faulkner, A.~J., Lyne, A.~G., {et~al.} 2005, \apj, 632, 1060

\bibitem[{{Strom}(2002)}]{str99}
{Strom}, R.~G. 2002, in The Universe at Low Radio Frequencies, ed. A.~{Pramesh
  Rao}, G.~{Swarup}, \& {Gopal-Krishna}, Vol. 199, 383

\bibitem[{{Tan}(1991)}]{tan91}
{Tan}, G.~H. 1991, in Astronomical Society of the Pacific Conference Series,
  Vol.~19, IAU Colloq. 131: Radio Interferometry. Theory, Techniques, and
  Applications, ed. T.~J. {Cornwell} \& R.~A. {Perley}, 42--46

\bibitem[{{Tarafdar} {et~al.}(2022){Tarafdar}, {Nobleson}, {Rana}, {Singha},
  {Krishnakumar}, {Joshi}, {Paladi}, {Kolhe}, {Batra}, {Agarwal}, {Bathula},
  {Dandapat}, {Desai}, {Dey}, {Hisano}, {Ingale}, {Kato}, {Kharbanda},
  {Kikunaga}, {Marmat}, {Pandian}, {Prabu}, {Srivastava}, {Surnis}, {Susarla},
  {Susobhanan}, {Takahashi}, {Arumugam}, {Bagchi}, {Banik}, {De}, {Girgaonkar},
  {Gopakumar}, {Gupta}, {Maan}, {Manoharan}, {Naidu}, \&
  {Pathak}}]{tarafdar+22}
{Tarafdar}, P., {Nobleson}, K., {Rana}, P., {et~al.} 2022, \pasa, 39, e053

\bibitem[{Taylor(1992)}]{tay92a}
Taylor, J.~H. 1992, Philos. Trans. Roy. Soc. London A, 341, 117

\bibitem[{{the EPTA and InPTA Collaborations}(2023{\natexlab{a}})}]{wm2}
{the EPTA and InPTA Collaborations}. 2023{\natexlab{a}}, \aa, this Volume

\bibitem[{{the EPTA and InPTA Collaborations}(2023{\natexlab{b}})}]{wm3}
{the EPTA and InPTA Collaborations}. 2023{\natexlab{b}}, \aa, this Volume

\bibitem[{{Tiburzi} {et~al.}(2021){Tiburzi}, {Shaifullah}, {Bassa}, {Zucca},
  {Verbiest}, {Porayko}, {van der Wateren}, {Fallows}, {Main}, {Janssen},
  {Anderson}, {Bak Nielsen}, {Donner}, {Keane}, {K{\"u}nsem{\"o}ller},
  {Os{\l}owski}, {Grie{\ss}meier}, {Serylak}, {Br{\"u}ggen}, {Ciardi},
  {Dettmar}, {Hoeft}, {Kramer}, {Mann}, \& {Vocks}}]{tsb+21}
{Tiburzi}, C., {Shaifullah}, G.~M., {Bassa}, C.~G., {et~al.} 2021, \aap, 647,
  A84

\bibitem[{{van Cappellen} {et~al.}(2022){van Cappellen}, {Oosterloo},
  {Verheijen}, {Adams}, {Adebahr}, {Braun}, {Hess}, {Holties}, {van der Hulst},
  {Hut}, {Kooistra}, {van Leeuwen}, {Loose}, {Morganti}, {Moss}, {Orr{\'u}},
  {Ruiter}, {Schoenmakers}, {Vermaas}, {Wijnholds}, {van Amesfoort}, {Arts},
  {Attema}, {Bakker}, {Bassa}, {Bast}, {Benthem}, {Beukema}, {Blaauw}, {de
  Blok}, {Bouwhuis}, {van den Brink}, {Connor}, {Coolen}, {Damstra}, {van
  Diepen}, {de Goei}, {D{\'e}nes}, {Drost}, {Ebbendorf}, {Frank}, {Gardenier},
  {Gerbers}, {Grange}, {Grit}, {Gunst}, {Gupta}, {Ivashina}, {J{\'o}zsa},
  {Janssen}, {Koster}, {Kruithof}, {Kuindersma}, {Kutkin}, {Lucero}, {Maan},
  {Maccagni}, {van der Marel}, {Mika}, {Morawietz}, {Mulder}, {Mulder},
  {Norden}, {Offringa}, {Oostrum}, {Overeem}, {Paragi}, {Pepping}, {Petroff},
  {Pisano}, {Polatidis}, {Prasad}, {de Reijer}, {Romein}, {Schaap},
  {Schoonderbeek}, {Schulz}, {van der Schuur}, {Sclocco}, {Sluman}, {Smits},
  {Stappers}, {Straal}, {Stuurwold}, {Verstappen}, {Vohl}, {Wierenga},
  {Woestenburg}, {Zanting}, \& {Ziemke}}]{cov22}
{van Cappellen}, W.~A., {Oosterloo}, T.~A., {Verheijen}, M.~A.~W., {et~al.}
  2022, \aap, 658, A146

\bibitem[{van Kerkwijk {et~al.}(1996)van Kerkwijk, Bergeron, \&
  Kulkarni}]{vbk96}
van Kerkwijk, M.~H., Bergeron, P., \& Kulkarni, S.~R. 1996, \apjl, 467, L89

\bibitem[{{van Straten}(2006)}]{vanStraten2006}
{van Straten}, W. 2006, \apj, 642, 1004

\bibitem[{van Straten {et~al.}(2012)van Straten, Demorest, \&
  Osłowski}]{psrchive}
van Straten, W., Demorest, P., \& Osłowski, S. 2012, Pulsar data analysis with
  PSRCHIVE

\bibitem[{{Verbiest} {et~al.}(2016){Verbiest}, {Lentati}, {Hobbs}, {van
  Haasteren}, {Demorest}, {Janssen}, {Wang}, {Desvignes}, {Caballero}, {Keith},
  {Champion}, {Arzoumanian}, {Babak}, {Bassa}, {Bhat}, {Brazier}, {Brem},
  {Burgay}, {Burke-Spolaor}, {Chamberlin}, {Chatterjee}, {Christy}, {Cognard},
  {Cordes}, {Dai}, {Dolch}, {Ellis}, {Ferdman}, {Fonseca}, {Gair},
  {Garver-Daniels}, {Gentile}, {Gonzalez}, {Graikou}, {Guillemot}, {Hessels},
  {Jones}, {Karuppusamy}, {Kerr}, {Kramer}, {Lam}, {Lasky}, {Lassus},
  {Lazarus}, {Lazio}, {Lee}, {Levin}, {Liu}, {Lynch}, {Lyne}, {Mckee},
  {McLaughlin}, {McWilliams}, {Madison}, {Manchester}, {Mingarelli}, {Nice},
  {Os{\l}owski}, {Palliyaguru}, {Pennucci}, {Perera}, {Perrodin}, {Possenti},
  {Petiteau}, {Ransom}, {Reardon}, {Rosado}, {Sanidas}, {Sesana}, {Shaifullah},
  {Shannon}, {Siemens}, {Simon}, {Smits}, {Spiewak}, {Stairs}, {Stappers},
  {Stinebring}, {Stovall}, {Swiggum}, {Taylor}, {Theureau}, {Tiburzi},
  {Toomey}, {Vallisneri}, {van Straten}, {Vecchio}, {Wang}, {Wen}, {You},
  {Zhu}, \& {Zhu}}]{vlh+16}
{Verbiest}, J.~P.~W., {Lentati}, L., {Hobbs}, G., {et~al.} 2016, \mnras, 458,
  1267

\bibitem[{Verbiest {et~al.}(2010)Verbiest, Lorimer, \& McLaughlin}]{vlm10}
Verbiest, J. P.~W., Lorimer, D.~R., \& McLaughlin, M.~A. 2010, \mnras, 405, 564

\bibitem[{{Verbiest} {et~al.}(2012){Verbiest}, {Weisberg}, {Chael}, {Lee}, \&
  {Lorimer}}]{vwc+12}
{Verbiest}, J.~P.~W., {Weisberg}, J.~M., {Chael}, A.~A., {Lee}, K.~J., \&
  {Lorimer}, D.~R. 2012, \apj, 755, 39

\bibitem[{Vigeland {et~al.}(2018)Vigeland, Deller, Kaplan, Istrate, Stappers,
  \& Tauris}]{vdk+18}
Vigeland, S.~J., Deller, A.~T., Kaplan, D.~L., {et~al.} 2018, The Astrophysical
  Journal, 855, 122

\bibitem[{{Zhu} {et~al.}(2019){Zhu}, {Desvignes}, {Wex}, {Caballero},
  {Champion}, {Demorest}, {Ellis}, {Janssen}, {Kramer}, {Krieger}, {Lentati},
  {Nice}, {Ransom}, {Stairs}, {Stappers}, {Verbiest}, {Arzoumanian}, {Bassa},
  {Burgay}, {Cognard}, {Crowter}, {Dolch}, {Ferdman}, {Fonseca}, {Gonzalez},
  {Graikou}, {Guillemot}, {Hessels}, {Jessner}, {Jones}, {Jones}, {Jordan},
  {Karuppusamy}, {Lam}, {Lazaridis}, {Lazarus}, {Lee}, {Levin}, {Liu}, {Lyne},
  {McKee}, {McLaughlin}, {Os{\l}owski}, {Pennucci}, {Perrodin}, {Possenti},
  {Sanidas}, {Shaifullah}, {Smits}, {Stovall}, {Swiggum}, {Theureau}, \&
  {Tiburzi}}]{zdw+19}
{Zhu}, W.~W., {Desvignes}, G., {Wex}, N., {et~al.} 2019, \mnras, 482, 3249

\end{thebibliography}


\appendix
\section{Dataset versions}\label{app:datasets}

The full EPTA DR2 dataset, hereafter referred to as ``DR2full'', is the primary source of data for the pulsar timing analyses presented in this paper, as well as the accompanying articles on the single-pulsar noise analysis \citep{wm2} and the search for correlated signals in \citep{wm3}. Apart from the DR2full, additional datasets were created, guided by iterative analyses carried out during the detailed investigation of single pulsar noise models in \citet{wm2}, as well as multiple analyses carried out to search for artefacts and spurious signals in the common signal searches carried out in \citet{wm3}. 

The first of these, the `DR2full+', consists of a combination of the DR2full dataset with the first data release of the InPTA \citep[henceforth InPTA-DR1][]{tarafdar+22} for an overlapping set of 10 pulsars, marked with an asterisk in Table~\ref{tab:dataoverview}. This dataset aims to improve the overall sensitivity of the DR2full to DM-variation linked noise-processes. The combination was performed using the `narrowband' TOAs from the InPTA-DR1 for two observing bands centred around \SIlist{500;1460}{\mega\hertz}, respectively. Unlike the EPTA data, the InPTA-DR1 data requires optimisation based on the flux density of the pulsar being observed, as well as, observing setup specific details leading to changes in the number of sub-bands generated, ranging from a minimum of \num{4} to a maximum of \num{16}. We refer interested readers to \citet{tarafdar+22} for the full details on the InPTA-DR1 data processing. To combine the InPTA-DR1 data with the EPTA DR2 data, for each pulsar we used the ephemeris produced from the EPTA DR2 dataset and followed the same steps as described in Section~\ref{sec:combination}. Specifically, we fitted for phase offset for each sub-band in the InPTA data individually with respect to the reference system in the EPTA data (see Section~\ref{sec:combination}). 

The single-pulsar noise analysis of these additional sets of data are presented in \citet{wm2}. For the `DR2full+' dataset, we performed single-pulsar Bayesian timing analysis following the same description as in Section~\ref{ssec:tnest}, using customised noise models determined by \citet{wm2}. The results of this analysis were later used for the gravitational wave searches presented in \citet{wm3}. 

\begin{table}[!h]
	\begin{center}
		\caption[]{Overview of the name designations for the EPTA dataset}
		\begin{tabular}{lcp{3cm}}
  \hline
  {Dataset} & {MJD range} & {Notes} \\
\hline \noalign {\smallskip}
DR2full & {50360.76} -- {59385.10} & Full EPTA DR2 data \\
DR2new & {55611.40} -- {59385.10} & Data only from the new backends collected in the past 10\,yr \\
DR2full+ &  {50360.76} -- {59644.16} & DR2full + InPTA DR1 for 10 overlap pulsars \\
DR2new+ &    {55611.40} -- {59644.16} & DR2new + InPTA DR1 for 10 overlap pulsars \\
			\hline \noalign {\smallskip}
			\label{tab:datasets}
		\end{tabular}
	\end{center}
\end{table}

A significant difference between the EPTA DR1 and DR2 datasets lies in the use of the technique of coherent dedispersion. This allows for a far more accurate modelling of the frequency-dependent DM delays, leading to the recovery of a sharper profile and thus, improved timing performance. To test the improvements derived from using only coherently dedispersed data, the `DR2new' dataset was created. This dataset spans, approximately, the final \num{10.3} years. Similar to the DR2full dataset, we also appended InPTA-DR1 data to the DR2new to produce the `DR2new+' dataset.

\section{Pulsar Ephemerides} \label{app:eph}
The measurements of the pulsar timing parameters obtained from the Bayesian timing analysis are presented  in the following pages, in Tables~\ref{tab:ephem1}--\ref{tab:ephem7} for all 25 EPTA DR2 pulsars. The measurement values are the medians of the posterior distributions and the errors denote the 1-$\sigma$ confidence intervals for the respective parameters. 


\begin{table*}[!htbp]
\centering
\caption{Measured timing model parameters for PSRs~J0030$+$0451, J0613$-$0200, J0751$+$1807 and J0900$-$3144. Here the epoch of spin frequency, sky position and DM are all set to MJD~55000. The definitions of the proper motion terms are: $\mu_\alpha = \dot{\alpha}\cos\delta$, $\mu_\delta=\dot{\delta}$, $\mu_\lambda = \dot{\lambda}\cos\beta$, $\mu_\beta=\dot{\beta}$. The components in the red-noise model used in the Bayesian analysis to obtain the timing solution are shown in the last row, where RN, DM, SV stand for achromatic red noise, chromatic red noise for DM variation, and scattering variation, respectively. All of the above apply to all tables in this section.}
\label{tab:ephem1}
\hspace*{-0.47cm}
\setlength{\tabcolsep}{4pt}
\begin{tabular}[c]{lllll}
\multicolumn{5}{l}{\rule{18.9cm}{0.5pt}} \\
Pulsar Jname  & J0030$+$0451 & J0613$-$0200 & J0751$+$1807 & J0900$-$3144 \\
\multicolumn{5}{l}{\rule{18.9cm}{0.5pt}} \\
Right ascension, $\alpha$ (J2000) & --- & 06:13:43.975688(1) & 07:51:09.155329(6) & 09:00:43.953105(8)\\
Declination, $\delta$ (J2000) & --- & $-$02:00:47.22541(4) & 18:07:38.4858(5) & $-$31:44:30.8950(1) \\
Ecliptic longitude $\lambda$ (deg) & 8.910356334(10)  & ---  & --- & --- \\
Ecliptic latitude $\beta$ (deg) & 1.4456958(4) & ---  & --- & --- \\
Spin frequency, $\nu$ (Hz) & 205.530695938456(2) & 326.6005620234831(4) & 287.457853995106(1) & 90.011841919354(1) \\
Spin frequency derivative, $\dot{\nu}$ (s$^{-2}$) & $-4.2977(1)\times10^{-16}$ & $-1.023017(8)\times10^{-15}$ & $-6.43455(6)\times10^{-16}$ & $-3.96012(7)\times10^{-16}$ \\
DM (cm$^{-3}$~pc) & 4.331(1) & 38.7759(7) & 30.2457(8) & 75.691(2) \\
DM1 (cm$^{-3}$~pc~yr$^{-1}$) & 0.0002(1) & $-8(2)\times10^{-5}$ & $-0.00046(7)$ & 0.0009(5) \\
DM2 (cm$^{-3}$~pc~yr$^{-2}$) & $-2(2)\times10^{-5}$ & $-1.8(6)\times10^{-5}$ & 10(20)$\times10^{-6}$ & $-0.00019(9)$ \\
Proper motion in $\alpha$, $\mu_{\rm\alpha}$ (mas\,yr$^{-1}$) & --- & 1.837(2) & $-2.70(1)$ & $-1.03(2)$ \\
Proper motion in $\delta$, $\mu_{\rm\delta}$ (mas\,yr$^{-1}$) & --- & $-10.359(6)$ & $-13.27(7)$ & 1.99(2) \\
Proper motion in $\lambda$, $\mu_{\rm\lambda}$ (mas\,yr$^{-1}$) & $-5.523(5)$ & --- & --- & ---  \\
Proper motion in $\beta$, $\mu_{\rm\beta}$ (mas\,yr$^{-1}$) & 3.1(2) & --- & --- & --- \\
Parallax, $\varpi$ (mas) & 3.09(6) & 1.00(5) & 0.85(4) & --- \\
Binary model & --- & T2  & T2  & T2  \\
Orbital period, $P_{\rm b}$ (d) & ---  & 1.198512575192(8) & 0.263144270793(3) & 18.7376360584(1) \\
Projected semi-major axis, $x$ (s) & ---  & 1.09144411(2) & 0.3966135(1) & 17.24880996(4) \\
Epoch of ascending node (MJD), $T_{\rm asc}$ & ---   & 53113.79635421(1) & 51800.21586830(2) & 52678.63028838(3) \\
$\hat{x}$ component of the eccentricity, $\kappa$ & ---   & 4.05(4)$\times10^{-6}$ & 2.9(1)$\times10^{-6}$ & 9.884(5)$\times10^{-6}$ \\
$\hat{y}$ component of the eccentricity, $\eta$ & ---    & 3.50(4)$\times10^{-6}$ & 3(1)$\times10^{-7}$ & 3.484(4)$\times10^{-6}$ \\
Orbital period derivative, $\dot{P}_{\rm b}$ & ---   & 3.5(2)$\times10^{-14}$ & $-3.50(5)\times10^{-14}$ & --- \\
Derivative of $x$, $\dot{x}$ & ---    & --- & 2(2)$\times10^{-16}$ & ---\\
3rd harmonic of Shapiro delay, $h_{\rm 3}$ (s) & ---   & 2.6(2)$\times10^{-7}$ & 1.9(2)$\times10^{-7}$ & ---\\
4th harmonic of Shapiro delay, $h_{\rm 4}$ (s) & ---  & --- & 4(23)$\times10^{-9}$ & ---\\
Ratio of harmonics amplitude, $\varsigma$ & ---    & 0.69(4) & --- & --- \\
Noise model &RN &RN, DM & DM &RN, DM \onehalfspacing \\
\multicolumn{5}{l}{\rule{18.9cm}{0.5pt}} \\
\end{tabular}
\end{table*}

\begin{table*}[!h]
\centering \caption{Measured timing model parameters for PSRs~J1012$+$5307, J1022$+$1001, J1024$-$0719, J1455$-$3330.
\label{tab:ephem2}}
\hspace*{-0.88cm}
\setlength{\tabcolsep}{4pt}
\begin{tabular}[c]{lllll}
\multicolumn{5}{l}{\rule{19.6cm}{0.5pt}} \\
Pulsar Jname  & J1012$+$5307 & J1022$+$1001 & J1024$-$0719 & J1455$-$3330 \\
\multicolumn{5}{l}{\rule{19.6cm}{0.5pt}} \\
Right ascension, $\alpha$ (J2000) & 10:12:33.437537(2) & ---  & 10:24:38.675394(3)  & 14:55:47.969872(8) \\
Declination, $\delta$ (J2000) & 53:07:2.30023(3) & --- & $-$07:19:19.43377(8) & $-$33:30:46.3803(2) \\
Ecliptic longitude $\lambda$ (deg) & ---  & 153.865866923(8)  & --- & --- \\
Ecliptic latitude $\beta$ (deg) & --- & $-$0.063926(8)  & --- & --- \\
Spin frequency, $\nu$ (Hz) & 190.2678344415654(2) & 60.7794479566973(2) & 193.715683448548(2) & 125.200243244993(2) \\
Spin frequency derivative, $\dot{\nu}$ (s$^{-2}$) & $-6.20041(2)\times10^{-16}$ & $-1.60094(1)\times10^{-16}$ & $-6.9593(2)\times10^{-16}$ & $-3.8097(1)\times10^{-16}$ \\
Second spin frequency derivative, $\ddot{\nu}$ (s$^{-3}$) & --- & --- & $-3.57(7)\times10^{-27}$ & --- \\
DM (cm$^{-3}$~pc) & 9.0211(4) & 10.2580(9) & 6.4885(10) & 13.569(3) \\
DM1 (cm$^{-3}$~pc~yr$^{-1}$) & 0.00012(1) & $-0.00016(6)$ & 0.0004(2) & 0.0004(3)  \\
DM2 (cm$^{-3}$~pc~yr$^{-2}$) & 1.6(3)$\times10^{-5}$ & 4.1(9)$\times10^{-5}$ & $-9(4)\times10^{-5}$ & $-4(4)\times10^{-5}$ \\
Proper motion in $\alpha$, $\mu_{\rm\alpha}$ (mas\,yr$^{-1}$) & 2.624(3) & --- & $-35.277(5)$ & 7.85(1) \\
Proper motion in $\delta$, $\mu_{\rm\delta}$ (mas\,yr$^{-1}$) & $-25.487(4)$ & --- & $-48.23(1)$ & $-1.98(4)$ \\
Proper motion in $\lambda$, $\mu_{\rm\lambda}$ (mas\,yr$^{-1}$) & --- & $-15.916(4)$ & --- & ---  \\
Proper motion in $\beta$, $\mu_{\rm\beta}$ (mas\,yr$^{-1}$) & --- & $-18(4)$ & --- & --- \\
Parallax, $\varpi$ (mas) & 0.90(8) & 1.16(8) & 1.01(4) & 1.3(1) \\
Binary model & T2 & DDH  & ---  & T2  \\
Orbital period, $P_{\rm b}$ (d) & 0.604672722921(3)  & 7.8051340(2) & --- & 76.17456861(2) \\
Projected semi-major axis, $x$ (s) & 0.58181715(6)  & 16.7654035(5) & --- & 32.3622232(5) \\
Longitude of periastron, $\omega$ (deg) & --- & 97.711(8) & --- & 223.458(1) \\
Epoch of periastron, $T_{\rm 0}$ (MJD) & --- & 50246.7172(2)  & ---  & 48980.1327(3) \\
Epoch of ascending node (MJD), $T_{\rm asc}$ & 50700.08174601(1) & --- & --- & --- \\
Orbital eccentricity, $e$ & --- & 9.697(1)$\times10^{-5}$ & ---  & 0.000169646(4) \\
$\hat{x}$ component of the eccentricity, $\kappa$ & 1.18(5)$\times10^{-6}$   & --- & --- & --- \\
$\hat{y}$ component of the eccentricity, $\eta$ & 9(5)$\times10^{-8}$ & --- & --- & --- \\
Advance of periastron, $\dot{\omega}$ (deg\,/\,yr) & --- & 0.0080(4)  & --- & --- \\
Orbital period derivative, $\dot{P}_{\rm b}$ & 5.46(6)$\times10^{-14}$ & 2.2(1)$\times10^{-13}$ & --- & 5(2)$\times10^{-12}$ \\
Derivative of $x$, $\dot{x}$ & 1.5(1)$\times10^{-15}$ & 1.39(2)$\times10^{-14}$ & --- & $-1.99(6)\times10^{-14}$ \\
3rd harmonic of Shapiro delay, $h_{\rm 3}$ (s) & 9.1(10)$\times10^{-8}$ & 6.5(2)$\times10^{-7}$ & --- & ---\\
4th harmonic of Shapiro delay, $h_{\rm 4}$ (s) & 5(1)$\times10^{-8}$  & --- & --- & --- \\
Ratio of harmonics amplitude, $\varsigma$ & --- & 0.54(1) & --- & --- \\
Solar wind electron density, $n_{\rm sw}$ (cm$^{-3}$) & --- & 11.1(3) & ---  & ---  \\
Noise model &RN, DM &RN, DM & DM &RN \\ 
\multicolumn{5}{l}{\rule{19.6cm}{0.5pt}} \\
\end{tabular}
\end{table*}

\begin{table*}[!h]
\centering \caption{Measured timing model parameters for PSRs~J1600$-$3053, J164$+$2224, J1713$+$0747 and J1730$-$2304.
\label{tab:ephem3}}
\hspace*{-0.75cm}
\setlength{\tabcolsep}{4pt}
\begin{tabular}[c]{lllll}
\multicolumn{5}{l}{\rule{19.4cm}{0.5pt}} \\
Pulsar Jname  & J1600$-$3053 & J1640$+$2224 & J1713$+$0747 & J1730$-$2304 \\
\multicolumn{5}{l}{\rule{19.4cm}{0.5pt}} \\
Right ascension, $\alpha$ (J2000) & 16:00:51.903339(2) & 16:40:16.744850(2)  & 17:13:49.5331917(3) & --- \\
Declination, $\delta$ (J2000) & $-$30:53:49.37555(7) & 22:24:08.84119(5) & 07:47:37.49258(1)  & --- \\
Ecliptic longitude $\lambda$ (deg)  & ---  & ---  & --- & 263.18603136(1) \\
Ecliptic latitude $\beta$ (deg) & --- & --- & --- & 0.188871(4) \\
Spin frequency, $\nu$ (Hz) & 277.9377069896062(8) & 316.123979331869(2) & 218.8118404171605(2) & 123.110287147370(2) \\
Spin frequency derivative, $\dot{\nu}$ (s$^{-2}$) & $-7.33880(3)\times10^{-16}$ & $-2.8156(1)\times10^{-16}$
 & $-4.08385(2)\times10^{-16}$ & $-3.05916(6)\times10^{-16}$ \\
DM (cm$^{-3}$~pc) &52.3243(4) & 18.426(1) & 15.9918(1) & 9.618(2) \\
DM1 (cm$^{-3}$~pc~yr$^{-1}$) & $-3(10)\times10^{-5}$ & 0.0003(1) & 1(14)$\times10^{-6}$ & 0.0004(2)  \\
DM2 (cm$^{-3}$~pc~yr$^{-2}$) & 4(3)$\times10^{-5}$ & $-5(2)\times10^{-5}$ & $-6(3)\times10^{-6}$ & $-2(40)\times10^{-6}$ \\
Proper motion in $\alpha$, $\mu_{\rm\alpha}$ (mas\,yr$^{-1}$) & $-0.943(3)$ & 2.102(4) & 4.9215(8) & --- \\
Proper motion in $\delta$, $\mu_{\rm\delta}$ (mas\,yr$^{-1}$) & $-6.92(1)$ & $-11.333(7)$ & $-3.920(2)$ & --- \\
Proper motion in $\lambda$, $\mu_{\rm\lambda}$ (mas\,yr$^{-1}$) & --- & --- & --- & 20.236(5)  \\
Proper motion in $\beta$, $\mu_{\rm\beta}$ (mas\,yr$^{-1}$) & --- & --- & --- & $-4.4(1.8)$ \\
Parallax, $\varpi$ (mas) & 0.72(2) & 0.8(2) & 0.88(1) & 2.08(6) \\
Binary model & T2 & DDH  & T2  & ---  \\
Orbital period, $P_{\rm b}$ (d) & 14.3484635(2) & 175.460664578(9) & 67.8251309746(7) & --- \\
Projected semi-major axis, $x$ (s) & 8.8016540(1) & 55.3297193(4) & 32.34241947(4) & --- \\
Longitude of periastron, $\omega$ (deg) & 181.819(3) & 50.7326(2) & 176.2000(4) & --- \\
Epoch of periastron, $T_{\rm 0}$ (MJD) & 52506.3733(1) & 51626.17953(9) & 48741.97387(7)  & --- \\
Orbital eccentricity, $e$ & 0.000173728(2) & 0.000797277(3) & 7.49405(2)$\times10^{-5}$  & --- \\
Advance of periastron, $\dot{\omega}$ (deg\,/\,yr) & 0.0036(1) & ---  & --- & --- \\
Orbital period derivative, $\dot{P}_{\rm b}$ & 3.6(3)$\times10^{-13}$ & 9(2)$\times10^{-12}$ & 2.6(7)$\times10^{-13}$ & --- \\
Derivative of $x$, $\dot{x}$ & $-3.55(6)\times10^{-15}$ & 1.12(4)$\times10^{-14}$ & --- & --- \\
Sine of inclination angle, $\sin{i}$ & 0.906(6) & --- & --- & ---\\
Companion mass, $M_{\rm c}$ ($M_{\odot}$) & 0.29(2) & --- & 0.296(3) & ---\\
3rd harmonic of Shapiro delay, $h_{\rm 3}$ (s) & --- & 3.8(2)$\times10^{-7}$ & --- & ---\\
Ratio of harmonics amplitude, $\varsigma$ & --- & 0.75(4) & --- & --- \\
Longitude of ascending node, $\Omega$ (deg) & --- & --- & 91.1(5)  & --- \\
Inclination angle, $i$ (deg) & --- & --- & 71.3(2)  & --- \\
Noise model &RN, DM, SV & DM & RN, DM & DM \\ 
\multicolumn{5}{l}{\rule{19.4cm}{0.5pt}} \\
\end{tabular}
\end{table*}

\begin{table*}[!h]
\centering 
\caption{Measured timing model parameters for PSRs~J1738$+$0333, J1744$-$1134, J1751$-$2857 and J1801$-$1417.
\label{tab:ephem4}}
\hspace*{-0.54cm} \vspace*{1cm}
\setlength{\tabcolsep}{4pt}
\begin{tabular}[c]{lllll}
\multicolumn{5}{l}{\rule{18.8cm}{0.5pt}} \\
Pulsar Jname & J1738$+$0333 & J1744$-$1134 & J1751$-$2857 & J1801$-$1417 \\
\multicolumn{5}{l}{\rule{18.8cm}{0.5pt}} \\
Right ascension, $\alpha$ (J2000) & 17:38:53.966386(6) & 17:44:29.4075472(8) & 17:51:32.69322(1) & 18:01:51.07336(2) \\
Declination, $\delta$ (J2000) & 03:33:10.8720(2) & $-$11:34:54.69423(6) & $-$28:57:46.519(2)  & $-$14:17:34.527(2) \\
Spin frequency, $\nu$ (Hz) & 170.937369887100(7) & 245.4261196898081(5) & 255.43611088568(2) & 275.85470899694(1) \\
Spin frequency derivative, $\dot{\nu}$ (s$^{-2}$) & $-7.0471(4)\times10^{-16}$ & $-5.38156(3)\times10^{-16}$
 & $-7.3239(8)\times10^{-16}$ & $-4.0361(7)\times10^{-16}$\\
DM (cm$^{-3}$~pc) &33.767(2) & 3.1379(4) & 42.81(1) & 57.24(1) \\
DM1 (cm$^{-3}$~pc~yr$^{-1}$) & $-0.0014(5)$ & $-7(3)\times10^{-5}$ & 3(99)$\times10^{-5}$ & 0.003(1) \\
DM2 (cm$^{-3}$~pc~yr$^{-2}$) & 1.0(8)$\times10^{-4}$ & 1.7(4)$\times10^{-5}$ & 0.0002(2) & $-0.0003(2)$ \\
Proper motion in $\alpha$, $\mu_{\rm\alpha}$ (mas\,yr$^{-1}$) & 7.09(1) & 18.806(2) & $-7.38(4)$ & $-10.85(4)$ \\
Proper motion in $\delta$, $\mu_{\rm\delta}$ (mas\,yr$^{-1}$) & 5.07(3) & $-9.386(10)$ & $-4.5(4)$ & $-1.9(3)$\\
Parallax, $\varpi$ (mas) & --- & 2.58(3) & 1.1(4) & 0.8(3) \\
Binary model & T2 & --- & T2  & ---  \\
Orbital period, $P_{\rm b}$ (d) & 0.35479073997(3) & --- & 110.74646085(1) & --- \\
Projected semi-major axis, $x$ (s) & 0.3434302(1) & --- & 32.5282233(8) & --- \\
Longitude of periastron, $\omega$ (deg) & --- & --- & 45.501(5) & --- \\
Epoch of periastron, $T_{\rm 0}$ (MJD) & --- & --- & 52491.572(2)  & --- \\
Epoch of ascending node (MJD), $T_{\rm asc}$ & 52500.1940104(2) & --- & --- & --- \\
Orbital eccentricity, $e$ & --- & --- & 0.00012792(1)  & --- \\
$\hat{x}$ component of the eccentricity, $\kappa$ & 1.0(7)$\times10^{-6}$ & --- & --- & --- \\
$\hat{y}$ component of the eccentricity, $\eta$ & 4(6)$\times10^{-7}$ & --- & --- & --- \\
Advance of periastron, $\dot{\omega}$ (deg\,/\,yr) & 0.0036(1) & ---  & --- & --- \\
Orbital period derivative, $\dot{P}_{\rm b}$ & $-3.0(7)\times10^{-14}$ & --- & --- & --- \\
Derivative of $x$, $\dot{x}$ & --- & --- & 3.7(2)$\times10^{-14}$ & --- \\
Noise model &RN & RN, DM & DM & DM \\ 
\multicolumn{5}{l}{\rule{18.8cm}{0.5pt}} \\
\end{tabular}
\end{table*}

\begin{table*}[!h]
\centering \caption{Measured timing model parameters for PSRs~J1804$-$2717, J1843$-$1113, J1857$+$0943 and J1909$-$3744.
\label{tab:ephem5}}
\hspace*{-0.78cm}
\setlength{\tabcolsep}{4pt}
\begin{tabular}[c]{lllll}
\multicolumn{5}{l}{\rule{19.34cm}{0.5pt}} \\
Pulsar Jname & J1804$-$2717 & J1843$-$1113 & J1857$+$0943 & J1909$-$3744 \\
\multicolumn{5}{l}{\rule{19.34cm}{0.5pt}} \\
Right ascension, $\alpha$ (J2000) & 18:04:21.13307(1) & 18:43:41.261937(7)
  & 18:57:36.390620(1) & 19:09:47.4335785(8) \\
Declination, $\delta$ (J2000) & $-$27:17:31.337(3) & $-$11:13:31.0684(5) & 09:43:17.20712(4) & $-$37:44:14.51579(3) \\
Spin frequency, $\nu$ (Hz) & 107.031649219533(4) & 541.809745036152(5) & 186.4940783779890(9) & 339.3156872184705(9)  \\
Spin frequency derivative, $\dot{\nu}$ (s$^{-2}$) & $-4.6812(2)\times10^{-16}$ & $-2.80559(3)\times10^{-15}$ & $-6.20522(4)\times10^{-16}$ & $-1.614806(7)\times10^{-15}$ \\
DM (cm$^{-3}$~pc) &24.688(4) & 59.962(2) & 13.2957(9) & 10.3925(2) \\
DM1 (cm$^{-3}$~pc~yr$^{-1}$) & 0.0005(7) & $-0.0009(3)$ & 0.00082(7) & $-0.00037(3)$ \\
DM2 (cm$^{-3}$~pc~yr$^{-2}$) & $-0.00012(9)$ & 5(8)$\times10^{-5}$ & $-0.00012(2)$ & 4.1(6)$\times10^{-5}$ \\
Proper motion in $\alpha$, $\mu_{\rm\alpha}$ (mas\,yr$^{-1}$) & 2.46(2) & $-1.99(2)$ & $-2.670(3)$ & $-9.523(1)$ \\
Proper motion in $\delta$, $\mu_{\rm\delta}$ (mas\,yr$^{-1}$) & $-16.9(4)$ & $-3.00(7)$ & $-5.428(6)$ & $-35.780(5)$ \\
Parallax, $\varpi$ (mas) & 1.1(3)  & --- & 0.89(6) & 0.94(2) \\
Binary model & T2 & --- & T2  & T2  \\
Orbital period, $P_{\rm b}$ (d) & 11.1287119652(3)  & --- & 12.32717138285(5) & 1.533449475874(1) \\
Projected semi-major axis, $x$ (s) & 7.2814511(1)  & --- & 9.23078029(8) & 1.89799110(1) \\
Epoch of ascending node (MJD), $T_{\rm asc}$ & 49610.1749842(2) & --- & 46423.31409197(5) & 56180.8496921865(6)\\
$\hat{x}$ component of the eccentricity, $\kappa$ & 1.219(3)$\times10^{-5}$ & --- & $-2.1565(9)\times10^{-5}$ & 5.4(7)$\times10^{-8}$ \\
$\hat{y}$ component of the eccentricity, $\eta$ & $-3.177(4)\times10^{-5}$ & --- & 2.454(5)$\times10^{-6}$ & $-1.07(4)\times10^{-7}$ \\
Orbital period derivative, $\dot{P}_{\rm b}$ & --- & --- & --- & 5.09(1)$\times10^{-13}$ \\
Derivative of $x$, $\dot{x}$ & --- & --- & --- & $-3.6(5)\times10^{-16}$ \\
Sine of inclination angle, $\sin{i}$ & --- & --- & 0.9989(2) & 0.99831(4) \\
Companion mass, $M_{\rm c}$ ($M_{\odot}$) & --- & --- & 0.258(5) & 0.2048(9) \\
Noise model & DM & DM & DM & RN, DM \\
\multicolumn{5}{l}{\rule{19.34cm}{0.5pt}} \\
\end{tabular}
\end{table*}

\begin{table*}[!h]
\centering \caption{Measured timing model parameters for PSRs~J1910$+$1256, J1911$+$1347, J1918$-$0642 and J2124$-$3358.
\label{tab:ephem6}}
\hspace*{-0.55cm}
\vspace*{1cm}
\setlength{\tabcolsep}{4pt}
\begin{tabular}[c]{lllll}
\multicolumn{5}{l}{\rule{18.9cm}{0.5pt}} \\
Pulsar Jname & J1910$+$1256 & J1911$+$1347 & J1918$-$0642 & J2124$-$3358 \\
\multicolumn{5}{l}{\rule{18.9cm}{0.5pt}} \\
Right ascension, $\alpha$ (J2000) & 19:10:9.701454(7) & 19:11:55.204694(3) & 19:18:48.033123(3) & 21:24:43.847830(7) \\
Declination, $\delta$ (J2000) & 12:56:25.4868(2) & 13:47:34.38383(6) & $-$06:42:34.8895(1) & $-$33:58:44.9196(2) \\
Spin frequency, $\nu$ (Hz) & 200.658802230113(7) & 216.171227371979(2) & 130.789514123371(1) & 202.793893746013(3)  \\
Spin frequency derivative, $\dot{\nu}$ (s$^{-2}$) & $-3.8969(3)\times10^{-16}$ & $-7.9086(1)\times10^{-16}$ & $-4.3947(6)\times10^{-16}$ & $-8.4590(1)\times10^{-16}$ \\
DM (cm$^{-3}$~pc) & 38.075(5) & 30.989(1) & 26.593(1) & 4.600(3) \\
DM1 (cm$^{-3}$~pc~yr$^{-1}$) & 0.0002(6) & $-0.0010(2)$ & $-0.0003(2)$ & $-0.0003(2)$ \\
DM2 (cm$^{-3}$~pc~yr$^{-2}$) & $-4(10)\times10^{-5}$ & 7(4)$\times10^{-5}$ & 2(3)$\times10^{-5}$ & 7(4)$\times10^{-5}$ \\
Proper motion in $\alpha$, $\mu_{\rm\alpha}$ (mas\,yr$^{-1}$) & 0.24(2) & $-2.900(5)$ & $-7.124(5)$ & $-14.09(1)$ \\
Proper motion in $\delta$, $\mu_{\rm\delta}$ (mas\,yr$^{-1}$) & $-7.10(3)$ & $-3.684(9)$ & $-5.96(2)$ & $-50.32(3)$ \\
Parallax, $\varpi$ (mas) & --- & 0.40(9) & 0.75(7) & 2.1(1) \\
Binary model & T2 & --- & DDH  & ---  \\
Orbital period, $P_{\rm b}$ (d) & 58.466742972(2) & --- & 10.91317774976(7) & --- \\
Projected semi-major axis, $x$ (s) & 21.1291048(4) & --- & 8.3504666(2) & --- \\
Longitude of periastron, $\omega$ (deg) & 106.005(3) & --- & 219.49(4) & --- \\
Epoch of periastron, $T_{\rm 0}$ (MJD) & 52968.4482(5) & --- & 51575.771(1) & --- \\
Orbital eccentricity, $e$ & 0.00023024(1) & --- & 2.032(1)$\times10^{-5}$ & --- \\
Orbital period derivative, $\dot{P}_{\rm b}$ & --- & --- & 2.6(7)$\times10^{-13}$ & --- \\
Derivative of $x$, $\dot{x}$ & $-1.5(1)\times10^{-14}$ & --- & --- & ---\\
3rd harmonic of Shapiro delay, $h_{\rm 3}$ (s) & --- & --- & 8.2(2)$\times10^{-7}$ & ---\\
Ratio of harmonics amplitude, $\varsigma$ & --- & --- & 0.918(9) & --- \\
Noise model & DM & DM & DM & DM \\
\multicolumn{5}{l}{\rule{18.9cm}{0.5pt}} \\
\end{tabular}
\end{table*}

\begin{table*}[!h]
\centering \caption{Measured timing model parameters for PSR~J2322$+$2057.
\label{tab:ephem7}}
\setlength{\tabcolsep}{4pt}
\begin{tabular}[c]{lllll}
\multicolumn{5}{l}{\rule{8.0cm}{0.5pt}} \\
Pulsar Jname & J2322$+$2057 \\
\multicolumn{5}{l}{\rule{8.0cm}{0.5pt}} \\
Right ascension, $\alpha$ (J2000) & 23:22:22.33517(2) \\
Declination, $\delta$ (J2000) & 20:57:02.6754(6) \\
Spin frequency, $\nu$ (Hz) & 207.96816335836(1)  \\
Spin frequency derivative, $\dot{\nu}$ (s$^{-2}$) & $-4.1769(7)\times10^{-16}$ \\
DM (cm$^{-3}$~pc) & 13.381(8) \\
DM1 (cm$^{-3}$~pc~yr$^{-1}$) & $-0.0002(10)$ \\
DM2 (cm$^{-3}$~pc~yr$^{-2}$) & 4(16)$\times10^{-5}$ \\
Proper motion in $\alpha$, $\mu_{\rm\alpha}$ (mas\,yr$^{-1}$) & $-18.30(5)$ \\
Proper motion in $\delta$, $\mu_{\rm\delta}$ (mas\,yr$^{-1}$) & $-14.9(1)$\\
Noise model & --- \\
\multicolumn{5}{l}{\rule{8.0cm}{0.5pt}} \\
\end{tabular}
\end{table*}

\section{Timing residuals} \label{app:timing rms}
In Figures~\ref{fig:res_p1}--\ref{fig:res_p5} we present the timing residuals of the 25 pulsars in the EPTA DR2. Residuals both before and after the application of red-noise subtraction are shown. 

\begin{figure*}[!htb]
\hspace*{-1.5cm}
\centering
  \centering
  \hspace*{-1cm}
  \includegraphics[scale=0.5]{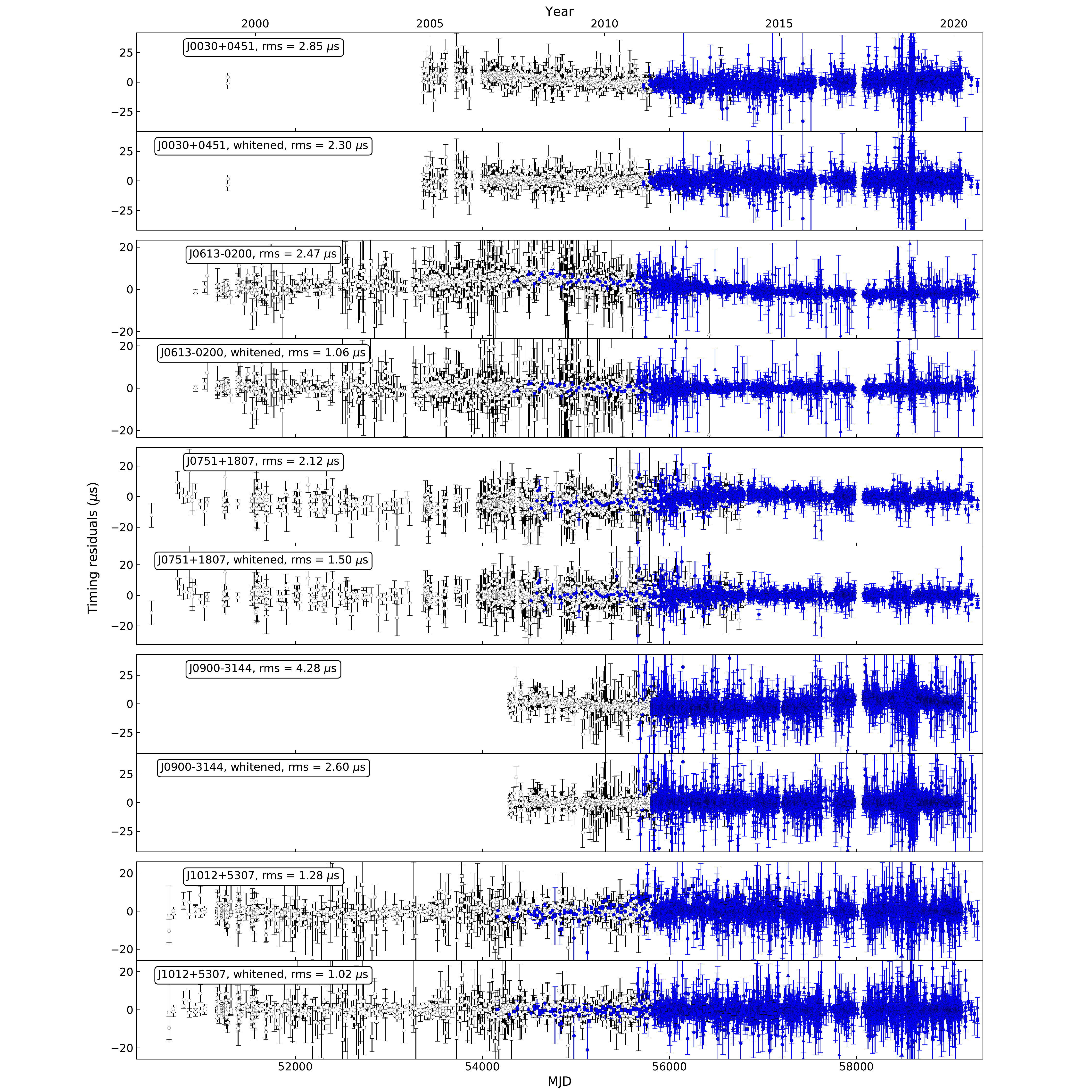}
  \vspace{-\baselineskip}
  \caption{Timing residuals of PSRs~J0030$+$0030, J0613$-$0200, J0751$+$1807, J0900$-$3144, J1012+5307. For each pulsar, the residuals before and after subtraction of DM and monochromatic red noise are shown. The squares, circles and triangles represent P-band, L-band and S/C-band observations, respectively (see Table~\ref{tab:source} for frequency coverage of each band). The blue/filled and black/unfilled symbols indicate the new backend data and those from EPTA DR1, respectively. \label{fig:res_p1}}
\end{figure*}

\begin{figure*}[!htb]
\hspace*{-1.5cm}
\centering
  \centering
    \hspace*{-1cm}
  \includegraphics[scale=0.5]{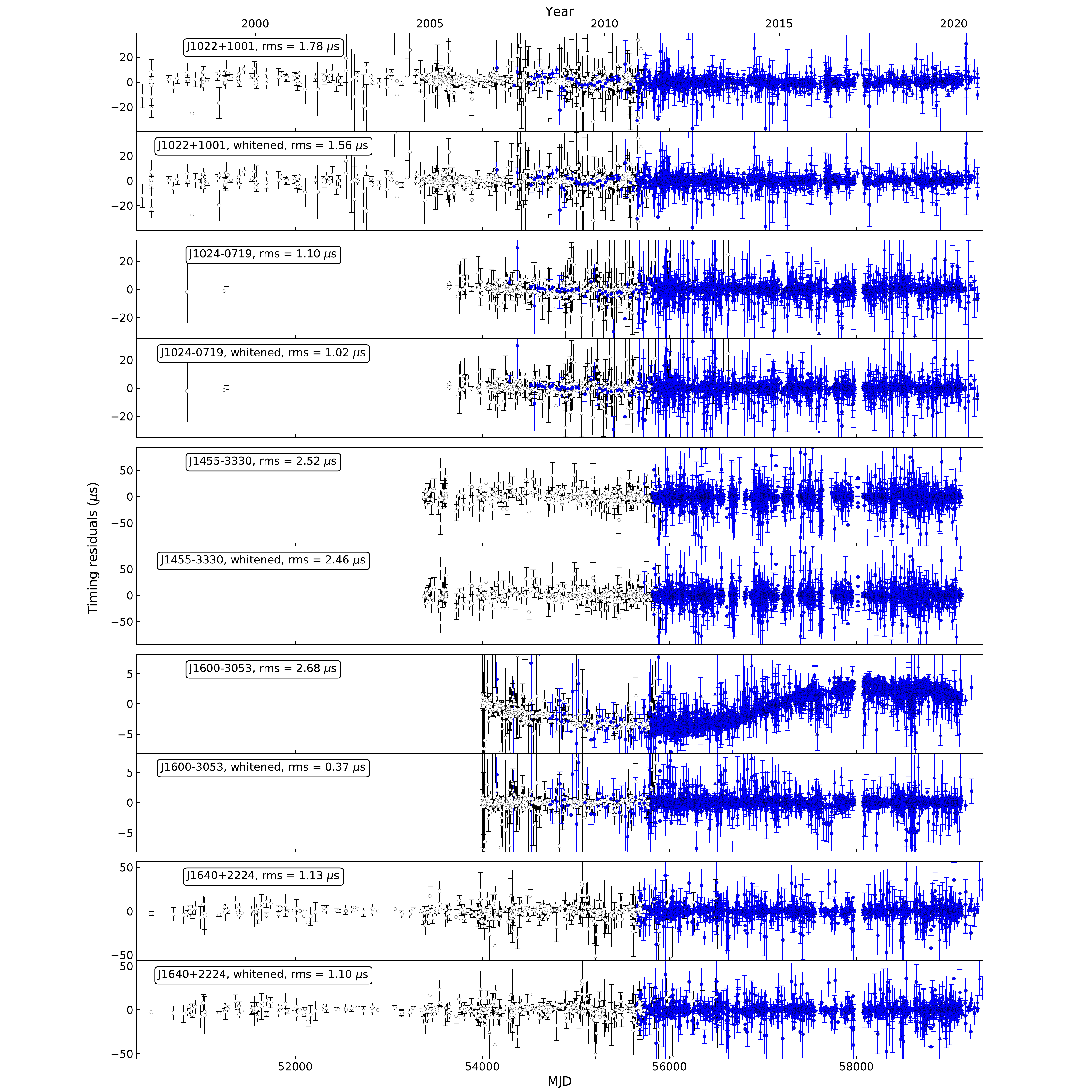}
  \caption{Timing residuals of PSRs~J1022$+$1001, J1024$-$0719, J1455$-$3330, J1600$-$3053, J1640$+$2224. Figure style is the same as Figure~\ref{fig:res_p1}. \label{fig:res_p2}}
\end{figure*}

\begin{figure*}[!htb]
\hspace*{-1.5cm}
\centering
  \centering
    \hspace*{-1cm}
  \includegraphics[scale=0.5]{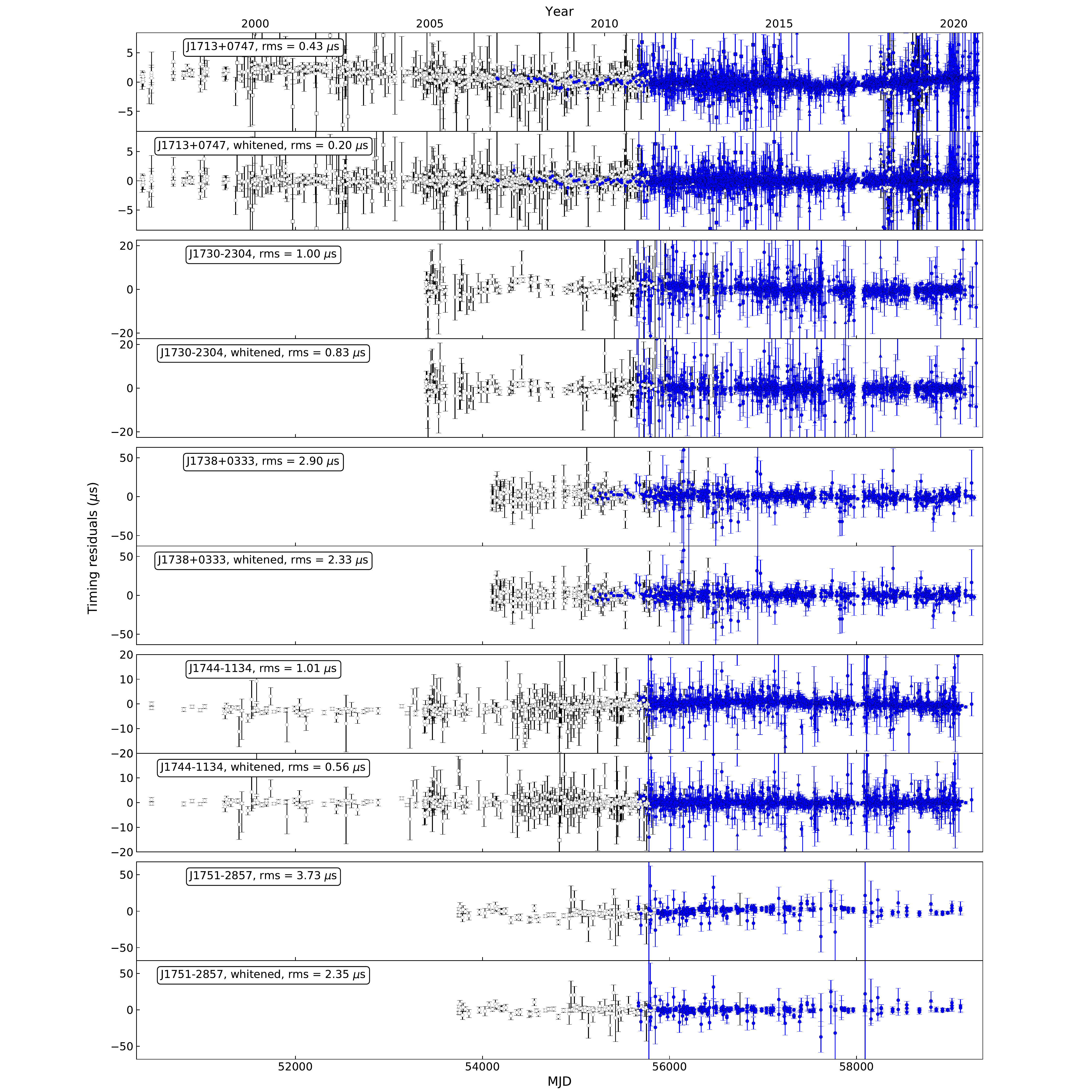}
  \caption{Timing residuals of PSRs~J1713$+$0747, J1730$-$2304, J1738$+$0333, J1744$-$1134, J1751$-$2857. Figure style is the same as Figure~\ref{fig:res_p1}. \label{fig:res_p3}}
\end{figure*}

\begin{figure*}[!htb]
\hspace*{-1.5cm}
\centering
  \centering
    \hspace*{-1cm}
  \includegraphics[scale=0.5]{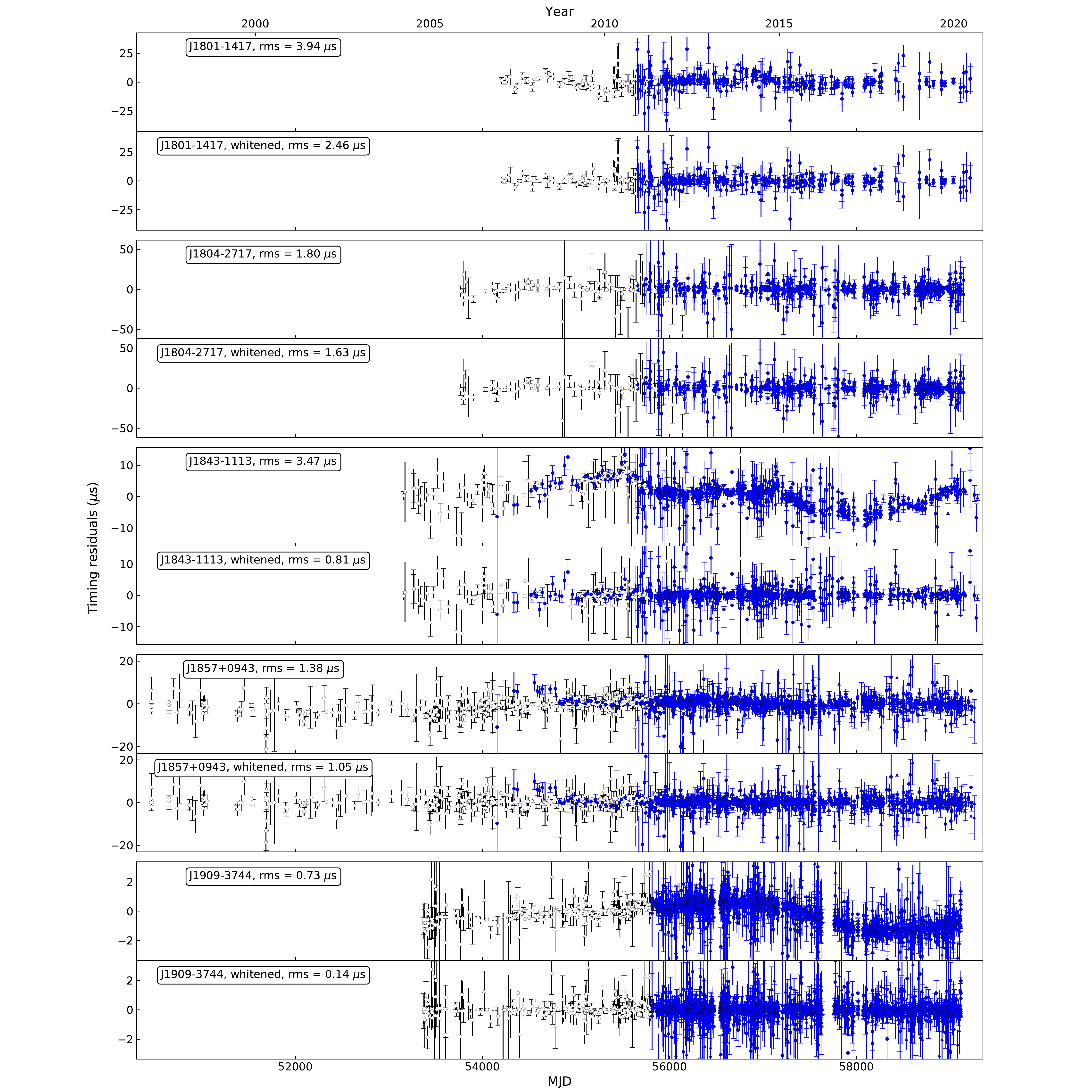}
  \caption{Timing residuals of PSRs~J1801$-$1417, J1804$-$2717, J1843$-$1113, J1857+0943, J1909$-$3744. Figure style is the same as Figure~\ref{fig:res_p1}. \label{fig:res_p4}}
\end{figure*}

\begin{figure*}[!h]
\hspace*{-1.5cm}
\centering
  \centering
    \hspace*{-1cm}
  \includegraphics[scale=0.5]{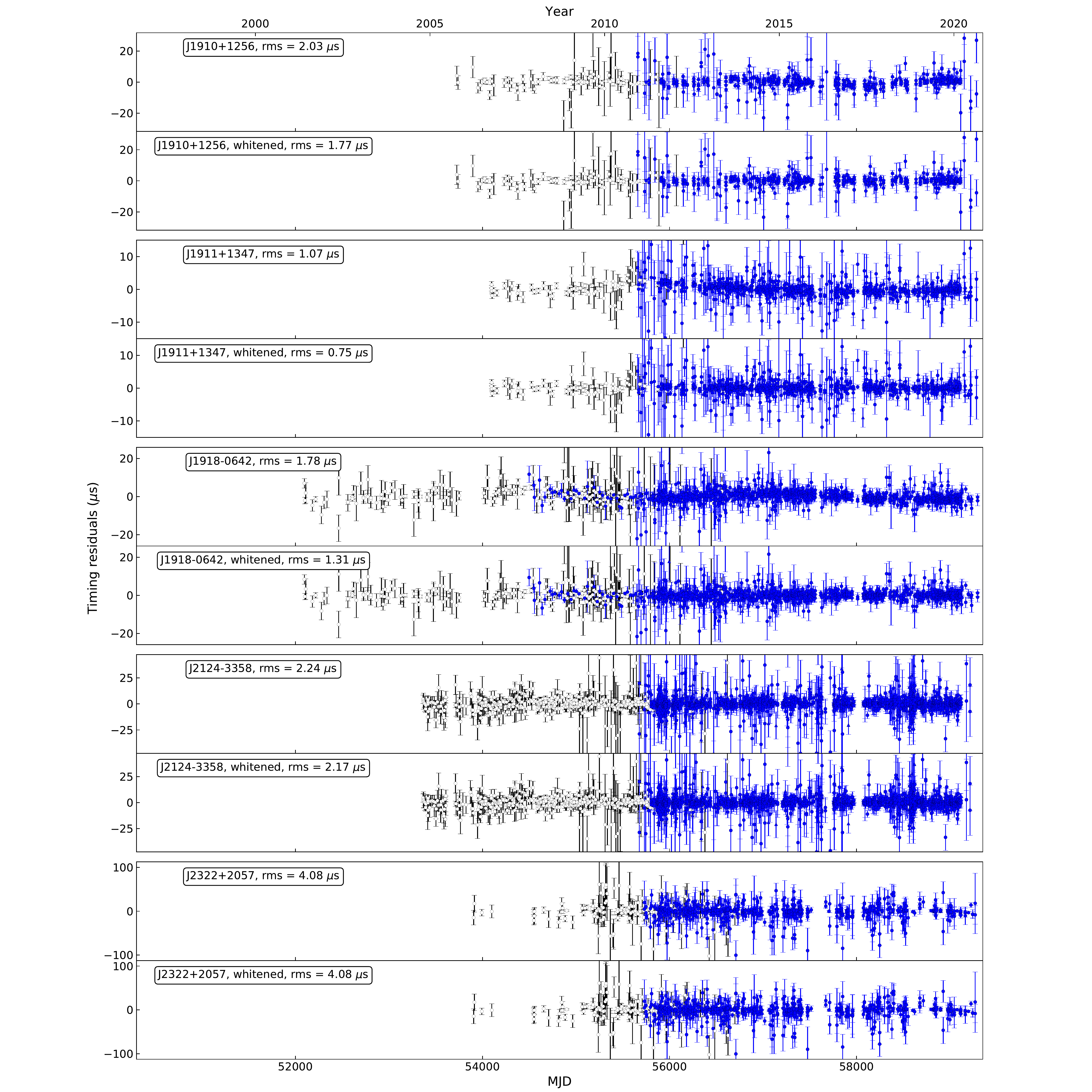}
  \caption{Timing residuals of PSRs~J1910+1256, J1911+1347, J1918$-$0642, J2124$-$3358, J2322+2057. Figure style is the same as Figure~\ref{fig:res_p1}. \label{fig:res_p5}}
\end{figure*}
\end{document}